\documentclass[11pt,oneside,openany]{book}

\usepackage{color}
\usepackage{graphics}
\usepackage{latexsym}

\newcommand{\MGIERG}{MGI-ERG}
\newcommand{\ERGE}{ERGE}
\newcommand{\ERG}{ERG}
\newcommand{\ibid}{\emph{ibid.}}
\newcommand{\viz}{viz}
\newcommand{\eg}{e.g.}
\newcommand{\ie}{i.e.}
\newcommand{\cf}{cf.}
\newcommand{\etc}{etc.}
\newcommand{\etcB}{etc}
\newcommand{\rhs}{r.h.s.}
\newcommand{\lhs}{l.h.s.}
\newcommand{\lhsB}{l.h.s}
\newcommand{\wrt}{with respect to}
\newcommand{\etal}{\emph{et al.}}
\newcommand{\aka}{a.k.a.}
\newcommand{\sic}{sic}
\newcommand{\QFT}{QFT}
\newcommand{\BS}{\Lambda_0}
\newcommand{\ES}{\Lambda}
\newcommand{\PV}{PV}
\newcommand{\match}{\eta}
\newcommand{\Proj}{\mathcal{P}}
\newcommand{\order}[1]{\mathcal{O}\left( #1 \right)}
\newcommand{\comm}[2]{\left[ #1,#2 \right]}

\newcommand{\SF}{\ensuremath{\mathcal{A}}}
\newcommand{\SH}{\ensuremath{\mathcal{C}}}

\newcommand{\mod}[1]{\left| #1 \right|}

\newcommand{\eq}[1]{(\ref{#1})}

\newcommand{\TwoByTwo}[4]
{
	\left(
		\begin{array}{cc}
			#1 & #2
		\\
			#3 & #4
		\end{array}
	\right)
}

\newcommand{\anticom}[2]
{
	\left\{
		#1, #2
	\right\}
}

\newcommand{\DAone}[1]{\Delta^{11}(#1)}
\newcommand{\EPAone}[2]{\frac{1}{#1^2}\frac{\alpha c_#2}{(\alpha+1) + c_#2(\alpha-1)}}
\newcommand{\Aone}[1]{\frac{\alpha c_#1}{\alpha+1 + c_#1(\alpha-1)}}
\newcommand{\Atwo}[1]{\frac{\alpha c_#1}{\alpha+1 + c_#1(1- \alpha)}}
\newcommand{\f}[1]{\frac{(1+\alpha) \tilde{c}_#1}{(1+\alpha)#1 \tilde{c}_#1 + 4 \alpha c_#1}}

\newcommand{\fB}[2]{\frac{(1+\alpha) \tilde{c}_#1}{(1+\alpha)#2 \tilde{c}_#1 + 4 \alpha c_#1}}
\newcommand{\gB}[2]{\frac{2 \alpha \tilde{c}_#1}{(1+\alpha)#2 \tilde{c}_#1 + 4 \alpha c_#1}}
\newcommand{\EP}[2]{\Delta^{#1}_{#2}}

\newcommand{\NewCO}{\mathsf{A}}
\newcommand{\NCO}[1]{\NewCO(#1)}
\newcommand{\NCOb}[1]{\NewCO_#1}
\newcommand{\InvCO}{\mathsf{B}}

\newcommand{\NCT}[1]{\tilde{\mathsf{A}}(#1)}

\newcommand{\NCOAlg}[1]{\frac{\alpha +1 + c_#1 (\alpha -1)}{\alpha c_#1}}
\newcommand{\NCTAlg}[1]{\frac{\alpha +1 + c_#1 (1-\alpha)}{\alpha c_#1}}

\newcommand{\DEP}[2]{\dot{\Delta}^{#1#2}}

\newcommand{\DB}[1]{\Delta^{B\bar{B}}(#1)}
\newcommand{\DD}[1]{\Delta^{D\bar{D}}(#1)}
\newcommand{\DCs}[1]{\Delta^{CC}(#1)}

\newcommand{\tr}{\ensuremath{\mathrm{tr}\,}}
\newcommand{\str}{\ensuremath{\mathrm{str}\,}}

\newcommand{\Int}[1]{\ensuremath{\int \!\! d^D \! #1 \,}}
\newcommand{\IntB}[1]{\ensuremath{\int \!\! \frac{d^D \! #1}{(2\pi)^D} \,}}
\newcommand{\volume}[1]{\ensuremath{d^D \! #1 \,}}

\newcommand{\AngVol}[1]{\ensuremath{\not{\!\Omega_#1}}}
\newcommand{\PowAngVol}[2]{\ensuremath{\not{\! \! \Omega_#1^#2}}}

\newcommand{\one}{\ensuremath{1\! \mathrm{l}}}

\newcommand{\pder}[2]{\ensuremath{\frac{\partial #1}{\partial #2}}}
\newcommand{\fder}[2]{\ensuremath{\frac{\delta #1}{\delta #2}}}

\newcommand{\flow}[1]{\ensuremath{\Lambda \partial_\Lambda #1}}
\newcommand{\flowConstAl}{\ensuremath{\Lambda \partial_\Lambda|_\alpha}}
\newcommand{\LConstAl}{\ensuremath{\left.\Lambda\right|_\alpha}}
\newcommand{\sym}{\ensuremath{\mathcal{S}}}
\newcommand{\GRk}{\ensuremath{\rhd}}
\newcommand{\CC}{CC}
\newcommand{\GRkp}{\ensuremath{\wedge}}

\renewcommand{\wedge}{>}

\newcommand{\ds}{\displaystyle}
\newcommand{\hf}{\ensuremath{\frac{1}{2}}}

\newcommand{\NUn}{\ensuremath{\mathsf{NC}}}

\newcommand{\strAA}{\ensuremath{\str A^1_\mu(p) A^1_\nu (-p)}}

\newcommand{\altoz}{\alpha \rightarrow 0}
\newcommand{\OLDs}{\mathcal{D}_1}
\newcommand{\TLDs}{\mathcal{D}_2}
\newcommand{\POLDs}{\mathcal{P}_1}
\newcommand{\TLDU}{\left.\mathcal{D}_2\right|_U}

\newcommand{\BPZ}{\ensuremath{\InvCO'(0)}}

\newcommand{\Op}[1]{\ensuremath{\mathcal{O}(p^{#1})}}
\newcommand{\Oq}[1]{\ensuremath{\mathcal{O}(q^{#1})}}
\newcommand{\Oep}{\ensuremath{\mathcal{O}(\epsilon)}}
\newcommand{\Oepz}{\ensuremath{\mathcal{O}(\epsilon^0)}}
\newcommand{\OepPow}[1]{\ensuremath{\mathcal{O}\left(\epsilon^{#1}\right)}}
\newcommand{\Omom}[1]{\ensuremath{\mathcal{O}\left({\mbox{mom}^{#1}}\right)}}
\newcommand{\Oone}{\ensuremath{\mathcal{O}(1)}}
\newcommand{\Oal}{\mathcal{O}(\alpha)}
\newcommand{\OalPow}[1]{\mathcal{O}\left(\alpha^{#1}\right)}

\newlength{\epminusone}
\newcommand{\emol}{\hspace{\epminusone}}
\newlength{\epzero}
\newcommand{\etozl}{\hspace{\epzero}}

\newlength{\NU}
\newcommand{\NUl}{\hspace{\NU}}

\newlength{\plus}
\newlength{\Sigmag}
\newcommand{\jhep}[3]{\emph{JHEP} #1 (#2) #3}
\newcommand{\NuclPhys}[4]{\emph{Nucl.\ Phys.\ }\textbf{#1 #2} (#3) #4}
\newcommand{\PhysRev}[4]{\emph{Phys.\ Rev.\ }\textbf{#1 #2} (#3) #4}
\newcommand{\IntJModPhys}[4]{\emph{Int.\ J.\ Mod.\ Phys.\ }\textbf{#1 #2} (#3) #4}
\newcommand{\PhysRep}[4]{\emph{Phys.\ Rep.\ }\textbf{#1 #2} (#3) #4}
\newcommand{\PhysRept}[3]{\emph{Phys.\ Rept.\ }\textbf{#1} (#2) #3}
\newcommand{\TheorMathPhys}[3]{\emph{Theor.\ Math.\ Phys.\ }\textbf{#1} (#2) #3}
\newcommand{\ModPhysLett}[4]{\emph{Mod.\ Phys.\ Lett.\ }\textbf{#1 #2} (#3) #4}
\newcommand{\AnnPhys}[3]{\emph{Ann.\ Phys.\ }\textbf{#1} (#2) #3}
\newcommand{\PhysLett}[4]{\emph{Phys.\ Lett.\ }\textbf{#1 #2} (#3) #4}
\newcommand{\ProgTheorPhys}[3]{\emph{Prog.\ Theor.\ Phys.\ }\textbf{#1} (#2) #3}
\newcommand{\Acta}[3]{\emph{Acta Phys.\ Slov.\ }\textbf{#1} (#2) #3}
\newcommand{\EurPhysJ}[3]{\emph{Eur.\ Phys.\ J.\ }\textbf{#1} (#2) #3}
\newcommand{\CEurJPhys}[3]{\emph{Central Eur.\ J.\ Phys.\ }\textbf{#1} (#2) #3}
\newcommand{\ZPhys}[4]{\emph{Z.\ Phys.\ } \textbf{#1 #2} (#3) #4}
\newcommand{\PhysRevLett}[3]{\emph{Phys.\ Rev.\ Lett.\ } \textbf{#1} (#2) #3}
\newcommand{\ClassQuantGrav}[3]{\emph{Class.\ Quant.\ Grav.\ }\textbf{#1} (#2) #3}
\newcommand{\hepth}[1]{hep-th/#1}
\newcommand{\hepph}[1]{hep-ph/#1}

\newcommand{\PoLep}{\ensuremath{\frac{p^{-2\epsilon}}{\Lambda^{-2\epsilon}}}}

\newcommand{\Pep}{\ensuremath{p^{-2\epsilon}}}

\newcommand{\eptoz}{\ensuremath{\epsilon \rightarrow 0}}

\newcommand{\EM}{\ensuremath{\gamma_\mathrm{EM}}}

\newcommand{\crit}{\ensuremath{J_\mathrm{max}}}

\newcommand{\nCr}[2]{\ensuremath{^{#1}C_{#2}\ }}

\newcommand{\norm}{\ensuremath{\Upsilon}}
\newcommand{\comp}[2]{\ensuremath{V^{#1,#2}}}
\newcommand{\CR}[2]{\ensuremath{V^{#1,#2;R}}}

\newcommand{\WBT}{\ensuremath{\ensuremath{\begin{array}{c}\begin{picture}(0,0)%
\includegraphics{pstex/WBT.pstex}%
\end{picture}%
\setlength{\unitlength}{3947sp}%
\begingroup\makeatletter\ifx\SetFigFont\undefined%
\gdef\SetFigFont#1#2#3#4#5{%
  \reset@font\fontsize{#1}{#2pt}%
  \fontfamily{#3}\fontseries{#4}\fontshape{#5}%
  \selectfont}%
\fi\endgroup%
\begin{picture}(405,126)(710,51)
\end{picture}
 \end{array}}}}
\newcommand{\BT}{\ensuremath{\ensuremath{\begin{array}{c}\begin{picture}(0,0)%
\includegraphics{pstex/BT.pstex}%
\end{picture}%
\setlength{\unitlength}{3947sp}%
\begingroup\makeatletter\ifx\SetFigFont\undefined%
\gdef\SetFigFont#1#2#3#4#5{%
  \reset@font\fontsize{#1}{#2pt}%
  \fontfamily{#3}\fontseries{#4}\fontshape{#5}%
  \selectfont}%
\fi\endgroup%
\begin{picture}(178,130)(937,47)
\end{picture}
 \end{array}}}}

\newcommand{\DiagSigma}{\ensuremath{\ensuremath{\begin{array}{c}\begin{picture}(0,0)%
\includegraphics{pstex/DiagSigma.pstex}%
\end{picture}%
\setlength{\unitlength}{3947sp}%
\begingroup\makeatletter\ifx\SetFigFont\undefined%
\gdef\SetFigFont#1#2#3#4#5{%
  \reset@font\fontsize{#1}{#2pt}%
  \fontfamily{#3}\fontseries{#4}\fontshape{#5}%
  \selectfont}%
\fi\endgroup%
\begin{picture}(101,101)(1005,-551)
\end{picture}
 \end{array}}}}
\newcommand{\ReplaceAwC}{\ensuremath{\ensuremath{\begin{array}{c}\begin{picture}(0,0)%
\includegraphics{pstex/ReplaceAwC.pstex}%
\end{picture}%
\setlength{\unitlength}{3947sp}%
\begingroup\makeatletter\ifx\SetFigFont\undefined%
\gdef\SetFigFont#1#2#3#4#5{%
  \reset@font\fontsize{#1}{#2pt}%
  \fontfamily{#3}\fontseries{#4}\fontshape{#5}%
  \selectfont}%
\fi\endgroup%
\begin{picture}(102,100)(1816,-713)
\end{picture}
 \end{array}}}}
\newcommand{\ReplaceAwSigma}{\ensuremath{\ensuremath{\begin{array}{c}\begin{picture}(0,0)%
\includegraphics{pstex/ReplaceAwSigma.pstex}%
\end{picture}%
\setlength{\unitlength}{3947sp}%
\begingroup\makeatletter\ifx\SetFigFont\undefined%
\gdef\SetFigFont#1#2#3#4#5{%
  \reset@font\fontsize{#1}{#2pt}%
  \fontfamily{#3}\fontseries{#4}\fontshape{#5}%
  \selectfont}%
\fi\endgroup%
\begin{picture}(118,114)(1926,-743)
\end{picture}
 \end{array}}}}
\newcommand{\DoubleCircle}{\ensuremath{\ensuremath{\begin{array}{c}\begin{picture}(0,0)%
\includegraphics{pstex/DoubleCircle.pstex}%
\end{picture}%
\setlength{\unitlength}{3947sp}%
\begingroup\makeatletter\ifx\SetFigFont\undefined%
\gdef\SetFigFont#1#2#3#4#5{%
  \reset@font\fontsize{#1}{#2pt}%
  \fontfamily{#3}\fontseries{#4}\fontshape{#5}%
  \selectfont}%
\fi\endgroup%
\begin{picture}(422,839)(2078,-1397)
\end{picture}
 \end{array}}}}
\newcommand{\DummyKernel}{\ensuremath{\stackrel{\bullet}{\mbox{\rule{1cm}{.2mm}}}}}

\newcommand{\Dal}{\ensuremath{\ensuremath{\begin{array}{c}\begin{picture}(0,0)%
\includegraphics{pstex/Dal.pstex}%
\end{picture}%
\setlength{\unitlength}{3947sp}%
\begingroup\makeatletter\ifx\SetFigFont\undefined%
\gdef\SetFigFont#1#2#3#4#5{%
  \reset@font\fontsize{#1}{#2pt}%
  \fontfamily{#3}\fontseries{#4}\fontshape{#5}%
  \selectfont}%
\fi\endgroup%
\begin{picture}(144,153)(416,370)
\put(452,427){\makebox(0,0)[lb]{\smash{\SetFigFont{8}{9.6}{\rmdefault}{\mddefault}{\updefault}{\color[rgb]{0,0,0}$\alpha$}%
}}}
\end{picture}
 \end{array}}}}

\newcommand{\COMP}{\mathsf{C}}
\newcommand{\Uni}[1]{\left. #1 \right|_\COMP}

\newcommand{\Cancel}[2]{
\begin{cancel}
Diagram~\ref{#1}  exactly cancels diagram~\ref{#2}.
\label{cancel:#1}
\end{cancel}
}

\newcommand{\CancelTmD}[2]{
\begin{cancel}
Diagram~\ref{#1}  cancels diagram~\ref{#2}, up to a discarded total momentum derivative.
\label{cancel:#1}
\end{cancel}
}

\newcommand{\CancelCom}[3]{
\begin{cancel}
Diagram~\ref{#1}  exactly cancels diagram~\ref{#2} #3
\label{cancel:#1}
\end{cancel}
}

\newcommand{\PartialCancelCom}[3]{
\begin{cancel}
Diagram~\ref{#1}  partially cancels diagram~\ref{#2}. #3
\label{cancel:#1}
\end{cancel}
}

\newcommand{\CancelOpCom}[3]{
\begin{cancel}
Diagram~\ref{#1} cancels diagram~\ref{#2} at $\Op{2}$#3
\label{cancel:#1}
\end{cancel}
}

\newcommand{\CancelRange}[4]{
\begin{cancel}
Diagrams~\ref{#1}--\ref{#2}  exactly cancel diagrams~\ref{#3}--\ref{#4}.
\label{cancel:#1}
\end{cancel}
}

\newcommand{\AGRO}[1]{\ensuremath{}\left[ #1 \right]_{ES}}
\newcommand{\AGROB}[1]{\ensuremath{}\left[ #1 \right]_{ES'}}

\newcommand{\dec}[3][0]{\ensuremath{\left[ #2 \hspace{#1in} \right]^{#3}}}

\newcommand{\decB}[3][0]{\ensuremath{\left[ \hspace{#1in} #2 \hspace{#1in} \right]^{#3}}}

\newcommand{\RO}[2][0]{\ensuremath{
\left[
#2 \hspace{#1in}
\right]_R
}} 

\newcommand{\LHlabs}[3][-2]{
\begin{array}{l}
\vspace{#1ex}
	#2\\
	#3		
\end{array}
}

\newcommand{\OLLabC}[4][0]
{
\LHlabs[#1]{\hspace{\scriptboldcurlybracket} \LabOLOb{#2}}{\CancelOLObject{#3}{#4}}
}

\newcommand{\COCO}[5][-2]{
	\LHlabs[#1]{\CancelOLObject{#2}{#3}}{\CancelOLObject{#4}{#5}}
}

\newcommand{\POCO}[5][-2]{
	\LHlabs[#1]{\hspace{\scriptboldcurlybracket} \Dprocess{#2}{#3}}{\CancelOLObject{#4}{#5}}
}

\newlength{\scriptboldcurlybracket}
\newcommand{\OLLabLBC}[4][0]
{
\LHlabs[#1]{\LabOLObLB{#2}}{\CancelOLObject{#3}{#4}}
}

\newcommand{\OLLabRBC}[4][0]
{
\LHlabs[#1]{\hspace{\scriptboldcurlybracket} \LabOLObRB{#2}}{\CancelOLObject{#3}{#4}}
}

\newcommand{\OLLabPro}[4][0]
{
\LHlabs[#1]{\LabOLOb{#2}}{\Dprocess{#3}{#4}}
}

\newcommand{\OLLabLab}[3][0]
{
\LHlabs[#1]{\LabOLOb{#2}}{\LabOLOb{#3}}
}

\newcommand{\LObLOb}[3][-2]
{
\OLLabLab[#1]{#2}{#3}
}

\newcommand{\OLLabLabLB}[3][-2]
{
\LHlabs[#1]{\hspace{\scriptboldcurlybracket} \LabOLOb{#2}}{\LabOLObLB{#3}}
}

\newcommand{\OLLabLabRB}[3][-2]
{
\LHlabs[#1]{\LabOLOb{#2}}{\LabOLObRB{#3}}
}

\newcommand{\LOLBLOLB}[3][-2]
{
\LHlabs[#1]{\LabOLObLB{#2}}{\LabOLObLB{#3}}
}

\newcommand{\OLLabRBLabRB}[3][-2]
{
\LHlabs[#1]{\LabOLObRB{#2}}{\LabOLObRB{#3}}
}

\newcommand{\LORBLORB}[3][-2]
{
\OLLabRBLabRB[#1]{#2}{#3}
}

\newcommand{\OLLabRBPro}[4][-2]
{
\LHlabs[#1]{\LabOLObRB{#2}}{\Dprocess{#3}{#4}}
}

\newlength{\strutheight}
\newcommand{\blankstrut}{\rule{0em}{\strutheight}}

\newcommand{\OLLabBlankLB}[2][-2]
{
\LHlabs[#1]{\LabOLObLB{#2}}{\blankstrut}
}

\newcommand{\LOBl}[2][-2]
{
\LHlabs[#1]{\LabOLOb{#2}}{\blankstrut}
}

\newcommand{\COBl}[3][-2]
{
\LHlabs[#1]{\CancelOLObject{#2}{#3}}{\blankstrut}
}

\newcommand{\sco}[3][0]{\ensuremath{
\begin{array}{c}
\vspace{#1in}
#2 \\
#3
\end{array}
}}

\newcommand{\Process}[1]{
\mbox{\scriptsize $\rightarrow \ref{#1}$}
}

\newcommand{\Discard}{
\mbox{\scriptsize $\rightarrow 0$}
}

\newcommand{\DiscardBlank}[1][-2]{
\begin{array}{c}
\vspace{#1ex}
	\mbox{\scriptsize $\rightarrow 0$} 
\\
	\blankstrut	
\end{array}
}

\newcommand{\ProPro}[3][-2]{
\begin{array}{c}
\vspace{#1ex}
	\mbox{\scriptsize $\rightarrow \ref{#2}$} 
\\
	\mbox{\scriptsize $\rightarrow \ref{#3}$}	
\end{array}
}

\newcommand{\ProBlank}[2][-2]{
\begin{array}{c}
\vspace{#1ex}
	\mbox{\scriptsize $\rightarrow \ref{#2}$} 
\\
	\blankstrut	
\end{array}
}

\newlength{\ProcessRefLength}
\newlength{\ProcessLength}
\newcommand{\Gprocess}[3][0]{
\ensuremath{
\settowidth{\LabLength}{\scriptsize\textbf{\ref{#2}}}
\settowidth{\ProcessRefLength}{\scriptsize\ref{#3}}
\addtolength{\LabLength}{\ProcessRefLength}
\addtolength{\LabLength}{\ProcessLength}
\addtolength{\LabLength}{0.8em}
\begin{minipage}{\LabLength}
\scriptsize
\begin{GD}\label{#2}$\rightarrow \ref{#3}$\end{GD}
\end{minipage}
}
}

\newcommand{\process}[3][0]{
\ensuremath{
\settowidth{\LabLength}{\scriptsize\textbf{\ref{#2}}}
\settowidth{\ProcessRefLength}{\scriptsize\ref{#3}}
\addtolength{\LabLength}{\ProcessRefLength}
\addtolength{\LabLength}{\ProcessLength}
\addtolength{\LabLength}{0.8em}
\begin{minipage}{\LabLength}
\scriptsize
\begin{ED}\label{#2}$\rightarrow \ref{#3}$\end{ED}
\end{minipage}
}
}

\newcommand{\Dprocess}[3][0]{
\settowidth{\LabLength}{\scriptsize\textbf{\ref{#2} }}
\settowidth{\ProcessRefLength}{\scriptsize\ref{#3}}
\addtolength{\LabLength}{\ProcessRefLength}
\addtolength{\LabLength}{\ProcessLength}
\addtolength{\LabLength}{0.8em}
\begin{minipage}{\LabLength}
\scriptsize
\begin{D}\label{#2} $\rightarrow \ref{#3}$ \end{D}
\end{minipage}
}

\newcommand{\Iprocess}[3][0]{
\settowidth{\LabLength}{\scriptsize\textbf{\ref{#2} }}
\settowidth{\ProcessRefLength}{\scriptsize\ref{#3}}
\addtolength{\LabLength}{\ProcessRefLength}
\addtolength{\LabLength}{\ProcessLength}
\addtolength{\LabLength}{0.8em}
\begin{minipage}{\LabLength}
\scriptsize
\begin{ID}\label{#2} $\rightarrow \ref{#3}$ \end{ID}
\end{minipage}
}

\newcommand{\PO}[2]{\Dprocess{#1}{#2}}

\newcommand{\discard}[2][0]{
\settowidth{\LabLength}{\scriptsize\textbf{\ref{#2}}}
\settowidth{\ProcessRefLength}{\scriptsize $\rightarrow 0$}
\addtolength{\LabLength}{\ProcessRefLength}
\addtolength{\LabLength}{\ProcessLength}
\addtolength{\LabLength}{0.8em}
\begin{minipage}{\LabLength}
\scriptsize
\begin{ED}\label{#2}$\rightarrow 0$\end{ED}
\end{minipage}}

\newcommand{\OLdiscard}[2][0]{
\settowidth{\LabLength}{\scriptsize\textbf{\ref{#2}}}
\settowidth{\ProcessRefLength}{\scriptsize $\rightarrow 0$}
\addtolength{\LabLength}{\ProcessRefLength}
\addtolength{\LabLength}{\ProcessLength}
\addtolength{\LabLength}{0.8em}
\begin{minipage}{\LabLength}
\scriptsize
\begin{D}\label{#2}$\rightarrow 0$\end{D}
\end{minipage}}

\newcommand{\Gdiscard}[2][0]{
\ensuremath{
\settowidth{\LabLength}{\scriptsize\textbf{\ref{#2}}}
\settowidth{\ProcessRefLength}{\scriptsize $\rightarrow 0$}
\addtolength{\LabLength}{\ProcessRefLength}
\addtolength{\LabLength}{\ProcessLength}
\addtolength{\LabLength}{0.8em}
\begin{minipage}{\LabLength}
\scriptsize
\begin{GD}\label{#2}$\rightarrow 0$\end{GD}
\end{minipage}}
}

\newlength{\CancelRefLength}
\newlength{\CancelRefLengthB}
\newlength{\CancelLength}
\newcommand{\CancelExObject}[3][0]{
\settowidth{\LabLength}{\scriptsize\textbf{\ref{#2}}}
\settowidth{\CancelRefLength}{\scriptsize$\ref{#3}$}
\addtolength{\LabLength}{\CancelRefLength}
\addtolength{\LabLength}{\CancelLength}
\begin{minipage}{\LabLength}
\scriptsize
\begin{EDC}\label{#2} \ref{#3} \textbf{\}}
\end{EDC}
\end{minipage}
}

\newcommand{\CancelGObject}[3][0]{
\settowidth{\LabLength}{\scriptsize\textbf{\ref{#2}}}
\settowidth{\CancelRefLength}{\scriptsize$\ref{#3}$}
\addtolength{\LabLength}{\CancelRefLength}
\addtolength{\LabLength}{\CancelLength}
\begin{minipage}{\LabLength}
\scriptsize
\begin{GDC}\label{#2} \ref{#3} \textbf{\}}
\end{GDC}
\end{minipage}
}

\newcommand{\CancelOLObject}[2]{
\settowidth{\LabLength}{\scriptsize\textbf{\ref{#2}}}
\settowidth{\CancelRefLength}{\scriptsize$\ref{#1}$}
\addtolength{\LabLength}{\CancelRefLength}
\addtolength{\LabLength}{\CancelLength}
\begin{minipage}{\LabLength}
\scriptsize
\begin{DC}\label{#1} \ref{#2} \textbf{\}}
\end{DC}
\end{minipage}
}

\newcommand{\CO}[2]{
	\CancelOLObject{#1}{#2}
}

\newcommand{\CancelOLObjectB}[2]{
\settowidth{\LabLength}{\scriptsize\textbf{\ref{#2}}}
\settowidth{\CancelRefLength}{\scriptsize$\ref{#1}$}
\addtolength{\LabLength}{\CancelRefLength}
\addtolength{\LabLength}{\CancelLength}
\begin{minipage}{\LabLength}
\scriptsize
\begin{DC}\label{#1} \ref{#2} \textbf{$|$}
\end{DC}
\end{minipage}
}

\newcommand{\CancelOLObjectC}[2]{
\settowidth{\LabLength}{\scriptsize\textbf{\ref{#2}}}
\settowidth{\CancelRefLength}{\scriptsize$\ref{#1}$}
\addtolength{\LabLength}{\CancelRefLength}
\addtolength{\LabLength}{\CancelLength}
\begin{minipage}{\LabLength}
\scriptsize
\begin{DC-B}\label{#1} \ref{#2} \textbf{\}}
\end{DC-B}
\end{minipage}
}

\newcommand{\COB}[2]{\CancelOLObjectB{#1}{#2}}
\newcommand{\COC}[2]{\CancelOLObjectC{#1}{#2}}

\newlength{\PartialCancelLength}
\newcommand{\PartialCancelExObject}[3][0]{
\settowidth{\LabLength}{\scriptsize\textbf{\ref{#3}}}
\settowidth{\CancelRefLength}{\scriptsize$\ref{#2}$}
\addtolength{\LabLength}{\CancelRefLength}
\addtolength{\LabLength}{\PartialCancelLength}
\begin{minipage}{\LabLength}
\scriptsize
\begin{EDC}\label{#2} \ref{#3}(P) \textbf{\}}
\end{EDC}
\end{minipage}
}

\newcommand{\PartialCancelGObject}[3][0]{
\settowidth{\LabLength}{\scriptsize\textbf{\ref{#3}}}
\settowidth{\CancelRefLength}{\scriptsize$\ref{#2}$}
\addtolength{\LabLength}{\CancelRefLength}
\addtolength{\LabLength}{\PartialCancelLength}
\begin{minipage}{\LabLength}
\scriptsize
\begin{GDC}\label{#2} \ref{#3}(P) \textbf{\}}
\end{GDC}
\end{minipage}
}

\newcommand{\labID}[2][0]{
\settowidth{\LabLength}{\scriptsize\textbf{\ref{#2}}}
\begin{minipage}{100em}
\scriptsize
\begin{ID}\label{#2}
\end{ID}
\end{minipage}
}

\newcommand{\DoubleCancelOLObject}[3]{
\settowidth{\LabLength}{\scriptsize\textbf{\ref{#1}}}
\settowidth{\CancelRefLength}{\scriptsize$\ref{#2}$}
\settowidth{\CancelRefLengthB}{\scriptsize$\ref{#3}$}
\addtolength{\LabLength}{\CancelRefLength}
\addtolength{\LabLength}{\CancelRefLengthB}
\addtolength{\LabLength}{\CancelLength}
\addtolength{\LabLength}{0.5em}
\begin{minipage}{\LabLength}
\scriptsize
\begin{DC}\label{#1} \ref{#2}, \ref{#3} \textbf{\}}
\end{DC}
\end{minipage}
}

\newcommand{\DCO}[3]{
	\DoubleCancelOLObject{#1}{#2}{#3}
}

\newlength{\DoubleCancelRefLengthA}
\newlength{\DoubleCancelRefLengthB}
\newlength{\DoublePartialCancelLength}
\newcommand{\PartialDoubleCancelExObject}[4][0]{
\settowidth{\LabLength}{\scriptsize\textbf{\ref{#2}}}
\settowidth{\DoubleCancelRefLengthA}{\scriptsize \textbf{\ref{#3}}}
\settowidth{\DoubleCancelRefLengthB}{\scriptsize \textbf{\ref{#4}}}
\addtolength{\LabLength}{\DoublePartialCancelLength}
\addtolength{\LabLength}{\DoubleCancelRefLengthA}
\addtolength{\LabLength}{\DoubleCancelRefLengthB}
\begin{minipage}{\LabLength}
\scriptsize
\begin{EDC}\label{#2} \ref{#3}, \ref{#4}(P) \textbf{\}}
\end{EDC}
\end{minipage}
}

\newlength{\DoublePartialUnRefCancelLength}
\newcommand{\PartialDoubleUnRefCancelExObject}[3][0]{
\settowidth{\LabLength}{\scriptsize\textbf{\ref{#2}}}
\settowidth{\DoubleCancelRefLengthA}{\scriptsize \textbf{\ref{#3}}}
\settowidth{\DoubleCancelRefLengthB}{\scriptsize \textit{??}}
\addtolength{\LabLength}{\DoublePartialCancelLength}
\addtolength{\LabLength}{\DoubleCancelRefLengthA}
\addtolength{\LabLength}{\DoubleCancelRefLengthB}
\begin{minipage}{\LabLength}
\scriptsize
\begin{EDC}\label{#2} ??, \ref{#3}(P) \textbf{\}}
\end{EDC}
\end{minipage}
}

\newcommand{\PartialDoubleCancelGObject}[4][0]{
\settowidth{\LabLength}{\scriptsize\textbf{\ref{#2}}}
\settowidth{\DoubleCancelRefLengthA}{\scriptsize \textbf{\ref{#3}}}
\settowidth{\DoubleCancelRefLengthB}{\scriptsize \textbf{\ref{#4}}}
\addtolength{\LabLength}{\DoublePartialCancelLength}
\addtolength{\LabLength}{\DoubleCancelRefLengthA}
\addtolength{\LabLength}{\DoubleCancelRefLengthB}
\begin{minipage}{\LabLength}
\scriptsize
\begin{GDC}\label{#2} \ref{#3}, \ref{#4}(P) \textbf{\}}
\end{GDC}
\end{minipage}
}

\newcommand{\PED}[5][0]{\ensuremath{
\sco[#1]{\process[#2]{#4}{#5}}{\ensuremath{\begin{array}{c}\input{pstex/#3.pstex_t} \end{array}}}
}}

\newcommand{\POLD}[5][0]{\ensuremath{
\sco[#1]{\Dprocess[#2]{#4}{#5}}{\ensuremath{\begin{array}{c}\input{pstex/#3.pstex_t} \end{array}}}
}}

\newcommand{\PGD}[5][0]{\ensuremath{
\sco[#1]{\Gprocess[#2]{#4}{#5}}{\ensuremath{\begin{array}{c}\input{pstex/#3.pstex_t} \end{array}}}
}}

\newcommand{\PID}[5][0]{\ensuremath{
\sco[#1]{\Iprocess[#2]{#4}{#5}}{\ensuremath{\begin{array}{c}\input{pstex/#3.pstex_t} \end{array}}}
}}

\newcommand{\DED}[4][0]{\ensuremath{
\sco[#1]{\discard[#2]{#4}}{\ensuremath{\begin{array}{c}\input{pstex/#3.pstex_t} \end{array}}}
}}

\newcommand{\DGD}[4][0]{\ensuremath{
\sco[#1]{\Gdiscard[#2]{#4}}{\ensuremath{\begin{array}{c}\input{pstex/#3.pstex_t} \end{array}}}
}}

\newcommand{\PEO}[5][0]{\ensuremath{
\sco[#1]{\process[#2]{#4}{#5}}{#3}
}}

\newcommand{\PGO}[5][0]{\ensuremath{
\sco[#1]{\Gprocess[#2]{#4}{#5}}{#3}
}}

\newcommand{\CEO}[5][0]{\ensuremath{
\sco[#1]{\CancelExObject[#2]{#4}{#5}}{#3}
}}

\newcommand{\CED}[5][0]{\ensuremath{
\sco[#1]{\CancelExObject[#2]{#4}{#5}}{\ensuremath{\begin{array}{c}\input{pstex/#3.pstex_t} \end{array}}}
}}

\newcommand{\CGD}[5][0]{\ensuremath{
\sco[#1]{\CancelGObject[#2]{#4}{#5}}{\ensuremath{\begin{array}{c}\input{pstex/#3.pstex_t} \end{array}}}
}}

\newcommand{\COLD}[4][0]{\ensuremath{
\sco[#1]{\CancelOLObject{#3}{#4}}{\ensuremath{\begin{array}{c}\input{pstex/#2.pstex_t} \end{array}}}
}}

\newcommand{\PCED}[5][0]{\ensuremath{
\sco[#1]{\PartialCancelExObject[#2]{#4}{#5}}{\ensuremath{\begin{array}{c}\input{pstex/#3.pstex_t} \end{array}}}
}}

\newcommand{\PCGD}[5][0]{\ensuremath{
\sco[#1]{\PartialCancelGObject[#2]{#4}{#5}}{\ensuremath{\begin{array}{c}\input{pstex/#3.pstex_t} \end{array}}}
}}

\newcommand{\PDCED}[6][0]{\ensuremath{
\sco[#1]{\PartialDoubleCancelExObject[#2]{#4}{#5}{#6}}{\ensuremath{\begin{array}{c}\input{pstex/#3.pstex_t} \end{array}}}
}}

\newcommand{\PDURCED}[4][0]{\ensuremath{
\sco[#1]{\PartialDoubleUnRefCancelExObject{#3}{#4}}{\ensuremath{\begin{array}{c}\input{pstex/#2.pstex_t} \end{array}}}
}}

\newcommand{\PDCGD}[6][0]{\ensuremath{
\sco[#1]{\PartialDoubleCancelGObject[#2]{#4}{#5}{#6}}{\ensuremath{\begin{array}{c}\input{pstex/#3.pstex_t} \end{array}}}
}}

\newcommand{\LabIllOb}[2][0]{
\settowidth{\LabLength}{\scriptsize \textbf{\ref{#2}}}
\addtolength{\LabLength}{0.8em}
\begin{minipage}{\LabLength}
\scriptsize
\begin{ID}\label{#2}
\end{ID}
\end{minipage}
}

\newcommand{\LabExOb}[2][0]{
\settowidth{\LabLength}{\scriptsize\textbf{\ref{#2}}}
\addtolength{\LabLength}{0.8em}
\ensuremath{
\begin{minipage}{\LabLength}
\scriptsize
\begin{ED}\label{#2}
\end{ED}
\end{minipage}
}
}

\newlength{\LabLength}
\newcommand{\LabGOb}[2][0]{
\settowidth{\LabLength}{\scriptsize \textbf{\ref{#2}}}
\addtolength{\LabLength}{0.8em}
\begin{minipage}{\LabLength}
\scriptsize
\begin{GD}\label{#2}
\end{GD}
\end{minipage}
}
\newcommand{\LabOLOb}[1]{
\settowidth{\LabLength}{\scriptsize \textbf{\ref{#1}}}
\addtolength{\LabLength}{0.8em}
\begin{minipage}{\LabLength}
\scriptsize
\begin{D}\label{#1}
\end{D}
\end{minipage}
}

\newcommand{\LE}[1]{
\settowidth{\LabLength}{\scriptsize \textbf{\ref{#1}}}
\addtolength{\LabLength}{0.8em}
\begin{minipage}{\LabLength}
\scriptsize
\begin{ED}\label{#1}
\end{ED}
\end{minipage}
}

\newcommand{\LO}[1]{\LabOLOb{#1}}
\newcommand{\LOLB}[1]{\LabOLObLB{#1}}
\newcommand{\LORB}[1]{\LabOLObRB{#1}}

\newcommand{\LabOLObLB}[1]{
\settowidth{\LabLength}{\scriptsize \textbf{\{ \ref{#1}}}
\addtolength{\LabLength}{0.8em}
\begin{minipage}{\LabLength}
\scriptsize
\begin{DC}\label{#1}
\end{DC}
\end{minipage}
}

\newcommand{\LabOLObRB}[1]{
\settowidth{\LabLength}{\scriptsize \textbf{\}\ref{#1}}}
\addtolength{\LabLength}{0.8em}
\begin{minipage}{\LabLength}
\scriptsize
\begin{D}\label{#1}
\textbf{\}}
\end{D}
\end{minipage}
}
\newcommand{\LID}[4][0]{\ensuremath{
\sco[#1]{\LabIllOb[#2]{#4}}{\ensuremath{\begin{array}{c}\input{pstex/#3.pstex_t} \end{array}}}
}}

\newcommand{\LED}[4][0]{\ensuremath{
\sco[#1]{\LabExOb[#2]{#4}}{\ensuremath{\begin{array}{c}\input{pstex/#3.pstex_t} \end{array}}}
}}

\newcommand{\LEO}[4][0]{\ensuremath{
\sco[#1]{\LabExOb[#2]{#4}}{#3}
}}

\newcommand{\LGD}[4][0]{\ensuremath{
\sco[#1]{\LabGOb[#2]{#4}}{\ensuremath{\begin{array}{c}\input{pstex/#3.pstex_t} \end{array}}}
}}

\newcommand{\LGO}[4][0]{\ensuremath{
\sco[#1]{\LabGOb[#2]{#4}}{#3}
}}

\newcommand{\LOLD}[3][0]{\ensuremath{
\sco[#1]{\LabOLOb{#3}}{\ensuremath{\begin{array}{c}\input{pstex/#2.pstex_t} \end{array}}}
}}

\newcommand{\scd}[3][0]{\ensuremath{
\begin{array}{c}
\vspace{#1in}
\ensuremath{\begin{array}{c}\input{pstex/#2.pstex_t} \end{array}} \\
\ensuremath{\begin{array}{c}\input{pstex/#3.pstex_t} \end{array}}
\end{array}
}}

\newcommand{\tco}[3][0]{\ensuremath{
\begin{array}{ccc}
#2 &  \hspace{#1in} & #3 
\end{array}
}}

\newcommand{\tcd}[3][0]{\ensuremath{
\hspace{#1in}
\tco[-0.2]{\ensuremath{\begin{array}{c}\input{pstex/#2.pstex_t} \end{array}}}{\ensuremath{\begin{array}{c}\input{pstex/#3.pstex_t} \end{array}}}
\hspace{#1in}
}}

\newcounter{GenDiaLab}
\Alph{GenDiaLab}
\newcounter{IllDiaLab}
\Alph{IllDiaLab}
\newcounter{ExDiaLab}
\Alph{ExDiaLab}
\newcounter{cancellation}
\addtocounter{cancellation}{1}
\newtheorem{D-ID}{Diagrammatic Identity}

\newtheorem{D}{}[chapter]
\newtheorem{DC}[D]{\{}
\newtheorem{DC-B}[D]{$|$}
\newtheorem{com}{Comment}
\newtheorem{ED}{}[chapter]
\newtheorem{Prop}{Proposition}[chapter]

\newtheorem{EDC}[ED]{\{}
\newtheorem{ID}{}[IllDiaLab]	
\newtheorem{GD}{}[GenDiaLab]	
\newtheorem{GDC}[GD]{\{}

\newlength{\PFheight}
\newcommand{\PF}[1]{
	 \stackrel{\rightarrow}{\vspace{\PFheight} #1}
}

\newcommand{\PB}[1]{
	\stackrel{\leftarrow}{\vspace{\PFheight} #1}
}

\newtheorem{cancel}{Cancellation}[chapter]

\newtheorem{CM}{Cancellation Mechanism}

\newlength{\dalpha}
\newlength{\Reduce}
\newcommand{\VertexSum}[2]{\ensuremath{
	\displaystyle \prod_{I=#1}^{#2} \sum_{V^{I_+}=0}^{V^I} 
}}

\newcommand{\JsumLim}{\ensuremath{2n+s}}

\newtheorem{assert}{Assertion}

\newlength{\hatwidth}
\newcommand{\TitlePage}[2]
{
\begin{center}
\large
	\textbf{UNIVERSITY OF SOUTHAMPTON}

	FACULTY OF SCIENCE

	School of Physics \& Astronomy

\vspace{2in}

\huge
	\textbf{The Manifestly Gauge Invariant Exact Renormalisation Group}

	\vspace{1ex}
	Volume #1 of II

\large

	\vspace{1ex}

	by

	\vspace{1ex}

	\textbf{Oliver Jacob Rosten}

	\vfill

Thesis for the degree of Doctor of Philosophy

\hspace{2ex}

#2
\end{center}
}

\newlength{\VertexWidth}

\newcommand{\Vertex}[1]
{
	\settowidth{\VertexWidth}{$#1$}
	\setlength{\unitlength}{1.2\VertexWidth}
	\begin{array}{c}
	\begin{picture}(1,1)(-0.5,-0.5)
		\put(0,0){\circle{1}}
		\put(-0.4,-0.1){$#1$}
	\end{picture}
	\end{array}
}

\newcounter{diagram}
\addtocounter{diagram}{1}

\begin{document}
\ifx\href\undefined\else\hypersetup{linktocpage=true}\fi 

\begin{titlepage}
	\TitlePage{I}{March 2005}
\end{titlepage}

\begin{titlepage}
	\begin{center}


UNIVERSITY OF SOUTHAMPTON

\vspace{1ex}

\underline{ABSTRACT}

\vspace{1ex}

FACULTY OF SCIENCE

\vspace{1ex}

SCHOOL OF PHYSICS \& ASTRONOMY

\vspace{1ex}

\underline{Doctor of Philosophy}

\vspace{1ex}

THE MANIFESTLY GAUGE INVARIANT EXACT RENORMALISATION GROUP

\vspace{1ex}

by Oliver Jacob Rosten
\end{center}
\vspace{1ex}

\noindent
We construct a manifestly gauge invariant Exact Renormalisation Group (ERG)
whose form is suitable for
computation in $SU(N)$ Yang-Mills theory, beyond one-loop. 
An effective cutoff is implemented by embedding the physical
$SU(N)$ theory in a 
spontaneously broken $SU(N|N)$ Yang-Mills theory.

To facilitate computations within this scheme, which proceed at every
step without fixing the gauge, we develop a set of diagrammatic
techniques. As an initial test of the formalism, the one-loop $SU(N)$ Yang-Mills
$\beta$-function, $\beta_1$, is computed, and the standard, universal answer
is reproduced.

It is recognised that the computational technique can be greatly
simplified. Using these simplifications, a partial proof
is given that, to all orders in perturbation theory,
the explicit dependence of perturbative $\beta$-function coefficients, $\beta_n$,
on certain non-universal elements of the manifestly gauge invariant
\ERG\ cancels out. This partial proof yields an extremely
compact, diagrammatic form for the surviving contributions to
arbitrary $\beta_n$, up to a set of terms which are
yet to be dealt with. The validity of the
compact expression is reliant on an unproven assertion
at the third loop order and above.

Starting from the compact expression for $\beta_n$, we 
specialise to $\beta_2$ and explicitly construct the
set of terms yet to be dealt with. From the resulting diagrammatic
expression for $\beta_2$, we extract a numerical coefficient which,
in the limit that the coupling of one of the unphysical
regulator fields is tuned to zero, yields the standard,
universal answer. Thus, we have performed the very first
two-loop, continuum calculation in Yang-Mills
theory, without fixing the gauge.

\end{titlepage}

\frontmatter

\tableofcontents
\listoffigures
\listoftables

\newpage
\begin{center}
	\textbf{DECLARATION OF AUTHORSHIP}
\end{center}

\hspace{3ex}

\noindent
I, Oliver Jacob Rosten
declare that the thesis entitled

\begin{center}
	The Manifestly Gauge Invariant Exact Renormalisation Group
\end{center}
and the work presented in it are my own. I confirm that:
\begin{itemize}
	\item 	this work was done mainly while in candidature for a research degree
			at this University;

	\item	no part of this thesis has previously been submitted for a degree or
			any other qualification at this University or any other institution;

	\item	where I have consulted the published work of others, this is always
			clearly attributed;

	\item	where I have quoted from the work of others, the source is always
			given. With the exception of such quotations, this thesis is
			entirely my own work;

	\item	I have acknowledged all main sources of help;

	\item	where the thesis is based on work done by myself jointly with others,
			I have made clear exactly what was done by others and what I have
			contributed myself;

	\item	none of this work has been published before submission.
\end{itemize}

\vfill

Signed:\dotfill

Date:\dotfill

\newpage
\chapter{Acknowledgements}

I would like to thank my supervisor,
Tim Morris, for friendship, help and
for setting me on the right path. I
acknowledge the Southampton University
Development Trust without the kind 
support of which,
this work would never have happened. 

Thanks to Stefano, for helping me
when I first started learning about
the \ERG\ and for bearing the
brunt of my enthusiasm. Thanks to Daniel
for much appreciated help with the 
bibliography.

I would like to thank the SHEP group for far too
many things to list but, most importantly,
for teaching me the value of intellectual
integrity.

In helping to relieve the immense stress of academic
work, I express my debt to 
the great composers but, most especially, to J.~S.~Bach, Beethoven
and the incomparable Schubert. 

My deepest thanks I save for last. First, to
my family, Esther, Edward and Judith whom
I love and appreciate,  very much.

Finally, I would like to thank Claire for her endless
love and support. Without her, and what she
has brought to my life, this piece of
work (much like myself) would be a pale imitation
of what it is now.

\newpage

\vspace*{3in}

\begin{center}
\Large 
To Harvey
\end{center}

\chapter{Motivation}

The original motivation for this work was the development
of a tool, formulated in the
continuum, for the non-perturbative study of gauge theories
and, in particular, Quantum Chromodynamics (QCD).

The idea of using the Exact Renormalisation Group (ERG)
in this context is
not a new one~\cite{Bergerhoff:1995zm}--\cite{Arnone:2004:SUSY}.
Compared to these
methods, the formalism that
we present in this thesis is in its formative stages;
indeed we are yet to even attempt a
non-perturbative calculation.

Nevertheless, we believe that our formalism---the Manifestly
Gauge Invariant ERG (\MGIERG)~\cite{TRM-Faro}--\cite{FurtherWork}---has great
potential. The fundamental reason for this confidence
is that the \MGIERG\ combines two of the most
powerful ideas in quantum field theory (QFT),
those of Wilsonian Renormalisation~\cite{Wilson} and gauge invariance,
treating them on an equal footing.
A novel consequence of this is that, within our 
framework, gauge invariance is \emph{manifest}. 
That there is no need to fix the gauge is, in 
itself, encouraging for non-perturbative applications, 
since
Gribov copies~\cite{Gribov}, and all associated problems,
are entirely avoided.

The formalism that this thesis builds upon (discussed in 
section~\ref{sec:Intro-GI-ERG-Reg}) has already been used to compute the 
perturbative one-loop
Yang-Mills $\beta$-function~\cite{GI-ERG-I, GI-ERG-II, YM-1-loop} (henceforth referred to as
$\beta_1$)---which, up until the work presented here, was
the only analytical calculation performed in 
continuum gauge theory without fixing the gauge.

Though an interesting result in it own right, the primary goal
of computing $\beta_1$ was to test the formalism. Given the 
successful outcome, it is an obvious question to ask why 
the formalism was not immediately applied
to non-perturbative problems.
The point is that whilst the calculation of $\beta_1$ was crucial
in establishing the consistency of the formalism, there were many
aspects of the formalism which remained un-tested, in this
comparatively simple piece of work.

The initial aim of this thesis was to compute the 
perturbative two-loop
Yang-Mills $\beta$-function (henceforth referred to as
$\beta_2$),
thereby providing a far more stringent test our
approach.\footnote{
Within a class of massless renormalisation schemes,
both $\beta_1$ and $\beta_2$ are universal, allowing
meaningful
comparison between our result and others.
} 
With this achieved, the hope was that
the stage would be set for application to the
non-perturbative domain.\footnote{Though we should note that the 
generalisation to
QCD, by the inclusion of
quarks, is not necessarily trivial---see section~\ref{sec:Conclusion:FutureWork}.}

Whilst we are now at this stage, the successful 
calculation of $\beta_2$ has also opened up entirely
unforeseen avenues of research. In its initial 
incarnation, the computation of $\beta_2$ was
almost prohibitively long; though demonstrating
the consistency of the \MGIERG\ beyond reasonable
doubt, the difficulty associated with this
was ominous. However, since completing the calculation
of $\beta_2$, a deep structure to the calculation
has been revealed.

At a stroke, this has removed almost all the 
difficulty associated with 
computing $\beta_2$ in the \MGIERG\@. 
Moreover, there are encouraging indications
that the extraction of numerical values for
$\beta_n$ may be easier than for competitive
techniques.\footnote{Of course, this is of real use only if either
the \MGIERG\ can be used to compute amplitudes, or a map
is made between our scheme and (say) $\overline{MS}$. We note
that such a map has been constructed for a different
gauge invariant ERG~\cite{Ellwanger:1997tp}.} (Currently, $\beta_4$ is
the highest loop $\beta$-function
coefficient to be computed using traditional techniques;
to appreciate the complexity of this calculation see~\cite{beta4-Verm, beta4-Other}.) Further investigation of the structure
of $\beta$-function coefficients is a piece
of research which follows directly. More interesting,
however, is to see whether the kind of simplifications
we see here are also manifest themselves in the computation of
perturbative amplitudes.

Thus, whilst the original motivation for this thesis
was simply to provide the groundwork for a formalism with
a view to  applying it to the non-perturbative study of
QCD, this emphasis has changed somewhat. In retrospect,
what this work has achieved is not only to demonstrate the consistency
of the \MGIERG, but also to go some way towards illuminating
its perturbative structure. This latter point has proven interesting
in its own right, thereby extending the usefulness of this work
and going beyond its original aims.

\chapter{Structure of Thesis}

This thesis is arranged into four parts.
The first part details the setup and breaks
down into two chapters. Chapter~\ref{ch:Intro}
introduces the work on which this thesis is based.
In particular, the Exact Renormalisation Group Equation
(\ERGE) of~\cite{YM-1-loop} is introduced, and its 
diagrammatic interpretation described.
In chapter~\ref{ch:NewFlowEquation} 
we modify the flow equation for subsequent
applications detailed in this thesis and
introduce new diagrammatics.

The second part details techniques developed for the new
flow equation. In chapter~\ref{ch:FurtherDiagrammatics}
we describe a new set of diagrammatic techniques. Their
application is illustrated
in chapter~\ref{ch:beta1} by applying them to the computation
of $\beta_1$. This allows us to reduce $\beta_1$ to a
set of terms differentiated \wrt\ the \ERG\ scale,
$\Lambda$.

In chapter~\ref{ch:LambdaDerivatives:Methodology}
we show how to evaluate these terms, developing the
formalism to the level suitable for extraction of $\beta_2$.

In the third part we investigate how the 
reduction of $\beta_1$ to a set of $\Lambda$-derivative
terms can be greatly simplified. We introduce and develop
the methodology in chapter~\ref{ch:bn:LambdaDerivsI}. 
This
guides us to a formula which automatically generates
(essentially) the $\Lambda$-derivative terms for a $\beta$-function
coefficient of arbitrary order, up to two types of unprocessed
terms. First, there are the `gauge remainders' which we  treat
in chapters~\ref{ch:bn:GRs}--\ref{ch:bn:Iterate}. Secondly,
there are the `$\Op{2}$' terms. A full treatment of
these is beyond the scope of this thesis, but we comment on
them in chapter~\ref{ch:Op2}. 

The fourth and final part is devoted to the computation of
$\beta_2$ and our conclusions. In chapter~\ref{ch:bulk},
we use the results of chapter~\ref{ch:bn:Iterate} to help us write
down an expression for $\beta_2$ in which all terms are 
differentiated \wrt\ $\Lambda$ (there is an additional set of terms,
the `$\alpha$-terms', which vanish in a certain limit). 
In chapter~\ref{ch:TotalLambdaDerivatives:Application}
we numerically evaluate this expression, using the results
of chapter~\ref{ch:LambdaDerivatives:Methodology}.
Finally, in chapter~\ref{ch:Conclusion}, we conclude.

During a first reading of this thesis, it is recommended
that the third part is omitted. As far as the rest
of the thesis is concerned, the main result of this part is to allow
us to jump straight to an expression for $\beta_2$ in
terms of (essentially) $\Lambda$-derivative diagrams.
It should be noted that this expression can, in
principle, be obtained with the methodology of part~II and
the diagrammatic identities introduced in part~III, alone.

The work of chapter~\ref{ch:Intro} is based on
existing, published work. The modification
of the flow equation of~\cite{YM-1-loop},
done in chapter~\ref{ch:NewFlowEquation}, 
is based on unpublished work by my collaborators,
Tim Morris and Stefano Arnone. 
The remaining work of chapter~\ref{ch:NewFlowEquation} 
is a mixture of
unpublished work by myself and Tim Morris.
All the remaining work is
my own, with the exception of that in section~\ref{sec:TLD:alpha},
which is based on unpublished work by Tim Morris.

\mainmatter

\part{Setup}

\chapter{Introduction}	\label{ch:Intro}

The history of the Exact Renormalisation Group dates
back to the seminal paper of  Wilson and Kogut~\cite{Wilson}
and to the contemporary work of Wegner and Houghton~\cite{W&H}.
Since then, it has gained a reputation as a powerful and
flexible technique for addressing a wealth of problems in
quantum field theory (\QFT) (see \eg~\cite{Wilson}--\cite{Litim:2003vp}).

In this thesis, we will be dealing with a particular
adaptation of the ERG for gauge theory~\cite{TRM-Faro}--\cite{FurtherWork}.
Our starting point, however, is the simpler scalar field
case~\cite{Wilson,W&H,Pol}, 
\cite{Wetterich:1993:Scalar}--\cite{scalar2loop} which, 
in section~\ref{sec:Intro:ScalarFT}, we use to illustrate some
general features of ERGs. This then allows a transparent
generalisation to the gauge theory case, done in section~\ref{sec:Intro:GIERG}.

\section{An ERG for Scalar Field Theory}	\label{sec:Intro:ScalarFT}

Though the original work on the ERG~\cite{Wilson,W&H} was
done in scalar field theory, we will utilise the equivalent
but simpler analysis due to Polchinski~\cite{Pol}. In all
that follows, we work in $D$ dimensional Euclidean space.

\subsection{The Polchinski Equation}

Given a scalar field theory, parametrised by $\phi$, we introduce
two scales, by hand. The first is the bare cutoff, $\BS$; physically, this
can be taken to represent the scale at which our \QFT\ 
ceases to be a good description of the phenomena we are trying to describe.
Secondly, we introduce a much lower scale, $\ES$,
henceforth referred to as the effective cutoff.

In scalar field theory, this procedure 
is essentially trivial, since there is no symmetry
forbidding the na\"ive implementation of a cutoff. Much
of the subtlety in the gauge theory case is due to the
fact that precisely the opposite of this is true 
(see section~\ref{sec:Intro-GI-ERG-Reg}).

The next step is to integrate out degrees of freedom between $\BS$ and $\ES$. 
The effect of this
procedure is to encode the information carried by the high energy
modes into the effective action, $S_\Lambda \equiv S$~\cite{Wilson, W&H, Pol}, \cite{Bagnuls:2001-Rep}--\cite{TRM-elements}.
Rather than supposing that the remaining modes are sharply cutoff
at $\Lambda$, a general cutoff function $c_p \equiv c(p^2/\ES^2)$ is introduced,
where $p$ denotes the momentum of some mode. Of $c_p$, we demand the following:
\begin{enumerate}
	\item	that it dies off quickly enough as its argument tends to infinity;
			this ensures that our theory is properly regulated;
	
	\item	that $c(0) = 1$, ensuring that the IR behaviour of our theory
			is unaffected by the precise details of the UV cutoff.
\end{enumerate}

The central ingredient for all that now follows  is
the Exact Renormalisation Group Equation (ERGE), or flow equation, which describes how
the effective action varies with $\Lambda$. 
To enable us to write this down we introduce the 
`seed action', $\hat{S}$, which we take to be equal to the regularised
kinetic term~\cite{Scalar-1-loop},
\[
\hat{S} = \frac{1}{2} \partial_\mu \phi \cdot c^{-1} \cdot \partial_\mu \phi,
\]
where, for the functions $f(x)$ and $g(y)$ and a momentum space kernel $W(p^2/\Lambda^2)$,
\begin{equation}
f \cdot W \cdot g = \int \!\! \int d^Dx \, d^Dy \, f(x) W_{xy} \, g(y),
\label{eq:f.W.g}
\end{equation}
with 
\begin{equation}
W_{xy} \equiv \IntB{p} W(p^2/\Lambda^2)e^{ip.(x-y)}.
\label{eq:W_xy}
\end{equation}

Defining $\Sigma = S-2\hat{S}$, the Polchinski equation is, up to a discarded vacuum 
energy~\cite{Pol, TRM-ERG+ApproxSolns, TRM-elements,Scalar-1-loop},
\begin{equation}
\flow{S} = -\frac{1}{\Lambda^2} \fder{S}{\phi} \cdot c' \cdot \fder{\Sigma}{\phi} + \frac{1}{\Lambda^2} \fder{}{\phi} \cdot c' \cdot \fder{\Sigma}{\phi},
\label{eq:pol}
\end{equation}
where a primed function is differentiated \wrt\ its argument.
For what follows, it is useful to recast this equation. First,
we define
\begin{equation}
	\dot{X} = -\flow{X}. 
\label{eq:dot(X)-OLD}
\end{equation}
Secondly, we introduce a new function,
\begin{equation}
	\Delta_p \equiv \frac{c_p}{p^2},
\label{eq:scalar-effprop}
\end{equation}
and now rewrite equation~(\ref{eq:pol}) as follows:
\begin{equation}
\dot{S} = \frac{1}{2} \fder{S}{\phi} \cdot \dot{\Delta} \cdot \fder{\Sigma}{\phi} - \frac{1}{2} \fder{}{\phi} \cdot \dot{\Delta} \cdot \fder{\Sigma}{\phi}.
\label{eq:pol2}
\end{equation}

This recasting is done in anticipation of the crucial role to
be played by $\Delta_p$. We notice from equation~(\ref{eq:scalar-effprop})
that this object is, in our scalar example at any rate, just the 
regularised propagator~\cite{TRM-ERG+ApproxSolns, Scalar-1-loop, scalar2loop}. 
That the regularised propagator appears in 
the flow equation is
not an automatic (or necessary) feature of an ERGE; in scalar field theory,
it generically\footnote{As opposed to in the special case of the Polchinski equation.} 
occurs as a consequence of a choice we are free to make for the solutions
of the flow equation~\cite{Scalar-1-loop, scalar2loop}. 
In Yang-Mills theory, things are more subtle. When we adapt our framework
do deal with Yang-Mills theory (see section~\ref{sec:Intro:YM}) 
we find that we need never fix the gauge~\cite{GI-ERG-I, GI-ERG-II}
and so cannot even define a propagator.
Nonetheless,
in this case, we will find that there is an object which looks
similar to and plays a role close to that of a regularised  propagator,
which does appear in the flow equation. Thus, we refer to $\Delta_p$ as an effective
propagator, mindful that we must not take its relationship to a \emph{bona fide}
propagator for granted.

\subsection{Properties of the Polchinski Equation}

There are four vital properties of the Polchinski equation,
which play a central role in extending the formalism to
encompass Yang-Mills theory:
\begin{enumerate}
	\item	All ingredients of the flow equation are infinitely differentiable in momenta. This property,
			referred to as quasilocality~\cite{TRM-MassiveScalar}, guarantees that
			each RG step $\Lambda \mapsto \Lambda - \delta \Lambda$
			does not generate IR singularities~\cite{Wilson}.

	\item	The flow corresponds to integrating out degrees of freedom;
			this simply follows from the derivation of the Polchinski
			equation given in~\cite{TRM-elements}.

	\item	The partition function is left invariant under the flow, since $e^{-S}$ transforms into a
			total derivative:
			\begin{equation}
				\flow e^{-S} = \frac{1}{2} \fder{}{\phi} \cdot \dot{\Delta} \cdot \left(\fder{\Sigma}{\phi} e^{-S} \right).
			\label{eq:InvariantUnderFlow}
			\end{equation}
			This ensures that, starting with some QFT at a high scale, 
			our effective action is guaranteed to still be describing the same physics
			at a very much lower scale (implicitly assuming the first point, above).

	\item	The flow is self-similar~\cite{SelfSimilarFlow}. This demands that the effective
			action depends only on the scale $\Lambda$ (this can be straightforwardly extended
			to massive theories~\cite{TRM-MassiveScalar}), and thus is a 
			`perfect action'~\cite{PerfectActions}. By definition, such actions lie on a
			renormalised trajectory guaranteeing, amongst other things, the existence of
			a continuum limit~\cite{TRM-elements}.

\end{enumerate}

An important consequence of these points is that locality, a property which is
generally not manifest in the Wilsonian effective action at some finite value of
$\Lambda$, is guaranteed~\cite{YM-1-loop}. Due to invariance of the partition function under the flow
(and the fact that the flow is free of IR divergences), we know that,
if we are dealing with a local action at some scale, then we must 
have been dealing with a local action at all higher scales and are
guaranteed to be dealing with a local action at all lower scales.
Now, we know that the flow lies on a
renormalised trajectory and that this trajectory is controlled by
the free-field (\aka\ Gaussian) fixed point. To obtain a continuum
limit, the flow is tuned such that it passes arbitrarily close
to the free-field fixed point. At this point, the action---being
that of a non-interacting field theory---is local. Hence, locality
is guaranteed at all other points along the flow.

It is worth commenting on the end of the renormalised 
trajectory, where $\ES \rightarrow \infty$.
The action here is just the bare action, whose form is determined by
the flow equation, but whose precise details amount to choices we are free to make.
As is generally our philosophy, we leave these choices either unmade or
at least implicit (see particularly section~\ref{sec:TLD:alpha}).

The crucial point recognised in~\cite{GI-ERG-I} is that the Polchinski equation is just
a special case of an equation with the above features; we now take these features
to \emph{define} what we mean by an ERGE. This idea is more rigorously 
explored in~\cite{TRM+JL-0,TRM+JL},
where it was realised that these generalised ERGs can be parametrised by a functional, 
$\Psi$, which induces a field redefinition along the flow.

With this in mind, we now make a non-trivial modification to the
Polchinski equation: we desist from restricting the seed action
to being just the kinetic term and allow it to become
general, insisting only on the following requirements:
first, that the vertices of $\hat{S}$ be infinitely differentiable and
secondly that $\hat{S}$ leads to convergent momentum integrals~\cite{Scalar-1-loop,scalar2loop,YM-1-loop}.
These properties which, generally speaking, we assume to be implicitly
fulfilled, are necessary for the consistency of the quantum theory (\ibid).

Having made this modification, we must now check that our
new flow equation satisfies the requisite requirements.

 From equation~(\ref{eq:InvariantUnderFlow}) it is clear
that the partition function will still be left invariant under the flow.
That the flow equation corresponds to integrating out can be argued as follows~\cite{GI-ERG-I}:
\begin{enumerate}
	\item	we can ensure that the flow equation is regularised, so all momentum
			integrals are bounded;

	\item	$\Lambda$ is the UV cutoff: momenta larger than some scale $q$ must vanish in the limit $q/\Lambda \rightarrow \infty$;

	\item	as $\Lambda \rightarrow 0$ all remaining contributions from any non-vanishing momentum scale disappear;

	\item	\emph{but} the physics is invariant under the flow.
			
	\item	Contributions from a given momentum scale must, therefore, be encoded in the effective action: we have integrated out!
\end{enumerate}

The self similarity of the flow is unchanged: the flow equation is written
only in terms of renormalised quantities, at the scale $\Lambda$.
Thus, as our new flow equation satisfies the requisite requirements, it is 
perfectly valid~\cite{Scalar-1-loop, scalar2loop, GI-ERG-I, YM-1-loop}. 

It is, of course, an obvious question as to what we have gained from this. Primarily,
this generalisation is necessary if we wish to be able to treat Yang-Mills theory,
as we will see in section~\ref{sec:Intro:YM}. However, the generalisation is
also of importance in its own right, as we now discuss.

The basic idea is that we do not expect the generalisation of
the seed action to change the result of a computation of some universal
quantity. Whilst this follows from the validity of the new flow equation,
there is also a useful physical interpretation: the freedom of $\hat{S}$
represents the continuum version of the choice of blocking 
transformations in the application of Wilsonian RG techniques to latticised
problems~\cite{Wilson, Scalar-1-loop}.

That a calculation must yield a result independent of the precise details
of many of the ingredients (we leave the cutoff functions general, too)
leads to a highly constrained calculational procedure. Generally speaking,
the only way to remove dependence on an instance of an unspecified 
(non-universal) component
of $\hat{S}$ is for there to be a second instance, of opposite sign. Indeed,
this is so constraining that calculations can be performed almost entirely
diagrammatically; in this way,
the (universal) scalar one-loop and two-loop $\beta$-functions
have
been correctly evaluated~\cite{Scalar-1-loop, scalar2loop}, as has the
one-loop Yang-Mills $\beta$-function~\cite{YM-1-loop}. One of the major results
of this thesis is a computation of the two-loop Yang-Mills $\beta$-function.

Now, whilst the freedom to leave the seed action largely unspecified has guided
us to an efficient calculational procedure we have, in some sense, complicated
the issue. Certainly, in scalar field theory where a general seed action is not
required, we have made life harder for ourselves by leaving it as general as
possible. The most obvious resolution to this is to simply regard the freedom
of the seed action as scaffolding: it has guided us to an efficient calculational
procedure, but now we can dispense with it, keeping the procedure but choosing
the simplest form for the seed action, consistent with our approach.

However, a consequence of one of the other major results of this thesis is that 
this point seems to be moot. Certainly, for the calculation of $\beta$-function
coefficients---\emph{to any order in perturbation theory}---it is possible to show that explicit dependence
on $\hat{S}$ is guaranteed to cancel out; indeed we give a compact expression
for the diagrammatic form of $\beta$-function coefficients, after
the cancellations have occurred (see part III).\footnote{Strictly,
what we present is an incomplete expression for Yang-Mills theory.
However, the scalar case is effectively contained within this
partial analysis.}
Thus, at least up until this stage of a
calculation, where any remaining seed action dependence is now implicit,
it makes no difference how complicated a seed action has been chosen; we
can simply bypass the entire procedure of cancelling explicit
instances of seed action components. Moreover, there are intriguing indications
that, at any loop order, even the implicit seed action dependence cancels out,
though the investigation of this will be saved for the future~\cite{FurtherWork}.

\section{The Manifestly Gauge Invariant Exact Renormalisation Group} \label{sec:Intro:GIERG}

With the review of ERGs for scalar field theory behind us,
we are now in a position to generalise the formalism for
the application to Yang-Mills theory. The central
issue is how to reconcile non-Abelian gauge theory
with the notion of a momentum cutoff, since a straightforward
implementation of the latter breaks the former. One
approach is to work in a gauge fixed formalism and
to allow this breaking to occur. The effects of the
breaking
can be kept track of with the `modified
Ward Identities' and it is hoped that they vanishes in the limit where
the momentum cutoff is removed~\cite{Bonini:1993}, \cite{Bonini:1993kt}--\cite{Litim:2002ce}.

The approach upon which this thesis is
based is one in which 
a regulator for Yang-Mills is built that
respects gauge invariance~\cite{SU(N|N)};
we review this in section~\ref{sec:Intro-GI-ERG-Reg}.\footnote{For
an alternative approach, based on the Vilkovisky-DeWitt~\cite{Vilkovisky:1984st,DeWitt}
formalism, see~\cite{Veneziano,Pawlowski-VDW}.}
Then, in section~\ref{sec:Intro:YM} we detail the
construction of a regulated, (manifestly) gauge invariant
flow equation. Some important properties of the flow
equation are discussed in 
sections~\ref{sec:Intro:Diagramatics}--\ref{sec:Intro:WeakCoupling}.
Much of the material in~\cite{YM-1-loop} is repeated in this
part of the thesis, as it forms the basis for everything that follows.
However, for the diagrammatic techniques in particular,
we try to simplify and clarify the exposition,
wherever possible.

\subsection{Regularisation} \label{sec:Intro-GI-ERG-Reg}

In this section, we review the regularisation of Yang-Mills
theory\footnote{Strictly speaking, this
regularisation is valid for $SU(N)$ Yang-Mills theory. Generalisation to
other gauge groups should be straightforward, though we make no such attempt to do so here.} 
introduced by Arnone \etal~\cite{SU(N|N)}. This
is a three-step procedure. First, we introduce the notion
of a gauge invariant cutoff function~\cite{GI-ERG-I}. However,
just as with gauge invariant higher derivative 
regularisation~\cite{Slavnov'sReg-A, Slavnov'sReg-B, HigherDerivReg-Falcetto} (to
which our approach is closely related),
this proves insufficient to completely regulate the theory~\cite{SU(N|N), One-loop-UVDivs}: 
certain
one-loop divergences survive. To cure these divergences, 
we introduce a set of gauge invariant Pauli-Villars (\PV)
fields. Whereas these were originally put in by hand~\cite{TRM-Faro, GI-ERG-I, GI-ERG-II}, 
it was later recognised that they arise in a natural way~\cite{pre-SU(N|N)-A, pre-SU(N|N)-B, SU(N|N)}.
Specifically, we embed our Yang-Mills theory in a certain graded Lie group which,
when broken in a particular manner, yields a set of heavy fields
which provide the necessary \PV\ regularisation. Finally,
a pre-regulator is required to consistently define the regularisation
scheme~\cite{SU(N|N)}.

Rather than following the historical approach of first introducing
the covariantized cutoff functions and then implementing the \PV\
regularisation, we will follow the approach of~\cite{SU(N|N)}, where the
embedding of $SU(N)$ gauge theory into the graded lie algebra, $SU(N|N)$,
is performed first. It is logical to proceed in this manner, not least because
the covariantised cutoff regularisation applies to the entire $SU(N|N)$
gauge theory, and not just the physical $SU(N)$ part. Indeed, this makes
it particularly clear that the problem of overlapping divergences,
commonly associated with \PV\ regularisation, is avoided in this
set-up: because the covariantised cutoff regularisation applies to all
fields, it applies to diagrams with external \PV\ fields, obviating
the problem.

\subsubsection{$SU(N|N)$ Regularisation}

A detailed discussion of the $SU(N|N)$ Lie superalgebra can be found in~\cite{Bars}.
However, we follow~\cite{SU(N|N)}, which focuses on the elements necessary
for our purposes.

We begin by looking at the graded Lie supergroup, $SU(N|M)$. The defining
representation is furnished by Hermitian $(N+M) \times (N+M)$ matrices,
\[
	\mathcal{H} = \TwoByTwo{H_N}{\theta}{\theta^{\dagger}}{H_M},
\]
where $H_N(H_M)$ is an $N\times N (M\times M)$ Hermitian matrix
with complex bosonic elements and $\theta$ is an $N\times M$ matrix
whose elements are complex Grassmann numbers.

The supertrace---which replaces
the trace as the natural cyclic invariant---is defined by
\[
	\str \mathcal{H} \equiv \tr \sigma \mathcal{H},
\]
where 
\[
	\sigma \equiv \sigma_3 =
	\left(
		\begin{array}{cc}
			\one_N	& 0 \\
			0		& -\one_N
		\end{array}
	\right).
\]

When expressing an element of the group in terms of
the generators and corresponding superangles,
we use the convention that only the superangles
carry Grassmann character. Hence, the generators
are taken to be Hermitian, complex valued matrices;
the superalgebra is thus constructed via both 
commutators and anticommutators.
The matrices that form the superalgebra of $SU(N|M)$ must
be supertraceless.

For what follows, it will be instructive to focus on
the bosonic part of the superalgebra. The traceless parts
of $H_N$ and $H_M$ furnish the defining representation
of the $SU(N)$ and $SU(M)$ groups, respectively. The
orthogonal, traceful part of $\mathcal{H}$ yields a $U(1)$.
Thus, the bosonic part of the $SU(N|M)$ superalgebra
forms an $SU(N)\times SU(M) \times U(1)$ subalgebra.

When we specialise $M=N$---as we now do---the $U(1)$ subalgebra
is generated by the identity, $\one_{2N}$. Note, though, that $SU(N|N)$
cannot be decomposed into the product of smaller algebras, since
the identity is generated by fermionic elements of the algebra. For example,
\[
	\anticom{\sigma_1}{\sigma_1} = 2\one_{2N},
\]
where
\[
	\sigma_1 = \TwoByTwo{0}{\one_N}{\one_N}{0}.
\]

We define the generators to be $S_\alpha \equiv \anticom{\one}{T_A}$,
where the rank of $\one$ is to be henceforth determined
by the context. The generator $\one$ has been separated out, in recognition
of the special role it will play. The remaining generators
are given by $T_A$, where $A$ runs over the $2(N^2-1)$
bosonic generators and the $2N^2$ fermionic generators.

An element of the $SU(N|N)$ superalgebra can thus be written
\begin{eqnarray*}
	\mathcal{H} & = & \mathcal{H}^\alpha S_\alpha = \mathcal{H}^0 \one + \mathcal{H}^A T_A,
\\
	\left( \mathcal{H} \right)^i_{\; j} & = & \mathcal{H}^0 \delta^i_{\; j} + \mathcal{H}^A \left( T_A \right)^i_{\; j}.
\end{eqnarray*}

The Killing supermetric is defined to be
\[
	h_{\alpha \beta} = 2 \str\left(S_\alpha S_\beta \right),
\]
where the generators are normalised such that
\begin{eqnarray}
	h_{\alpha \beta} & = &	\left(	
							\begin{array}{c|ccc|ccc|ccccc}
								0 	&&&&&&&&&&& 			\\ \hline
									& \ \; 1 &&&&&&&&&& 			\\
									&&  \ \; 1 &&&&&&&&& 			\\
									&&& \ddots &&&&&&&& 	\\ \hline
									&&&& -1 &&&&&&& 		\\
									&&&&& -1 &&&&&& 		\\
									&&&&&& \ddots &&&&& 	\\ \hline
									&&&&&&& 0 & i &&& 		\\
									&&&&&&& -i & 0 &&& 		\\
									&&&&&&&&& 0 & i & 		\\
									&&&&&&&&& -i & 0 		\\
									&&&&&&&&&&& \ddots		\\
							\end{array}
							\right). \label{eq:supermetric}
\\
						&&	\hspace{0.5em} \underbrace{\hspace{0.9em}}_{U(1)}\,
							\underbrace{\hspace{5.8em}}_{SU_1(N)}\,
							\underbrace{\hspace{6em}}_{SU_2(N)}\,
							\underbrace{\hspace{9em}}_{\mathrm{Fermionic}} \nonumber
\end{eqnarray}

The supermetric is not invertible, on account of the $U(1)$
generator commuting with everything. However, we can usefully
define the restriction of $h_{\alpha \beta}$ to the $T_A$ space:
\[
	g_{AB} = 2\str \left(T_A T_B \right) = h_{AB},
\]
which does have an inverse:
\[
	g_{AB}g^{BC}  = g^{CB}g_{BA} = \delta_A^C.
\]
This metric can be used to raise and lower indices:
\[
	X_A \equiv g_{A B} X^B \Rightarrow X^A = X_B g^{AB} \neq X_B g^{BA};
\]
the inequality following from~(\ref{eq:supermetric}).
The requirement that $X^A T_A = X_A T^A$ guides us to the dual
relations for raising and lowering indices on the generators:
\[
	T^A \equiv T_B g^{BA}.
\]
The completeness relation for the $T^A$ is
\begin{equation}
(T^A)^i_{\; j} (T_A)^k_{\ l} = \frac{1}{2} \delta^i_{\; l}\, (\sigma_3)^k_{\ j} 
									- \frac{1}{4N}\left[\delta^i_{\; j}\,(\sigma_3)^k_{\ l} + (\sigma_3)^i_{\; j}\,\delta^k_{\ l}\right].
\label{eq:Completeness}
\end{equation}

Having introduced the $SU(N|N)$ algebra, we now demonstrate
how we can utilise it to construct a set of regulating fields
for our physical $SU(N)$ gauge theory. The physical gauge theory
has 
connection $A^1_\mu(x) \equiv A^1_{a\mu} \tau_1^a$,
where $\tau_1^a$ are the generators of $SU(N)$, 
normalised to $\tr (\tau_1^a \tau_1^b) = \delta^{ab}/2$.

We now embed $A^1_\mu$ into the superfield $\SF_\mu \equiv \SF_\mu^\alpha S_\alpha$:
\[
	\SF_\mu = 
	\left(
		\begin{array}{cc}
			A_\mu^1 	& B_\mu 
		\\
			\bar{B}_\mu & A_\mu^2
		\end{array} 
	\right) + \SF_\mu^0 \one.
\]
$A^2$ is an unphysical copy of our original $SU(N)$ gauge  theory.
$B$ and $\bar{B}$ are unphysical, wrong statistics (\ie\ fermionic)
gauge fields; it is these that will acquire mass and provide \PV\
regularisation. Lastly, there is a central term, corresponding to the 
field $\SF^0_\mu$. In chapters~\ref{ch:NewFlowEquation} 
and~\ref{ch:FurtherDiagrammatics}
we will,
for the first time, arrive at a full understanding of the role that
this field plays, within the \MGIERG.

For much of this chapter,
it will prove useful to treat the block diagonal and block off-diagonal
components of $\SF$ together, which we
represent by $A$ and $B$, respectively. Whether $B$ refers
to both block-off diagonal components of just to the top-most one
depends on the context; we hope this is not confusing.

Construction of an $SU(N|N)$ gauge theory proceeds in the usual manner. First,
we define the covariant derivative
\begin{equation}
	\nabla_\mu = \partial_\mu - i \SF_\mu
\label{eq:Intro:CovarDeriv}
\end{equation}
where, for future convenience, we have scaled out the coupling $g$. The superfield
strength is now just $\mathcal{F}_{\mu\nu} = i[ \nabla_\mu, \nabla_\nu ]$. In 
anticipation of the covariantised cutoff regularisation, the kinetic
term will be of the form
\begin{equation}
	\str \mathcal{F}_{\mu\nu} \left(\frac{\nabla}{\Lambda}  \right)^n \cdot \mathcal{F}_{\mu\nu} 
\label{eq:KineticTerm}
\end{equation}
where the dot, in this context, tells us that $\nabla$ acts via commutation.

There are a number of important things to note about the kinetic term. The
first is that, since it is built out of commutators, it does not
have an $\SF^0$ component. Thus, if $\SF^0$, were to appear anywhere else in
the action, it would act as a Lagrange multiplier. To avoid the unwanted
constraints that this would entail, we demand that the action is independent of $\SF^0$.\footnote{
We cannot, however, simply get rid of $\SF^0$, since it is generated by (fermionic)
gauge transformations~\cite{SU(N|N)}. It is possible to modify the representation of the Lie bracket~\cite{Bars, SU(N|N)}
to remove $\SF^0$, but we do not pursue this any further as it seems to complicate
matters.
}
Hence, in addition to being invariant under supergauge transformations, parameterised
by $\Omega$, the action is also invariant under `no-$\SF^0$' 
symmetry~\cite{SU(N|N), YM-1-loop, Antonio'sThesis}: 
$\delta \SF^0_\mu(x) = \lambda_\mu(x)$. Hence, the action is invariant under the
shift
\begin{equation}
	\delta \SF_\mu = \nabla_\mu \cdot \Omega + \lambda_\mu \one.
\label{eq:Intro:GT}
\end{equation}

The second point about the kinetic term~(\ref{eq:KineticTerm}) is that the supertrace
structure causes $A^2$ to come with the wrong sign action,
leading to a violation of unitarity~\cite{SU(N|N)}. We will return to this
issue after a discussion of the breaking of $SU(N|N)$, to which we now turn.

To break $SU(N|N)$ down to its bosonic subgroup $SU(N) \times SU(N) \times U(1)$,
we introduce the superscalar field $\SH$ which 
transforms under the adjoint representation of $U(N|N)$~\cite{SU(N|N)}:
\begin{equation}
	\SH =
	\left(
		\begin{array}{cc}
			C^1		& D
		\\
			\bar{D}	& C^2
		\end{array}
	\right).
\label{eq:SH-defn}
\end{equation}
Note that we have not factored out $\SH^0$: $C^1$, unlike $A^1$,
is not identified with a physical field and so it is not an
inconvenience to allow it to 
contain a $\SH^0$ component.
Under supergauge transformations,
\[
	\delta \SH = -i [\SH, \Omega].
\]

The idea now is to allow $\SH$ to develop a vacuum expectation value along
the $\sigma$ direction, thereby spontaneously breaking the $SU(N|N)$
invariance, via the Higgs mechanism. The fermionic fields, $B$ and $D$,
acquire masses of order $\Lambda$. If we were to work in the unitarity gauge,
then  $B$ eats the Goldstone fermion, $D$. However, we do
not fix the gauge and so these two fields mix, under gauge transformations.
This observation underpins the diagrammatic techniques of~\cite{YM-1-loop}
and, as a consequence, is of vital importance to much of the work
that follows.

One of the necessary properties that our regularisation scheme must obey
is that any unphysical regulating fields decouple, in the limit that
the symmetry breaking scale tends to infinity. The fermionic fields 
and the bosonic components of the superhiggs field
can
be given an arbitrarily high mass, decoupling as required. The field
$A^2$, on the other hand, remains massless. However, it only communicates
with the physical $A^1$ field via the heavy fermionic fields; the lowest
dimension effective interaction,
\[
	\frac{1}{\Lambda^4}\;  \tr \left(F_{\mu \nu}^1 \right)^2 \tr \left(F_{\mu \nu}^2 \right)^2,
\]
 is irrelevant guaranteeing that,
if our theory is renormalisable, then $A^2$ will decouple, courtesy
of the Appelquist-Carazzone theorem~\cite{A&C}. Once our covariant cutoff
regularisation is in place, our theory will indeed be renormalisable, so 
long as $D \leq 4$. Thus, in the limit that $\Lambda \rightarrow \infty$,
we recover our physical, unitary, $SU(N)$ gauge theory.

We conclude this section by describing how to perform functional derivatives
\wrt\ $\SF$ and $\SH$. This is straightforward in the latter case:
$\SH$ is an unconstrained superfield ($\SF$, on the other hand, must be
supertraceless which leads to complications) and so~\cite{GI-ERG-II, YM-1-loop, SU(N|N)}:
\[
	\fder{}{\SH} \equiv
	\left(
		\begin{array}{cc}
			
			\ds \fder{}{C^1}	& \ds -\fder{}{\bar{D}}
		\\ [2ex]
			\ds \fder{}{D}		& \ds -\fder{}{C^2}
		\end{array}
	\right).
\]

Following~\cite{GI-ERG-II, YM-1-loop} we introduce the constant
supermatrices $X$ and $Y$ and, neglecting the irrelevant spatial 
dependence,  schematically represent
first supersowing, whereby two supertraces are sown together:
\begin{equation}
	\pder{}{\SH} \str \SH Y = Y \ \ \Longrightarrow \ \ \str X \pder{}{\SH} \str \SH Y = \str XY,
\label{eq:SuperSowing-C}
\end{equation}
and secondly supersplitting, whereby a single supertrace is split into two supertraces:
\begin{equation}
	\str \pder{}{\SH} X \SH Y = \str X \str Y.
\label{eq:SuperSplitting-C}
\end{equation}

For $\SF$, however, we define~\cite{SU(N|N)}
\begin{equation}
	\fder{}{\SF_\mu} \equiv 2T_A \fder{}{A_{A \mu}} + \frac{\sigma}{2N} \fder{}{\SF^0_\mu},
\label{eq:fder-SF}
\end{equation}
where $T_A$, as before, are the traceless generators of $SU(N|N)$
\ie\
\[
	\SF_\mu = \SF_\mu^0 \one + \SF_\mu^A T_A.
\]

It is not surprising that, in this context, we do not have exact
supersowing and supersplitting as we had above; rather, there will
be corrections,
\begin{eqnarray}
	\str X \pder{}{\SF} \str \SF Y	& = & \str XY - \frac{1}{2N} \str X \tr Y \label{eq:SuperSowing-A}
\\ [1ex]
	\str \pder{}{\SF} X \SF Y 		& = & \str X \str Y - \frac{1}{2N} \tr XY \label{eq:SuperSplitting-A},
\end{eqnarray}
which follow from the completeness relation~(\ref{eq:Completeness}).

Under supergauge transformations, the functional derivatives transform as
follows~\cite{YM-1-loop}:
\begin{eqnarray}
	\delta \left(\fder{}{\SH} \right)		& =	& -i \comm{\fder{}{\SH}}{\Omega}
\label{eq:Intro:GT-ddC}
\\[1ex]
	\delta \left(\fder{}{\SF_\mu} \right)	& =	& -i \comm{\fder{}{\SF_\mu}}{\Omega} + i \frac{\one}{2N} \tr \comm{\fder{}{\SF_\mu}}{\Omega}.
\label{eq:Intro:GT-ddA}
\end{eqnarray}
Notice that $\delta / \delta \SF$ does not transform homogeneously:
there is a correction term which ensures that $\delta / \delta \SF$
remains traceless. This proves important when we come to construct
our flow equation: the structure of the flow equation must be
such that the correction term vanishes.

\subsubsection{Covariant Cutoff Regularisation}		\label{sec:Intro:CovCutoffRegn}

The key to incorporating a covariantised cutoff
is the appropriate generalisation of equations~(\ref{eq:f.W.g})
and~(\ref{eq:W_xy}). Thus, we consider replacing $f$ and $g$
with two supermatrix representations, $u$ and $v$. Knowing
that our aim is to create a supergauge invariant, we use this
to define what we mean by the covariantisation of the kernel,
$W$~\cite{GI-ERG-I, YM-1-loop}:
\begin{eqnarray}
\lefteqn{
		u\left\{W\right\}_\SF v = 
		\sum_{n=0}^{\infty} \Int{x}\volume{y}\volume{x_1}\!\!\cdots\volume{x_n} 
		\overline{W}_{\mu_1\cdots\mu_n}(x_1,\ldots,x_n;x,y)
	} \nonumber
\\ 
	&	& \hspace{6em} \str \left[u(x)\SF_{\mu_1}(x_1)\cdots \SF_{\mu_n}(x_n) \cdot v(y) \right] \hspace{6em}
\label{eq:Intro:Wine:Commutator}
\end{eqnarray}
where, without loss of generality, we can demand that
$u \left\{ W \right\}_\SF v = v \left\{ W \right\}_\SF u$. The subscript $\SF$ 
tells us that the covariantisation involves only the supergauge field, $\SF$;
we will involve the superscalar $\SH$, shortly.
The $\overline{W}_{\mu_1\cdots\mu_n}(x_1,\ldots,x_n;x,y)$ are called the vertices
of the `wine' $\left\{ W \right\}_\SF$.
Each of the $\SF$s in~\eq{eq:Intro:Wine:Commutator} acts via commutation
on the expression to its right, as indicated by
the dot between $\SF_{\mu_n}$ and $v(y)$ .\footnote{In determining what, precisely,
each $\SF$ acts on, we
arrange for the argument of the (cyclically invariant) 
supertrace to be of the form
indicated in~\eq{eq:Intro:Wine:Commutator} \ie\ if the fields have
been cycled around, we bring them back to the form of~\eq{eq:Intro:Wine:Commutator}.}

By demanding this commutator structure, we ensure that the wines
respect no-$\SF^0$ symmetry. Moreover, since $\one$ commutes
with everything, if $v = \one g(y)$ for all $y$, only
the term corresponding to the zero-point wine survives:
\begin{equation}
	u \left\{ W \right\}_\SF \one g(y) = (\str{u})\cdot W \cdot g.
\label{eq:CoincidentLine}
\end{equation}

This identity is of particular importance. Indeed, in section~\ref{sec:Intro:YM}, following~\cite{YM-1-loop}, 
we will show that it is actually
required if our flow equation is to be supergauge invariant.
The relationships between the wine vertices arising from supergauge
invariance are discussed in sections~\ref{sec:Intro:BrokenPhase} 
and~\ref{sec:GR-Wines}.

Rather than working with the wine vertices, $\overline{W}$, of~\eq{eq:Intro:Wine:Commutator},
it is more convenient to expand out the commutators. Doing so, we
relabel the supergauge fields that appear on the 
\rhs\ of $v(y)$ in terms of $\SF_{\nu_i}(y_i)$~\cite{GI-ERG-II}:
\begin{eqnarray}
	\lefteqn{
		u\left\{W\right\}_\SF v = 
		\sum_{m,n=0}^{\infty} \Int{x}\volume{y}\volume{x_1}\!\!\cdots\volume{x_n}\volume{y_1}\!\!\cdots\volume{y_m} 
	} \nonumber 
\\
	& 	& W_{\mu_1\cdots\mu_n,\nu_1\cdots\nu_m}(x_1,\ldots,x_n;y_1,\ldots,y_m;x,y) \nonumber 
\\
	& 	& \str \left[u(x)\SF_{\mu_1}(x_1)\cdots \SF_{\mu_n}(x_n)v(y)\SF_{\nu_1}(y_1)\cdots \SF_{\nu_m}(y_m) \right].
\label{eq:Wine}
\end{eqnarray}
In the case where both $m,n=0$,
we drop the arguments $x_1 \cdots x_n$ and $y_1 \cdots y_m$. The resulting vertex,
which we denote simply by $W_{xy}$, will commonly be referred to as a `zero-point' wine.
A wine vertex with an additional argument is a `one-point' wine, and so forth.

The new wine vertices, $W_{\mu_1\cdots\mu_n,\nu_1\cdots\nu_m}(x_1,\ldots,x_n;y_1,\ldots,y_m;x,y)$,
are related to the $\overline{W}$; from these relationships there follows a set
of identities between the various 
new vertices---the
`coincident line identities' of~\cite{GI-ERG-I} and~\cite{GI-ERG-II}. We
give these in momentum space:
\begin{eqnarray}
	 W_{\mu_1\cdots\mu_n,\nu_1\cdots\nu_m}(p_1,\ldots,p_n;q_1,\ldots,q_m;r,s) \nonumber
\\
	= (-)^m \sum_{\mathrm{interleaves}} W_{\lambda_1 \cdots \lambda_{m+n}} (k_1, \ldots, k_{m+n};;r,s),
\label{eq:Interleaves}
\end{eqnarray}
where the sum runs over all interleaves of the sequences $p_1^{\mu_1}, \ldots, p_n^{\mu_n}$
and $q_m^{\nu_m},\ldots,q_1^{\nu_1}$. In other words, the ordered 
sequence $k_1^{\lambda_1}, \ldots, k_{m+n}^{\lambda_{m+n}}$
comprises all arrangements of $p^\mu$s and $q^\nu$s such that 
the $p^\mu$s are ordered \wrt\ each other
and the $q^\nu$s are in reverse order.

For our purposes, we will require not~(\ref{eq:Interleaves}), but a
relationship which follows:\footnote{This relationship is most easily
understood diagrammatically---see figure~\ref{fig:CoincidentLine}.}
\begin{eqnarray}
	W_{\lambda_1 \cdots \lambda_{n}} (k_1, \ldots, k_{n};;r,s)  + W_{;\lambda_1 \cdots \lambda_{n}} (;k_1, \ldots, k_{n};r,s) \nonumber
\\	
	+ \sum_{m=1}^{n-1} W_{\lambda_1 \cdots \lambda_{m}, \lambda_{m+1} \cdots \lambda_{n}} (k_1, \ldots, k_{m};k_{m+1}, \ldots k_{n};r,s) = 0.
\label{eq:WineRelation}
\end{eqnarray}

To introduce $\SH$s is straightforward and can be done in the
most convenient manner. In~\cite{YM-1-loop} the strategy used
was to restrict instances of $\SH$s to both ends of a wine,
and demand that they act via commutation. Thus, the `full wine',
$\{W\}$ was defined
\begin{equation}
	u\{W\}v = u \left\{ W \right\}_\SF v - \frac{1}{4} \SH \cdot u  \left\{ W_m \right\}_\SF \SH \cdot v,
\label{eq:fullwine-Old}
\end{equation}
where $W_m(p,\Lambda)$ is just some new kernel.

We will essentially use equation~(\ref{eq:fullwine-Old}) but choose
to lift
the restriction that instances of $\SH$ occur only at the ends
of the wine. We do this because, as we will see in section~\ref{sec:NFE:NewDiags-II}, 
it is natural in the diagrammatic framework we develop. Moreover,
it will not increase the amount of work we have to do in an
actual calculation. Nonetheless,
we demand that these additional instances of $\SH$ also
act via commutation, meaning that~(\ref{eq:CoincidentLine}) is preserved
for the full wine and that~(\ref{eq:Interleaves}) is appropriately
generalised to include fields which are Lorentz scalars.

We should note that, in chapter~\ref{ch:NewFlowEquation},
the flow equation is modified such that there are terms
which generate instances of $\SH$ to which wines attach.
Though these $\SH$s are not decorations of the wine
in the sense we have just discussed, it is often useful
to treat them as such. However, these $\SH$s act via
anticommutation, rather than commutation. When treating
them as part of a wine, they can always be `unpackaged',
so that all remaining decorative $\SH$s do, indeed,
act via commutation.

When viewed as part of the wine, these new instances of
$\SH$ give rise to multi-supertrace components of the wine
which, in our current set up do not exist. Again, there
is nothing to stop us from introducing such terms in 
complete generality since, in the diagrammatic framework
of section~\ref{sec:NFE:NewDiags-II}, they are
incorporated  naturally.

In addition to modifying~(\ref{eq:Wine}) to include $\SH$s,
we will anticipate the form of the flow equation by supposing
that $u$ and $v$ involve functional derivatives. If these derivatives
are \wrt\ the superfields $Z_1$ and $Z_2$, then the structure
of the flow equations turns out to be such that we can label
the different wines by these fields. Hence, we have~\cite{YM-1-loop}:
\begin{eqnarray}
	\lefteqn{
		\fder{}{Z_1^c} \left\{W^{Z_1Z_2}\right\} \fder{}{Z_2^c} = 
		\sum_{m,n=0}^{\infty} \Int{x}\volume{y}\volume{x_1}\!\!\cdots\volume{x_n}\volume{y_1}\!\!\cdots\volume{y_m} 
	} \nonumber 
\\ [1ex]
	& 	& W^{X_1 \cdots X_n, Y_1 \cdots Y_m, Z_1 Z_2}_{\, a_1 \; \cdots \, a_n,\, b_1\, \cdots b_m}(x_1,\ldots,x_n;y_1,\ldots,y_m;x,y) \nonumber 
\\ [1ex]
	& 	& \str \left[\fder{}{Z_1^c(x)}X^{a_1}_1(x_1)\cdots X^{a_n}_n(x_n) \fder{}{Z_2^c(y)} Y^{b_1}_1 (y_1) \cdots Y^{b_m}_m(y_m) \right],
\label{eq:Wine-B}
\end{eqnarray}
where the superfields $X_i$, $Y_i$ and $Z_i$ are either $\SF$s or $\SH$s.
The indices $a_i$, $b_i$ and $c$ are, respectively,
equal to $\mu_i$, $\nu_i$ and $\rho$ in the case that the field
which they label is a $\SF$ and are null when they label a $\SH$.
The wine vertices of~(\ref{eq:Wine-B}) satisfy coincident line
identities~(\ref{eq:CoincidentLine}) and~(\ref{eq:Interleaves});
the commutators arising from the action of the superfields have
been expanded out.

We now have all the ingredients we need to regularise $SU(N)$ gauge theory,
via a (covariantised) cutoff~\cite{SU(N|N)}. Having embedded our gauge theory
in the supergauge field, $\SF$, we apply higher derivative
regularisation to the $\SF$ kinetic term, via the cutoff function $c$:
\[
	\frac{1}{2} \mathcal{F}_{\mu\nu} \left\{ c^{-1} \right\} \mathcal{F}_{\mu\nu}.
\]
Similarly, we can use the cutoff function $\tilde{c}$ to sufficiently
improve the UV behaviour of the superscalar kinetic term. In~\cite{SU(N|N)},
it was assumed that the asymptotic behaviour of the cutoff functions is
governed by a power law\footnote{There
is no reason to suspect that regularisation cannot be achieved with
different cutoff profiles.}:
\[
	c_p \sim \left(\frac{p^2}{\Lambda^2}\right)^{-r}, \ \ \tilde{c}_p \sim \left(\frac{p^2}{\Lambda^2}\right)^{-\tilde{r}}.
\]
It was then proven that, provided the following constraints are satisfied
\begin{equation}
	r -\tilde{r} > 1, \ \ \tilde{r} > 1,
\label{eq:CutoffConstraints}
\end{equation}
the physical $SU(N)$ gauge theory is, indeed, properly regulated in $D \leq 4$.

It should be noted that 
to unambiguously
define  contributions which are finite only by virtue of the \PV\ regularisation, a
pre-regulator~\cite{Slavnov'sReg-B, HigherDerivReg-Falcetto, PreReg-Warr, TRM-Faro} 
must be used, in $D=4$. To understand this, consider some
\PV\ regulated loop integral. Gauge invariance demands that such an integral be
invariant under shifting the loop momenta. However, such a shift can generate
a non-vanishing surface term. The effect of our pre-regulator is to discard such
terms; for the purposes of this thesis---where we can perform calculations in
$D = 4 -2 \epsilon$---we use dimensional regularisation to automatically
remove such contributions. We emphasise that, assuming we are investigating
phenomena that can be studied in an intermediate dimension,
there is no barrier to utilising dimensional regularisation as a 
pre-regulator, even non-perturbatively~\cite{SU(N|N)}. Interestingly, as we
will discuss in the conclusion, it seems that there may be a diagrammatic
recipe for implementing pre-regularisation,  which one might hope would be
applicable to phenomena for which one must strictly work in $D=4$.

Finally, it has been traditional to impose the following constraints~\cite{TRM-Faro, GI-ERG-II, YM-1-loop}:
\begin{equation}
	\fder{}{\SF_\mu} \left\{ W \right\} = 0,\ \ \fder{}{\SH} \left\{ W \right\} =0.
\label{eq:WBT(SF)}
\end{equation}
As we will see, when we introduce the diagrammatics for our flow
equation, this restriction forbids certain types of diagram---the diagrams
in which `a wine bites its own tail'~\cite{TRM-Faro, GI-ERG-I}. The
problem with these diagrams is that they are not properly UV regularised;
though our gauge theory \emph{is} properly regularised, this does not
necessarily imply that the flow equation is properly regularised. This
is a deep issue: the constraint~(\ref{eq:WBT(SF)}) can be implemented
by an implicit choice of covariantisation but, strictly, this requires
that we temporarily \emph{break} gauge invariance, restoring it only
after an appropriate limit is taken (\ibid). For much of this
thesis, we will use this prescription, and thus simply never draw the
diagrams in which the wine bites its own tail. However, the work
of part~III implies a more satisfactory resolution to this problem,
which is discussed in the conclusion.

\subsection{The Flow Equation}	\label{sec:Intro:YM}

Having introduced both the \ERG\ and a regularisation scheme
for $SU(N)$ Yang-Mills theory, based on a cutoff, we can now 
combine the two. The regularised,\footnote{Up to
the subtlety of wines biting their tails.} manifestly supergauge invariant flow
equation of~\cite{YM-1-loop} is:
\begin{equation}
\Lambda \partial_\Lambda S = -a_0[S,\Sigma_g] + a_1[\Sigma_g],
\label{eq:Flow:General}
\end{equation}
where
\begin{equation}
a_0[S,\Sigma_g] = \frac{1}{2} \fder{S}{\SF_\mu} \{\dot{\Delta}^{\SF \SF}\} \fder{\Sigma_g}{\SF_\mu}
	+  \frac{1}{2} \fder{S}{\SH} \{\dot{\Delta}^{\SH \SH}\} \fder{\Sigma_g}{\SH}
\label{eq:flow:old:a_0}
\end{equation}
and
\begin{equation}
a_1[\Sigma_g] = \frac{1}{2} \fder{}{\SF_\mu} \{\dot{\Delta}^{\SF \SF}\} \fder{\Sigma_g}{\SF_\mu}
	+  \frac{1}{2} \fder{}{\SH} \{\dot{\Delta}^{\SH \SH}\} \fder{\Sigma_g}{\SH}.
\label{eq:flow:old:a_1}
\end{equation}

We immediately note the similarity of the above to the scalar
equation~(\ref{eq:pol2}).
Superficially, the only differences are as follows. First, we have the two fields, $\SF$ and $\SH$,
introduced in a symmetric fashion, rather than the single field, $\phi$. Secondly, we have
implemented the covariantisation of the cutoff functions. However, there is a more subtle difference.
Recall that the scalar equation involves the functional $\Sigma = S - 2\hat{S}$. The above
equations, though, involve  a slightly different quantity,
\[
	\Sigma_g = g^2 S - 2\hat{S}.
\]

Let us examine where this difference comes from. Consider performing
a straightforward generalisation of the scalar equation to a
form suitable for Yang-Mills theory. This is most easily
done by using Abelian gauge theory, as an intermediate step~\cite{GI-ERG-I}.
In this case, there is no need to covariantise the cutoff
and so we simply let $\phi \rightarrow A_\mu$ in our scalar
flow equation, where
the covariant derivative is defined to be $D_\mu = \partial_\mu - i g A_\mu$.

However, when we generalise to non-Abelian gauge theory, it is
desirable to scale $g$ out of the covariant derivative \viz\
equation~(\ref{eq:Intro:CovarDeriv}). The exact preservation of
relationship~(\ref{eq:Intro:GT}) now implies that, given the form of
the covariant derivative~(\ref{eq:Intro:CovarDeriv}), $\SF$ cannot
renormalise: it is only $g$ that runs (\ibid).

Let us consider making the change $A_\mu \rightarrow A_\mu/g$
in our \ERG\ for Abelian gauge theory; we choose also to let 
$\hat{S} \rightarrow \hat{S}/g^2$. The flow equation is
\[
	\flow{S} + \frac{\beta}{g} \Int{x} A_\mu(x) \fder{S}{A_\mu(x)}= 
			-\frac{1}{2} \fder{S}{A_\mu}\cdot  \dot{\Delta} \cdot \fder{\Sigma_g}{A_\mu} 
			+ \frac{1}{2} \fder{}{A_\mu} \cdot \dot{\Delta} \cdot \fder{\Sigma_g}{A_\mu},
\]
where $\beta \equiv \flow g$. The presence of the second term on the left
is undesirable, since it destroys manifest gauge invariance:
gauge transformations generate a term containing
$\partial_\mu (\delta S / \delta A_\mu)$.
We can demonstrate
that this term vanishes, using gauge invariance,
but that it does so is not manifestly the case.

The solution is simply to drop the $\beta$-term (\ibid).
Thus, equations~(\ref{eq:Flow:General})--(\ref{eq:flow:old:a_1}) do not
represent a trivial generalisation of the scalar flow equation. However, our
non-Abelian flow equation
satisfies the necessary requirements 
and so is perfectly valid.

At this point it is worth pausing to repeat the argument of~\cite{YM-1-loop}
that our non-Abelian flow equation is actually gauge invariant. Invariance
of the $\delta / \delta \SH$ terms follows directly from the homogeneous
transformation of $\delta / \delta \SH$ (equation~(\ref{eq:Intro:GT-ddC})),
enforcing constraints on the wine vertices. These constraints 
will be discussed in sections~\ref{sec:Intro:BrokenPhase}
and~\ref{sec:GR-Wines}.

In the $A$-sector, however, matters
are complicated by 
the inhomogeneous transformation of
$\delta / \delta \SF$, given by equation~(\ref{eq:Intro:GT-ddA}).
Under a supergauge transformation, we can use equations~(\ref{eq:Intro:GT-ddA})
and~(\ref{eq:CoincidentLine}) to give:
\[
	\delta
	\left(	
		\fder{S}{\SF_\mu} \{\dot{\Delta}^{\SF \SF}\} \fder{\Sigma_g}{\SF_\mu}
	\right)
	= \frac{i}{2N} \tr \comm{\fder{S}{\SF_\mu}}{\Omega} \cdot \dot{\Delta}^{\SF \SF} \cdot \str \fder{\Sigma_g}{\SF_\mu} + (S \leftrightarrow \Sigma_g).
\]
By equation~(\ref{eq:fder-SF}) and no-$\SF^0$ symmetry,
\[
	\str \fder{\Sigma_g}{\SF_\mu} = \fder{\Sigma_g}{\SF_\mu^0} = 0
\]
and likewise for the term with $S \leftrightarrow \Sigma_g$. Gauge
invariance of the quantum term follows, similarly.

We emphasise that the demonstration of gauge invariance depends, crucially,
on two ingredients: equation~(\ref{eq:CoincidentLine}) and no-$\SF^0$
symmetry. We recall that equation~(\ref{eq:CoincidentLine}) is
intimately connected with the coincident line identities; we will
need to use this later, when examining gauge invariance in the broken
phase.

That our flow equation satisfies the requisite 
requirements ensures that two possible
sources for concern are alleviated. On the one hand, we have discussed
already that locality is guaranteed. On the other hand, we might wonder whether
it is \emph{really} $SU(N)$ Yang-Mills theory that we are describing at low energies;
after all, we expect a hugely complicated action depending on all
the component fields of $\SF$ and $\SH$. However, we know that the partition
function, and hence the physics we are describing, is invariant under the flow.
Since, in $D\leq 4$, the regulator fields decouple---as they must---as $\Lambda \rightarrow \infty$,
we are describing $SU(N)$ Yang-Mills theory at the top end of the renormalised trajectory,
and thus everywhere else along it.\footnote{As pointed out in~\cite{YM-1-loop}, asymptotic freedom
ensures that strong corrections do not affect 
the conclusions we draw as $\Lambda \rightarrow \infty$.}

The next point to make about the flow equation is the novel mass dimension of $\SH$:
whereas we might expect $\SH$ to come with mass dimension one, we choose instead for
it to have mass dimension zero~\cite{YM-1-loop}. The reason relates to properly implementing
the regularisation into the ERG. From~\cite{SU(N|N)}, the higher derivative regularisation
scale and the effective spontaneous symmetry breaking scale must be tied together,
both flowing with $\Lambda$. In general, this is not guaranteed within the \ERG,
but occurs only if the seed seed action is suitably constrained; these constraints
are particularly simple if $\SH$ is dimensionless.

Let us define $V(\SH)$  to be the effective superhiggs potential 
in $S$, minimised when $\SH$ acquires its effective vacuum expectation value 
$\bar{\SH}$. We also introduce a separate---and generally different---potential, $\hat{V}(\SH)$, for 
$\hat{S}$. Now, the nice thing about $\SH$ being dimensionless is
that, classically, we can insist that the minima of both $V$ and $\hat{V}$ lie at $\SH = \sigma$~\cite{YM-1-loop}.
This immediately implies that neither action, at the classical level, has a one-point $\SH$ vertex
in the broken phase.

Protecting $\bar{\SH} = \sigma$ from loop corrections can be done by
tuning $\hat{S}$. In section~\ref{sec:NFE:ND-II:App} we will see explicitly
how this is done.

The set-up of the flow equation is completed by defining $g$ through
a renormalisation condition. As in~\cite{GI-ERG-I,YM-1-loop}, $g$ is the
coupling of the physical $SU(N)$ field, $A^1$, and so:
\begin{equation}
	S[\SF = A^1, \SH = \bar{\SH}] =\frac{1}{2g^2} \, \tr \!\! \Int{x} \left(F_{\mu\nu}^1 \right)^2 + \cdots,
\label{eq:Intro:RenormCondition}
\end{equation}
where the ellipsis denotes higher dimension operators and ignored vacuum energy.

\subsection{Diagrammatics}	\label{sec:Intro:Diagramatics}

\subsubsection{Diagrammatics for the Action}

The flow equations have a particularly simple diagrammatic interpretation,
as a consequence of the supersowing and supersplitting relations~(\ref{eq:SuperSowing-C}),
(\ref{eq:SuperSplitting-C}), (\ref{eq:SuperSowing-A}) and~(\ref{eq:SuperSplitting-A}). To
describe these diagrammatics, we must first give a superfield expansion for the action. This
is simple: supergauge invariance demands that this expansion be in terms of supertraces and
products of supertraces~\cite{YM-1-loop}:
\begin{eqnarray}
	S	& =	& 	\sum_{n=1}^\infty \frac{1}{s_n} \Int{x_1} \!\! \cdots \volume{x_n} 
				S^{X_1 \cdots X_n}_{\,a_1 \, \cdots \, a_n}(x_1, \cdots, x_n) \str X_1^{a_1}(x_1) \cdots X_n^{a_n} (x_n)
	\nonumber \\
		& +&	\frac{1}{2!} \sum_{m,n=0}^{\infty} \frac{1}{s_n s_m} \Int{x_1} \!\!\cdots \volume{x_n}  \volume{y_1} \!\! \cdots \volume{y_m}
	\nonumber \\
		&	&
				\hspace{4em} S^{X_1 \cdots X_n, Y_1 \cdots Y_m}_{\, a_1 \, \cdots \, a_n \, , \, b_1\cdots \, b_m}(x_1, \cdots, x_n; y_1 \cdots y_m)
	\nonumber \\
		&	&	
				\hspace{6em} \str X_1^{a_1}(x_1) \cdots X_n^{a_n} (x_n) \, \str Y_1^{b_1}(y_1) \cdots Y_m^{b_m} (y_m)
	\nonumber \\
		&+	& \ldots \label{eq:Intro:ActionExpansion}
\end{eqnarray}
where, in the symmetric phase, the $X^{a_i}$ are $\SF_{\mu_i}$ or $\SH$ and the $Y^{b_j}_j$ are $\SF_{\nu_i}$ or $\SH$
and the vacuum energy is ignored. Due to the invariance of the supertrace under cyclic permutations of its
arguments, we take only one cyclic ordering for the lists $X_1 \cdots X_n$, $Y_1 \cdots Y_m$ in the sums over $n,m$.

Given that the arguments of each element of the lists $X_1 \cdots X_n$, $Y_1 \cdots Y_m$
are dummy arguments, we now want to check if any terms in our sums are invariant under
non-trivial cyclic permutations of the fields. The coefficients $s_n$ ($s_m$) represent
the order of any such cyclic subgroup. Thus, for the single supertrace term 
\[
	\frac{1}{s_n} S^{\SF \SH \SF \SH} \str \SF \SH \SF \SH,
\]
in which we have suppressed position arguments and Lorentz indices,
we take $s_n = 2$.

We write the momentum space vertices as
\begin{eqnarray*}
	\lefteqn{
		S^{X_1 \cdots X_n}_{\,a_1 \, \cdots \, a_n}(p_1, \cdots, p_n) \left(2\pi \right)^D \delta \left(\sum_{i=1}^n p_i \right)
	}
\\
	& &
	=
	\Int{x_1} \!\! \cdots \volume{x_n} e^{-i \sum_i x_i \cdot p_i} S^{X_1 \cdots X_n}_{\,a_1 \, \cdots \, a_n}(x_1, \cdots, x_n),
\end{eqnarray*}
where all momenta are taken to point into the vertex. We will employ the shorthand
\[
	S^{X_1 X_2}_{\, a_1 \, a_2}(p) \equiv S^{X_1 X_2}_{\, a_1 \, a_2}(p,-p).
\]

\begin{center}
\begin{figure}[h]
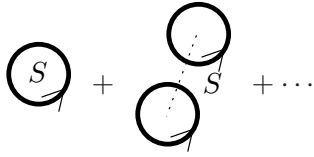

	\[
		\ensuremath{\begin{array}{c}\input{pstex/GeneralSingleSupertrace.pstex_t} \end{array}} +  \ensuremath{\begin{array}{c}\input{pstex/GeneralDoubleSupertrace.pstex_t} \end{array}} + \cdots
	\]
\caption{Diagrammatic representation of the action.}
\label{fig:Intro:Diagrammatics:Action}
\end{figure}
\end{center}

The diagrammatic representation of the action is given in figure~\ref{fig:Intro:Diagrammatics:Action}.

Each circle stands for a single supertrace containing any number of fields. The $S$
is to remind us that this is an expansion of the action, since the seed action $\hat{S}$
has a similar expansion.
The dotted line reminds us that the two supertraces in the double supertrace term are
part of the same vertex.
The arrows
indicate the cyclic sense in which fields should be read off; henceforth, this
will always be done in the counterclockwise sense and so these arrows will generally
be dropped.\footnote{Arrows can always be dropped in complete diagrams formed by
the flow equation. However, if we look at the diagrammatic representation of a wine,
in isolation, then it will be necessary to keep these arrows.} 
In turn, each of
these supertraces can now be expanded in terms of a sum over all explicit, cyclically
independent combinations of fields, as in figure~\ref{fig:Intro:Diagrammatics:SingleStr},
where closed circles represent $\SF$s and open circles represent $\SH$s.

\begin{center}
\begin{figure}[h]
	\[
		\ensuremath{\begin{array}{c}\input{pstex/GeneralSingleSupertrace-B.pstex_t} \end{array}} =
		\left[
		\ensuremath{\begin{array}{c}\input{pstex/Vertex-SH.pstex_t} \end{array}}
		+
		\scd[0.1]{Vertex-SFSF}{Vertex-SHSH}
		+
		\begin{array}{c}
			\ensuremath{\begin{array}{c}\input{pstex/Vertex-SFSFSF.pstex_t} \end{array}}	
		\\
			\ensuremath{\begin{array}{c}\input{pstex/Vertex-SFSFSH.pstex_t} \end{array}}
		\\
			\ensuremath{\begin{array}{c}\input{pstex/Vertex-SFSHSH.pstex_t} \end{array}}
		\\
			\ensuremath{\begin{array}{c}\input{pstex/Vertex-SHSHSH.pstex_t} \end{array}}
		\end{array}
		+\cdots
		\right.
	\]
\caption{Expansion of a single supertrace in powers of $\SF$ and $\SH$.}
\label{fig:Intro:Diagrammatics:SingleStr}
\end{figure}
\end{center}

In this case, we have chosen not to indicate whether the supertraces we
are considering come from the Wilsonian effective action, the
seed action, or some linear combination of the two.

Explicit instances of fields are referred to as decorations. Note that there are
no supertraces containing a single $\SF$, since $\str \SF =0$. The
 $\str \SF \SH$ vertex vanishes by charge conjugation invariance,
under which $\SF \rightarrow -\SF^T$ and $\SH \rightarrow \SH^T$~\cite{YM-1-loop};
the transpose of a supermatrix being defined as in~\cite{GI-ERG-II}:
\[
\mathrm{for} \ X = \TwoByTwo{X^{11}}{X^{12}}{X^{21}}{X^{22}}, \ \ 
	X^T \equiv \TwoByTwo{\left. X^{11} \right.^T}{-\left. X^{21} \right.^T}{\left. X^{12} \right.^T}{\left. X^{22} \right.^T}.
\]
The extra sign
in the top right element ensures that we can rewrite
\begin{equation}
	\str X^T Y^T \cdots Z^T = \str Z\cdots YX,
\label{eq:str-Transpose}
\end{equation}
since this sign compensates for the sign picked up upon the
anticommutation of fermionic fields in the above re-ordering.\footnote{
The definition of the transpose, though it ensures~(\ref{eq:str-Transpose}),
is not unique~\cite{GI-ERG-II}.}

As it stands, figures~\ref{fig:Intro:Diagrammatics:Action} and~\ref{fig:Intro:Diagrammatics:SingleStr}
provide representations of the action~(\ref{eq:Intro:ActionExpansion}). However, as we will
see shortly, it is often useful to interpret the explicitly decorated terms as just the
vertex coefficient functions $S^{X_1 \cdots X_n}_{\,a_1 \, \cdots \, a_n}(p_1, \cdots, p_n)$ etc.,
the accompanying supertrace(s) and cyclic symmetry factors having been stripped off.

\subsubsection{Diagrammatics for the Wines}

\begin{center}
\begin{figure}[h]
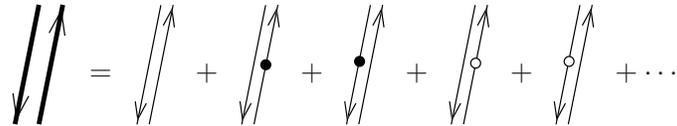

	\[
		\ensuremath{\begin{array}{c}\input{pstex/GeneralWine.pstex_t} \end{array}} = \ensuremath{\begin{array}{c}\input{pstex/Wine-0pt.pstex_t} \end{array}} + \ensuremath{\begin{array}{c}\input{pstex/Wine-SF-R.pstex_t} \end{array}} +\ensuremath{\begin{array}{c}\input{pstex/Wine-SF-L.pstex_t} \end{array}} + \ensuremath{\begin{array}{c}\input{pstex/Wine-SH-R.pstex_t} \end{array}} +\ensuremath{\begin{array}{c}\input{pstex/Wine-SH-L.pstex_t} \end{array}} +\cdots
	\]
\caption{Diagrammatic representation of the wines.}
\label{fig:Intro:Diagrammatics:Wines}
\end{figure}
\end{center}

The wines have a very similar diagrammatic expansion~\cite{GI-ERG-II, YM-1-loop}
to the action,
as indicated in figure~\ref{fig:Intro:Diagrammatics:Wines}.

\begin{center}
\begin{figure}[h]
	\[
		\ensuremath{\begin{array}{c}\input{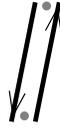} \end{array}} 
	\]
\caption{True diagrammatic representation of equation~(\ref{eq:Wine-B}).}
\label{fig:Intro:Diagrammatics:Wines-True}
\end{figure}
\end{center}

The ellipsis represents terms with any number of fields distributed in 
all possible ways between the two sides of the wine.
We should note that figure~\ref{fig:Intro:Diagrammatics:Wines} is not
strictly a representation  of equation~(\ref{eq:Wine-B}).
Equation~(\ref{eq:Wine-B}) involves not
only the wine vertices and the associated decorative fields, 
but also two functional derivatives which sit at the ends of
the wine. For the purposes of this chapter, we can just
directly include these functional derivatives in our
diagrammatics; they would act as unambiguous labels for
the different wines in the flow equation, as shown in
figure~\ref{fig:Intro:Diagrammatics:Wines-True}.

The grey circles can be either both $\SF$s or both $\SH$s;
since they sit at the end of the wine
they represent functional derivatives and label the wine.

We are almost ready to give a diagrammatic representation of
the flow equation, in the symmetric phase. Indeed, if
we were to allow only decorations by $\SH$, we could do
this immediately, by virtue of the exact supersowing 
relation~(\ref{eq:SuperSowing-C})
and exact supersplitting relation~(\ref{eq:SuperSplitting-C}).

\subsubsection{Diagrammatic Flow Equation in the $\SH$-sector}

For that which follows in this section, we assume all fields, including
those that have been differentiated, to be in the $C$-sector.
In this case, we note that the exact supersowing
relation applies to the classical part of the flow equation~(\ref{eq:flow:old:a_0}).
This is easy to see from equation~(\ref{eq:Wine}) (with the restriction to
decoration by $\SF$ removed) if we take $u = \delta S / \delta \SH$ and 
$v = \delta \Sigma_g / \delta \SH$. From the \lhs\ of equation~(\ref{eq:SuperSowing-C}),
both of these derivatives produce just a string of superfields; everything
is tied back into a supertrace by the covariantisation.

Similarly, the supersplitting relation
applies to the quantum part of the flow equation~(\ref{eq:flow:old:a_1}).
Thus, in this scenario, we can represent the flow equation as shown in 
figure~\ref{fig:Intro:OldFlow:SH-only}~\cite{YM-1-loop}; the ellipses
represent terms with additional supertraces.
\begin{center}
\begin{figure}[h]
	\[
		\dec{\ensuremath{\begin{array}{c}\input{pstex/Action.pstex_t} \end{array}} + \ensuremath{\begin{array}{c}\input{pstex/Action-DS.pstex_t} \end{array}} +\cdots}{\bullet} = 
		\frac{1}{2} 
		\left[
			\ensuremath{\begin{array}{c}\input{pstex/ClassicalTerm.pstex_t} \end{array}} - \ensuremath{\begin{array}{c}\input{pstex/QuantumTerm.pstex_t} \end{array}} - \ensuremath{\begin{array}{c}\input{pstex/WineBiteTail.pstex_t} \end{array}}
		\right] + \cdots
	\]
\caption{Diagrammatic representation of the flow equation, assuming that all fields are $\SH$s.}
\label{fig:Intro:OldFlow:SH-only}
\end{figure}
\end{center}

There are a number of important points to make about this figure. The first is that
the final diagram is the term for which the wine bites its own tail. As discussed
already, we will henceforth discard such terms.

Let us now focus
on the dumbbell-like structure, which represents the classical term.
The lobe at the top of the diagram (which encloses $S$) was spawned by
$\delta S / \delta \SH$. This derivative breaks $S$ open, by knocking out a
$\SH$, represented by the open circle sitting at the bottom of
the lobe.
The interpretation of the bottom lobe is much the same, though
this time it is $\Sigma_g$ that has been differentiated. The two broken supertraces
are now joined back together by the wine. Note that we need not put arrows on the wine:
the counterclockwise sense in which we read off fields is determined by the lobes.

At the moment, we are taking the \lhs\ of the equation to directly
represent the action---symmetry factors and all---and so we must
be quite careful to interpret figure~\ref{fig:Intro:OldFlow:SH-only}
in a way that is consistent with both figure~\ref{fig:Intro:Diagrammatics:Action}
and equation~(\ref{eq:Intro:ActionExpansion}). (This is not entirely
clear in either~\cite{GI-ERG-I} or~\cite{YM-1-loop}.)
We will shortly see that
there is a prescription which enables us to remove all symmetry factors,
as we would hope.

To be specific, suppose that the vertex
whose flow we are computing has $m$ supertraces, the first
of which is decorated by $s_1$ $\SH$s and the last of which is
decorated by $s_m$ $\SH$s. Hence, the overall symmetry factor is
\begin{equation}
	\frac{1}{m!\prod_{i=1}^m s_i}.
\label{eq:SymmetryFactor}
\end{equation}

Whereas the vertex whose flow we are computing comes
with this overall factor, the diagrams on the \rhs\ of
figure~\ref{fig:Intro:OldFlow:SH-only} come with
additional factors, due to the derivatives breaking
open $S$ and $\Sigma_g$.

Let us illustrate this with an example. We consider the classical
part of the flow
of all double supertrace terms with one $\SH$ on one supertrace
and two $\SH$s on the other. Thus we combine 
$S^{\SH \SH,\SH} \str \SH \SH \str \SH$
with $S^{\SH, \SH \SH}  \str \SH \str \SH \SH$. This has
the effect of killing the $1/m!$ in equation~(\ref{eq:SymmetryFactor}).

Figure~\ref{fig:Intro:ClassicalFlowEx}
depicts the flow, where the ellipsis denotes un-drawn terms
in which $\SH$s decorate the wine and the missing quantum term.
\begin{center}
\begin{figure}[h]
	\[
	\dec{
		\ensuremath{\begin{array}{c}\input{pstex/Vertex-SHSH-SH.pstex_t} \end{array}}
	}{\bullet}
	= \frac{1}{2} 
	\left[
		\begin{array}{l}
			4 \ensuremath{\begin{array}{c}\input{pstex/Dumbbell-SHx2-W-SHx2-SH.pstex_t} \end{array}} + 4 \ensuremath{\begin{array}{c}\input{pstex/Dumbbell-SHx2-SH-W-SHx2.pstex_t} \end{array}}
		\\
			+ 2 \ensuremath{\begin{array}{c}\input{pstex/Dumbbell-SHx2-W-SH-SHx2.pstex_t} \end{array}} + 2 \ensuremath{\begin{array}{c}\input{pstex/Dumbbell-SH-SHx2-W-SHx2.pstex_t} \end{array}}

		\\
			+2 \ensuremath{\begin{array}{c}\input{pstex/Dumbell-SHx2-DW-SHL-SH-SH.pstex_t} \end{array}} + \cdots
		\end{array}
	\right]
	\]
\caption{Example of the classical part of the flow.}
\label{fig:Intro:ClassicalFlowEx}
\end{figure}
\end{center}

Notice that on the \rhs, as with the \lhs, each $S^{\SH, \SH \SH}$
vertex is counted twice, since we do not care which of the supertraces
comes `first'. Again, this has the effect of killing the $1/m!$
coming from equation~(\ref{eq:SymmetryFactor}).

A factor of two arises each time we differentiate a supertrace containing two $\SH$s.
The final diagram has a factor of two due to the symmetry of the
double supertrace vertex: either of the $\str \SH$s could be differentiated,
with the same result.

Let us now consider changing the interpretation of the diagrammatics: we will
strip off the common supertrace structure and extract the symmetry factors $1/s_i$
and $1/m!$
from the vertices. Note that only the final diagram has a surviving
factor of $1/m!$.
This yields the diagrams of figure~\ref{fig:Intro:ClassicalFlowEx-B},
where we note that we have \emph{not} changed the diagrammatic notation
and so must be careful with our interpretation.
\begin{center}
\begin{figure}[h]
	\[
	\frac{1}{2}
	\dec{
		\ensuremath{\begin{array}{c}\input{pstex/Vertex-SHSH-SH.pstex_t} \end{array}}
	}{\bullet}
	= \frac{1}{2} 
	\left[
		\begin{array}{l}
			\ensuremath{\begin{array}{c}\input{pstex/Dumbbell-SHx2-W-SHx2-SH.pstex_t} \end{array}} + \ensuremath{\begin{array}{c}\input{pstex/Dumbbell-SHx2-SH-W-SHx2.pstex_t} \end{array}}
		\\
			+ \frac{1}{2} \ensuremath{\begin{array}{c}\input{pstex/Dumbbell-SHx2-W-SH-SHx2.pstex_t} \end{array}} + \frac{1}{2} \ensuremath{\begin{array}{c}\input{pstex/Dumbbell-SH-SHx2-W-SHx2.pstex_t} \end{array}}

		\\
			+ \ensuremath{\begin{array}{c}\input{pstex/Dumbell-SHx2-DW-SHL-SH-SH.pstex_t} \end{array}} + \cdots
		\end{array}
	\right]
	\]
\caption{Reinterpretation of figure~\ref{fig:Intro:ClassicalFlowEx}, where now the
diagrammatic elements comprise just the vertex coefficient functions and
the wine vertex coefficient functions.}
\label{fig:Intro:ClassicalFlowEx-B}
\end{figure}
\end{center}

This is almost what we want. 
Finally, to kill the factor of $1/2$ on the \lhs, we multiply through by two.
 Taking this factor inside the square brackets, the diagrams
with a factor of two are those for which the dumbbell structure has a total of two fields sitting
on the outside (we do not count the fields which have been hit by  derivatives). Thus,
to compute the flow of a \emph{vertex coefficient function}, we can use the following mnemonic,
for the classical term:
\begin{enumerate}
	\item	take an overall factor of $1/2$, corresponding to the $1/2$ associated
			with $a_0$ (see equation~(\ref{eq:flow:old:a_0}));

	\item	draw all dumbbell structures with the same implied supertrace structure as the
			vertex whose flow has been computed (remember, the actual supertraces have
			been stripped off) where, \emph{on the dumbbell structure, we sum over all
			cyclic permutations of the fields on the outside}.
\end{enumerate}

The latter point makes sense, since it is necessary to ensure that the 
vertex coefficient function whose flow we
are computing is cyclically invariant.

This recipe clearly works in our example: the diagrams with a relative factor of two are
precisely those for which we can cyclically permute the fields on the dumbbell structure.
In a very real sense, we can think of stripping off the supertraces as promoting the dummy
arguments carried by the fields to actual arguments. In this sense, the fields each effectively
carry a label. Thus, when we draw all dumbbell
structures, we count differently the various cyclic permutations of the fields.

Returning now to figure~\ref{fig:Intro:OldFlow:SH-only}, we look at the 
quantum term, which we note looks somewhat like a padlock.
The treatment of this term is much the same as with the classical term,
in that we can remove all symmetry factors by computing the flow
of vertex coefficient functions. However, things are complicated
by the fact that, unlike the dumbbell term, the quantum term has
a minimum of two supertraces (the inner circuit is not connected to the
outer circuit). If either of these supertraces is empty, the diagram
will vanish since $\str \one = 0$. Consequently, the quantum part of the
flow relates 
$m$ supertrace vertices to $m-1$ supertrace
vertices (this receives corrections in the broken phase).  
To compute the flow of a vertex coefficient function, we can use the following mnemonic,
for the quantum term:
\begin{enumerate}
	\item	take an overall factor of $1/2$, corresponding to the $1/2$ associated
			with $a_1$ (see equation~(\ref{eq:flow:old:a_1}));

	\item	draw all quantum terms with the same implied supertrace structure as the
			vertex whose flow has been computed, where we sum over all cyclic permutations
			of the fields on both the inner and outer circuits of the padlock structure.
			Note also that we independently count the case where a given set of fields
			are on the inner / outer circuit.
\end{enumerate}

\subsubsection{The Full Diagrammatic Flow Equation}	\label{sec:Intro:FullDiagrammatics}

Having described the diagrammatics of the flow equation with
all fields in $\SH$-sector, we now generalise to include $\SF$s.
The subtlety occurs due to the corrections to both 
supersowing~(\ref{eq:SuperSowing-A}) and 
supersplitting~(\ref{eq:SuperSplitting-A}), when the differentiated
fields are $\SF$s.

Since we will be applying an extension of such considerations in
chapter~\ref{ch:NewFlowEquation}, we review the methodology for
computing the corrections given in~\cite{YM-1-loop}. Moreover,
the intuition we will gain about the source of corrections to
the diagrammatic flow equation will prove invaluable.

The idea is a simple one: if the superfield $\SF$ were unconstrained,
rather than restricted to being supertraceless, then supersplitting
and supersowing would be exact, just as in the $\SH$ sector. Thus
we imagine promoting $\SF$ to a full superfield, $\SF^e$  via the map:
\[
	\SF_\mu \mapsto \SF_\mu^e \equiv \SF_\mu + \sigma \SF^\sigma_\mu.
\]
We choose to take $\SF^\sigma_\mu$ arbitrary, so the map is not
unique; similarly, with the extension of all functionals:
\[
	S^e[\SF^e, \SH] \equiv S[\SF \mapsto \SF^e, \SH] \   \mathrm{\etc}
\]

On the other hand, the reverse procedure, in which we project back to the supertraceless
space is unique:
\[
	\pi \SF_\mu^e = \SF_\mu, \ \ \pi S^e = S \ \  \mathrm{\etc}
\]
Functional derivatives \wrt\ $\SF^e$ can be written~\cite{YM-1-loop}
\begin{equation}
	\fder{}{\SF^e_\mu} = \fder{}{\SF_\mu} + \frac{\one}{2N} \fder{}{\SF_\mu^\sigma},
\label{eq:ExtendedFunctionalDeriv}
\end{equation}
or, equivalently (\ibid)
\begin{equation}
	\fder{}{\SF_\mu} = \fder{}{\SF^e_\mu} - \frac{\one}{2N} \tr \fder{}{\SF_\mu^e}.
\label{eq:ExtendedFunctionalDeriv-B}
\end{equation}

Noting that $\pi$ and $\delta / \delta \SF^\sigma$ do not commute, our strategy
is as follows. We trivially rewrite both terms in the flow equation (equations~(\ref{eq:flow:old:a_0})
and~(\ref{eq:flow:old:a_1})) by extending all functionals and then projecting out.
Under the projection, we cannot simply exchange derivatives \wrt\ $\SF$
with derivatives \wrt\ $\SF^e$ but we can substitute for the former,
in terms of the latter, 
using equation~(\ref{eq:ExtendedFunctionalDeriv}) or~(\ref{eq:ExtendedFunctionalDeriv-B}). 
Then
we collect together terms into those for which supersowing / supersplitting is
exact and those which are corrections.

Let us see how this works for the quantum term~(\ref{eq:flow:old:a_1}).
\begin{equation}
	\fder{}{\SF_\mu} \{ \dot{\Delta}^{\SF \SF} \} \fder{\Sigma_g}{\SF_\mu} 
	= \pi 
	\left\{ 
		\fder{}{\SF_\mu} \{ \dot{\Delta}^{\SF \SF} \}^e \fder{\Sigma_g^e}{\SF_\mu} 
	\right\}.
\label{eq:Extended-a_1}
\end{equation}
Focusing on the term on which the projection operator acts, we substitute
for unconstrained derivatives, using equation~(\ref{eq:ExtendedFunctionalDeriv}):
\begin{eqnarray*}
	\lefteqn{
		\fder{}{\SF^e_\mu} \{ \dot{\Delta}^{\SF \SF} \}^e \fder{\Sigma_g^e}{\SF^e_\mu}
	-\frac{\one}{2N} \fder{}{\SF_\mu^\sigma} \{ \dot{\Delta}^{\SF \SF} \}^e \fder{\Sigma_g^e}{\SF_\mu^e}
	} 
\\
	&&
	-\fder{}{\SF_\mu^e} \{ \dot{\Delta}^{\SF \SF} \}^e \frac{\one}{2N} \fder{\Sigma_g^e}{\SF_\mu^\sigma}
	+\frac{\one}{2N} \fder{}{\SF_\mu^\sigma}  \{ \dot{\Delta}^{\SF \SF} \}^e \frac{\one}{2N} \fder{\Sigma_g^e}{\SF_\mu^\sigma}.
\end{eqnarray*}

The next step is to use equation~(\ref{eq:CoincidentLine}). Immediately, this tells us
that the final term vanishes. The second and third terms will involve
\[
	\str \fder{}{\SF_\mu^e} = \fder{}{\SF^0_\mu},
\]
by equations~(\ref{eq:ExtendedFunctionalDeriv}) and~(\ref{eq:fder-SF}). Thus, equation~(\ref{eq:Extended-a_1})
becomes:
\begin{equation}
	\fder{}{\SF_\mu} \{ \dot{\Delta}^{\SF \SF} \} \fder{\Sigma_g}{\SF_\mu} 
	= 
	\pi 
	\left\{ 
		\fder{}{\SF^e_\mu} \{ \dot{\Delta}^{\SF \SF} \}^e \fder{\Sigma_g^e}{\SF^e_\mu} 
	\right\}
	-\frac{1}{N}
	\pi
	\left\{ 
		\fder{}{\SF^\sigma_\mu} \cdot \dot{\Delta}^{\SF \SF} \cdot \fder{\Sigma_g^e}{\SF_\mu^0}
	\right\},
\label{eq:a_1-Corrs}
\end{equation}
as in~\cite{YM-1-loop}.

The first term represents exact supersplitting; the second term is
the correction. The crucial point, emphasised by the presence of
$\delta /\delta \SF^0_\mu$, is that the correction term is tied up with
no-$\SF^0$ symmetry. Indeed, if the $\Sigma_g^e$ of the last term
were not extended, then the correction would just vanish, precisely
by no-$\SF^0$ symmetry.

The question is, does the extension of $\Sigma_g$ lead to any surviving terms?
Consider an unextended vertex containing an arbitrary number of supertraces,
and suppose that one of these supertraces is just $\str \SF \SF$.
This part of the vertex is automatically no-$\SF^0$ symmetric,
which follows by recognising that the lowest order constraints
implied by no-$\SF^0$ symmetry can be deduced by making the 
transformation $\delta \SF_\mu = \lambda_\mu \one$ in 
equation~(\ref{eq:Intro:ActionExpansion})~\cite{YM-1-loop, Antonio'sThesis}.

However, it is only $\str \SF \SF$ terms that vanish when we
make this transformation. For the action as a whole to be no-$\SF^0$
symmetric, the survival of terms from this transformation implies 
relationships between
the vertex coefficient functions.
Now, these constraints
are there, irrespective of whether or not we extend the action (and seed
action).

Thus, when we do extend the actions, $\delta \Sigma_g^e / \delta \SF^0$
will vanish unless $\delta \SF^0$ strikes a $\str \SF^e \SF^e$. In
this extended case, such a contribution now survives since $\str \SF^e \neq 0$.
Returning to equation~(\ref{eq:a_1-Corrs}), the $\delta / \delta \SF^\sigma$
must strike  the $\str \SF^e$, since otherwise this contribution will
vanish, upon projection to the restricted space. This yields a 
(super)group theory
factor of $-\frac{1}{N} \str \sigma = -2$.

Given that similar arguments tell us that the classical part of
the flow equation~(\ref{eq:flow:old:a_0}) does not, in fact, receive
any corrections to supersowing~\cite{YM-1-loop}, we are ready to
write down the full diagrammatic flow equation. We simply reproduce
figure~\ref{fig:Intro:OldFlow:SH-only} (discarding the final diagram), 
where the differentiated
fields (along with all decorations) can now be either both $\SH$s or
both $\SF$s (the mixed case does not exist, since there is no 
$\dot{\Delta}^{\SH \SF}$), so long as we use the following prescription:
\emph{if, in the quantum term, an undecorated wine attaches at
both ends to a $\str \SF \SF$ term, then we supplement the usual
group theory factor with an additional $-2$}.

\begin{center}
\begin{figure}[h]
	\[
		\ensuremath{\begin{array}{c}\input{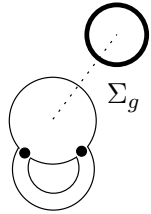} \end{array}}
	\]
\caption{A term which survives, despite na\"ively vanishing by group theory considerations.}
\label{fig:Intro:SuperSplittingCorrection}
\end{figure}
\end{center}

In the unbroken phase, this actually causes a class of
diagram which would na\"ively vanish, to survive. These
diagrams can possess any number of supertraces;  the
double supertrace case is shown in figure~\ref{fig:Intro:SuperSplittingCorrection}.

Now, the expected group theory factor associated with this diagram arises from the two empty
circuits in the padlock structure. These yield $(\str \one)^2 = 0$. However, according
to our prescription, the group theory factor of this diagram is supplemented by $-2$,
causing it to survive.

\subsubsection{The Coincident Line Identities}

We can cast equation~(\ref{eq:WineRelation})---which we recall
follows from the coincident line identities~(\ref{eq:Interleaves})---in 
a particularly simple diagrammatic form. This is 
shown in figure~\ref{fig:CoincidentLine}.
\begin{center}
\begin{figure}[h]
	\begin{equation}
		\ensuremath{\begin{array}{c}\input{pstex/Wine-XXXX.pstex_t} \end{array}} \hspace{1em} + \hspace{1em} \ensuremath{\begin{array}{c}\input{pstex/Wine-XXX-X.pstex_t} \end{array}} \hspace{1em} + \cdots + \hspace{1em} \ensuremath{\begin{array}{c}\input{pstex/Wine-X-XXX.pstex_t} \end{array}} \hspace{1em} + \hspace{1em} \ensuremath{\begin{array}{c}\input{pstex/Wine--XXXX.pstex_t} \end{array}} = 0
	\label{eq:WineRelation-Diag}
	\end{equation}
\caption{Diagrammatic representation of equation~(\ref{eq:WineRelation}).}
\label{fig:CoincidentLine}
\end{figure}
\end{center}

Notice that equation~(\ref{eq:WineRelation-Diag}) implies equation~(\ref{eq:CoincidentLine}),
as it should. If the top (bottom) end of the wine is plugged by $\one$, rather than a functional
derivative, then no matter what the bottom (top) end attaches to, we can cycle the decorations
of the wine around, without changing the supertrace structure of the diagram as a whole.
Hence, the only class of diagram to  survive, in this case, is that for which the wine is undecorated.

\subsection{The Broken Phase}	\label{sec:Intro:BrokenPhase}

\subsubsection{Using Partial Supermatrices}

To work in the broken phase, in which we recall that $\SH$ fluctuates around $\sigma$, is easy. We decompose:
\begin{equation}
	\SF_\mu = A_\mu + B_\mu;  \hspace{2em} \SH = C + D + \sigma.
\label{eq:BrokenPhase}
\end{equation}

Now, instead of decorating  vertices with $\SF$s and $\SH$s, we
decorate instead with the fields $A,B,C,D$ and $\sigma$. 
We can simplify appearances of $\sigma$ by noting that it (anti)commutes
with $(B,D)$ $A,C$. Thus, we can take the convention~\cite{YM-1-loop}
that all instances of
$\sigma$ are consecutive, upon which we use $\sigma^2 = \one$ to 
remove all pairs. Consequently, vertices have at most one instance
of $\sigma$, which we note does not come with a position label
and can be ignored when determining symmetry factors~\cite{YM-1-loop}.
Vertices must contain an even number of fermionic
fields, else they can be cast as the supertrace of a block off-diagonal
matrix, which vanishes.

To deal with the broken sector wines, let us return to equation~(\ref{eq:Wine-B}).
The fields $X$ and $Y$ can be any of $A$--$D$ and $\sigma$, whereas $Z$
is restricted to just the dynamical fields. Wines (assuming the 
diagrammatic interpretation 
of figure~\ref{fig:Intro:Diagrammatics:Wines-True}, 
as opposed to figure~\ref{fig:Intro:Diagrammatics:Wines}), like vertices, must be net
bosonic, but we note that the functional derivatives can be \wrt\
opposite statistics partners (this then implies that there must be an odd
number of fermionic decorations). Once again, we can simplify instances of
$\sigma$ by noting that it (anti)commutes with not only
the decorations, but also the functional derivatives.

It is instructive to compare the zero-point wines in the symmetric phase
to those in the broken phase. Let us start in the
symmetric phase. From equation~(\ref{eq:fullwine-Old}),
we see that the second term does not contribute to
$\dot{\Delta}^{\SF \SF}$, since it
 has at least two decorative $\SH$s and so cannot
contribute to a zero-point wine. 

In the broken phase, we
note that $\sigma$ commutes with $\delta / \delta A$, and
so the second term does not contribute to $\dot{\Delta}^{AA}$.
Therefore,
\[
	\dot{\Delta}^{AA} = \dot{\Delta}^{\SF \SF}.
\]
However, $\sigma$ anticommutes with $\delta / \delta B$
and so, in this case, there is an additional contribution:
\[
	\dot{\Delta}^{BB} = \dot{\Delta}^{\SF \SF} + \dot{\Delta}^{\SF \SF}_m.
\]
Similarly,
\begin{eqnarray*}
	\dot{\Delta}^{CC} & = & \dot{\Delta}^{\SH \SH},
\\ 
	\dot{\Delta}^{DD} & = & \dot{\Delta}^{\SH \SH} + \dot{\Delta}^{\SH \SH}_m.
\end{eqnarray*}

Still following~\cite{YM-1-loop}, we can reduce the number of fields
we have to deal with in the broken sector by defining\footnote{Actually,
this definition differs from that in~\cite{YM-1-loop} in which
$F = (B_\mu, D \sigma)$; the new definition will remove some annoying signs, later.}
the Euclidean five vector
\begin{equation}
	F_M = (B_\mu, \sigma D).
\label{eq:Intro:F}
\end{equation}
Putting $B$ and $D$ on the same footing makes perfect sense since,
as we have already noted, these fields gauge transform into each other,
in the broken phase.

As one might hope, the $B$
and $D$ sector wines can be collected together into a composite $F$
wine. For the zero-point kernel, we expect to find an object like
$\dot{\Delta}^{\sigma D \sigma D}$. However, we can simplify this.
Anticommuting one of the $\sigma$s through the functional derivative,
we can combine it with the other, at the expense of a minus sign.
Therefore, the zero-point $F$-wine looks like~\cite{YM-1-loop}
\begin{equation}
	\dot{\Delta}^{F\, F}_{MN}(p) =
	\left(
		\begin{array}{cc}
			\dot{\Delta}^{BB}_p \delta_{\mu \nu}	& 0
		\\
						0							& -\dot{\Delta}^{DD}_p
		\end{array}
	\right).
\label{eq:F-Wine}
\end{equation}

We note that there is no such object as a mixed $BD$ wine, since
this would require the existence of an $\SF\SH$ wine in the
unbroken phase.

The diagrammatic flow equation receives correction, since we are no longer
differentiating \wrt\ $\SF$ and $\SH$, but only \wrt\ either their
block diagonal or block off-diagonal components. The corrections are easy
to deduce by introducing the projectors $d_\pm$ on to the block (off) diagonal
components~\cite{GI-ERG-II}:
\begin{equation}
	d_\pm X = \frac{1}{2} (X \pm \sigma X \sigma).
\label{eq:Block(off)DiagonalProjectors}
\end{equation}

\begin{center}
\begin{figure}[h]
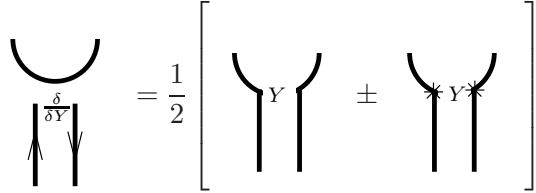

	\[
		\ensuremath{\begin{array}{c}\input{pstex/DiffY.pstex_t} \end{array}} = \frac{1}{2}
		\left[
			\ensuremath{\begin{array}{c}\input{pstex/DiffY-A.pstex_t} \end{array}} \pm \ensuremath{\begin{array}{c}\input{pstex/DiffY-B.pstex_t} \end{array}}
		\right]
	\]
\caption{Diagrammatic representation of differentiation \wrt\ a partial
supermatrix, $Y$.}
\label{fig:Intro:DiffPartialSuper}
\end{figure}
\end{center}

Hence, differentiating \wrt\ partial superfields yields insertions
of $\sigma$. Representing these insertions by \DiagSigma, we see
the diagrammatic interpretation in figure~\ref{fig:Intro:DiffPartialSuper},
where we differentiate \wrt\ the partial supermatrix, $Y$.

Clearly, instances of $\sigma$, whether they come from vertices containing
a $\SH$ in the unbroken phase or from differentiation \wrt\ partial
supermatrices, affect the group theory structure of diagrams. For circuits
containing fields, we may expect to find (implied) supertrace structures like 
$\str X Y \cdots \sigma$. For empty circuits, the effect is more profound.
In the unbroken phase, if a circuit is devoid of dynamical fields, then
it must yield $\str \one =0$. However, in the broken phase we can 
have a circuit devoid of dynamical fields but which contains an insertion
of $\sigma$. Such a circuit now yields $\str \sigma = 2N$.

\subsubsection{Symmetries}

We have already discussed no-$\SF^0$ symmetry to the desired
depth, for this section of the thesis. We will return to
it, however, in chapters~\ref{ch:NewFlowEquation} and~\ref{ch:FurtherDiagrammatics}. The other
symmetries we will make use of are the residual $SU(N|N)$
symmetries left over, after symmetry breaking, and
charge conjugation (\CC) invariance, which we have already
mentioned in the symmetric phase.\footnote{In~\cite{YM-1-loop},
where both actions were restricted to single supertrace terms,
use was made of the symmetry under $C,D,\sigma \leftrightarrow -C,-D,-\sigma$.
In the current treatment, with this restriction lifted, the symmetry is
of no use, since terms which appear to violate this symmetry
\eg\ $S^{AAC}$ in fact do not. This is because we can always arrange
for there to be an implicit $\sigma$
hiding on a second supertrace, devoid of any dynamical
fields  \eg\ $S^{AAC, \sigma}$.}

The residual $SU(N|N)$ symmetries follow from
decomposing $\Omega$ into bosonic and fermionic parts, which 
we denote by $\omega = d_+\Omega$ and $\tau = d_- \Omega$, respectively. Defining
$D_\mu = \partial_\mu - i A_\mu$, the unbroken bosonic gauge transformations give us~\cite{YM-1-loop}
\begin{equation}
	\begin{array}{rcl}
		\delta A_\mu	& =	& D_\mu \cdot \omega
	\\
		\delta B_\mu	& =	& -i B_\mu \cdot \omega
	\\
		\delta C		& = & -i C \cdot \omega
	\\
		\delta D		& = & -i D \cdot \omega,
	\end{array}
\label{eq:Intro:UnBroken-GT}
\end{equation}
whereas the broken fermionic gauge transformations yield
\begin{equation}
	\begin{array}{rcl}
		\delta B_\mu	& =	& D_\mu \cdot \tau
	\\
		\delta A_\mu	& =	& -i B_\mu \cdot \tau
	\\
		\delta D		& = & -i C \cdot \tau + 2i\tau \sigma
	\\
		\delta C		& = & -i D \cdot \tau.	
\end{array}
\label{eq:Intro:Broken-GT}
\end{equation}

Likewise, we can deduce the gauge transformation of 
functional derivatives \wrt\ the broken sector fields
by decomposing equations~(\ref{eq:Intro:GT-ddC})
and~(\ref{eq:Intro:GT-ddA}). Doing so, it is apparent
that $\delta / \delta A$ does not transform homogeneously;
gauge invariance of our broken phase flow
equation is guaranteed by the coincident line identities.

Two Ward identities now follow from applying~(\ref{eq:Intro:UnBroken-GT}),
(\ref{eq:Intro:Broken-GT}) and the broken phase form of~(\ref{eq:Intro:GT-ddC})
and~(\ref{eq:Intro:GT-ddA})
to the field expansions~(\ref{eq:Wine-B}) 
and~(\ref{eq:Intro:ActionExpansion}).
The (unbroken) bosonic gauge transformations~\cite{GI-ERG-I, YM-1-loop} yield
\begin{equation}
	q_\nu U^{\cdots X \! A Y \cdots}_{\cdots \, a \; \nu \, b \ \cdots} (\ldots, p,q,r, \ldots) 
	=
	U^{\cdots XY \cdots}_{\cdots \, a \; b \ \cdots}( \ldots, p, q+r, \ldots) - 	
	U^{\cdots XY \cdots}_{\cdots \, a \; b \ \cdots}( \ldots, p+q, r, \ldots),
\label{eq:WID-Unbroken}
\end{equation}
where $U$ can be a vertex from either action or any wine and $X,Y$ can
be any of $A,C,F$. There is an appealing geometrical picture of
this~\cite{TRM-Faro,GI-ERG-I,YM-1-loop}: we can view the momentum of the 
field $A_\nu$ as being pushed forward
(diagrammatically, this is in the counterclockwise sense)
on to the next obstruction with a plus and pulled back on to the previous obstruction
with a minus. An obstruction is any dynamical field ($\sigma$ commutes with
$\omega$, and so is blind to this process), or a derivative \wrt\ any of the
(dynamical) fields. Thus, in the case that the momentum of $A_\nu$ hits the end of a
wine, we take either $X$ or $Y$ above to be $Z_1$ or $Z_2$ in~(\ref{eq:Wine-B}), as
appropriate.

The second Ward identity follows from  the (broken)
fermionic gauge transformations and is most neatly written in terms of the
composite field $F$, rather than its components $B$ and $\sigma D$. To this end, we define
a Euclidean five momentum~\cite{YM-1-loop}\footnote{Again, this definition is different 
from~\cite{YM-1-loop}, where the two comes with a minus sign.}
\begin{equation}
	q_M = (q_\mu, 2).
\label{eq:Intro:q_M}
\end{equation}
The Ward identities corresponding to the broken, fermionic gauge transformations
can now be written in the following compact form:
\begin{equation}
	q_N U^{\cdots X \! F Y \cdots}_{\cdots \, a  N  b \ \cdots} (\ldots, p,q,r, \ldots) 
	=
	U^{\cdots X\PF{Y} \cdots}_{\cdots \, a \; b \ \cdots}( \ldots, p, q+r, \ldots) - 	
	U^{\cdots \PB{X}Y \cdots}_{\cdots \, a \; b \ \cdots}( \ldots, p+q, r, \ldots),
\label{eq:WID-Broken}
\end{equation}
where $X$ and $Y$ are as before.\footnote{$\sigma$ anticommutes with $\tau$ and so
if the momentum of a fermionic field is pushed through a $\sigma$, we pick up a minus sign.} 
For $X = A,C$, $\PF{X} = \PB{X}$ is the corresponding
opposite statistics partner.
$\PF{F}_M = (A_\mu, -C\sigma)$ and $\PB{F}_M = (A_\mu, C \sigma)$.\footnote{The signs
of the $C$-components are opposite to~\cite{YM-1-loop}.}

The final symmetry we look at is charge conjugation (\CC) invariance, which will play
a central role in the techniques developed in 
chapter~\ref{ch:FurtherDiagrammatics}. Our $SU(N|N)$ gauge theory
in the broken phase is invariant under~\cite{YM-1-loop}:
\begin{eqnarray}
	A_\mu 	& \rightarrow & - A^T,
\label{eq:CC-A}
\\
	C		& \rightarrow & C^T,	
\label{eq:CC-C}
\\
	F_M		& \rightarrow & - F_M^T.
\label{eq:CC-F}
\end{eqnarray}

In chapter~\ref{ch:FurtherDiagrammatics} we will see that both the
Ward identities and \CC\
have an intuitive and appealing diagrammatic interpretation.

\subsection{The Weak Coupling Expansion}	\label{sec:Intro:WeakCoupling}

All calculations performed in this thesis are
done in the perturbative regime, and so we now
examine the form that the flow equations take,
in this limit.

Following~\cite{GI-ERG-I, YM-1-loop}, the action has the weak
coupling expansion
\begin{equation}
	S = \sum_{i=0}^\infty \left( g^2 \right)^{i-1} S_i = \frac{1}{g^2}S_0 + S_1 + \cdots,
\label{eq:WeakCouplingExpansion-Action}
\end{equation} 
where $S_0$ is the classical effective action and the $S_{i>0}$
the $i$th-loop corrections. The seed action has a similar expansion:
\begin{equation}
	\hat{S} = \sum_{i=0}^\infty  g^{2i}\hat{S}_i.
\label{eq:WeakCouplingExpansion-SeedAction}
\end{equation} 
Note that these definitions are consistent with $\Sigma_g = g^2 S -2\hat{S}$; 
identifying powers of $g$ in the flow equation, it is clear that
$S_i$ and $\hat{S}_i$ will always appear together.
With this in mind, we now define
\begin{equation}
	\Sigma_i = S_i - 2\hat{S}_i.
\label{eq:WeakCouplingSigma}
\end{equation}
The $\beta$-function takes the usual form:
\begin{equation}
	\beta \equiv \flow g = \sum_{i=1}^\infty  g^{2i+1} \beta_i.
\label{eq:WeakCouplingExpansion-beta}
\end{equation}

We now substitute expansions~(\ref{eq:WeakCouplingExpansion-Action}), 
(\ref{eq:WeakCouplingExpansion-SeedAction}) and~(\ref{eq:WeakCouplingExpansion-beta})
into equation~(\ref{eq:Flow:General}) and, using equation~(\ref{eq:WeakCouplingSigma}),
obtain the weak coupling expansion of the flow equation:
\begin{equation}
	\flow S_n = -2\sum_{r=1}^n (n-r-1) \beta_r S_{n-r} - \sum_{r=0}^n a_0 \left[ S_{n-r}, \Sigma_r \right] + a_1 \left[\Sigma_{n-1} \right].
\label{eq:WeakCouplingExpansion-oldFlow}
\end{equation}

To simplify the diagrammatics, in particular, we introduce the following shorthands:
\begin{equation}
	n_r \equiv n-r, \ \ n_+ \equiv n+1, \ \ n_- \equiv n-1.
\label{eq:LoopOrderShorthands}
\end{equation}
Furthermore, for a Wilsonian effective action vertex of loop order
$m$, we will often refer to the vertex just as $m$, rather than $S_m$.
Similarly, we will use $\widehat{m}$ to refer to $\hat{S}_m$. For tree level
seed action vertices, we may abbreviate $\hat{0}$ to just $\widehat{\hspace{\hatwidth}}$.

In figure~\ref{fig:Intro:Diagrammatics:n-loopOLD} we give a
diagrammatic representation of equation~(\ref{eq:WeakCouplingExpansion-oldFlow})
where, for the fields which have been differentiated,
we sum over all dynamical, broken phase fields,
discarding any combinations which correspond to wines that
do not exist (\eg\ $AC$). For all other fields,
we also allow instances of $\sigma$.
\begin{center}
\begin{figure}[h]
	\[
	\dec{
		\ensuremath{\begin{array}{c}\input{pstex/FullVertex-S_n.pstex_t} \end{array}}
	}{\bullet}
	= 2 \sum_{r=1}^n (n-r-1) \beta_r \ensuremath{\begin{array}{c}\input{pstex/FullVertex-S_n_r.pstex_t} \end{array}}
	+\frac{1}{2} \sum_{r=0}^n \ensuremath{\begin{array}{c}\input{pstex/FullDumbbell-S_n_r-Sig_r.pstex_t} \end{array}} -\frac{1}{2} \ensuremath{\begin{array}{c}\input{pstex/FullPadlock-Sig_n.pstex_t} \end{array}}
	\]
\caption{Diagrammatic representation of equation~(\ref{eq:WeakCouplingExpansion-oldFlow}),
where diagrams in which the wine bites its own tail have been discarded.}
\label{fig:Intro:Diagrammatics:n-loopOLD}
\end{figure}
\end{center}

\chapter{The New Flow Equation} \label{ch:NewFlowEquation}

\section{The Need for a New Flow Equation}		\label{sec:NFE:TheNeed}

Given the successful computation of $\beta_1$~\cite{YM-1-loop} using the flow
equation of the previous chapter, 
(equations~(\ref{eq:Flow:General})--(\ref{eq:flow:old:a_1})),
one may very well ask why it is necessary for the flow
equation to be modified. The key point is that we expect
the (unphysical) field $A^2$ to come with its own coupling, $g_2$,
which is distinct from $g$ in the broken phase. In our
computation of $\beta_1$, this consideration plays no role but,
for higher loop calculations, it is of crucial importance.

To understand this, we recall the standard argument as to why
the coefficients $\beta_1$ and $\beta_2$ are guaranteed to
agree, between certain renormalisation schemes~\cite{Weinberg, YM-1-loop}. 
Let us consider
relating our coupling, $g(\Lambda)$, to some coupling 
$\tilde{g}(\mu \mapsto \Lambda)$ which corresponds to a different
renormalisation scheme.

Given the dimensionless
one-loop matching coefficient, $\match$, we can perturbatively
match the two couplings:
\[
	\frac{1}{\tilde{g}^2} = \frac{1}{g^2} + \match + \mathcal{O}(g^2).
\]
Using the definition of the $\beta$-function (equation~(\ref{eq:WeakCouplingExpansion-beta})) and its
analogue for $\tilde{g}$, we obtain:
\begin{equation}
	\tilde{\beta}_1 + g^2 \tilde{\beta}_2 = \beta_1 + g^2 \beta_2 + \flow \match + \mathcal{O}(g^4).
\label{eq:Universal-beta}
\end{equation}
Thus, 
if $\flow \match = 0$ (or, at any rate, does not contribute until $\mathcal{O}(g^4)$), 
we will obtain $\tilde{\beta}_1 = \beta_1$
and $\tilde{\beta}_2 = \beta_2$. 

We must now examine under what conditions $\flow \match$ vanishes. Clearly,
if there is some finite physical scale other than $\Lambda$, then we will
be able to construct an $\match$ whose flow does not vanish.
Similarly, this
can be achieved by the presence of dimensionless running couplings, other
than $g$.\footnote{As noted in~\cite{YM-1-loop} this second point is equivalent to 
the first since dimensional transmutation leads to the introduction of
some finite physical scale, other than $\Lambda$.}

Within the \MGIERG, there are two potential ways in which a running $\match$
can arise. The
first is through a generic choice of $\hat{S}$, which can lead to even tree level
running of $\match$, destroying the agreement between the one-loop $\beta$-functions. 
In~\cite{YM-1-loop}, it was demonstrated that such running can
removed by a suitable choice of $\hat{S}$. 
In section~\ref{sec:TLD:Couplings} we extend this argument
to show that such running can in fact be removed to all orders.

However, this still leaves us with the dimensionless
coupling $g_2$ which, like $g$, we know to run at one-loop.
There is no reason not to expect this to spoil agreement between
our value of $\beta_2$ and the standard one.

At this juncture, it is worth emphasising that a disagreement between
our first two $\beta$-function coefficients and the standard 
values is not necessarily a signature of a sick formalism. Although often referred
to as universal, $\beta_1$ and $\beta_2$ are not universal in the sense that a
physically observable scattering amplitude is. They are better
described as being pseudo-universal, since 
it can be \emph{arranged}
for them to depend only on universal details.
Hence, there is nothing fundamentally wrong with 
the matching coefficient $\eta$ running, even at tree level.

However, the whole point of computing $\beta_2$ within the \MGIERG\ is
to provide a stringent test of the formalism. This is most easily done
by arranging things so our value should coincide with the standard value
and so we will loosely refer to our $\beta$-function coefficients
as universal. 

Let us
return to equation~(\ref{eq:Universal-beta}) and 
consider $\match = \match(g_2)$. Now,
\[
	\flow \match  = \flow g_2 \, \pder{\match}{g_2}.
\]
Perturbatively, we know that $\flow g_2 \sim g_2^3$ and
thus, if $g_2$ were to vanish, then so does its flow.
Hence, our strategy is to tune $g_2 \rightarrow 0$, at the end of our calculation.
So long as $\partial \match / \partial g_2$ does not diverge in
this limit, then we can hope to recover the usual $\beta_2$.

This is all very well and good, but we have still not said
why we need to modify the flow equation. The trouble is that,
in its current form, the flow equation cannot 
conveniently distinguish
between the fields $A^1$ and $A^2$. 
In order for us to isolate
the effects of $g_2$, supplementary terms must be added, so
that the necessary distinction can be made.

For convenience, we now define a new variable,
\begin{equation}
	\alpha \equiv \frac{g_2^2}{g^2}.
\label{eq:Definition:alpha}
\end{equation}
We also now change notation slightly such
that
\begin{equation}
	\dot{X} \equiv -\flowConstAl X;
\label{eq:dot(X)-NEW}
\end{equation}
this replacing our old definition~(\ref{eq:dot(X)-OLD}).

\section{Modifying the Flow Equation}

Our aim is to modify the flow equation, so that
it treats $A^1$ and $A^2$ differently. It is not
hard to guess how to go about doing this.
Currently, the flow equation involves objects
like $\delta \Sigma_g / \delta \SF_\mu$. In addition
to terms such as this, we now want to try and include 
terms containing things
like $\sigma \delta \Sigma_g / \delta \SF_\mu$; the presence
of the $\sigma$ will ensure that, in our new terms, 
$A^2$ fields pick up 
a sign, relative to $A^1$ fields. Consequently, $A^1$ and $A^2$
will be treated differently by the flow equation.
The difficulty is ensuring that any terms
we add to the flow equation respect no-$\SF^0$
symmetry.

To this end, we introduce an operator, $\Proj$,
which, when acting on the superfield $X$, removes
the $X^0$ component:
\begin{equation}
	\Proj(X) = X - \frac{\one}{2N} \str (\sigma X).
\label{eq:A^0-Projector-BPhase}
\end{equation}

This then naturally leads us to the symmetric
phase version of this operator, $\Proj_\SH$, where
\begin{equation}
	\Proj_\SH(X) = X - \one \frac{\str (\SH X)}{\str \SH}.
\label{eq:A^0-Projector-Sphase}
\end{equation}
We note that $\Proj_\SH$ is a projector, since it
satisfies $\Proj_\SH^2 = \Proj_\SH$.

Now, we can use our projector to implement $\sigma \delta \Sigma_g / \delta \SF_\mu$
terms in our flow equation. In the broken phase, rather than multiplying
$\sigma$ by $\delta \Sigma_g / \delta \SF_\mu$, we want to first project
out the $\SF^0$ part. Then, in the symmetric phase, we
replace all $\sigma$s by $\SH$s. Thus, the building blocks for our
new terms look like
\begin{equation}
	\SH \Proj_\SH \left( \fder{\Sigma_g}{\SF_\mu} \right).
\label{eq:BuildingBlock}
\end{equation}

However, this is not quite what we want. If we are to introduce terms
like
\[
	\SH \fder{\Sigma_g}{\SF_\mu},
\]
then, to preserve \CC\ invariance, we must also introduce terms like
\[
	\fder{\Sigma_g}{\SF_\mu} \SH.
\]
To do this, we simply let the first $\SH$ of equation~(\ref{eq:BuildingBlock})
act via anti-commutation. 

The final change we must make to the building block~(\ref{eq:BuildingBlock}),
so that it is suitable for use in the flow equation, is to multiply
though by $\str \SH$. If we were not to do this, then our flow equation
would contain terms with an overall factor of $1/\str{\SH}$.

With these points in mind, we introduce a new wine, $\{\dot{\Delta}^{\SF \SF}_\sigma \}$
and write down a new flow equation,
which can distinguish between $A^1$ and $A^2$, whilst
respecting no-$\SF^0$ symmetry.

\begin{equation}
-\flow S = a_0[S,\Sigma_g] - a_1[\Sigma_g],
\label{eq:Flow:General-New}
\end{equation}
where
\begin{eqnarray}
a_0[S,\Sigma_g] &= &	\frac{1}{2} \fder{S}{\SF_\mu} \{\dot{\Delta}^{\SF \SF}\} \fder{\Sigma_g}{\SF_\mu}
						+\frac{1}{2} \fder{S}{\SH} \{\dot{\Delta}^{\SH \SH}\} \fder{\Sigma_g}{\SH}
\nonumber
\\[1ex]
				& +&	\frac{1}{16N} \fder{S}{\SF_\mu} \{\dot{\Delta}^{\SF \SF}_\sigma\}
							\left\{
								\SH \str{\SH}, \Proj_\SH 
								\left(\fder{\Sigma_g}{\SF_\mu} \right) 
							\right\} 
\nonumber
\\[1ex]
				& + &	\frac{1}{16N}
							\left\{
								\SH \str{\SH} , \Proj_\SH \left(\fder{S}{\SF_\mu}\right) 
							\right\} 	
						\{\dot{\Delta}^{\SF \SF}_\sigma\} \fder{\Sigma_g}{\SF_\mu}
\label{eq:flow:new:a_0}
\end{eqnarray}
and
\begin{eqnarray}
a_1[\Sigma_g]	&=	& 	\frac{1}{2} \fder{}{\SF_\mu} \{\dot{\Delta}^{\SF \SF}\} \fder{\Sigma_g}{\SF_\mu}
					+  	\frac{1}{2} \fder{}{\SH} \{\dot{\Delta}^{\SH \SH}\} \fder{\Sigma_g}{\SH}.
\nonumber
\\[1ex]
				&+	& 	\frac{1}{16N} \fder{}{\SF_\mu} \{\dot{\Delta}^{\SF \SF}_\sigma\}
							\left\{
								\SH	\str \SH ,  \Proj_\SH \left(\fder{\Sigma_g}{\SF_\mu} \right)
							\right\}
\nonumber
\\[1ex]
				& + & 	\frac{1}{16N}
							\left\{
								\SH	\str \SH ,  \Proj_\SH \left(\fder{}{\SF_\mu}\right) 
							\right\}
							\{\dot{\Delta}^{\SF \SF}_\sigma\} 
						\fder{\Sigma_g}{\SF_\mu}.
\label{eq:flow:new:a_1}
\end{eqnarray}

The are several things to note about the new flow equation. First,
it is indeed gauge invariant: it is straightforward to demonstrate
that, for the new terms, the inhomogeneous part of $\delta (\delta / \delta \SF)$
(see equation~(\ref{eq:Intro:GT-ddA})) cancels out.  Secondly, we note that
the new terms come with an overall factor of $1/16N$. This
is just a convenient normalisation factor, as we will see shortly.
Thirdly, we have been completely diplomatic in the way the new terms have
been introduced which is why, in addition to the anti-commutator
structure, $a_0$ (and $a_1$) has two new parts, rather than one.
This diplomacy is down to choice. We do this to maintain the
simple diagrammatic rule that we decorate all structures
in all possible ways.

Needless to say, our new flow equation satisfies the usual
requirements we demand of a flow equation and so is perfectly
valid. However, we do emphasise that our flow equation is by no means
unique: there are an infinite class of good flow equations
which distinguish between $A^1$ and $A^2$ whilst respecting no-$\SF^0$
symmetry. We will return to this point in the conclusion.

\section{The New Diagrammatics-I}	\label{sec:NFE:NewDiags-I}

\subsection{The Symmetric Phase}

By supplementing the flow equation with new terms,
it is inevitable that our diagrammatics must change.
The obvious expectation is that it must become more 
complicated, which is borne out in this section. 
However, this complication turns out to be
a temporary evil, as  we will be guided
to  diagrammatics which
are actually simpler than those of the previous 
chapter!

The first thing to note in deriving diagrammatic
rules for the new terms in the flow equation is
that, for these new terms, supersowing and supersplitting
are exact (this is straightforward to check, using the
techniques of section~\ref{sec:Intro:FullDiagrammatics}).

Let us examine the new classical
terms.
Expanding out the projector,
we have things like
\[
\left[
	\left\{
		\SH, \fder{\Sigma_g}{\SF_\mu} 
	\right\} \str{\SH} 
	- 2 \SH \str 
	\left(
		\SH \fder{\Sigma_g}{\SF_\mu} 
	\right)
\right].
\]

In the first term, we will be tying either
$\SH \, \delta \Sigma / \delta \SF_\mu$ 
or $ \delta \Sigma / \delta \SF_\mu \, \SH$
back into a supertrace. In the second term above,
the $ \delta \Sigma / \delta \SF_\mu$ is already
part of a supertrace; it is the  $\SH$ sitting
outside the supertrace that must be tied up.

Now, either way, there is
an instance of $\SH$ that is tied up by the covariantisation.
Referring to the superfield expansion
of the wine~(\ref{eq:Wine}), it is clear that this
$\SH$  sits just past the end of one side
of the wine. Thus, as an intermediate step, we
can represent the classical part of the flow
equation as in figure~\ref{fig:NewFlowEquation:Classical}.
The double circle notation will be interpreted shortly,
but we note that the double circles attach only
to $\dot{\Delta}^{\SF \SF}_\sigma$ wines (a corollary
of which is that all functional derivatives are \wrt\ $\SF$,
in the new terms) and that their position arguments are
the same as those of the functional derivatives they sit
next to.
\begin{center}
\begin{figure}[h]
	\[
	a_0[\Sigma_g] 	= \frac{1}{2} \ensuremath{\begin{array}{c}\input{pstex/ClassicalTerm-B.pstex_t} \end{array}} 
					+ \frac{1}{16N} 
					\left[
						\ensuremath{\begin{array}{c}\input{pstex/NewClassicalTerm-A.pstex_t} \end{array}}
						+\ensuremath{\begin{array}{c}\input{pstex/NewClassicalTerm-B.pstex_t} \end{array}}
						+\ensuremath{\begin{array}{c}\input{pstex/NewClassicalTerm-C.pstex_t} \end{array}}
						+\ensuremath{\begin{array}{c}\input{pstex/NewClassicalTerm-D.pstex_t} \end{array}}
					\right]
	\]
\caption{Diagrammatics for the classical part of the new flow equation.}
\label{fig:NewFlowEquation:Classical}
\end{figure}
\end{center}

\begin{center}
\begin{figure}[h]
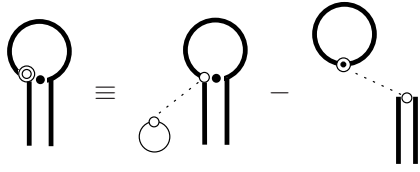

	\[
	\ensuremath{\begin{array}{c}\input{pstex/DoubleCircle.pstex_t} \end{array}} \equiv \ensuremath{\begin{array}{c}\input{pstex/DoubleCircle-A.pstex_t} \end{array}} - \ensuremath{\begin{array}{c}\input{pstex/DoubleCircle-B.pstex_t} \end{array}}
	\]
\caption{Interpretation of the double circle notation.}
\label{fig:NewFlowEquation:Details}
\end{figure}
\end{center}

The interpretation of the double circles is simple. For the
component which looks like
\[
	\SH \fder{\Sigma_g}{\SF_\mu} \str{\SH},
\]
we have a dumbbell structure as usual but with 
\begin{enumerate}
	\item	an extra $\SH$ sitting
			between one end of the wine 
			and one of the lobes,
	\item	an accompanying
			$\str \SH$. 
\end{enumerate}
For the component looking like
\[
	\SH \str
	\left(
		\SH \fder{\Sigma_g}{\SF_\mu} 
	\right),
\]
the dumbbell structure now terminates in a $\SH$, rather than the usual
lobe. However, there is an accompanying supertrace, in which we
can think of one of the $\SF$s of $\Sigma_g$ having been replaced by
a $\SH$. The diagrammatic interpretation is shown in 
figure~\ref{fig:NewFlowEquation:Details}.

In both diagrams, the dotted line
is a `false' kernel which, if joining two
fields with position arguments $x$ and $y$,
is taken as $\delta(x-y)$. 
 The interpretation of the first diagram
is straightforward. In the second diagram, the \ReplaceAwC\
tells us to replace an $\SF$ with a $\SH$. As for the wine, it
is `plugged' by a $\SH$, rather than ending in a functional
derivative; the functional derivative being linked to this $\SH$
via a false kernel.
In both diagrams of figure~\ref{fig:NewFlowEquation:Details}, the wine
is of type $\dot{\Delta}^{\SF \SF}_\sigma$. Note that one the $\SF$s
which labels such wines may now be linked to the end of the wine by
a false kernel.

The diagrammatics for the quantum term, shown in 
figure~\ref{fig:NewFlowEquation:Quantum}, is obvious 
(we ignore diagrams in which the wine bites its own tail).
\begin{center}
\begin{figure}[h]
	\[
	a_0[\Sigma_g] 	= \frac{1}{2} \ensuremath{\begin{array}{c}\input{pstex/QuantumTerm-B.pstex_t} \end{array}} 
					+ \frac{1}{16N} 
					\left[
						\ensuremath{\begin{array}{c}\input{pstex/NewQuantumTerm-A.pstex_t} \end{array}}
						+\ensuremath{\begin{array}{c}\input{pstex/NewQuantumTerm-C.pstex_t} \end{array}}
						+\ensuremath{\begin{array}{c}\input{pstex/NewQuantumTerm-D.pstex_t} \end{array}}
						+\ensuremath{\begin{array}{c}\input{pstex/NewQuantumTerm-B.pstex_t} \end{array}}
					\right]
	\]
\caption{Diagrammatics for the quantum part of the new flow equation.}
\label{fig:NewFlowEquation:Quantum}
\end{figure}
\end{center}

\subsection{The Broken Phase}	\label{sec:NFE:BrokenPhase}

Adapting the symmetric phase diagrammatics
to the broken phase is harder than
it was for the old flow equation.

With the old flow equation, we started off with
fields $\SF$ and $\SH$ and broke the supersymmetry
to give the dynamical fields $A,C,F$ and insertions
of $\sigma$. The flow equation was then written naturally
in terms of vertices and wines decorated by these fields.

Now, however, whilst the symmetry breaking is still the same,
the flow equation allows us to sensibly
`look inside' $A$ and compute the flow of (say) just $A^1$.
The difficulty comes because $A$ decomposes into 
not just $A^1$ and $A^2$, but also $\SF^0$; the 
presence of this latter field is awkward. Nonetheless,
we will find that there is a prescription we can
use to automatically remove $\SF^0$ from the diagrammatics,
so long as we accept some further corrections to
supersowing and supersplitting.

\begin{center}
\begin{figure}[h]
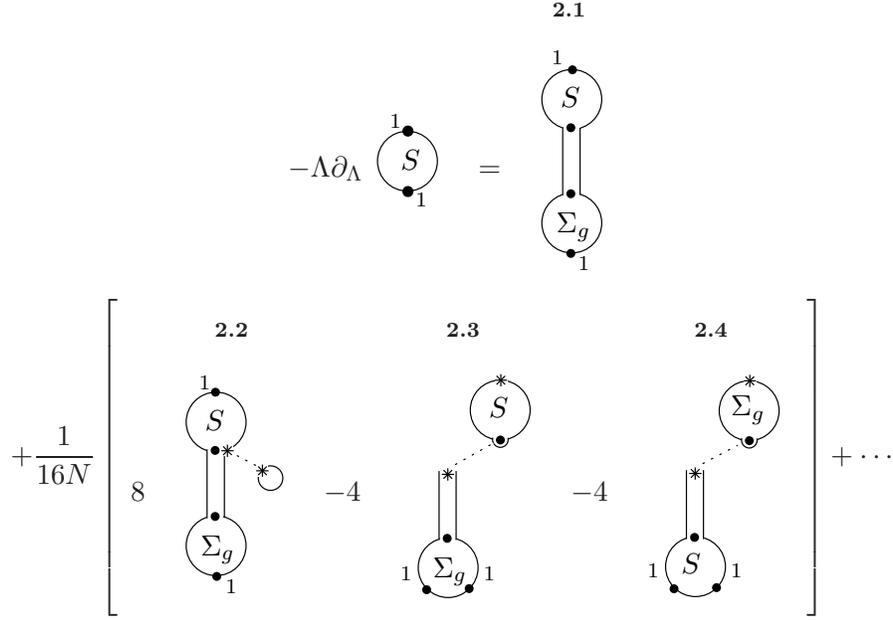

	\[	
	\begin{array}{c}
	\vspace{1ex}
		\begin{array}{ccc}
		\vspace{0.1in}
									&	& \LabExOb{Dumbbell-A1A1-Bars}
		\\
			-\flow \ensuremath{\begin{array}{c}\input{pstex/Vertex-A1A1.pstex_t} \end{array}} & = & \ensuremath{\begin{array}{c}\input{pstex/Dumbbell-A1A1-Bars.pstex_t} \end{array}}
		\end{array}
	\\
		\ds
		+
		\frac{1}{16N}
		\left[
			\begin{array}{cccccc}
			\vspace{0.1in}
					& \LabExOb{Dumbbell-A1A1-Bars-New-A}&	&\LabExOb{Bottle-A1A1-Unity}&	&\LabExOb{Bottle-A1A1-Unity-B}
			\\
				8	&\ensuremath{\begin{array}{c}\input{pstex/Dumbbell-A1A1-Bars-New-A.pstex_t} \end{array}}		&-4	&\ensuremath{\begin{array}{c}\input{pstex/Bottle-A1A1-Unity.pstex_t} \end{array}}		&-4	&\ensuremath{\begin{array}{c}\input{pstex/Bottle-A1A1-Unity-B.pstex_t} \end{array}}
			\end{array}
		\right]
		+ \cdots
	\end{array}
	\]
\caption{Classical part of the flow of a vertex decorated
by two $A^1$s.}
\label{fig:NFE:ClassicalFlow-A1A1}
\end{figure}
\end{center}

Before doing this, let us look at the diagrammatics
as would naturally follow from the previous chapter.
As an illustration, we will consider the
flow of a vertex decorated by two $A^1$s. 
In figure~\ref{fig:NFE:ClassicalFlow-A1A1}
we focus on the classical part of the flow equation.
Filled circles represent $A$s and, if they are
tagged with a `1', then they are restricted to
the $A^1$ sector. 
We have suppressed the Lorentz
indices of the decorative fields. Note also that we
have attached our false kernel to instances of $\sigma$,
which do not carry a position argument. If the $\sigma$
has replaced an $A$, then the position argument of
the $A$ is the information carried by the appropriate end
of the false kernel. However, the other end may attach to
a $\sigma$ with no associated field to provide a position
argument. In this case, 
the false
kernel does not carry a position space $\delta$-function,
but just serves to remind us how the diagrams of
which they comprise a part were formed.
The ellipsis represents the un-drawn diagrams spawned
by the quantum term.

There are a number of important points to make about the
diagrams of figure~\ref{fig:NFE:ClassicalFlow-A1A1}.

Diagram~\ref{Dumbbell-A1A1-Bars} is the only diagram
produced by a term from the original flow equation.
The factor of $1/2$ associated
with the original $a_0$ terms has been killed by the factor
of two coming from summing over the two possible locations of
the fields which decorate the dumbbell.
Notice that the differentiated fields are in the $A$-sector
only. Generically, the differentiated fields can also
be $F$s and $C$s. However, both of these
are forbidden in this case by our choice of
external fields: in the former case because $S^{AF}$
vertices do not exist and, in the latter case, because the
$S^{AC}$ vertex vanishes by \CC.

Now, given that the external fields are both $A^1$s,
it would perhaps be expected that the differentiated
fields are $A^1$s, also. To investigate this,
it is useful to construct the projectors
\begin{equation}
	\sigma_{+} \equiv
	\left(
		\begin{array}{cc}
			\one	&	0
		\\
				0	&	0
		\end{array}
	\right), \ \ 
		\sigma_{-} \equiv
	\left(
		\begin{array}{cc}
			0	&	0
		\\
			0	&	\one
		\end{array}
	\right).
\label{eq:NFE:Projectors}
\end{equation}
As we would expect from a pair of projectors,
$\sigma_+^2 = \sigma_+$, $\sigma_-^2 = \sigma_-$
and $\sigma_+\sigma_- = \sigma_-\sigma_+ = 0$.

We cannot take the differentiated field to
be just an $A^1$, since this would throw away the effects of $\SF^0$
which, for the time being, we must explicitly keep.
In other words, we can think of the differentiated
field as being $\sigma_+ \SF \sigma_+$, but note that
this has an $\SF^0$ component. The corrections
to supersowing and splitting arising from differentiating
the partial supermatrix $\sigma_+ \SF \sigma_+$ are
just insertions of $\sigma_+$ which, for classical
terms at any rate, can just be ignored, since they
do not change the supertrace structure.

Let us now focus on the wine in this diagram. Coming
from one of the original terms in the flow equation,
the wine must be just $\dot{\Delta}^{AA}$. 
The labels of the wine imply that it attaches at
each end to $\delta / \delta A$. There is no
problem with this as we can take the functional
derivative to act \wrt\ a full $A$, the unwanted
components being killed by the $\sigma_+$s. Finally,
we note that there cannot be any instances of $\sigma$ on the wine: since
such $\sigma$s act by commutation, the wine dies by virtue of the fact that
$[\sigma, \delta / \delta A] =0$.

Let us now turn to diagram~\ref{Dumbbell-A1A1-Bars-New-A}. This
is a combination of four diagrams produced by the additions
to the flow equation, 
corresponding to the four ends of the wine on which the
$\sigma$ can sit. Since these diagrams are equivalent,
we can add them. Summing over the two locations of the decorative fields
gives an overall relative factor of eight, as shown. The wine is  
$\dot{\Delta}^{AA}_\sigma$. That it must be this, and not $\dot{\Delta}^{AA}$,
can be deduced by the presence of the $\sigma$ at one of its ends.
The final point to note about this diagram is the extra supertrace, which
is just $\str \sigma = 2N$. Thus we see that this diagram has
the same overall factor as the first diagram.

Diagram~\ref{Bottle-A1A1-Unity} necessitates some new notation. The vertex
with argument $S$ has a feature looking like \ReplaceAwSigma. This
tells us that an $A$ has been differentiated and replaced with
a $\sigma$. However, the vertex on which this feature is present already
has a $\sigma$ and so we combine them, to leave $\str \one =0$. 
In fact, we need never have drawn this diagram. The $S$ vertex
coefficient function is $S^{A \sigma}$; single $A$ vertices 
vanish by both \CC\ and Lorentz invariance. In anticipation of what is to come, though, we note
that multiple supertrace
terms can have separate $\str A \sigma$ factors. This will
play a key role in what follows, since $\str A \sigma$
is none other than $2N \SF^0$.

Diagram~\ref{Bottle-A1A1-Unity-B} also
vanishes for the reasons just discussed.

Thus, to recap, only the first two diagrams survive. 
They come with the same relative factor, have exactly
the same vertices and so can be combined. The result
will be that the two vertices are now joined by the
sum of $\dot{\Delta}^{AA}$ and $\dot{\Delta}^{AA}_\sigma$.
It seems natural to define
\begin{equation}
	\dot{\Delta}^{A^1A^1} \equiv \dot{\Delta}^{11} \equiv \dot{\Delta}^{AA} + \dot{\Delta}^{AA}_\sigma.
\label{eq:Wine-A1A1:Defn}
\end{equation}

We can do an analogous analysis in the $A^2$ sector. The key difference
here is that the embedded $\sigma$ of the analogue of diagram~\ref{Dumbbell-A1A1-Bars-New-A}
gives rise to a minus sign since, whilst $\str A^1 A^1 \sigma = \str A^1 A^1$,
$\str A^2 A^2 \sigma = -\str A^2 A^2$. In this case, we are led
to the definition
\begin{equation}
	\dot{\Delta}^{A^2A^2} \equiv \dot{\Delta}^{22} \equiv \dot{\Delta}^{AA} - \dot{\Delta}^{AA}_\sigma.
\label{eq:Wine-A2A2:Defn}
\end{equation}

However, neither~(\ref{eq:Wine-A1A1:Defn}) or~(\ref{eq:Wine-A2A2:Defn})
work quite as we would like: the wines $\dot{\Delta}^{AA}$ and $\dot{\Delta}^{AA}_\sigma$
do not attach to just $A^1$ or $A^2$, as the labels of $\dot{\Delta}^{11}$ 
and $\dot{\Delta}^{22}$ seem to imply; 
we know that there is also attachment to $\SF^0$.

Nonetheless, both the physics of the situation and the diagrammatics seem to be guiding us to a formalism
where we work with $A^1$s and $A^2$s. It turns out that the most
efficient way to proceed is to follow this lead. Recalling
that
\[
	\fder{}{\SF_\mu} = 2 T_A \fder{}{\SF_{A\mu}} + \frac{\sigma}{2N} \fder{}{\SF_\mu^0}
\]
we now split up the first term into derivatives \wrt\ $A^1$ and $A^2$:
\begin{equation}
	\fder{}{A_\mu} = 2 \sigma_+ \tau_1^{a} \fder{}{A_{\mu a}^1}  - 2 \sigma_- \tau_2^a \fder{}{A_{\mu a}^2}  + \frac{\sigma}{2N} \fder{}{\SF_\mu^0},
\label{eq:SF-deriv-cmpts}
\end{equation}
where $\tau_i^a$ are the generators of the $SU(N)$ gauge theories carried by $A^i$, for $i=1,2$,
normalised to $\tr \tau_i^a \tau_j^b = 1/2 \delta^{ab} \delta_{ij}$.
We now exploit no-$\SF^0$ symmetry and so use the prescription that, since all 
\emph{complete} functionals (as opposed to individual contributions---see below)
are independent of $\SF^0$, we can take $\delta/ \delta \SF^0$ not to act.

At this point it is worth pausing to consider this in more detail.
Let us return to figure~\ref{fig:NFE:ClassicalFlow-A1A1}. We noted
in the first diagram that, tempting as it was to take both
vertices to be $S^{A^1 A^1}$, we should instead take
them to be $S^{A^1 A}$ vertices. The thing stopping us 
from working with just $A^1$s is
the differentiated field: we are of course at liberty to specialise
the decorative fields to be whatever we want. Now, however, we
are supposing the $\delta / \delta \SF^0$ does not act; in this
case, we really can take the vertices to be $S^{A^1 A^1}$.

In this picture, diagrams~\ref{Dumbbell-A1A1-Bars} and~\ref{Dumbbell-A1A1-Bars-New-A}
still combine but now, with all fields in the $A^1$ sector,
it is natural to talk of the vertices being joined by $\dot{\Delta}^{11}$.

What of the final two diagrams? We have already noted that they vanish
anyway, but this is only an accidental feature of the particular
vertex whose flow we chose to compute. The real point is
that they both contain a $\str \sigma A = \str \SF^0$, which is
attached to a wine. In our new picture, we simply never draw such
diagrams: we are not interested
in external $\SF^0$s and need not consider internal
ones because we do not allow $\delta / \delta \SF^0$ to act.

It is important to realise, though, that this does not mean that the action does
not contain $\str \sigma A$ components; indeed, precisely the opposite is necessary
for no-$\SF^0$ symmetry to be preserved. To see this, consider part of the action:
\[
	\cdots + \frac{1}{2} S^{AA\sigma} \str{AA\sigma} + \frac{1}{2!} S^{A\sigma, A\sigma} \str A \sigma \str A \sigma + \cdots.
\]
No-$\SF^0$ symmetry, whose lowest order effects can be deduced by letting $\delta A_\mu = \lambda \one$~\cite{YM-1-loop, Antonio'sThesis}
tells us that
\begin{equation}
	 S^{AA\sigma} \str{A\sigma} + S^{A\sigma, A\sigma} \str \sigma \str A \sigma =0,
\label{eq:Intro:No-A0-Ex}
\end{equation}
\ie\ that $S^{AA\sigma} = -2N  S^{A\sigma, A\sigma}$.

Thus, the presence of the $S^{A\sigma, A\sigma}$ vertex in the action is necessary to 
ensure no-$\SF^0$ symmetry; the point is that, when using the flow equations,
we can choose never to attach to such objects.

Moreover, there is an additional simplification we find when
working with $A^1$s and $A^2$s: all diagrams involving
\ReplaceAwSigma vanish. This is because the differentiated field
is now restricted to being an $A^1$ or $A^2$. However,
\[
	\str \sigma \fder{}{A} = \tr \sigma_+ \fder{}{A^1} - \tr \sigma_- \fder{}{A^2} \equiv 0.
\]
Had we still been using exact supersowing and supersplitting, $\str \sigma \delta / \delta A$
terms can survive, via the extension of $\SF$ to $\SF^e$ (\cf\ equation~(\ref{eq:ExtendedFunctionalDeriv})). 

\begin{center}
\begin{figure}[h]
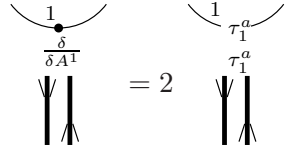

	\[
		\ensuremath{\begin{array}{c}\input{pstex/Attach-A1.pstex_t} \end{array}} = 2 \ensuremath{\begin{array}{c}\input{pstex/Attach-A1-B.pstex_t} \end{array}}
	\]
\caption{Attachment of a wine via $\delta / \delta A^1$.}
\label{fig:NFE:Attachement-A1}
\end{figure}
\end{center}

There is, though, a price to pay for all this: supersowing and supersplitting
both receive new corrections, which we henceforth refer to
as attachment corrections. To compute these corrections, consider
attachment of a wine via $\delta / \delta A^1$, as shown in
figure~\ref{fig:NFE:Attachement-A1}. 

Note that, in the second diagram, we have retained the `1'
which previously labelled the $A^1$, to remind us that
the flavours of any fields which followed or preceded
the $A^1$ are restricted. 

Next, we use the completeness relation for $SU(N)$,
\begin{equation}
	2 \left( \tau_1^a \right)^i_{\ j} \left( \tau_1^a \right)^k_{\ l} = \delta^i_{\ l} \delta^k_{\ j} - \frac{1}{N} \delta^i_{\ j} \delta^k_{\ l},
\label{eq:SU(N)-Completeness}
\end{equation}
to obtain figure~\ref{fig:NFE:Attachement-A1-B}, where attachments like those in
the first column of diagram on the \rhs\ will be henceforth
referred to as direct.
\begin{center}
\begin{figure}[h]
	\begin{eqnarray*}
		\ensuremath{\begin{array}{c}\input{pstex/Attach-A1.pstex_t} \end{array}} 	& = 	& \ensuremath{\begin{array}{c}\input{pstex/Attach-A1-C.pstex_t} \end{array}} - \frac{1}{N} \ensuremath{\begin{array}{c}\input{pstex/Attach-A1-D.pstex_t} \end{array}}
	\\
						& \equiv& \ensuremath{\begin{array}{c}\input{pstex/Attach-A1-C.pstex_t} \end{array}} -\frac{1}{N} \ensuremath{\begin{array}{c}\input{pstex/Attach-A1-E.pstex_t} \end{array}}
	\end{eqnarray*}
\caption{A re-expression of figure~\ref{fig:NFE:Attachement-A1}.}
\label{fig:NFE:Attachement-A1-B}
\end{figure}
\end{center}

There is a very similar expression for the $A^2$ sector. Now, however, $\sigma_+$s
are replaced with $\sigma_-$s and the sign of the $1/N$ contribution flips, which
can be traced back to~(\ref{eq:supermetric}). We can also understand the 
sign flip heuristically because we are tying everything
back into supertraces and not traces; we recall that the supertrace 
yields the trace of the bottom block-diagonal of a supermatrix but picks
up a minus sign (see \eg\ equation~(\ref{eq:SF-deriv-cmpts})).

In the $B, D$ and $C$ sectors we do not get any $1/N$ 
attachment corrections. Derivatives \wrt\
the fields $B$ and $\bar{B}$,
can simply be written
\begin{eqnarray}
	\fder{}{B}		& = & \sigma_- \fder{}{\SF}\sigma_+,
\label{eq:SigmaAlg-B}
\\
	\fder{}{\bar{B}}& = & \sigma_+ \fder{}{\SF}\sigma_-,
\label{eq:SigmaAlg-Bbar}
\end{eqnarray}
and so derivatives \wrt\ these partial superfields just yield
insertions of $\sigma_\pm$.

In the $C$-sector, $\SH^0$ does not play a privileged role and
so has not been factored out of the definition of $C^1$ and $C^2$
(see equation~(\ref{eq:SH-defn})). Hence, derivatives \wrt\ the components
of $\SH$ simply yield insertions of $\sigma_\pm$. 

Let us now reconsider the classical part of the flow of
a vertex decorated by two fields. These fields can be any
of $A^1, \ A^2, \ C^1, \ C^2, \ F$ and $\bar{F}$ (though
certain choices \eg\ $A^1A^2$ correspond to a vanishing vertex).
The diagrammatics is shown in figure~\ref{fig:NFE:ClassicalFlow-TP},
where we have neglected not only the quantum terms but also
any diagrams in which the wine is decorated.\footnote{
Such diagrams only exist in the $C$-sector, in this case, since
this is the only sector for which one-point (seed action) vertices
exist.}
\begin{center}
\begin{figure}[h]
	\[
	\begin{array}{cccc}
	\vspace{0.1in}
								&	& \LabExOb{Dumbbell-TP} &
	\\
		-\flow\ensuremath{\begin{array}{c}\input{pstex/Vertex-TP.pstex_t} \end{array}} 	&=	& \ensuremath{\begin{array}{c}\input{pstex/Dumbbell-TP.pstex_t} \end{array}} 		&+ \cdots
	\end{array}
	\]
\caption{Classical part of the flow of a two-point vertex decorated
by the fields $A^1, \ A^2, \ C^1, \ C^2, \ F$ and $\bar{F}$.}
\label{fig:NFE:ClassicalFlow-TP}
\end{figure}
\end{center}

The differentiated fields now label the wines, where
we take $\DEP{C^1}{C^1} = \DEP{C^2}{C^2} = \DEP{C}{C}$
and $\DEP{\bar{F}}{F} = \DEP{F}{\bar{F}}$. The only subtlety
comes in the $A^{1,2}$ sectors, where we know that
there are corrections, which have been implicitly absorbed
into the Feynman rules. We will be more specific about
the corrections in this case---\ie\ where the wine is 
undecorated.

From figure~\ref{fig:NFE:Attachement-A1-B} we obtain 
the relation of figure~\ref{fig:NFE:Attachment-A1A1}.
\begin{center}
\begin{figure}[h]
	\[
		\ensuremath{\begin{array}{c}\input{pstex/Attach-A1A1.pstex_t} \end{array}} =	 \ensuremath{\begin{array}{c}\input{pstex/Attach-A1A1-A.pstex_t} \end{array}} -\frac{1}{N} \ensuremath{\begin{array}{c}\input{pstex/Attach-A1A1-B.pstex_t} \end{array}} -\frac{1}{N} \ensuremath{\begin{array}{c}\input{pstex/Attach-A1A1-C.pstex_t} \end{array}} +\frac{1}{N^2} \ensuremath{\begin{array}{c}\input{pstex/Attach-A1A1-D.pstex_t} \end{array}}
	\]
\caption{The attachment corrections for
 $\DEP{A^1}{A^1}$.}
\label{fig:NFE:Attachment-A1A1}
\end{figure}
\end{center}

The expression in figure~\ref{fig:NFE:Attachment-A1A1} now simplifies. The loop in
the middle of the final diagram is decorated only by $\sigma_+^2$ and so
yields $\str \sigma_+ = N$. This diagram then cancels either of those
with factor $-1/N$. We now redraw the remaining diagram with factor $-1/N$,
as shown in figure~\ref{fig:NFE:Attachment-A1s,A2s}, together
with a similar expression in the $A^2$ sector.
\begin{center}
\begin{figure}[h]
	\begin{eqnarray}
		\ensuremath{\begin{array}{c}\input{pstex/Attach-A1A1.pstex_t} \end{array}} &=	& \ensuremath{\begin{array}{c}\input{pstex/Attach-A1A1-A.pstex_t} \end{array}} -\frac{1}{N} \ensuremath{\begin{array}{c}\input{pstex/Attach-A1-A1-N.pstex_t} \end{array}}
	\label{eq:Attach-A1A1}
	\\[2ex]
		\ensuremath{\begin{array}{c}\input{pstex/Attach-A2A2.pstex_t} \end{array}} &=	& \ensuremath{\begin{array}{c}\input{pstex/Attach-A2A2-A.pstex_t} \end{array}} +\frac{1}{N} \ensuremath{\begin{array}{c}\input{pstex/Attach-A2-A2-N.pstex_t} \end{array}}
	\label{eq:Attach-A2A2}
	\end{eqnarray}	
\caption{The attachment corrections for
$\DEP{A^1}{A^1}$ and $\DEP{A^2}{A^2}$.}
\label{fig:NFE:Attachment-A1s,A2s}
\end{figure}
\end{center}

Then meaning of the double dotted lines and the associated field
hiding behind the line of the supertrace should be clear: 
the double dotted lines stand for $\DEP{A^i}{A^i}$, and the sub-sector of the
associated fields tells us whether $i=1$ or $2$. (We will use
the term sub-sector whenever we want to specify that some 
unlabelled field is a component of the one of the fields $A$, $C$
or $F$.)

Returning now to diagram~\ref{Dumbbell-TP} (figure~\ref{fig:NFE:ClassicalFlow-TP})
we should interpret the
wine when the dummy fields are in the $A^1$ or $A^2$ sectors
according to~(\ref{eq:Attach-A1A1}) and~(\ref{eq:Attach-A2A2}).

In our analysis of the terms spawned by the classical part of the flow
equation, we have so far restricted ourselves to diagrams in which the
wine is undecorated and for which there are no insertions of the dynamical
components of $\SH$ (recall that such 
insertions occur as in  figure~\ref{fig:NewFlowEquation:Details}).

Let us first relax just the former restriction and analyse
the mapping into the $A^1$, $A^2$ basis.  
The simplest case to deal with is where the 
decorations of the wines $\{\DEP{\SF}{\SF}\}$ and $\{\dot{\Delta}_\sigma^{\SF\SF}\}$
are, on each side \emph{independently}, net bosonic. In this
scenario, we are just mapped into the $A^1$, $A^2$ basis,
as before. 

The next case to examine is where the decorations on one side
of the wine are net bosonic but those on the other side are net
(anti) fermionic. This immediately tells us that one of the functional
derivatives sitting at the end of the wine must be 
(anti) fermionic. Now, as before, we would like to pair up diagrams
with a $\{\DEP{\SF}{\SF}\}$ wine with those with a $\{\dot{\Delta}_\sigma^{\SF\SF}\}$ wine.
However, there is a subtlety: $\sigma$ anticommutes with net fermionic
structures. Thus, of the four terms generated
 with a $\sigma$ at the end of one side of the wine
attached to a separate $\str \sigma$, two cancel.
Hence, the vertex from $\{\dot{\Delta}_\sigma^{\SF\SF}\}$
has a factor of half relative to the vertex from
$\{\DEP{\SF}{\SF}\}$. 

We can still choose a prescription to map us into the $A^1$, $A^2$ basis
by absorbing this factor into our definition of 
\eg\ $\dot{\Delta}^{1B,F \cdots,\cdots}$, where both
ellipses denote net bosonic decorations.\footnote{
Note that the $B$ denotes a functional derivative \wrt\
$B$; since this removes a fermion, the wine decorations
must be net fermionic.}

The final case to examine is where the decorations on 
one side of the wine are net fermionic whilst those on the
other side are net anti-fermionic.\footnote{If the decorations
on both sides are (anti) fermionic, then the vertex belongs
to $\{\DEP{F}{\bar{F}} \}$.}
The functional 
derivatives must both be bosonic and in 
separate sub-sectors.\footnote{
It is straightforward to check that one cannot construct a legal
wine of this type for which the functional derivatives
are fermionic and anti-fermionic.
}
There are no 
surviving contributions involving $\{\dot{\Delta}_\sigma^{\SF\SF}\}$
vertices.
Nonetheless, we can still define an object $\{\dot{\Delta}^{A^1 A^2}\}$
though we note that it must have net fermionic decorations on both sides.

It is important to realise that the broken phase form of
equation~\eq{eq:WineRelation-Diag}
(which we recall follows from the  coincident line identities)
for wines possessing fermionic decorations mixes contributions
from different wines. This is because moving a fermionic decoration
from one side of the wine, though a bosonic functional derivative, to the other 
side 
changes the sub-sector of the functional derivative. This is illustrated
in figure~\ref{fig:NFE:CLI-FlavourChange}, where we explicitly indicate
which portions of the wine are unchanged by insertions of $\sigma_\pm$.
Fermionic fields are denoted by diamonds.
\begin{center}
\begin{figure}[h]
	\[
		\ensuremath{\begin{array}{c}\input{pstex/Wine-1BF.pstex_t} \end{array}} \hspace{2em} + \hspace{1.5em} \ensuremath{\begin{array}{c}\input{pstex/Wine-2B-F.pstex_t} \end{array}} \hspace{2em} = 0
	\]
\caption{Example of how the broken phase form of~\eq{eq:WineRelation-Diag} mixes different wines.}
\label{fig:NFE:CLI-FlavourChange}
\end{figure}
\end{center}

We might worry that this picture is incompatible with gauge invariance:
$\delta / \delta A^1$ and $\delta / \delta A^2$ necessarily
strike different vertices and so it is perhaps not 
immediately obvious why the  inhomogeneous part of $\delta (\delta / \delta A)$
will cancel.

However, under gauge transformations, we recall that the inhomogeneous
part of $\delta (\delta / \delta A) \sim \one \tr \comm{\fder{}{\SF}}{\Omega}$.
In both diagrams of figure~\ref{fig:NFE:CLI-FlavourChange}, the $\one$
plugs the top of the wine (it does not matter that the $\one$ reduces
to $\sigma_+$ in one case and $\sigma_-$ in the other, since there are other
fields on the same portion of supertrace). The associated
functional derivative is inside a separate trace, and so can strike the same thing in both diagrams.
Thus, under gauge transformations of $\delta / \delta A^1$ and  $\delta / \delta A^2$,
the non-homogeneous parts still vanish for the usual reasons, but we must be
careful to recognise that the broken phase form of~\eq{eq:WineRelation-Diag}
involves vertices from different wines.

Let us now relax the second restriction above and so allow insertions
of the dynamical components of $\SH$.
We have seen how diagrams with insertions of $\sigma$---linked with
a separate $\str \sigma$---arising from the new
terms in the flow equation, can be combined with  diagrams 
spawned by the original terms in the flow equation, in certain
cases. However, diagrams with
insertions of the dynamical components of $\SH$ can never be similarly combined.
Hence, such diagrams occur only with $\dot{\Delta}_\sigma^{\SF\SF}$ wines
and, moreover, there is no natural way of mapping such terms 
into the $A^1$, $A^2$ basis. There
is, of course, nothing to stop us performing functional derivatives
\wrt\ $A^1$ and $A^2$ only---indeed, this is what we will do. However,
the  differentiated fields do not now label the wines in a natural
way.

In fact, as we will see, considerations such as this
ultimately do not bother us: we will find that, in our calculation
of $\beta$-function coefficients, all these awkward terms 
cancel, between themselves. Hence, our strategy is to lump
all the unpleasant terms into our diagrammatic rules, so that
they essentially become hidden. We can, at any stage, unpack them
if necessary but, generally speaking, we will find that we do not
need to do so. 

To this end, we note that insertions of $\SH$ always occur at
the ends of wines, and so include them in our diagrammatic
rules for the wines. The actual interpretation is easy, but must
be done with care.
As a first example, consider unpacking a wine
with two dummy fields at the end, `decorated' by a $C^1$ and a $\sigma$. This is
shown in figure~\ref{fig:NFE:Unpack-C1}, where the ellipsis denotes wines 
arising from the new terms in the flow equation whose associated
functional derivatives are fermionic.
\begin{center}
\begin{figure}[h]
	\begin{eqnarray*}
		\lefteqn{\ensuremath{\begin{array}{c}\input{pstex/Wine-Dummy-C1.pstex_t} \end{array}} \rightarrow \ensuremath{\begin{array}{c}\input{pstex/Wine-A1A1-C.pstex_t} \end{array}} \ + \ensuremath{\begin{array}{c}\input{pstex/Wine-C1C1-C1.pstex_t} \end{array}} \ + \ensuremath{\begin{array}{c}\input{pstex/Wine-FF-C1.pstex_t} \end{array}}}
	\\
	&&
		+\frac{1}{32N}	
		\left[
			\begin{array}{c}
			+ 2N \ensuremath{\begin{array}{c}\input{pstex/Wine-A1-A1-Embed-C1.pstex_t} \end{array}} + 2N \ensuremath{\begin{array}{c}\input{pstex/Wine-A1-A1-Embed-C1-B.pstex_t} \end{array}} + \ensuremath{\begin{array}{c}\input{pstex/Wine-A1-A1-Embed-sigma--C.pstex_t} \end{array}} + \ensuremath{\begin{array}{c}\input{pstex/Wine-A1-A1-Embed-sigma--Cb.pstex_t} \end{array}}
		\\[10ex]
			- \ensuremath{\begin{array}{c}\input{pstex/Wine-PlugSig-T-C1.pstex_t} \end{array}} \hspace{3em} - \ensuremath{\begin{array}{c}\input{pstex/Wine-PlugSig-T-C1-B.pstex_t} \end{array}} \hspace{3em} +\cdots 
		\end{array}
		\right]
	\end{eqnarray*}
\caption{Unpacking a wine, `decorated' by a $C^1$ and a $\sigma$.}
\label{fig:NFE:Unpack-C1}
\end{figure}
\end{center}

It is thus apparent that the decoration of the parent wine by $C^1$ 
and $\sigma$
is not a decoration of a wine in the usual sense. Whilst the parent has
components in which $C^1$ and $\sigma$ decorate the wine in the manner
we are used to (the first three
diagrams on the \rhs), it is also clear that it has components in
which either the $C^1$ or $\sigma$ is embedded at the end of the wine.
However, it makes sense to recognise these embedded fields as decorations. By doing
this, we are essentially generalising the notion of a wine to 
include multiple supertraces. Nonetheless, we are mindful that
these new decorations behave differently from the ones
we have encountered so far. In particular, since embedded components
of $\SH$ do not act via commutation, wine vertices involving
these embedded fields do not participate in relations such as~\eq{eq:WineRelation-Diag}.

The factor of $1/32N$ deserves comment, since one would expect it to
be a factor of two larger, from the new flow equation. However,
the terms with the $\dot{\Delta}^{AA}_\sigma$ wine can also 
be generated from a parent diagram in which the decorative $C^1$
and $\sigma$ are in a different order. Hence, we define the diagrammatic
rules to compensate for this.

It may seem like we have missed two diagrams, in which $C^1$ plugs one end of the wine.
However, this leaves
us with a $\sigma \delta / \delta A$ which vanishes, in our picture,
as we have already discussed.

Whilst our example involves $C^1$ and $\sigma$, it is clear that
we could instead have taken any pair of 
fields belonging to $\SH$,
though we must be very careful taking two instances
of $\sigma$. If one of them is embedded at the end of the wine
and the other is on a lone supertrace,
then the corresponding diagram has been used, already,
to map us into the $A^1$-$A^2$ basis. However,
if one of the instances is a `normal' decoration, then such
diagrams should be counted. 
If there are two instances of $\sigma$ on a wine, it would
be natural to eliminate them via $\sigma^2 = \one$. Now,
however, it is not necessarily correct to do this,
as one could be on a different supertrace.

The interpretation of additional decorations of the parent is straightforward. 
Any instances of
the components of $\SF$ are just treated like normal decorations of the wine.
However, instances of the components of $\SH$ can act as either
normal wine decorations or can be embedded at the ends. In the latter
case, they must be accompanied by an additional component of $\SH$,
to create a valid diagram. 

This completes the
analysis of the diagrammatics we have performed for the new flow equation.
We have not explicitly looked at the quantum term, but there are no new
considerations in this case. Thus, we can summarise the prescription
that we use in the broken phase.
\begin{enumerate}
	\item	All decorative fields are instances of
			$A^1$, $A^2$, $F$, $\bar{F}$, $C^1$, $C^2$ and $\sigma$;
			$\SF^0$ is excluded.

	\item	Differentiation is \wrt\ all dynamical fields, above, where:
			\begin{enumerate}
				\item	differentiation \wrt\ $A^1$ or $A^2$ leads to 
						attachment corrections of the type
						shown in figure~\ref{fig:NFE:Attachement-A1-B};
				
				\item	differentiation \wrt\ all other fields just involves
						insertions of $\sigma_\pm$;

				\item	diagrams involving $\str \sigma \delta / \delta A^{1,2}$
						vanish, identically.
			\end{enumerate}

	\item	Full diagrams without any insertions of the components of $\SH$
			at the ends of the wine are naturally written in terms of
			the above fields and their corresponding kernels \ie\ $A^1$s
			attach to a (decorated) $\DEP{1}{1}$ etc.
		
	\item	Full diagrams with insertions of the components of $\SH$ are restricted
			to those for which at least one insertion is not a $\sigma$. These
			diagrams involve the wine $\dot{\Delta}^{AA}_\sigma$ but, for convenience,
			are packaged together with the decorated wines of the previous item.
\end{enumerate}

\section{The New Diagrammatics-II}	\label{sec:NFE:NewDiags-II}

\subsection{Construction}	\label{sec:NFE:DiagsII-Construction}

The work of the previous section now guides us to
a more compact and intuitive diagrammatics, which
is considerably easier to deal with. By packaging up
the remaining $\dot{\Delta}^{AA}_\sigma$ wines
with decorated instances of the other wines, 
we have taken a step in the right
direction. In anticipation that these compact, packaged
objects cancel in their entirety when we perform actual
calculations, it clearly makes sense
to bundle together wines of a different flavour. However,
we have really done more than that, as we now point out.

Considering a full diagram with genuine decorations on one 
side of its wine, we know that
these decorations  are all on the same supertrace. However, 
fields embedded at the ends of a wine attach, via a false
kernel, to fields which can be on an entirely
different supertrace. Thus, by packaging together terms 
as we did in the previous section, we have started to
combine diagrams with differing supertrace structures.
In this section, we extend this to its natural 
conclusion.

The basic idea is that, rather than considering diagrams
with a specific supertrace structure, we instead sum 
over all legal supertrace structures, consistent with
the decorative fields. Thus, let us suppose
that we wish to compute the flow of all vertices
which can be decorated by the set of fields $\{f\}$.
The new flow equation takes a very simple,
intuitive form, as shown in figure~\ref{fig:NFE:New-Diags-1}.
\begin{center}
\begin{figure}[h]
	\[
	-\flow 
	\dec{
		\ensuremath{\begin{array}{c}\input{pstex/Vertex-S.pstex_t} \end{array}}
	}{\{f\}}
	=
	\frac{1}{2}
	\dec{
		\ensuremath{\begin{array}{c}\input{pstex/Dumbbell-S-Sigma_g.pstex_t} \end{array}} - \ensuremath{\begin{array}{c}\input{pstex/Vertex-Sigma_g.pstex_t} \end{array}}
	}{\{f\}}
	\]
\caption{The diagrammatic form of the flow equation, when we treat
single and multiple supertrace terms together.}
\label{fig:NFE:New-Diags-1}
\end{figure}
\end{center}

Let us now analyse each of the elements of figure~\ref{fig:NFE:New-Diags-1} in turn.
On the \lhs, we have the set of vertices whose flow we are computing.
This set comprises all cyclically independent arrangements of the fields $\{f\}$,
over all possible (legal) supertrace structures.
When we specify
the fields $\{f\}$, we use a different notation from before. As an example,
consider $\{f\} = \{A^1_\mu, A^1_\nu, C^1\}$, shown in figure~\ref{fig:NFE:NewVertices-Ex}.
Note that we have not drawn any vertices comprising a supertrace decorated only by an $A^1$,
since these vanish.
\begin{center}
\begin{figure}[h]
	\[
	\dec{
		\ensuremath{\begin{array}{c}\input{pstex/Vertex-S.pstex_t} \end{array}}
	}{A^1_\mu A^1_\nu C^1}
	\equiv
	\ensuremath{\begin{array}{c}\input{pstex/Vertex-S-A1A1C1.pstex_t} \end{array}}
	=
	\ensuremath{\begin{array}{c}\input{pstex/Vertex-S-A1_muA1_nuC1.pstex_t} \end{array}} + \ensuremath{\begin{array}{c}\input{pstex/Vertex-S-A1_nuA1_muC1.pstex_t} \end{array}} + \ensuremath{\begin{array}{c}\input{pstex/Vertex-S-A1_nuA1_mu-C1.pstex_t} \end{array}}
	\]
\caption{A new style vertex decorated by two $A^1$s and a $C^1$.}
\label{fig:NFE:NewVertices-Ex}
\end{figure}
\end{center}

It is apparent that we denote $A$s by wiggly lines and $C$s by dashed lines. 
A wildcard field will be denoted by a solid line.

Notice how, in the new style diagram, we explicitly indicate the sub-sector of
all the fields. This is because there is no need for them to be on the same supertrace
and so, for example, there is nothing to prevent an $A^1A^1C^2$ vertex. In the old
notation, however, all fields on the same circle are on the same supertrace and
so, once we know the sub-sector of one field, the sub-sectors of the remaining fields
on the same circle follow, uniquely.

To symbolically represent the new vertices, we will somewhat loosely write \eg\
$S_{\mu \nu}^{1 \, 1 C^{1}}$. If we need to emphasise that we are using the
new style diagrammatics, as opposed to the old style diagrammatics, then
we will write $S_{\ \, \mu \nu}^{\{ 1 \, 1 C^{1} \}}$, reminding us that the
fields are arranged in all cyclically independent ways
over all possible (legal) supertrace structures.

With these points in mind, let us return to figure~\ref{fig:NFE:New-Diags-1}. The diagrams
on the \rhs\ both involve the structure \DummyKernel. This is a dummy kernel which
attaches, at either end, to dummy fields. 
The fields at the ends can be any of $A^1$, $A^2$, $C^1$,
$C^2$, $\bar{F}$ or $F$, so long as the corresponding diagram actually exists. 
The dummy
kernel can be decorated by any subset of the fields $\{f\}$ where, if a
pair of decorative fields are both components of $\SH$ (and one of them is dynamical),
then we include the possibility that the kernel can be of the type $\dot{\Delta}^{\SF\SF}_\sigma$.
In this case, we note that there are implicit factors of $1/16N$.

The relationship between the new diagrammatics for the wines and the old diagrammatics
is straightforward, and is illustrated in figure~\ref{fig:NFE:NewKernel-Ex}
for the case of a new-style wine decorated by a single $A^1$.
\begin{center}
\begin{figure}[h]
	\[
	\dec{
		\ensuremath{\begin{array}{c}\input{pstex/DummyKernel.pstex_t} \end{array}}
	}{A^1}
	\equiv
	\ensuremath{\begin{array}{c}\input{pstex/DummyKernel-A1.pstex_t} \end{array}}
	=
	\ensuremath{\begin{array}{c}\input{pstex/DummyKernel-A1-R.pstex_t} \end{array}} + \ensuremath{\begin{array}{c}\input{pstex/DummyKernel-A1-L.pstex_t} \end{array}}
	\]
\caption{A new style (dummy) wine decorated by a single $A^1$.}
\label{fig:NFE:NewKernel-Ex}
\end{figure}
\end{center}

Having described the new diagrammatics for vertices and kernels,
we are nearly ready to complete our interpretation of figure~\ref{fig:NFE:New-Diags-1}.
Before we do so, however, we use our new notation to hide one further detail:
instances of $\sigma$.

When using the old flow equation, instances of $\sigma$ mixed 
fields from different sub-sectors \eg\ $\sigma C = C^1\sigma_+ - C^2\sigma_-$.
Now, however, instances of $\sigma$ yield a constant factor of 
either $\pm \one$ or $2N$. Thus, instances of $\sigma$
can be replaced by numerical factors accompanying wines / vertices.
We need only remember that the multiple supertrace decorations
of wines arising from the new terms in the flow equation exist.
Such terms require two instances of $\SH$ (at least one of
which we take to be dynamical). 
In the case that one of these fields is a $\sigma$,
we must remember that it is now hidden.

With these implicit instance of $\sigma$ in mind, the interpretation
of the \rhs\ of figure~\ref{fig:NFE:New-Diags-1} is simple:
the decorative fields $\{f\}$ are distributed around the two diagrams
in all possible, independent ways.

\subsection{Identical Fields}	\label{sec:NFE:IdenticalFields}

Let us reconsider the flow of a vertex in our
new diagrammatic approach by looking again at
figure~\ref{fig:NFE:New-Diags-1}. Suppose that we wish to
explicitly decorate with the set of fields $\{f\}$.
In the case that some of these fields are identical---we
will be more precise about the meaning of this shortly---the
set of fully decorated diagrams on the \rhs\ can be simplified.

When explicitly decorating a diagram,
we sum over all independent permutations of the fields. The
individual components of each diagram---vertices and wines---automatically
include all permutations of any decorations. So, the rule is
that we must divide the fields $\{f\}$ in all possible ways between
the number of structures. Suppose we have two identical fields, $X$ and
$Y$, and suppose further that we have drawn the
diagram in which $X$ decorates structure $U$ and $Y$ decorates structure $V$.
Rather than also counting the diagram in which $X$ decorates $V$ and
$Y$ decorates $U$, we can take just the first case and multiply by two.

As an example, we will compute the flow of a vertex decorated
by $A^1_\mu(p)$ and $A^1_\nu(-p)$. These fields are counted as identical:
by Bose symmetry, the diagrams which they 
decorate are invariant under $p_\mu \leftrightarrow -p_\nu$. 
Thus, we obtain the diagrams of figure~\ref{fig:NFE:flow-A1A1}, where we have
suppressed both Lorentz indices and sub-sector labels.
\begin{center}
\begin{figure}[h]
	\begin{eqnarray*}
		-\flow \ensuremath{\begin{array}{c}\input{pstex/Vertex-S-AA.pstex_t} \end{array}} & = & - \ensuremath{\begin{array}{c}\input{pstex/Dumbbell-S-AA-hatS.pstex_t} \end{array}} -2 \ensuremath{\begin{array}{c}\input{pstex/Dumbbell-S-A-W-A-hatS.pstex_t} \end{array}} + \ensuremath{\begin{array}{c}\input{pstex/Dumbbell-S-A-Sigma_g-A.pstex_t} \end{array}}
	\\
								&	& -\frac{1}{2} 
		\left[ 
			\ensuremath{\begin{array}{c}\input{pstex/Sigma_g-DEP.pstex_t} \end{array}} - 2\ensuremath{\begin{array}{c}\input{pstex/Sigma_g-A-W-A.pstex_t} \end{array}} - \ensuremath{\begin{array}{c}\input{pstex/Sigma_g-W-AA.pstex_t} \end{array}}
		\right]
	\end{eqnarray*}
\caption{The flow of a vertex decorated by two $A^{1}$s (or $A^2$s), using the
new diagrammatics.}
\label{fig:NFE:flow-A1A1}
\end{figure}
\end{center}

Since the fields $A^1_\mu(p)$ and $A^1_\nu(-p)$
are never differentiated, we call them external fields---as
opposed to the internal fields sitting at the ends of the wines.

Let us start by looking at the dumbbell terms. The first diagram contains
a one-point vertex. One-point vertices do not exist in the $A$ or $F$-sectors,
and we demand that Wilsonian effective action one-point vertices
vanish, in order that the superhiggs potential is not shifted by
quantum corrections~\cite{YM-1-loop}. Hence, the one-point vertex
must be a seed action vertex in either of
the $C$ sub-sectors. Therefore, the external fields must decorate the $S$-lobe,
rather than the $\Sigma_g$ lobe. Since both of these fields are on the
same vertex, there is no factor of two. However, by taking only the
seed action part of the $\Sigma_g$ vertex, we get a factor
of $-2$. (Recall $\Sigma_g = g^2 S -2\hat{S}$.)

The next dumbbell term comes with an additional factor of two, compared
to the first, since the two external fields are now on different structures. We
note that this diagram in fact vanishes, since if the wine is in the
$C$-sector (as in the first diagram) then the bottom vertex is an $AC$
vertex, which does not exist. The final dumbbell contains a `full' $\Sigma_g$.
This diagram picks up a factor of two in recognition of the indistinguishability
of the $A^1_\mu(p)$ and $A^1_\nu(-p)$.

The first and third quantum terms do not pick up a factor of two,
since the external fields decorate the same structure. The middle
quantum term, on the other hand, comes with a relative factor of
two.

When we come to use the \MGIERG\ for computation, we will
compute the flow of vertices that are part of some complete
diagram. Such vertices will generically contain fields which are external
\wrt\ the diagram as a whole and some which are internal to
the diagram as a whole (these latter fields can, of course,
be external \wrt\ some sub-diagram). Whether or not
internal fields---whose flavour, we recall, is summed over---can 
be treated as identical depends
on the topology of the diagram in question.
This will become clear when we start to manipulate
complete diagrams in chapter~\ref{ch:beta1}.

\subsection{Additional Notation}

\begin{center}
\begin{figure}[h]
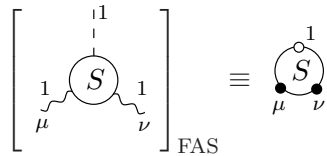

	\[
		\left[\ensuremath{\begin{array}{c}\input{pstex/Vertex-S-A1A1C1.pstex_t} \end{array}}\right]_\mathrm{FAS} \equiv \ensuremath{\begin{array}{c}\input{pstex/Vertex-S-A1_muA1_nuC1.pstex_t} \end{array}}
	\]
\caption{An example of the meaning of FAS.}
\label{fig:NFE:FAS}
\end{figure}
\end{center}

Whilst we will generally use the
new diagrammatics thus described, from
now on, it is occasionally useful to
flip back to the old style mentality
of specifying the supertrace structure.
It turns out that, in this thesis, we
only ever have recourse to do
this for single supertrace terms and
so introduce the notation `Fields as Shown'
or FAS. An example of this is illustrated
in figure~\ref{fig:NFE:FAS}.

The second new piece of notation
allows us to perform an intermediate step
between going from figure~\ref{fig:NFE:New-Diags-1}---where
all the decorations are implicit---to a set of explicitly
decorated diagrams. 

\begin{center}
\begin{figure}[h]
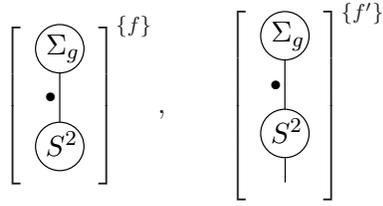

	\[
		\dec{
		\ensuremath{\begin{array}{c}\input{pstex/Dumbbell-S2-Sigma_g.pstex_t} \end{array}}
	}{\{f\}}, \hspace{2em}
	\dec{
		\ensuremath{\begin{array}{c}\input{pstex/Dumbbell-S2-Explicit-Sigma_g.pstex_t} \end{array}}
	}{\{f'\}}
	\]
\caption{Forcing a vertex on an implicitly decorated diagram
to have precisely two decorations.}
\label{fig:SpecificNumDecs}
\end{figure}
\end{center}

Suppose that we have a diagram for which we want to focus 
on the components possessing a two-point vertex. So long
as there is still at least one field sitting as
an implicit decoration, we must specify that our vertex
has precisely two decorations. We use a superscript on
the vertex argument to denote this. Two
examples are shown in figure~\ref{fig:SpecificNumDecs}.

In the first diagram, we see that we must use one 
and only one element of the set
$\{f\}$ to decorate the bottom vertex. In
the second diagram, the elements of $\{f'\}$ must all
decorate some structure other than the bottom vertex.

\subsection{Constraints in the $C$-sector} \label{sec:NFE:ND-II:App}

Recall from section~\ref{sec:Intro:YM}
that, in order to ensure quantum corrections do not
shift the minimum of the Higgs potential from the classical
choice, $\sigma$, $\hat{S}$ is constrained. 
Assuming that these quantum corrections vanish means that
all Wilsonian effective action one-point $C$ vertices
vanish. By computing the flow of an object we know to
vanish, we arrive at the desired constraint equation,
as shown in figure~\ref{fig:NFE:C-secConstraint}. Note that
the external fields can be in either the $1$ or $2$ sub-sector.
\begin{center}
\begin{figure}[h]
	\begin{eqnarray}
	-\flow \ensuremath{\begin{array}{c}\input{pstex/Vertex-S-C.pstex_t} \end{array}} = 0 	&=				& \frac{1}{2} \left[ \ensuremath{\begin{array}{c}\input{pstex/Dumbbell-S-C-Sigma_g.pstex_t} \end{array}} + \ensuremath{\begin{array}{c}\input{pstex/Dumbbell-S-Sigma_g-C.pstex_t} \end{array}} - \ensuremath{\begin{array}{c}\input{pstex/Sigma_g-C-DEP.pstex_t} \end{array}} - \ensuremath{\begin{array}{c}\input{pstex/Sigma_g-W-C.pstex_t} \end{array}} \right] 
\nonumber
\\[2ex]
								&\Rightarrow 	& -2 \ensuremath{\begin{array}{c}\input{pstex/Dumbbell-S-C-Sh.pstex_t} \end{array}} = \ensuremath{\begin{array}{c}\input{pstex/Sigma_g-C-DEP.pstex_t} \end{array}} + \ensuremath{\begin{array}{c}\input{pstex/Sigma_g-W-C.pstex_t} \end{array}}
\label{eq:NPC-sec-Constraint}
	\end{eqnarray}
\caption{The constraint arising from ensuring that
the position of the minimum of the Higgs potential is
unaffected by quantum corrections.}
\label{fig:NFE:C-secConstraint}
\end{figure}
\end{center}

To go from the first line to the second line, we have 
used $\Sigma_g = g^2 S - 2\hat{S}$ and have
discarded all one-point, Wilsonian effective action vertices.
To satisfy equation~(\ref{eq:NPC-sec-Constraint}), we tune the
one-point, seed action vertex in the first diagram, which is something
we are free to do.

\section{The Weak Coupling Expansion}

\subsection{The flow Equation}

By isolating the effects of $g_2$ (equivalently, $\alpha$),
the weak coupling expansion of the flow equation, (\ref{eq:WeakCouplingExpansion-oldFlow}), changes
since we now write\footnote{We avoid writing $\partial / \partial \alpha$ as $\partial_\alpha$
to avoid confusion later, when we will have momentum derivatives which are written \eg\
$\partial_\alpha^k$.}
\[
	\flow S = \flowConstAl S + \flow \alpha \ \pder{S}{\alpha}.
\]

To obtain our new weak coupling expansion of the flow equation, it
is necessary that we understand $\flow \alpha$, in the perturbative
domain. The first thing we note is that we have $\beta$-functions for
both $g$ and $g_2$, where the coefficients generically depend on $\alpha$:
\begin{eqnarray}
	\flow \frac{1}{g^2} 	& = & -2 \sum_{i=1}^\infty \beta_i(\alpha) g^{2(i-1)} 
\label{eq:NFE:WCE:flow-g}
\\
	\flow \frac{1}{g^2_2} 	& = & -2 \sum_{i=1}^\infty \tilde{\beta}_i(1/\alpha) g^{2(i-1)}_2.
\label{eq:NFE:WCE:flow-g_2}
\end{eqnarray}

Now, we know already the value of $\beta_1$ and, moreover, that it is independent of $\alpha$~\cite{YM-1-loop}.
The coefficient $\beta_2(\alpha)$ is what we will compute in this thesis. For generic
$\alpha$, we expect it to disagree with the standard value but, as discussed earlier,
hope that agreement is reached for $\beta_2(0)$.

What of the coefficients $\tilde{\beta}_i(1/\alpha)$?  These are determined
through the renormalisation condition
\begin{equation}
	S[\SF = A^2, \SH = \bar{\SH}] = -\frac{1}{2\alpha g^2} \, \tr \!\! \Int{x} \left(F_{\mu\nu}^2 \right)^2 + \cdots,
\label{eq:Intro:RenormCondition-g_2}
\end{equation}
where the ellipsis denotes higher dimension operators and ignored vacuum energy. The minus
sign arises because
the $SU(N)$ Yang-Mills theory carried by $A^2$ comes with the wrong
sign kinetic term; if we were to absorb this sign into the coupling, $g_2^2$,
then we could think of our unphysical Yang-Mills theory as having
imaginary coupling. Hence, $\tilde{\beta_1} = -\beta_1$
and $\tilde{\beta_2}(0) = \beta_2(0)$.\footnote{We note that there is
no reason to expect $\tilde{\beta}_2(1/\alpha) = \beta_2(\alpha)$
since $g$ and $g_2$ are not treated symmetrically in the flow equation.}

Utilising equations~(\ref{eq:NFE:WCE:flow-g}) and~(\ref{eq:NFE:WCE:flow-g_2})
and rewriting $g_2^2$ in terms of $\alpha g_1^2$, it is apparent
that $\flow \alpha$ has the following weak coupling expansion:
\begin{equation}
	\flow \alpha = \sum_{i=1}^{\infty} \gamma_i g^{2i},
\label{eq:flow-alpha}
\end{equation}
where
\[
	\gamma_i = -2 \alpha \left(\beta_i(\alpha) - \alpha^i \tilde{\beta}_i(1/\alpha) \right).
\]
Knowing that $\tilde{\beta}_1 = -\beta_1$, we have
\begin{equation}
	\gamma_1 = -2 \beta_1 \alpha (1+\alpha),
\label{eq:gamma_1}
\end{equation}
which it turns out we will need, for our computation of $\beta_2$.

Now, the weak coupling expansion reads:
\begin{equation}
	\dot{S}_n = \sum_{r=1}^n \left[ 2(n-r-1) \beta_r S_{n-r} + \gamma_r \pder{S_{n-r}}{\alpha} \right]
		+ \sum_{r=0}^n a_0 \left[ S_{n-r}, \Sigma_r \right] -  a_1 \left[\Sigma_{n-1} \right],
\label{eq:WeakCouplingExpansion-NewFlow}
\end{equation}
where we recall that $\dot{S}$ has been redefined (equation~(\ref{eq:dot(X)-NEW})) to mean $\flowConstAl{S}$.
Incidentally, we note that the kernels $\DEP{f}{f}$ appearing in the flow equation are defined
according to the new definition of $\dot{X}$ \ie\ they are differentiated at constant $\alpha$.
This is a choice we are free to make about the flow equation and do so, since it makes life
easier.

The diagrammatics for the new weak coupling flow equation follow, directly.
However, we note that the classical term can be brought 
into a more symmetrical form. This follows from the invariance of
$a_0[S_{n-r}, \Sigma_r] + a_0[S_{r}, \Sigma_{n-r}]$ 
under $r \rightarrow n-r$. We exploit this by recasting
the classical term as follows:
\begin{eqnarray}
	a_0[\bar{S}_{n-r}, \bar{S}_r] 	& \equiv 	& a_0[S_{n-r}, S_r] - a_0[S_{n-r}, \hat{S}_r] - a_0[\hat{S}_{n-r}, S_r].
\label{eq:NFE:BarNotation}
\\
									& =			& 	\left\{
														\begin{array}{cc}
															\frac{1}{2} \left(a_0[S_{n-r}, \Sigma_r] + a_0[S_{r}, \Sigma_{n-r}]\right) 	& n-r \neq r
														\\
															a_0[S_r, \Sigma_r]														& n-r =r.
														\end{array}
													\right.
\nonumber
\end{eqnarray}

Hence, we can rewrite the flow equation in the following form:
\begin{equation}
	\dot{S}_n = \sum_{r=1}^n \left[ 2(n-r-1) \beta_r S_{n-r} + \gamma_r \pder{S_{n-r}}{\alpha} \right]
		+ \sum_{r=0}^n a_0 \left[\bar{S}_{n-r}, \bar{S}_r \right] -  a_1 \left[\Sigma_{n-1} \right].
\label{eq:WeakCouplingExpansion-NewFlow-B}
\end{equation}
The diagrammatic version is shown in figure~\ref{fig:NFE:NewDiagFE}, where we have used the various
shorthands described in section~\ref{sec:Intro:WeakCoupling}.
\begin{center}
\begin{figure}[h]
	\[
	\dec{
		\ensuremath{\begin{array}{c}\input{pstex/Vertex-n-LdL.pstex_t} \end{array}} 
	}{\{f\}}
	= 
	\dec{
		2 \sum_{r=1}^n \left[\left(n_r -1 \right) \beta_r +\gamma_r \pder{}{\alpha} \right]\ensuremath{\begin{array}{c}\input{pstex/Vertex-n_r-B.pstex_t} \end{array}} 
		+ \frac{1}{2} \sum_{r=0}^n \ensuremath{\begin{array}{c}\input{pstex/Dumbbell-n_r-r.pstex_t} \end{array}} -\frac{1}{2} \ensuremath{\begin{array}{c}\input{pstex/Vertex-Sigma_n_-B.pstex_t} \end{array}}
	}{\{f\}}
	\]
\caption{The new diagrammatic form for the weak coupling flow equation.}
\label{fig:NFE:NewDiagFE}
\end{figure}
\end{center}

Terms like the one on the \lhs, in which the entire diagram is struck by
$\flowConstAl$, are referred to as $\Lambda$-derivative terms. On the \rhs,
in addition to the usual classical and quantum terms, we have the 
so called $\beta$ and $\alpha$ terms.

It is worth remarking that the diagrammatics of figure~\ref{fig:NFE:NewDiagFE}
can be applied to the old flow equation as well: we simply take the fields $\{f\}$
to be $A$s, $C$s and $F$s and interpret the wines accordingly. 
Notice, indeed, that the old flow equation actually
provides a very strong clue as to the (algebraic) form the new flow equation must take.
Consider taking $\{f\} = AA$ (\ie\ we are using the old flow equation)
and, as in~\cite{YM-1-loop}, working only with single supertrace terms.\footnote{
This prevents vertices  containing $\sigma$ and an even number of
the dynamical components of $\SH$, since such terms
are not invariant under $\SH \rightarrow -\SH$. This restriction
simplifies the following analysis which, anyway, is intended for illustrative
purposes only.
}

Now, something odd is going on with the quantum term. Na\"ively,
this goes as $\str AA \str \one = 0$. However, we know from 
section~\ref{sec:Intro:BrokenPhase} that, since we are differentiating
\wrt\ only partial supermatrices, we must perform the $\sigma$ algebra,
as in figure~\ref{fig:Intro:DiffPartialSuper}. Thus, the quantum
term survives, going as $\str AA \sigma \str \sigma$, which
introduces a relative minus
sign in the $A^2$-sector. In the 
computation of $\beta_1$ in~\cite{YM-1-loop}, this minus sign is telling
us that  $\tilde{\beta}_1 = -\beta_1$.

Thus, the flow equation is guiding us to the sort of new terms
we need in order to properly distinguish $A^1$ and $A^2$: namely
terms possessing a wine which can be decorated by a single $\sigma$.
Returning to our example of the flow of $AA$, we will have a new
quantum term which, after performing the $\sigma$ algebra, will go as
$\str AA \str \sigma$, where the $\sigma$ from the wine has
combined with a $\sigma$ from the $\sigma$ algebra. Moreover,
the existence of a wine decorated by a single $\sigma$
necessitates that the restriction to single supertrace terms
is lifted (we can now hide instances of $\sigma$
on otherwise empty supertraces, allowing us 
to preserve $\SH \rightarrow -\SH$ symmetry). This ineluctably
guides us to the current formalism.

\subsection{The Two-Point, Tree Level Vertices}	\label{sec:NFE:TPTL}

In preparation for the calculation of $\beta$-function
coefficients, we now determine the two-point, tree
level vertices. Of the two-point, tree level vertices,
we demand that:
\begin{enumerate}
	\item	they are consistent with the renormalisation 
			conditions~(\ref{eq:Intro:RenormCondition}) and~(\ref{eq:Intro:RenormCondition-g_2});

	\item	full $SU(N|N)$ invariance is recovered in the $\SF$ sector
			and full $U(N|N)$ invariance is recovered in the $C$-sector, for
			sufficiently high energies;

	\item	the high energy behaviour is consistent with the
			constraints~(\ref{eq:CutoffConstraints}) (or their appropriate generalisation), to ensure
			that the regularisation does its job.
\end{enumerate}

These three points are not sufficient to uniquely determine
the two-point, tree level vertices and, indeed, it is not strictly
necessary to do so. Nonetheless, we have found it useful to 
have concrete algebraic expressions. We save for the future
a more general treatment which, with the understanding of
the \MGIERG\ we now have, should be straightforward~\cite{FurtherWork}.

To determine the two-point, tree level vertices, our starting
point is the tree level, classical flow equation,
obtained by specialising equation~(\ref{eq:WeakCouplingExpansion-NewFlow})
or~(\ref{eq:WeakCouplingExpansion-NewFlow-B})
to $n=0$:
\begin{equation}
	\dot{S}_0 = a_0[S_0, \Sigma_0].
\label{eq:TreeLevelFlow}
\end{equation}
We now further specialise, to consider the flow of all two-point vertices,
as shown in figure~\ref{fig:NFE:TLTP-flow}. Recall that the solid
lines represent dummy fields, which we choose to be instances
of $A^1$, $A^2$, $C^1$, $C^2$, $\bar{F}$ and $F$.
\begin{center}
\begin{figure}[h]
	\begin{equation}
		\ensuremath{\begin{array}{c}\input{pstex/Vertex-TLTP-LdL.pstex_t} \end{array}} = \ensuremath{\begin{array}{c}\input{pstex/Dumbbell-S_0-Sigma_0.pstex_t} \end{array}}
	\label{eq:TLTP-flow}
	\end{equation}
\caption{Flow of all possible two-point, tree level vertices.}
\label{fig:NFE:TLTP-flow}
\end{figure}
\end{center}

There are a number of  things to note.
First, the overall half associated with $a_0$ terms has been killed
by the two ways in which the dumbbell structure can be decorated
by the two wildcard fields. Secondly, there is no possibility
of embedding components of $\SH$ at the ends  of the wines and
so the $\dot{\Delta}^{AA}_\sigma$ kernel does not appear.\footnote{Up to
instances which have been used to map is into the $A^1$, $A^2$ basis,
in the first place.}
Lastly, diagrams containing one-point, tree level vertices 
have not been drawn, since these vertices do not exist in any sector,
for either action.\footnote{Recall that one-point, seed action vertices
exist only from the one-loop level.}

We can now solve equation~(\ref{eq:TLTP-flow}) (with appropriate
boundary conditions---see section~\ref{sec:NFE:Universality}), giving an
expression for the Wilsonian effective action, two-point,
tree level vertices in terms of the seed action two-point,
tree level vertices and the zero-point kernels.

The next step to make is to follow~\cite{TRM-Faro, GI-ERG-I, GI-ERG-II, YM-1-loop}
and utilise the freedom inherent in $\hat{S}$ 
by choosing the two-point,
tree level (single supertrace) components of $\hat{S}$ to be equal to the two-point, tree
level (single supertrace) components of $S$ \ie\
\[
	\hat{S}^{\ \,f\, f}_{0 RS}(k) = S^{\ \,f\, f}_{0 RS}(k).
\]
We emphasise that this is simply a choice we make, since it
turns out to be helpful to do so.

We now choose a form for the two-point, tree level $\hat{S}$ vertices, 
based on the general
properties they must possess. In~\cite{YM-1-loop}, this has been 
done already for the special case of $\alpha=1$ and we summarise
the results below: 
\begin{eqnarray}
	\hat{S}^{\ AA}_{0 \mu \, \nu}(p)	& = & \frac{2}{c_p}\Box_{\mu \nu}(p),
\label{eq:OLD-TLTP-A1A1}
\\
	\hat{S}^{\ BB}_{0 \mu \, \nu}(p)	& = & \frac{2}{c_p}\Box_{\mu \nu}(p) + \frac{4 \Lambda^2}{\tilde{c}_p} \delta_{\mu \nu},
\label{eq:OLD-TLTP-BB}
\\
	\hat{S}^{\ B \sigma D}_{0 \mu}(p)	& = & \frac{2 \Lambda^2 p_\mu}{\tilde{c}_p},
\label{eq:OLD-TLTP-BD}
\\
	\hat{S}^{\ DD}_0(p)					& = & \frac{\Lambda^2 p^2}{\tilde{c}_p},
\label{eq:OLD-TLTP-DD}
\\
	\hat{S}^{\ CC}_0(p)					& = & \frac{\Lambda^2 p^2}{\tilde{c}_p} + 2 \lambda \Lambda^4,
\label{eq:OLD-TLTP-CC}
\end{eqnarray}
where $\lambda >0$ is an undetermined constant. Since the actions in~\cite{YM-1-loop}
were restricted to single supertrace terms, $\hat{S}^{AA\sigma}$ and $\hat{S}^{CC\sigma}$
vertices
were forbidden, under $\SH \rightarrow -\SH$ symmetry.\footnote{In our treatment,
these restrictions vanish, since we can always hide instances of $\sigma$
on additional, otherwise empty, supertraces.} Note, though, that  the
$\hat{S}^{DD\sigma}$ vertex
is  excluded more generally: cyclicity of the supertrace tells us that
$ 2\str DD\sigma = \str D \{D,\sigma\}$, but this vanishes, since $D$ and $\sigma$ anti-commute;
similarly for $\hat{S}^{BB\sigma}$.
Due to the restriction to single supertrace terms, the vertices $\hat{S}^{C,C}$, 
$\hat{S}^{C,C\sigma}$ and $\hat{S}^{C\sigma,C\sigma}$ were never considered.

It will be instructive to understand where expressions~(\ref{eq:OLD-TLTP-A1A1})--(\ref{eq:OLD-TLTP-CC}) 
come from, and so we repeat the analysis of~\cite{YM-1-loop}. 
First, let us focus on the $A$-sector. The Ward identity~(\ref{eq:WID-Unbroken})
tells us that $p_\mu\hat{S}^{\ AA}_{0 \mu \, \nu}(p) = 0$. Given this and that the vertex is of mass dimension two
it must, by Lorentz invariance, take the form given by equation~(\ref{eq:OLD-TLTP-A1A1}). 
Now, since we have identified the two-point, tree level, seed action vertices with the corresponding
Wilsonian effective action vertices, they are subject to the renormalisation 
conditions~(\ref{eq:Intro:RenormCondition}) and~(\ref{eq:Intro:RenormCondition-g_2}) (recall that we are
working with $\alpha=1$).\footnote{Generally, of course, seed action vertices are independent
of the renormalisation conditions.}
These conditions, together with equation~(\ref{eq:WeakCouplingExpansion-Action}),
tell us that $c(0) = 1$.

Next, we look at the pure $D$ vertex. Goldstone's theorem~\cite{Goldstone} tells us that $D$
is massless and so~(\ref{eq:OLD-TLTP-DD}) follows by Lorentz invariance and dimensions. We can
think of equation~(\ref{eq:OLD-TLTP-DD}) as providing a definition of what we mean by $\tilde{c}$.

The $B\sigma D$ vertex is determined by the Ward identity~(\ref{eq:WID-Broken}). Recalling
that $F = (B, \sigma D)$, we have
\[
	p_\mu \hat{S}_{0 \mu}^{\ B\sigma D}(p) = 2\hat{S}^{DD}(p).
\]
Equation~(\ref{eq:OLD-TLTP-BD}) therefore follows by Lorentz invariance. The longitudinal part
of the pure $B$ vertex now follows also from the Ward identity~(\ref{eq:WID-Broken}), since
\[
	p_\mu \hat{S}_{0 \mu\, \nu} ^{\ BB}(p) = 2\hat{S}^{B\sigma D}_\nu(p).
\]
As for the transverse part of the pure $B$ vertex, by Lorentz invariance and dimensions
it must look like $2\Box_{\mu \nu}(p)/ f(p)$, for the dimensionless function, $f$. However,
to recover $SU(N|N)$ invariance at high energies, $f(p)$ must coincide with $c(p)$,
in this regime. For the sake of simplicity, we choose $f(p) = c(p)$.

Finally, consider the $C$-sector vertex. Whereas the Goldstone mode $D$ is
massless, $C$ is not and so we expect the kinetic terms of
these two fields to differ by a mass term. In order that
we recover $U(N|N)$ invariance in the $\SH$-sector at sufficiently
high energies, this mass term must be sub-leading, in this regime.
We thus introduce the function $\lambda_p$ of which we demand that it must not
vanish at $p=0$ and that it ensures the mass term is sub-leading
at high energies. The simplest choice is to take $\lambda$
to be a positive constant.

Having justified the forms of the two-point, tree level vertices for when $\alpha=1$
and the actions are restricted to single supertrace terms, we must now generalise
the analysis, for our current purposes.

We begin in the $A$-sector, where now we have two independent vertices to deal with,
$\hat{S}^{\ \, 1 1}_{0 \mu \nu}(p)$ and $\hat{S}^{\ \, 2 2}_{0 \mu \nu}(p)$.\footnote{
These are linear combinations of $\hat{S}_{0 \mu \, \nu}^{\ A A}(p)$
and $\hat{S}_{0 \mu \, \nu}^{\ A A \sigma}(p)$.}
As before,
gauge invariance, Lorentz invariance and dimensions tell us that
\begin{eqnarray}
	\hat{S}^{\ \, 1 1}_{0 \mu \nu}(p) & = & \NCO{p^2/\Lambda^2, \alpha} \Box_{\mu \nu}(p),
\label{eq:NFE:S11-General}
\\
	\hat{S}^{\ \, 2 2}_{0 \mu \nu}(p) & = & \NCT{p^2/\Lambda^2, \alpha} \Box_{\mu \nu}(p),
\label{eq:NFE:S22-General}
\end{eqnarray}
where $\NCO{p^2/\Lambda^2, \alpha}$ and $\NCT{p^2/\Lambda^2, \alpha}$ are to be determined. Of
these two functions, we know the following:
\begin{enumerate}
	\item	$\NCO{0,\alpha} = 2$, due to the renormalisation condition~(\ref{eq:Intro:RenormCondition});

	\item	$\NCT{0,\alpha} = 2/\alpha$, due to the renormalisation condition~(\ref{eq:Intro:RenormCondition-g_2});

	\item	$\NCO{p^2/\Lambda^2, 1} = \NCT{p^2/\Lambda^2, 1} = 2/c_p$, for consistency with~\cite{YM-1-loop};

	\item	At sufficiently high energies, $SU(N|N)$ invariance is recovered and
			so the two functions must coincide, in this regime.
\end{enumerate}
As mentioned already, these points do not unambiguously determine 
$\NCO{p^2/\Lambda^2, \alpha}$ and $\NCT{p^2/\Lambda^2, \alpha}$. However,
for definiteness, we make the following choices:
\begin{eqnarray}
	 \NCO{p^2/\Lambda^2, \alpha} 	& = &  \NCOAlg{p},
\label{eq:NFE-A1-c^-1}
\\
	 \NCT{p^2/\Lambda^2, \alpha}	& = &  \NCTAlg{p},
\label{eq:NFE-A2-c^-1}
\end{eqnarray}
which gives 
\begin{eqnarray}
	\hat{S}^{\ \, 1 1}_{0 \mu \nu}(p) & = &  \NCOAlg{p} \Box_{\mu \nu}(p),
\label{eq:NFE-TLTP-A1A1}
\\
	\hat{S}^{\ \, 2 2}_{0 \mu \nu}(p) & = &  \NCTAlg{p}	\Box_{\mu \nu}(p).
\label{eq:NFE-TLTP-A2A2}
\end{eqnarray}
We will henceforth use the shorthand
\[
	\NCO{p} \equiv \NCOb{p} \equiv \NCO{p^2/\Lambda^2, \alpha}
\]
where we note that
\begin{equation}
	\NCO{0} = 2.
\label{eq:NFE:A(0)}
\end{equation}
For future convenience, we define 
\begin{equation}
\InvCO_p \equiv \NewCO^{-1}_p.
\label{eq:def-B}
\end{equation}

Now we move on to the pure $D$-sector. The only change we need make is that,
rather than looking at both block off-diagonal components of $\SH$ together---what
we have been calling $D$---we look at the two components $\bar{D}$ and $D$, separately.
\begin{equation}
	\hat{S}^{\ D\bar{D}}_0(p) = \hat{S}^{\ \bar{D} D}_0(p)  = \frac{\Lambda^2 p^2}{\tilde{c}_p}.
\label{eq:NFE-TLTP-DD}
\end{equation}

Once again, the mixed $BD$ vertex is now uniquely determined by gauge invariance. Hence,
\begin{equation}
	\hat{S}^{\ D\bar{B}}_{0 \ \  \mu} (p) = - \hat{S}^{\ \bar{D} B}_{0 \ \ \mu}(p)  = \frac{2 \Lambda^2 p_\mu}{\tilde{c}_p}.
\label{eq:NFE-TLTP-BD}
\end{equation}
In turn, this now uniquely determines the longitudinal part of the pure $B$ vertex,
as before. For the 
transverse part of the pure $B$ vertex, we note that $SU(N|N)$ invariance demands that
it is the same as the $A^1$ and $A^2$ vertices at sufficiently high energies. Thus, for simplicity,
we choose
\begin{equation}
	\hat{S}^{\ B\bar{B}}_{0 \mu \, \nu}(p) 	=  	\hat{S}^{\ \bar{B} B}_{0 \mu \, \nu}(p)
											=	\frac{\alpha + 1}{\alpha c_p}\Box_{\mu \nu}(p) 
												+ \frac{4 \Lambda^2}{\tilde{c}_p} \delta_{\mu \nu}.
\label{eq:NFE-TLTP-BB}
\end{equation}

Finally, we deal with the $C$-sector. For the case where both fields are
on the same supertrace we now have a new vertex, namely $\hat{S}_0^{\ C C \sigma}$.
Linear combinations of $\hat{S}_0^{\ C C}$ and $\hat{S}_0^{\ C C \sigma}$
define the vertices $\hat{S}^{\ C^1C^1}_0(p)$ and $\hat{S}^{\ C^2C^2}_0(p)$,
which we take to be equal. As before, we take these
vertices to be like the pure $D$ vertices, plus a mass term. Thus we have
\begin{equation}
	\hat{S}^{\ C^1C^1}_0(p)	= \hat{S}^{\ C^2C^2}_0(p) = \frac{\Lambda^2 p^2}{\tilde{c}_p} + 2 \lambda \Lambda^4,
\label{eq:NFE-TLTP-CC}
\end{equation}
where we note that $\lambda$ can now depend on $\alpha$.

Since our current analysis includes multi-supertrace terms, there is nothing to
prevent us from having $S^{C^{1,2},C^{1,2}}(p)$ and $\hat{S}^{C^{1,2},C^{1,2}}(p)$
vertices (we treat the $A$-sector shortly). 
At tree level, at any rate, it is useful to exclude these;
for the seed action vertices, we simply set them to zero.
To see whether the Wilsonian effective action vertices vanish,
we must compute their flow, 
as shown 
in figure~\ref{fig:NFE:S^C,C}. We have used the
old-style notation, as we wish to be specific about the supertrace
structure.
\begin{center}
\begin{figure}[h]
	\begin{equation}
	\begin{array}{ccccccc}
										&	&\LE{Dumbbell-CC-C,C}	&	&\LE{Dumbbell-C,C,C,C}	&	&\LE{Dumbbell-C,C,C,C-hat}
	\\[1ex]
		\dec{\ensuremath{\begin{array}{c}\input{pstex/Vertex-C,C.pstex_t} \end{array}}}{\bullet}	&=	&-\ensuremath{\begin{array}{c}\input{pstex/Dumbbell-CC-C,C.pstex_t} \end{array}} 	&+	& \ensuremath{\begin{array}{c}\input{pstex/Dumbbell-C,C,C,C.pstex_t} \end{array}} &- 2&\ensuremath{\begin{array}{c}\input{pstex/Dumbbell-C,C,C,C-hat.pstex_t} \end{array}}
	\end{array}
	\label{eq:NFE-C,C}
	\end{equation}
\caption{The flow of $S^{\ C^{1,2}, C^{1,2}}_0(p)$.}
\label{fig:NFE:S^C,C}
\end{figure}
\end{center}

Setting $\hat{S}^{\ C^{1,2},C^{1,2}}_0(p) =0$
causes diagrams~\ref{Dumbbell-CC-C,C} and~\ref{Dumbbell-C,C,C,C-hat}
disappear. Equation~(\ref{eq:NFE-C,C}) simplifies and we choose the solution
$S^{\ C^{1,2},C^{1,2}}_0(p) =0$.

Though we will return to the two-point, tree level vertices in
more generality in section~\ref{sec:NFE:Universality},
we will henceforth always take the tree level $C^{1,2}, C^{1,2}$
vertices to vanish. 

Now, we can also have multi-supertrace terms in the $A$-sector,
which must take the form $\str \sigma \SF \; \str \sigma \SF$. These
cannot be set to zero because they are related, by no-$\SF^0$ symmetry,
to single supertrace two-point vertices (see equation~(\ref{eq:Intro:No-A0-Ex}).
However, since we are working in the scheme where we cannot attach to $\SF^0$,
these terms will interest us no further.
Thus, when we now refer to two-point,
tree level vertices, we strictly mean only those with a single
supertrace.

For ease of reference, we collect together all the new two-point, tree level
vertices in appendix~\ref{app:TLTPs}.

It is worth remarking that we have checked that 
unbroken actions can be constructed which yield
the above vertices, in the broken phase.

\subsection{The Zero-Point Kernels and Effective Propagators} \label{sec:NFE:Zero-PointKernels+EffProps}

Having set the Wilsonian effective action and seed action two-point, tree level vertices
equal to each other and having chosen their algebraic form,
let us now return to equation~(\ref{eq:TLTP-flow}). We can
simplify this, since $\Sigma_{0 M N}^{\ \; X\, X}(p) = S_{0 M N}^{\ \; X\, X}(p) - 2 \hat{S}_{0 M N}^{\ \; X\, X}(p)$
can be replaced by $- S_{0 M N}^{\ \; X\, X}(p)$. 
This makes it clear that we have an equation involving
only two-point, tree level vertices---which we know---and zero-point
wines. We can thus determine these latter objects.

This is straightforward, and is presented in~\cite{YM-1-loop} for the special
case of $\alpha=1$. In the current case, we will just quote the
new results. To this end, we define $x = p^2 / \Lambda^2$ and 
introduce two new functions:
\begin{eqnarray}
	f_p & = & \fB{p}{x},
\label{eq:NFE-f}
\\[1ex]
	g_p	& = & \gB{p}{x}.
\label{eq:NFE-g}
\end{eqnarray}
These functions reduce to the definitions in~\cite{YM-1-loop} for $\alpha=1$
and, crucially, still satisfy the relationship
\begin{equation}
	xf_p + 2g_p =1.
\label{eq:NFE-xf+2g}
\end{equation}

As will become clear later, when we start to perform diagrammatic
calculations, it is this relationship that is important
to us, rather than the precise algebraic forms of $f$ and $g$,
which contain some degree of choice.

Denoting differentiation \wrt\ $x$ by a prime, we are now ready to
give the zero-point kernels:
\begin{eqnarray}
	\DEP{1}{1}(p)		& = &	\frac{2}{\Lambda^2}	\left[ \Aone{p} \right]',
\label{eq:NFE-DEP-11}
\\[1ex]
	\DEP{2}{2}(p)		& = &	\frac{2}{\Lambda^2}	\left[ \Atwo{p} \right]',
\label{eq:NFE-DEP-22}
\\[1ex]
	\DEP{B}{\bar{B}}(p)	& = &	\frac{1}{\Lambda^2} \left[ x \tilde{c}_p g_p \right]',
\label{eq:NFE-DEP-BB}
\\[1ex]
	\DEP{D}{\bar{D}}(p) & = &	\frac{2}{\Lambda^4}\frac{1}{x} \left[ x^2 \tilde{c}_p f_p \right]',
\label{eq:NFE-DEP-DD}
\\[1ex]
	\DEP{C^1}{C^1}(p)	& = &	\frac{1}{\Lambda^4}\frac{1}{x} \left[ \frac{2x^2 \tilde{c}_p}{x + 2\lambda \tilde{c}_p} \right]',
\label{eq:NFE-DEP-C1C1}
\\[1ex]
	\DEP{C^2}{C^2}(p)	& = &	\frac{1}{\Lambda^4}\frac{1}{x} \left[ \frac{2x^2 \tilde{c}_p}{x + 2\lambda \tilde{c}_p} \right]'.
\label{eq:NFE-DEP-C2C2}
\end{eqnarray}
Note that, in the $B$ and $D$-sectors, it does not make any
difference whether the barred or unbarred field comes first.

The next step we perform is absolutely central to the all
that follows, and its importance cannot be over emphasised.
We now determine the integrated kernels, where we note
that their appearance has been anticipated by the notation
$\dot{\Delta} \equiv -\flowConstAl \Delta$. The
integrated kernels are as follows:
\begin{eqnarray}
	\EP{11}{}(p)		& = & \frac{1}{p^2} \Aone{p},
\label{eq:NFE:EP-11}
\\[1ex]
	\EP{22}{}(p)		& = & \frac{1}{p^2} \Atwo{p},
\label{eq:NFE:EP-22}
\\[1ex]
	\EP{B\bar{B}}{}(p)	& = & \frac{1}{2 \Lambda^2} \tilde{c}_p g_p,
\label{eq:NFE:EP-BB}
\\[1ex]
	\EP{D\bar{D}}{}(p)	& = & \frac{1}{\Lambda^4} \tilde{c}_p f_p,
\label{eq:NFE:EP-DD}
\\[1ex]
	\EP{C^1C^1}{}(p)	& = & \frac{1}{\Lambda^4}	\frac{\tilde{c}_p}{x+ 2\lambda \tilde{c}_p},
\label{eq:NFE:EP-C1C1}
\\[1ex]
	\EP{C^2C^2}{}(p)	& = & \frac{1}{\Lambda^4}	\frac{\tilde{c}_p}{x+ 2\lambda \tilde{c}_p},
\label{eq:NFE:EP-C2C2}
\end{eqnarray}
and are equivalent to those of~\cite{YM-1-loop} for $\alpha=1$.

These objects have a very close relationship to the two-point, 
tree level vertices. In the $C$-sector, the integrated
kernels are just the inverse of the 
corresponding 
two-point, tree level
vertices:
\begin{equation}
	S_0^{C^1 C^1}(p) \EP{C^1 C^1}{}(p) =  S_0^{C^2 C^2}(p) \EP{C^2 C^2}{}(p) = 1
\label{eq:NFE-EffProp-C}
\end{equation}

In the $A^{1,2}$ sectors, the integrated kernels are the inverses
of the corresponding two-point, tree level vertices only in
the transverse space:
\begin{eqnarray}
	S^{\ \, 1 1}_{0 \mu \nu}(p) \EP{11}{}(p) & = & \delta_{\mu \nu} - \frac{p_\mu p_\nu}{p^2},
\label{eq:NFE-EffProp-A1}
\\ [1ex]
	S^{\ \, 22}_{0 \mu \nu}(p) \EP{22}{}(p) & = & \delta_{\mu \nu} - \frac{p_\mu p_\nu}{p^2}.
\label{eq:NFE-EffProp-A2}
\end{eqnarray}

In the $B$ and $D$ sectors, the results are most naturally
phrased in terms of the composite field $F$. Following
equation~(\ref{eq:F-Wine}) we define
\begin{equation}
	\EP{\, F\, \bar{F}}{MN}(p) =
	\left(
		\begin{array}{cc}
			\EP{B\bar{B}}{}(p) \delta_{\mu \nu}		& 0
		\\
						0							& -\EP{D\bar{D}}{}(p)
		\end{array}
	\right),
\label{eq:NFE-EP-F}
\end{equation}
giving us
\begin{equation}
	S_{0 MS}^{\ \, F \, \bar{F}}(p) \EP{\bar{F}\, F}{SN}(p) = S_{0 MS}^{\ \, \bar{F} \, F}(p) \EP{F\, \bar{F}}{SN}(p) = \delta_{MN} - p'_M p_N
\label{eq:NFE-EffProp-F}
\end{equation}
where, as in the introduction (equation~(\ref{eq:Intro:q_M})),
\begin{equation}
	p_N = (p_\nu, 2)
\label{eq:NFE:k}
\end{equation}
and we define 
\begin{equation}
	p'_M = \left( \frac{f_p p_\mu}{\Lambda^2}, g_p \right).
\label{eq:NFE:k'}
\end{equation}
Note that equation~(\ref{eq:NFE:k'})
differs from the analogous definition of~\cite{YM-1-loop} by a sign in the fifth
component. This can be traced back to our definition of $F$ (see equation~(\ref{eq:Intro:F})).

We note that, when looking at components of the fields
in equation~(\ref{eq:NFE-EffProp-F}) labelled by $M$
and $N$, we must take particular care: whilst
$F= (B,D)$, $\bar{F} = (\bar{B}, -\bar{D})$. This
extra sign can be easily forgotten.

Now, just as the integrated kernels in the $A^{1,2}$ sectors
are the inverses of the corresponding two-point, tree level
vertices in the transverse space, so too in the $F$-sector. This
follows upon contracting equation~(\ref{eq:NFE-EffProp-F}) with $p_M$:
\[
	p_M (\delta_{MN} - p'_M p_N) = p_N\left[1 - (p^2 f_p /\Lambda^2 + 2g_p)\right] = 0,
\]
where we have used the definitions~(\ref{eq:NFE:k}) and~(\ref{eq:NFE:k'}),
have recognised that $x = p^2 / \Lambda^2$ and have used equation~(\ref{eq:NFE-xf+2g}).

In summary, we see that the $C$-sector integrated kernels
are the inverses of the two-point, tree level vertex whereas,
in the $A$ and $F$ sectors, the integrated kernels
are the inverses of the two-point, tree level vertices in
a restricted sense only. In recognition of the similarity of
the integrated kernels to propagators, we will henceforth refer
to them as effective propagators~\cite{YM-1-loop}. However,
despite their similarities in both form and the role they play,
we emphasise that they are not propagators in the usual sense.
Indeed, since we never fix the gauge, we know that we cannot
even define a propagator for the $\SF$-sector fields.

The zero-point kernels and effective propagators are collected
together in appendix~\ref{app:Kernels}.

As a final remark, we note that we have not explicitly
mentioned $\dot{\Delta}^{AA}_\sigma$ in this analysis.
Of course, it can be obtained from $\DEP{1}{1}$ and $\DEP{2}{2}$,
but this is not really the point. As we will see in 
part~III, diagrams containing explicit instances of
$\dot{\Delta}^{AA}_\sigma$ (as opposed to implicit instances,
which have been absorbed when mapping to the $A^1$-$A^2$ basis)
will always cancel and so the details of 
$\dot{\Delta}^{AA}_\sigma$ need never concern us.

\subsection{Diagrammatic Identities}	\label{sec:NFE:DiagrammaticIdentities}

We can phrase the relationship between the
two-point, tree level vertices and the
effective propagators in particularly concise form. In
preparation, we consider more carefully the objects left
over when a two-point, tree level vertex is contracted into an
effective propagator. We use the $F$-sector relationship~(\ref{eq:NFE-EffProp-F})
as a template for \emph{all} sectors, which we can do if we 
employ a prescription for allowing $p_M'$ and $p_N$ to depend
on the sector of the calculation.

To be specific, we can take
\begin{equation}
		S_{0 MS}^{\ \, X \, Y}(p) \EP{Y Z}{SN}(p) = \delta_{MN} - p'_M p_N
\label{eq:NFE:EffPropReln}
\end{equation}
where the fields $X,Z$ are any broken phase fields, the field
$Y$ is summed over and
we identify the components of the \rhs\ according to table~\ref{tab:NFE:k,k'}.
(Note that the field $Z$ must, in fact, be the same as $Y$, since `mixed'
effective propagators do not exist.)
\begin{center}
\begin{table}[h]
	\[
	\begin{array}{c|ccc}
				& \delta_{MN}		& p_M'							& p_N
	\\ \hline
		F		& \delta_{MN}		& (f_k p_\mu / \Lambda^2, g_k)	& (p_\nu, 2)
	\\
		A		& \delta_{\mu \nu}	& p_\mu / p^2					& p_\nu
	\\
		C		& \one				& \mbox{---}					& \mbox{---}
	\end{array}
	\]
\caption{Prescription for unifying the relationships~(\ref{eq:NFE-EffProp-C}), (\ref{eq:NFE-EffProp-A1}), (\ref{eq:NFE-EffProp-A2}) and~(\ref{eq:NFE-EffProp-F}).}
\label{tab:NFE:k,k'}
\end{table}
\end{center}

Henceforth, we refer to $p'_M p_N$ as a `gauge remainder'~\cite{YM-1-loop}. This
captures the notion that the effective propagators generally leave something
behind (other than a Kronecker $\delta$-function) when contracted into
a two-point, tree level vertex. Moreover, if vertices or wines
are struck by these objects, then they can be processed
via the gauge invariance 
identities~(\ref{eq:WID-Unbroken}) and~(\ref{eq:WID-Broken}). 
The algebraic details of the gauge remainders are reproduced
in appendix~\ref{app:GRs}.

We now give a diagrammatic form for equation~(\ref{eq:NFE:EffPropReln}),
using the convention that the diagrammatic form of the two-point,
tree level vertices represents only the single supertrace components.
The two supertrace contributions either vanish automatically ($F$
sector) or by choice ($C$ sector) or are not attached to ($A$-sector).

All diagrammatic identities are reproduced in appendix~\ref{app:D-IDs},
for easy reference.

\begin{D-ID}
Consider a two point, tree level (single supertrace) vertex, 
attached to an effective propagator which,
only ever appearing as an internal line,
attaches to an arbitrary structure. This is shown
diagrammatically in figure~\ref{fig:NFE:EffProp}, where the effective propagator
is
denoted by a long, solid line.
\begin{center}
\begin{figure}[h]
	\begin{equation}
		\ensuremath{\begin{array}{c}\input{pstex/EffPropReln.pstex_t} \end{array}}	\equiv \ensuremath{\begin{array}{c}\input{pstex/K-Delta.pstex_t} \end{array}} - \ensuremath{\begin{array}{c}\input{pstex/FullGaugeRemainder.pstex_t} \end{array}} 
							\equiv \ensuremath{\begin{array}{c}\input{pstex/K-Delta.pstex_t} \end{array}} - \ensuremath{\begin{array}{c}\input{pstex/DecomposedGR.pstex_t} \end{array}}
	\label{eq:NFE:EffPProp}
	\end{equation}
\caption{The effective propagator relation.}
\label{fig:NFE:EffProp}
\end{figure}
\end{center}

The object $\ensuremath{\begin{array}{c}\begin{picture}(0,0)%
\includegraphics{pstex/FullGaugeRemainder-B.pstex}%
\end{picture}%
\setlength{\unitlength}{3947sp}%
\begingroup\makeatletter\ifx\SetFigFont\undefined%
\gdef\SetFigFont#1#2#3#4#5{%
  \reset@font\fontsize{#1}{#2pt}%
  \fontfamily{#3}\fontseries{#4}\fontshape{#5}%
  \selectfont}%
\fi\endgroup%
\begin{picture}(174,174)(2239,-823)
\end{picture}
 \end{array}}$ decomposes into two constituents:
\begin{eqnarray}
	\ensuremath{\begin{array}{c}\input{pstex/GaugeRemainder-ppr.pstex_t} \end{array}} & \equiv & p'_M,
\label{eq:NFE:Diag:p'}
\\[1ex]
	\ensuremath{\begin{array}{c}\input{pstex/GaugeRemainder-p.pstex_t} \end{array}} & \equiv & p_N.
\label{eq:NFE:Diag:p}
\end{eqnarray}

\label{D-ID:NFE:EffPropReln}
\end{D-ID}

Diagrammatic identity~\ref{D-ID:NFE:EffPropReln} is one of the foundations
of our entire diagrammatic approach and so we will give it 
a special name: the effective propagator relation.
However, on closer inspection, it looks like we may have missed
something. We know that, in the $A$-sector, there are attachment corrections. 
Thus, we might expect there to be
$1/N$ corrections to the effective propagator relation, arising
from where the effective propagator attaches to the vertex (see
figure~\ref{fig:NFE:Attachement-A1-B}). In practise, though,
these corrections always vanish. This is a consequence of
the diagrammatic structure shown in figure~\ref{fig:NFE:EffProp}
only ever appearing as an internal structure, in some larger diagram.

\begin{center}
\begin{figure}[h]
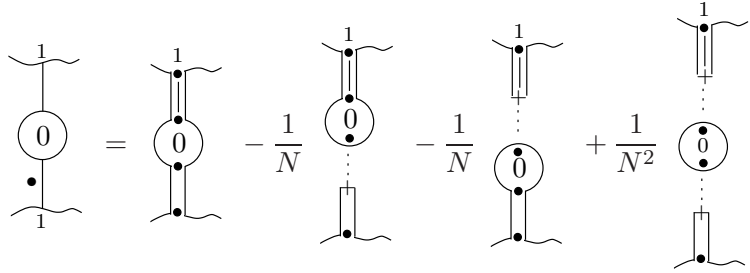

	\[
	\ensuremath{\begin{array}{c}\input{pstex/EffPropCorr-Pack.pstex_t} \end{array}} = \ensuremath{\begin{array}{c}\input{pstex/EffPropCorr-UnPack-A.pstex_t} \end{array}} -\frac{1}{N} \ensuremath{\begin{array}{c}\input{pstex/EffPropCorr-UnPack-C.pstex_t} \end{array}} -\frac{1}{N} \ensuremath{\begin{array}{c}\input{pstex/EffPropCorr-UnPack-B.pstex_t} \end{array}}
							+\frac{1}{N^2} \ensuremath{\begin{array}{c}\input{pstex/EffPropCorr-UnPack-D.pstex_t} \end{array}}
	\]
\caption{Showing how $1/N$ corrections to the effective propagator relation vanish.}
\label{fig:NFE:EffPropCorrs}
\end{figure}
\end{center}

To understand why this is the
case, consider a two-point, tree level vertex in the $A^1$
sector (the same analysis can be repeated in the $A^2$ sector).
To one field we attach an effective propagator and to the other we attach
a wine. The other ends of the effective 
propagator and wine are somehow tied up; we are not interested
in these details or in any corrections to the corresponding attachments.
This scenario is depicted in figure~\ref{fig:NFE:EffPropCorrs}. On the \lhs,
we use compact notation, where any corrections to the attachments
are packaged up. On the \rhs, we make the corrections to the 
vertex attachments explicit. Note that, as in~\cite{YM-1-loop},
an old-style wine with a line down its spine denotes an effective propagator.

The crucial point is that, in the final diagram, the vertex---which attaches to other structures
exclusively via false kernels---yields a factor of $\str \sigma_+ = N$.
Hence, the final diagram cancels either the second or third diagrams. Taking it to cancel
the third diagram makes it clear that what we are effectively left with is just the $1/N$ correction
due to the attachment of the wine to the vertex. In other words, we can use the effective
propagator relation as advertised.

Had we attached a second effective propagator to the vertex, rather than a wine, our
argument still works. Though it is up to us to choose, we can only apply the effective
propagator relation to one of the effective propagators. The remaining effective propagator
plays the role of the wine, in the above argument. Thus, it is not true to say
that effective propagators only attach to two-point vertices directly (\ie\ without any corrections).
What is true is that the effective propagator relationship can be applied na\"ively,
without any consideration to possible $1/N$ corrections.

Finally, notice how in diagrammatic identity~\ref{D-ID:NFE:EffPropReln} we have used
the equivalence symbol in equation~(\ref{eq:NFE:EffPProp}). 
In other words, we are taking the \rhs\ of equation~(\ref{eq:NFE:EffPProp})
to define what we mean by the \lhs.

There now follows a second diagrammatic identity, which follows
from the properties of $p'_M$ and $p_N$.

\begin{D-ID}
Using the definitions~(\ref{eq:NFE:Diag:p'}) and~(\ref{eq:NFE:Diag:p}),
together with the information of table~\ref{tab:NFE:k,k'} and equation~(\ref{eq:NFE-xf+2g})
gives us the diagrammatic identity below.
\begin{equation}
	\ensuremath{\begin{array}{c}\begin{picture}(0,0)%
\includegraphics{pstex/GR-relation.pstex}%
\end{picture}%
\setlength{\unitlength}{3947sp}%
\begingroup\makeatletter\ifx\SetFigFont\undefined%
\gdef\SetFigFont#1#2#3#4#5{%
  \reset@font\fontsize{#1}{#2pt}%
  \fontfamily{#3}\fontseries{#4}\fontshape{#5}%
  \selectfont}%
\fi\endgroup%
\begin{picture}(314,174)(2239,-821)
\end{picture}
 \end{array}} = 1
\label{eq:GR-relation}
\end{equation}
The index at the apex of $\GRk$ is contracted into the index carried
by $\GRkp$.
Note that this relationship exists only in the $A^{1,2}$ and $F$-sectors.

A number of corollaries follow from equation~(\ref{eq:GR-relation}) and the
fact that $p_N$ is independent of both $\Lambda$ and $\alpha$,
in all sectors.

\paragraph{Corollaries}
\begin{eqnarray*}
	\dec{\ensuremath{\begin{array}{c}\begin{picture}(0,0)%
\includegraphics{pstex/GR-relation.pstex}%
\end{picture}%
\setlength{\unitlength}{3947sp}%
\begingroup\makeatletter\ifx\SetFigFont\undefined%
\gdef\SetFigFont#1#2#3#4#5{%
  \reset@font\fontsize{#1}{#2pt}%
  \fontfamily{#3}\fontseries{#4}\fontshape{#5}%
  \selectfont}%
\fi\endgroup%
\begin{picture}(314,174)(2239,-821)
\end{picture}
 \end{array}}}{\bullet}					& = 			& 0	
\\
													& \Rightarrow 	& \ensuremath{\begin{array}{c}\begin{picture}(0,0)%
\includegraphics{pstex/kkpr-LdL.pstex}%
\end{picture}%
\setlength{\unitlength}{3947sp}%
\begingroup\makeatletter\ifx\SetFigFont\undefined%
\gdef\SetFigFont#1#2#3#4#5{%
  \reset@font\fontsize{#1}{#2pt}%
  \fontfamily{#3}\fontseries{#4}\fontshape{#5}%
  \selectfont}%
\fi\endgroup%
\begin{picture}(314,293)(2239,-821)
\put(2433,-612){\makebox(0,0)[lb]{\smash{\SetFigFont{8}{9.6}{\rmdefault}{\mddefault}{\updefault}{\color[rgb]{0,0,0}$\bullet$}%
}}}
\end{picture}
 \end{array}} = 0	
\\
	\pder{}{\alpha} \left[ \ensuremath{\begin{array}{c}\begin{picture}(0,0)%
\includegraphics{pstex/GR-relation.pstex}%
\end{picture}%
\setlength{\unitlength}{3947sp}%
\begingroup\makeatletter\ifx\SetFigFont\undefined%
\gdef\SetFigFont#1#2#3#4#5{%
  \reset@font\fontsize{#1}{#2pt}%
  \fontfamily{#3}\fontseries{#4}\fontshape{#5}%
  \selectfont}%
\fi\endgroup%
\begin{picture}(314,174)(2239,-821)
\end{picture}
 \end{array}} \right] & = 			& 0
\\
													& \Rightarrow	& \ensuremath{\begin{array}{c}\begin{picture}(0,0)%
\includegraphics{pstex/k.pstex}%
\end{picture}%
\setlength{\unitlength}{3947sp}%
\begingroup\makeatletter\ifx\SetFigFont\undefined%
\gdef\SetFigFont#1#2#3#4#5{%
  \reset@font\fontsize{#1}{#2pt}%
  \fontfamily{#3}\fontseries{#4}\fontshape{#5}%
  \selectfont}%
\fi\endgroup%
\begin{picture}(174,174)(2239,-821)
\end{picture}
 \end{array}} \pder{}{\alpha} \ensuremath{\begin{array}{c}\begin{picture}(0,0)%
\includegraphics{pstex/kpr.pstex}%
\end{picture}%
\setlength{\unitlength}{3947sp}%
\begingroup\makeatletter\ifx\SetFigFont\undefined%
\gdef\SetFigFont#1#2#3#4#5{%
  \reset@font\fontsize{#1}{#2pt}%
  \fontfamily{#3}\fontseries{#4}\fontshape{#5}%
  \selectfont}%
\fi\endgroup%
\begin{picture}(174,174)(2379,-821)
\end{picture}
 \end{array}} \equiv \ensuremath{\begin{array}{c}\begin{picture}(0,0)%
\includegraphics{pstex/GRk-dal-GRkpr.pstex}%
\end{picture}%
\setlength{\unitlength}{3947sp}%
\begingroup\makeatletter\ifx\SetFigFont\undefined%
\gdef\SetFigFont#1#2#3#4#5{%
  \reset@font\fontsize{#1}{#2pt}%
  \fontfamily{#3}\fontseries{#4}\fontshape{#5}%
  \selectfont}%
\fi\endgroup%
\begin{picture}(314,357)(2239,-821)
\put(2445,-560){\makebox(0,0)[lb]{\smash{\SetFigFont{8}{9.6}{\rmdefault}{\mddefault}{\updefault}{\color[rgb]{0,0,0}$\alpha$}%
}}}
\end{picture}
 \end{array}} =0
\end{eqnarray*}
We have introduced \Dal\ to represent $\partial / \partial \alpha$.
\label{D-ID:GaugeRemainderRelation}
\end{D-ID}

In retrospect, diagrammatic identity~\ref{D-ID:GaugeRemainderRelation}
arises as a consequence of the effective propagator relation
and gauge invariance. To see this, we note from the 
Ward identities~(\ref{eq:WID-Unbroken}) and~(\ref{eq:WID-Broken})
that
\begin{equation}
	p_M S_{0 MS}^{\ \, X  X}(p) = 0.
\label{eq:GR-TLTP}
\end{equation}
This follows directly in the $A$-sector, since 
one-point $A$-vertices do not exist.
In the $F$-sector, though,
we are left with a one-point $C$-vertex. However,
such vertices do not exist at tree level. 

Now consider attaching an
effective propagator to the vertex $S_{0 MS}^{\ \, X  X}(p)$
in equation~(\ref{eq:GR-TLTP}). Clearly, we must
still be left with zero, but we have the option of utilising the effective propagator
relation before the $p_M$ has acted. Diagrammatically,
this gives:
\[
	\ensuremath{\begin{array}{c}\begin{picture}(0,0)%
\includegraphics{pstex/GR-TLTP-EP.pstex}%
\end{picture}%
\setlength{\unitlength}{3947sp}%
\begingroup\makeatletter\ifx\SetFigFont\undefined%
\gdef\SetFigFont#1#2#3#4#5{%
  \reset@font\fontsize{#1}{#2pt}%
  \fontfamily{#3}\fontseries{#4}\fontshape{#5}%
  \selectfont}%
\fi\endgroup%
\begin{picture}(1081,306)(2490,-1356)
\put(2776,-1250){\makebox(0,0)[lb]{\smash{\SetFigFont{11}{13.2}{\rmdefault}{\mddefault}{\updefault}{\color[rgb]{0,0,0}0}%
}}}
\end{picture}
 \end{array}} = 0 = \ensuremath{\begin{array}{c}\begin{picture}(0,0)%
\includegraphics{pstex/k.pstex}%
\end{picture}%
\setlength{\unitlength}{3947sp}%
\begingroup\makeatletter\ifx\SetFigFont\undefined%
\gdef\SetFigFont#1#2#3#4#5{%
  \reset@font\fontsize{#1}{#2pt}%
  \fontfamily{#3}\fontseries{#4}\fontshape{#5}%
  \selectfont}%
\fi\endgroup%
\begin{picture}(174,174)(2239,-821)
\end{picture}
 \end{array}} - \ensuremath{\begin{array}{c}\begin{picture}(0,0)%
\includegraphics{pstex/kkprk.pstex}%
\end{picture}%
\setlength{\unitlength}{3947sp}%
\begingroup\makeatletter\ifx\SetFigFont\undefined%
\gdef\SetFigFont#1#2#3#4#5{%
  \reset@font\fontsize{#1}{#2pt}%
  \fontfamily{#3}\fontseries{#4}\fontshape{#5}%
  \selectfont}%
\fi\endgroup%
\begin{picture}(499,174)(2239,-820)
\end{picture}
 \end{array}},
\]
which implies diagrammatic identity~\ref{D-ID:GaugeRemainderRelation}.

We conclude this section with a wonderful relationship between
the effective propagators and the gauge remainders. To motivate
this, consider an $A^1$-sector effective propagator attached to
the $p'_M$ part of a gauge remainder. Explicitly keeping track of
which end of the effective propagator carries momentum $p$ and
which carries $-p$, we can write:
\[
	\EP{11}{}(-p,p) \frac{p_\mu}{p^2} = (-p)_\mu \tilde{\Delta}^{11}(-p,p)
\]
where, trivially, $\tilde{\Delta}^{11}(-p,p) = -\EP{11}{}(-p,p)/p^2 = -\EP{11}{}(p)/p^2$. The
point of doing this is that we have converted a $p'_M$ like part
of a gauge remainder attached to one end of an effective
propagator into a $p_M$ like part attached to the other end. The minus
sign is inserted so that we have a consistent picture of the
momentum flow.

We can thus think of components of gauge remainders at one end of
an effective propagator as begin able to pass through the effective
propagator, but inducing a change, in the process.
We call the object $\tilde{\Delta}$
a \emph{pseudo effective propagator}. 

The remarkable thing is that an equivalent relationship holds in the $F$-sector,
too:
\[
	\EP{\, F\, \bar{F}}{MN}{(-p,p)} p'_N = (-p)_M \tilde{\Delta}^{F \bar{F}}(-p,p),
\]
where $\tilde{\Delta}^{F \bar{F}}(p) = -\tilde{c}_p f_p g_p/ 2\Lambda^4$ and
we note that $-p_M \neq (-p)_M$. This leads us to another diagrammatic
identity.

\begin{D-ID}
Consider an effective propagator attached, at one end
to the $p'_M$ part of a gauge remainder. This can be
redrawn as follow:
\[
	\ensuremath{\begin{array}{c}\begin{picture}(0,0)%
\includegraphics{pstex/EP-GRpr.pstex}%
\end{picture}%
\setlength{\unitlength}{3947sp}%
\begingroup\makeatletter\ifx\SetFigFont\undefined%
\gdef\SetFigFont#1#2#3#4#5{%
  \reset@font\fontsize{#1}{#2pt}%
  \fontfamily{#3}\fontseries{#4}\fontshape{#5}%
  \selectfont}%
\fi\endgroup%
\begin{picture}(628,174)(1926,-821)
\end{picture}
 \end{array}} = \ensuremath{\begin{array}{c}\begin{picture}(0,0)%
\includegraphics{pstex/GR-PEP.pstex}%
\end{picture}%
\setlength{\unitlength}{3947sp}%
\begingroup\makeatletter\ifx\SetFigFont\undefined%
\gdef\SetFigFont#1#2#3#4#5{%
  \reset@font\fontsize{#1}{#2pt}%
  \fontfamily{#3}\fontseries{#4}\fontshape{#5}%
  \selectfont}%
\fi\endgroup%
\begin{picture}(786,174)(2089,-821)
\end{picture}
 \end{array}},
\]
where the dash-dot line represents the pseudo effective propagators.

\upshape
\paragraph{Corollary}
\[
	\ensuremath{\begin{array}{c}\begin{picture}(0,0)%
\includegraphics{pstex/EP-FullGR.pstex}%
\end{picture}%
\setlength{\unitlength}{3947sp}%
\begingroup\makeatletter\ifx\SetFigFont\undefined%
\gdef\SetFigFont#1#2#3#4#5{%
  \reset@font\fontsize{#1}{#2pt}%
  \fontfamily{#3}\fontseries{#4}\fontshape{#5}%
  \selectfont}%
\fi\endgroup%
\begin{picture}(778,174)(1926,-826)
\end{picture}
 \end{array}} = \ensuremath{\begin{array}{c}\begin{picture}(0,0)%
\includegraphics{pstex/GR-PEP-GR.pstex}%
\end{picture}%
\setlength{\unitlength}{3947sp}%
\begingroup\makeatletter\ifx\SetFigFont\undefined%
\gdef\SetFigFont#1#2#3#4#5{%
  \reset@font\fontsize{#1}{#2pt}%
  \fontfamily{#3}\fontseries{#4}\fontshape{#5}%
  \selectfont}%
\fi\endgroup%
\begin{picture}(929,174)(2089,-824)
\end{picture}
 \end{array}} = \ensuremath{\begin{array}{c}\begin{picture}(0,0)%
\includegraphics{pstex/FullGR-EP.pstex}%
\end{picture}%
\setlength{\unitlength}{3947sp}%
\begingroup\makeatletter\ifx\SetFigFont\undefined%
\gdef\SetFigFont#1#2#3#4#5{%
  \reset@font\fontsize{#1}{#2pt}%
  \fontfamily{#3}\fontseries{#4}\fontshape{#5}%
  \selectfont}%
\fi\endgroup%
\begin{picture}(778,174)(1965,-825)
\end{picture}
 \end{array}}
\]

\label{D-ID:PseudoEffectivePropagatorRelation}
\end{D-ID}

In retrospect, diagrammatic identity~\ref{D-ID:PseudoEffectivePropagatorRelation}
can be deduced from the previous two diagrammatic identities.
Diagrammatic identities~\ref{D-ID:NFE:EffPropReln}
and~\ref{D-ID:GaugeRemainderRelation} imply that
\[
	\ensuremath{\begin{array}{c}\begin{picture}(0,0)%
\includegraphics{pstex/TLTP-EP-GR.pstex}%
\end{picture}%
\setlength{\unitlength}{3947sp}%
\begingroup\makeatletter\ifx\SetFigFont\undefined%
\gdef\SetFigFont#1#2#3#4#5{%
  \reset@font\fontsize{#1}{#2pt}%
  \fontfamily{#3}\fontseries{#4}\fontshape{#5}%
  \selectfont}%
\fi\endgroup%
\begin{picture}(1059,306)(2512,-1356)
\put(2776,-1250){\makebox(0,0)[lb]{\smash{\SetFigFont{11}{13.2}{\rmdefault}{\mddefault}{\updefault}{\color[rgb]{0,0,0}0}%
}}}
\end{picture}
 \end{array}} = \ensuremath{\begin{array}{c}\begin{picture}(0,0)%
\includegraphics{pstex/kpr.pstex}%
\end{picture}%
\setlength{\unitlength}{3947sp}%
\begingroup\makeatletter\ifx\SetFigFont\undefined%
\gdef\SetFigFont#1#2#3#4#5{%
  \reset@font\fontsize{#1}{#2pt}%
  \fontfamily{#3}\fontseries{#4}\fontshape{#5}%
  \selectfont}%
\fi\endgroup%
\begin{picture}(174,174)(2379,-821)
\end{picture}
 \end{array}} - \ensuremath{\begin{array}{c}\begin{picture}(0,0)%
\includegraphics{pstex/kprkkpr.pstex}%
\end{picture}%
\setlength{\unitlength}{3947sp}%
\begingroup\makeatletter\ifx\SetFigFont\undefined%
\gdef\SetFigFont#1#2#3#4#5{%
  \reset@font\fontsize{#1}{#2pt}%
  \fontfamily{#3}\fontseries{#4}\fontshape{#5}%
  \selectfont}%
\fi\endgroup%
\begin{picture}(508,174)(2379,-817)
\end{picture}
 \end{array}} = 0.
\]
In other words, the (non-zero) structure $\ensuremath{\begin{array}{c}\begin{picture}(0,0)%
\includegraphics{pstex/EP-GR.pstex}%
\end{picture}%
\setlength{\unitlength}{3947sp}%
\begingroup\makeatletter\ifx\SetFigFont\undefined%
\gdef\SetFigFont#1#2#3#4#5{%
  \reset@font\fontsize{#1}{#2pt}%
  \fontfamily{#3}\fontseries{#4}\fontshape{#5}%
  \selectfont}%
\fi\endgroup%
\begin{picture}(406,118)(3165,-1234)
\end{picture}
 \end{array}}$ kills
a two-point, tree level vertex. But, by equation~(\ref{eq:GR-TLTP}), 
this implies that the structure $\ensuremath{\begin{array}{c}\begin{picture}(0,0)%
\includegraphics{pstex/EP-GR.pstex}%
\end{picture}%
\setlength{\unitlength}{3947sp}%
\begingroup\makeatletter\ifx\SetFigFont\undefined%
\gdef\SetFigFont#1#2#3#4#5{%
  \reset@font\fontsize{#1}{#2pt}%
  \fontfamily{#3}\fontseries{#4}\fontshape{#5}%
  \selectfont}%
\fi\endgroup%
\begin{picture}(406,118)(3165,-1234)
\end{picture}
 \end{array}}$
must be equal, up to some factor, to $\ensuremath{\begin{array}{c}\begin{picture}(0,0)%
\includegraphics{pstex/k-mirror.pstex}%
\end{picture}%
\setlength{\unitlength}{3947sp}%
\begingroup\makeatletter\ifx\SetFigFont\undefined%
\gdef\SetFigFont#1#2#3#4#5{%
  \reset@font\fontsize{#1}{#2pt}%
  \fontfamily{#3}\fontseries{#4}\fontshape{#5}%
  \selectfont}%
\fi\endgroup%
\begin{picture}(174,174)(2089,-821)
\end{picture}
 \end{array}}$.

\subsection{Universality}	\label{sec:NFE:Universality}

As we have pointed out already, the precise algebraic
forms of the two-point, tree level vertices,
effective propagators and gauge remainders have
an element of choice in them. In this section, we
wish to clarify precisely which parts are universal,
which parts are forced upon us by the regularisation
and which parts are purely down to choice. Similarly,
we comment on whether or not we expect our diagrammatic
identities to be generally true.

The entire purpose of our setup is ultimately
to provide a means for computing physically observable quantities.
Thus, despite the elaborate structure of the \MGIERG,
our real interest is in the physical field $A^1$.

All physically observable quantities must
be controlled by the renormalisation condition
for $A^1$ (equation~(\ref{eq:Intro:RenormCondition})),
which is where the physics feeds in.
Consequently, the single universal element of all
that we have discussed in this section is the $\Op{2}$
part of the $S_{0 \mu \nu}^{\ 1\, 1}(p)$ vertex:
\[
	S_{0 \mu \nu}^{\ 1\, 1}(p) = 2\Box_{\mu \nu}(p) + \Op{4}.
\]

Now consider the actual computation of a (pseudo)-universal
quantity within our framework. Whilst it is true that
the final answer will be controlled by the renormalisation
condition for $A^1$, the extraction of a meaningful answer
implicitly assumes that the framework is consistent. Thus,
whilst a universal quantity cannot depend on the fact that
we are using a spontaneously broken
$SU(N|N)$ regularisation scheme, within
the \ERG, the fact that we have chosen compute in this way
makes certain demands. We have seen a manifestation of these
demands in the Ward identities and
in constraints on the UV behaviour of the Wilsonian
effective action and seed action, two-point,
tree level vertices.

Generally speaking, there is no reason why the Wilsonian
effective action and seed action two-point, tree level
vertices need be identified with each other and so the
fact that they are is purely down to choice. 
Having made this choice, we are still at liberty to choose
the precise algebraic form of the two-point, tree level
vertices, so long as this choice is consistent
with the renormalisation condition, gauge invariance
and preserves the regularisation.
Irrespective  of how we do this, 
we can ensure that the diagrammatic identities all still hold. We have
seen that, given the effective propagator relationship
and gauge invariance, all the other diagrammatic identities follow.
However, the effective propagator relationship itself
follows once we have set the Wilsonian effective action and seed action
two-point, tree level vertices equal to each other.

Let us examine this statement in more detail. Having
set the Wilsonian effective action and seed action two-point,
tree level vertices equal we have, from equation~(\ref{eq:TLTP-flow})
\[
	 \flowConstAl S_0^{\ C^1 C^1}(p) =  S_0^{\ C^1 C^1}(p) \DEP{C^1}{C^1}(p) S_0^{\ C^1 C^1}(p).
\]
Lorentz invariance demands that $S_0^{\ C^1 C^1}(p)$ is an even function
of $p$ and so
\[
	\DEP{C^1}{C^1}(p) = -\flowConstAl \left[S_0^{\ C^1 C^1}(p)\right]^{-1}.
\]
Integrating up, we have
\[
	\EP{C^1 C^1}{}(p) = \left[S_0^{\ C^1 C^1}(p)\right]^{-1} + \frac{\mbox{const}}{p^4}.
\]
Let us examine the integration constant, in more detail. The
only dimensionful quantities available are $\Lambda$ and $p$.
Since the integration constant is
necessarily independent  of the former it must go as  $1/p^4$,
by dimensions.
The dimensionless constant which multiplies this, being
independent of $\Lambda$, must be independent of $p$
and so only function of $\alpha$. 

Now, $C^1$ is a massive field, so demanding that its effective
propagator is well behaved as $p \rightarrow 0$ uniquely
fixes the integration constant to zero. The effective
propagator relation follows, directly.

In the $A^1$-sector, equation~(\ref{eq:TLTP-flow}) tells us that
\[
	\flowConstAl S^{\ 1 \, 1}_{0 \mu \nu}(p) = S^{\ 1 \, 1}_{0 \mu \alpha}(p) \DEP{1}{1}(p) S^{\ 1 \, 1}_{0 \alpha \nu}(p).
\]
Using equation~(\ref{eq:NFE:S11-General}) and the fact that 
$\Box_{\mu \alpha}(p) \Box_{\alpha \nu}(p) = p^2 \Box_{\mu \nu}(p)$
we have:
\[
	-\flowConstAl \frac{1}{\NCO{p}} = p^2 \DEP{1}{1}(p)
\]
and, therefore, that
\[
	\frac{1}{\NCO{p}} = p^2 \EP{11}{}(p) + p^2 \times \mbox{const}.
\]
Demanding that the effective propagator vanishes
in the limit $p \rightarrow \infty$~\cite{YM-1-loop} 
uniquely fixes the  constant, which
is dimensionless and independent of $\Lambda$, to be zero;
the effective propagator relation follows.

Similar arguments to those above yield the
effective propagator relation in the $A^2$, $C^2$
and $F$-sectors.

Now, given that we work in the scheme where the Wilsonian
effective action and seed action two-point, tree level
vertices are equal, there is a second place in
which the $A^1$ renormalisation condition is forced to
feed in. Since,
\[
	S^{\ 1 \, 1}_{0 \mu \nu}(p)\EP{11}{}(p) = \delta_{\mu \nu} - p_\mu p_\nu / p^2
\] 
and the $\Op{2}$ part of the vertex is universal, this implies
that the $\Op{-2}$ part of the effective propagator is universal,
also.

To conclude this section, we introduce some nomenclature. Since we have
furnished the two-point,
tree level vertices and effective propagators with algebraic
realisations, we will refer to such objects as `algebraic'.
We note that components of these objects may be either
universal or non-universal. These objects are
the only ones for which we will ever introduce an
algebraic realisation.\footnote{We do not count the
zero-point wines separately, since they are just differentiated
effective propagators.} 

Although we never introduce algebraic realisations for
any other objects (\ie\
higher point / higher loop order vertices
and decorated wines) we note that they may have components
related, via gauge invariance, to our algebraic terms.
In turn, this implies that such objects can have, buried
in them, universal components.

\subsection{Constraints in the $C$-sector}

We conclude our discussion of the weak coupling flow
equations by giving the weak coupling version
of the constraint~(\ref{eq:NPC-sec-Constraint})
which, we recall, arises from demanding that
all Wilsonian effective action one-point $C^{1,2}$
vertices vanish.

Using the diagrammatic weak coupling flow equation
(figure~\ref{fig:NFE:NewDiagFE}) we specialise
$\{f\}$ to $C^{1,2}$. Immediately, we find a number
of simplifications, since all terms containing
a vertex with a one-point Wilsonian effective
action vertex (after decoration) vanish. Hence,
the $\beta$ and $\alpha$ terms vanish.
Expanding out the bar notation according to
equation~(\ref{eq:NFE:BarNotation}), the
$a_0[S_{n-r}, S_r]$ term dies, since one of
the vertices is compelled to be a Wilsonian
effective action one-point $C^{1,2}$ vertex.
The remaining two terms combine, and we note that
here we are compelled to leave both the seed action
vertex and wine undecorated.

There is an additional simplification. Since each diagram
is decorated by a single field, this field must carry
zero momentum. Now,  it follows by dimensions (and is
trivial to confirm) that
\[
	\DEP{C^1}{C^1}(0) = 4 \EP{C^1C^1}{}(0),
\]
and similarly for $C^2$.
If we make this replacement when the wine attaches
to a two-point, tree level vertex, then we can use
the effective propagator relation. We thus
arrive at the constraint for the $n\geq 1$-loop seed action,
one-point vertices,
shown in figure~\ref{fig:NFE:C-Constraint:Weak}.\footnote{We recall
that, at tree level, the one-point, seed action vertex is guaranteed
to vanish
because $\SH$ is dimensionless and the classical
minimum of the Wilsonian effective action Higgs potential lies at $\sigma$~\cite{YM-1-loop}.}
\begin{center}
\begin{figure}[h]
	\begin{equation}
		4 \ensuremath{\begin{array}{c}\input{pstex/Vertex-n-S-C.pstex_t} \end{array}} = -\sum_{r=1}^{n-1} \ensuremath{\begin{array}{c}\input{pstex/Dumbbell-n_r-S-r-C.pstex_t} \end{array}} - \frac{1}{2} \left[ \ensuremath{\begin{array}{c}\input{pstex/Sigma_n-1-C-DEP.pstex_t} \end{array}} + \ensuremath{\begin{array}{c}\input{pstex/Sigma_n-1-W-C.pstex_t} \end{array}}\right] 
	\label{eq:NFE:C-Constraint:Weak}
	\end{equation}
\caption{The weak coupling regime constraint to ensure that the minimum of the superhiggs
potential is not shifted by quantum corrections.}
\label{fig:NFE:C-Constraint:Weak}
\end{figure}
\end{center}

Note that the $n$-loop seed action vertex is related to objects of
loop order $n-1$ or lower. Thus, by appropriately tuning $\hat{S}$,
we can  satisfy the constraint order by order in perturbation
theory.

\part{Techniques}

\chapter{Further Diagrammatic Techniques} \label{ch:FurtherDiagrammatics}

Having introduced both the new flow equation and the corresponding diagrammatic interpretation,
we devote this chapter to describing a further set of diagrammatic techniques. The current
state of affairs is that we can diagrammatically represent the flow of a vertex, and can
process the result further by utilising the effective propagator relation.

The effective propagator relation allows us to replace a two-point vertex connected to
an effective propagator with a Kronecker $\delta$ and a `gauge remainder'. In section~\ref{sec:GRs}, 
we will see how we can deal with these remainders, diagrammatically. This will
require that we broaden our understanding of both the Ward identities 
and no-$\SF^0$ symmetry.

In section~\ref{sec:MomentumExpansions}, we utilise the insights gained from the treatment
of the gauge remainders to develop a diagrammatic technique for Taylor expanding vertices and wines in momenta.

We round off the diagrammatic techniques with a further set of 
diagrammatic identities that
will be heavily used in the computation of 
$\beta$-function coefficients.

\section{Gauge Remainders} \label{sec:GRs}

Up until now, we have refered to the composite
object $\GRkp \!\! \GRk$ as a gauge remainder.
Henceforth, we will often loosely refer to the
individual components as gauge remainders. To
make an unambiguous reference, we call $\GRk$
an active gauge remainder, $\GRkp$ a processed
gauge remainder and $\GRkp \!\! \GRk$ a full
gauge remainder.

\subsection{Vertices}

\begin{center}
\begin{figure}[h]
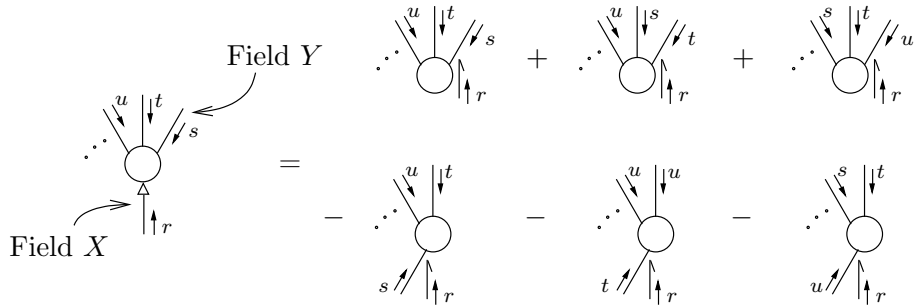

	\[
	\ensuremath{\begin{array}{c}\input{pstex/GR-Defining.pstex_t} \end{array}} = \ensuremath{\begin{array}{c}\input{pstex/GR-Defining-B.pstex_t} \end{array}}
	\]
\caption[Diagrammatics for the Ward Identities.]{The \lhs\ shows the contraction of an arbitrary vertex with one of its momenta. On the \rhs, the first row of diagrams
shows all possible pushes forward on to the explicitly drawn fields and the second row shows all possible pulls back.}
\label{fig:GR:Defining}
\end{figure}
\end{center}

We begin by considering an arbitrary vertex, which is contracted with
the momentum carried by one of its fields, $X$, as shown on the \lhs\ of figure~\ref{fig:GR:Defining}.
All of the fields shown are wildcards, though the field $X$ has no support in the $C$-sector.
 To proceed, we use
the  Ward identities~(\ref{eq:WID-Unbroken}) and~(\ref{eq:WID-Broken}), which
tell us that we either push forward or pull back (with a minus sign) the momentum of $X$ to the next
field on the vertex. We recall from section~\ref{sec:Intro:Diagramatics} that fields are read off
a vertex in the counterclockwise sense; hence, we push forward counterclockwise and pull back clockwise.

Since the vertex contains all possible (cyclically independent) orderings of the fields, spread 
over all (legal) combinations of supertraces, any of the fields could precede or follow $X$. Hence, we must sum
over all possible pushes forward and pulls back, as shown on the \rhs\ of figure~\ref{fig:GR:Defining}.

It is clear from the Ward identities~(\ref{eq:WID-Unbroken}) and~(\ref{eq:WID-Broken}) that
the diagrams on the \rhs\ of figure~\ref{fig:GR:Defining} have no explicit
dependence on the field $X$.
Nonetheless, to interpret the diagrams on the \rhs\ unambiguously, without reference to the parent,
we must retain information about $X$. This is achieved by keeping the line which used to represent $X$ but which is
now terminated by a half arrow, rather than entering the vertex. This line carries information about the flavour of $X$
and its momentum, whilst  indicating which field it is that has been pushed forward / pulled back on to. 
The half arrow can go on either side of the line.\footnote{This should be borne in 
mind when we encounter pseudo effective propagators, attached to $\GRk$.}

The diagrammatics we have been using has been, up until now, completely blind to details concerning the ordering of fields
and the supertrace structure. If we are to treat gauge remainders diagrammatically, we can no longer exactly 
preserve these features.
Let us suppose that we have pushed forward the momentum of $X$ on to the field $Y$, as depicted in the first diagram on the \rhs\ of 
figure~\ref{fig:GR:Defining}. Clearly, it must be the case that $X$ and $Y$
are on the same supertrace and that $Y$ is immediately after $X$, in the counter-clockwise sense.
The other fields on the vertex---which we will call spectator fields---can be in any order and distributed 
amongst any  number of supertraces,
up to the requirement that they do not come between the fields $X$ and $Y$. To deduce the momentum flowing into the vertex
along $Y$, we simply follow the indicated momentum routing. Hence, momentum $r+s$ enters the vertex along $Y$, in the case that
it is the field $Y$ that has been pushed forward (pulled back) on to. Similarly, if we push forward on to the field carrying momentum $t$,
then momentum $r+t$ enters the vertex, along this field.
We must also take into account that the flavour of $Y$ will change if $X$ is fermionic;
these changes are given beneath~(\ref{eq:WID-Broken}). 

However, our current form for the
the Ward identities are in terms of the fields $A, \ C$ and $F$, whereas we
want to be working with $A^1$s, $A^2$s and \etcB. This is easy enough to 
achieve: we can always project down to the required sub-sector using
$\sigma_\pm$ (see equation~(\ref{eq:NFE:Projectors})). There is, though, one
major subtlety: the fermionic gauge transformations can generate $\SF^0$.
Though we will see that, for the calculations we wish to perform,
we can adopt a prescription whereby we effectively
remove $\SF^0$ once and for all, as an intermediate step 
we define
\begin{eqnarray*}
	\tilde{A}^1 & \equiv & A^1 + \sigma_+ \SF^0 
\\ 
	\tilde{A}^2 & \equiv & A^2 + \sigma_- \SF^0.
\end{eqnarray*}

The flavour changes arising from pushing forward/ pulling back the (anti) fermionic field, $X$
on to the field, $Y$,
are summarised in table~\ref{tab:GR:Flavour}. Bosonic fields have
been treated together.
We note that the entries in the table contain some annoying signs. There is nothing that can be done about these; we 
simply incorporate them into the diagrammatic rules. Before any momenta have been contracted into vertices, the wildcard 
fields have neat interpretations in terms of $F$, $\bar{F}$, $(A^1,C^1)$ and $(A^2,C^2)$. After such contractions have been performed,
though, the precise meaning of the wildcard fields must be deduced from the field structure of the diagram.
\begin{center}
\begin{table}[h]
\[
	\begin{array}{l|l|c|c}
		X 		& 	Y 		& \PF{Y} 			& \PB{Y}
\\ \hline\hline
		\bar{F} & (A^1,C^1) &(\bar{B}, \bar{D})	& \mbox{---}
\\
  				& (A^2,C^2) &\mbox{---} 		& (\bar{B},\bar{D}) 
\\
  				& F 		&(\tilde{A}^2, C^2) & (\tilde{A}^1,C^1) 
\\ \hline
		F 		& (A^2,C^2) & F 				& \mbox{---}
\\ 
  				& (A^1,C^1) & \mbox{---} 		& F 
\\
  				& \bar{F} 	&(\tilde{A}^1,-C^1) & (\tilde{A}^2,-C^2) 
\end{array}
\]
\caption[The flavour changing effect of pushing forward and / or pulling back the momentum of a fermionic field on to some other field.
]{The flavour changing effect of pushing forward and / or pulling back the momentum of the 
fermionic field $X$ on to the field $Y$. If the field $Y$ is bosonic, then
the flavour of $X$ and $Y$ uniquely determine whether $X$ must follow $Y$, or vice-versa. The empty entries correspond to the case
where the required ordering of fields does not exist. The entries in this table do not include the relative minus sign always present 
between pushes forward and pulls back.}
\label{tab:GR:Flavour}
\end{table}
\end{center}

Let us suppose that we have some vertex which
is decorated by, \eg, an $F$ and a $\bar{B}$ and that
these fields are on the same supertrace. Now consider
contracting the vertex with the momentum of the $F$.
There are two cases to analyse. The first is where there
are other fields on the same supertrace as the $F$ and
$\bar{B}$. For argument's sake, we will take them to be such
that the $\bar{B}$ follows the $F$ (in the counterclockwise
sense). Now the $F$ can push forward on to the $\bar{B}$,
generating a $\tilde{A}^1$. 
In this case,
the vertex \emph{coefficient function} is blind to whether its argument
involves $\tilde{A}^1$ or $A^1$: starting with a vertex containing $\tilde{A}^1$,
we can always remove the $\SF^0$ part by no-$\SF^0$ symmetry.

The second case to look at is where there are no other fields
on the same supertrace as $F$ and $\bar{B}$. Now the $F$ can
both push forward and pull back on to the $\bar{B}$ generating,
respectively, $\tilde{A}^1$ and $\tilde{A}^2$. However,
we cannot rewrite $\tilde{A}^{1,2}$ as $A^{1,2}$ since,
whereas $\str \tilde{A}^{1,2} \neq 0$, $\str A^{1,2} =0$.
Our strategy is to rewrite vertices involving a separate $\str \tilde{A}^{1,2}$
factor via no-$\SF^0$ symmetry.

We have encountered a no-$\SF^0$ relation already---see equation~(\ref{eq:Intro:No-A0-Ex}).
Now we will generalise this relationship, which is most readily done by example.
Consider the following part of the action, where we remember that all position arguments
are integrated over:
\begin{eqnarray*}
	&& \cdots + \frac{1}{3} S^{1 \, 1 \, 1}_{\alpha \beta \gamma}(x,y,z) \str \tilde{A}^1_\alpha (x) \tilde{A}^1_\beta (y) \tilde{A}^1_\gamma (z)
\\
	&&
	+\frac{1}{2}  S^{1 \, 1, A \sigma}_{\alpha \beta,  \gamma}(x,y;z) \str \tilde{A}^1_\alpha (x) \tilde{A}^1_\beta (y) \; \str A_\gamma (z) \sigma
	+\cdots
\end{eqnarray*}
We note that, in the second term, we have combined $S^{AA,A}$ with $S^{A,AA}$, thereby killing the factor of $1/2!$
associated with each of these vertices.

To determine the no-$\SF^0$ relationship between these vertices, we shift  $A$: $\delta A_\mu(x) = \lambda_\mu(x) \one$,
and collect together terms with the same supertrace structure and the same dependence on $\lambda_\mu$.
By restricting ourselves to the portion of the action shown, we only find common terms which
depend on a single power of $\lambda_\mu$. By
using a larger portion of the action we can, of course, obtain higher order relationships, though
we will not require these. In the single
supertrace vertex, this operation simply kills the factor of $1/3$; in the double supertrace term, we
focus on shifting the lone $A$ which yields:
\begin{eqnarray*}
	&&  \cdots + \lambda_\gamma(z)  S^{1 \, 1 \, 1}_{\alpha \beta \gamma}(x,y,z) \str \tilde{A}^1_\alpha (x) \tilde{A}^1_\beta (y)
\\
	&&
	+\frac{1}{2}  \lambda_\gamma(z) S^{1 \, 1, A \sigma}_{\alpha \beta, \gamma}(x,y;z) \str \tilde{A}^1_\alpha (x) \tilde{A}^1_\beta (y) \; \str \sigma
	+\cdots
\end{eqnarray*}

Notice that we can rewrite the first term to recognise the invariance of the
supertrace under cycling the remaining fields: we replace the existing vertex
coefficient function
by
\[
	\frac{1}{2} \left[ S^{1 \, 1 \, 1}_{\alpha \beta \gamma}(x,y,z) + S^{1 \, 1 \, 1}_{\alpha \gamma \beta}(x,z,y) \right].
\]
Our no-$\SF^0$ relation is, therefore:
\[
	S^{1 \, 1 \, 1}_{\alpha \beta \gamma}(x,y,z) + S^{1 \, 1 \, 1}_{\alpha \gamma \beta}(x,z,y) + 2N S^{1 \, 1, A \sigma}_{\alpha \beta, \gamma}(x,y;z) = 0.
\]
We now recast the final term, so that we work with $\tilde{A}^1$s and $\tilde{A}^2$s,
rather than $A \sigma$. This is a little counterintuitive. At first, we recognise
that $A \sigma = \tilde{A}^1 - \tilde{A}^2$. However, $\tilde{A}^1$ and
$\tilde{A}^2$ are not independent. Specifically,
\[
	\str \tilde{A}^1 = N\SF^0 = - \str \tilde{A}^2.
\]
Consequently, we need to be careful what we mean by the vertex
coefficient functions $S^{\cdots, \tilde{A}^{1,2}}$. If, as we will do,
we treat the vertex coefficient functions $S^{\cdots, \tilde{A}^1}$
and $S^{\cdots, \tilde{A}^2}$ as independent then, by recognising
that
\[
	S^{\cdots, \tilde{A}^1} (\str \cdots) \str \tilde{A}^1 + S^{\cdots, \tilde{A}^2} (\str \cdots) \str \tilde{A}^2
\]
is equivalent to
\[
	S^{\cdots, A\sigma} (\str \cdots) \str A \sigma
\]
and writing out the explicitly indicated supertraces in terms of $\SF^0$ and $N$, we find
that
\[
	S^{\cdots, \tilde{A}^1} - S^{\cdots, \tilde{A}^2}  = 2S^{\cdots, A\sigma}.
\]
The factor of two on the \rhs\ is, perhaps, unexpected; we emphasise that it
comes from splitting up the variable $\SF^0$ between two other variables. In figure~\ref{fig:noA0}
we give a diagrammatic form for the subset of first order no-$\SF^0$ relations which relate single
supertrace vertices to two supertrace vertices. 
\begin{center}
\begin{figure}[h]
	\begin{eqnarray}
		&	& \ensuremath{\begin{array}{c}\input{pstex/Vertex-WW-A1-Full.pstex_t} \end{array}} \hspace{1em} + \cdots + \ensuremath{\begin{array}{c}\input{pstex/Vertex-A1-WW-Full.pstex_t} \end{array}} \hspace{1em} + \ensuremath{\begin{array}{c}\input{pstex/Vertex-WA1W-Full.pstex_t} \end{array}} \nonumber
	\\[1ex]
		& + & \ensuremath{\begin{array}{c}\input{pstex/Vertex-WW-A2-Full.pstex_t} \end{array}} \hspace{1em} + \cdots + \ensuremath{\begin{array}{c}\input{pstex/Vertex-A2-WW-Full.pstex_t} \end{array}} \hspace{1em} + \ensuremath{\begin{array}{c}\input{pstex/Vertex-WA2W-Full.pstex_t} \end{array}} \nonumber
	\\[1ex]
		& +	& N \left[ \ensuremath{\begin{array}{c}\input{pstex/Vertex-WW-Full--A1.pstex_t} \end{array}} \hspace{1em} -\hspace{0.5em} \ensuremath{\begin{array}{c}\input{pstex/Vertex-WW-Full--A2.pstex_t} \end{array}} \right] \nonumber
	\\[1ex]
		& =	& 0
	\label{eq:NoA0}
	\end{eqnarray}
\caption{Diagrammatic form of the first order no-$\SF^0$ relations.}
\label{fig:noA0}
\end{figure}
\end{center}

A number of comments are in order. First,
this relationship is trivially generalised to include terms
with additional supertraces. Secondly, if we restrict the action to single supertrace terms,
as in~\cite{YM-1-loop, Antonio'sThesis}, then the first two lines reproduce
the no-$\SF^0$ relations of~\cite{Antonio'sThesis}. Thirdly, the Feynman rules
are such that some of the diagrams of figure~\ref{fig:noA0} can be set to zero,
for particular choices of the fields which decorate the vertex. For example,
if all decorative fields are $A^1$s or $C^1$s, then the second row
of diagrams effectively vanishes. Note, though, that if there are fermionic 
decorations, then there are guaranteed to be contributions from all rows.

In the calculations performed in this thesis,
the only place we generate $\tilde{A}^1$s
is as internal fields. 
Since internal fields are always attached to wines (or effective propagators),
we can absorb the factors of $N$ appearing in equation~(\ref{eq:NoA0}) into our 
rules for attaching
wines / effective propagators to vertices.
This is illustrated in
figure~\ref{fig:GR:WinePrescription}, where the ellipsis denotes
un-drawn pushes forward and pulls back, on to the remaining fields, $\{f\}$, which also
decorate the vertex.
\begin{center}
\begin{figure}[h]
	\begin{eqnarray*}
		\ensuremath{\begin{array}{c}\input{pstex/Struc-W-F-FGR.pstex_t} \end{array}}	& =		& \ensuremath{\begin{array}{c}\input{pstex/Struc-W-F-FGR-M.pstex_t} \end{array}} - \ensuremath{\begin{array}{c}\input{pstex/Struc-W-F-FGR-M-B.pstex_t} \end{array}} + \cdots
	\\
							&\equiv	& \left[\ensuremath{\begin{array}{c}\input{pstex/Struc-Thick-W-F-FGR-M.pstex_t} \end{array}} - \frac{1}{N} \ensuremath{\begin{array}{c}\input{pstex/Struc-Thick-W-F-FGR-M-Corr.pstex_t} \end{array}}\right] 
									-\left[ \ensuremath{\begin{array}{c}\input{pstex/Struc-Thick-W-F-FGR-M-B.pstex_t} \end{array}} + \frac{1}{N} \ensuremath{\begin{array}{c}\input{pstex/Struc-Thick-W-F-FGR-M-Corr-B.pstex_t} \end{array}} \right] 
	\\[1ex]
							&	+	& \cdots
	\end{eqnarray*}
\caption{Prescription adopted for internal fermionic fields decorating a vertex
struck by
a gauge remainder.
}
\label{fig:GR:WinePrescription}
\end{figure}
\end{center}

There are a number of things to note. First, the gauge remainder strikes the
vertex and not the base of the wine; this is ambiguous from the way
in which we have drawn the diagrams though this ambiguity is, in fact, 
deliberate, as we will discuss shortly. Thus, whilst the vertex now possesses
an $A^{1,2}$ field, the wine is still labelled by $\bar{B}$. Secondly---and this
is the whole point of our prescription---the field struck by the gauge remainder becomes
an $A^{1,2}$, and not an $\tilde{A}^{1,2}$; the missing contributions to the vertex
have been effectively absorbed into the wine Feynman rule. Thirdly, we implicitly
sum over all possible ways in which the un-drawn fields, $\{f\}$, decorate the 
vertex in all diagrams (including those with the old-style notation). This ensures
that all diagrams in the no-$\SF^0$ relationship~\eq{eq:NoA0} are included.
Lastly, we note once more that the Feynman rules are such that certain
terms can be set to zero when we look at particular realisations of $\{f\}$.

\begin{center}
\begin{figure}[h]
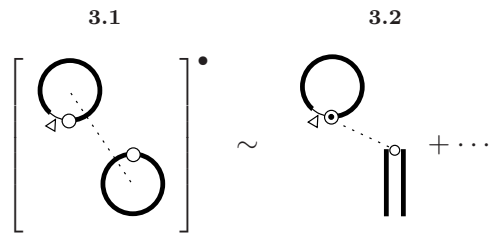

	\[
	\begin{array}{ccc}
		\LE{GR-Replace-AwC-Ex-pre}					& 		& \LE{GR-Replace-AwC-Ex} 
	\\[1ex]
		\dec{\ensuremath{\begin{array}{c}\input{pstex/GR-Replace-AwC-Ex-pre.pstex_t} \end{array}}}{\bullet}	&\sim 	& \ensuremath{\begin{array}{c}\input{pstex/GR-Replace-AwC-Ex.pstex_t} \end{array}} + \cdots
	\end{array}
	\]
\caption{A gauge remainder strikes a vertex in which a component of $\SF$ has been 
replaced by a component of $\SH$.}
\label{fig:GR:Replace-A-w-C}
\end{figure}
\end{center}

We conclude our discussion of the effect of gauge remainders
on vertices by considering diagrams generated
by the new terms in the flow equation in which a component of $\SF$ decorating
a vertex is replaced by a component of $\SH$.\footnote{
Though not necessary for the following analysis, we recall from section~\ref{sec:NFE:BrokenPhase}
that, in the $A^1$, $A^2$
basis, components of $\SF$ must be replaced by dynamical components of $\SH$.
}
The situation is illustrated in figure~\ref{fig:GR:Replace-A-w-C} where, for reasons that
will become apparent, we schematically indicate the type of vertex whose flow generates the terms we
are interested in.

The effect of the gauge remainder requires a little thought.
In diagram~\ref{GR-Replace-AwC-Ex-pre}, the gauge remainder can
clearly strike the $\SH$. However, in diagram~\ref{GR-Replace-AwC-Ex}
the $\SH$ is not part of the vertex coefficient function and so
is blind to the effects of the gauge remainder.

Allowing the $\SH$ of diagram~\ref{GR-Replace-AwC-Ex}
to strike the $\SF$, which labels the vertex coefficient function,
the vertex coefficient 
function changes. Now we have a strange situation:
looking just at the coefficient functions of the diagrams (\ie\ ignoring
the implied supertrace structure), diagrams~\ref{GR-Replace-AwC-Ex-pre}
and~\ref{GR-Replace-AwC-Ex}  are consistent, after the action of the gauge
remainder. However, the implied supertrace structures of the two diagrams
seems to differ, because the $\SH$ of diagram~\ref{GR-Replace-AwC-Ex} is
blind to the gauge remainder.

The solution is simple: we allow the effect of the gauge remainder
striking the $\SF$ in diagram~\ref{GR-Replace-AwC-Ex} to induce
a similar change in the $\SH$. This amounts to a diagrammatic prescription
which ensures that all our diagrams continue to represent both
numerical coefficients and implied supertrace structure. The key point
is that we are free to do this with the $\SH$ since, not being part
of a vertex (or, strictly, part of a wine) it does not contribute to the
numerical value of the diagram but serves only to keep track of the
supertraces which have been implicitly stripped off from the vertex whose
flow we are computing. 

Finally, we should take account of attachment corrections, if we
are to work in the $A^1$, $A^2$ basis. Attachment corrections
effectively detach the embedded component of $\SH$ from the 
vertex, causing it to become an isolated $\str \SH$. 
In diagram~\ref{GR-Replace-AwC-Ex-pre}, this field cannot now be
struck by the gauge remainder. In diagram~\ref{GR-Replace-AwC-Ex},
when the gauge remainder acts, it no longer induces a change in the embedded 
component of $\SH$.
In fact, such contributions must cancel against other terms formed by
the action of the gauge remainder; this is discussed further in
section~\ref{sec:Diags:GI}.

\subsection{Wines} \label{sec:GR-Wines}

Thus far, we have been considering the effects of gauge remainders on vertices. It is straightforward to generalise this
analysis to the effect on wines; the generic case is shown in  Figure~\ref{fig:GR:kernels}.
\begin{center}
\begin{figure}[h]
	\[
	\ensuremath{\begin{array}{c}\input{pstex/GR-kernels.pstex_t} \end{array}} = \ensuremath{\begin{array}{c}\input{pstex/GR-kernels-M.pstex_t} \end{array}}
	\]
\caption[Contraction of an arbitrary wine with one of its momenta.]{Contraction of an arbitrary wine with one of its momenta. The 
sense in which we will take pushes forward and pulls back
is as in figure~\ref{fig:NFE:NewKernel-Ex}.}
\label{fig:GR:kernels}
\end{figure}
\end{center}

If the field whose momentum is contracted into the wine is fermionic, then pushes 
forward and pulls back will involve flavour changes.
Let us begin by supposing that one of the fields hit decorates the wine (as opposed
to being a derivative sitting at the end); in this case, the
flavour changes are given by table~\ref{tab:GR:Flavour}. Note that instances of $\SH$
embedded at the ends of the wine behave like normal wine decorations, as 
far as gauge remainders are concerned. This follows from the gauge invariance 
of the flow equation and is natural if
we view these embedded fields as behaving just like multi-supertrace
components of the wine. Of course, if the gauge remainder does strike a 
component of $\SH$ which is really an embedded $\SH$ then it must be that
this component of $\SH$ is forced to be on the same portion of
supertrace as the rest of the wine. This is just a manifestation of
the statement that the action of gauge remainders necessitates
partial specification of the supertrace structure (\cf\ our 
treatment of vertices).

When we generate internal $\tilde{A}^{1,2}$s, we would like
to attach to them according to the prescription of figure~\ref{fig:GR:WinePrescription}
\ie\ we wish to extend this prescription such that the structure to which the
wine attaches is generic, as opposed to being just a vertex. We can and will do
this, though note that whether or not the $1/N$ corrections actually survive
depends on whether or not we endow our wines with completely general supertrace structure.
If we do allow completely general wine decorations then the $1/N$ corrections 
arise---as they did before---by combining terms with a lone $\str \tilde{A}^1$ (or $\str \tilde{A}^2$)
with those without. If, however, we take the only multi-supertrace terms of
the wine to be those involving embedded $\SH$s, then our no-$\SF^0$ relations
for the wine will cause the $1/N$ corrections to vanish. Since these corrections
are to be hidden in our Feynman rules, it does not matter which scheme we
employ.

The next task is to consider what happens when we push forward (pull back) on to
the end of a wine. This is straightforward: we know that $\delta / \delta \SF^0$
never attaches, irrespective of whether it has been generated by a fermionic
gauge transformation. Using the prescription that we discard $\delta / \delta \SF^0$,
the flavour
changes involved when a functional derivative sitting at the end of the wine
is hit 
are summarised in table~\ref{tab:GR:WineFlavourChanges}.

\begin{center}
\begin{table}[h]
\[
\begin{array}{l|l|c|c}
X & Y & \PF{Y} & \PB{Y}\\ \hline\hline
\bar{F} & \left(\fder{}{A^1},\fder{}{C^1}\right) & \fder{}{F} & \mbox{---} \\
  & \left(\fder{}{A^2},\fder{}{C^2}\right) & \mbox{---} & \fder{}{F} \\
  & -\fder{}{\bar{F}} & \left(\fder{}{\tilde{A}^2},-\fder{}{C^2}\right) & \left(\fder{}{\tilde{A}^1},-\fder{}{C^1}\right) \\ \hline
F & \left(\fder{}{A^2},\fder{}{C^2}\right) & -\left(\fder{}{\bar{B}},\fder{}{\bar{D}} \right) & \mbox{---}\\ 
  & \left(\fder{}{A^1},\fder{}{C^1}\right) & \mbox{---} &  -\left(\fder{}{\bar{B}},\fder{}{\bar{D}} \right) \\
  & \fder{}{F} & \left(\fder{}{\tilde{A}^1},\fder{}{C^1}\right) &  \left(\fder{}{\tilde{A}^2},\fder{}{C^2}\right)
\end{array}
\]
\caption[The flavour changing effect of pushing forward and / or pulling back the momentum of a fermionic field
on to the end of a wine.]{
The flavour changing effect of pushing forward and / or pulling back the momentum of the fermionic field $X$ on to the derivative, $Y$, at the end of a wine.}
\label{tab:GR:WineFlavourChanges}
\end{table}
\end{center}

\subsection{Gauge Invariance}	\label{sec:Diags:GI}

We mentioned under figure~\ref{fig:GR:WinePrescription} that
the diagrams on the \rhs\ of the figure are ambiguous: if
we ignore the \lhs, it is
not clear whether we have pushed forward around the bottom structure
or pulled back down the wine. In this section we will argue that
the two must be equivalent, by gauge invariance, and then
demonstrate this to be the case.

Consider the flow of some vertex decorated by the fields $f_1 \cdots f_n$.
Using the form of the flow equation given in figure~\ref{fig:NFE:New-Diags-1},
we explicitly decorate with $f_1$, but leave the other fields as
unrealised decorations (see part~III for much more detail concerning
this procedure). This yields the diagrams of figure~\ref{fig:GR:GaugeInvariance}.
\begin{center}
\begin{figure}[h]
	\[
	-\flow 
	\dec{
		\ensuremath{\begin{array}{c}\input{pstex/Vertex-S-f_1.pstex_t} \end{array}}
	}{f_2 \cdots f_n}
	=
	\frac{1}{2}
	\dec{
		\ensuremath{\begin{array}{c}\input{pstex/Dumbbell-S-f_1-W-Sigma_g.pstex_t} \end{array}} + \ensuremath{\begin{array}{c}\input{pstex/Dumbbell-S-W-f_1-Sigma_g.pstex_t} \end{array}} \hspace{0.75em} +  \ensuremath{\begin{array}{c}\input{pstex/Dumbbell-S-W-Sigma_g-f_1.pstex_t} \end{array}}  -  \ensuremath{\begin{array}{c}\input{pstex/Vertex-Sigma_g-f1-W.pstex_t} \end{array}} - \ensuremath{\begin{array}{c}\input{pstex/Vertex-Sigma_g-W-f1.pstex_t} \end{array}}
	}{f_2 \cdots f_n}
	\]
\caption{Flow of a vertex decorated by the fields $f_1 \cdots f_n$.}
\label{fig:GR:GaugeInvariance}
\end{figure}
\end{center}

Now consider contracting each of the diagrams of figure~\ref{fig:GR:GaugeInvariance}
with the momentum of $f_1$. On the \lhs, this generates the flow of a set of 
vertices decorated by $m-1$ fields. Amongst the diagrams generated on the \rhs\
are those for which we push forward / pull back on to fields to which the wine
attaches. For each of these diagrams, there is then a corresponding diagram
(with opposite sign) where we have pulled back / pushed forward on to the end of the wine.
Such diagrams, in which we push forward / pull back on to an internal field cannot be
generated by the \lhs; thus as a consequence of gauge invariance, it must be
that they cancel amongst themselves.

\begin{Prop}

Consider the set of diagrams generated by the flow equation,
each of which
necessarily possesses a single wine that we will take
to attach to the fields $X_1$ and $X_2$.
Suppose that we contract each of these diagrams with
the momentum of one of the (external) fields, $Y$.

Of the resultant diagrams, we collect together those for which the momentum of $Y$
is pushed forward and / or pulled back round a vertex, on to $X_1$ ($X_2$).
We add to this set of diagrams all those for which the momentum of
$Y$ is pushed forward and / or pulled back along the wine on to
the end attaching to $X_1$ ($X_2$).

We now split these sets into subsets, where the elements of each subset have
exactly the same supertrace structure. The elements of each of these
subsets cancel, amongst themselves.

\label{Prop-GI}
\end{Prop}

The proof of proposition~\ref{Prop-GI} is essentially trivial. If
we choose to work in the picture where we retain $\SF^0$, then we know
that there are no attachment corrections.\footnote{We recall from 
figure~\ref{fig:Intro:SuperSplittingCorrection}
that, in this picture, there is a correction to supersplitting.
However, the structure which receives the correction
is devoid of external fields and so cannot be struck by
a gauge remainder.} In this case, we can simply use 
tables~\ref{tab:GR:Flavour}
and~\ref{tab:GR:WineFlavourChanges} to demonstrate that proposition~\ref{Prop-GI}
is true.

Transferring to the $A^1$, $A^2$ basis complicates matters by generating
attachment corrections. However, we know that this picture is equivalent
to the one in which we retain $\SF^0$, and so the effects of
these corrections cannot spoil the truth of proposition~\ref{Prop-GI}.

It is, though, instructive to see how proposition~\ref{Prop-GI}
can be demonstrated in the $A^1$, $A^2$ basis, directly. However, rather than
giving a complete exposition, we will demonstrate the truth of
proposition~\ref{Prop-GI} in the $A^1$, $A^2$ basis for the original flow
equation only (\ie\ we neglect the effects of embedded components of $\SH$).

We begin by supposing that
the field we are pushing forward / pulling back is in the $A$-sector.
In this case, the effects of the attachment corrections are straightforward: irrespective of
whether the field decorates the structure to which the wine attaches, or the wine
itself, the flavour of the fields to which the wine attaches are the same, 
and so the attachment corrections are the same in both cases. Hence,
proposition~\ref{Prop-GI} is clearly true, in this case.
The subtlety
comes when the field we are pushing forward / pulling back is fermionic.

\begin{center}
\begin{figure}[h]
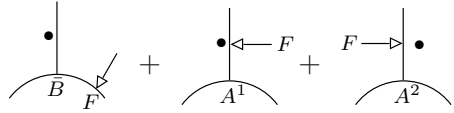

	\[
		\ensuremath{\begin{array}{c}\input{pstex/Struc-W-F-FGR.pstex_t} \end{array}} + \ensuremath{\begin{array}{c}\input{pstex/Struc-W-1F-GR.pstex_t} \end{array}} +  \ensuremath{\begin{array}{c}\input{pstex/Struc-W-2F-GR.pstex_t} \end{array}}
	\]
\caption{The push forward and / or pull back round the vertex cancels the pull back and / or push forward down the wine.}
\label{fig:GR:GI-A}
\end{figure}
\end{center}

The first case that we will examine is shown in figure~\ref{fig:GR:GI-A}. 

The potential
problem arises because attachments to $A^{1,2}$ come with a correction,
whereas attachments to $\bar{B}$ do not.
However, the first diagram has been discussed earlier in this chapter
(see figure~\ref{fig:GR:WinePrescription}): when we push forward
and / or pull back
on to the field to which the wine attaches, this generates an effective
attachment correction. These corrections are precisely the same as those present in the second
and third diagrams (see figure~\ref{fig:NFE:Attachement-A1-B}) and so the cancellation
required for proposition~\ref{Prop-GI} to be true goes ahead as required!

\begin{center}
\begin{figure}[h]
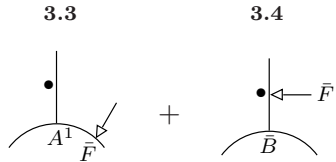

	\[
	\begin{array}{ccc}
			\LE{Diags-W-1-FGR}	&	& \LE{Diags-W-BF-GR}
	\\[1ex]
			\ensuremath{\begin{array}{c}\input{pstex/Struc-W-1-FGR.pstex_t} \end{array}}	& +	& \ensuremath{\begin{array}{c}\input{pstex/Struc-W-BF-GR.pstex_t} \end{array}}
	\end{array}
	\]
\caption{The push forward round the vertex does not quite cancel the pull back down the wine.}
\label{fig:GR:GI-B}
\end{figure}
\end{center}

The second case to deal with is shown in figure~\ref{fig:GR:GI-B} (an identical analysis can
be performed with $A^2$ instead of $A^1$s).

This is the most subtle case and so must be examined in some detail. 
The potential problem is clearly  that there is an attachment correction 
where the bottom end of the wine attaches in
diagram~\ref{Diags-W-1-FGR} but not in diagram~\ref{Diags-W-BF-GR}.

Let us start
by supposing that the wine of diagram~\ref{Diags-W-1-FGR} is undecorated and
that the wine of diagram~\ref{Diags-W-BF-GR} has no additional decorations. The 
top ends of each wine must now attach to an $A^1$. However, we know that when an
undecorated wine attaches at both ends to an $A^1$, there is effectively only
an attachment correction at one end (see figures~\ref{fig:NFE:Attachment-A1A1} and~\ref{fig:NFE:Attachment-A1s,A2s}).
Hence, in this case, the push forward round the vertex cancels the pull back down the wine.

Next, let us add an arbitrary number of bosonic decorations to the wines
of each diagram (in diagram~\ref{Diags-W-BF-GR} we are clearly not
interested in the case where any of the decorations come between the existing decoration
and the bottom end of the wine). Now, let us focus on the correction term in
at the bottom end of the wine in
diagram~\ref{Diags-W-1-FGR}. Since we plug the end of the wine with a $\sigma_+$,
we can cycle the bosonic decorations around the end of the wine (remember that
we are neglecting embedded components of $\SH$). The sum of all diagrams
thus obtained 
then vanishes by the coincident line identities, as is clear by considering
the broken phase version of figure~\ref{fig:CoincidentLine}.

\begin{center}
\begin{figure}[h]
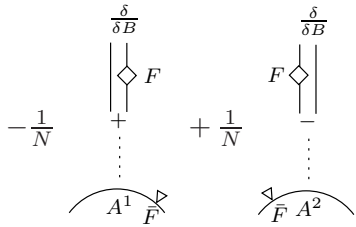

	\[
	\begin{array}{ccc}
		
	\\[1ex]
		-\frac{1}{N} \ensuremath{\begin{array}{c}\input{pstex/Corrs-FermionicDec-A.pstex_t} \end{array}} + \frac{1}{N} \ensuremath{\begin{array}{c}\input{pstex/Corrs-FermionicDec-B.pstex_t} \end{array}}
	\end{array}
	\]
\caption{Two diagrams which, after the gauge remainders strike the explicitly drawn fields,
cancel by the coincident line identities.}
\label{fig:GR:GI-Bb}
\end{figure}
\end{center}

Having dealt with bosonic decorations, we now consider the case where the wines
of diagrams~\ref{Diags-W-1-FGR} and~\ref{Diags-W-BF-GR} are decorated
by a single additional (anti)fermion.
In this case, it is not obvious that the
attachment correction in diagram~\ref{Diags-W-1-FGR} vanishes by the coincident
line identities: we cannot cycle the (anti)fermionic decoration around the $\sigma_+$
plugging the wine. However, we have missed something. Consider the diagrams 
of figure~\ref{fig:GR:GI-Bb}.

The pull back on to $A^2$ in the second diagram generates the same field as the
push forward on to $A^1$ in the first diagram (see table~\ref{tab:GR:Flavour}).
However, pulls back come with a relative minus sign, compared to pushes forward.
This sign
cancels the existing relative minus between the two diagrams and now the
two diagrams cancel,
courtesy of the coincident line identity shown in figure~\ref{fig:NFE:CLI-FlavourChange}.
This argument can be repeated for any number of fermionic decorations.

We have therefore demonstrated the truth of 
proposition~\ref{Prop-GI}, working explicitly in the $A^1$, $A^2$ basis,
for the original flow equation. It is straightforward, if tedious, to
extend this proof to encompass the new terms in the flow equation, and
so we omit the details; rather we argue that proposition~\ref{Prop-GI}
must be true, due to the equivalence of the $A^1$, $A^2$ basis and
the basis in which we retain $\SF^0$ (or simply as a consequence of
supergauge invariance).

\subsection{Cancellations Between Pushes forward / Pulls back} \label{sec:GRs:bosonic-cancellations}

We saw in the previous section how, in certain circumstances, a push forward round 
a vertex could be cancelled by a pull back, also round the vertex 
(see figure~\ref{fig:GR:GI-Bb}). In this section, such 
cancellations are treated in complete generality.

Referring back to figure~\ref{fig:GR:Defining}, we now ask when it 
is possible for the pulls back of
the second row to cancel the pushes forward of the first row (this argument can
be repeated for wines). It is clear that, if the field structure of
corresponding terms is exactly the same, then they will cancel, due to the relative minus sign. 
For the purposes of this section, we wish to consider the case where any cancellations occur independently of the
spectator fields. In other words, we will not consider cancellations which involve changing the ordering, flavour or indices
of the spectator fields; this is delayed until the next section. Furthermore, whilst all the wildcard fields we are considering
include all possible field choices, we do not sum over these choices, but consider each independently.

Let us temporarily
suppose that $X$  is in the $A$-sector and focus on the case where its momentum is pushed forward on to field $Y$. If both $X$ and $Y$ are
bosonic, then the flavours of $X$ and $Y$ are independent of which field precedes the other. Moreover, for a given field arrangement,
the flavours of the other fields will not change if the order of $X$ and $Y$ is swapped. In this case, the push forward
on to $Y$ will be exactly cancelled by the corresponding pull back.

However, if either $X$ or $Y$ is fermionic, then interchanging their order will necessarily change the field content of the vertex.
This follows because a bosonic field in the 1-sector precedes an $F$ and follows an $\bar{F}$, whereas a bosonic field 
in the 2-sector follows an $F$ and precedes an $\bar{F}$.
As an example, consider $p_\mu S^{1F\bar{F},\ldots}_{\mu R S\ldots}(p,r,s,\ldots)$. To cancel the push forward on to $F_R$
would require us to change the flavour of $X$ to $A^2$: $p_\mu S^{F 2 \bar{F}\ldots}_{R \mu S\ldots}(r,p,s,\ldots)$. 
Instead, we could try and cancel the
push forward on to $F_R$ by constructing the term $p_\mu S^{F\bar{F}1\ldots}_{S R \mu\ldots}(s,r,p,\ldots)$, 
but now it is the $\bar{F}$ carrying the 
index $R$, rather than the $F$. As we will see in the next section such a term can, in general, either cancel or double the original
push forward. However, for the purposes of this section, we note that the spectator field $\bar{F}_S$ has suffered a change
and so we do not consider this further.

Similarly, if both $X$ and $Y$ are fermionic, then interchanging them will alter the field content of the vertex,  if other fields
are present on the same supertrace.
Then, we have the choice of altering the spectators or letting
 $X,Y\rightarrow \bar{X}, \bar{Y}$. The former case will be dealt with in the next section. In the latter case, we note from
table~\ref{tab:GR:Flavour} that pushing forward the momentum of
$F_R$ on to $\bar{F}_S$ yields $(A^1,C^1)_S$ whereas the pulling back the momentum of 
$\bar{F}_R$ into $F_S$ yields $-(A^1,-C^1)_S$. These contribution do
produce a cancellation over the first four indices, but the fifth index contributions add.

In conclusion, when dealing with a single vertex, a push forward can only completely 
cancel a pull back, independently of
the spectator fields, when both fields
involved are bosonic. When we generalise this analysis to 
full diagrams, rather than individual vertices, we might 
expect this constraint to be relaxed: all internal fields will be 
summed over and we have seen how, for example, pushing forward 
the momentum of an $A^1$ on to an $F$ could be can be cancelled by 
pulling back the momentum of an $A^2$. Thus, if the $A$ field is internal,
then we will be including both cases, automatically.
However, when dealing with full diagrams,
we must be aware that interchanging fields can alter the supertrace 
structure of the diagram and so we will actually find
that the conditions for cancellation between pushes forward and 
pulls back are even more stringent (see section~\ref{sec:GR:CompleteDiagrams}).

\subsection{Charge Conjugation}	\label{sec:Diagrammatics:CC}

In the previous section we looked at whether pushes forward could cancel pulls back, independently of the spectator fields.
If the properties of the spectator fields are allowed to change, then we find that every push forward is related to a pull back,
by \CC.

Referring back to equations~(\ref{eq:CC-A})--(\ref{eq:CC-F}) we can construct the following
diagrammatic recipe for \CC:
\begin{enumerate}
	\item	reverse the sense in which we read fields off from the vertices / wines,

	\item 	pick up a minus sign for each field in the $A$-sector;

	\item	let $\bar{F} \leftrightarrow -F$;
\end{enumerate}
where we must remember that fields pushed forward / pulled back on to may have changed flavour and that the wildcard fields 
should be interpreted according to table~\ref{tab:GR:Flavour}.
Rather than having to
specify the sense in which fields are to be read off, we can instead replace a given diagram with its mirror image, whilst obeying
points two and three above~\cite{TRM-Faro,GI-ERG-II,YM-1-loop}.

Now let us return to figure~\ref{fig:GR:Defining} and consider taking the mirror image of the bottom row of diagrams.
Since the location and order of the spectator fields is unspecified we see that, up to a possible sign, the first and 
second rows are actually identical! However, whether corresponding entries in the two rows add or cancel, depends
on the whether the original vertex is even or odd under \CC. In the former case, pushes forward and pulls back will add;
in the latter case they will cancel. 

It is important to note that this argument is separate from that of the previous section, as illustrated in figure~\ref{fig:GR:CC-Ex}.
The diagrams of the first and fourth rows are equal, by \CC, as are the diagrams of the second and third rows. However, 
diagrams~\ref{Ex:CC-1} and~\ref{Ex:CC-2}, whose parents are not related by \CC, cancel, as do diagrams~\ref{Ex:CC-3} and~\ref{Ex:CC-4}.
\begin{center}
\begin{figure}[h]
\[
	\ensuremath{\begin{array}{c}\input{pstex/GR-CC-Ex.pstex_t} \end{array}}
\]
\caption[An example of applying \CC\ to lone vertices struck by a gauge remainder.
]{Rows one and four are equal by \CC; likewise for rows two and three. Independently of this, there are cancellations
between diagrams of rows one and two and also diagrams of rows three and four.}
\label{fig:GR:CC-Ex}
\end{figure}
\end{center}

\subsection{Complete Diagrams} \label{sec:GR:CompleteDiagrams}

So far, we have just been concerned with isolated vertices and so now turn to full diagrams. We still wish to combine
pushes forward and pulls back using \CC\ but, to do so, we must look at the \CC\ properties of whole diagrams,
rather than the properties of individual vertices.

\begin{center}
\begin{figure}[h]
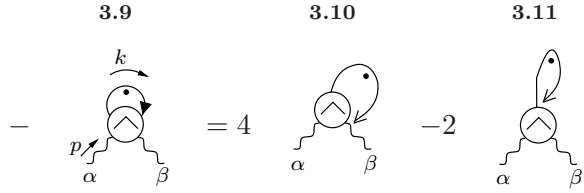

	\[
	\begin{array}{cccccc}
			& \LE{ExD:d1:1.6}	&	& \LE{ExD:d1:2.1}	&	&\LE{ExD:d1:2.2}
	\\[1ex]
		- 	& \ensuremath{\begin{array}{c}\input{pstex/GR-CC-1.6.pstex_t} \end{array}}	& =4& \ensuremath{\begin{array}{c}\input{pstex/Beta1-2.1-lab.pstex_t} \end{array}}& -2&\ensuremath{\begin{array}{c}\input{pstex/Beta1-2.2-lab.pstex_t} \end{array}}
	\end{array} 
	\]
\caption[Example of a gauge remainders in a complete diagram.
]{Example of a gauge remainders in a complete diagram. The dummy index $R$ is given by $\rho$, if restricted to the first four indices.}
\label{fig:GR:d1:1.6}
\end{figure}
\end{center}

We begin by looking at the
example illustrated in figure~\ref{fig:GR:d1:1.6}. Each diagram has two external fields, which we will choose to be $A^1$s,
carrying indices $\alpha$ and $\beta$ and momenta $p$ and $-p$. 
By Bose symmetry, the diagrams are symmetric under $p_\alpha \leftrightarrow -p_\beta$.

The first comment to make is that the diagrammatics is slightly different from the previous case.
Rather than terminating the pushed forward / pulled back field-line with a half arrow, we just utilise the fact
that the corresponding field line already ends in a $\wedge$ and use this to indicate the field hit.

Returning to the diagrams of figure~\ref{fig:GR:d1:1.6}
we see that, not only can we collect pushes forward and pulls back, but we can also exploit any symmetries of
the diagrams to collect terms. Looking at diagram~\ref{ExD:d1:1.6}, it makes no difference whether the gauge remainder hits the field carrying
$\alpha$ or the field  carrying $\beta$. Since we can push forward or pull back on to either of these fields, this accounts for the
factor of four multiplying diagram~\ref{ExD:d1:2.1}.

Diagram~\ref{ExD:d1:2.2} is interesting. Having used \CC\ to collect the push forward
and pull back, let us now suppose that all fields leaving the three-point vertex are in the $A$-sector. 
We note that the field struck by the gauge remainder has an $\SF^0$ component,
but suppose that this has been absorbed into an attachment correction. In this
picture, we cannot have an $A^1$ alone on a supertrace and so all three fields
must be on the same supertrace. However, we are still free to interchange
 $A_\alpha(p)$ and $A_\beta(-p)$ and, summing over the two possible locations
of these fields, we have:
\[
\hat{S}^{1\, 1\, 1}_{\alpha \beta \rho}(p,-p,0) + \hat{S}^{1\, 1\, 1}_{\beta \alpha \rho}(-p,p,0).
\]
These two terms cancel, as a result of \CC.

\begin{center}
\begin{figure}[h]
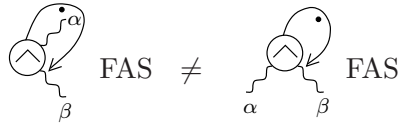

	\[
		\ensuremath{\begin{array}{c}\input{pstex/Beta1-2.1-lab-B.pstex_t} \end{array}} \mathrm{FAS} \ \ \neq \ \ \ensuremath{\begin{array}{c}\input{pstex/Beta1-2.1-lab.pstex_t} \end{array}} \mathrm{FAS} 
	\]
\caption{Two components of diagram~\ref{ExD:d1:2.1} which are not equal, due to their differing
supertrace structure.}
\label{fig:SupertraceStrucure-Ex}
\end{figure}
\end{center}

We might wonder if \CC\ causes components of diagram~\ref{ExD:d1:2.1} to cancel. However, 
attachment corrections aside, the index structure $\beta \alpha R$
corresponds to a different supertrace structure from the index structure $\alpha \beta R$. In the former case, the two $A^1$s are
on different supertraces whereas, in the latter case, they are on the same supertrace. This is illustrated
in figure~\ref{fig:SupertraceStrucure-Ex}.

If we include  attachment corrections (see figure~\ref{fig:NFE:Attachement-A1-B}), then the wine
of diagram~\ref{ExD:d1:2.1}
can attach to the vertex via a false kernel. In this case, there is only one supertrace, and so
all fields are necessarily on it. Such components of diagram~\ref{ExD:d1:2.1} do cancel
amongst themselves. However, these cancellations are generally hidden by our notation
and are of no practical importance anyway, until we come to extracting numerical contributions to
$\beta$-function coefficients.

Returning to diagram~\ref{ExD:d1:2.2}, for the diagram to survive, 
the field carrying momentum $k$ must be in the $F$-sector. In this case, the gauge remainder
can produce a $C$-sector field. Under interchange of $\alpha$ and $\beta$, such a vertex is even and so survives.

 This
serves to illustrate a general feature of these diagrammatics, alluded to at the end of 
section~\ref{sec:GRs:bosonic-cancellations}. Suppose that we are pushing forward the momentum of a field $X$ on to the 
field $Y$. If we can rearrange the diagram such that, leaving all other fields alone, we can place $X$ on the other side of
$Y$, then the resulting pull back on to $Y$ will cancel the push forward, so long as no flavours  or indices have changed in the
rearrangement and the supertrace structure is still the same. 

Here is how this applies to our examples.
In the case of diagram~\ref{ExD:d1:2.1}, to convert a pull back on to $A_\beta^1$ into a push forward, we must change the location
of $A_\alpha^1$, to maintain the same supertrace structure (up to attachment corrections). 
The resulting term can then just be collected with the pull back, by
\CC. Hence the push forward on to $A_\beta^1$ can never be completely cancelled by a pull back.

In the case of diagram~\ref{ExD:d1:2.2} we can convert a push forward into a pull back without changing the locations
of the spectator fields and without having to change the supertrace structure.
If the fields carrying momentum $k$ are in the $A$-sector, then interchanging them does not result in any flavour
changes, and so the push forward cancels the pull back. However, if the fields carrying momentum $k$ are fermionic, then interchanging
them requires us to replace $\bar{F} \leftrightarrow F$. This constitutes a change of flavour and we find that the push forward does not
completely cancel the pullback, since we are left with a contribution arising from the $C$-sector.

We note that, just as we can use \CC\ to
redraw vertices struck by gauge remainders, so too can we use \CC\ to
redraw entire diagrams. Given some diagram, the diagrammatic effect of \CC\
is to replace a diagram by its mirror image, letting $\bar{F} \leftrightarrow F$ (\sic)
and picking up a minus sign for each gauge remainder that has been 
performed.\footnote{So far, we have only encountered diagrams in which a single
gauge remainder has acted, but we will come across more general cases later.}
Picking up a sign in this manner automatically keeps track
of the signs associated with the rules of section~\ref{sec:Diagrammatics:CC}.

\begin{center}
\begin{figure}[h]
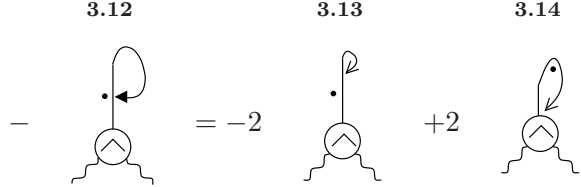

	\[
	\begin{array}{cccccc}
			& \LE{ExD:d1:1.10}	&	& \LE{ExD:d1:2.3(S)}&	&\LE{ExD:d1:2.4(S)}
	\\[1ex]
		-	& \ensuremath{\begin{array}{c}\input{pstex/GR-CC-1.10.pstex_t} \end{array}} 	&=-2& \ensuremath{\begin{array}{c}\input{pstex/Beta1-2.3S.pstex_t} \end{array}} &+2 & \ensuremath{\begin{array}{c}\input{pstex/Beta1-2.2.pstex_t} \end{array}}
	\end{array}
	\]
\caption{Example of a gauge remainder on a wine, in a full diagram.}
\label{fig:GR:ExD:1.10}
\end{figure}
\end{center}

Let us now examine a second example, as shown in figure~\ref{fig:GR:ExD:1.10}.

It is crucially important to recognise that, whilst diagram~\ref{ExD:d1:1.10}
may superficially look like a diagram in which the wine bites its own tail, it is
very different. The difference arises due to the gauge remainder, and means that
such diagrams cannot be discarded on account of~(\ref{eq:WBT(SF)}). (We can view
the gauge remainders as being some non-trivial kernel $K(x,y)$ sitting between
the functional derivatives in~(\ref{eq:WBT(SF)})---which we take to carry position
argument $x$---and $\{W\}$, which we take to carry position argument $y$. Only if
the kernel reduces to $\delta(x-y)$, which it does not, can the constraints~(\ref{eq:WBT(SF)})
contribute with non-zero measure.)

Comparing with figure~\ref{fig:GR:d1:1.6}, 
we see that diagram~\ref{ExD:d1:2.4(S)} has exactly the same 
structure as diagram~\ref{ExD:d1:2.2}. Although the former diagram
involves a pull back along the wine and the latter case involves a push forward around a vertex,
we know from section~\ref{sec:Diags:GI} that these two diagrams are identical. Taking into
account the relative sign, it is clear that they cancel.

This cancellation leads us to a cancellation mechanism, which 
simply recognises that the cancellation we have seen between
diagrams~\ref{ExD:d1:2.2} and~\ref{ExD:d1:2.4(S)} is just a specific example of a cancellation
that is always guaranteed to occur between two diagrams of a certain general structure.

\begin{CM}
Consider the two sub-diagrams in figure~\ref{fig:CM:VertexJoinWine}. There are no restrictions
on either the arbitrary set of decorative fields, $\{f\}$, or
the form of the complete diagrams of which these two sub-diagrams are part. 
The vertex argument, $v$, is arbitrary.
\begin{center}
\begin{figure}[h]
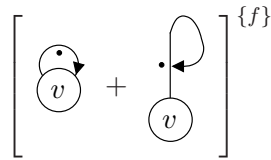

	\[
	\dec{
		\ensuremath{\begin{array}{c}\input{pstex/v-W-GR.pstex_t} \end{array}} + \ensuremath{\begin{array}{c}\input{pstex/v-WGR.pstex_t} \end{array}}
	}{\{f\}}
	\]
\caption{Two sub-diagrams between which, after the action of the gauge remainders,
certain contributions are guaranteed to cancel.}
\label{fig:CM:VertexJoinWine}
\end{figure}
\end{center}

Allowing the gauge remainders to act,
the push forward (pull back) around the vertex on to the field to which the wine attaches 
exactly cancels
the pull back (push forward) along the wine on to the end to which the vertex attaches.
\label{CM:VertexJoinWine}
\end{CM}

Returning now to figure~\ref{fig:GR:ExD:1.10}, it is worth making some comments
about the structure at the top of
diagram~\ref{ExD:d1:2.3(S)}. First, the line segment which joins the top
of the wine to the $\GRkp$---thereby forming a `hook'---performs no role other 
than to make this join. In other
words, it is neither a section of wine nor an effective propagator. We could imagine
deforming this line segment so that the hook becomes arbitrarily large. Despite
appearances, we must always remember that this line segment simply performs the role
of a Kronecker delta. When part of a complete diagram, this line
segment can always be distinguished from an effective propagator,
to which it can be made to look identical, by the context. This
follows because hooks in which the line segment is a Kronecker $\delta$
only ever attach to effective propagators or wines, whereas hook-like structures
made out of an effective propagator only ever attach to vertices
(this is particularly clear from the perspective of part~III, but see 
diagram~\ref{ExD:d1:2.4(S)} for an example).
When viewed in isolation, we will always take the hook structure to
comprise just a line segment and so will draw the hook as tightly as
possible.

Secondly, this hook structure---around which we will assume flows momentum 
$l$---is purely algebraic.
To have any chance of surviving, the hook must be
 in either the $D$ or $\bar{D}$-sector. This follows from referring to the algebraic
form of the gauge remainders (appendix~\ref{app:GRs}) and noting that these are the only sectors
for which $l'$ is not an odd function of momentum. This is entirely consistent
with the structure of the rest of the diagram: to prevent the
diagram from vanishing when we sum over independent permutations of fields
on the vertex, the wine must be in the $C$-sector. Since gauge remainders have no
support in the $C$-sector, and $AC$ wines do not exist, this means that the field
decorating the wine in the parent diagram~\ref{ExD:d1:1.10} and, by the same reasoning,
the hook must be fermionic.

When determining the algebraic form of the hook one must be very careful, due
to the prevalence of awkward signs. The simplest way to get the right answer is to
suppose that the hook is in either the full $F$ or $\bar{F}$ sector and project out later.
With this way of looking at things, the gauge remainder forming the hook
is given by $g_l$: since we have been using the full fermionic sector, there
were no signs picked up when $l'$ was originally generated by application of
the effective propagator relation (see section~\ref{sec:NFE:Zero-PointKernels+EffProps}). There may, however, be signs picked
up at the end of the wine, which can be determined via table~\ref{tab:GR:WineFlavourChanges}.

Appearance of these signs depends on
whether the hook is in the $F$ or $\bar{F}$ sector which, in turn, depends on whether the field on the vertex to
which the wine attaches
is in the $C^1$ or $C^2$ sector, respectively.\footnote{Remember that the vertex could be,
for example, $\hat{S}_{0 \mu \nu}^{\ 1 \, 1, C^2}(p,-p;0)$.}
We choose to use the prescription that whenever these signs appear, we
will absorb them into the definition of the hook. This then allows us to
take the wine to which the hook attaches to be just $\dot{\Delta}^{C^1 C^1}(0)$
or $\dot{\Delta}^{C^2 C^2}(0)$, without the need to worry about signs.

The final ingredient we need to obtain the algebraic form for the hook is to
realise that the inside of the hook constitutes an empty loop and so
gives a group theory factor of $\pm N$, depending on flavour. Thus we have:
\begin{eqnarray}
	\ensuremath{\begin{array}{c}\begin{picture}(0,0)%
\includegraphics{pstex/hook-R-1.pstex}%
\end{picture}%
\setlength{\unitlength}{3947sp}%
\begingroup\makeatletter\ifx\SetFigFont\undefined%
\gdef\SetFigFont#1#2#3#4#5{%
  \reset@font\fontsize{#1}{#2pt}%
  \fontfamily{#3}\fontseries{#4}\fontshape{#5}%
  \selectfont}%
\fi\endgroup%
\begin{picture}(153,283)(159,346)
\put(159,346){\makebox(0,0)[lb]{\smash{\SetFigFont{8}{9.6}{\rmdefault}{\mddefault}{\updefault}{\color[rgb]{0,0,0}1}%
}}}
\end{picture}
 \end{array}} & = & (-N) \int_l g_l \label{eq:hook-R-1} \\
	\ensuremath{\begin{array}{c}\begin{picture}(0,0)%
\includegraphics{pstex/hook-R-2.pstex}%
\end{picture}%
\setlength{\unitlength}{3947sp}%
\begingroup\makeatletter\ifx\SetFigFont\undefined%
\gdef\SetFigFont#1#2#3#4#5{%
  \reset@font\fontsize{#1}{#2pt}%
  \fontfamily{#3}\fontseries{#4}\fontshape{#5}%
  \selectfont}%
\fi\endgroup%
\begin{picture}(153,283)(159,346)
\put(159,346){\makebox(0,0)[lb]{\smash{\SetFigFont{8}{9.6}{\rmdefault}{\mddefault}{\updefault}{\color[rgb]{0,0,0}2}%
}}}
\end{picture}
 \end{array}} & = & - (N) \int_l g_l. \label{eq:hook-R-2}
\end{eqnarray}
The numbers at the base of the hook in the above equations indicate the sector
of the wine to which the hooks attach. The overall minus sign in the second equation
can be traced back to table~\ref{tab:GR:WineFlavourChanges}, whereas the sign 
in the first equation is due to the group theory factor.

We now see a further advantage to having absorbed signs into the definition of
the hook: we can trivially extend the diagrammatic effect of \CC\ to cover
its action on a disconnected hook.
Quite simply, we have that
\begin{equation}
	\ensuremath{\begin{array}{c}\begin{picture}(0,0)%
\includegraphics{pstex/hook-R.pstex}%
\end{picture}%
\setlength{\unitlength}{3947sp}%
\begingroup\makeatletter\ifx\SetFigFont\undefined%
\gdef\SetFigFont#1#2#3#4#5{%
  \reset@font\fontsize{#1}{#2pt}%
  \fontfamily{#3}\fontseries{#4}\fontshape{#5}%
  \selectfont}%
\fi\endgroup%
\begin{picture}(130,178)(182,451)
\end{picture}
 \end{array}} = - \ensuremath{\begin{array}{c}\begin{picture}(0,0)%
\includegraphics{pstex/hook-L.pstex}%
\end{picture}%
\setlength{\unitlength}{3947sp}%
\begingroup\makeatletter\ifx\SetFigFont\undefined%
\gdef\SetFigFont#1#2#3#4#5{%
  \reset@font\fontsize{#1}{#2pt}%
  \fontfamily{#3}\fontseries{#4}\fontshape{#5}%
  \selectfont}%
\fi\endgroup%
\begin{picture}(130,178)(288,451)
\end{picture}
 \end{array}},
\label{eq:hookCC}
\end{equation}
consistent with our previous definition that we take the mirror image, picking
up a sign for every performed gauge remainder. Of course, had we not absorbed signs
coming from the end of the wine to which the hook attaches, then it would only have
made sense to apply our previous \CC\ recipe
to an entire diagram possessing a hook, rather than factorisable components.

We conclude this section by discussing a particular scenario---which will
crop up repeatedly in our computation of $\beta$-function coefficients---in
which it is possible to neglect attachment corrections.

\begin{center}
\begin{figure}[h]
	\[
	\dec{
		\ensuremath{\begin{array}{c}\input{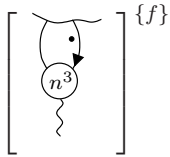} \end{array}}
	}{\{f\}}
	\]
\caption{A diagram for which corrections to supersplitting / supersowing
are restricted.}
\label{fig:Diags:GR-SuperCorrs}
\end{figure}
\end{center}

Consider some complete diagram possessing
a three-point vertex which is decorated by an external $A^1$-sector
field and is struck by a gauge remainder. 
The type of
diagram we are
considering is represented in figure~\ref{fig:Diags:GR-SuperCorrs},
where the fields $\{f\}$ can attach anywhere except the bottom vertex, which must
be three-point. If any of these fields are internal fields, then we take
pairs of them to be connected by an effective propagator.

We now argue that we can 
forget about any attachment corrections.
Let us start by supposing that the gauge remainder
is in the $F$-sector. If the gauge remainder strikes
the internal field, then it can generate an effective
attachment correction. However this correction isolates the newly
formed two-point vertex from the rest of the diagram,
leaving us with $\str A^1 = 0$.

Next, suppose that the gauge remainder is in the $A$-sector.
Since two
of the three fields entering the vertex are now in the $A$-sector,
the third must also be bosonic. Moreover, 
the final field must be in the $A$-sector also, else the action
of the gauge remainder will produce an $AC$ vertex, which does not
exist. Now, any attachment corrections 
would mean that all fields on the vertex are guaranteed
to be on the same portion of supertrace, \wrt\ the diagram
as a whole, irrespective of location. Summing over the independent
locations of the fields causes the diagram to vanish, by \CC.

Henceforth, whenever we deal with a three-point vertex
decorated by an external field and struck by a gauge remainder,
we will automatically discard all attachment corrections.

\subsection{Socket Notation} \label{sec:socket}

\begin{center}
\begin{figure}[h]
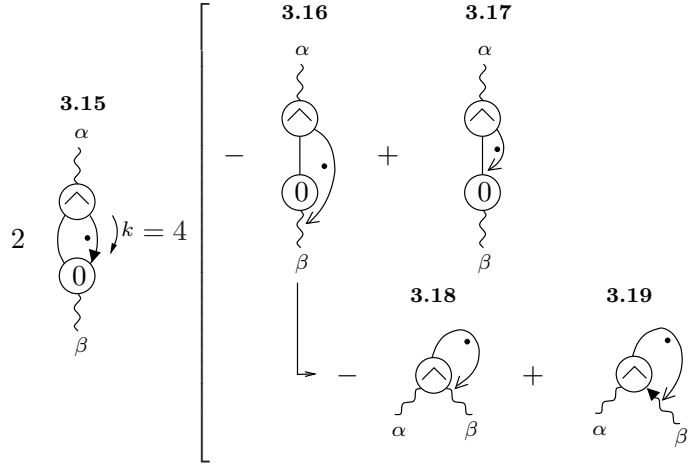

	\[
	\ensuremath{\begin{array}{c}\input{pstex/GR-CC-1.20.pstex_t} \end{array}}	
	= 4
	\left[
		\ensuremath{\begin{array}{c}\input{pstex/GR-CC-1.20M.pstex_t} \end{array}}
	\right]
	\]
\caption{Gauge remainder which produces a diagram in which the effective propagator relation can be applied.
}
\label{fig:GR:nested}
\end{figure}
\end{center}

If a gauge remainder hits a three-point vertex, then it will produce a two-point vertex. If one (or both) of the fields
leaving the resulting two-point vertex are internal fields attached to effective propagators, then we will be able to
use the effective propagator relation once more. An example is shown in figure~\ref{fig:GR:nested}. 

Referring also to 
figure~\ref{fig:GR:d1:1.6}, 
diagram~\ref{ExD:d1:2.10} straightforwardly cancels diagram~\ref{ExD:d1:2.1}.
Note that, this cancellation would have failed if the effective
propagator relation were supplemented by $1/N$ corrections 
(see section~\ref{sec:NFE:DiagrammaticIdentities}).

We are now interested in generalising this cancellation. In essence, what
we are trying to demonstrate is that it does not matter how a diagram
in which a gauge remainder has acted is formed: we should not be able to tell
whether it has been formed directly or only after the application of the effective propagator
relation.

This leads us to the socket notation: rather than specifying which field a gauge
remainder has hit, instead we state that it has hit some socket which can be filled
by any available field.\footnote{One may very well ask why we do not simply
take gauge remainders to strike an arbitrary field, rather than an empty
socket, which can be filled by an arbitrary field. We will see
the value of the socket notation in part~III.} This is illustrated in figure~\ref{fig:socket}. Note that
whilst we allow sockets to represent fields which decorate wines, we do not
take them to represent the derivatives at the end of a wine. 
\begin{center}
\begin{figure}[h]
	\[
	\begin{array}{c}
	\vspace{0.2in}
		\dec{
			\LED[0.1]{}{v-GR}{v-GR}
		}{\{f\}} 
		=2
		\dec{
			 \LED[0.1]{}{v-socket}{v-socket}
		}{\{f\}}
	\\
		\dec{
			\LED[0.1]{}{W-GR}{W-GR}
		}{\{f\}} 
		=2
		\dec{
			 \LED[0.1]{}{W-socket}{W-socket}
			+\LED[0.1]{}{W-BE}{W-BE}
			+\LED[0.1]{}{W-BB}{W-BB}
		}{\{f\}}
	\end{array}
	\]
\caption{Basic idea of socket notation.}
\label{fig:socket}
\end{figure}
\end{center}	

Guided by the cancellation of diagrams~\ref{ExD:d1:2.10} and~\ref{ExD:d1:2.1},
we are now led to consider the following diagrams: taking the parents in
figure~\ref{fig:socket}, we detach the gauge remainder from the structure it
strikes and attach it instead to a three-point, tree level vertex. This vertex
is now attached to the structure previously hit by the gauge remainder. Allowing the
gauge remainder to act will produce the diagrams of figure~\ref{fig:socket-B}.
\begin{center}
\begin{figure}[h]
	\[
	2
	\dec{
		\ensuremath{\begin{array}{c}\input{pstex/v-TLTP-socket.pstex_t} \end{array}}
	}{\{f\}}, \ \ 
	2
	\dec{
		\ensuremath{\begin{array}{c}\input{pstex/W-TLTP-socket.pstex_t} \end{array}}
	}{\{f\}}
	\]
\caption{Partner diagrams to  diagrams~\ref{v-socket} and~\ref{W-socket}.}
\label{fig:socket-B}
\end{figure}
\end{center}	

Using table~\ref{tab:GR:Flavour} and recalling that $F=(B,D)$ and $\bar{F} = (\bar{B},-\bar{D})$,
it is straightforward to show that we can utilise the effective propagator relation
to demonstrate that, up to nested gauge remainders, the diagrams of figure~\ref{fig:socket-B}
do, indeed, reduce to diagrams~\ref{v-socket} and~\ref{W-socket}, irrespective of what we
use to fill the sockets.

\subsection{Nested Contributions} \label{sec:GRs-Nested}

The basic methodology for nested gauge remainders is  similar to the methodology just presented, but
we must take account of the fact that the supertrace structure is now partially specified. In particular,
this will generally mean that we cannot use \CC\ to collect nested pushes forward and pulls back (the exception 
being if a gauge remainder produced in one factorisable sub-diagram hits a separate factorisable sub-diagram)
and so must count them separately. Indeed, nested pushes forward and pulls back can have different supertrace 
structures as illustrated by considering the result of processing diagram~\ref{ExD:d1:2.11}, as shown in figure~\ref{fig:GR:nested-M}.
\begin{center}
\begin{figure}[h]
	\[
	4
	\left[
		\ensuremath{\begin{array}{c}\input{pstex/GR-CC-1.20M-B.pstex_t} \end{array}}
	\right]
	\]
\caption{Result of processing diagram~\ref{ExD:d1:2.11}.}
\label{fig:GR:nested-M}
\end{figure}
\end{center}

We begin our analysis of 
the diagrams of figure~\ref{fig:GR:nested-M}
by noting that there are no attachment corrections. The first gauge remainder
struck a three-point (tree level) vertex decorated by an
external $A$ to generate the nested gauge remainder. In turn, the nested
gauge remainder has struck a three-point (tree level) vertex decorated
by an external $A$. From our
discussion at the end of section~\ref{sec:GR:CompleteDiagrams} we thus 
deduce that all attachment corrections in the diagrams of figure~\ref{fig:GR:nested-M}
can be neglected.

Diagrams~\ref{ExD:d1:2.17a} and~\ref{ExD:d1:2.18a} have supertrace structure $\pm N\str{A^1_\alpha A^1_\beta}$, with
the plus or minus depending on the sector of the wildcard fields. On the other hand, diagrams~\ref{ExD:d1:2.17b} and~\ref{ExD:d1:2.18b}
have supertrace structure $\str{A^1_\alpha}\str{A^1_\beta} =0$.
Note in the latter case that the particular supertrace structure puts constraints
on the field content of the diagram. Specifically, the wine in 
diagrams~\ref{ExD:d1:2.17b} and~\ref{ExD:d1:2.18b} cannot be fermionic. Let us suppose
that it is. Then, the end which attaches to the vertex must be an $F$ and so
the end which attaches to the $\wedge$ must be an $\bar{F}$. However, referring
to table~\ref{tab:GR:Flavour} we see that an $\bar{F}$ cannot pull back on to bosonic
fields in the 1-sector. There is an inconsistency in such a diagram and so our
original supposition that it exists must be wrong.

\begin{center}
\begin{figure}[h]
	\[
	\ensuremath{\begin{array}{c}\input{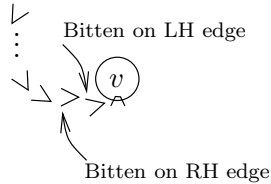} \end{array}}
	\]
\caption{An arbitrarily nested gauge remainder bites a socket on a vertex.}
\label{fig:GRstring}
\end{figure}
\end{center}

The diagrams of figure~\ref{fig:GR:nested-M} are one-loop diagrams and it is clear that
there is no way in which we can generate further gauge remainders. For
calculations of arbitrary loop order, though, we can imagine generating arbitrarily
nested diagrams. It would then be useful for us to know if there is any simple way of
keeping track of which gauge remainders have pushed forward and which have pulled back.
Of course, with sufficient thought, it is always possible to stare at a complicated
diagram and deduce the pattern of pushes forward and pulls back. However, there is
an easier way. Note in diagrams~\ref{ExD:d1:2.17a}--\ref{ExD:d1:2.18b}
that the nested gauge remainder is bitten on its right-hand edge by the original
gauge remainder. Had we pushed forward, rather than pulled back with the initial
gauge remainder, then this bite would have been to the left. With this in mind, consider a string
of gauge remainders which bite a socket decorating either a wine or a vertex; the latter
case is shown in figure~\ref{fig:GRstring}.

\begin{center}
\begin{figure}[h]
	\[
	\ensuremath{\begin{array}{c}\input{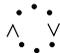} \end{array}}
	\]
\caption{An arbitrarily nested version of the hook.}
\label{fig:nestedhook}
\end{figure}
\end{center}

The sense in which the socket on the vertex is bitten is obvious. To determine
in which sense the nested gauge remainders are bitten, we simply equate bites
on the left with pushes forward and bites on the right with pulls back. There is only
one case where we must take a little care. Consider a nested version of the hook
discussed in section~\ref{sec:GR-Wines}. It is not surprising that we must take
care here, as this is an example where the final gauge remainder in the string has bitten
the end of a wine, rather than a socket. We can consider the gauge remainder that forms
the un-nested hook of equations~(\ref{eq:hook-R-1}) and~(\ref{eq:hook-R-2})
to have bitten itself on the right. This, however, corresponds to a push forward and
\emph{not} a pull back. Thus, for the arbitrarily nested hook shown in figure~\ref{fig:nestedhook},
the number of pushes forward is equal to the number of bites on the left, plus one and the number of pulls
back is equal to the number of bites on the right, minus one.

\subsection{Double Gauge Remainders}	\label{sec:Diagrammatics:DoubleGRs}

\begin{center}
\begin{figure}[h]
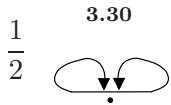

	\[
	\frac{1}{2} 
	\LED[0.1]{}{Struc-WGRx2-B}{Diags-WGRx2-B}
	\]
\caption{An example of a double gauge remainder diagram.}
\label{fig:DoubleGR}
\end{figure}
\end{center}

We conclude our discussion of the gauge remainders by examining diagrams possessing
two active gauge remainders. The techniques we use to process such terms are
exactly the same as  the ones detailed already in this chapter. There is,
however, a qualitatively new type of diagram that arises, together with a source
of possible ambiguity. To investigate both of these issues, we focus on a
two-loop diagram, in which both gauge remainders bite the wine.\footnote{
Double gauge remainder diagrams are not restricted to those in which both gauge
remainders bite a wine; one or both of the gauge remainders can bite a vertex.}
This
is shown in figure~\ref{fig:DoubleGR}.

\begin{center}
\begin{figure}[h]
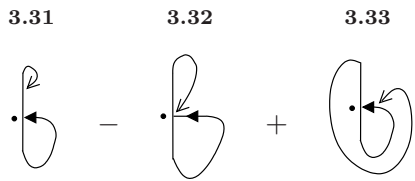

	\[
	\begin{array}{ccccc}
		\LE{Diags-WGRx2-Top}	&	& \LE{Diags-WGRx2-Trap}	&	& \LE{Diags-WGRx2-Bottom}
	\\[1ex]
		\ensuremath{\begin{array}{c}\input{pstex/Struc-WGRx2-Top.pstex_t} \end{array}}	& -	& \ensuremath{\begin{array}{c}\input{pstex/Struc-WGRx2-Trap.pstex_t} \end{array}}	& +	& \ensuremath{\begin{array}{c}\input{pstex/Struc-WGRx2-Bottom.pstex_t} \end{array}}
	\end{array}
	\]
\caption{Result of allowing one gauge remainder in diagram~\ref{Diags-WGRx2-B} to act.}
\label{fig:DoubleGR-A}
\end{figure}
\end{center}

We proceed by allowing first one gauge remainder to act and then, if possible,
by allowing the second gauge remainder to act. The qualitatively new type of diagram
arises because one of the
effects of the first gauge remainder can be to `trap' the second gauge remainder,
by biting the field on the wine to which the second gauge remainder attaches.
The active gauge remainder now does not have the same momentum flowing through
it as the field it is trying to bite, and so it cannot act.
The other diagrams generated are those in
which the processed gauge remainder bites one of the
ends of the wine. The result of allowing the first gauge remainder to
act is shown in figure~\ref{fig:DoubleGR-A}, where we have collected pulls
back and pushes forward, as usual.

Diagram~\ref{Diags-WGRx2-Trap} possesses the trapped gauge remainder. In diagram~\ref{Diags-WGRx2-Bottom}
we have recognised that the wine ends where it attaches to the active gauge remainder
and so this is where the processed gauge remainder bites. Note that
we can trivially redraw this diagram, as shown in figure~\ref{fig:DoubleGR-B}.\footnote{
Diagram~\ref{Diags-WGRx2-Bottom-B} highlights how one must be very careful when drawing
which end of an active gauge remainder a processed gauge remainder bites; diagrams~\ref{Diags-WGRx2-Bottom-B}
and~\ref{Diags-WGRx2-Trap} are clearly different.
}
\begin{center}
\begin{figure}[h]
	\[
	\LED[0.1]{}{Struc-WGRx2-Bottom-B}{Diags-WGRx2-Bottom-B}
	\]
\caption{A trivial redrawing of diagram~\ref{Diags-WGRx2-Bottom}.}
\label{fig:DoubleGR-B}
\end{figure}
\end{center}

\begin{center}
\begin{figure}[h]
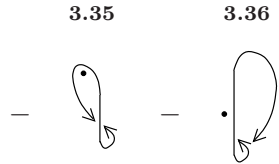

	\[
	\begin{array}{cccc}
			& \LE{Diags-WGRx2-L-Bottom}	&	& \LE{Diags-WGR-N1-Bottom-PF}
	\\[1ex]
		-	& \ensuremath{\begin{array}{c}\input{pstex/Struc-WGRx2-L-Bottom.pstex_t} \end{array}}	& -	& \ensuremath{\begin{array}{c}\input{pstex/Struc-WGR-N1-Bottom-PF.pstex_t} \end{array}}
	\end{array}
	\]
\caption{Two of the terms produced by processing diagrams~\ref{Diags-WGRx2-Top}
and~\ref{Diags-WGRx2-Bottom-B}.}
\label{fig:DoubleGR-C}
\end{figure}
\end{center}

Finally, consider allowing the active gauge remainder to act in diagrams~\ref{Diags-WGRx2-Top}
and~\ref{Diags-WGRx2-Bottom-B}. We are not interested in all the contributions. Rather,
we just want to focus on
\begin{enumerate}
	\item	the term produced by diagram~\ref{Diags-WGRx2-Top} where the
			gauge remainder pulls back along the wine, to the same end as the hook;

	\item	the term produced by diagram~\ref{Diags-WGRx2-Bottom-B} in which the
			gauge remainder pulls back to the bottom of the wine.
\end{enumerate}
Reflecting the former diagram about a horizontal line, we arrive at the two diagrams
of figure~\ref{fig:DoubleGR-C}.\footnote{The reflection does not yield a net sign as
we pick up one for each of the processed gauge remainders.} 

\begin{center}
\begin{figure}[h]
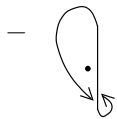

	\[
	-\LED[0.1]{}{Struc-WGR-N1-Bottom-PF-B}{Diags-WGR-N1-Bottom-PF-B}
	\]
\caption{A trivial redrawing of diagram~\ref{Diags-WGR-N1-Bottom-PF}.}
\label{fig:DoubleGR-D}
\end{figure}
\end{center}

From the way in which these two diagrams have been drawn, it is clear that
they are distinct and that they must be treated as such. We can, however, redraw
diagram~\ref{Diags-WGR-N1-Bottom-PF} by sliding the outer gauge remainder round the hook
to where the inner gauge remainder bites the wine. This is shown in figure~\ref{fig:DoubleGR-D}.

Diagram~\ref{Diags-WGR-N1-Bottom-PF-B} is, of course, still the same as
diagram~\ref{Diags-WGR-N1-Bottom-PF} but it is starting to look very similar
to diagram~\ref{Diags-WGRx2-L-Bottom}. Indeed, we must make sure that we never
slide the gauge remainder so far round the hook that it appears to bite the wine
at the same point as the gauge remainder which forms the hook. If this
were to happen, then such a diagram would be ambiguous.

\section{Momentum Expansions} \label{sec:MomentumExpansions}

The computation of $\beta$-function coefficients involves working at fixed order in external momentum. 
If a diagram contains a structure that is already manifestly of the desired order, then it is useful to Taylor expand 
at least some of the remaining structures in the external momentum. Vertices can always be expanded in momentum,
as it is a requirement of the setup that such a step is possible~\cite{GI-ERG-I, GI-ERG-II, TRM-elements}. 
Whereas wines, too, can always be Taylor expanded in momentum it is not necessarily
possible to do so with effective propagators that form part of a diagram, as this step can introduce IR divergences.
This will be discussed in detail in 
chapter~\ref{ch:LambdaDerivatives:Methodology}.

In addition to allowing us to deal with diagrams possessing a structure already of the desired
order in external momentum,
the methodology of this chapter will also be of use in the extraction of universal coefficients from 
$\Lambda$-derivative terms. To this end, we will provide explicit lowest order expansions
of certain vertices and discuss features of particular illustrative diagrams.

The key idea in what follows is that, if an $A$-field decorating an $n$-point vertex (wine) carries zero momentum, then we can relate
this vertex (wine) to the momentum derivative of a set of $(n-1)$-point vertices (wines). This relation arises as a consequence of
the Ward identity~(\ref{eq:WID-Unbroken}) and so it is no surprise that the diagrammatics of this chapter is very similar to those of the last.

\subsection{Basics}

\begin{center}
\begin{figure}[h]
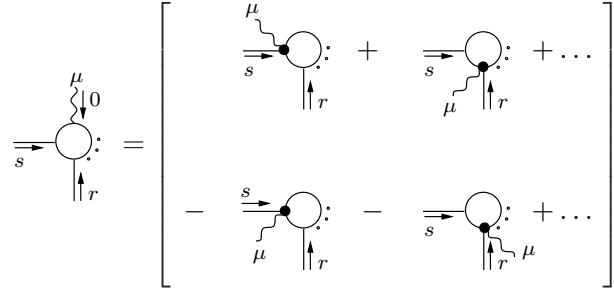

\[
\ensuremath{\begin{array}{c}\input{pstex/MomExp-dummy.pstex_t} \end{array}} =
\left[
\ensuremath{\begin{array}{c}\input{pstex/MomExp-dummyM.pstex_t} \end{array}}
\hspace{0.3in}
\right]
\]
\caption[Diagrammatics expression for a vertex decorated by an $A$-field carrying zero momentum.
]{Diagrammatics expression for a vertex decorated by an $A$-field carrying zero momentum. The filled circle attached to 
the $A$-field line tells us to first replace all momenta with dummy momenta; then
to differentiate \wrt\ the dummy momenta of the field hit, holding
all other momenta constant and finally to replace the dummy momenta with the original momenta.}
\label{fig:MomExp:dummy}
\end{figure}
\end{center}

Consider the structure, $U$, which can be either or vertex or wine, the decorations of which include an $A$-field carrying zero momentum.
Let us begin by supposing that this $A$-field is sandwiched between the fields $X$ and $Y$. By the Ward Identity~(\ref{eq:WID-Unbroken}), we have:
\begin{eqnarray}
\lefteqn{
\epsilon_\mu U^{\cdots X A Y \cdots}_{\cdots R \, \mu \, S \cdots}(\ldots,r,\epsilon,s-\epsilon,\ldots) =} \nonumber \\
& &
U^{\cdots X Y \cdots}_{\cdots R \, S \cdots}(\ldots,r,s,\ldots) - U^{\cdots X Y \cdots}_{\cdots R \, S \cdots}(\ldots,r+\epsilon,s-\epsilon,\ldots).
\end{eqnarray}
Taylor expanding both sides in $\epsilon$ and equating the $\Oep$ terms yields:
\begin{equation}
 U^{\cdots X A Y \cdots}_{\cdots R \, \mu \, S \cdots}(\ldots,r,0,s,\ldots) = \left.\left(\partial_\mu^{s'} - \partial_\mu^{r'} \right)
 U^{\cdots X Y \cdots}_{\cdots R \, S \cdots}(\ldots,r',s',\ldots)\right|_{r'=r,s'=s}.
\label{eq:MomeExp:Defining}
\end{equation}
This equation, for the case of a vertex, is represented diagrammatically in figure~\ref{fig:MomExp:dummy}, which highlights the similarity between
the  momentum expansions and gauge remainders. The top row on the \rhs\ correspond to `push forward like' terms, whereas those
on the second row correspond to `pull back like' terms. (As with the gauge remainders, pushes forward are performed
in the counterclockwise sense.)

As with the gauge remainders, we must consider all possible independent locations of the $A$-field \wrt\ the other fields.
Hence, terms between the first and second rows can cancel, if the field hit is bosonic.

We now want to convert derivatives \wrt\ the dummy momenta to derivatives \wrt\ the original momenta. There are two cases to deal with.
The first---in which we shall say that the momenta are paired---is where there are a pair of fields, carrying equal and opposite momentum. 
The second---in which we shall say that the momenta are coupled---is where there are three fields carrying, say, $(r, s, -s-r)$.

\subsubsection{Case I: Paired Momenta}

This is the simplest case to deal with. If the momentum $r$ has been replaced with dummy momentum $r'$ and $-r$ has been
replaced with dummy momentum $s'$ then
\[
\left(\partial_\mu^{r'} - \partial_\mu^{s'} \right) \rightarrow \partial_\mu^r.
\]
Hence, we can collect together a push forward like diagram with a pull back like diagram to give a derivative \wrt\ one of 
the original momenta. An example of this is shown in figure~\ref{fig:MomExp:Defining}, for a field-ordered three-point vertex.
\begin{center}
\begin{figure}[h]
\[
\ensuremath{\begin{array}{c}\input{pstex/MomExp-Defining.pstex_t} \end{array}}
\]
\caption[Re-expressing a three-point vertex with zero momentum entering along an $A$-field.]
{A field ordered three-point vertex with zero momentum entering along an $A$-field can be expressed as 
the momentum derivative of a two-point vertex. The open circle attached to the $A$-field line represents a derivative
\wrt\ the momentum entering the vertex along the field hit.}
\label{fig:MomExp:Defining}
\end{figure}
\end{center}

\subsubsection{Case II: Coupled Momenta}

The structures in this section contain momentum arguments of the form $(r,s,-s-r)$.
Referring back to equation~(\ref{eq:MomeExp:Defining}), we will denote the dummy momenta
by $(r',s',t')$. We can make progress by noting that:
\begin{eqnarray}
\left(\partial_\mu^{r'} - \partial_\mu^{s'} \right) & \rightarrow & \left(\left.\partial_\mu^{r}\right|_s - 
\left.\partial_\mu^{s}\right|_r\right) \label{eq:dummy-original}
\end{eqnarray}
and likewise, for all other combinations of $(r',s',t')$.
Thus, as with the previous case, we need to combine a pair of terms differentiated \wrt\ dummy momenta
to obtain a structure which is differentiated \wrt\ its original momenta. The difference is that, whilst in the
previous case the pair combined into one diagram, in this case they remain as a pair. An example is
shown in figure~\ref{fig:MomExp:Coupled}.
\begin{center}
\begin{figure}[h]
\[
\ensuremath{\begin{array}{c}\input{pstex/MomExp-Coupled.pstex_t} \end{array}}
\]
\caption[Re-expressing a four-point vertex with zero momentum entering along an $A$-field.]
{A field ordered four-point vertex with zero momentum entering along an $A$-field can be expressed as 
the momentum derivative of two three-point vertices. 
}
\label{fig:MomExp:Coupled}
\end{figure}
\end{center}

The open circle attached to the $A$-field line represents a derivative
\wrt\ the momentum entering the vertex along the field hit. However, this derivative is performed holding the
momentum of the field hit in the partner diagram constant. Hence, the final two diagrams 
of figure~\ref{fig:MomExp:Coupled} must be interpreted as a pair.
The difference between this
and the paired momentum case highlights the care that
must be taken interpreting the new diagrammatics.

\subsection{Wines}

When we come to deal with wines, we must adapt the diagrammatic notation slightly.
If the momentum derivative strikes a field decorating a wine, then we just use the
current notation. However, it is desirable to change the notation when the
momentum derivative strikes one of the ends a wine.
In complete diagrams, placing the
diagrammatic object representing a momentum derivative at the end of the wine becomes confusing; rather we place the object in middle 
and use an arrow to indicate which end of the wine it acts on, as shown in figure~\ref{fig:MomExp:wine}.

\begin{center}
\begin{figure}[h]
\[
\ensuremath{\begin{array}{c}\input{pstex/MomExp-wine.pstex_t} \end{array}}
\]
\caption{A field ordered one-point wine with zero momentum entering along a decorative $A$-field can be expressed as
the momentum derivative of a zero-point wine.}
\label{fig:MomExp:wine}
\end{figure}
\end{center}

Hence, the second diagram denotes a derivative \wrt\ $+k$, whereas the third diagram denotes a derivative \wrt\ $-k$.

\subsection{Algebraic Expansions}

In this section we provide the lowest order momentum expansions for the following:
\begin{enumerate}
	\item	a three-point, tree level vertex for which all fields are in the $A^1$ sector;

	\item	a four-point, tree level vertex for which all fields are in the $A^1$ sector.
\end{enumerate}

The first case has already been partially examined in figure~\ref{fig:MomExp:Defining} and, algebraically,
this reads:
\begin{equation}
S^{AXY}_{\mu\, RS}(0,s,-s) = \partial^s_\mu S^{XY}_{RS}(s,-s)
\label{eq:MomExp:3pt}
\end{equation}
Using equation~(\ref{eq:app:TLTP-11}) and recalling that $c(0)=1$, we obtain
\begin{equation}
\partial_\mu^k S_{0 \alpha \beta}^{\ 1 \, 1}(k,-k) = 2(2k_\mu \delta_{\alpha \beta} - k_\alpha \delta_{\beta \mu} - 
k_{\beta} \delta_{\alpha \mu}) + \mathcal{O}(k^3).
\label{eq:MomExp:3pt-algebra}
\end{equation}

The second case has already been partially examined in figure~\ref{fig:MomExp:Coupled} and, algebraically,
this reads:
\[
S^{1\, 1  \, 1 \, 1}_{\alpha \beta \rho \, \sigma}(l,-l-k,0,k) = \left.\partial_\rho^k\right|_l
S^{1\, 1\, 1}_{\alpha \beta \sigma}(l,-l-k,k).
\]
Expanding both sides to zeroth order in momenta and utilising equation~(\ref{eq:MomExp:3pt-algebra}) yields:
\begin{equation}
S^{\ 1\, 1\, 1\, 1}_{0 \alpha \beta \rho \, \sigma}(0,0,0,0)
\equiv S^{\ 1\, 1\, 1\, 1}_{0 \alpha \beta \rho \, \sigma}(\underline{0})
= -2(2\delta_{\alpha \rho} \delta_{\beta \sigma}
- \delta_{\beta \rho} \delta_{\alpha \sigma} - \delta_{\sigma \rho} \delta_{\alpha \beta}).
\label{eq:MomExp:4pt}
\end{equation}

\subsection{Complete Diagrams} \label{sec:MomExp:CompleteDiagrams}

The diagrams we deal with fall into two classes, depending on whether
or not they possess a structure which is manifestly of the desired
order in external momentum to which we wish to work. In the case that
such a structure is present, we will Taylor expand all components of
the  rest of the diagram that we are allowed to, in external momentum.
The other case will arise in our discussion of $\Lambda$-derivatives
(see chapters~\ref{ch:LambdaDerivatives:Methodology} 
and~\ref{ch:TotalLambdaDerivatives:Application}) and here,
rather than being forced to Taylor expand certain structures, we choose
which structures to Taylor expand.

\subsubsection{Forced Expansions}

\begin{center}
\begin{figure}[h]
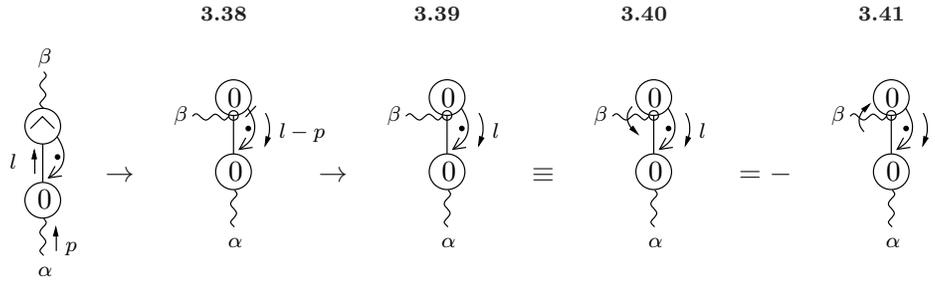

	\[
	\begin{array}{ccccccccc}
							&			&\LE{ExD:d1:2.12M}	&			& \LE{ExD:d1:2.12Mb}&		&\LE{ExD:d1:2.12Mc}	&	&\LE{ExD:d1:2.12Md}
	\\[1ex]
		\ensuremath{\begin{array}{c}\input{pstex/Beta1-2.12-lab.pstex_t} \end{array}}	&\rightarrow&\ensuremath{\begin{array}{c}\input{pstex/ExD-d12.12M.pstex_t} \end{array}}	&\rightarrow&\ensuremath{\begin{array}{c}\input{pstex/ExD-d12.12Mb.pstex_t} \end{array}}	&\equiv	& \ensuremath{\begin{array}{c}\input{pstex/ExD-d12.12Mc.pstex_t} \end{array}}	&=-	&\ensuremath{\begin{array}{c}\input{pstex/ExD-d12.12Md.pstex_t} \end{array}}
	\end{array}
	\]
\caption{Manipulation of a diagram at $\Op{2}$. Discontinuities in momentum flow are indicated by a bar.}
\label{fig:MomExp:Ex}
\end{figure}
\end{center}

The generalisation to complete diagrams is straightforward and, as an example, 
figure~\ref{fig:MomExp:Ex} shows how we can manipulate diagram~\ref{ExD:d1:2.12}.

The two-point vertex at the base of the diagram is $\Op{2}$, which is the order in $p$ to
which we wish to work. We call this base structure an `$\Op{2}$ stub'.
The first step is to Taylor expand the three-point vertex to zeroth order in $p$,
as shown in diagram~\ref{ExD:d1:2.12M}. There is now a discontinuity in momentum arguments, since
although momentum $l$ flows into and out of the differentiated two-point vertex, this vertex is attached to an
effective propagator 
carrying momentum $l$ and a wine carrying momentum $l-p$. 
This discontinuity is indicated by the bar between the vertex and the wine.
We can Taylor expand the wine to zeroth order in momentum, too, and this
is done in diagram~\ref{ExD:d1:2.12Mb}. Since the discontinuity in momentum has now vanished, the bar is
removed. We will encounter terms shortly in which
the bar must be retained. Treatment of diagrams with an $\Op{2}$ stub in which the wine is decorated by
an external field is exactly the same.

In diagram~\ref{ExD:d1:2.12Mc} we have introduced an arrow on the diagrammatic representation
of the derivative. We have come across this arrow already the context of wines but have not
yet required it for vertices. Indeed, in the current example, it is effectively redundant notation.
We note, though, that we can reverse the direction of the arrow, at the expense of a minus sign,
as in diagram~\ref{ExD:d1:2.12Md}. By reversing the direction of the arrow, we are now differentiating \wrt\
the momentum leaving the vertex along the struck field, rather than the momentum thus entering. We will
require this notation shortly.

\subsubsection{Unforced Expansions}

\begin{center}
\begin{figure}[h]
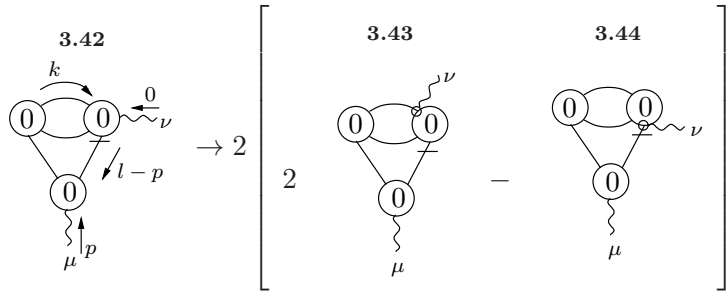

	\[
	\LED{}{MomExp-D28.1}{ExD:d:28.1} \rightarrow
	2
	\left[
		\begin{array}{cccc}
				& \LE{ExD:d:28.1:sub1}	&	& \LE{ExD:d:28.1:sub2}
		\\[1ex]
			2	& \ensuremath{\begin{array}{c}\input{pstex/MomExp-D28.1Ma.pstex_t} \end{array}}	& - & \ensuremath{\begin{array}{c}\input{pstex/MomExp-D28.1Mb.pstex_t} \end{array}}
		\end{array}
	\right]
	\]
\caption{Example of a two-loop diagram in which we choose to Taylor expand in momenta, rather
than this choice being forced upon us.}
\label{fig:MomExp:D28.1}
\end{figure}
\end{center}

To illustrate the kind of diagrams in which we \emph{choose} to Taylor expand structures,
consider figure~\ref{fig:MomExp:D28.1}. The right-most vertex of the first diagrams is not forced to carry zero momentum 
along its external field; rather we construct this diagram, for reasons which will become apparent in 
section~\ref{sec:TLD:SS-Mirror-ComplementaryTerms}.
In direct analogy with the gauge remainders, the overall factor of two on the \rhs\ arises from collecting together \CC\ pairs,
and the additional factor of two in front of diagram~\ref{ExD:d:28.1:sub1} arises from invariance under swapping the two $k$-dependent internal 
fields. Now, however, we arrive at a potential source of confusion. We know that two diagrams on the \rhs\ must be interpreted as a pair. But,
we are free to route momenta such that, in the first diagram, whilst holding $l$ constant,
 we are either differentiating \wrt\ $k$ or \wrt\ $(l-k)$. This choice, in turn, determines what must be held constant whilst differentiating
\wrt\ $l$ in the second diagram.

Noting that the relative factor of two between the two diagrams on the \rhs\ disappears if we choose an order for the interchangeable internal 
fields in the second diagram, the effect of the momentum derivatives on the right-most vertex is summarised in table~\ref{tab:MomExp:D28.1}.
In the second column, we include the overall minus sign of the diagram (which arises from pulling back the momentum derivative).
\begin{center}
\begin{table}[h]
\[
\begin{array}{l|l|l}
\mathrm{Diagram~\ref{ExD:d:28.1:sub1}} & \mathrm{Diagram~\ref{ExD:d:28.1:sub2}} & \mathrm{Net \ Effect} \\ \hline
\left.\partial^k_\nu\right|_l & -\left.\partial^{-l}_\nu\right|_k & \left.\partial^k_\nu\right|_l  +\left.\partial^l_\nu\right|_k \\
\left.\partial^{(l-k)}_\nu\right|_l & - \left.\partial^{-l}_\nu\right|_{(l-k)} & \left.\partial^l_\nu\right|_k
\end{array}
\]
\caption{Algebraic realisation of the derivative striking the four-point vertices of 
diagrams~\ref{ExD:d:28.1:sub1} and~\ref{ExD:d:28.1:sub2},
for different momentum routing.}
\label{tab:MomExp:D28.1}
\end{table}
\end{center}

It seems that the net effect on the right-most vertex depends on the choice of momentum routing. 
However, we do not expect individual elements of a diagram to be invariant under momentum
rerouting; rather it is the diagram as a whole which is invariant under such a change. Thus,
we can we choose to route $k$ however we please in diagram~\ref{ExD:d:28.1} and need only
ensure that diagrams~\ref{ExD:d:28.1:sub1} and~\ref{ExD:d:28.1:sub2} are consistent with this.

Note, though, that by performing the unforced Taylor expansion, diagram~\ref{ExD:d:28.1} is
no longer invariant under rerouting the loop momentum $l$. As we will see in 
section~\ref{sec:TLD:Subtractions}, diagrams such as~\ref{ExD:d:28.1} are in fact
constructed in  actual calculations to facilitate extraction
of numerical answers, rather than arising as a direct consequence of
the flow equations. Hence, to incorporate such diagrams, we must add and subtract them
from a calculation (\ie\ we merely conveniently re-express zero). 
Diagrams genuinely
generated by the flow equations are always invariant under momentum rerouting;
indeed this is necessary for gauge invariance~\cite{SU(N|N), TRM-Faro}.

There is a further subtlety concerning the treatment of diagrams in which we perform an unforced Taylor expansion,
which can also be illustrated by our current example.
When we come to extract the universal contribution from the $\Lambda$-derivative terms, it turns out the four-point vertex in
diagram~\ref{ExD:d:28.1} will be Taylor expanded to zeroth order in all momenta, not just in $p$. 
 Let us denote the field ordered four-point, tree level vertex
(with all fields in the $A^1$ sector) by:
\begin{equation}
S^{\ 1\, 1\, 1\, 1}_{0 \alpha \nu \rho \, \sigma}(p-l,-p,k,l-k)
\label{eq:MomExp:4pt:4A}
\end{equation}
where $\rho, \sigma$ are the indices corresponding to the indistinguishable internal fields.
Given that, as a result of momentum rerouting invariance, the parent diagram is invariant under the interchange of these two internal fields
and that the left-hand three-point vertex changes sign if we let $\rho \leftrightarrow \sigma$, 
 the diagram is invariant under the addition  of 
\begin{equation}
a\left(\delta_{\nu \rho} \delta_{\alpha \sigma} + \delta_{\alpha \rho} \delta_{\nu \sigma}\right) 
+ b \delta_{\sigma \rho} \delta_{\alpha \nu},
\label{eq:MomExp:4pt-vanishing}
\end{equation}
to the four-point vertex, where $a$ and $b$ are arbitrary coefficients. 
This means that, when we Taylor expand the four-point vertex to zeroth
order in momenta, there is a spectrum of equivalent forms that we can take.

In particular, we will obtain different forms depending on the order we choose for $\rho$ and $\sigma$. These differences,
of course, just give rise to a contribution of the form given by equation~(\ref{eq:MomExp:4pt-vanishing}). However, to actually evaluate 
the $\Lambda$-derivative terms generally involves a complete loss
of momentum rerouting invariance at a certain stage of the calculation (\ie\ we ultimately lose
the freedom to reroute $k$, as well as $l$). At this point, 
the specific choice for a four-point vertex will become important.

\begin{center}
\begin{figure}[h]
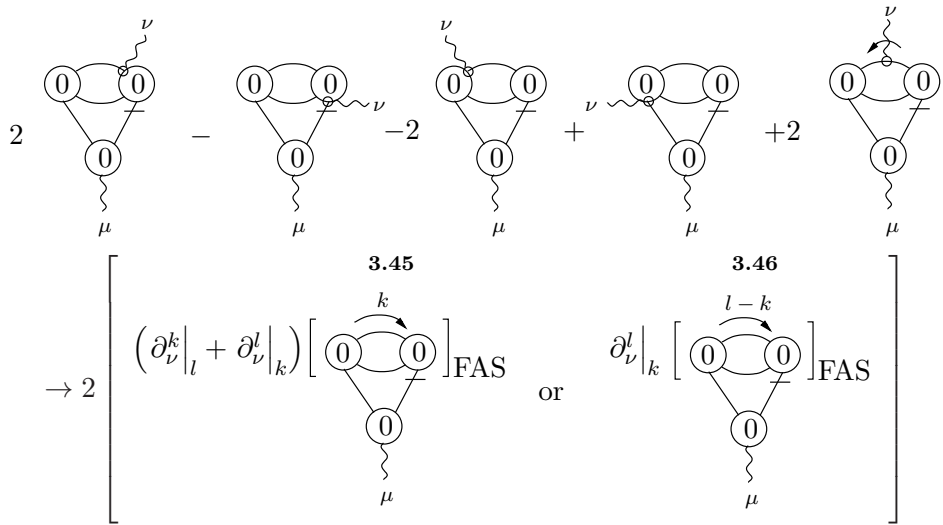

\[
\begin{array}{c}
	\ensuremath{\begin{array}{c}\input{pstex/MomExp-D28.1-subs.pstex_t} \end{array}}
\\
	\rightarrow 2
	\left[
		\ensuremath{\begin{array}{c}\input{pstex/MomExp-D28.1-subsM.pstex_t} \end{array}}
	\hspace{0.3in}
	\right]
\end{array}
\]
\caption{A set of diagrams which naturally combine.}
\label{fig:MomExp:D28.1:subs}
\end{figure}
\end{center}

We conclude this section by illustrating how certain groups of diagrams
in which an unforced Taylor expansion has been performed can be
combined. In chapter~\ref{ch:TotalLambdaDerivatives:Application},
we will be lead to consider the set of diagrams 
shown in the first row of figure~\ref{fig:MomExp:D28.1:subs}.

We now fix the order of the interchangeable fields, choosing to do so in such a way that the momentum routing between pairs of diagrams
is the same.
Referring to the $k$-dependent sub-diagram, all structures are hit by a momentum derivative with the exception of the
inner most effective propagator. We now analyse the two different momentum routings.

In the first instance, we will take momentum $k$ to travel around the upper-most effective propagator. Noting that
the inner-most effective propagator carries momentum $l-k$ and that 
$\left(\left.\partial^k_\nu\right|_l +\left.\partial^l_\nu\right|_k\right) F(l-k) = 0$,
for any function $F$, we can take the momentum derivatives outside the $k$-dependent sub-diagram, to yield the first
diagram~\ref{ExD:d:28.1:subM1}.

In the second instance, we will take momentum $(l-k)$ to flow around the upper-most effective propagator. The inner-most 
effective propagator now carries $k$ and, referring to table~\ref{tab:MomExp:D28.1}, both vertices are hit by
$\left.\partial^l\right|_k$. Given that $\left(\partial^{(l-k)}_\nu-\left.\partial^{l}_\nu\right|_k\right) F(l-k) = 0$
 we can take the momentum derivatives outside the $k$-dependent sub-diagram, to yield the second
diagram~\ref{ExD:d:28.1:subM2}.

\begin{center}
\begin{figure}[h]
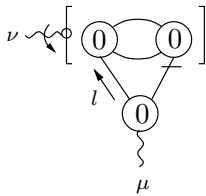

	\[
	\LED[0.1]{}{MomExp-D28.1-TMD}{DerivSqBracket-Ex}
	\]
\caption{Completely diagrammatic representation of diagrams~\ref{ExD:d:28.1:subM1} and~\ref{ExD:d:28.1:subM2}.}
\label{fig:MomExp:D28.1:TMD}
\end{figure}
\end{center}

It is apparent that these two diagrams differ by a total derivative \wrt\ $k$.
 This gives rise to 
a surface term which must be thrown away, for the consistency of the regularisation scheme~\cite{SU(N|N)}; thus, in this thesis,
calculations are performed in general dimension $D$, so that such terms are automatically discarded. We can redraw both terms
as the diagram shown in figure~\ref{fig:MomExp:D28.1:TMD}.

Now we see why we introduced arrows on momentum derivative striking vertices
in figure~\ref{fig:MomExp:Ex}. Let us return to diagram~\ref{DerivSqBracket-Ex}.
Focusing on the 
structure enclosed by 
square brackets, we ignore its internal structure and note only that
it carries external momentum $l$. The arrow on the derivative
tells us to differentiate \wrt\ $+l$ (with $l$ routed as shown). This rule is consistent with
the way in which we treat momentum derivatives of plain vertices.

\subsubsection{Charge Conjugation}

We conclude our discussion of momentum expansions by commenting on how
we can redraw a diagram using \CC. For any diagram, we use the following recipe:
\begin{enumerate}
	\item	take the mirror image (this includes reflecting
			any arrows accompanying derivative symbols);

	\item	pick up a minus sign for each performed gauge remainder;

	\item	pick up a minus sign for each derivative symbol.
\end{enumerate}

\section{More Diagrammatic Identities} \label{sec:D-ID}

In this section we detail a number of diagrammatic identities involving 
a selection of the following:
two-point tree-level vertices, (un)decorated wines, effective propagators,
gauge remainders and momentum derivatives

\begin{D-ID}
A sub-diagram comprising a two-point, tree level vertex, differentiated \wrt\ momentum and attached to an effective propagator can, using the effective
propagator relation,
 be redrawn, in the following way:
\[
\ensuremath{\begin{array}{c}\input{pstex/D-ID-OneWine.pstex_t} \end{array}}.
\]
The final diagram is interpreted as the derivative \wrt\ the momentum entering the encircled structure from the left. Had the
arrow pointed the other way, then the derivative would be \wrt\ the momentum entering the structure from the right.
\label{D-ID:OneWine}
\end{D-ID}

\begin{D-ID}
A sub-diagram comprising a two-point, tree level  vertex, differentiated \wrt\ momentum and attached to two effective
propagators can, using diagrammatic 
identity~\ref{D-ID:OneWine} and the effective
propagator relation, be redrawn in the following way:
\[
\ensuremath{\begin{array}{c}\input{pstex/D-ID-TwoWines-A.pstex_t} \end{array}} 
+ \frac{1}{2}
\left[
\ensuremath{\begin{array}{c}\input{pstex/D-ID-TwoWines-B.pstex_t} \end{array}} 
\right].
\]
We could re-express this in a less symmetric form by taking either of the two rows in the square brackets and removing the
factor of half.
\label{D-ID:TwoWines}
\end{D-ID}

\begin{D-ID}
A sub-diagram comprising a two-point, tree level vertex, struck by $\GRk$
vanishes:
\[
	\ensuremath{\begin{array}{c}\begin{picture}(0,0)%
\includegraphics{pstex/TLTP-GR.pstex}%
\end{picture}%
\setlength{\unitlength}{3947sp}%
\begingroup\makeatletter\ifx\SetFigFont\undefined%
\gdef\SetFigFont#1#2#3#4#5{%
  \reset@font\fontsize{#1}{#2pt}%
  \fontfamily{#3}\fontseries{#4}\fontshape{#5}%
  \selectfont}%
\fi\endgroup%
\begin{picture}(454,238)(1438,-114)
\put(1648,-57){\makebox(0,0)[lb]{\smash{\SetFigFont{11}{13.2}{\rmdefault}{\mddefault}{\updefault}{\color[rgb]{0,0,0}0}%
}}}
\end{picture}
 \end{array}} = 0.
\]
\paragraph{Corollaries}
\begin{eqnarray*}
	\ensuremath{\begin{array}{c}\begin{picture}(0,0)%
\includegraphics{pstex/dTLTP-GR.pstex}%
\end{picture}%
\setlength{\unitlength}{3947sp}%
\begingroup\makeatletter\ifx\SetFigFont\undefined%
\gdef\SetFigFont#1#2#3#4#5{%
  \reset@font\fontsize{#1}{#2pt}%
  \fontfamily{#3}\fontseries{#4}\fontshape{#5}%
  \selectfont}%
\fi\endgroup%
\begin{picture}(457,396)(1432,-114)
\put(1648,-57){\makebox(0,0)[lb]{\smash{\SetFigFont{11}{13.2}{\rmdefault}{\mddefault}{\updefault}{\color[rgb]{0,0,0}0}%
}}}
\end{picture}
 \end{array}} 		& \equiv & - \ensuremath{\begin{array}{c}\begin{picture}(0,0)%
\includegraphics{pstex/TLTP-dGR.pstex}%
\end{picture}%
\setlength{\unitlength}{3947sp}%
\begingroup\makeatletter\ifx\SetFigFont\undefined%
\gdef\SetFigFont#1#2#3#4#5{%
  \reset@font\fontsize{#1}{#2pt}%
  \fontfamily{#3}\fontseries{#4}\fontshape{#5}%
  \selectfont}%
\fi\endgroup%
\begin{picture}(550,431)(1438,-114)
\put(1648,-57){\makebox(0,0)[lb]{\smash{\SetFigFont{11}{13.2}{\rmdefault}{\mddefault}{\updefault}{\color[rgb]{0,0,0}0}%
}}}
\end{picture}
 \end{array}}
\\
	\ensuremath{\begin{array}{c}\input{pstex/LdL-TLTP-GR.pstex_t} \end{array}} 	& \equiv & -\ensuremath{\begin{array}{c}\input{pstex/TLTP-LdL-GR.pstex_t} \end{array}} = 0
\\
	\ensuremath{\begin{array}{c}\input{pstex/dal-TLTP-GR.pstex_t} \end{array}}	& \equiv & -\ensuremath{\begin{array}{c}\input{pstex/TLTP-dal-GR.pstex_t} \end{array}} = 0
\end{eqnarray*}

Recall that \Dal\ represents differentiation \wrt\ $\alpha$. We have used the
independence of $\GRk$ on both $\Lambda$ and $\alpha$ (in all sectors).
\label{D-ID:TLTP-GR}
\end{D-ID}

\begin{D-ID}
A sub-diagram comprising a two-point, tree level vertex, differentiated \wrt\ momentum and attached to two effective propagators, 
one of which terminates with a $\wedge$ can,
using diagrammatic 
identities~\ref{D-ID:PseudoEffectivePropagatorRelation} and~\ref{D-ID:OneWine}, be redrawn in the following form:
\[
\ensuremath{\begin{array}{c}\input{pstex/D-ID-TwoWinesGR.pstex_t} \end{array}}.
\]
\label{D-ID:TwoWinesGR}
\end{D-ID}

\begin{D-ID}
A sub-diagram comprising a two-point, tree level vertex, attached to an un-decorated wine
which terminates in a $\wedge$ can be redrawn, using diagrammatic identities~\ref{D-ID:PseudoEffectivePropagatorRelation} 
and~\ref{D-ID:GaugeRemainderRelation},
the effective propagator relation and the tree level flow equation:
\begin{eqnarray*}
\ensuremath{\begin{array}{c}\input{pstex/D-ID-VWG-A.pstex_t} \end{array}} 
& \equiv &
\left[
\ensuremath{\begin{array}{c}\begin{picture}(0,0)%
\includegraphics{pstex/D-ID-VWG-B.pstex}%
\end{picture}%
\setlength{\unitlength}{3947sp}%
\begingroup\makeatletter\ifx\SetFigFont\undefined%
\gdef\SetFigFont#1#2#3#4#5{%
  \reset@font\fontsize{#1}{#2pt}%
  \fontfamily{#3}\fontseries{#4}\fontshape{#5}%
  \selectfont}%
\fi\endgroup%
\begin{picture}(656,238)(376,-654)
\put(537,-596){\makebox(0,0)[lb]{\smash{\SetFigFont{11}{13.2}{\rmdefault}{\mddefault}{\updefault}{\color[rgb]{0,0,0}0}%
}}}
\end{picture}
 \end{array}} 
\right]^\bullet -
\ensuremath{\begin{array}{c}\input{pstex/D-ID-VWG-C.pstex_t} \end{array}} 
-
\ensuremath{\begin{array}{c}\input{pstex/D-ID-VWG-D.pstex_t} \end{array}} 
\\
& \equiv & 
\ensuremath{\begin{array}{c}\begin{picture}(0,0)%
\includegraphics{pstex/D-ID-VWG-E.pstex}%
\end{picture}%
\setlength{\unitlength}{3947sp}%
\begingroup\makeatletter\ifx\SetFigFont\undefined%
\gdef\SetFigFont#1#2#3#4#5{%
  \reset@font\fontsize{#1}{#2pt}%
  \fontfamily{#3}\fontseries{#4}\fontshape{#5}%
  \selectfont}%
\fi\endgroup%
\begin{picture}(155,227)(878,-583)
\put(935,-440){\makebox(0,0)[lb]{\smash{\SetFigFont{8}{9.6}{\rmdefault}{\mddefault}{\updefault}{\color[rgb]{0,0,0}$\bullet$}%
}}}
\end{picture}
 \end{array}} 
\end{eqnarray*}
\label{D-ID:VertexWineGR}.

\end{D-ID}

\begin{D-ID}
Consider the diagram shown on the \lhs\ of figure~\ref{fig:D-ID-GR-W-dGR-A}.
We can redraw this as shown on the \rhs\ since,
if the newly drawn $\GRk$ hits the external field, then the diagram
vanishes.
\begin{center}
\begin{figure}[h]
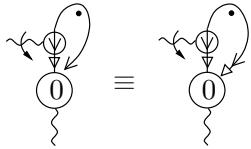

	\[
	\ensuremath{\begin{array}{c}\input{pstex/Beta1-2.16-A.pstex_t} \end{array}} \equiv \ensuremath{\begin{array}{c}\input{pstex/Beta1-2.16-B.pstex_t} \end{array}}
	\]
\caption{The first step in redrawing a diagram to yield
a diagrammatic identity.}
\label{fig:D-ID-GR-W-dGR-A}
\end{figure}
\end{center}

We now allow the original $\GRk$ to act. It must trap
the remaining $\GRk$. Taking the \CC\ of the resultant
diagram, we reflect it but pick up no net sign, since we
have both one processed gauge remainder and one momentum
derivative. Having done this, we then reverse the direction
of the arrow on the derivative, which does yield a minus sign.
This gives us the diagrammatic identity, shown in figure~\ref{fig:D-ID-GR-W-dGR}.
\begin{center}
\begin{figure}[h]
	\[
	\ensuremath{\begin{array}{c}\input{pstex/Beta1-2.16-A.pstex_t} \end{array}} \equiv \ensuremath{\begin{array}{c}\input{pstex/Beta1-2.16-C.pstex_t} \end{array}}
	\]
\caption{A diagrammatic identity.}
\label{fig:D-ID-GR-W-dGR-pre}
\end{figure}
\end{center}

We can, however, generalise this diagrammatic identity by stripping
off the two-point, tree level vertices. We are justified in doing this:
the integral over loop momentum will, by Lorentz invariance, yield
a Kronecker $\delta$ for both diagrams of figure~\ref{fig:D-ID-GR-W-dGR}.
In other words, we need not worry that stripping off the two-point,
tree level vertices is valid only up to some term which kills the vertex.
\begin{center}
\begin{figure}[h]
	\[
	\ensuremath{\begin{array}{c}\input{pstex/Beta1-2.16-A-NoV.pstex_t} \end{array}} \equiv \ensuremath{\begin{array}{c}\input{pstex/Beta1-2.16-C-NoV.pstex_t} \end{array}}
	\]
\caption{Diagrammatic identity~\ref{D-ID-GR-W-dGR}.}
\label{fig:D-ID-GR-W-dGR}
\end{figure}
\end{center}

Note that this identity holds if we replace the wine by an effective propagator.

\label{D-ID-GR-W-dGR}
\end{D-ID}

\chapter{One Loop Diagrammatics} \label{ch:beta1}

In this chapter, we present the entire diagrammatics for the computation of $\beta_1$,
arriving at a manifestly gauge invariant, diagrammatic expression, from which the universal value in $D=4$ can be 
immediately extracted. 
This computation of $\beta_1$ not only serves as an illustration of the diagrammatic techniques 
of chapters~\ref{ch:NewFlowEquation} and~\ref{ch:FurtherDiagrammatics}, but is a necessary
intermediate step in the computation of $\beta_2$.
Much of the work presented in this chapter overlaps with the computation
of $\beta_1$ presented in~\cite{YM-1-loop}. However, there are a number of important differences, which we now outline.

First, the computation here is done for an unrestricted Wilsonian effective action. Previously, the action was restricted to
just single supertrace terms; a consequence of which is that the single supertrace terms $S^{\ AAC}_{0 \mu \nu}(p,q,r)$ 
and $S^{\ AAC\sigma}_{0 \mu \nu}(p,q,r)$
can be set to zero (\emph{ibid}). The second major difference is that the diagrammatics are no longer terminated after the use of the
effective propagator relation. Gauge remainders and $\Op{2}$ manipulations are dealt with in an entirely diagrammatic fashion.
Moreover, the diagrammatics are utilised even at the point where all quantities are
`algebraic' (see section~\ref{sec:NFE:Universality}).

We also choose to use a completely general $\hat{S}$, thereby demonstrating complete scheme independence. In fact, the inclusion
of $\hat{S}_1$ (higher loop vertices do not occur in the calculation) actually leads only to a trivial extension of the scheme
independence. The instances of $\hat{S}$ beyond tree level are restricted to those of the form $\hat{S}_1^C$
(at $\Op{2}$), and are only ever involved
in cancellations via the constraint equation~(\ref{eq:NFE:C-Constraint:Weak}).\footnote{
In the one loop calculation we choose to set ${S}_1^C=0$ at the beginning of the calculation,
rather than at the end. This means that we never compute the flow of this vertex in
the current calculation and so must resort to the previously obtained constraints.}
Nonetheless, it is instructive to see this occurring and to confirm that
$\beta_1$ is universal.

\section{A Diagrammatic Expression for $\beta_1$}

\subsection{The Starting Point}

The key to extracting $\beta$-function coefficients from the weak coupling flow equations~(\ref{eq:WeakCouplingExpansion-NewFlow-B}) is to
use the renormalisation condition~(\ref{eq:Intro:RenormCondition}), which places a constraint on  the vertex
$S^{1\, 1}_{\mu \nu}(p)$. From equations~(\ref{eq:WeakCouplingExpansion-Action}) and~(\ref{eq:app:TLTP-11}), this constraint is saturated at tree level:
\[
S^{1\, 1}_{\mu \nu}(p) = \frac{2}{g^2}\Box_{\mu \nu}(p) + \Op{4} = \frac{1}{g^2} \NCO{0}\Box_{\mu \nu}(p) + \Op{4} =  \frac{1}{g^2} S^{\ 1\, 1}_{0 \mu \nu}(p),
\]
where we have used equation~(\ref{eq:NFE:A(0)}).
Hence, all higher loop two-point vertices, $S_{n\geq 1 \mu \nu}^{\ \ \ \ 1\, 1}(p)$ vanish at $\Op{2}$.

To utilise this information, we
begin by specialising equation~(\ref{eq:WeakCouplingExpansion-NewFlow-B}) to compute the flow of $S_{1 \mu \nu}^{\ 1 \, 1}(p)$:
\begin{equation}
	\flow{S_{1 \mu \nu}^{\ 1 \, 1}(p)} = 2\beta_1 S_{0 \mu \nu}^{\ 1 \, 1}(p) - \gamma_1 \pder{S_{0 \mu \nu}^{\ 1 \, 1}(p)}{\alpha}
	- \sum_{r=0}^1 a_0[\bar{S}_{1-r},\bar{S}_r]^{1\, 1}_{\mu \nu}(p) +  a_1[\Sigma_0]^{1\, 1}_{\mu \nu}(p).
\label{eq:Beta1-Defining-pre}
\end{equation}

The $a_0$ term can be simplified. Defining $\Pi_{RS}^{XY}(k) = S_{RS}^{XY}(k) - \hat{S}_{RS}^{XY}(k)$
and using the definition of the barred vertices, (\ref{eq:NFE:BarNotation}), we can write
\[
	- \sum_{r=0}^1 a_0[\bar{S}_{1-r},\bar{S}_r]^{1\, 1}_{\mu \nu}(p) = -2a_0[\Pi_0, S_1]^{1\, 1}_{\mu \nu}(p) + 2a_0[S_0, \hat{S}_1]^{1\, 1}_{\mu \nu}(p).
\]

All the $a_0$ terms generate two vertices, joined by a wine. Unless one of these vertices is decorated by a single field, both
vertices must be decorated by an internal field and one of the external fields. Now, one-point $\Pi_0$ vertices do
not exist and two-point $\Pi_0$ vertices vanish, since we have identified the two-point, tree level Wilsonian effective
action vertices with the corresponding seed action vertices. We choose to discard one-point $S_1^C$ vertices at this
stage of the calculation and so $a_0[\Pi_0, S_1]^{1\, 1}_{\mu \nu}(p)$ does not contribute.

The next step is to focus on the $\Op{2}$ part of equation~(\ref{eq:Beta1-Defining-pre}). Noting that $S_{1 \mu \nu}^{\ 1 \, 1}(p)$
is at least $\Op{4}$, and that $\NCO{0}$ is independent of $\alpha$,
we arrive at
an algebraic expression for $\beta_1$:
\begin{equation}
-4 \beta_1 \Box_{\mu \nu}(p) + \Op{4} = a_1[\Sigma_0]^{1\, 1}_{\mu \nu}(p) + 2a_0[S_0,\hat{S}_1]^{1\, 1}_{\mu \nu}(p),
\label{eq:Beta1-Defining}
\end{equation}
which is shown diagrammatically in figure~\ref{fig:Beta1-LevelZero}. It is implicit in all that follows that, unless otherwise stated,
 the external indices are $\mu$ and $\nu$ and
we are working at $\Op{2}$.  
\begin{center}
\begin{figure}[h]
	\[
	\begin{array}{c}
		-4 \beta_1 \Box_{\mu \nu}(p) + \Op{4} = 
	\\[2ex]
		\ds
		\frac{1}{2}
		\left[
			\begin{array}{c}
				\begin{array}{ccccc}
					\POCO{d1:1.2(W)}{fig:Beta1:FirstManip}{d1:1.2(S)}{d1:1.5}	&	& \COCO{d1:1.3(W)}{d1:1.22}{d1:1.3(S)}{d1:1.8}	&	& \COBl{d1:1.4}{d1:1.12}
				\\[1ex]
					\ensuremath{\begin{array}{c}\input{pstex/d1-1.2.pstex_t} \end{array}}													& +2& \ensuremath{\begin{array}{c}\input{pstex/d1-1.3.pstex_t} \end{array}}									&-	& \ensuremath{\begin{array}{c}\input{pstex/d1-1.4.pstex_t} \end{array}}
				\end{array}
			\\
				\begin{array}{cccc}
						& \DCO{d1:C-1}{d1:1.16}{d1:1.16b}	&	&\OLdiscard{d1:C-2}
				\\[1ex]
					+2	& \ensuremath{\begin{array}{c}\input{pstex/d1-C-1.pstex_t} \end{array}}						&+4	& \ensuremath{\begin{array}{c}\input{pstex/d1-C-2.pstex_t} \end{array}}
				\end{array}
			\end{array}
		\right]	
	\end{array}
	\]
\caption[A diagrammatic representation of the equation for $\beta_1$.]{
A diagrammatic representation of the equation for $\beta_1$. On the \rhs, we implicitly take the indices to be $\mu$ and $\nu$ and
work at $\Op{2}$.}
\label{fig:Beta1-LevelZero}
\end{figure}
\end{center}

The diagrams are labelled in boldface as follows: any diagram containing $\Sigma$ as a vertex
argument has two labels: the first corresponds to the Wilsonian effective action part and
the second corresponds to the seed action part. Diagrams with only a Wilsonian
effective action vertex or a seed action vertex have a single label.
If the reference number of a diagram
is followed by an arrow, it can mean one of two things:
\begin{enumerate}
	\item 	$\rightarrow 0$ denotes that the corresponding diagram can be set to zero, for some reason;

	\item 	$\rightarrow$ followed by a number (other than zero)
			indicates the number of the figure in which the corresponding
			diagram is processed.
\end{enumerate}
If a diagram is cancelled, then its reference number is enclosed in curly braces,
together with the reference number of the diagram against which it cancels.
A diagram will not be taken as cancelled until the diagram against
which it cancels has been explicitly generated. Thus, at various stages
of the calculation where we collate surviving terms, we include those diagrams
whose cancelling partner does not yet exist.

Returning to figure~\ref{fig:Beta1-LevelZero}, 
the first three diagrams  are formed by the $a_1[\Sigma_0]$ term and the last two are formed by the 
$a_0[S_0,\hat{S}_1]$ term.  
We do not draw any diagrams possessing either a one-point, tree level vertex
or a wine which bites its own tail (see sections~\ref{sec:Intro-GI-ERG-Reg}, \ref{sec:Intro:Diagramatics} and~\ref{sec:Conclusion:Discussion}).
In the third diagram, we have used the equality between Wilsonian effective action and seed action two-point, tree
level vertices
to replace $\Sigma_{0 RS}^{\ XX}(k)$ with $-S_{0 RS}^{\ XX}(k)$. The final diagram vanishes: the one-point
vertex must be in the $C$-sector but, since an $\dot{\Delta}^{AC,A}$ wine does not exist, it is not possible to form a
legal diagram.

Note that we have not included the diagram which can be obtained from diagram~\ref{d1:C-2}
by taking the field on the wine and placing it on the top-most vertex, since such a term vanishes
at $\Op{2}$: the vertex $S_{0 \mu \alpha}^{\ 1 \, 1}(p)$ is, as we know already, at least $\Op{2}$;
the same too applies to $\hat{S}_{1 \mu \alpha}^{\ 1 \, 1}(p)$, as a consequence of the
Ward identity~(\ref{eq:WID-Unbroken}).\footnote{There is no argument that the $\Op{2}$ 
part of $\hat{S}_{n\geq 1 \mu \nu}^{\ \ \ \ 1\, 1}(p)$ vanishes,
since the renormalisation condition does not apply to the seed action.}

\subsection{Diagrammatic Manipulations}

As it stands, we cannot directly extract a value for $\beta_1$ from equation~(\ref{eq:Beta1-Defining}).
The \rhs\ is phrased in terms of non-universal objects. Whilst one approach would be to choose
a particular scheme in which to compute these objects~\cite{GI-ERG-II} we know from~\cite{YM-1-loop}
that this is unnecessary: owing to the universality of $\beta_1$, all
non-universalities must somehow cancel out. To proceed, we utilise the flow equations.

Our aim is to try and reduce the expression for $\beta_1$ to a set of $\Lambda$-derivative terms---terms where
the entire diagram is hit by $\Lambda \partial_\Lambda|_\alpha$---since, as we will see in chapter~\ref{ch:LambdaDerivatives:Methodology},
such terms either vanish directly or combine to give only universal contributions (in the limit that $D \rightarrow 4$).

The approach we use is to start with the term containing the highest point Wilsonian
effective action vertex.
By focusing on the term with the highest point vertex,
we guarantee that the wine in the diagram is un-decorated. Now,
we know that an un-decorated wine is $-\flowConstAl$ of an effective propagator.
Hence, up to a term in which the entire diagram is hit by $-\flowConstAl$,
we can move the $-\flowConstAl$ from the effective propagator to the vertex.
This step is only useful if the vertex is a Wilsonian effective action vertex,
for now it can be processed, using the flow equations.

From figure~\ref{fig:Beta1-LevelZero} it is clear that the highest point Wilsonian effective
action vertex in our calculation of $\beta_1$
is the four-point, tree level vertex contained in 
diagram~\ref{d1:1.2(W)}. The manipulation of this diagram is shown in figure~\ref{fig:Beta1:FirstManip}. 
For the time being,
we will always take the $\Lambda$-derivative to act before we integrate over loop momenta (this is fully 
discussed in chapter~\ref{ch:LambdaDerivatives:Methodology}).
\begin{center}
\begin{figure}[h]
	\[
	\frac{1}{2} \ensuremath{\begin{array}{c}\input{pstex/Beta1-1.2.pstex_t} \end{array}} 
	= 
	\frac{1}{2}
	\dec{
		\POLD[0.1]{}{Beta1-LdL-A}{d1:1.4b}{fig:beta1}
	}{\bullet}	
	-\frac{1}{2}
	\POLD[0.1]{}{Beta1-1.2c}{d1:1.2(W):Manip}{fig:beta1:LevelOne}
	\]
\caption[The manipulation of diagram~\ref{d1:1.2(W)}.]{
The manipulation of diagram~\ref{d1:1.2(W)}. In the final diagram, the $\Lambda$-derivative operates on just the four-point vertex.}
\label{fig:Beta1:FirstManip}
\end{figure}
\end{center}

We can now use the tree-level flow equation~(\ref{eq:TreeLevelFlow}) to process the $\Lambda$-derivative of the four-point vertex. 
The flow of a four-point vertex with two $A^1$ fields and two wildcards is shown in figure~\ref{fig:App:TLFP-A1A1XX}. Throwing
away all terms which vanish at $\Op{2}$ and joining the wildcards
together with an effective propagator, we arrive at 
figure~\ref{fig:beta1:LevelOne}. 
\begin{center}
\begin{figure}
	\[
	\frac{1}{2}
	\left[
		\begin{array}{cccccccc}
				&\PO{d1:1.5,6}{fig:beta1:LevelOne:Manip}&	&\PO{d1:1.8,9}{fig:beta1:LevelOne:Manip}&	&\PO{d1:1.12:pre}{fig:beta1:LevelOne:Manip}	
				&	&\PO{d1:1.15}{fig:beta1:LevelTwo-B}
		\\[1ex]
			2	& \ensuremath{\begin{array}{c}\input{pstex/d1-1.5,6.pstex_t} \end{array}}							&+4	&\ensuremath{\begin{array}{c}\input{pstex/d1-1.8,9.pstex_t} \end{array}}							&+	&\ensuremath{\begin{array}{c}\input{pstex/d1-1.12-pre.pstex_t} \end{array}}							
				&-2	&\ensuremath{\begin{array}{c}\input{pstex/d1-1.15.pstex_t} \end{array}}
		\\[1ex]
				&\CO{d1:1.15b}{d1:1.19b}&	&\PO{d1:1.10:pre}{fig:beta1:LevelOne:Manip}	&	&\PO{d1:1.13:pre}{fig:beta1:LevelOne:Manip}	&	&\CO{d1:1.11}{d1:1.24}
		\\[1ex]
			+4	&\ensuremath{\begin{array}{c}\input{pstex/d1-1.15b.pstex_t} \end{array}}			&+2	&\ensuremath{\begin{array}{c}\input{pstex/d1-1.10-pre.pstex_t} \end{array}}							&+4	&\ensuremath{\begin{array}{c}\input{pstex/d1-1.13-pre.pstex_t} \end{array}}							&+4	&\ensuremath{\begin{array}{c}\input{pstex/d1-1.11.pstex_t} \end{array}}
		\\[1ex]
				&\PO{d1:1.7}{fig:beta1:c0}	&	&\PO{d1:1.14}{fig:beta1:LevelTwo-A}	&	&\CO{d1:1.14b}{d1:1.15d}&	&\CO{d1:1.14c}{d1:1.15e}
		\\[1ex]
			+2	&\ensuremath{\begin{array}{c}\input{pstex/d1-1.7.pstex_t} \end{array}}				&-	&\ensuremath{\begin{array}{c}\input{pstex/d1-1.14.pstex_t} \end{array}}						&+	&\ensuremath{\begin{array}{c}\input{pstex/d1-1.14b.pstex_t} \end{array}}			&+	&\ensuremath{\begin{array}{c}\input{pstex/d1-1.14c.pstex_t} \end{array}}
		\end{array}
	\right]
	\]
\caption{The re-expression of diagram~\ref{d1:1.2(W):Manip}, using the tree-level flow equation.}
\label{fig:beta1:LevelOne}
\end{figure}
\end{center}

The diagrams have been arranged in a very specific way. 
The first eight diagrams  have all had the loop of the parent diagram split open and joined
back to the rest of the diagram with some other structure(s). The very first diagram contains the highest-point
vertex of this subset. Moving along the first five diagrams, we take the basic structure of the first diagram
and distribute the external fields in all possible ways, being careful to
insert the correct vertex arguments. Having exhausted all possibilities, we move to the
next three diagrams, where we take the first five diagrams  but now demand that one end of the effective propagator
attaches to the wine, rather than the bottom vertex (we discard any terms possessing one-point, tree level
vertices). 
Finally, the bottom row contains all diagrams (which survive at $\Op{2}$) in which
the loop of the parent diagram has not been split open.

Diagrams~\ref{d1:1.5,6}--\ref{d1:1.12:pre}, \ref{d1:1.10:pre} and~\ref{d1:1.13:pre}
 can be 
further manipulated using the effective propagator relation, 
since they contain a two-point, tree level vertex contracted into an effective propagator. The results of this procedure are shown in 
figure~\ref{fig:beta1:LevelOne:Manip}. Diagrams in which a wine bites its own tail have been discarded as have those in which 
a gauge remainder strikes a two-point vertex. Henceforth, we will assume that such terms have always
been discarded.
\begin{center}
\begin{figure}[h]
	\[
	\frac{1}{2}
	\left[
		\begin{array}{c}
			\begin{array}{cccccccc}
					&\CO{d1:1.5}{d1:1.2(S)}	&	&\PO{d1:1.6}{fig:beta1:GR1}	&	&\CO{d1:1.8}{d1:1.3(S)}	&	& \CO{d1:1.9}{d1:1.23}
			\\[1ex]
				2	& \ensuremath{\begin{array}{c}\input{pstex/d1-1.5.pstex_t} \end{array}}			&-2	&\ensuremath{\begin{array}{c}\input{pstex/d1-1.6.pstex_t} \end{array}}				&+4	&\ensuremath{\begin{array}{c}\input{pstex/d1-1.8.pstex_t} \end{array}}			&-4	&\ensuremath{\begin{array}{c}\input{pstex/d1-1.9.pstex_t} \end{array}}
			\end{array}
		\\[1ex]
			\begin{array}{cccccc}
					&\CO{d1:1.12}{d1:1.4}	&	&\PO{d1:1.10}{fig:beta1:GR1}&	&\PO{d1:1.13}{fig:beta1:GR1}
			\\[1ex]
				+	&\ensuremath{\begin{array}{c}\input{pstex/d1-1.4.pstex_t} \end{array}}			&-2	&\ensuremath{\begin{array}{c}\input{pstex/d1-1.10.pstex_t} \end{array}}				&-4	&\ensuremath{\begin{array}{c}\input{pstex/d1-1.13.pstex_t} \end{array}}
			\end{array}
		\end{array}
	\right]	
	\]
\caption{Manipulation of diagrams~\ref{d1:1.5,6}--\ref{d1:1.12:pre}, 
\ref{d1:1.10:pre} and~\ref{d1:1.13:pre}.}
\label{fig:beta1:LevelOne:Manip}
\end{figure}
\end{center}

Three of the diagrams generated exactly cancel
the contributions in the first row of figure~\ref{fig:Beta1-LevelZero} 
containing seed action vertices (or Wilsonian effective action two-point, tree level vertices).

\Cancel{d1:1.5}{d1:1.2(S)}
\Cancel{d1:1.8}{d1:1.3(S)}
\Cancel{d1:1.12}{d1:1.4}

Other than the $\Lambda$-derivative term, diagram~\ref{d1:1.4b}, there are now only two diagrams left which contain four-point vertices.
The first of these, diagram~\ref{d1:1.7}, can be manipulated at $\Op{2}$ since the bottom vertex is at least $\Op{2}$ and
the rest of the diagram
can be Taylor expanded in $p$. Taking the zeroth order
contribution from the differentiated effective
propagator yields, by equation~(\ref{eq:app:NFE-DEP-11}) (\cf\ equation~(\ref{eq:def-B})), 
the non-universal contribution \BPZ. If we were to take
the zeroth order contribution from the four-point vertex\footnote{
As we will see in section~\ref{sec:beta1:A'(0)} this manipulation
is not necessary as there is a more elegant way to deal with the diagram.},
 it would reduce to the (double) momentum derivative of a two-point vertex.
In the second of the diagrams containing a four-point vertex, diagram~\ref{d1:1.6}, the vertex is hit by a gauge remainder and so will automatically be reduced
to a three-point vertex. Thus the effect of our manipulations is to ensure that all occurrences of the highest-point vertex in 
the calculation occur only in a $\Lambda$-derivative term.

Before moving on to the next stage of the calculation, we compare our current expression to that of reference~\cite{YM-1-loop}. Ignoring the
multiple supertrace terms contained within each of our diagrams, the two expressions are 
superficially
the same, up to diagrams~\ref{d1:1.10:pre} and~\ref{d1:1.14}--\ref{d1:1.14c}.
In each of these terms, the internal field joining the two three-point vertices must be in the $C$-sector. If it is in the $F$-sector, then
each diagram vanishes because net fermionic vertices vanish. If it is in the $A$-sector, then charge conjugation invariance causes the diagrams to 
vanish when we sum over permutations of the bottom vertex (see section~\ref{sec:Diagrammatics:CC}). Looking a little
harder, we see a related difference between the current expression and that of reference~\cite{YM-1-loop}:
amongst the components of diagrams~\ref{d1:1.15} and~\ref{d1:1.15b} are diagrams possessing $AAC$ vertices.

Now, with the aim of removing all three-point vertices from the calculation (up to $\Lambda$-derivative terms), we iterate the procedure.
Referring to figure~\ref{fig:beta1:LevelOne}, only diagrams~\ref{d1:1.14} and~\ref{d1:1.15} possess exclusively Wilsonian effective
action vertices and an un-decorated wine
and so
it is these which we manipulate. 

Figure~\ref{fig:beta1:LevelTwo-A} shows the manipulation of diagram~\ref{d1:1.14} which proceeds
along exactly the same lines as the manipulations of figure~\ref{fig:Beta1:FirstManip}. This time, however, we utilise figure~\ref{fig:App:TLThP-WildCardx3}
for the flow of a three-point vertex with three wildcard fields and figure~\ref{fig:App-TLThP-A1A1X} for the flow of
a three-point vertex containing two external $A^1$s (which carry moment $p$ and $-p$). In this latter case, we discard all terms which
vanish at $\Op{2}$.
\begin{center}
\begin{figure}
	\[
	\begin{array}{c}
	\vspace{2ex}
		\ds 
		-\frac{1}{2}
		\dec{
			\POLD[0.1]{}{Beta1-LdL-D}{d1:1.15c}{fig:beta1}
		}{\bullet}
		-\frac{1}{2}
		\left[
			\begin{array}{ccc}
				\CO{d1:1.15d}{d1:1.14b}	&	& \CO{d1:1.15e}{d1:1.14c}
			\\[1ex]
				\ensuremath{\begin{array}{c}\input{pstex/d1-1.15d.pstex_t} \end{array}}			& +	& \ensuremath{\begin{array}{c}\input{pstex/d1-1.15e.pstex_t} \end{array}}
			\end{array}
		\right]
	\\
		\ds
		+\frac{1}{2}
		\left[
			\begin{array}{ccccccc}
				\COB{d1:1.16}{d1:C-1}	&	&\COC{d1:1.16b}{d1:C-1}	&	&\PO{d1:1.17}{fig:beta1:GR1}&	&\PO{d1:1.18}{fig:beta1:GR1}
			\\[1ex]
				\ensuremath{\begin{array}{c}\input{pstex/d1-1.16.pstex_t} \end{array}}			&-	&\ensuremath{\begin{array}{c}\input{pstex/d1-1.16b.pstex_t} \end{array}}			&+2	&\ensuremath{\begin{array}{c}\input{pstex/d1-1.17.pstex_t} \end{array}}				&+2	&\ensuremath{\begin{array}{c}\input{pstex/d1-1.18.pstex_t} \end{array}}
			\end{array}
		\right]
	\end{array}
	\]
\caption{Manipulation of diagram~\ref{d1:1.14} using the tree level flow equations.}
\label{fig:beta1:LevelTwo-A}
\end{figure}
\end{center}

As with diagrams~\ref{d1:1.2(S)}, \ref{d1:1.3(S)} and~\ref{d1:1.4},
we find that diagrams of the same structure as the parent but containing a seed action
vertex are cancelled. 
\Cancel{d1:1.15d}{d1:1.14b}
\Cancel{d1:1.15e}{d1:1.14c}

We also find, as promised that the sole instance of a one-point, seed action vertex
is cancelled.
\begin{cancel}
Diagrams~\ref{d1:1.16}  and~\ref{d1:1.16b} exactly cancel diagram~\ref{d1:C-1}
by virtue of the constraint equation~(\ref{eq:NFE:C-Constraint:Weak}).
\label{cancel:d1:1.16}
\end{cancel}

In this context, the notation used in figure~\ref{fig:beta1:LevelTwo-A} to 
describe the cancellation of diagram~\ref{d1:C-1} has an obvious
interpretation. Note that 
the only surviving terms from figure~\ref{fig:beta1:LevelTwo-A} both contain gauge remainders.

Figure~\ref{fig:beta1:LevelTwo-B} shows the manipulation of diagram~\ref{d1:1.15} (which first
appears in figure~\ref{fig:beta1:LevelOne}), where the overall factor of $1/2$ arises from
the symmetry of the $\Lambda$-derivative term~\ref{d1:1.19} under rotations by $\pi$ (alternatively, the indistinguishability of the
two internal fields). When we compute the flow of the vertices of diagram~\ref{d1:1.19}, we can 
utilise this symmetry to remove the
factor of $1/2$.

\begin{center}
\begin{figure}[h]
	\[
	\begin{array}{c}
	\vspace{2ex}
		\ds
		-\frac{1}{2}
		\dec{
			\POLD[0.1]{}{Beta1-LdL-B}{d1:1.19}{fig:beta1}
		}{\bullet} +
	\\
		\left[
			\begin{array}{cccccccc}
					&\CO{d1:1.19b}{d1:1.15b}&	&\PO{d1:1.20}{fig:beta1:GR1}&	&\PO{d1:1.21}{fig:beta1:c0}	&	&\CO{d1:1.22}{d1:1.3(W)}
			\\[1ex]
				-2	&\ensuremath{\begin{array}{c}\input{pstex/d1-1.15b.pstex_t} \end{array}}			&+2	&\ensuremath{\begin{array}{c}\input{pstex/d1-1.20.pstex_t} \end{array}}				&-	&\ensuremath{\begin{array}{c}\input{pstex/d1-1.21.pstex_t} \end{array}}				&-	&\ensuremath{\begin{array}{c}\input{pstex/d1-1.22.pstex_t} \end{array}}
			\\[1ex]
					&\CO{d1:1.23}{d1:1.9}	&	&\PO{d1:1.24b}{fig:beta1:GR1}	&	&\CO{d1:1.24}{d1:1.11}	&	&\PO{d1:1.25}{fig:beta1:GR1}
			\\[1ex]
				+2	&\ensuremath{\begin{array}{c}\input{pstex/d1-1.9.pstex_t} \end{array}}			&-	&\ensuremath{\begin{array}{c}\input{pstex/d1-1.24b.pstex_t} \end{array}}					&-2	&\ensuremath{\begin{array}{c}\input{pstex/d1-1.11.pstex_t} \end{array}}			&+2	&\ensuremath{\begin{array}{c}\input{pstex/d1-1.25.pstex_t} \end{array}}
			\end{array}
		\right]
	\end{array}
	\]
\caption{Manipulation of diagram~\ref{d1:1.15} using the tree level flow equation.}
\label{fig:beta1:LevelTwo-B}
\end{figure}
\end{center}

We find a number of cancellations. The first of these is the expected cancellation of
the partner of the parent diagram, possessing a seed action vertex.
\Cancel{d1:1.19b}{d1:1.15b}

The next cancellation completes the removal of all terms from figure~\ref{fig:Beta1-LevelZero}
formed by the action of $a_0[\Sigma_0]$.

\Cancel{d1:1.22}{d1:1.3(W)}

Of the remaining cancellations, one involves two 
diagrams, each possessing active gauge remainders,
which we notice can be cancelled without the need to perform the
gauge remainders. The final cancellation occurs only at $\Op{2}$.

\CancelCom{d1:1.23}{d1:1.9}{ since when a three-point tree level vertex is struck by 
a gauge remainder, it makes no difference whether it is a Wilsonian effective
action vertex or a seed action vertex.}

\CancelOpCom{d1:1.24}{d1:1.11}{. In each case, we Taylor expand the three-point, tree level vertex
to zeroth order in external momentum, reducing it to the momentum derivative of
a two-point, tree level vertex.
It is then of no consequence that one of the three-point, tree level vertices was a 
seed action vertex whereas the other was a Wilsonian effective action vertex,
since the two-point, tree level Wilsonian effective action vertices are identified
with the corresponding seed action vertices.}

At this stage, up to diagrams in which
the sole three-point vertex is hit by a gauge remainder, 
we have removed all three-point, tree level vertices
from the calculation with the following exceptions:
\begin{enumerate}
	\item 	the $\Lambda$-derivative terms, diagrams~\ref{d1:1.19} and~\ref{d1:1.15c};

	\item 	the \BPZ\ term, diagram \ref{d1:1.21};

	\item 	three diagrams~\ref{d1:1.10}, \ref{d1:1.17} and~\ref{d1:1.18}, 
			which are each left with a three-point, tree level vertex,
			even after the action of the gauge remainders.
\end{enumerate}

The last three terms, which  all possess an
$S_{0 \mu \nu}^{\ 1 \, 1 C^{1,2}}$ vertex, have no analogue in
the version of the calculation presented in~\cite{YM-1-loop}.
To make further progress, we must process the gauge remainders.
In figure~\ref{fig:beta1:GR1}, we utilise the techniques of section~\ref{sec:GRs} to 
manipulate the gauge remainders, stopping after the use of the
effective propagator relation. We discard all terms which vanish due to their supertrace structure
being $\str A^1_\mu \, \str A^1_\nu$ and neglect attachment corrections to
three-point vertices which are decorated by an external field and struck by a gauge remainder.
\begin{center}
\begin{figure}
	\[
	\begin{array}{l}
	\vspace{0.2in}
		\ref{d1:1.6} = 2 
		\left[
			\begin{array}{cccc}
					&\CO{d1:2.1}{d1:2.10}	&	&\CO{d1:2.2}{d1:2.4(S)}
			\\[1ex]
				2	&\ensuremath{\begin{array}{c}\input{pstex/d1-2.1.pstex_t} \end{array}}			&-	&\ensuremath{\begin{array}{c}\input{pstex/d1-2.2.pstex_t} \end{array}}
			\end{array}
		\right]
	\\
	\vspace{0.2in}
		\ref{d1:1.10} + \ref{d1:1.18} = 2 
		\left[
			\begin{array}{ccc}
				\COBl{d1:2.3}{d1:2.15b}	&	&\COCO{d1:2.4(W)}{d1:2.8}{d1:2.4(S)}{d1:2.2}
			\\[1ex]
				\ensuremath{\begin{array}{c}\input{pstex/d1-2.3.pstex_t} \end{array}}				&-	&\ensuremath{\begin{array}{c}\input{pstex/d1-2.4.pstex_t} \end{array}}
			\end{array}
		\right]
	\\
	\vspace{0.2in}
		 \ref{d1:1.17} = 2
		\left[
			\begin{array}{ccc}
				\CO{d1:2.8}{d1:2.4(W)}	&	&\PO{d1:2.9}{fig:beta1:LdLConversion}
			\\[1ex]
				\ensuremath{\begin{array}{c}\input{pstex/d1-2.8.pstex_t} \end{array}}				&-	&\ensuremath{\begin{array}{c}\input{pstex/d1-2.9.pstex_t} \end{array}}
			\end{array}
		\right]
	\\
	\vspace{0.2in}
		\ref{d1:1.20} = -4 
		\left[
			\begin{array}{ccccc}
				\CO{d1:2.10}{d1:2.1}&	& \PO{d1:2.11}{fig:beta1:GR2}	&	&\PO{d1:2.12}{fig:beta1:Op2}
			\\[1ex]
				\ensuremath{\begin{array}{c}\input{pstex/d1-2.1.pstex_t} \end{array}}			& -	& \ensuremath{\begin{array}{c}\input{pstex/d1-2.11.pstex_t} \end{array}}					&-	&\ensuremath{\begin{array}{c}\input{pstex/d1-2.12.pstex_t} \end{array}}
			\end{array}
		\right]
	\\
	\vspace{0.2in}
		\ref{d1:1.13} = 4 
		\left[
			\begin{array}{ccccc}
				\PO{d1:2.7-pre}{fig:beta1:Op2}	&	&\CO{d1:2.5}{d1:2.13}	&	& \OLdiscard{d1:2.6}
			\\[1ex]
				\ensuremath{\begin{array}{c}\input{pstex/d1-2.7-pre.pstex_t} \end{array}}					& +	& \ensuremath{\begin{array}{c}\input{pstex/d1-2.5.pstex_t} \end{array}}			& - & \ensuremath{\begin{array}{c}\input{pstex/d1-2.6.pstex_t} \end{array}}
			\end{array}
		\right]
	\\
		\ref{d1:1.24b} = 2 
		\left[
			\POLD[0.1]{}{Beta1-GR-G}{d1:2.16:pre}{fig:beta1:Op2}
		\right]
	\hspace{3em}
		\ref{d1:1.25} = -4  
		\left[
			\begin{array}{ccc}
				\CO{d1:2.13}{d1:2.5}	&	&\PO{d1:2.14}{fig:beta1:GR2}
			\\[1ex]
				\ensuremath{\begin{array}{c}\input{pstex/d1-2.5.pstex_t} \end{array}}			& -	& \ensuremath{\begin{array}{c}\input{pstex/d1-2.14.pstex_t} \end{array}}
			\end{array}
		\right] 
	\end{array}
\]
\caption{Terms arising from processing the gauge remainders
of the diagrams in figures~\ref{fig:beta1:LevelTwo-A} and~\ref{fig:beta1:LevelTwo-B}.}
\label{fig:beta1:GR1}
\end{figure}
\end{center}

 Note that diagram~\ref{d1:2.16:pre}
is our first example of a diagram possessing a trapped gauge remainder.
There is no corresponding diagram in which the gauge remainder
bites the external field because then we are left with an \emph{active}
gauge remainder striking a two-point, tree level vertex.
Note also that all diagrams which either cannot be processed further
or do not contain an $S_{0 \mu \nu}^{\ 1 \, 1 C^{1,2}}$ vertex cancel,
amongst themselves.
\Cancel{d1:2.4(W)}{d1:2.8}
\Cancel{d1:2.4(S)}{d1:2.2}
\Cancel{d1:2.10}{d1:2.1}
\Cancel{d1:2.5}{d1:2.13}

Whilst these cancellations are very encouraging, it is not
clear that we are any closer to solving the mystery of
the diagrams containing $S_{0 \mu \nu}^{\ 1 \, 1 C^{1,2}}$ vertices.
We will, however, persevere and 
process the nested gauge remainders arising from the
previous procedure. 
The result of this is shown in figure~\ref{fig:beta1:GR2}.

\begin{center}
\begin{figure}[h]
	\[
	\begin{array}{c}
		\ref{d1:2.11} = 4
		\left[
			\begin{array}{ccc}
				\CO{d1:2.17}{d1:2.20}	&	&\PO{d1:2.18}{fig:beta1:LdLConversion}
			\\[1ex]	
				\ensuremath{\begin{array}{c}\input{pstex/d1-2.17.pstex_t} \end{array}}			& -	&\ensuremath{\begin{array}{c}\input{pstex/d1-2.18.pstex_t} \end{array}}
			\end{array}
		\right] 
	\\[10ex]
		\ref{d1:2.14} = 4
		\left[
			\begin{array}{ccc}
				\PO{d1:2.19}{fig:beta1:c0}	&	&\CO{d1:2.20}{d1:2.17}
			\\[1ex]	
				\ensuremath{\begin{array}{c}\input{pstex/d1-2.19.pstex_t} \end{array}}				& -	&\ensuremath{\begin{array}{c}\input{pstex/d1-2.17.pstex_t} \end{array}}
			\end{array}
		\right]
	\end{array}
\]
\caption{Diagrams arising from processing the nested gauge remainders of figure~\ref{fig:beta1:GR1}.}
\label{fig:beta1:GR2}
\end{figure}
\end{center}

Once again, we find a cancellation between a pair of the terms
generated by this procedure.

\Cancel{d1:2.20}{d1:2.17}

This exhausts the active gauge remainders and so is a good point to pause and
collate the surviving terms. These fall into five sets:
\begin{enumerate}
	\item 	The $\Lambda$-derivative terms, diagrams~\ref{d1:1.4b}, \ref{d1:1.15c} and~\ref{d1:1.19};
	
	\item 	The \BPZ\ terms, diagrams~\ref{d1:1.7}, \ref{d1:1.21} and~\ref{d1:2.19}. Notice
			that the last of these has been formed via the action of a nested gauge
			remainder;
	
	\item	Terms possessing an $\Op{2}$ stub formed by the action of a gauge remainder,
			diagrams~\ref{d1:2.16:pre}, \ref{d1:2.7-pre} and~\ref{d1:2.12}. Notice that
			the former of these has a trapped gauge remainder;

	\item	Terms possessing a $S_{0 \mu \nu}^{\ 1 \, 1 C^{1,2}}$ vertex,
			diagrams~\ref{d1:2.9} and~\ref{d1:2.3};

	\item	Diagram~\ref{d1:2.18}.
\end{enumerate}

We will leave the first three sets of diagrams, for the time being,
and focus on the final two. Remarkably, diagram~\ref{d1:2.9} from the fourth
set and diagram~\ref{d1:2.18} share a common feature! Both can
be re-drawn via diagrammatic identity~\ref{D-ID:VertexWineGR}.
Having thus re-drawn these diagrams, we then find that they can be converted
into $\Lambda$-derivative terms. This whole procedure is shown 
in figure~\ref{fig:beta1:LdLConversion}, where we have discarded
any terms which vanish at $\Op{2}$.
\begin{center}
\begin{figure}[h]
	\[
	\begin{array}{lcccl}
		\ref{d1:2.9}	& =	& 2 \ensuremath{\begin{array}{c}\input{pstex/Beta1-2.9.pstex_t} \end{array}} 		& = & 
		2 
		\dec{
			\ensuremath{\begin{array}{c}\input{pstex/Beta1-2.15a.pstex_t} \end{array}}
		}{\bullet} 
		-\ensuremath{\begin{array}{c}\input{pstex/d1-2.15b.pstex_t} \end{array}} - \ensuremath{\begin{array}{c}\input{pstex/d1-2.15c.pstex_t} \end{array}}
	\\[10ex]
						&	&						& =	&
		2 
		\dec{
			\POLD[0.1]{}{Beta1-2.15a}{d1:2.15a}{fig:beta1}
		}{\bullet}
		- 2 \COLD[0.1]{d1-2.3}{d1:2.15b}{d1:2.3} + \Op{4}
	\\[10ex]	
		\ref{d1:2.18}	& =	&4 \ensuremath{\begin{array}{c}\input{pstex/Beta1-2.25pre.pstex_t} \end{array}}	& =	& 
		2 
		\dec{
			\POLD[0.1]{}{Beta1-LdL-C}{d1:2.25}{fig:beta1}
		}{\bullet}
	\end{array}
	\]
\caption{Re-drawing of diagrams~\ref{d1:2.9} and~\ref{d1:2.18}
using diagrammatic identity~\ref{D-ID:VertexWineGR} and their subsequent
conversion into $\Lambda$-derivative terms.}
\label{fig:beta1:LdLConversion}
\end{figure}
\end{center}

Note how diagram~\ref{d1:2.25} has a factor of $1/2$, relative to the parent diagram.
This recognises the indistinguishability of the two processed gauge remainders
possessed by this diagram.

Finally, we find the cancellation of the remaining diagrams possessing
a $S_{0 \mu \nu}^{\ 1 \, 1 C^{1,2}}$ vertex, up to those which are cast
as $\Lambda$-derivative terms.

\Cancel{d1:2.15b}{d1:2.3}

Our next task is to analyse the surviving diagrams possessing an $\Op{2}$ stub formed by
the action of a gauge remainder. This is a two-step process. First, we Taylor
expand each of the sub-diagrams attached to the stub to zeroth order in $p$.\footnote{As
will become clear in chapter~\ref{ch:LambdaDerivatives:Methodology}, this step can potentially
generate IR divergence, in which case it is not legal. At the one-loop level, though, this
never happens.}
We then re-draw them, if possible, using various diagrammatic identities. The results of
the complete procedure are shown in figure~\ref{fig:beta1:Op2}.
\begin{center}
\begin{figure}[h]
	\[
	\begin{array}{lcl}
	\vspace{0.2in}
		\ref{d1:2.7-pre} 	&\rightarrow& 4
		\COLD[0.1]{Beta1-2.24}{d1:2.7}{d1:2.24}
	\\
	\vspace{0.2in}
		\ref{d1:2.12} 		&\rightarrow& 4 \ensuremath{\begin{array}{c}\input{pstex/Beta1-Op2-A.pstex_t} \end{array}}
		= -4 
		\left[
			\begin{array}{ccccc}
				\PO{d1:2.21}{fig:beta1:FinalManip}	&	&\PO{d1:2.16b}{fig:beta1:FinalManip}&	&\CO{d1:2.16}{d1:2.22a}
			\\[1ex]
				\ensuremath{\begin{array}{c}\input{pstex/d1-2.21.pstex_t} \end{array}}						& +	& \ensuremath{\begin{array}{c}\input{pstex/d1-2.16b.pstex_t} \end{array}}						& +	&\ensuremath{\begin{array}{c}\input{pstex/d1-2.16.pstex_t} \end{array}}
			\end{array}
		\right] 
	\\
		\ref{d1:2.16:pre} 	&\rightarrow& -2 \ensuremath{\begin{array}{c}\input{pstex/Beta1-Op2-C.pstex_t} \end{array}}
		= 2
		\left[
			\begin{array}{ccc}
				\PO{d1:2.22b}{fig:beta1:FinalManip}	&	&\CO{d1:2.22a}{d1:2.16}
			\\[1ex]
				\ensuremath{\begin{array}{c}\input{pstex/d1-2.16b.pstex_t} \end{array}}						&+2	& \ensuremath{\begin{array}{c}\input{pstex/d1-2.16.pstex_t} \end{array}}
			\end{array}
		\right]
\end{array}
\]
\caption{Manipulations at $\Op{2}$, followed by a re-expression of the resulting diagrams.}
\label{fig:beta1:Op2}
\end{figure}
\end{center}

Some comments are in order. Having Taylor expanded diagram~\ref{d1:2.12},
we obtain the final set of diagrams by means of diagrammatic identity~\ref{D-ID:OneWine}.
In the case of diagram~\ref{d1:2.16:pre} the procedure after Taylor expansion
is different: we re-express it as a total momentum derivative---which we discard---plus
supplementary terms. These supplementary terms come with a relative minus sign and
correspond to the momentum derivative hitting all structures other than the wine.
We then note that the two contributions in which the momentum derivative strikes
a $\wedge$ can be combined, via diagrammatic identity~\ref{D-ID-GR-W-dGR}.
This gives a cancellation.
\Cancel{d1:2.22a}{d1:2.16}

The four surviving diagrams possessing an $\Op{2}$ stub formed by the
action of a gauge remainder can now be combined into $\Lambda$-derivatives.
Diagram~\ref{d1:2.21} can be re-drawn via diagrammatic 
identity~\ref{D-ID:VertexWineGR} and then, together with diagram~\ref{d1:2.7}
converted into a $\Lambda$-derivative term.
Diagram~\ref{d1:2.22b} precisely halves the overall factor
of diagram \ref{d1:2.16b}. The resultant diagram is then re-drawn
via diagrammatic identity~\ref{D-ID-GR-W-dGR} which, upon
inspection, is actually a $\Lambda$-derivative term (where we exploit
the fact that $\flow$ kills \GRk). This is all
shown in figure~\ref{fig:beta1:FinalManip}. Note that we have taken
the $\Lambda$-derivative to strike entire diagrams rather than
just the sub-diagram attached to the stub. This step is
valid at $\Op{2}$.
\begin{center}
\begin{figure}[h]
	\[
	\begin{array}{l}
	\vspace{0.2in}
		\ref{d1:2.21} = 4 \ensuremath{\begin{array}{c}\input{pstex/Beta1-2.23pre.pstex_t} \end{array}}
		\rightarrow 4
		\dec{
			\POLD[0.1]{}{Beta1-LdL-F}{d1:2.23}{fig:beta1}
		}{\bullet}
		-4 \COLD[0.1]{Beta1-2.24}{d1:2.24}{d1:2.7}
	\\
		\ref{d1:2.16b} + \ref{d1:2.22b} = -2 \ensuremath{\begin{array}{c}\input{pstex/Beta1-2.26pre.pstex_t} \end{array}}
		\rightarrow 2 
		\dec{
			\POLD[0.1]{}{Beta1-LdL-G}{d1:2.26}{fig:beta1}
		}{\bullet}
\end{array}
\]
\caption{The final conversion into $\Lambda$-derivative terms.}
\label{fig:beta1:FinalManip}
\end{figure}
\end{center}

In diagram~\ref{d1:2.26} we have moved the momentum derivative from
the $\GRk$ onto the pseudo effective propagator, discarding a total
momentum derivative, in the process.

\Cancel{d1:2.24}{d1:2.7}

We now find that all terms, other than the $\Lambda$-derivatives, have cancelled, with the sole exception of the \BPZ\ terms---which
have been collected together in figure~\ref{fig:beta1:c0}.
\begin{center}
\begin{figure}[h]
\[
\ensuremath{\begin{array}{c}\input{pstex/./Beta1-c0.pstex_t} \end{array}}
\]
\caption{The set of \BPZ\ terms
(which do not manifestly vanish at $\Op{2}$).}
\label{fig:beta1:c0}
\end{figure}
\end{center}

 In anticipation of the cancellation of the \BPZ\ terms, the
$\Lambda$-derivative terms have been collected together in figure~\ref{fig:beta1} to give, for the first time, 
an entirely diagrammatic expression for $\beta_1$,
in terms of $\Lambda$-derivatives. 
\begin{center}
\begin{figure}[h]
	\[
	4 \beta_1 \Box_{\mu \nu} (p) = 
	-\frac{1}{2}
	\dec{
		\begin{array}{c}
			\begin{array}{ccccccc}
				\LO{Beta1-LdL-A}	&	& \LO{Beta1-LdL-B}	&	& \LO{Beta1-LdL-C} 	&	&\LO{Beta1-LdL-D}
			\\[1ex]
				\ensuremath{\begin{array}{c}\input{pstex/Beta1-LdL-A.pstex_t} \end{array}}	& -	& \ensuremath{\begin{array}{c}\input{pstex/Beta1-LdL-B.pstex_t} \end{array}}	&+4	& \ensuremath{\begin{array}{c}\input{pstex/Beta1-LdL-C.pstex_t} \end{array}}	& -	&\ensuremath{\begin{array}{c}\input{pstex/Beta1-LdL-D.pstex_t} \end{array}}
			\end{array}
		\\[10ex]
			\begin{array}{cccccc}
					& \LO{Beta1-LdL-E}	&	& \LO{Beta1-LdL-F}	&	& \LO{Beta1-LdL-G}
			\\[1ex]
				+4	& \ensuremath{\begin{array}{c}\input{pstex/Beta1-LdL-E.pstex_t} \end{array}}	&+8	& \ensuremath{\begin{array}{c}\input{pstex/Beta1-LdL-F.pstex_t} \end{array}}	&+4	& \ensuremath{\begin{array}{c}\input{pstex/Beta1-LdL-G.pstex_t} \end{array}}
			\end{array}
		\end{array}
	}{\bullet}
	\]
\caption{Diagrammatic, gauge invariant expression for $\beta_1$, phrased entirely in terms of $\Lambda$-derivatives.}
\label{fig:beta1}
\end{figure}
\end{center}

The diagrams in this expression will arise so many times in future that we will name them. 
The complete set of diagrams inside the square brackets will be referred to as $\OLDs$. The
The first
three of these are henceforth referred to as the standard set. The last two will be known as the little set.

\subsection{The \BPZ\ Terms} \label{sec:beta1:A'(0)}

The first thing to notice about the \BPZ\ terms is that they are very similar
to the first three diagrams of figure~\ref{fig:beta1}. Indeed, the \BPZ\ terms
are very nearly just the standard set joined to an $\Op{2}$ stub, via an un-decorated
wine. The only difference is that the standard set contains exclusively Wilsonian
effective action vertices, whereas the \BPZ\ terms do not. However, we know
that the \BPZ\ terms can be manipulated, at $\Op{2}$. Doing this, we can replace the
four-point (three-point) seed action vertex with a double (single) momentum
derivative of a two-point, tree level vertex. Now, rather than making this
replacement, we use the equality of the two-point, tree level Wilsonian effective
action and seed action vertices to realise that, at $\Op{2}$, we can trade the seed
action vertices of the un-processed \BPZ\ terms for Wilsonian effective action vertices.
Now the \BPZ\ terms take the form of the standard set attached to an $\Op{2}$ stub, 
via an un-decorated wine.

At this stage, we might wonder why the set of \BPZ\ terms does not contain diagrams
like the fourth and fifth of figure~\ref{fig:beta1}. The answer is that we have discarded
these terms already, on the basis that they vanish at $\Op{2}$.\footnote{
\BPZ\ terms corresponding to the little set do not occur at all.}

There are several strategies to demonstrate that the \BPZ\ terms vanish. In 
reference~\cite{YM-1-loop}, it was demonstrated algebraically
that the \BPZ\ terms cancel, at $\Op{2}$: Taylor expanding the standard
set sub-diagrams to zeroth order in $p$, 
we can algebraically substitute for all constituent structures. There is, however, a much
more elegant way to proceed which minimises the algebra and is more intuitive.

Let us assume for the moment that  our calculation of $\beta_1$
is consistent (of course, part of the purpose of having performed this calculation
is to demonstrate this). Then we know that the set of diagrams contributing to $\beta_1 \Box_{\mu\nu}(p)$
must be transverse in $p$. The \BPZ\ terms are automatically transverse and so the only
diagrams not manifestly transverse are those constituting the standard set. For the calculation to be
consistent, then, the standard set must be transverse in $p$ and hence at least $\Op{2}$.
This immediately tells us that the \BPZ\ terms are at least $\Op{4}$ and so can be discarded.

Hence, our task is to demonstrate the transversality of the standard set.
Figure~\ref{fig:SS-Transversality} shows the result of contracting one of the free indices of the standard set with its external
momentum where, as usual, we have used the techniques of section~\ref{sec:GRs}. 
\begin{center}
\begin{figure}[h]
	\[
	\begin{array}{c}
		\ensuremath{\begin{array}{c}\input{pstex/SS-Transversality.pstex_t} \end{array}}
	\\[1ex]
		= 4
		\left[
			\ensuremath{\begin{array}{c}\input{pstex/SS-Transversality-B.pstex_t} \end{array}}
		\right]
	\end{array}
	\]
\caption{The result of contracting the standard set with its external momentum. The first three diagrams on the \rhs\ cancel and the
fourth vanishes by Lorentz invariance.}
\label{fig:SS-Transversality}
\end{figure}
\end{center}

Now we analyse the diagrams on the right hand side.
Algebraically, the first three terms go like:
\begin{eqnarray*}
\mbox{$A$-sector} & = &
4N\int_l \frac{l_\alpha}{l^2}\left(-1 + \frac{l\cdot(l+p)}{(l+p)^2} +  \frac{p\cdot(l+p)}{(l+p)^2} \right) \\
\mbox{$F$-sector} & = &
4(-N)
\int_l \frac{f_l l_\alpha}{\Lambda^2}\left(-1 + \frac{l\cdot(l+p)f_{l+p}}{\Lambda^2} + 2g_{l+p} + \frac{p\cdot(l+p)f_{l+p}}{\Lambda^2} \right),
\end{eqnarray*}
where the $UV$ finite sum vanishes after using the relationship~(\ref{eq:app-xf+2g}) and shifting momenta.
As we will see in section~\ref{sec:D-ID-New}, this is a special case of
diagrammatic identity~\ref{D-ID-Alg-2}.
The final term of figure~\ref{fig:SS-Transversality} vanishes by Lorentz invariance: the $l$-integral contains a single index and
the only momentum available to carry this index, after integration over $l$, is $p$. However, this index is contracted into a vertex
transverse in $p$ and so the diagram vanishes.

Therefore, contracting the standard set with its external momenta yields zero. 
Since the standard set carries two Lorentz indices we have proven that it must
be transverse in external momenta, as predicted. This, then, guarantees that the \BPZ\
terms vanish, at $\Op{2}$, and also confirms the consistency of the calculation. 
At the two-loop level, we will find that the analogue of the
\BPZ\ terms vanish for much the same reason.

\section{Conclusions}

The diagrammatic expression for $\beta_1$ is one of the major results of this thesis
and there are a number of comments worth making.

The first thing to do is to compare it to the expression
for $\beta_1$ obtained in~\cite{YM-1-loop}. Of those diagrams
present in figure~\ref{fig:beta1}, only the single supertrace
components of the first two, not possessing $AAC$ vertices,
are present in~\cite{YM-1-loop}. Diagrams~\ref{Beta1-LdL-C},
\ref{Beta1-LdL-F} and~\ref{Beta1-LdL-G} are present in 
algebraic form and diagrams~\ref{Beta1-LdL-D} and~\ref{Beta1-LdL-E}
are completely absent.

This is precisely what we expect: the treatment in~\cite{YM-1-loop}
of gauge remainders was algebraic and
single supertrace terms only were included; a consequence of this latter point being
that
$AAC$ vertices---which diagrams~\ref{Beta1-LdL-D} and~\ref{Beta1-LdL-E} (and also
components of~\ref{Beta1-LdL-B}) 
possess---can be set to zero.

Given the differences between the new approach and the old approach, an
obvious question to ask is whether or not we obtain the expected universal
coefficient in $D=4$. The answer turns out to be trivially in the
affirmative, as we will discuss in chapter~\ref{ch:LambdaDerivatives:Methodology}.
This is very encouraging, giving us faith that
the changes we have made to the flow equation (see chapter~\ref{ch:NewFlowEquation})
are consistent.\footnote{Of course, given that we can obtain the universal
one-loop coefficient using the old flow equation, the true test of
the new flow equation is the computation of $\beta_2$.}

One of the primary benefits of re-computing $\beta_1$ has been the
diagrammatic techniques which have been developed, as a by-product.
The new techniques fall into two classes. First, there are those
which enable us to process gauge remainders and to manipulate
diagrams at $\Op{2}$. Although not used originally, these methods 
can be applied to the old calculation of $\beta_1$, in exactly
the same way as we have just applied them in the new calculation.
Were we to return to the computation of $\beta_1$ using the
old flow equation, it would indeed be desirable to use the new techniques:
the new approach is both considerably more efficient and more elegant than the old one.

Let us suppose that we were to re-do the old calculation in this way. The truly
remarkable thing is that, up to diagrams possessing $AAC$ vertices, the
old calculation could be \emph{exactly} mapped onto the new one: all
we need do is replace the old style diagrams with the `supertrace blind' diagrams.
This use of supertrace blind diagrammatics represents the second class of
new techniques. By allowing many diagrams to be processed in parallel,
it hugely increases the efficiency of computations.

It is thus unarguably the case that much has been gained in the exercise of
re-computing $\beta_1$. Indeed, such a step must be taken if we wish to
compute $\beta_2$:
though universal only in $D=4$, the diagrammatic
expression for $\beta_1$ is valid in $D=4-2\epsilon$ dimensions and, as such, is necessary
for the computation of $\beta_2$.

Whilst the exercise of re-computing $\beta_1$ is necessary
both to prepare for
further calculations and to test the formalism, we should
ask whether it has any merits on its own. Though we have made much
of the efficiency and elegance of the approach, it perhaps does not compare
all that well to traditional methods of computing $\beta_1$. Though, as we
will see, the universal coefficient is extremely easy to extract from
the final diagrammatic expression, we had to spend quite some time obtaining
this expression. Nonetheless, once one knows the rules,
the steps that take us from the initial diagrammatic expression
of figure~\ref{fig:Beta1-LevelZero} to the final one are algorithmic.
This suggests that our methodology is ideally suited for automation
on a computer. We can, however, do very much better than this.

Re-examining the calculation, it becomes clear that, at each stage,
we repeat the same steps numerous times. For example,
the conversion of diagrams~\ref{d1:1.14} and~\ref{d1:1.15} into
$\Lambda$-derivative terms mirror each other exactly 
(see figures~\ref{fig:beta1:LevelTwo-A} and~\ref{fig:beta1:LevelTwo-B}). Both diagrams
possess two three-point, tree level Wilsonian effective action vertices;
upon their manipulation, the partner diagrams possessing seed action
vertices are exactly cancelled. Indeed, thinking about this more carefully,
if we take two three-point, tree level vertices, two effective propagators and
two external fields, the only $\Lambda$-derivative terms we can construct
are precisely diagrams~\ref{d1:1.15c} and~\ref{d1:1.19}. This is starting
to suggest that perhaps the generation of these $\Lambda$-derivative terms
can be done in parallel.

These ideas will be developed in part~III. Essentially, what
we will find is that all the manipulations and cancellations seen in this
chapter are merely special cases of more general mechanisms. Understanding
these mechanisms completely will allow us to dispense with all the intermediate
stages we have laboured through in going from our initial diagrammatic expression
to the final one! The ramifications of this can hardly be over emphasised. If we could jump
straight from equation~(\ref{eq:Beta1-Defining})
to the diagrammatic expression of figure~\ref{fig:beta1}, then suddenly
our methodology becomes competitive with traditional ways of computing $\beta_1$.

\chapter{$\Lambda$-Derivative Terms} \label{ch:LambdaDerivatives:Methodology}

In this chapter we detail the methodology used for treating 
$\Lambda$-derivative terms.
Following a statement of the basic idea in section~\ref{sec:LD:M:intro}, 
the methodology is developed in section~\ref{sec:TLD:Methodology:1-loop} 
by looking at the 1-loop integrals which contribute to $\beta_1$. 

In section~\ref{sec:TLD:Methodology:2-loop}, we build on the 1-loop case 
to obtain the expected general form for 2-loop integrals.
As preparation for the evaluation of $\beta_2$, we introduce an intermediate step
which will allow us to explicitly demonstrate how 
non-universal contributions can cancel between diagrams. 
The methodology for this, which also has applications to
terms which require manipulation at \Op{2}, is presented 
in section~\ref{sec:TLD:Subtractions}.

We conclude the chapter by showing that certain running couplings which
can spoil the universality of $\beta_2$ can always be removed by a suitable
choice of the seed action.
This is an extension of work done in~\cite{YM-1-loop}.

\section{Introduction} \label{sec:LD:M:intro}

The simplest $\Lambda$-derivative terms we will encounter are those
contributing to $\beta_1$. From figure~\ref{fig:beta1} we know that
\[
	4 \beta_1 \Box_{\mu \nu} (p) = -\frac{1}{2} \dec{\OLDs}{\bullet}.
\]
We now want to make the integral over loop momentum (which
we will take to be $k$) to be explicit and so, mindful that
we will need to be more precise about the \rhs, write
\begin{equation}
	4 \beta_1 \Box_{\mu \nu} (p) = -\frac{1}{2} \int_k \dec{\OLDs(k)}{\bullet}.
\label{eq:TDM:Interchange}
\end{equation}

The next step that we wish to perform is to interchange the order of the $\Lambda$-derivative
and the momentum integral. This step is trivial only if the integral is convergent, even after this
change. We temporarily ignore this subtlety and so now have
\[
	4 \beta_1 \Box_{\mu \nu} (p) = -\frac{1}{2} \dec{\int_k \OLDs(k)}{\bullet}.
\]

Since the \lhs\ of this equation comprises a number times $\Op{2}$, it follows
that the coefficient multiplying the $\Op{2}$ part of the \rhs\ must be
dimensionless. Consequently, we can schematically write
\[
	\beta_1 = \flowConstAl \left( \mbox{Dimensionless Quantity} \right).
\]
For the \rhs\ to survive differentiation \wrt\ $\Lambda|_\alpha$, 
it must either depend on some dimensionless running 
coupling---other than $g$ and $\alpha$---or there must be some scale, other than $\Lambda$,
available for the construction of dimensionless quantities.  
We show how we can avoid introducing such
couplings in section~\ref{sec:TLD:Couplings}.

One scale which is available is $p$ and so we can envisage contributions to $\beta_1$ of 
the form (in $D=4$)
\[
	\flowConstAl \ln p^2 / \Lambda^2.
\]
Indeed, the standard set gives rise to contributions precisely of this type (see section~\ref{sec:TLD-RegByp}). However,
as we will see in section~\ref{sec:TDM:Indep-p}, we cannot form contributions
of this type from the little set---but we know from~\cite{YM-1-loop} that the little set does
contribute to $\beta_1$.

For the little set, then, what scales are there, other than $\Lambda$, available
for the construction of dimensionless quantities? Courtesy of the $SU(N|N)$ regularisation,
we know that there are no scales in the UV.\footnote{
Although we will find a subtle interplay between the IR and UV, which is commented on at
the end of 
section~\ref{sec:TLD-RegByp}.} 
Na\"ively, we would not expect a scale to arise
in the IR, either. However, we stated
that interchanging the order of differentiation \wrt\ $\flowConstAl$ and loop integration
in~(\ref{eq:TDM:Interchange}) is trivial only if the integral is convergent, even after
this step. The key point is that performing this step has the capacity to introduce IR divergences.
This is most easily appreciated by noting that $\Delta^{11}_k \sim 1/k^2$, 
whereas $\dot{\Delta}^{11}_k \sim 1/\Lambda^2$.
Thus, to legally move $\flowConstAl$ outside of the loop integral, we must introduce some IR regulator,
which then provides the scale necessary to form dimensionless quantities. After allowing 
$\flowConstAl$ to act, this unphysical scale will disappear. IR divergences introduced in 
this way will be called \emph{pseudo}-divergences, since they are an artifact of the way
in which we have chosen to perform the calculation.

Noting that in the case of the standard set it is effectively $p$ which is 
providing the IR regularisation,
our strategy for evaluating loop integrals is to look at the IR end. Scanning through the
list of effective propagators~(\ref{eq:app:NFE:EP-11})--(\ref{eq:app:NFE:EP-C2C2}), it is apparent that
the leading contributions occur when all effective propagators are in the $A$-sector;
likewise for any instances of $\wedge$. Gauge invariance and considerations
as to the supertrace structure will eventually determine that all contributions to
$\beta_1$ and $\beta_2$ are ultimately
limited to the lowest order momentum contributions from the 
$A^1$ sector; it is precisely this regime---and this regime alone---that is universal.

\section{1-loop Integrals}	\label{sec:TLD:Methodology:1-loop}

\subsection{Vanishing Diagrams}

We begin our analysis by looking at diagrams~\ref{Beta1-LdL-D} and~\ref{Beta1-LdL-E} 
which, up to an overall factor, are reproduced in figure~\ref{fig:TLD:1:Vanishing}.
Recall that these two diagrams have no analogue in the computation of $\beta_1$
presented in~\cite{YM-1-loop} and so we had better find that they vanish.
\begin{center}
\begin{figure}[h]
\[
	\dec{
		\begin{array}{ccc}
		\vspace{0.1in}
			\LabOLOb{Ill-Beta1-LdL-D}	&	&\LabOLOb{Ill-Beta1-LdL-E}
		\\
			\ensuremath{\begin{array}{c}\input{pstex/Beta1-LdL-D.pstex_t} \end{array}}			&-4	&\ensuremath{\begin{array}{c}\input{pstex/Beta1-LdL-E.pstex_t} \end{array}}			
		\end{array}
	}{\bullet}
\]
\caption{Two diagrams which give a vanishing contribution to $\beta_1$ in $D=4$.}
\label{fig:TLD:1:Vanishing}
\end{figure}
\end{center}

We now focus on the IR end of the loop integral.
The bottom vertex does not carry the loop momentum, nor does the effective
propagator leaving it. However, we know that this effective propagator must
be in the $C$-sector (else the bottom vertex vanishes
by \CC, when we sum over permutations of the fields). 
This immediately tells us that diagram~\ref{Ill-Beta1-LdL-E}
is IR safe, even if we interchange the order the $\Lambda$-derivative
and the momentum integral: performing this step, the loop integral just goes as
\[
	\int_k g_k.
\]

Let us now focus on diagram~\ref{Ill-Beta1-LdL-D}.
To try and find IR divergences, we take the fields involved in the loop integral to be in the $A$-sector.
To deal with the top vertex, we recall that it is 
Taylor expandable in momenta~\cite{GI-ERG-I, GI-ERG-II, TRM-elements}. Hence, to try and isolate
the most IR divergent contribution from the loop integral, we take the minimum number of
powers of momenta from the top vertex consistent with  Lorentz invariance. Given that the field entering
this vertex from beneath is in the $C$-sector, we can take \Omom{0}.

Diagram~\ref{Ill-Beta1-LdL-D} is clearly IR safe in $D=4$ since the loop integral 
looks at worst---without even taking into account the  $\Lambda$-derivative---like
\[
\int_k 1/k^2,
\]
in the IR. 

Thus we can safely interchange the order of integration and 
differentiation \wrt\ \LConstAl\  for both diagrams~\ref{Ill-Beta1-LdL-D} and~\ref{Ill-Beta1-LdL-E}.
Having done so,
we know that (the $\Op{2}$ parts of) both diagrams will vanish, when computed in $D=4$. Bearing in mind 
that we must pre-regularise~\cite{TRM-Faro,YM-1-loop} (see section~\ref{sec:Intro:CovCutoffRegn}) 
and in preparation for the 2-loop calculation, 
we ultimately want to
be doing our computations in $D=4-2\epsilon$ dimensions. Rescaling our loop momenta $k \rightarrow k/\Lambda$ we
see that, at $\Op{2}$, the diagrams acquire an overall factor of $\Lambda^{-2 \epsilon}$. Working in this
scheme, the $\Lambda$-derivative of the diagrams now $\sim \epsilon$ which, of course, vanishes in the $D=4$ limit.

Note that diagram~\ref{Beta1-LdL-A} (the first element of the standard set)
also vanishes in $D=4$, after differentiation
\wrt\ \LConstAl. However, we will always
keep this term together with the other elements of the standard set. This is done because
we  often exploit the fact that the set is transverse, and want to be able to do
this irrespective both of $D$
and whether or not it is struck by a $\Lambda$-derivative.

\subsection{IR Regularisation Provided by $p$} \label{sec:TLD-RegByp}

In this section, we will encounter diagrams that survive differentiation \wrt\ $\Lambda$ in $D=4$
and for which $p$ plays a role in the IR regularisation. The only diagrams
that fall into this class are diagrams~\ref{Beta1-LdL-B} and~\ref{Beta1-LdL-C}. However, 
as just mentioned, diagram~\ref{Beta1-LdL-A} will come along for the ride;
a full understanding of the standard set is crucial for the 2-loop calculation.
These three diagrams
are reproduced, up to an overall factor, in figure~\ref{fig:TLD:StandardSet}. We henceforth take
their loop momentum to be $k$.

\begin{center}
\begin{figure}[h]
\[
	\dec{
		\begin{array}{ccccc}
		\vspace{0.1in}
			\LabOLOb{Ill-Beta1-LdL-A}	&	&\LabOLOb{Ill-Beta1-LdL-B}	&	&\LabOLOb{Ill-Beta1-LdL-C}
		\\
			\ensuremath{\begin{array}{c}\input{pstex/Beta1-LdL-A.pstex_t} \end{array}}			&-	&\ensuremath{\begin{array}{c}\input{pstex/Beta1-LdL-B.pstex_t} \end{array}}			&+4	&\ensuremath{\begin{array}{c}\input{pstex/Beta1-LdL-C.pstex_t} \end{array}}
		\end{array}
	}{\bullet}
\]
\caption{The standard set struck by $\flowConstAl$.}
\label{fig:TLD:StandardSet}
\end{figure}
\end{center}

Diagram~\ref{Ill-Beta1-LdL-A} is IR safe. The second and third diagrams, before
differentiation \wrt\ \LConstAl, 
have the IR structure
\begin{equation}
\int_k \frac{\mathcal{O}(p^2, p.k, k^2)}{k^2 (k-p)^2}, \label{eq:TLD:GeneralIRStructure}
\end{equation}
where we have taken a single power of momentum from each of the three-point vertices, have chosen all effective propagators
to be in the $A$-sector and have evaluated any cutoff functions at zero
momentum. 

Note that choosing the effective propagators to be in the $A$-sector 
constrains diagram~\ref{Ill-Beta1-LdL-B}, considerably. For three-point vertices decorated
exclusively by $A$-fields, it must be the case that all fields are on the same supertrace
and hence in the same sub-sector. Consequently, for the contributions to diagram~\ref{Ill-Beta1-LdL-B}
with the severest IR behaviour, all fields must be in the $A^1$-sector.

Returning to equation~(\ref{eq:TLD:GeneralIRStructure}), 
it is clear that the presence of $p$ in the denominator is required to regulate the integral in the IR, at
least when we choose to take $\mathcal{O}(p^2)$ from the vertices. Performing the integral in $D=4$ will
then give us something of the form
\[
\mathcal{O}(p^2)(a \ln (p^2/\Lambda^2) + b).
\]
When this is hit by the $\Lambda$ derivative---which we can move outside the integral---only the 
first term will survive and so we will be left
with a (universal) coefficient multiplying two powers of $p$.

That the final answer is Taylor expandable in $p$ gives us an alternative way in which to 
evaluate $\Lambda$-derivative terms.
Having moved the $\Lambda$-derivative outside the integral,
we Taylor expand the denominator in $p$. Doing this, $p$ will no longer act as a regulator
and so we will then generate IR divergences, when we perform the integral. However, all divergences will 
be killed by the $\Lambda$-derivative. 
To parameterise these pseudo-divergences, we must introduce an IR regulator; it is natural to use 
dimensional regularisation.

It may, at this stage, seem a little perverse to have traded one IR regulator, $p$---which
occurs naturally---for
another. However, even for diagrams which are not Taylor expandable in $p$, it will
turn out that we are often interested in the Taylor expandable part. By Taylor expanding
in $p$, the resulting integrals tend to be easier to perform. 

We now discuss how this
procedure works, in more detail:
\begin{enumerate}
\item Take $\mathcal{O}$(mom) from each of the vertices;

\item Taylor expand the denominator in $p$;

\item replace the upper limit of the radial integral with $\Lambda$, thereby cutting off modes above this scale;

\item rescale $k \rightarrow k/\Lambda$, so that the diagram acquires an overall factor of $\Lambda^{-2\epsilon}$.
\end{enumerate}

Having done the angular integral, we are left with an expression of the form:
\[
 \frac{(\Lambda^{-2\epsilon})^\bullet}{(4\pi)^{D/2}} \int_0^1 \frac{k^{D-1}}{k^4}dk  \, \mathcal{O}(p^2).
\]
Performing the integral gives us a factor of $1/\epsilon$, as expected. This is killed by a factor of $\epsilon$
arising from differentiation \wrt\ \LConstAl, confirming the consistency of the approach.

Before moving on, we must justify the validity of the third step. We know that the integral we are dealing
with has support only in the IR. However, $\int_k 1/k^4$ is not UV regulated and so we must incorporate the 
effects of the UV regularisation. Since the details of the regularisation will not affect the IR, at leading
order, we choose the simplest form that cuts off momentum modes above the scale $\Lambda$. The non-universal
corrections to this will necessarily remove any IR divergence, even before differentiation \wrt\ $\LConstAl$,
 and so vanish in the limit that $\epsilon \rightarrow 0$.
An implicit part of this step is that we now throw away all F-sector diagrams and evaluate any cutoff 
functions at zero momentum.

In readiness for the 2-loop calculation, we will now extend our analysis of the standard set, and will give its general
form. We know the following facts about the standard set:
\begin{enumerate}
	\item 	the sum of the diagrams is transverse;

	\item 	when struck by \flowConstAl, the coefficient of the $\Op{2}$ term is universal,
			up to \Oep\ corrections (this follows from~\cite{YM-1-loop});

	\item 	in $D=4$, the diagrams have the structure
			\[
			\Op{2} \ln (p^2/\Lambda^2) + \Op{2} +\cdots,
			\]
			where the ellipsis denotes terms which are higher order in $p$.
\end{enumerate}
From these three points and dimensions it follows that, in $D=4-2\epsilon$, the standard set takes the algebraic form
\[
\frac{N\Lambda^{-2\epsilon}}{(4\pi)^{D/2}} \left[ a_0 \left( 1 - \frac{p^{-2\epsilon}}{\Lambda^{-2\epsilon}} \right) 
\frac{1}{\epsilon} + \ldots \right]
\Box_{\alpha \beta}(p) + \order{p^4, p^{4-2\epsilon}},
\]
where $a_0$ is a universal coefficient and the ellipsis denotes terms which are higher order in $\epsilon$.

We have no further interest in the $\order{p^4, p^{4-2\epsilon}}$ terms and so turn to the ellipsis.
There are two ways in which we expect these terms to arise. On the one hand, 
compared to equation~(\ref{eq:TLD:GeneralIRStructure}),
we can take additional 
powers of $k$ in the numerator of the integrand, giving us non-universal contributions which are Taylor expandable in $p$.
On the other hand, we generically expect the coefficient $a_0$ to have arisen from some function of $D$ in which we have
taken $D=4$. Expanding this function in $\epsilon$ will
give rise to sub-leading contributions.
These contributions will be called computable. 
At the two-loop level, we will see how computable
contributions can combine to give universal quantities.

Just as we talk of computable parts of
some diagram, so too will we talk of the complimentary
non-computable parts. We emphasise that by non-computable
we really mean that the corresponding coefficients
cannot be computed without specifying non-universal details
of our set-up (\ie\ cut-off functions and the precise form 
of the covariantisation); it is not that we cannot, in principle,
calculate them.

Hence, the standard set takes the following form
\begin{equation}
\frac{N\Lambda^{-2\epsilon}}{(4\pi)^{D/2}} \left[ 
\sum_{i=0}^\infty \left(a_i + b_i\frac{p^{-2\epsilon}}{\Lambda^{-2\epsilon}} \right)\epsilon^{i-1} \right]
\Box_{\alpha \beta}(p) + \order{p^4, p^{4-2\epsilon}}, \label{eq:TLD:StandardSetGeneralForm}
\end{equation}
where $b_0 = -a_0$,  
the $a_{i>0}$ are a mixture of computable and non-computable contributions and the $b_i$ are entirely computable.

Notice that 1-loop computations are insensitive to the $p^{-2\epsilon}/\Lambda^{-2\epsilon}$ terms: taking
into account the additional factor of $\Lambda^{-2\epsilon}$ sitting outside, such terms are independent of $\Lambda$
and so will be killed by the $\Lambda$-derivative. At two-loops, where the standard set can occur
as a sub-diagram, we expect such contributions to survive.

We now discuss how to calculate the coefficients $b_i$. 
It is convenient to begin by contracting the standard set with
$\delta_{\alpha \beta}$. The $b_i$ arise from integrals of the form
\[
	\int_k \frac{\order{p^2, p.k, k^2}}{k^2(k-p)^2}.
\]
Note that the $\order{k^2}$ term does not contribute to the $b_i$: the denominator
becomes just $1/(k-p)^2$ and so, by shifting momentum, we can remove $p$
from the denominator entirely. It is now not possible to generate a power of $p^{-2\epsilon}$.

The next step is to combine denominators, using Feynman parameterisation~\cite{P&S}:
\[
\int_0^1 da h(a) \int_k \frac{\mathcal{O}(p^2)}{(k^2 + K^2)^2}, 
\]
where $K^2 = a(1-a)p^2$ and $h(a)$ is some function of $a$. At this stage, it is now tempting to proceed as before and restrict the range of the
radial integral. However, this leaves us with an unpleasant calculation 
as we cannot use standard dimensional regularisation formulae. Besides, there is a much simpler way
to proceed: we differentiate twice, with respect to $p^2$.

The effect on the integral is to ensure that it is UV regulated by the denominator of the integrand---without the need
for any cutoff regularisation. We call this 
automatic UV regularisation (which will play an important role at two loops). 
Since we are  interested in the part of the integral which has support in the IR, 
there is no need for us to  restrict the range of integration, as doing so would only serve to
make the calculation harder.
Retaining just A-sector diagrams, we evaluate all cutoff functions at zero momentum, leaving us
with an integral we can do using standard dimensional regularisation techniques (\ibid).

We perform the integral and compare it to the second derivative \wrt\ $p^2$ of equation~(\ref{eq:TLD:StandardSetGeneralForm}),
contracted with $\delta_{\alpha \beta}$. This is one place 
where the value of keeping the standard set together manifests itself:
because we know the standard set to be transverse, we know the effect of contracting with $\delta_{\alpha \beta}$.
Equating powers of $\epsilon$ allows us to determine the $b_i$. The first two coefficients are:
\begin{eqnarray}
b_0 & = & -20/3   \label{eq:SS:b_0} \\
b_1 & = & -124/9 + 20 \EM /3 \label{eq:SS:b_1}
\end{eqnarray}
where $\EM$ is the Euler-Mascheroni constant.

At this point it is worth pausing to consider why the coefficients
$a_0$ and $b_i$ have no dependence on $N$. We note in 
equation~(\ref{eq:TLD:StandardSetGeneralForm}) that we have extracted
an overall factor of $N$, but we might suspect that the $a_i$ and $b_i$
incorporate attachment corrections.\footnote{The overall
factor of $N$ is what we expect for the diagrams if there are no attachment 
corrections. In this case, each diagram comprises one loop
decorated by the two external fields and one empty loop. The empty loop
yields $\str \sigma_+ = N$ if the internal fields are
bosonic and $\str \sigma_- = -N$ if the internal fields are fermionic.}
 Let us look
first at diagram~\ref{Ill-Beta1-LdL-C}. Since this is formed by
the action of gauge remainders on three-point (tree level) vertices
decorated by an external field, we know from the end of
section~\ref{sec:GR:CompleteDiagrams} that we can discard all $1/N$
corrections. 

Now consider diagram~\ref{Ill-Beta1-LdL-B}. The contributions to $a_0$
and $b_i$ come when all fields are in the $A$-sector. If either
of the effective propagators attaches via a $1/N$ correction,  
then the external fields are always guaranteed
to be on the same supertrace, irrespective of location: the diagram vanishes by \CC.
However, we expect the $a_{i\geq1}$, to be non-trivial functions of $1/N$,
since these coefficients receive contributions in which the vertices
of diagram~\ref{Ill-Beta1-LdL-B} are each \CC\ even (\ie\ $AAC$ vertices)
and from diagram~\ref{Ill-Beta1-LdL-A}.

We conclude our analysis of the standard set by
noting a beautiful interplay between the IR and the UV, illustrated 
by diagram~\ref{Ill-Beta1-LdL-C}. The strategy we have used
is to pull the $\Lambda$-derivative outside of the momentum 
integral and then focus on the IR end. 
Focusing on the IR end allows us
to throw away the regulating diagram. However, the regulating
diagram was required to
define the $A$-sector diagram when we interchanged the
order of differentiation \wrt\ $\Lambda$ and loop integration.

Now, suppose that we 
had left the $\Lambda$-derivative inside the integral. Then 
the $A$-sector diagram actually dies, since $A$-sector (processed)
gauge remainders are independent 
of $\Lambda$. We are left with the $B$-sector diagram, which 
provides the same leading order contribution as the $A$-sector 
diagram, but arising from the UV!

Interplay such as this will only arise when the components
of some diagram which gives a contribution in the IR are not
regulated by cutoff functions, alone.

\subsection{Loop Integrals Independent of $p$}	\label{sec:TDM:Indep-p}

We conclude our survey of 1-loop integrals by looking at the final two diagrams which contribute to $\beta_1$,
as reproduced in figure~\ref{fig:TLD:LittleSet}, up to an overall factor. 
\begin{center}
\begin{figure}[h]
\[
	\dec{
		\begin{array}{cccc}
		\vspace{0.1in}
				&\LabOLOb{Ill-Beta1-LdL-F}	&	&\LabOLOb{Ill-Beta1-LdL-G}
		\\
			2	&\ensuremath{\begin{array}{c}\input{pstex/Beta1-LdL-F.pstex_t} \end{array}}			&+	&\ensuremath{\begin{array}{c}\input{pstex/Beta1-LdL-G.pstex_t} \end{array}}
		\end{array}
	}{\bullet}
\]
\caption{The little set struck by $\flowConstAl$.}
\label{fig:TLD:LittleSet}
\end{figure}
\end{center}

The difference between these diagrams and the ones just analysed is that
$p$ is not involved in the IR regularisation. The integrals just go like
\[
\int_k \frac{1}{k^4}
\]
in the IR and we can use the techniques of the previous section to evaluate such terms. We thus expect the complete
set of 
diagrams contributing to $\beta_1$ to take the following form, before differentiation \wrt\ $\LConstAl$:
\begin{equation}
\frac{N\Lambda^{-2\epsilon}}{(4\pi)^{D/2}} \left[
\sum_{i=0}^\infty \left( A_{i} + B_{i}\frac{p^{-2\epsilon}}{\Lambda^{-2\epsilon}} \right)\epsilon^{i-1} \right]
\Box_{\mu \nu}(p) + \order{p^4, p^{4-2\epsilon}}, \label{eq:TLD:1-loopGeneralForm}
\end{equation}
where the $B_i$ are computable and the $A_i$ generally contain both computable and non-computable parts. The universal coefficient $A_0$
yields the sole contribution to $\beta_1$, in the $\epsilon \rightarrow 0$ limit.

\section{2-loop Integrals} \label{sec:TLD:Methodology:2-loop}

In this section, we develop the machinery of the previous section to deal with two-loop
$\Lambda$-derivative terms.
The integrals we have to deal with fall into two classes: factorisable and non-factorisable.
In the former case---dealt with in section~\ref{sec:TLD:Methodology:Factorisable}---the loop-integrals are independent, whereas, in the 
latter case---dealt with in section~\ref{sec:TLD:Methodology:NFactorisable}---they are not.

Following on from the one-loop case we expect and, indeed,
find---see chapters~\ref{ch:bulk} and~\ref{ch:TotalLambdaDerivatives:Application}---that we can
write\footnote{In the limit that $\alpha \rightarrow 0$.}
\[
	4\beta_2 \Box_{\mu \nu} (p) = -\frac{1}{2} \dec{\TLDs}{\bullet}.
\]
One of the main sources of complication in the two-loop case is that, even
after differentiation \wrt\ $\flowConstAl$, individual elements of $\TLDs$
can still possess IR divergences. It is only the sum of diagrams contributing
to $\TLDs$ that we expect to give a finite (universal) contribution after
differentiation \wrt\ $\flowConstAl$.

Since we are interested in two-loop integrals which contribute to $\beta_2$ we will,
up to factors of $\Pep$, work at $\Op{2}$.

\subsection{The Factorisable Case} \label{sec:TLD:Methodology:Factorisable}

To understand the algebraic form of two-loop diagrams, we need first to understand their
structure. Since we are dealing with factorisable terms, we expect them to comprise
two one-loop sub-diagrams, each of which carries external momentum $p$. These sub-diagrams must
be connected to each other, and so we predict that they will be joined together by an effective propagator
(for explicit examples, see chapter~\ref{ch:bulk}). This effective propagator
just contributes powers of the external momentum and so we take 
the general form of a factorisable two-loop integral to be:
\begin{equation}
\frac{N^2\Lambda^{-4\epsilon}}{(4\pi)^D} \sum_{i=0}^\infty \left(c_i + d_i \frac{p^{-2\epsilon}}{\Lambda^{-2\epsilon}}
+ e_i \frac{p^{-4\epsilon}}{\Lambda^{-4\epsilon}} \right) \frac{1}{\epsilon^{i-2}}\Op{2},
\label{eq:TLD:TwoLoopGenericForm}
\end{equation}
where we obtain a power of $p^{-2\epsilon}/\Lambda^{-2\epsilon}$ for each loop in which $p$ provides IR regularisation.

\subsection{The Non-Factorisable Case} \label{sec:TLD:Methodology:NFactorisable}

To understand the non-factorisable case will require a little more work.
We begin by trying to construct the most divergent possible
type of diagram. To do this, we can adapt our study of factorisable diagrams.
We know that factorisable diagrams constitute two one-loop sub-diagrams joined
by an effective propagator. For a diagram constructed entirely of vertices
and effective propagators 
this means that our most divergent (factorisable) two-loop diagram will comprise four three-point,
tree level vertices and five effective propagators.
(We can, of course, construct diagrams using processed gauge remainders,
in addition to the other ingredients, but this will not tell us anything
extra about the IR structure.)	

We expect the most divergent non-factorisable diagrams to possess exactly the same
ingredients, but joined together in a different way. Taking the loop momenta to be $l$
and $k$, we know that one of the effective propagators must $\sim 1/(l-k)^2$ (assuming
it to be in the $A$-sector).
Conservation of four-momentum at a vertex then implies that there must be at least one
effective propagator carrying $l$ and at least one carrying $k$.
Knowing that, at $\Op{2}$ the integrand must be of mass dimension $-8$, we
expect the most divergent type of diagram we can
construct (assuming no IR regularisation is provided by $p$) to take the following
form in the IR:
\[
\Op{2}\int_{l,k} \frac{\mathcal{O}(\mathrm{mom}^2)}{k^2 (l-k)^2 l^6}.
\]
In fact, we will see in section~\ref{sec:TLD:SS-Mirror} that, for the set of diagrams contributing to
$\beta_2$, gauge invariance prevents the appearance of $1/l^6$ and so we would find that,
taking the above form, we would be forced to have $l^2$ in the numerator. 
Hence, the most divergent type of integral we find has the following structure in the IR:
\begin{equation}
\Op{2}\int_{l,k} \frac{1}{k^2 (l-k)^2 l^4} \label{eq:TLD:StandardTwoLoopIntegral}.
\end{equation}
To evaluate the contribution coming from the IR, we observe that the $l$-integral is automatically UV regulated. Thus, using dimensional
regularisation,
we perform the $l$-integral first, with unrestricted
range of integration, and perform the $k$-integral second, with the range of radial integration restricted to $\Lambda$. We obtain one power of
$1/\epsilon$ from the Feynman parameter integral and a further power from the radial $k$-integral. 

Doing the integrals the other way around would be awkward, 
as we 
cannot then use an un-restricted range of integration for the inner integral.

Had there been $p$s present in the denominators, providing regularisation, we would expect accompanying factors of $p^{-2\epsilon}/\Lambda^{-2\epsilon}$.
Consequently,  equation~(\ref{eq:TLD:TwoLoopGenericForm}) is the form for a generic 2-loop integral.

\subsection{Considerations for $\beta_2$} \label{sec:TLD:Methodology:beta2}

For the actual computation of $\beta_2$, we can
constrain some of the coefficients in equation~(\ref{eq:TLD:TwoLoopGenericForm}). Since the terms corresponding
 to the coefficients $e_i$ are independent of $\Lambda$, we
can drop them as they will vanish after differentiation.
Given that we are comparing our final answer with the Taylor expandable, finite expression
$
\beta_2 \Box_{\mu \nu} (p),
$
it must be that the coefficients $c_0$ and $d_i$ vanish. 

The coefficients $c_0$ and $d_0$ are entirely computable. The coefficient $d_1$,
on the other hand, comprises both a purely computable part and a non-computable part
multiplied by a computable coefficient. 
The computable part and the computable coefficient must both be zero.

Ultimately, we will be left with the coefficient $c_1$ being the
only contribution to the final answer. One of the primary tasks ahead is to show that the non-universal contributions to
$c_1$ cancel between diagrams.

This problem really has two sides. First, we 
must show that non-computable contributions from vertices etc.\ cancel out. Then
we must show that the computable contributions to $c_1$ combine to give 
the standard, universal answer.

\section{Subtraction Techniques} \label{sec:TLD:Subtractions}

\subsection{Basics}	\label{sec:TLD:Subtractions:Basics}

Rather than attempting to process 2-loop diagrams directly, we perform an intermediate step whereby we 
add and subtract a set of terms designed to remove all non-computable contributions from the calculation.
We illustrate this technique with a simple example. Consider the two-loop integral arising
from the computation of the scalar two-loop $\beta$-function, within the ERG~\cite{scalar2loop}.
\[
	\int_{l,k} 
	\dec{
		\Delta^2_l\Delta_{l-k}\Delta_k - \frac{1}{2}\Delta^2_l\Delta^2_k 
	}{\bullet},
\]
where $\Delta_l = c(l)/l^2$. We can trivially rewrite this as
\[
	\int_{l,k} 
	\dec{
		\Delta^2_l\Delta_{l-k}\Delta_k - \Delta^2_l\Delta^2_k + \Delta^2_l\Delta^2_k - \frac{1}{2}\Delta^2_l\Delta^2_k 
	}{\bullet},
\]
where we call the second term a subtraction and the third term its corresponding addition. The addition
trivially combines with the final term, though this is of no particular significance.

Now focus on the $k$ integral, in the first term. Following~\cite{scalar2loop, Bonini},
we know that we can set $c(l-k) = c(k)$, as contributions higher order in $l$ are
killed in the \eptoz\ limit.
Next, use the by now familiar prescription for the cutoff functions: if the integral is regulated without the cutoff
functions, then we simply evaluate them at zero momentum, leaving the domain of integration unrestricted. 
If the integral requires the cutoff functions for regularisation, then restrict the
domain of integration and evaluate the cutoff function at zero momentum. The sub-leading
corrections to this will manifest themselves
as additional powers of momenta, in the numerator.

If we were to start
taking such sub-leading (non-computable) contributions under the $k$-integral, then this will allow us to
 Taylor expand the $k$-integral in $l$. For the $l$-integral to still diverge in the IR---and
thus survive differentiation \wrt\ \LConstAl(in the limit that $\epsilon 
\rightarrow 0$)---we must take $l^0$. However, such non-computable contributions will be cancelled by exactly the same 
contributions coming from the second term, above. Hence we have, up to \Oep\ corrections:
\[
\int_{l,k} \left[ \Delta^2_l(\Delta_{l-k}\Delta_k)_\COMP - \Delta^2_l(\Delta^2_k)_\COMP + \frac{1}{2}\Delta^2_l\Delta^2_k \right]^\bullet,
\]
where $\COMP$ tells us that, when considered as a pair, the first two $k$-integrals yield a computable contribution.

For the first term not to die in the $\epsilon 
\rightarrow 0$ limit, we must set $c(l) \rightarrow 1$ and so can
 extend the `$\COMP$' to cover the whole term.
We have:
\begin{equation}
	\int_{l,k}
	\left( 
		\left[
			\Delta^2_l\Delta_{l-k}\Delta_k
		\right]^\bullet_\COMP 
		- 
		\left[\Delta^2_l
		\right]^\bullet 
		\left[\Delta^2_k
		\right]_\COMP
 		- \Delta^2_l 
		\left[
			\Delta^2_k
		\right]_\COMP^\bullet
		+ \Delta^2_l
		\left[
			\Delta^2_k 
		\right]^\bullet 
	\right) \label{eq:TLD:SubtractionSet2},
\end{equation}
where, in the final term, we have used the freedom to interchange $l$ and $k$.
Terms like the second and third, comprising a computable, factorisable sub-diagram, are to be 
 called semi-computable.
The third and forth terms combine to give the non-computable ($\NUn$) contribution $\Delta^2_l[\Delta^2_k ]^\bullet_{\NUn}$ which,
up to $\Oep$ corrections, allows us to write
\[
	\int_{l,k} 
	\left( 
		\left[
		\Delta^2_l\Delta_{l-k}\Delta_k
		\right]^\bullet_\COMP
		- 
		\left[
			\Delta^2_l
		\right]^\bullet 
		\left[
			\Delta^2_k
		\right]_\COMP
		+ 
		\left[
			\Delta^2_l
		\right]_\COMP
		\left[
			\Delta^2_k 
		\right]^\bullet_{\NUn} 
	\right). 
\]

Exploiting the freedom to interchange $l,k$, 
the second and third terms combine to leave us with a purely computable contribution:
\[
	\int_{l,k} 
	\left[
		\Delta^2_l\Delta_{l-k}\Delta_k - \frac{1}{2}\Delta^2_l\Delta^2_k 
	\right]^\bullet_\COMP.
\]
We have demonstrated that the original integral does, indeed, give something which is computable.

To calculate this integral, we can use a mixture of the techniques already discussed. In the factorisable case,
we simply restrict the ranges of the integrals. In the non-factorisable case, we note that the $l$-integral is automatically
UV regulated. Hence, we perform this integral first with an unrestricted domain of integration but then restrict the domain
for the remaining $k$-integral. It is straightforward to confirm 
that we reproduce the expected, universal answer
\[
	-\frac{17}{3} \frac{1}{(4 \pi)^4}.
\]

\subsection{Generalisation to the Gauge Case} \label{sec:TLD:Subtractions:Gauge}

Constructing subtractions in the gauge case is exactly analogous to the simpler case just analysed. In the same way,
the subtractions are constructed such as to remove  non-computable contributions, by noting that denominators can be
Taylor expanded in momenta if sufficient powers of momentum are present in the numerator.

As the whole formalism is based around Taylor expansion, it is not surprising that we will need to use the techniques of
section~\ref{sec:MomentumExpansions} to Taylor expand vertices in momenta. We know that the lowest order terms constitute
derivatives of lower order vertices, and that the sign of this derivative depends on whether we have had  to push forward or
pull back. We define the subtraction to be the term which removes non-computable components from the parent diagram, and not by its sign.
Hence, a subtraction involving a pull back will come with a positive sign.

The real subtlety in the gauge case comes from the diagrammatic 
manipulations that will be performed, subsequent to the construction
of the subtractions. Recall from  equation~(\ref{eq:TLD:SubtractionSet2}) that, after the 
construction of the subtraction, we arrive at  a
non-factorisable, computable term; two semi-computable, factorisable terms and a  factorisable term.
That this plain factorisable term is a combination of the addition and another term is besides the point; the point is that the
addition is never under the influence of $\COMP$. However, the semi-computable terms, which are of exactly the same form as the addition,
are under the influence of $\COMP$.

In the gauge case, we will encounter examples where the additions can be manipulated, using the diagrammatic techniques of
section~\ref{ch:FurtherDiagrammatics}. We must now answer the question of whether the semi-computable terms---which are of exactly the 
same form---can be manipulated in the same way. As we will see, the answer depends on a choice we are free to make and, given that,
on the precise structure of the diagrams in question.

The key to this problem is understanding exactly what it is that is meant by $\COMP$. 
Let us suppose that we have taken a non-factorisable two-loop diagram and constructed a factorisable
subtraction. We will suppose that the sub-diagram of this latter term, to which we apply
$\COMP$, is just diagram~\ref{Ill-Beta1-LdL-F}.

The algebraic form of this diagram is (without the restriction to $\COMP$)
\begin{equation}
\Box_{\mu \beta}(p) (\Lambda^{-2\epsilon})^\bullet \int_k \partial_\alpha^k \left[\frac{\InvCO_k}{k^2} \right] \frac{k_\beta}{k^2}.
\label{eq:TLD:keepcutoff}
\end{equation}

To compute the part of this diagram left over, after combining with our non-factorisable
term we do the following: evaluate the cutoff function at zero momentum, restrict the range
of the integral, and proceed as usual.

Next, suppose that we were to move the momentum derivative from the $\InvCO_k/k^2$ term to 
the $k_\beta/k^2$ term, throwing away the total derivative, in the process. This yields
\begin{equation}
-\Box_{\mu \beta}(p) (\Lambda^{-2\epsilon})^\bullet \int_k \frac{\InvCO_k}{k^2}  \partial_\alpha^k \left[\frac{k_\beta}{k^2}\right] .
\label{eq:TLD:keepcutoff-B}
\end{equation}

Evaluating this integral in the usual way gives a different contribution, at sub-leading order,
than the integral of equation~(\ref{eq:TLD:keepcutoff}). What is going on?
The point is, that whilst going from equation~(\ref{eq:TLD:keepcutoff})
to equation~(\ref{eq:TLD:keepcutoff-B}) is usually a perfectly valid step, it breaks down when these
terms are under the influence of $\COMP$. Specifically, because the effect of $\COMP$ has been
to replace cutoff functions with a restricted range of integration,
we are no longer justified in throwing away what would previously have been total
momentum derivative terms.

Nonetheless, it is technically useful to be able to move the momentum derivative around,
whilst being able to forget about any total momentum
derivatives. We can do this if we employ
the correct prescription:  reinstate a term
to both the parent and subtraction (after they are under the influence of $\COMP$)
such that, at sub-leading order, we can move
momentum derivatives around with impunity. To understand what this term must be,
let us return to equation~(\ref{eq:TLD:keepcutoff}). 
Rather than discarding the cutoff function
straight away, we will first allow the momentum derivative to act.
 
Doing so, averaging over angle and substituting
$x = k^2$ yields:
\[
\frac{N \! \AngVol{D}}{D} \Box_{\mu \alpha}(p) (\Lambda^{-2\epsilon})^\bullet \int dx x^{1-\epsilon}
\left(\frac{\InvCO'_x}{x} - \frac{\InvCO_x}{x^2} \right).
\]
Integrating the first term by parts, and discarding the resulting surface term gives:
\[
\frac{N \!\AngVol{D}}{D} \Box_{\mu \alpha}(p) \int dx \frac{\InvCO_x}{x^{1+\epsilon}}(\epsilon-1).
\]
Now if we remove the cutoff function and restrict the range of integration, 
it is apparent that the $\InvCO'$ term has provided a sub-leading contribution;
indeed, this is precisely the sub-leading contribution we are after!

We have seen how, when cutoff functions are necessary for UV regularisation,
their derivatives can supply sub-leading contributions in the IR. This can be
rephrased by saying that, under the influence of $\COMP$, total momentum derivative
contributions can no longer be discarded, at sub-leading order, 
unless we reinstate terms to parent
and subtraction.

When dealing with
automatically regulated integrals, however, total momentum derivatives can be thrown away,
even under the influence of $\COMP$. This follows because the range of integration need not be restricted,
even after we have evaluated any cutoff functions at zero momentum. Equivalently,
in this case, we can move momentum derivatives around without the need to reinstate the derivative
of cutoff functions.

Returning to diagrams under the influence of $\COMP$
for which total momentum derivatives cannot be discarded,
we note that there is a downside to the prescription that we reinstate
derivatives of cutoff functions to both parent and subtraction.
Let us suppose that the subtraction involves
a momentum derivative striking an effective propagator
and that it happens to be manipulable. 
If, under the influence of $\COMP$, we do not reinstate any terms to the
parent and subtraction, then we can perform all diagrammatic manipulations,
since both the effective propagator relation
and the gauge invariance identities hold. However, if we do reinstate terms, then we would
have to supplement the usual 
diagrammatic identities
with corrections.
Generally, we will not perform any manipulations if this
correction is present but will choose to do so, if it is not.

\subsection{Application to Terms Manipulable at \Op{2}}		\label{sec:TotalDerivativeMethods:Op2-NTE}

The application of the subtraction techniques we have described is not limited
to $\Lambda$-derivative terms, but is useful for any class of terms in which 
we wish to perform Taylor expansions. We have already encountered such manipulations
when we dealt with the \Op{2}\ terms in the $\beta_1$ diagrammatics. For example,
the elements of the little set were derived from diagrams which were Taylor expanded
in $p$. Let us return to this by considering  one of the parents of the
little set, diagram~\ref{d1:2.12}, which is
reproduced in figure~\ref{fig:TLM-M:TE-Ex-A}.
\begin{center}
\begin{figure}[h]
	\[
	4 \LOLD[0.1]{Diagram2.12-Lab}{d1:2.12-Lab}
	\]
\caption{Reproduction of diagram~\ref{d1:2.12}, with explicit momentum routing.}
\label{fig:TLM-M:TE-Ex-A}
\end{figure}
\end{center}

Due to the fact that the structure carrying momentum $l-p$ is a zero-point wine,
rather than an effective propagator, it is clear that $p$ is not required to 
regulate this diagram in the IR; hence we can expand not only the vertex but also
the wine to zeroth order in $p$. For this diagram, there is no need to construct
a subtraction. Indeed, at one-loop, there is never any need to construct subtractions for diagrams
manipulable at $\Op{2}$. 

At two-loops, however, the situation is different. Consider the first diagram shown 
in figure~\ref{fig:TLM-M:TE-Ex-B}. In anticipation of what follows, we have constructed a
subtraction.
\begin{center}
\begin{figure}[h]
	\[
		\begin{array}{ccc}
		\vspace{0.1in}
			\LOBl{d:111b.11}	&		& \LObLOb{d:111b.11-s}{d:111b.11-a}
		\\
			\ensuremath{\begin{array}{c}\input{pstex/Diagram111b.11.pstex_t} \end{array}} & \mp 	& \ensuremath{\begin{array}{c}\input{pstex/Diagram111b.11-s.pstex_t} \end{array}}
		\end{array}
	\]
\caption{A two-loop diagram with an $\Op{2}$ stub, which cannot be Taylor
expanded in $p$, and its subtraction.}
\label{fig:TLM-M:TE-Ex-B}
\end{figure}
\end{center}

Note that the second diagram comes with two labels, the first of which refers
to the subtraction and the second of which refers to the addition.

We begin by focusing on diagram~\ref{d:111b.11}.
Since we can always Taylor expand vertices in momenta, let us suppose that
we take a power of $l$ from the top-most vertex (we cannot take any powers of $p$, at \Op{2}) and
let us choose to take a power of $k$ from the other vertex. The leading IR behaviour
of the $l$-integral is now
\[
	\int_l \frac{1}{l^2 (l-p)^2};
\]
this is not Taylor expandable in $p$. Note that had we taken a power of $l$ from
the right-hand vertex, rather than a power of $k$, then the extra power of $l$
in the integrand would render the diagram Taylor expandable in $p$.

Now let us consider the subtraction and addition. The addition (diagram~\ref{d:111b.11-a})
 is manipulated in the
usual way; this is basically what we would like to have done with diagram~\ref{d:111b.11},
in the first place. The effect of the subtraction on the parent is to cancel
all those components which are Taylor expandable in $p$. This immediately tells us
the following about any surviving contributions to diagram~\ref{d:111b.11}:
\begin{enumerate}
	\item	all fields carrying momentum $l$ must be in the $A$-sector;

	\item	we must take $\mathcal{O}(l^0)$ from the $k$-integral (note that
			the $k$-integral is Taylor expandable in $l$);

	\item	we must discard any remaining contributions to the $l$
			integral which do not $\sim \Pep$.
\end{enumerate}

The contributions to diagram~\ref{d:111b.11} not removed by diagram~\ref{d:111b.11-s} are shown in figure~\ref{fig:TLM-M:TE-Ex-C}.
\begin{center}
\begin{figure}[h]
	\[
		\left[
			\LOLD[0.1]{Diagram111b.11-Rem}{d:111b.11-Rem}
		\right]_{\Pep}
	\]
\caption{The contribution to diagram~\ref{d:111b.11} not removed by its subtraction.}
\label{fig:TLM-M:TE-Ex-C}
\end{figure}
\end{center}

As required, we have taken the $\mathcal{O}(l^0)$ from the $k$-integral. 
Though drawn as a dummy field, the field attached to the circle representing
the derivative is implicitly in the $A$-sector; this follows because the only
derivatives of vertices we consider are those which arise from a field in the $A$-sector
carrying zero momentum.
The tag \Pep\
demands that we take the \Pep\ component of the diagram.
Note, of course, that this tag implicitly assumes that we are using dimensional
regularisation. However, were we to use some other means of
regulating IR divergences, diagrams such as~\ref{d:111b.11-Rem} would
still exist, but the tag would be appropriately generalised.

\section{Ensuring Universality} \label{sec:TLD:Couplings}

The central tenet of our analysis of the $\Lambda$-derivative terms
has been that the derivative \wrt\ \LConstAl\
of a dimensionless integral must vanish, unless
there is a scale other than $\Lambda$, from which we can construct dimensionless
quantities. Implicit in this is that there are no dimensionless 
running\footnote{Where, strictly, we mean running \wrt\ \LConstAl.} 
couplings,
hidden in the integrand. 

The most obvious candidates for dimensionless running couplings can immediately be discounted:
$g$ counts the loop order, and so never appears in loop integrals and the presence of
$\alpha$ is irrelevant, since it is held constant, when differentiating \wrt\ $\Lambda$.
The first question we must address is whether there are
actually any other candidates for dimensionless running couplings.

To see how they could arise, in principle, consider the flow of any
vertex, with mass dimension $\geq 0$. Now Taylor expand in momenta
and focus on the term which is the same order in momenta as the 
mass dimension of the vertex. The coefficient of this term must be
dimensionless; if this coefficient flows, then we have 
found what we are looking for.

As a first example, let us consider the flow of an $m$-loop
vertex, decorated by an arbitrary number, $q$, of $C$s. We take the $C$s to
carry momenta $r_i$. Recalling
that $C$s are of mass dimension zero~\cite{YM-1-loop,GI-ERG-II},
all such vertices are of mass dimension four. Hence, we are interested
in the $\Omom{4}$ component of each of the vertices.

The crucial point for what follows is that, no matter what the value of $m$,
the flow is guaranteed to produce a certain type of term: specifically,
we will always have a dumbbell structure consisting of a two-point, tree
level vertex, joined by an un-decorated wine to a seed action vertex.
This is illustrated in figure~\ref{fig:TLD:FlowC^q}.
\begin{center}
\begin{figure}[h]
\[
	\begin{array}{ccccccc}
	\vspace{0.1in}
						&	&\LabOLOb{Dumbbell-TLTP-C-UW-Cq}&									&\LabOLOb{Dumbbell-TLTP-C-UW-Cq-B}	&
	\\
		\ensuremath{\begin{array}{c}\input{pstex/Vertex-Cq.pstex_t} \end{array}} 	&= -&\ensuremath{\begin{array}{c}\input{pstex/Dumbbell-TLTP-C-UW-Cq.pstex_t} \end{array}}		&- \hspace{1em}\cdots \hspace{1em} -	& \ensuremath{\begin{array}{c}\input{pstex/Dumbbell-TLTP-C-UW-Cq-B.pstex_t} \end{array}} 		& -\cdots
	\end{array}
\]
\caption{The flow of a vertex decorated by an arbitrary
number of $C$s.}
\label{fig:TLD:FlowC^q}
\end{figure}
\end{center}

The first ellipsis denotes diagrams of the same structure as~\ref{Dumbbell-TLTP-C-UW-Cq} 
and~\ref{Dumbbell-TLTP-C-UW-Cq-B} but for which a different $C$ decorates the two-point,
tree level vertex. Each of these diagrams possess a 
seed action vertex. These seed action vertices are the highest loop vertices which appear;
moreover, all other vertices generated at this loop order possess fewer legs.
The second  ellipsis denotes the remaining terms generated by the flow.

Focusing on the \Omom{4}\ components of all diagrams generated by the flow,
we now tune  the $m$-loop, $q$-point, seed action  vertices to exactly
cancel the remaining terms. This choice of seed action is one we are entirely at
liberty to make; it ensures that there are no hidden running couplings in 
this sector of the calculation. It perhaps seems a little artificial that
we only ensure universality after some (implicit) choice for the seed action.
We must remember, though, that $\beta$-function coefficients are not strictly
universal, and that scheme dependence even at one-loop is not necessarily 
a sign of a sick formalism. Our choice of seed action is merely done to allow
comparison of the values we compute for $\beta_1$ and $\beta_2$ with
those computed in, say, $\overline{MS}$.

In anticipation of what follows, we emphasise that the crucial ingredient in
what we have just done is
that the flow of an $m$-loop, $q$-point vertex generates an $m$-loop, $q$-point 
seed action vertex. Moreover, there are no vertices generated with higher loop
order, and for the rest of the same loop order, the number of legs is $< q$.
Consequently,
for each Wilsonian effective action vertex whose
flow we compute, it is a different seed action vertex we tune. This ensures
that we are never in the situation where we have to try and
tune the same seed action vertex in two different directions.

Let us now move on to consider an $m$-loop vertex decorated by $q$ 
$C$s and also by a single $A$ (we do not care which sub-sector this
field is in). The vertex is now of mass dimension three and
so it is the \Omom{3}\ part we are interested in. This time, we are guaranteed
to generate  $m$-loop, $q+1$-point
seed action vertices, joined to a  two-point, tree-level vertex by an undecorated
wine. Now, however, we see a difference between this case and the previous one:
the tree-level, two-point can be decorated by either an $A$ or a $C$. We illustrate
this in figure~\ref{fig:TLD:FlowAC^q}, where we take the $A$ to carry momentum $p$
and the $C$s to carry momenta $r_i$.
\begin{center}
\begin{figure}[h]
\[
	\begin{array}{ccccccc}
	\vspace{0.1in}
						&	&\LabOLOb{Dumbbell-TLTP-C-UW-A-Cq}	&									&\LabOLOb{Dumbbell-TLTP-A-UW-A-Cq}	&
	\\
		\ensuremath{\begin{array}{c}\input{pstex/Vertex-A-Cq.pstex_t} \end{array}}&= -&\ensuremath{\begin{array}{c}\input{pstex/Dumbbell-TLTP-C-UW-A-Cq.pstex_t} \end{array}}		&- \hspace{1em}\cdots \hspace{1em}-	& \ensuremath{\begin{array}{c}\input{pstex/Dumbbell-TLTP-A-UW-A-Cq.pstex_t} \end{array}}		& -\cdots
	\end{array}
\]
\caption{The flow of a vertex decorated by a single $A$ and an arbitrary
number of $C$s.}
\label{fig:TLD:FlowAC^q}
\end{figure}
\end{center}

The first ellipsis denotes diagrams of the same structure 
as~\ref{Dumbbell-TLTP-C-UW-A-Cq} but for which a different $C$ 
decorates the two-point, tree level vertex. The final
ellipsis denotes the remaining terms generated by the flow equation.

Having decorated with an $A$, we must now take account of gauge invariance.
The most obvious effect is that the two-point, tree level vertex of 
diagram~\ref{Dumbbell-TLTP-A-UW-A-Cq}, unlike that of 
diagram~\ref{Dumbbell-TLTP-C-UW-A-Cq}, is forced to be $\Op{2}$. Given that
we are working at \Omom{3}, this means that we must take a single power
of momentum from the corresponding $m$-loop, seed action vertex.
Now, if this single power of momentum is $p$ then, by Lorentz invariance,
its index must be contracted with the two-point, tree level vertex---killing
it. Hence, this single power of momentum must be one of the $r_i$. In turn,
this means that we can Taylor expand the $m$-loop, seed action vertex 
of diagram~\ref{Dumbbell-TLTP-A-UW-A-Cq} to zeroth
order in $p$; the effect of this is to reduce it to the derivative of
an $m-1$-point vertex. This means that gauge invariance has caused us to lose
one of our $m$-loop, $q+1$-point seed action vertices. We can still tune the flow
of our Wilsonian effective action vertex to zero, though, by virtue of the
presence of diagram~\ref{Dumbbell-TLTP-C-UW-A-Cq} and the diagrams represented by
the first ellipsis.

Let us examine the tuning of the seed action vertices in a little more
detail. To do this, consider contracting  the diagrams of figure~\ref{fig:TLD:FlowAC^q}
with the momentum carried by the $A$. On the \lhs\ of the equation, we now have
the flow of (a set of) $m$-loop vertices, decorated by $q$ $C$s. Since we were
working at \Omom{3} but have contracted with a power of momentum, we should now
be looking at \Omom{4}. However, we know from our work on pure-$C$ vertices that
the flow of such terms has already been tuned to zero. Thus, returning
to the diagrams of figure~\ref{fig:TLD:FlowAC^q}, gauge invariance ensures
that we need only tune the seed action vertices so as to remove those \Omom{3}\
contributions transverse in $p$! Note that this reproduces the conclusions of~\cite{YM-1-loop},
in which the special case of the flow of a tree level $ACC$ vertex was considered.

Next, we extend our
analysis to a vertex decorated by two $A$s and $q$ $C$s. We now work at \Omom{2}.
If we take $q>0$, then our analysis just mirrors what we have done:
to avoid dimensionless,  running couplings, gauge invariance ensures that we need
only tune the seed action vertices  transverse in the momenta of the $A$s. What if $q$=0? Now the 
renormalisation condition guarantees that the flow of the vertex vanishes;
this is, of course, exactly what we have been utilising to compute $\beta$-function
coefficients.

In the case of a vertex decorated by three $A$s and $q$ $C$s, we work
at \Omom{1}. Any terms involving two-point, tree level vertices decorated by
$A$s vanish, at the desired order in momentum. Consequently, irrespective of
the value of $q$, gauge invariance---in conjunction with what we
have just done---ensures that the flow of our vertex 
vanishes at \Omom{1}, without the need for any further tuning. Similarly, 
this result implies that gauge invariance 
can be used to demonstrate that the flow
of a vertex decorated by four $A$s and any number of $C$s vanishes,
without the need for further tuning.

Our final task is to extend this analysis to include fermionic fields.
To do this, we will treat $B$s and $D$s separately, due to their
differing mass dimensions. The point here is that neither
$S_{0 \alpha \beta}^{\ B \bar{B}}(k)$ nor $S_0^{\ D\bar{D}}(k)$ vanishes at 
zero momentum. Thus, whereas gauge invariance can force some or
all of the highest loop order seed action vertices with the maximum number of
legs
to be written as derivatives of lower point vertices when $A$s are
amongst the decorative fields, no such thing happens here.
Thus, both $B$s and $D$s behave essentially like the $C$s, of the previous analysis.

As a final point, we might worry about the vertex 
$S_{0 \alpha}^{\ B \bar{D}}(k)$---which vanishes at zero momentum. 
However, supposing that it is the $B(\bar{D})$
that is the internal field, this vertex will always be accompanied by a term
in which there is a $D(\bar{B})$ as an internal field. It is the seed action
vertex at the other end of the corresponding dumbbell which is the one we tune.

We have thus demonstrated that all dimensionless couplings, other than $g$
and $\alpha$, can be prevented from running by a suitable choice of the seed
action.

\newpage
\pagestyle{empty}
	\TitlePage{II}{March 2005}
\newpage
\pagestyle{plain}

\part{$\beta_{n_+}$ Diagrammatics}

\chapter{Introduction} \label{ch:bn:Intro}

In chapter~\ref{ch:beta1}, we saw how it is possible to derive an expression for
$\beta_1$ in terms of $\Lambda$-derivatives, with all other diagrams 
cancelling out.  We noted how the structure of the calculation suggested
that many of these cancellations could be done in parallel.

We now explore this idea in much greater depth,
viewing the one-loop cancellations
as a subset of the cancellations that take place in the
computation of the arbitrary $\beta$-function coefficient, $\beta_{n+1} \equiv \beta_{n_+}$.

The major difference between the diagrammatic methodology we will employ
for the computation of $\beta_{n_+}$ compared to that
used for $\beta_1$ is the use
of the generalised notation of figure~\ref{fig:NFE:New-Diags-1}.
The idea behind this notation is to exploit one of the diagrammatic
effects of manifest gauge invariance. When we compute the flow of
a vertex decorated by some set of fields, $\{f\}$, manifest gauge invariance
tells us that we can represent the resulting diagrams by some base structure
which is then decorated in all possible ways by the fields $\{f\}$.
In chapter~\ref{ch:beta1}, we performed all these decorations explicitly,
generating a large number of diagrams.
In the subsequent chapters, however, we will leave these decorations largely
implicit and doing this will save us a huge amount of time.

Indeed, we will find that partial decoration, at most, is necessary
for diagrammatic manipulations to be performed. As anticipated, this
allows us to process large numbers of terms in parallel. Moreover, when
we come to cancel these terms, we will find that we are able to do so
without further decoration; the entire diagrammatic procedure is
radically simplified.

However, in its current form, the generalised methodology is not
sufficiently developed, for our purposes. 
Hence, in the initial 
stages, 
our approach is pedagogical. 
We start our computation of $\beta_{n_+}$ in chapter~\ref{ch:bn:LambdaDerivsI},
where we begin by 
explicitly converting diagrams
into $\Lambda$-derivative terms, plus corrections. The first purpose
this serves is to allow us to furnish the generalised methods with
the desired set of features. Having done this, we then
proceed with the usual diagrammatic procedure: we identify two-point,
tree level vertices and apply the effective propagator relation, as
appropriate. This then allows us to identify cancellations, 
a subset of which we could map back to the special case of $n=0$.

However, as we iterate the procedure, it becomes clear that the
cancellations we identify within our generalised framework
are simply repeated with each iteration. This
guides us from explicit cancellations to cancellation mechanisms, which 
guarantee that certain types of diagram will be cancelled
in a calculation of $\beta_{n_+}$. Once we have identified these mechanisms,
we can dispense with explicit computation; rather, we can immediately write down 
diagrammatic expressions, valid to all orders in perturbation theory,
which give the result of iterating
the diagrammatic procedure, until exhaustion.

In section~\ref{sec:bn:critical} we arrive at 
a diagrammatic form for $\beta_{n_+}$ in terms of $\Lambda$-derivative,
$\alpha$ and $\beta$-terms, up to gauge remainders and diagrams which
require manipulation at $\Op{2}$. 

The next task is to examine the gauge remainder
diagrams. Once again, the heavily compacted notation utilised in this
chapter makes the problem tractable. The initial manipulation of
the gauge remainders is done in chapter~\ref{ch:bn:GRs}. Having found that many
of the terms generated cancel amongst themselves, we then introduce a set of
diagrammatic identities in chapter~\ref{ch:FurtherD-ID} to facilitate further progress.
The validity of the final identity, which applies at three loops
and beyond, is only partially demonstrated and so we make an unproven
assertion that it is true in general.
In chapter~\ref{ch:bn:Iterate} we find that, aided by these identities, we
can iterate the diagrammatic procedure, converting the surviving gauge remainder
terms into further $\Lambda$-derivative, $\alpha$ and $\beta$ terms. This allows us
to supplement our previous expression for $\beta_{n_+}$, to explicitly include
the effects of processing all gauge remainders. 

The final thing to be done is to
process those terms which require manipulation at $\Op{2}$. Unfortunately, it
is beyond the scope of this thesis to do this, though we comment on
some of the associated issues in chapter~\ref{ch:Op2}.

Before embarking on the analysis, proper, some preliminary remarks 
are in order. We will encounter three types of labelled
diagram in this
chapter. The first type are those arising from the initial, 
step by step reduction of $\beta_{n_+}$ to $\Lambda$-derivatives
and etc. These diagrams
will be labelled 
\textbf{chapter.\#}. 

The second type are so called general
diagrams which can arise from any level of
iteration; special cases of these diagrams reduce to the explicitly
generated
diagrams. General diagrams are labelled
\textbf{G.\#}. 

The third type are the  illustrative diagrams and are 
labelled \textbf{I.\#}.

In addition to diagrams being processed, discarded and cancelled
we will encounter a further option: partial cancellation.
If diagram~$Y$ is partially cancelled by diagram~$X$,
then the label above diagram~$X$ would be
\[
\mbox{\textbf{\{$\mathbf{X}$}}
\ Y(P) 
\mbox{\textbf{\}}}.
\]
If the remaining part of diagram $Y$ partially
cancels a further diagram, $W$, then the label above
diagram $Y$ would be
\[
\mbox{\textbf{\{$\mathbf{Y}$}}
\ X, W(P) 
\mbox{\textbf{\}}}.
\]

Unlabelled diagrams are those which are never referred to
explicitly, though they are implicitly referred
to in the generalised  formalism which we develop.

\chapter{Initial Recasting} \label{ch:bn:LambdaDerivsI}

In this chapter we will generate an expression for
$\beta_{n_+}$ in  terms of $\Lambda$-derivative, $\alpha$
and $\beta$-terms, up to gauge remainders and  diagrams with
an $\Op{2}$ stub.
The methodology we employ will be developed
as we proceed; consequently, the ideas presented
at the beginning of the chapter are refined, as we
get deeper into the calculation.

Indeed, to begin with, we will mirror the calculation of
$\beta_1$: starting from the flow equation~(\ref{eq:WeakCouplingExpansion-NewFlow-B}),
we write down a diagrammatic expression for $\beta_{n_+}$. We call
this the zeroth level of the calculation. At the first level,
we perform the usual diagrammatic procedure of converting manipulable
diagrams into $\Lambda$-derivative terms, plus corrections. Iterating the procedure,
we present the results of the second and third level manipulations. This
is sufficient to enable us to refine the formalism and
to deduce the result of manipulations performed at an arbitrary level.

\section{Level-Zero Manipulations} \label{sec:bn:L0}

To compute $\beta_{n_+}$, we start with flow equation~(\ref{eq:WeakCouplingExpansion-NewFlow-B})
specialised to $\{f\} = A^1_\mu(p), A^1_\nu(-p)$ and $n=n+1$ and work at $\Op{2}$:
\begin{equation}
-4 \beta_{n_+} \Box_{\mu \nu}(p) + \Op{4} = a_1[\Sigma_n]^{1\, 1}_{\mu \nu}(p) - \sum_{r=0}^{n_+} a_0[S_r,\Sigma_{{n_+}_r}]^{1\, 1}_{\mu \nu}(p)
\label{eq:Beta_n+:Defining}
\end{equation}
This is represented in figure~\ref{fig:beta_n+:Defining}. 
Unless stated 
otherwise, we will take the indices of the two external fields to be $\mu,\nu$ and work at $\Op{2}$. In preparation for later,
when we will encounter diagrams containing many vertices, it will prove useful to have a tag for each vertex, independent of the vertex argument.
Henceforth, we label vertices alphabetically, in the counter-clockwise sense; the bottom left vertex of a diagram being $A$.
\begin{center}
\begin{figure}[h]
	\[
	-4 \beta_{n_+} \Box_{\mu \nu}(p) + \Op{4} = 
	\frac{1}{2}
	\dec{
		\PED[0.1]{4}{Sigma_n}{bn:L0:a1}{fig:bn+:L0:a1:sep}
		-
		\sum_{r=0}^{n_+}
		\PED[0.1]{6}{Beta-n+L0-a0}{bn:L0:a0}{fig:Beta-n+L1:a0:pre}
	}{11} 
	\]
\caption{A diagrammatic representation of the equation for $\beta_{n_+}$. Vertices are tagged alphabetically.}
\label{fig:beta_n+:Defining}
\end{figure}
\end{center}

Although we will not perform any decorations at this stage, it is worth noting that
the decoration of
diagrams~\ref{bn:L0:a1} and~\ref{bn:L0:a0} is straightforward. Following
section~\ref{sec:NFE:IdenticalFields}, if we were to place the two 
identical external fields on different structures we would pick up
a factor of two whereas, if we were to place them on the same structure, we would not. 

It is also worth examining the behaviour of the decorated diagrams at $\Op{2}$.

Examining diagram~\ref{bn:L0:a1} at $\Op{2}$
is trivial: none of the component diagrams vanish at $\Op{2}$ and so we need do nothing further. 
Diagram~\ref{bn:L0:a0}, however,
is more subtle. If we attach one external field to each of the vertices, then the resulting diagram vanishes at $\Op{2}$. Whilst it
is completely legitimate to discard such terms, we choose to retain (some of) them. The reason for this is as follows.

The components of diagram~\ref{bn:L0:a0} containing Wilsonian effective action vertices joined by an undecorated wine
can be manipulated, using the flow equations. The salient point is that a set of the diagrams generated do not,
individually, vanish at $\Op{2}$. Of course, it must be the case the sum 
over the elements of this set has no $\Op{2}$ component. 
Crucially, we will see that these diagrams cancel diagrams
arising from the manipulation of diagram~\ref{bn:L0:a1}. 
For the computation of $\beta_2$, at any rate, this
allows us to bring the diagrammatic expression into a form from which we can extract the numerical value in $D=4$.

These considerations are not restricted to the zeroth level of the calculation. At each level we will be confronted
with terms that vanish at $\Op{2}$ and, rather than throwing them away, we choose to process
the manipulable parts.
The only exception to this rule concerns diagrams containing two-point, tree level vertices. Whilst it
is possible to manipulate such diagrams, it is not useful to do so. The reason for this is easily seen by considering
the $r=0$ component of diagram~\ref{bn:L0:a0}, with an external field attached to each of the vertices.
If we were to manipulate this diagram, we just end up computing the flow of $S^{\ \ \ 11}_{n_{+} \mu \alpha}(p)$ which is,
of course, what we did to obtain equation~(\ref{eq:Beta_n+:Defining}), in the first place! Hence, this manipulation
will just increase our work, without giving anything new.

Note that any terms which manifestly vanish at $\Op{2}$ and which cannot be manipulated can 
be discarded. However, we choose to retain a subset of 
them as it turns out that they are cancelled
in parallel with terms that do not vanish in this way. Hence, it would
be inefficient to try and separate them out.

Returning to diagram~\ref{bn:L0:a0},
if we do not attach any external
fields to one (both) of the vertices, then the corresponding vertex 
is a one-point vertex and so has no Wilsonian effective action contribution~\cite{YM-1-loop}.
However, just as we choose to manipulate the terms which vanish at
$\Op{2}$, so too do we choose to manipulate one-point vertices. Again, it is perfectly valid to
discard such terms---indeed, this is the strategy employed in the one-loop
calculation. On the other hand, by manipulating them we  
avoid having to enforce
the constraint~(\ref{eq:NFE:C-Constraint:Weak});
rather, \emph{all} diagrams involved in this constraint equation will be
automatically
cancelled by diagrams generated by the manipulation of one-point
Wilsonian effective action vertices. When the dust has settled,
all that we are left with is $\Lambda$-derivative terms containing
one-loop Wilsonian effective action vertices. Now we utilise
the fundamental constraint that such vertices vanish to remove
these terms.

\section{Level-One Manipulations} \label{sec:bn:L1}

\subsection{Diagram~\ref{bn:L0:a1}} \label{sec:bn:L0:a1}

We know from the one-loop calculation that if, in diagram~\ref{bn:L0:a1}, we take the Wilsonian effective action contribution
and attach both external fields to the vertex, then we can process this diagram along the lines of figure~\ref{fig:Beta1:FirstManip}.

The first step of this procedure is to isolate the manipulable component. Taking only the Wilsonian effective action contributions to 
the vertex is trivial. However, we now need to introduce some notation to denote whether or not the wine is decorated.
To indicate an undecorated wine,
we enclose the $\flowConstAl$ in a box, which we take to mean that we decorate \emph{first}
and then differentiate. Since it is illegal to  decorate effective propagators, this has
the desired effect. 

To indicate a decorated wine, we define an object called the \emph{reduced} wine:
\begin{equation}
	\left(\dot{\Delta}\right)_R \equiv \stackrel{\circ}{\Delta} =  \dot{\Delta} - \dot{\Delta}^{XY}_{ST}(k),
\label{eq:RW}
\end{equation}
where we have suppressed all arguments of the generic wine, $\dot{\Delta}$, and its reduction.
The isolation of the manipulable component of diagram~\ref{bn:L0:a1} is shown in figure~\ref{fig:bn+:L0:a1:sep}.
\begin{center}
\begin{figure}[h]
	\[
		\frac{1}{2}
		\dec{
			\PED[0.1]{4}{Sigma_n-M}{bn:L0:a1:M}{fig:beta_n+:Level1} + 
			\PDCED[0.1]{}{Sigma-n-DW}{bn:L0:a1:DW}{bn:L0:a1:C-W}{bn:L0:a1:S}
			-2 \PDCED[0.1]{}{Sigma_n-S}{bn:L0:a1:S}{bn:L1:a0-Both2ptI-A}{bn:L0:a1:DW}
		}{11}
	\]
\caption{Isolation of the manipulable component of diagram~\ref{bn:L0:a1}.}
\label{fig:bn+:L0:a1:sep}
\end{figure}
\end{center}

We now process diagram~\ref{bn:L0:a1:M},
as shown in figure~\ref{fig:beta_n+:Level1}. The key thing
to notice is that we have promoted the effective propagator
of the $\Lambda$-derivative term, and the associated internal
fields to which it attaches, to an unrealised  decoration. This is denoted by
$\{I^2, \Delta\}$. 
For the $\Lambda$-derivative
term, decoration is performed \emph{before} the diagram is hit by $\flowConstAl$. Indeed, we remind ourselves of
this by the square brackets around the $\Lambda$-derivative term. In the absence of these, we would decorate
$\emph{after}$ the vertex is hit by $\flowConstAl$.
Henceforth, we will often loosely refer to the promotion of objects to unrealised decorations
simply as promotion to decorations.
\begin{center}
\begin{figure}[h]
	\[
	\frac{1}{2}
	\dec{
		\begin{array}{c}
		\vspace{0.2in}
			\displaystyle
			\dec{\ensuremath{\begin{array}{c}\input{pstex/./Beta-n+L1-LdL.pstex_t} \end{array}}}{\bullet}
			-\sum_{r=1}^{n}
			\left[
				2(n_r-1)\beta_r \ensuremath{\begin{array}{c}\input{pstex/Vertex-n_r.pstex_t} \end{array}}
				+\gamma_r \pder{}{\alpha} \ensuremath{\begin{array}{c}\input{pstex/Vertex-n_r.pstex_t} \end{array}}
			\right]
		\\
			\displaystyle 
			+\frac{1}{2}\sum_{r=0}^{n} \PED[0.1]{4}{Beta-n+L1-a0}{bn:L1:a1:a0}{fig:beta_n+:L1:a0:tree}
			+\frac{1}{2} \PED[0.1]{4}{Beta-n+L1-a1}{bn:L1:a1:a1}{fig:Beta-n+L2:a1:a1}
		\end{array}
	}{11\{I^2,\Delta\}}
\]
\caption{The result of processing  diagram~\ref{bn:L0:a1:M}.}
\label{fig:beta_n+:Level1}
\end{figure}
\end{center}

Comparing with the one-loop calculation this figure is, up to the
$\Lambda$-derivative term, the analogue
of figure~\ref{fig:beta1:LevelOne}. Note, though, that when we specialise
the diagrams of figure~\ref{fig:beta_n+:Level1} to $n=0$, the second, third
and fifth diagrams must be discarded, else they have an (illegal) negative
vertex argument. Thus, all the diagrams of figure~\ref{fig:beta1:LevelOne}
are generated entirely by decorating diagram~\ref{bn:L1:a1:a0} (for $n=0$).
This huge simplification has come about not just through leaving the external
fields as unrealised decorations but also due to the similar promotion
of the effective propagator. This latter step will prove central
to the following analysis.

Our first task in analysing the diagrams of figure~\ref{fig:beta_n+:Level1}
is to determine the combinatorics associated with attaching the various fields
to the diagram. We have already discussed the attachment of external fields, in the previous section. When we decorate
with internal fields, what we will ultimately be doing is joining them together with an effective propagator. The rule
is simply that, if the effective propagator starts on one structure and ends on another, then we must pick up
a factor of two which recognises that we can choose which end of the effective propagator to attach to which structure.
If both ends of the effective propagator attach to the same structure, then there is no such factor.

As an example of how this works, we will reproduce the factors of the diagrams~\ref{d1:1.14} and~\ref{d1:1.15}, by
attaching fields to (the Wilsonian effective action component of) diagram~\ref{bn:L1:a1:a0}.
To reproduce diagram~\ref{d1:1.14}, we must attach the external fields to the same vertex and the effective 
propagator must start and end on the same vertex.
However, we could have decorated either of the vertices
with the external (internal) fields and so this gives us a factor of two. Taking into account the overall
factor of $-1/4$ in front of diagram~\ref{bn:L1:a1:a0} exactly reproduces diagram~\ref{d1:1.14}.

Now, to reproduce diagram~\ref{d1:1.15}, we must attach one external field and one internal field to each vertex.
This gives a factor of four. However, due to the symmetry of the diagram, this exhausts all possibilities, yielding
an overall factor of $-1$, as required.

We must now consider the implicit decoration of diagrams~\ref{bn:L1:a1:a0} and~\ref{bn:L1:a1:a1}
with the two external fields and an effective propagator. The resultant terms will split into four classes:
\begin{enumerate}
\item terms which can be manipulated by applying the effective propagator relation;

\item terms which can be manipulated at $\Op{2}$;

\item terms which can be  processed using the flow equations;

\item terms with which we can do nothing.
\end{enumerate}

From our experiences with the one-loop calculation, we expect terms in the first class to cancel
a subset of the un-manipulated contributions to diagram~\ref{bn:L0:a1}. We will see this explicitly
below.  Terms in the second class  will be commented on in section~\ref{ch:Op2}.
Terms in the third class
will be dealt with in section~\ref{sec:bn:L2}, yielding
diagrams which either directly, or through further manipulation, cancel
the terms of the fourth class.

To prepare for the iteration of the diagrammatic procedure, we first want to split off all
terms which can be manipulated directly \ie\ without the need for using the flow equations.
These terms are all those which contain a two-point, tree level  vertex. If this vertex is
decorated solely by
internal fields, then the resulting diagram is in the first class above; if the vertex is decorated
by an external field, then it is in the second class. All such terms are a subset of the components
of diagram~\ref{bn:L1:a1:a0} that occur when the summed variable $r$ takes its boundary values, zero or $n$.

To facilitate this separation, we define the reduced, $n$-loop vertex thus:
\[
v^{R}_n =
\left\{
\begin{array}{cc}
v_n & \mathrm{n>0} \\
v_0 - v^{\ XX}_{0 RS}(k) & \mathrm{n=0}
\end{array}
\right.
\]
where we have suppressed all arguments of the generic vertex $v_n$ and its reduction.
By definition, the reduced vertex does not contain a two-point, tree level component. 

We must take care rewriting diagram~\ref{bn:L1:a1:a0}
in terms of reduced vertices since the $n=0$ case behaves slightly differently from the $n>0$ case. 
When $n>0$, it is not possible for both vertices to be simultaneously tree level; hence we get only three terms.
One contains two reduced vertices and two contain one reduced $n$-loop vertex and a two-point,
tree level vertex. These latter two terms are identical and so can be combined. Moreover, in this case, the reduced $n$-loop vertex can 
be replaced by just an $n$-loop vertex. This step cannot be performed when $n=0$ since it 
will over-count the diagram comprising two two-point, tree level 
vertices---which do exist in this case---by a factor of
two. 

However, it is undesirable to explicitly draw the term comprising two two-point, tree level vertices since it is only present in 
the case when $n=0$. Rather, we introduce the  variable $1/\vartheta_n$,
which takes the value of $1/2$ when the following
is satisfied: the $n$-loop vertex is both tree level and two-point. Now, in the terms containing one two-point, tree level
vertex and one reduced vertex, we can replace the reduced vertex by a normal vertex for all $n$, so long as we multiply
by a factor of $1/\vartheta_n$.\footnote{As the calculation develops, we will
find that there is a way to isolate all two-point, tree level vertices, irrespective
of the value of $n$, in a unified manner. However, this relies on terms which have not
yet been generated.}

Finally, then, the separation of terms is shown in figure~\ref{fig:beta_n+:L1:a0:tree}
\begin{center}
\begin{figure}[h]
	\[
	-\frac{1}{4}
	\dec{
		\displaystyle \sum_{r=0}^{n}
		\PED[0.1]{4}{./Beta-n+L1-a0-Re}{bn:L1:a1:a0-Re}{fig:Beta-n+L1:a1:a0:sep}
		+ \frac{2}{\vartheta_n}
		\PED[0.1]{4}{./Beta-n+L1-a0-2pt}{bn:L1:a0-One2pt}{fig:beta_n+:L1:a0:tree:A}
	}{11\{I^2,\Delta\}}
	\]
\caption{Splitting off the tree level two-point, tree level vertices from diagram~\ref{bn:L1:a1:a0}.}
\label{fig:beta_n+:L1:a0:tree}
\end{figure}
\end{center}

We now partially decorate diagram~\ref{bn:L1:a0-One2pt} by specifying which field attaches to the two-point, tree level vertex.
To partially decorate with an external field, we simply attach one of them to the two-point, tree level vertex with a
factor of unity. We will only pick up the factor 
of two arising from the attachment of the remaining external field\footnote{
Since one of the external fields has attached to the two-point, tree level vertex, the other must necessarily attach to a different structure.}
when the attachment is 
actually made.\footnote{This prescription is designed to make 
partial decoration consistent with full decoration. When decorating an arbitrary diagram
with identical fields, we determine the associated factor only after completing the decoration.}

 When we partially decorate with
internal fields, what we choose to  do is attach one end of the effective propagator. When we decorate with the 
other internal field---which in this case necessarily attaches to a different structure---we will pick up a factor of two, which recognises the
indistinguishability of the internal fields. As with the external fields, this factor is suppressed until the final decoration is actually made. 
Figure~\ref{fig:beta_n+:L1:a0:tree:A} shows the result of this procedure, where
we note that a dumbbell with barred arguments simplifies if one of vertices
is a two-point, tree level vertex. This follows on account of
\begin{eqnarray*}
	a_0[\bar{S}_m, \bar{S}_{0RS}^{\ XY}(k)] 	& = & a_0[S_m, S_{0RS}^{\ XY}(k)] - a_0[S_m, \hat{S}_{0RS}^{\ XY}(k)] - a_0[\hat{S}_m, {S}_{0RS}^{\ XY}(k)]
\\
												& = & -a_0[\hat{S}_m, {S}_{0RS}^{\ XY}(k)],
\end{eqnarray*}
where we have used the equality of the Wilsonian effective action and seed action two-point, tree level
vertices.

In the case where we have attached one end of the effective propagator, the remaining decorative
fields include the internal field to which the other end must attach. This is denoted by $\{I\}$. However, after this section we 
will drop this notation since, if there is a loose end of an effective propagator present, it is implicit that it must attach elsewhere.
Moreover, when we decorate with an effective propagator, we will now drop the associated internal fields, since their presence is
implied.
\begin{center}
\begin{figure}[h]
	\[
	\frac{1}{2 \vartheta_n}
	\left[
		\dec{
			\PED[0.1]{4}{./Beta-n+L1-a0-2ptI}{bn:L1:a0-2ptI}{fig:beta_n+:L1:a0:EP}
		}{11\{I\}}
		+
		\dec{
			\ensuremath{\begin{array}{c}\input{pstex/./Beta-n+L1-a0-2ptE.pstex_t} \end{array}}
		}{1\{I^2,\Delta\}}
	\right]
\]
\caption{Partial decoration of diagram~\ref{bn:L1:a0-One2pt} with either an external field
or an effective propagator.}
\label{fig:beta_n+:L1:a0:tree:A}
\end{figure}
\end{center}

The next step is to attach the loose end of the effective propagator in
diagram~\ref{bn:L1:a0-2ptI}, as shown on the  LHS of figure~\ref{fig:beta_n+:L1:a0:EP}. 
Since the effective propagator is guaranteed to
start and end on a different structure, this will necessarily yield a factor
of two.
On the RHS of figure~\ref{fig:beta_n+:L1:a0:EP}, we apply the effective propagator relation,
discarding the diagram in which the wine bites its own tail. Henceforth, we will automatically
discard such diagrams.
\begin{center}
\begin{figure}[h]
	\[
	\frac{1}{\vartheta_n}
	\dec{
		\ensuremath{\begin{array}{c}\input{pstex/./Beta-n+L1-a0-2ptI-MA.pstex_t} \end{array}}
	}{11}
	=
	\frac{1}{\vartheta_n}
	\dec{
		\PCED[0.1]{}{Struc-n-S-W}{bn:L1:a0-Both2ptI-A}{bn:L0:a1:S}
		-\ensuremath{\begin{array}{c}\input{pstex/Struc-n-S-W-GR.pstex_t} \end{array}}
		-\ensuremath{\begin{array}{c}\input{pstex/Struc-n-S-WGR.pstex_t} \end{array}}
	}{11}
\]
\caption{Result of tying up the loose end in diagram~\ref{bn:L1:a0-2ptI}, followed by
subsequent application of the effective propagator relation.}
\label{fig:beta_n+:L1:a0:EP}
\end{figure}
\end{center}

We now encounter our first cancellation.  
Having split diagram~\ref{bn:L0:a1} into non-manipulable and
manipulable parts we expect, from our experiences
of the one-loop calculation, the former to be cancelled by manipulations of the latter.

\begin{cancel}
Diagram~\ref{bn:L1:a0-Both2ptI-A} has the same structure as diagram~\ref{bn:L0:a1:S}.
Irrespective of the value of $n$, diagram~\ref{bn:L1:a0-Both2ptI-A} exactly cancels the  contributions
to diagram~\ref{bn:L0:a1:S}, for which at least one of the external fields decorates the vertex. The
only remaining contribution to diagram~\ref{bn:L0:a1:S} is where both external fields decorate the wine
\ie\ the vertex is a two-point vertex. If $n>0$, this contribution is exactly cancelled by diagram~\ref{bn:L1:a0-Both2ptI-A}.
However, when $n=0$, the factor of $1/\vartheta_n$ means that only half the contribution is removed.
Hence, diagram~\ref{bn:L1:a0-Both2ptI-A} only partially cancels diagram~\ref{bn:L0:a1:S}.
\label{C:L0:a1:S}
\end{cancel}

This first cancellation removes those components of the parent diagram whose vertices are uniquely seed 
action. (The only surviving component, occurring for $n=0$, is a two-point, tree level vertex, for which the
seed action contribution is just equal to the Wilsonian effective action contribution. Hence, we
take such a
contribution not to possess a uniquely seed action vertex.) As we will see 
throughout this chapter, this is a general feature of the diagrammatic method: starting from some parent
diagram, we split off the non-manipulable terms from the manipulable ones. Diagrams in the former
class can be further sub-divided, depending on whether or not they possess any seed action vertices.
Upon manipulation of the parent diagram, these seed action vertices will be removed.

To cancel the remaining contributions to the parent diagram---\ie\ those non-manipulable components
possessing only uniquely Wilsonian effective action vertices---we will have to proceed to the next level 
of manipulation. The diagram which cancelled diagram~\ref{bn:L0:a1:S} was one the
components of diagram~\ref{bn:L1:a1:a0} which could be processed by using the effective propagator
relation. It is the contributions to diagram~\ref{bn:L1:a1:a0} that must be manipulated via the 
flow equations which will generate the terms necessary to complete the cancellation (see section~\ref{sec:bn:L1:a1:a0-Re})
of the non-manipulable components of diagram~\ref{bn:L0:a1}.

Again, we will find that this is a general feature of the diagrammatics. The non-manipulable
contributions to some parent diagram will generally be completely removed after two levels of manipulation.
At the first level, the uniquely seed action contributions will be cancelled whereas
the uniquely Wilsonian effective action contributions will be cancelled at the
second level.\footnote{In the case that there are any remaining contributions \ie\
those containing a two-point, tree level vertex then, in the case where this vertex
is decorated
solely by internal fields, the corresponding diagrams will cancel among themselves.}

\subsection{Diagram~\ref{bn:L0:a0}} \label{sec:bn:L0:a0}

We now process the components of diagram~\ref{bn:L0:a0} containing Wilsonian
effective action vertices and undecorated wines. 
As discussed in section~\ref{sec:bn:L0}
we do not manipulate any terms containing two-point, tree level vertices.
Indeed, we can discard all terms containing such a vertex. The two-point, tree level vertex 
is necessarily decorated by an external field and so the diagram is
at least $\Op{2}$. Since any two-point vertex decorated by an external field is at 
least $\Op{2}$,
the diagram vanishes at $\Op{2}$ if the remaining external field decorates the 
other vertex. If, instead, the remaining external field
decorates the wine, then the diagram vanishes in exactly the same way that the one-loop
diagram~\ref{d1:C-2} vanished.
 
Discarding all terms containing a two-point, tree level vertex, we isolate the remaining manipulable components,
as shown in figure~\ref{fig:Beta-n+L1:a0:pre}.
\begin{center}
\begin{figure}
	\[
	-\frac{1}{2} \sum_{r=0}^{n_+}
	\dec{
		\PED[0.1]{4}{Beta-n+L0-a0-M}{bn:L0:a0:M}{fig:Beta-n+L1:a0}
		+\CED[0.1]{4}{Beta-n+L0-a0-DW}{bn:L0:a0:DW}{bn:L2:a0:a0:a0:2ptx2} 
		-2\CED[0.1]{5}{Beta-n+L0-a0-S}{bn:L0:a0:S}{bn:L1:a0:a0-S}	
	}{11}
	\]
\caption{Isolation  of those components of 
 diagram~\ref{bn:L0:a0} which can be usefully processed.}
\label{fig:Beta-n+L1:a0:pre}
\end{figure}
\end{center}

The next step is to convert diagram~\ref{bn:L0:a0:M} into a $\Lambda$-derivative term. This procedure
will turn the wine into an effective propagator and generate two correction terms, in which the
$\Lambda$-derivative strikes each of the vertices.
Due to the invariance of the parent diagram under interchange of the vertex arguments, we choose to
compute the flow of only one of these vertices, 
but multiply by two.

Following our treatment of diagram~\ref{bn:L0:a1} we want to convert the newly formed
effective propagator into a decoration. However, when we come to reattach it,
our current rule for attaching effective propagators will give us an extra factor 
of two.\footnote{This follows because our current rule allows effective propagators joining two different
structures to attach either way round. In the case under consideration, though, the
effective propagator started out life as a wine, originally part of a dumbbell structure
formed by the flow.}
We could
demand that, for two unjoined structures, the first join does not come with a factor of two. Alternatively,
as we choose to do, we simply allow all joins to come with a factor of two but include an extra factor
of $1/2$ for each wine, originally part of a dumbbell structure,
which becomes transformed into a decorative effective propagator.

To proceed further, we must know how to 
 compute the flow of a reduced vertex.
Consider for a moment the flow of a two-point, tree level  vertex. From section~\ref{sec:NFE:TPTL}, we know that this
would produce a pair of two-point, tree level  vertices joined together by an undecorated wine. 
This is precisely the diagram that must be excluded from the flow of a reduced vertex. 
To exclude such diagrams,
we can extend the action of $R$ such that it is defined to operate on a dumbbell
structure in the following way. The action of $R$ removes all terms for which
the following are satisfied: both vertices are two-point, tree level vertices \emph{and} the wine is undecorated. 
If a quantum term is generated by the action
of the flow on a reduced vertex we can drop the $R$, as such terms appear only in the flow of $(n>0)$-loop vertices.

The result of both promoting the effective propagator to a decorative field and computing the flow of
the relevant vertex is shown in figure~\ref{fig:Beta-n+L1:a0}. Note that we adopt the rule
that, upon decoration, there must be no disconnected structures.
\begin{center}
\begin{figure}[h]
	\[
	\displaystyle
	-\frac{1}{4} \sum_{r=0}^{n_+}
	\left[
		\begin{array}{c}
		\vspace{0.2in}
			\displaystyle
			\dec{
				\scd{Vertex-r-R}{Vertex-n+_r-R}
			}{\bullet}
			-2\sum_{t=1}^r
				\left[
					2(r_t-1)\beta_t \dec{\scd[0.1]{Vertex-r_t}{Vertex-n+_r-R}}{}
					+\gamma_t \dec{\sco[0.1]{\pder{}{\alpha}\ensuremath{\begin{array}{c}\input{pstex/Vertex-r_t.pstex_t} \end{array}}}{\ensuremath{\begin{array}{c}\input{pstex/Vertex-n+_r-R.pstex_t} \end{array}}}}{}
				\right]
		\\
			\displaystyle -\sum_{t=0}^{r}
			\left(
				\sco[0.2]{
					\RO{\PED[0.1]{4}{Beta-n+L1-a0-r}{bn:L1:a0:a0-r}{fig:Beta-n+L1:a0-2pts}}
				}{\ensuremath{\begin{array}{c}\input{pstex/Vertex-n+_r-R.pstex_t} \end{array}}}
			\right)
			+ 
			\dec{
				\PEO[0.1]{4}{\scd{Sigma_r_-}{Vertex-n+_r-R}}{bn:L1:a0:a1}{fig:bn:L1:a0:a1-B}
			}{}
		\end{array}
	\right]^{11\Delta}
\]
\caption{Result of processing diagram~\ref{bn:L0:a0:M} by converting it into a $\Lambda$-derivative term, plus corrections.}
\label{fig:Beta-n+L1:a0}
\end{figure}
\end{center}

The combinatorics associated with decorating the diagrams of figure~\ref{fig:Beta-n+L1:a0} is very simple.
We have a single effective propagator with which we must join the bottom vertex to some other structure.
This will produce a factor of two. Decoration with the external fields is equally straightforward though,
of course, 
they can both attach to the same structure.

Before moving on, we note that we can rewrite diagram~\ref{bn:L1:a0:a1}. Since we cannot have a vertex
argument less than zero the sum over $r$, for this term, must start from one. Letting $r \rightarrow r+1$,
we can redraw diagram~\ref{bn:L1:a0:a1}, as shown in figure~\ref{fig:bn:L1:a0:a1-B}.
\begin{center}
\begin{figure}[h]
	\[
		-\frac{1}{4} \sum_{r=0}^n 
		\dec{		
			\CEO[0.1]{5}{\scd[0.1]{Sigma_r}{Vertex-n_r-R}}{bn:L1:a0:a1-B}{bn:L1:a0:a1-Cancel}
		}{11 \Delta}
	\]
\caption{A re-expression of diagram~\ref{bn:L1:a0:a1}.}
\label{fig:bn:L1:a0:a1-B}
\end{figure}
\end{center}

As in section~\ref{sec:bn:L0:a1}, we now remove from diagram~\ref{bn:L1:a0:a0-r} all those components which can
be processed without using the flow equations \ie\ all those containing two-point, tree level vertices.
These terms occur when the summed variable $t$ takes its boundary values; the isolation of two-point, tree level
vertices is shown in figure~\ref{fig:Beta-n+L1:a0-2pts}.
Note that, 
unlike the analogous case from the previous section, we do not need to include something
like $1/\vartheta_r$: since the variable $r$, unlike $n$, is summed over, diagrams containing two two-point, tree level 
vertices necessarily occur and so can be included naturally.

In those diagrams for which one of the vertices attached to the dumbbell structure is reduced, we need not include
the overall $R$
(note, though, that it would be illegal to retain the overall $R$ and remove the $R$ from the vertices). Diagrams
containing a single two-point, tree level vertex have been partially decorated.
\begin{center}
\begin{figure}
	\[
	\begin{array}{c}
	\vspace{0.2in}
		\displaystyle
		+\frac{1}{4}\sum_{r=0}^{n_+}
		\left[
			\begin{array}{c}
				\vspace{0.2in}
				-2 \dec{
					\sco[0.2]{\PED[0.1]{4}{./Beta-n+L1-a0-Attach-r}{bn:L1:a0:a0-n-TP-I}{fig:Beta-n+L1-Join}}{\ensuremath{\begin{array}{c}\input{pstex/Vertex-n+_r-R.pstex_t} \end{array}}}
				}{11}	
				-2 \dec{
					\sco[0.2]{\DED[0.1]{4}{./Beta-n+L1-a0-Attach-r-1}{bn:L1:a0:a0-n-TP-E}}{\ensuremath{\begin{array}{c}\input{pstex/Vertex-n+_r-R.pstex_t} \end{array}}}
				}{1\Delta}
			\\
				\displaystyle
				+ \sum_{t=0}^r 	
				\dec{
					\sco[0.2]{\PED[0.1]{4}{./Beta-n+L1-a0-r-R}{bn:L1:a0:a0-r-R}{fig:Beta-n+L2:a0:a0}}{\ensuremath{\begin{array}{c}\input{pstex/Vertex-n+_r-R.pstex_t} \end{array}}}
				}{11\Delta}
			\end{array}
		\right]
	\\
		\displaystyle
		-\frac{1}{4} 	
		\dec{
			\sco[0.2]{
				\PED[0.1]{5}{./Beta-n+L1-a0-2-TP}{bn:L1:a0:a0-2-TP}{fig:Beta-n+L1-Join}
			}{\ensuremath{\begin{array}{c}\input{pstex/Vertex-n+R.pstex_t} \end{array}}}
		}{11\Delta}
	\end{array}
	\]
\caption[Isolation of the components of diagram~\ref{bn:L1:a0:a0-r} containing a two-point, tree level  vertex.]
{Isolation of the components of diagram~\ref{bn:L1:a0:a0-r} containing a two-point, tree level  vertex.
Dumbbells with a single two-point, tree level vertex have been partially
decorated.}
\label{fig:Beta-n+L1:a0-2pts}
\end{figure}
\end{center}

Referring to the diagrams of figure~\ref{fig:Beta-n+L1:a0-2pts}, there are a number of comments to
make.
First, we note that we can discard diagram~\ref{bn:L1:a0:a0-n-TP-E}. If we attach the remaining external field
to the wine, we know that such a diagram vanishes at $\Op{2}$. If, instead, we attach the remaining external field
to one of the vertices, then this vertex must either be of the form $S^{\ \; 1\ 1}_{m \alpha \beta}(p)$ or
$S^{\ \{1\ 1 C\}}_{m \, \alpha \beta}(p,-p,0)$ (or the hatted versions of these vertices), for some loop order $m$. 
Either way, the vertex is $\Op{2}$ and so the diagram as whole is at least $\Op{4}$. Since the diagram
contains a seed action vertex, we cannot manipulate it---however we choose to decorate it---and so we can 
discard it.

Next, we further process diagrams~\ref{bn:L1:a0:a0-n-TP-I} and~\ref{bn:L1:a0:a0-2-TP}. In the former case, we tie up the loose
end noting that, to create a legal diagram, it must attach to the bottom vertex. This attachment yields a factor of
two. In the latter case, we note that we must attach the effective propagator to one of the two-point, tree level vertices.
If, instead, we attach an external field to each of these vertices then we can discard this diagram as it vanishes at
$\Op{2}$ and cannot be usefully
manipulated.\footnote{In this particular case, the diagram can not be manipulated at all. The effective propagator is forced to decorate
the wine. However, even if this were not the case, our prescription for treating such diagrams would demand that
we throw it away, anyway.} Having decorated one of the two-point, tree level  vertices with the effective propagator,
we now have no choice but to decorate both the wine and the remaining two-point, tree level  vertex with external fields.
This yields an additional factor of two.

Figure~\ref{fig:Beta-n+L1-Join} shows how we tie up the loose end of the propagator in both of these cases and the subsequent
application of the effective propagator relation. We are free to discard all gauge remainders, since they
either strike a two-point vertex or have no support---as is the case when they attach to a one-point vertex.
We can discard diagram~\ref{bn:L1:a0:a0-J-n-Re1}
\begin{center}
\begin{figure}
	\[
	\displaystyle -\sum_{r=0}^{n_+} 
	\dec{\CED[0.1]{5}{Beta-n+L0-a0-S}{bn:L1:a0:a0-S}{bn:L0:a0:S}}{11} -2
	\dec{
		\DED[0.1]{4}{./Beta-n+L1-a0-Join-Re-1}{bn:L1:a0:a0-J-n-Re1}
	}{}
	\]
\caption[Decoration of the two-point, tree level vertices of diagrams~\ref{bn:L1:a0:a0-n-TP-I} and~\ref{bn:L1:a0:a0-2-TP} and
application of the effective propagator relation.]
{Decoration of the two-point, tree level vertices of diagrams~\ref{bn:L1:a0:a0-n-TP-I} and~\ref{bn:L1:a0:a0-2-TP} and
application of the effective propagator relation. In the final diagram, we complete the decoration as the
remaining field is forced to decorate the wine.}
\label{fig:Beta-n+L1-Join}
\end{figure}
\end{center}

As expected, we now find another cancellation. Having split the parent diagram into manipulable
and non-manipulable components (see figure~\ref{fig:Beta-n+L1:a0:pre}), the seed action
contributions to the latter are now removed.

\begin{cancel}
Diagram~\ref{bn:L1:a0:a0-S} exactly cancels diagram~\ref{bn:L0:a0:S}. 
\label{C:L0:a0:S}
\end{cancel}

Note that, in this case, the issue of whether or not vertices of the parent diagram are uniquely seed
action or Wilsonian effective action vertices never arises.
 
\section{Level-Two Manipulations} \label{sec:bn:L2}

By processing diagrams~\ref{bn:L0:a1} and~\ref{bn:L0:a0}, we have performed all possible level-one manipulations.
At the second level, though, we must be more discerning about what we choose to manipulate. Ignoring gauge remainders
and $\Op{2}$ manipulations, which will be dealt with later, there are four diagrams, components of which 
are candidates for
manipulation: diagrams~\ref{bn:L1:a1:a1}, \ref{bn:L1:a1:a0-Re}, \ref{bn:L1:a0:a1-B} and~\ref{bn:L1:a0:a0-r-R}.

These diagrams are ones which have been formed by the action of either $a_0$ or $a_1$, at each level of
manipulation. The history of each of the diagrams is summarised in table~\ref{tab:L2-candidates-history}.
\begin{center}
\begin{table}
	\[
	\begin{array}{ccc}
		\mathrm{Diagram} 	& \mathrm{Level-0} & \mathrm{Level-1} \\ \hline
		\ref{bn:L1:a1:a1} 	& a_1 & a_1 \\
		\ref{bn:L1:a1:a0-Re}& a_1 & a_0 \\
		\ref{bn:L1:a0:a1-B}	& a_0 & a_1 \\
		\ref{bn:L1:a0:a0-r-R}&a_0 & a_0
	\end{array}
	\]
\caption{The history of the four diagrams which are candidates for level-two manipulations.}
\label{tab:L2-candidates-history}
\end{table}
\end{center}

Let us consider the effect of manipulating diagram~\ref{bn:L1:a1:a0-Re}. Suppose that we use the decorative
effective propagator to form a loop on one of the vertices---in the case that $n=0$,
this would correspond to diagram~\ref{d1:1.14}. We henceforth refer to loops formed by effective
propagators attaching at both ends to the same vertex as simple loops.
When we manipulate diagram~\ref{bn:L1:a1:a0-Re}, one of the diagrams
produced will correspond to the $\Lambda$-derivative striking this simple loop. This diagram has the same
topology as diagram~\ref{bn:L1:a0:a1-B}---which is one of the other candidates for manipulation---and, as we will see,
actually cancels part of it. Thus it is clear that we should not be manipulating both diagrams~\ref{bn:L1:a1:a0-Re}
and~\ref{bn:L1:a0:a1-B}. The best way to proceed is to manipulate the first of these diagrams. 
We choose to do this because diagram~\ref{bn:L1:a1:a0-Re} also
contains a manipulable component with no simple loops---in the case that $n=0$, 
this would be diagram~\ref{d1:1.15}---which must be manipulated, come what may.

As we iterate the diagrammatic procedure, we will generally find that diagrams with a structure formed by the
action of $a_1$ will cancel diagrams in which the $\Lambda$-derivative has been moved from 
effective propagators joining different vertices to a simple loop. There is, of course, one exception:
diagrams formed by the (repeated) action of $a_1$ possess only a single vertex and so cannot
possess any effective propagators which join different vertices. These terms must be manipulated;
indeed, we have already done so with diagram~\ref{bn:L0:a1}. Likewise, we must manipulate diagram~\ref{bn:L1:a1:a1}.

Our final candidate for manipulation, diagram~\ref{bn:L1:a0:a0-r-R}, is formed by the repeated action of $a_0$.
This should be manipulated.

Upon manipulation of the diagrams in this section, we find, as expected, a number of cancellations.
For some of these, a pattern is apparent, even at this level of manipulation and
this leads us to the first generic cancellation mechanisms, of this chapter. As we build
up the set of cancellation mechanisms, we can start to ask what happens when, starting with
a certain diagram, we iterate the diagrammatic procedure, until exhausted. We find that,
for diagram~\ref{bn:L0:a0}, we can describe the final result of this procedure in a remarkably
compact form.

\subsection{Diagram~\ref{bn:L1:a1:a0-Re}} \label{sec:bn:L1:a1:a0-Re}

We now iterate the diagrammatic procedure by processing the manipulable component of diagram~\ref{bn:L1:a1:a0-Re}.
The isolation  of those terms containing only Wilsonian effective action vertices and whose wine is undecorated
is shown in figure~\ref{fig:Beta-n+L1:a1:a0:sep}.
\begin{center}
\begin{figure}[h]
	\[
	-\frac{1}{4} \sum_{r=0}^{n}
	\dec{
		\PED[0.1]{4}{Dumbell-n_r-r-M}{bn:L1:a1:a0-M}{fig:Beta-n+L2}
		+ \CED[0.1]{4}{Dumbell-n_r-r-DW}{bn:L1:a1:a0-DW}{bn:L3:a1:a0:a0-a0-W}
		-2 \CED[0.1]{5}{Dumbell-n_r-r-S}{bn:L1:a1:a0-S}{bn:L2:a1:a0:a0-J-n-Re1}
	}{11\Delta}
	\]
\caption{Isolation of the manipulable component of diagram~\ref{bn:L1:a1:a0-Re}.}
\label{fig:Beta-n+L1:a1:a0:sep}
\end{figure}
\end{center}

The manipulations to be performed on diagram~\ref{bn:L1:a1:a0-M}
are similar to those of the previous section but are complicated by the fact that the parent diagram
is decorated by an effective propagator.

The first consideration is how this effective propagator attaches. In the current
case of the wine being undecorated, the effective propagator can either join the two vertices or form a simple loop.
This is shown in figure~\ref{fig:Beta-n+L2-a1-a0-pre}.
\begin{center}
\begin{figure}[h]
	\[
	-\frac{1}{2}\sum_{r=0}^{n}
	\dec{
		\LID[0.1]{1}{Beta-n+L2-a1-a0-pre}{bn:L2:a1:a0:pre1} + 
		\LID[0.1]{1}{Beta-n+L2-a1-a0-pre2}{bn:L2:a1:a0:pre2}
	}{11}
	\]
\caption{Attachment of the effective propagator to diagram~\ref{bn:L1:a1:a0-M}.}
\label{fig:Beta-n+L2-a1-a0-pre}
\end{figure}
\end{center}

In both cases, the attachment of the effective propagator yields a factor of two. In the former case,
this is because  we can choose which end of the effective propagator to attach to which vertex; in the
latter case the factor arises since we can attach the simple loop to either vertex.

We denote the total number of simple loops by $L$ (which, in this case is either zero or unity) which
we decompose into the sum $L_A + L_B$, by looking at the number of simple loops on each vertex, separately.
The quantity $J_A$ denotes the total number effective propagators leaving the bottom vertex
which attach to a different structure. In this case, this reduces just to the total number of joins between
the two vertices; being either one or two.

When we re-express our manipulable term as a $\Lambda$-derivative plus corrections, we must include an overall
factor of  $1/J_A$ since the $\Lambda$-derivative can reproduce the parent diagram by striking any one
of $J_A$ indistinguishable effective propagators. Thus we see that diagram~\ref{bn:L2:a1:a0:pre1} will pick up an extra
factor of $1/2$, compared with diagram~\ref{bn:L2:a1:a0:pre2}. However, by choosing the canonical form
for the $\Lambda$-derivative term we will find that, remarkably, this relative factor can be incorporated automatically!
Consider the form for the $\Lambda$-derivative term, shown in figure~\ref{fig:Beta-n+L2-a1-a0-preLdL}.
\begin{center}
\begin{figure}[h]
	\[
	-\frac{1}{16}\sum_{r=0}^{n}
	\dec{
		\dec{
			\LID[0.1]{}{Beta-n+L2-LdL}{ID-Beta-n+L2-LdL}
		}{\bullet}
	}{11\Delta^2}
	\]
\caption{The canonical form of the $\Lambda$-derivative term for diagrams~\ref{bn:L2:a1:a0:pre1} and~\ref{bn:L2:a1:a0:pre2}.}
\label{fig:Beta-n+L2-a1-a0-preLdL}
\end{figure}
\end{center}

The first thing we must do is define the diagrammatic rules for attachment of the effective propagators. In accord
with the discussion of section~\ref{sec:bn:L0:a1}, each join between the two vertices yields a factor of two.
Additionally, if we make $J_A$ joins from a total of $J$ effective propagators, then we
will take there to be $^J C_{J_A}$
different ways in which we can do this. 

These rules now allow us to deduce the overall factor of diagram~\ref{ID-Beta-n+L2-LdL}.
Starting 
from the factor of $-1/4$ coming from the parent diagram~\ref{bn:L1:a1:a0-M},
there are two additional contributions. First, the promotion of an effective
propagator which was previously a wine gives a factor of $1/2$.
Secondly, if we make just the single join between the vertices, we now have a choice of $J=2$ effective propagators with which
to do this; we must compensate for this with a factor of $1/J$.
 If we make two joins between the vertices, then (up to
interchanging the ends of each propagator), there is only one way we can do this. However, as we have just
discussed, such a term
must come with an additional factor of $1/J$.

Thus by giving the $\Lambda$-derivative term a overall factor of $1/16$, the
contribution corresponding to diagram~\ref{bn:L2:a1:a0:pre1} acquires a factor of $1/2$, relative to the one corresponding
to diagram~\ref{bn:L2:a1:a0:pre2}.

Having arrived at a prescription for converting diagram~\ref{bn:L1:a1:a0-M}
into a $\Lambda$-derivative term, we can now understand what we have done in a slightly different way, but one that will generalise
readily to more complex diagrams. 

Referring back to figure~\ref{fig:Beta-n+L2-a1-a0-preLdL}
let us now suppose that, having made the $J_A$ joins between the two vertices, we allow the $\Lambda$-derivative to strike one of these
$J_A$ indistinguishable effective propagators. The resultant diagram, which possesses one wine and $J_A-1$ effective propagators,  has an overall factor of
\[
\frac{J!}{(J-J_A)!(J_A-1)!} \times 2^{J_A}.
\]
The next step is to `un-decorate' by removing the $J_A-1$ effective propagators. This reduces the overall factor of
the diagram by $^{J-1} C_{J_A -1} \times 2^{J_A-1}$ which leaves
\[
2 J.
\]
Crucially, the resultant diagram is of exactly the same topology as the parent diagram. In order for it
to reproduce the parent diagram, then, we must simply compensate by a factor of $1/2J$.

The final ingredient we need to fully process diagram~\ref{bn:L1:a1:a0-M} is the correction terms
to the $\Lambda$-derivative. These fall into two classes depending on whether the $\Lambda$-derivative
strikes one of the vertices or one of the simple loops. In both cases, note that the $\Lambda$-derivative
term is invariant under interchange of the vertex arguments.

In the latter case, let us suppose that we decorate one of the vertices---say vertex-$B$---with $L_B$ simple loops.
This can be done in $^J C_{L_B}$ different ways. Allowing the $\Lambda$-derivative to strike one
of these simple loops and un-decorating the diagram leaves over a factor of $J$. This is a very
nice result, since it means that this correction term does not include the diagram specific term $L_B$.
Since the $\Lambda$-derivative term is invariant under the interchange of its vertex arguments,
we need only consider decorating one of the vertices with simple loops, so long as we multiply by two.
The final point about diagrams in which the $\Lambda$-derivative strikes a simple loop
is that the effective propagator, forming the simple loop
was, by definition, undecorated before being hit by the $\Lambda$-derivative. Consequently, it must
be undecorated after differentiation, which we denote by enclosing the $\Lambda$-derivative in a box.
 
Finally, we process  diagram~\ref{bn:L1:a1:a0-M}
in figure~\ref{fig:Beta-n+L2}. 
\begin{center}
\begin{figure}[h]
	\[
	\begin{array}{c}
 		\displaystyle 
		-\frac{1}{16}\sum_{r=0}^{n}
		\left[
		\begin{array}{c}
		\vspace{0.2in}
			\displaystyle
			\dec{
				\ensuremath{\begin{array}{c}\input{pstex/./Beta-n+L2-LdL.pstex_t} \end{array}}
			}{\bullet}
			-2\sum_{t=1}^r
				\left[
					2(r_t-1)\beta_t \dec{\scd[0.1]{Vertex-r_t}{Vertex-n_r-R}}{}
					+\gamma_t \dec{\sco[0.1]{\pder{}{\alpha}\ensuremath{\begin{array}{c}\input{pstex/Vertex-r_t.pstex_t} \end{array}}}{\ensuremath{\begin{array}{c}\input{pstex/Vertex-n_r-R.pstex_t} \end{array}}}}{}
				\right]

		\\
			\displaystyle 
			-\sum_{t=0}^{r}
			\left(
				\sco[0.2]{
					\RO{
						\PED[0.1]{4}{Beta-n+L2-a0-r}{bn:L2:a1:a0:a0-r}{fig:Beta-n+L2-2pts}
					}				
				}{\ensuremath{\begin{array}{c}\input{pstex/Vertex-n_r-R.pstex_t} \end{array}}}
			\right)
			+
			\left[
				\sco[0.2]{
					\LED[0.1]{}{Beta-n+L2-a1-r}{bn:L2:a1:a0:a1-r}
				}{\ensuremath{\begin{array}{c}\input{pstex/Vertex-n_r-R.pstex_t} \end{array}}}
			\right]
		\end{array}
		\right]^{11\Delta^2} \\
	\\
		\displaystyle
		+ \frac{1}{4} \sum_{r=0}^{n}
		\dec{
			\sco[0.2]{
				\PED[0.1]{}{Beta-n+L2-r-loop}{bn:L2:a1:a0:loop-r}{fig:CancelLoop}
			}{\ensuremath{\begin{array}{c}\input{pstex/Vertex-n_r-R.pstex_t} \end{array}}}
		}{11\Delta}
	\end{array}
\]
\caption{Manipulation of diagram~\ref{bn:L1:a1:a0-M}.}
\label{fig:Beta-n+L2}
\end{figure}
\end{center}

Our next task is to understand the combinatorics associated with decorating the diagrams of
figure~\ref{fig:Beta-n+L2}. This feature of the first three diagrams and diagram~\ref{bn:L2:a1:a0:loop-r} 
has just been discussed, and leads us to the combinatorics for
the attachment of effective propagators to diagrams~\ref{bn:L2:a1:a0:a0-r} and~\ref{bn:L2:a1:a0:a1-r}.
We will not explicitly analyse the latter case, since the arguments are just a simplified version 
of those which apply to the former.

Referring to diagram~\ref{bn:L2:a1:a0:a0-r}, let us consider making $J_A$ joins between the dumbbell
structure and the bottom vertex. For this to be consistent with the diagrammatics of the parent diagram,
we know that this must come with a factor of $^J C_{J_A}$. However, within the dumbbell structure, 
we now have a choice of three individual structures to which we can attach the effective propagators. 

Figure~\ref{fig:DumbellExternalFields}
shows the attachment of the $J_A$ identical effective propagators to the dumbbell structure. The number of effective propagators
on each of the structures
is denoted by $m_1 \cdots m_3$, where $m_1+m_2+m_3 = J_A$. We sum over all possible values of the $m_i$.
\begin{center}
\begin{figure}[h]
	\[
	\sum_{m_1,m_2,m_3}
	\dec[0.2]{\scd{DumbellExternalFields}{Vertex-n_r-R}}{\Delta^{J-J_A}}
	\]
\caption{Attachment of $J_A$ effective propagators, the loose ends of which all attach to the bottom vertex.}
\label{fig:DumbellExternalFields}
\end{figure}
\end{center}

To find the combinatoric factor we must multiply $^J C_{J_A}$ by the number of ways of
sub-dividing the $J_A$ effective propagators into three groups, containing the number of elements
indicated in the diagram. This is just
\begin{equation}
^J C_{J_A} \ ^{J_A}C_{m_1}\  ^{J_A-m_1}C_{m_2} = \frac{J!}{(J-J_A)!} \times \frac{1}{m_1!m_2!m_3!}.
\label{eq:DumbellExternalFields}
\end{equation}

If we were to decorate just one of the structures---say the one decorated with $m_1$ effective 
propagators---leaving the decoration of the others until later,
then the combinatoric factor is just 
\begin{equation}
^J C_{m_1}.
\label{eq:SingleExternalAttached}
\end{equation} 

It is easy to confirm the consistency of this
with equation~(\ref{eq:DumbellExternalFields}): if we were to decorate one structure at a time, then 
the combinatoric factor would be
\[
^J C_{m_1} \ ^{J-m_1} C_{m_2} \ ^{J-m_1-m_2} C_{m_3}
\]
which reduces to the above expression.

Now we turn to the case of effective propagators which join constituent parts of the dumbbell structure.
Numbering the three dumbbell structures $1,2,3$---as implied in figure~\ref{fig:DumbellExternalFields}---we denote
the number of joins between the $i^{\mathrm{th}}$ and $j^{\mathrm{th}}$ structures by $J_{ij}$. Note
that $J_{ii}$ corresponds to a simple loop and that $J_{ij} = J_{ji}$. 
The sum over the total number of joins, $\sum_{i\leq j} J_{ij}$ must equal $L_B$.
We denote
\[
\sum_{i < j} J_{ij} = L_{B'},
\]
which represents the number of joins between separate structures on the dumbbell.
The combinatoric factor for decorating the dumbbell structure with the $L_B$ fields is
\begin{eqnarray*}
\lefteqn{
\nCr{J}{L_B}  \nCr{L_B}{L_{B'}} \nCr{L_{B'}}{J_{12}} \nCr{L_{B'}-J_{12}}{J_{13}} \nCr{L_{B'}-J_{12}-J_{13}}{J_{23}}
} \\
& & 
\times \nCr{L_B-L_{B'}}{J_{11}} \nCr{L_B-L_{B'}-J_{11}}{J_{22}} \nCr{L_B-L_{B'}-J_{11}-J_{22}}{J_{33}}  \\
& & = \frac{J!}{(J-L_B)!} \frac{1}{\Pi_{i\leq j} (J_{ij}!)},
\end{eqnarray*}
where we pick up an additional factor of
\[
2^{L_{B'}}
\]
from the freedom to interchange the ends of the effective propagators joining different structures.
Here, too, if we were to attach just the set of effective propagators corresponding to $J_{ij}$,
then this would come with a combinatoric factor of
\[
\nCr{J}{J_{ij}}.
\]


Now that we understand the combinatorics associated with decorating the diagrams
of figure~\ref{fig:Beta-n+L2} we wish to identify cancellations.
As in section~\ref{sec:bn:L0:a0}, we begin by
isolating two-point, tree level vertices, as shown in figure~\ref{fig:Beta-n+L2-2pts}. 

\begin{center}
\begin{figure}[h]
	\[
	\displaystyle
	+\frac{1}{16}\sum_{r=0}^{n}
	\begin{array}{c}
	\vspace{0.2in}
		\left[
			\begin{array}{c}
				\vspace{0.2in}
				-4 	
				\dec{
					\sco[0.2]{\PED[0.1]{4}{Beta-n+L2-a0-Attach-r}{bn:L2:a1:a0:a0-n-TP-I}{fig:Beta-n+L2-Join}
					}{\ensuremath{\begin{array}{c}\input{pstex/Vertex-n_r-R.pstex_t} \end{array}}}		
				}{11\Delta}
				-2 \dec{\scd[0.2]{./Beta-n+L2-a0-Attach-r-1}{Vertex-n_r-R}}{1\Delta^2}
			\\
				\displaystyle
				+ \sum_{t=0}^r 
				\dec{
					\PEO[0.1]{4}{\scd{./Beta-n+L2-a0-r-R}{Vertex-n_r-R}}{bn:L2:a1:a0:a0-r-R}{fig:Beta-n+L3:a1:a0:a0}
				}{11\Delta^2}
			\end{array}
		\right]
	\\
		\displaystyle
		-\frac{1}{16} 
		\dec{
			\sco[0.2]{\RO{\PED[0.1]{5.5}{Beta-n+L2-a0-2-TP}{bn:L2:a1:a0:a0-2-TP}{fig:Beta-n+L2-Join^2}}
			}{\ensuremath{\begin{array}{c}\input{pstex/Vertex-n-R.pstex_t} \end{array}}}
		}{11\Delta^2}
	\end{array}
	\]
\caption{Isolation of those components of diagram~\ref{bn:L2:a1:a0:a0-r} containing a two-point, tree level  vertex.}
\label{fig:Beta-n+L2-2pts}
\end{figure}
\end{center}

Diagram~\ref{bn:L2:a1:a0:a0-n-TP-I}
is an example of the by now familiar case in which an effective propagator attaches to a two-point, tree level
vertex. By equation~(\ref{eq:SingleExternalAttached}) it comes with a combinatoric factor of $\nCr{2}{1}$, which
has been incorporated.

We now attach the loose end
of the effective propagator, which can be done in one of two ways:
either we can attach it to the dumbbell structure,
or we can attach it to the bottom vertex. Either way, we pick up a factor of two.
The results of attaching the loose end, together with application of the effective propagator relation, 
are shown in figure~\ref{fig:Beta-n+L2-Join}. 
\begin{center}
\begin{figure}[h]
	\[
	\displaystyle
	-\frac{1}{2} \sum_{r=0}^{n}
	\left[
	\begin{array}{c}
	\vspace{0.2in}
		\CED[0.1]{5}{Dumbell-n_r-r-S}{bn:L2:a1:a0:a0-J-n-Re1}{bn:L1:a1:a0-S}
		-\ensuremath{\begin{array}{c}\input{pstex/Dumbell-n_r-r-S-GR.pstex_t} \end{array}}
	\\
	\vspace{0.2in}
		\displaystyle
		+
		\dec{
			\sco{
				\PED[0.1]{}{Struc-r-S-R-W}{bn:L2:a1:a0:EP-loop-r}{fig:CancelLoop}
				-\ensuremath{\begin{array}{c}\input{pstex/Struc-r-S-R-W-GR.pstex_t} \end{array}}
				-\ensuremath{\begin{array}{c}\input{pstex/Struc-r-S-R-WGR.pstex_t} \end{array}}
			}{\ensuremath{\begin{array}{c}\input{pstex/Vertex-n_r-R.pstex_t} \end{array}}}
		}{}
	\end{array}
	\right]^{11\Delta}
\]
\caption{Tying up the loose ends of diagram~\ref{bn:L2:a1:a0:a0-n-TP-I}, together with application of the effective propagator 
relation.}
\label{fig:Beta-n+L2-Join}
\end{figure}
\end{center}

Having performed  manipulation via the flow equations and having processed 
one of the generated terms using the effective propagator relation, we
find a cancellation, as expected. 

\begin{cancel}
Diagram~\ref{bn:L2:a1:a0:a0-J-n-Re1} exactly cancels diagram~\ref{bn:L1:a1:a0-S}.
\label{C:L1:a1:a0:S}
\end{cancel}

Now we return to 
diagram~\ref{bn:L2:a1:a0:a0-2-TP}. 
Upon partial decoration, this will give us new types of diagram. Whilst we have encountered dumbbell
structures comprising two two-point, tree level vertices, this is the
first time we have also had more than just one effective propagator.

We follow the usual
recipe: partially decorate the diagram by attaching fields / effective propagators
to the two-point, tree level vertices; tie up all loose ends and finally 
 apply the effective propagator
relation.
The result of this procedure is shown in figure~\ref{fig:Beta-n+L2-Join^2}. 
Note that we do not include the diagram in which both two-point, tree level
vertices are decorated by external fields, since this vanishes at $\Op{2}$
and cannot be usefully manipulated.
Nor have we drawn those terms in which each of the two-point, tree level vertices is decorated by an
effective propagator, the loose ends of which are both attached to the wine. Whilst there is nothing wrong with
this, in principle, in this particular case such attachments will use up all the effective propagators
but leave us with a loose vertex.
\begin{center}
\begin{figure}[h]
\[
	\begin{array}{c}
	\vspace{0.2in}
		\displaystyle
		-\frac{1}{2} 
		\dec{
			\ensuremath{\begin{array}{c}\input{pstex/Dumbell-n-R-TLTP.pstex_t} \end{array}}
			-\ensuremath{\begin{array}{c}\input{pstex/Dumbell-n-R-TLTP-GR.pstex_t} \end{array}}
			-\left[
			\sco[0.2]{
					\ensuremath{\begin{array}{c}\input{pstex/Beta-n+L2-a0-EI.pstex_t} \end{array}}
					}{\ensuremath{\begin{array}{c}\input{pstex/Vertex-n-R.pstex_t} \end{array}}}
			\right]
		}{1\Delta} 
		-\frac{1}{4} 
		\dec{
			\CEO[0.1]{5}{\scd[0.1]{./Beta-n+L2-a0-EE}{Vertex-n-R}}{bn:L2:a1:a0:TLTP-loop}{bn:L1:a0:a1-Cancel-Correction}
		}{11\Delta} 
	\\
		\displaystyle
		-\frac{1}{2} 
		\dec{
			\PCED[0.1]{}{Struc-n-R-RW}{bn:L0:a1:C-W}{bn:L0:a1:DW}
			-2\ensuremath{\begin{array}{c}\input{pstex/Struc-n-R-RW-GR.pstex_t} \end{array}}
			+\ensuremath{\begin{array}{c}\input{pstex/Struc-n-R-GR-RW-GR.pstex_t} \end{array}}
			-2 \left( \ensuremath{\begin{array}{c}\input{pstex/./Beta-n+L2-a0-II-B.pstex_t} \end{array}} \right)
		}{11}
	\end{array}
\]
\caption[Partial decoration of diagram~\ref{bn:L2:a1:a0:a0-2-TP}.]{Partial decoration of diagram~\ref{bn:L2:a1:a0:a0-2-TP}. All loose ends are tied up
and the effective propagator relation is applied, as appropriate.}
\label{fig:Beta-n+L2-Join^2}
\end{figure}
\end{center}

The terms generated fall into three classes, depending on how many times the effective propagator
relation has been applied. Ignoring gauge remainders, we are interested in the terms for which
it has been applied more than once. We will return to diagram~\ref{bn:L2:a1:a0:TLTP-loop},
in which the relation has been applied once, in section~\ref{sec:bn:L2:a1:a0:a0-r-R}. In the
meantime, we focus on diagram~\ref{bn:L0:a1:C-W}, for which the relation has been applied twice.

\begin{cancel}
Diagram~\ref{bn:L0:a1:C-W} has the same structure as diagram~\ref{bn:L0:a1:DW}.
Irrespective of the value of $n$, diagram~\ref{bn:L0:a1:C-W} exactly cancels the  contributions
to diagram~\ref{bn:L0:a1:DW}, for which at least one of the external fields decorates the vertex.
The
only remaining contribution to diagram~\ref{bn:L0:a1:DW} is where both external fields decorate the wine
\ie\ the vertex is a two-point vertex. If $n>0$, this contribution is exactly cancelled by diagram~\ref{bn:L0:a1:C-W}.
However, when $n=0$, this contribution survives, due to the reduction of the vertex in diagram~\ref{bn:L0:a1:C-W}.
Hence, diagram~\ref{bn:L0:a1:C-W} only partially cancels diagram~\ref{bn:L0:a1:DW}.
\label{C:L0:a1:DW}
\end{cancel}

This is the cancellation promised in section~\ref{sec:bn:L0:a1} in which the non-manipulable components
of diagram~\ref{bn:L0:a1} containing uniquely Wilsonian effective action vertices are
removed. That the cancellation is between terms generated two levels apart makes perfect
sense. We will assume---as will later be proven to be the case---that the cancellation
of non-manipulable components of diagrams possessing vertices of unique character occurs due the generation
of terms formed by the (repeated) action of $a_0$ on the partner manipulable component. 

Any term generated by the action of $a_0$ possesses one more vertex than the parent diagram.
If the dumbbell structure generated by the action of $a_0$ 
contains a two-point, tree level vertex, then the other vertex---assuming, for the time being
that it is not a two-point, tree level vertex---must be a seed action vertex.
We can, of course, get rid of the two-point, tree level vertex by attaching it to 
another structure with
an effective propagator.
However, the remaining term necessarily contains a seed action vertex; indeed, it is
this contribution that cancels those partners of the parent diagram possessing a uniquely seed action 
vertex.

If we wish to cancel a diagram containing uniquely Wilsonian effective action vertices,
then we must iterate the diagrammatic procedure. Focusing on those terms formed by
the action of $a_0$ at each level, we now arrive at diagrams with two more vertices
than the original parent. We can choose these extra vertices to both be two-point, tree
level vertices, joined by a wine. The  remaining loose vertices are Wilsonian effective action
vertices. We can now remove both two-point, tree level vertices by 
attaching different effective propagators to each. One of the resultant terms will
cancel the non-manipulable, uniquely Wilsonian effective action contributions to the parent
diagram.

We are now left to deal with the cancellation of those terms whose vertex does not
have a unique character.

\begin{cancel}
We have seen how diagrams~\ref{bn:L0:a1:S} and~\ref{bn:L0:a1:DW} have been 
entirely cancelled 
by diagrams~\ref{bn:L1:a0-Both2ptI-A} and~\ref{bn:L0:a1:C-W} for $n>0$ but only
partially cancelled for $n=0$. In the latter case, the surviving contributions
to each diagram contain a two-point (tree level) vertex. In diagram~\ref{bn:L0:a1:DW},
the whole of this contribution survives; in diagram~\ref{bn:L0:a1:S} only half of
it survives. Combining the two terms, we have $\Pi_{0 RS}^{\ XX}(k)$ which vanishes!
Hence, diagram~\ref{bn:L0:a1:S} partially cancels diagram~\ref{bn:L0:a1:DW}.

In this way, we have removed all non-manipulable contributions to diagram~\ref{bn:L0:a1}.
\label{C:L0:a1:TLTP}
\end{cancel}

\subsection{Diagram~\ref{bn:L1:a1:a1}} \label{sec:bn:L1:a1:a1}

Having processed the term formed by the action of $a_1$ at the zeroth level and $a_0$
at the first level, we now process the partner term, formed by the action
of $a_1$, at the first level. The manipulable
contribution to this diagram is, of course, the Wilsonian effective action
contribution for which the wine is undecorated. This latter constraint forces the
unattached effective propagator to form a simple loop, giving two simple loops, in total. 
Hence, when we re-express this
term as a $\Lambda$-derivative term plus corrections, we must include an 
overall factor of $1/2$, to recognise the indistinguishability of the two simple loops.
The conversion of the manipulable component of diagram~\ref{bn:L1:a1:a1} into a
$\Lambda$-derivative term plus corrections is shown in figure~\ref{fig:Beta-n+L2:a1:a1}.
However, since we have not explicitly isolated the manipulable component from the non-manipulable components, these
latter diagrams must be included.
\begin{center}
\begin{figure}[h]
	\[
	\begin{array}{c}
	\vspace{0.2in}
		\displaystyle
		\frac{1}{8}
		\dec{
			\begin{array}{c}
			\vspace{0.2in}
				\displaystyle
				\dec{\ensuremath{\begin{array}{c}\input{pstex/Beta-n+L2-a1-a1-LdL.pstex_t} \end{array}}}{\bullet}	
				-\sum_{r=1}^{n_-}
				\left[
				2(n_r-2) \beta_{n_-} \ensuremath{\begin{array}{c}\input{pstex/Vertex-n-r-1.pstex_t} \end{array}}
				+ \gamma_r \pder{}{\alpha} \ensuremath{\begin{array}{c}\input{pstex/Vertex-n-r-1.pstex_t} \end{array}}
				\right]
			\\
				\displaystyle 
				-\frac{1}{2} \sum_{r=0}^{n_-}
				\PED[0.1]{4}{./Beta-n+L2-a1-a1-a0}{bn:L2:a1:a1:a0}{fig:Beta-n+L2:a1:a1-Re}
				+\frac{1}{2}
				\LED[0.1]{}{Beta-n+L2-a1-a1-a1}{bn:L2:a1:a1:a1}
			\end{array}
		}{11\Delta^2}
	\\
		\displaystyle
		+ \frac{1}{4}
		\dec{
			\PDURCED[0.1]{Sigma_n_-DW}{bn:L1:a1:a1:DW}{bn:L1:a1:a1:S}
			-2 \PDCED[0.1]{8}{Sigma_n_-S}{bn:L1:a1:a1:S}{bn:L2:a1:a1:a0:2ptI-A}{bn:L1:a1:a1:DW}
		}{11\Delta}
	\end{array}
\]
\caption[Result of processing the manipulable component of diagram~\ref{bn:L1:a1:a1}.]
{Result of processing the manipulable component of diagram~\ref{bn:L1:a1:a1}. The
incomplete cancellation of diagram~\ref{bn:L1:a1:a1:DW} is due to the termination of
the explicit computation of $\beta_{n_+}$ before the diagram required to complete
the cancellation is generated.}
\label{fig:Beta-n+L2:a1:a1}
\end{figure}
\end{center}

We see that this figure is similar to figure~\ref{fig:beta_n+:Level1}; the only differences being
the vertex arguments, the number of decorative effective propagators and the relative factor between
the diagrams and their corresponding parent. Nonetheless, we can split
off the two-point, tree level vertices from diagram~\ref{bn:L2:a1:a1:a0} in exactly the same way
as we did with diagram~\ref{bn:L1:a1:a0}. This procedure is shown in figure~\ref{fig:Beta-n+L2:a1:a1-Re},
where we have also used the effective propagator relation, wherever possible, and have
tied up any loose ends.

Compared with the similar procedure performed on the previous level, there is one major difference:
when we come to attach an effective propagator to
the two-point, tree level vertex, there are two such objects from which to choose. The resulting factor of 
two will exactly cancel the factor
of $1/2$ arising from the indistinguishability of the simple loops of the parent diagram.
\begin{center}
\begin{figure}[h]
\[
	\begin{array}{c}
	\vspace{0.2in}
		\displaystyle 
		-\frac{1}{16}\sum_{r=0}^{n_-}
		\dec{
			\ensuremath{\begin{array}{c}\input{pstex/Beta-n+L2-a1-a1-a0-Re.pstex_t} \end{array}}
		}{11\Delta^2}
		+\frac{1}{8 \vartheta_{n_-}}
		\dec{
			\ensuremath{\begin{array}{c}\input{pstex/Beta-n+L2-a1-a1-a0-2ptE.pstex_t} \end{array}}
		}{1\Delta^2}
	\\
		\displaystyle
		+\frac{1}{2 \vartheta_{n_-}}
		\dec{
			\CED[0.1]{}{Struc-n_-S-W}{bn:L2:a1:a1:a0:2ptI-A}{bn:L1:a1:a1:S}
			-\ensuremath{\begin{array}{c}\input{pstex/Struc-n_-S-W-GR.pstex_t} \end{array}}
			-\ensuremath{\begin{array}{c}\input{pstex/Struc-n_-S-WGR.pstex_t} \end{array}}
		}{11\Delta}
\end{array}
\]
\caption{Result of splitting off two-point, tree level vertices from diagram~\ref{bn:L2:a1:a1:a0}.}
\label{fig:Beta-n+L2:a1:a1-Re}
\end{figure}
\end{center}

We now find that all components of the parent diagram for which the vertex is uniquely seed action
are removed.

\begin{cancel}
Diagram~\ref{bn:L2:a1:a1:a0:2ptI-A} partially cancels diagram~\ref{bn:L1:a1:a1:S}.
The cancellation is complete for 
$n_- >0$. In the case where $n_- =0$ the contribution containing
a two-point vertex is only half removed; all other contributions
being completely cancelled.
\label{C:L1:a1:a1:S}
\end{cancel}

Note that, although we have been performing second level manipulations,
we only find cancellations that go back a single level. This is because 
the parent diagram was generated by the repeated action of $a_1$ alone.
Only by generating terms with $a_0$ do we find cancellations and,
since the terms in this section have a history involving at most one instance
of $a_0$, the cancellations go back just a single level.

Cancellation~\ref{C:L1:a1:a1:S} is simply a generalisation of cancellation~\ref{C:L0:a1:S}
and  leads us to the first of the general cancellation mechanisms,
of this chapter. Our aim is to show how the non-manipulable components of a diagram
formed by the exclusive and repeated action of $a_1$ cancel. In the process,
we will also see how the seed action contributions to a diagram comprising two
vertices joined by a wine and decorated by arbitrary number of effective propagators
cancel.

To begin, we consider the formation of a level-$J$ term by the repeated action of $a_1$.
The very first instance of $a_1$ in this calculation forms the diagram~\ref{bn:L0:a1}
which comprises a single vertex and a wine.
When we convert this into a $\Lambda$-derivative term, this wine is promoted to an effective
propagator. The daughter diagram formed by the action of $a_1$, diagram~\ref{bn:L1:a1:a1},
has the same structure as the parent but is decorated by an additional
effective propagator. The loop order of the vertex argument has reduced by one.

Iterating this procedure, it is clear that a level-$J$ term, formed by
the repeated action of $a_1$ comprises a single vertex, a wine and is decorated by
the usual external fields and by $J$ effective propagators. The loop order of this
term is $n-J$. Consequently, $J \leq n$ and so the generation of terms
by the repeated action of $a_1$ must ultimately terminate.

Figure~\ref{fig:Beta-n+:LJ:a1...a1} depicts the separation of the manipulable component of
a level-$J$ term formed by the repeated action of $a_1$. The curious looking factor, $\norm_{J,0}$, is 
an overall normalisation factor. 
The reason it is written in this apparently baroque form will become
apparent later.
\begin{center}
\begin{figure}[h]
	\[
	\begin{array}{l}
	\vspace{0.2in}
		-\norm_{J,0}
		\dec{
			\LGD[0.1]{1}{Sigma_n_J}{LJ-1V}
		}{11\Delta^J}	
		=
	\\
		-\norm_{J,0}
		\dec{
			\PGD[0.1]{4}{Sigma_n-J-M}{bn:LJ:a1...a1}{fig:Beta-n+:LJ:a1...a1-M}
			+ \PDCGD[0.1]{6}{Sigma_n-J-DW}{bn:LJ:a1...a1-DW}{bn:LJ+2:a0-a0-loop}{bn:LJ:a1...a1-S}
			-2 \PDCGD[0.1]{6}{Sigma_n-J-S}{bn:LJ:a1...a1-S}{LJ+1-1V-a0-2ptI}{bn:LJ:a1...a1-DW}
		}{11\Delta^{J}}
	\end{array}
	\]
\caption{Isolation of the manipulable component of a level-$J$ diagram, formed by the
repeated action of $a_1$.}
\label{fig:Beta-n+:LJ:a1...a1}
\end{figure}
\end{center}

We now manipulate diagram~\ref{bn:LJ:a1...a1}, turning it into a $\Lambda$-derivative
term plus corrections, as shown in figure~\ref{fig:Beta-n+:LJ:a1...a1-M}.
\begin{center}
\begin{figure}[h]
	\[
	-\frac{\norm_{J,0}}{J+1}
	\dec{
		\begin{array}{c}
		\vspace{0.2in}
			\displaystyle
			\dec{
				\ensuremath{\begin{array}{c}\input{pstex/Vertex-n_J.pstex_t} \end{array}}
			}{\bullet}
			- \sum_{r=1}^{n_J}
			\left[
				2(n_J - r -1) \beta_r - \gamma_r \pder{}{\alpha}
			\right] 
			\ensuremath{\begin{array}{c}\input{pstex/Vertex-n_J+r.pstex_t} \end{array}}		
		\\
			\displaystyle
			-\frac{1}{2} \sum_{r=0}^{n_J}
			\PGD[0.1]{4}{LJ+1-1V-a0}{LJ+1-1V-a0}{fig:Beta-n+:LJ:a1...a1-Split}
			+\frac{1}{2}
			\LGD[0.1]{2}{Sigma_n-J+1}{LJ+1-1V-a1}
	\end{array}
	}{11\Delta^{J+1}}
	\]
\caption{The result of processing diagram~\ref{bn:LJ:a1...a1}.}
\label{fig:Beta-n+:LJ:a1...a1-M}
\end{figure}
\end{center}

Comparing diagrams~\ref{LJ-1V} and~\ref{LJ+1-1V-a1} gives a recursion relation for the normalisation factor:
\[
\norm_{J+1,0} = \frac{1}{2}\frac{\norm_{J,0}}{J+1}.
\]
Since the zeroth level diagram formed by $a_1$, diagram~\ref{bn:L0:a1}, has an overall factor of
$1/2$, this fixes $\norm_{0,0} = -1/2$, from which we can  determine 
the normalisation factor:
\begin{equation}
	\norm_{J,0} = -\frac{1}{J!} \left(\frac{1}{2}\right)^{J+1}.
\label{eq:norm_(J,0)}
\end{equation}

We now proceed in familiar fashion and strip off the two-point, tree level
vertices from diagram~\ref{LJ+1-1V-a0}. Having done this, we partially decorate these vertices,
tying up all loose ends and applying the effective propagator relation, as appropriate. The
final result of this procedure is shown in figure~\ref{fig:Beta-n+:LJ:a1...a1-Split}
\begin{center}
\begin{figure}[h]
	\[
	\norm_{J+1,0} 
	\left[
	\begin{array}{c}
	\vspace{0.2in}
		\displaystyle
		\sum_{r=0}^{n_J}
		\dec{
			\PGD[0.1]{4}{Dumbell-n_J+rR-rR}{LJ+1-1V-a0-R}{fig:Beta-n+:LJ:a1...a1-a0-M}
		}{11\Delta^{J+1}}
		-\frac{2}{\vartheta_{n_J}}
		\dec{
			\ensuremath{\begin{array}{c}\input{pstex/LJ+1-1V-a0-2ptE.pstex_t} \end{array}}
		}{1\Delta^{J+1}}
	\\
	\displaystyle
	- \frac{4(J+1)}{\vartheta_{n_J}}
	\dec{
		\PCGD[0.1]{}{Struc-n_J-S-W}{LJ+1-1V-a0-2ptI}{bn:LJ:a1...a1-S}
		-\ensuremath{\begin{array}{c}\input{pstex/Struc-n_J-S-W-GR.pstex_t} \end{array}}
		-\ensuremath{\begin{array}{c}\input{pstex/Struc-n_J-S-WGR.pstex_t} \end{array}}
	}{11\Delta^J}
	\end{array}
	\right]
	\]
\caption{Isolation of the two-point, tree level vertices of diagram~\ref{LJ+1-1V-a0} and etc.}
\label{fig:Beta-n+:LJ:a1...a1-Split}
\end{figure}
\end{center}

Noting that $2(J+1)\norm_{J+1,0} = \norm_{J,0}$, we find a cancellation.

\PartialCancelCom{LJ+1-1V-a0-2ptI}{bn:LJ:a1...a1-S}{The cancellation
is complete for $n_J>0$. In the case where $n_J =0$ the contribution
to diagram~\ref{bn:LJ:a1...a1-DW} containing a two-point, tree level vertex
is only half removed.}

Since this cancellation is guaranteed to happen, 
irrespective of the level, $J$, this gives our second cancellation mechanism
(see section~\ref{sec:GR:CompleteDiagrams} for the first).

\begin{CM}
Consider a level-$J$ diagram created by the repeated action of $a_1$.
The component of this diagram for which the vertex is a Wilsonian effective
action vertex and the wine is undecorated can be further manipulated,
using the flow equations. This manipulation
is guaranteed to produce a further set of terms generated by $a_0$.
A subset of these terms contain a dumbbell structure for
which at least one of the vertices is a two-point, tree level vertex.
Attaching the two-point, tree level vertex to the other vertex
generates a diagram which, up to gauge remainders,
will partially cancel the seed action 
contributions to the parent diagram.

Ignoring these gauge remainders, the cancellation is complete for 
$n_J >0$. In the case where $n_J =0$, the contribution containing
a two-point vertex is only half removed; all other contributions
being completely cancelled.
\label{CM:a1-S}
\end{CM}

In order to complete the cancellation of the non-manipulable components
of diagram~\ref{LJ-1V}, we must manipulate diagram~\ref{LJ+1-1V-a0-R}.
The first stage of this is shown in figure~\ref{fig:Beta-n+:LJ:a1...a1-a0-M}
in which, having separated the manipulable part, 
we perform the conversion into a $\Lambda$-derivative, plus corrections.
\begin{center}
\begin{figure}
	\[
	\frac{\norm_{J+1,0}}{2(J+2)}
	\sum_{r=0}^{n_J}
	\begin{array}{c}
	\vspace{0.2in}
		\displaystyle
		\dec{
			\begin{array}{c}
			\vspace{0.2in}
				\displaystyle
				\dec{\scd[0.1]{Vertex-r-R}{Vertex-n_J+r-R}}{\bullet}
			\\
			\vspace{0.2in}
				\displaystyle
				-2\sum_{t=0}^r
				\left[
					2(r_t-1)\beta_t \left[\scd[0.1]{Vertex-r_t}{Vertex-n_J+r-R}  \right] 
					+\gamma_t \left[\sco[0.1]{\pder{}{\alpha}{\ensuremath{\begin{array}{c}\input{pstex/Vertex-r_t.pstex_t} \end{array}}}}{\ensuremath{\begin{array}{c}\input{pstex/Vertex-n_J+r-R.pstex_t} \end{array}}} \right]
				\right]
			\\
				\displaystyle
				-\sum_{t=0}^{r}
				\left(
				\sco[0.1]{
					\RO{
						\PGD[0.1]{4}{Beta-n+L2-a0-r}{LJ+2-1V-a0-a0}{fig:Beta-n+:LJ:a1...a1-a0-a0-Split}
					}				
				}{\ensuremath{\begin{array}{c}\input{pstex/Vertex-n_J+r-R.pstex_t} \end{array}}}
			\right)
			+
			\left[
				\scd[0.1]{Vertex-r-R}{Sigma_n_J+r+1}
			\right]
			\end{array}
		}{11\Delta^{J+2}}
	\\
		\displaystyle
		-
		\norm_{J+1,0} \sum_{r=0}^{n_J}
		\dec{
			\left[\scd[0.1]{Vertex-r-R}{Vertex-n_J+r-R-Dloop} \right]
			- \ensuremath{\begin{array}{c}\input{pstex/Dumbell-n_J+rR-rR-DW.pstex_t} \end{array}}
			+2 \CGD[0.1]{5}{Dumbell-n_J+rR-rR-S}{bn:LJ+1-1V-a0-S}{bn:LJ+2-1V-a0-a0-S}
		}{11\Delta^{J+1}}
	\end{array}
	\]
\caption{Manipulation of diagram~\ref{LJ+1-1V-a0-R}.}
\label{fig:Beta-n+:LJ:a1...a1-a0-M}
\end{figure}
\end{center}

The final stage is to strip off the two-point, tree level vertices from diagram~\ref{LJ+2-1V-a0-a0},
perform the partial decoration of these vertices, tie up all loose ends and apply the 
effective propagator relation, as appropriate. This is shown in figure~\ref{fig:Beta-n+:LJ:a1...a1-a0-a0-Split}
where, other than the diagram which does not possess any two-point, tree level vertices,
we have retained only those diagrams which render diagrams generated earlier in this
section completely cancelled.
\begin{center}
\begin{figure}
	\[
	\begin{array}{c}
	\vspace{0.2in}
		\displaystyle 
		- \norm_{J+2,0} \sum_{r=0}^{n_J} \sum_{t=0}^r
		\dec{
			\scd[0.1]{Dumbell-tR-r_tR-R}{Vertex-n_J+r-R}
		}{11\Delta^{J+2}}
	\\
		\displaystyle
		+2\norm_{J+1,0} \sum_{r=0}^{n_J}
		\dec{
			\CGD[0.1]{5}{Dumbell-n_J+rR-rR-S}{bn:LJ+2-1V-a0-a0-S}{bn:LJ+1-1V-a0-S}
		}{11\Delta^{J+1}}
		+\norm_{J,0}
		\dec{
			\PCGD[0.1]{5}{Sigma_n_J-R-DW}{bn:LJ+2:a0-a0-loop}{bn:LJ:a1...a1-DW}
		}{11\Delta^J} + \cdots
	\end{array}
	\]
\caption{Isolation of selected components of diagram~\ref{LJ+2-1V-a0-a0}.}
\label{fig:Beta-n+:LJ:a1...a1-a0-a0-Split}
\end{figure}
\end{center}

\Cancel{bn:LJ+2-1V-a0-a0-S}{bn:LJ+1-1V-a0-S}

\PartialCancelCom{bn:LJ+2:a0-a0-loop}{bn:LJ:a1...a1-DW}{The cancellation
is complete for $n_J>0$. In the case where $n_J =0$ the contribution
to diagram~\ref{bn:LJ:a1...a1-DW} containing a two-point, tree level vertex
survives.}

\begin{cancel}
The components of diagrams~\ref{bn:LJ:a1...a1-DW} and~\ref{bn:LJ:a1...a1-S} which survive
from cancellations~\ref{cancel:LJ+1-1V-a0-2ptI} and~\ref{cancel:bn:LJ+2:a0-a0-loop} exactly cancel.
\label{cancel:bn:LJ:a1...a1-DW,bn:LJ:a1...a1-S}
\end{cancel}

The cancellations shown in figure~\ref{fig:Beta-n+:LJ:a1...a1-a0-a0-Split}
give us further cancellation mechanisms. The first cancellation we will
promote to a cancellation mechanism is the partial cancellation
of diagram~\ref{bn:LJ:a1...a1-DW} by diagram~\ref{bn:LJ+2:a0-a0-loop}.

\begin{CM}
Consider a level-$J$ diagram created by the repeated action of $a_1$.
Iterating the diagrammatic procedure twice creates a set of terms
generated by two instances of $a_0$ which possess a total of three vertices---
two more
than the parent diagram. Those terms for which the two extra vertices
are both two-point, tree level vertices, each joined to the original
vertex by an effective propagator, partially cancel the Wilsonian 
effective action contributions to the parent diagram, for which the 
wine is decorated, up to gauge remainders.

The cancellation is complete for $n_J >0$. In the case where $n_J=0$,
the contribution to the parent diagram containing a two-point,
tree level vertex survives; all other contributions being completely
cancelled.
\label{CM:a1-DW}
\end{CM}

This cancellation mechanism, combined with cancellation mechanism~\ref{CM:a1-S}
leads to yet another cancellation mechanism.
\begin{CM}
Consider a level-$J$ diagram created by the repeated action of $a_1$. By
cancellation mechanisms~\ref{CM:a1-S} and~\ref{CM:a1-DW}, all
non-manipulable components of this diagram are guaranteed to be
cancelled
for $n_J>0$. For $n_J=0$, two contributions survive, both
of which possess a two-point, tree level vertex. One
comes from the Wilsonian effective action contribution to
the parent; the other is half of the seed action contribution
to the parent. Given that the original vertex argument of the
parent is $\Sigma_{n_J}$, these contributions are
guaranteed to cancel since $\Pi_{0 RS}^{\ XX}(k) = 0$.
\label{CM:a1-TLTP}
\end{CM}

Cancellation mechanisms \ref{CM:a1-S}--\ref{CM:a1-TLTP}
thus guarantee that all non-manipulable components of a diagram
formed by the repeated action of $a_1$ are guaranteed to be
cancelled, in the computation of $\beta_{n_+}$.

The final cancellation mechanism of this section arises
from the cancellation of diagram~\ref{bn:LJ+1-1V-a0-S} by
diagram~\ref{bn:LJ+2-1V-a0-a0-S}. This cancellation
is for the seed action contribution to a diagram 
containing a dumbbell structure decorated by any
number of effective propagators, provided this number is
greater than zero.\footnote{This restriction
arises because of the way in which we have generated the diagrams:
their history involves the action of a single instance of $a_1$.}
However, by combining this general
cancellation with the specific cancellation~\ref{C:L0:a0:S},
we can remove this restriction on the number of effective 
propagators.
\begin{CM}
Consider a level-$J$ diagram formed by the action of a single
instance of $a_0$ preceded by $J$
instances of $a_1$.
Stripping off all two-point, tree level components of the
vertices, we arrive
at the parent diagram for what follows.

 The component of this diagram for which both vertices are  Wilsonian effective
action vertices and the wine is undecorated can be further manipulated,
using the flow equations. This manipulation
is guaranteed to produce a further set of terms generated by $a_0$.
A subset of these terms  contain a dumbbell comprising a single 
two-point, tree level vertex and thus, necessarily, also a reduced
seed action vertex. If we join this two-point, tree level vertex 
to the loose Wilsonian effective action vertex
then the resulting diagram
will exactly cancel the seed action 
contributions to the parent diagram, up to gauge remainders.
\label{CM:a1...a0-S}
\end{CM}

\subsection{Diagram~\ref{bn:L1:a0:a0-r-R}} \label{sec:bn:L1:a0:a0-r-R}

Processing the  manipulable component of diagram~\ref{bn:L1:a0:a0-r-R} provides us with the first---and simplest---example
of the manipulation of a diagram containing more than two vertices.
That this diagram is the simplest of its type follows directly from its history: since
it is formed by the repeated action of $a_0$, it has, upon decoration, the minimum number of joins between structures allowed. 
 Since we will be focusing
on the manipulable components of this diagram we do not decorate the wine and so the effective
propagator must join one of the wines from the dumbbell structure to the bottom
vertex. Although we will ultimately undo
this step, we will make this join, to get a better picture of the issues we will face in converting our
diagram into a $\Lambda$-derivative term.

Due to the invariance
of the dumbbell structure under the interchange of vertex arguments, we need only consider joining one
of the constituent vertices to vertex-$A$, so long as we multiply by two. Once the attachment is made,
we will have an additional factor of two coming, as usual, from joining two different structures
with an effective propagator. This is shown in figure~\ref{fig:Beta-n+L2:a0:a0-pre}.
\begin{center}
\begin{figure}[h]
	\[
	\displaystyle
	\sum_{r=0}^{n_+} \sum_{t=0}^r
	\dec{
		\LID[0.1]{1}{Beta-n+L2-a0-a0-pre}{bn:L2:a0:a0-pre}
	}{11}
	\]
\caption{Attaching the effective propagator to the component of
diagram~\ref{bn:L1:a0:a0-r-R} with an undecorated wine.
}
\label{fig:Beta-n+L2:a0:a0-pre}
\end{figure}
\end{center}

It is not immediately clear whether processing this term is of any use since, if we were to cast
the manipulable term as a $\Lambda$-derivative, we would generate a term in which the $\Lambda$-derivative
hits the effective propagator joining the bottom vertex to the middle vertex. This would just be of the
same topology as the parent and so we would not have gained anything. 

The solution is to consider two explicit pairs of values for $\{r,t\}$. Figure~\ref{fig:Beta-n+L2:a0:a0-pre2}
shows diagram~\ref{bn:L2:a0:a0-pre} specialised to:
\[
\{r,t\} = \{r',t'\}, \ \{n_+-r'+t',t'\}.
\]
where the constraints $r\geq t$ and $n_+ \geq r$ simply imply that $r'\geq t'$ and that $n_+ \geq r'$.
We assume, for the time being, that $r' \neq n_+-r'+t'$.
\begin{center}
\begin{figure}
	\[
	\dec{\ensuremath{\begin{array}{c}\input{pstex/./Beta-n+L2-a0-a0-pre2.pstex_t} \end{array}}}{11}
	\]
\caption{Specialisation of diagram~\ref{bn:L2:a0:a0-pre} to two specific realisations of $\{r,t\}$,
which we assume to be different.}
\label{fig:Beta-n+L2:a0:a0-pre2}
\end{figure}
\end{center}

Rotating diagram~\ref{bn:L2:a0:a0-pre2B} by $\pi$, it is clear that we can process the manipulable parts
of diagrams~\ref{bn:L2:a0:a0-pre2A} and~\ref{bn:L2:a0:a0-pre2B}, together. 
Note that if 
$r' = n_+-r'+t'$ then matters are more complicated. In this case, the two diagrams of figure~\ref{fig:Beta-n+L2:a0:a0-pre2}
are identical and, to prevent over-counting, we must multiply by a factor of $1/2$.
Having included this factor, as appropriate, the manipulable components of 
diagrams~\ref{bn:L2:a0:a0-pre2A} and~\ref{bn:L2:a0:a0-pre2B} can be processed,
irrespective of whether $r' = n_+-r'+t'$. This is shown in figure~\ref{fig:Beta-n+L2:a0:a0-LdL-Ex},
where $S_{AC}$ is a function of the arguments of the top and bottom vertices; being $1/2$ when they are
the same and unity otherwise.
\begin{center}
\begin{figure}[h]
	\[
	S_{AC} 	\dec{
				\LID[0.1]{1}{./Beta-n+L2-a0-a0-LdL-Ex}{bn:L2:a0:a0-LdL-Ex}
			}{11 \bullet} - \mbox{corrections}
	\]
\caption{Conversion of diagrams~\ref{bn:L2:a0:a0-pre2A} and~\ref{bn:L2:a0:a0-pre2B} into a $\Lambda$-derivative term.}
\label{fig:Beta-n+L2:a0:a0-LdL-Ex}
\end{figure}
\end{center}

Note that we can interpret $S_{AC}$ as  arising
due to symmetry properties of diagram~\ref{bn:L2:a0:a0-LdL-Ex} \emph{before we decorate
with the external fields}. Diagram~\ref{bn:L2:a0:a0-LdL-Ex} is only invariant under rotations
by $\pi$ when  $r' = n_+-r'+t'$ and this is, of course, precisely when we must take $S_{AC} = 1/2$.

We can, however, be much more intelligent about $S_{AC}$ and remove it completely by
choosing the canonical form for the $\Lambda$-derivative term. Let us begin by considering
the form for the $\Lambda$-derivative term shown in figure~\ref{fig:Beta-n+L2:a0:a0-LdL-Ex2}.
\begin{center}
\begin{figure}[h]
	\[
	\displaystyle
	\frac{1}{16} \sum_{r=0}^{n_+} \sum_{t=0}^r
	\sym 
	\dec{
		\dec{
			\sco{
				\LID[0.1]{}{Beta-n+L2-a0-a0-LdL}{ID-Beta-n+L2-a0-a0-LdL}
			}{\ensuremath{\begin{array}{c}\input{pstex/Vertex-n+_r-R.pstex_t} \end{array}}}
		}{\bullet}
	}{11\Delta^2}
	\]
\caption{A trial form for the $\Lambda$-derivative term arising from diagram~\ref{bn:L1:a0:a0-r-R}.}
\label{fig:Beta-n+L2:a0:a0-LdL-Ex2}
\end{figure}
\end{center}

The operator $\sym$ is defined to operate 
after the attachment of all effective propagators joining different vertices but
\emph{before} the attachment of external fields or simple loops. For the time being,
we leave it undefined but note that its action can, in principle, depend on any symmetries
of the diagram.
The overall factor of the diagram requires comment. Starting from the factor of $1/4$ coming from the parent diagram,
there are two additional contributions. First, the promotion of an effective
propagator which was previously a wine gives a factor of $1/2$, as we know from section~\ref{sec:bn:L0:a0}.
Secondly, having chosen any pair of vertices, we now have a choice of two effective propagators with
which to join them; we must compensate for this with a factor of $1/2$.

Let us now examine the diagrams corresponding to  the following set of values for $\{r,t\}$.
\begin{equation}
\{r,t\} = \left(\{r',t'\};\ \{n_+-r'+t',t'\}; \ \{n_+-t',n_+-r'\} \right) + t'\rightarrow r'-t'. 
\label{eq:ValuesForThreeVertexTerm}
\end{equation}
If any particular ordered pair occurs more than once, we discard all but the first instance, 
to avoid over-counting.
It is easy to check that these six realisations correspond to the six permutations of the ordered
triplet
\[
\{n_r',t',r'_{t'}\}.
\]
where the condition that we do not over-count demands that any particular ordered triplet occurs only once.
Hence, if $m$ of the vertices have the same argument, there are a total of
$3!/m!$ distinct ordered triplets.

The next step is to join together the vertices in figure~\ref{fig:Beta-n+L2:a0:a0-LdL-Ex2}.
There are $3!$ ways to order the vertices in a line. Joining the vertices together, this
over-counts the number of diagrams by a factor of two, since ($A$---$B$---$C$) is
the same as ($C$---$B$---$A$). We must now ascertain whether any of the remaining
three diagrams are indistinguishable. If a diagram has $m$ vertices with identical argument, then
permuting these vertices will produce a set of $m!$ indistinguishable diagrams.
However, if any of these permutations corresponds to rotating the diagram by $\pi$,
then this permutation has already been counted. Consequently, diagrams not invariant under rotations
by $\pi$ come with a factor of two, relative to those which are thus invariant.
Hence, we see that by representing the $\Lambda$-derivative
term in the form of diagram~\ref{fig:Beta-n+L2:a0:a0-LdL-Ex2}, we can completely eliminate $S_{AC}$:
$\sym$ just reduces to a constant factor, $2/3!$, since any symmetries of the diagrams are automatically
taken care of!

Once again, having arrived at a prescription for converting diagram~\ref{bn:L1:a0:a0-r-R} into a $\Lambda$-derivative
term, we can interpret the result in a more generalisable manner. To reproduce the
manipulable component of diagram~\ref{bn:L1:a0:a0-r-R}
from diagram~\ref{ID-Beta-n+L2-a0-a0-LdL} (with $\sym = 2/3!$), consider joining any pair of vertices together with either
of the effective propagators. There are $^3 C_2 \times 2$ different ways to do this, and each of them is 
identical. Up to an additional factor of two coming from the actual attachment, this is the factor we 
must take if we desire the $\Lambda$-derivative to strike the first effective propagator with which we decorate.
Hence, compared to diagram~\ref{bn:L1:a0:a0-r-R}, diagram~\ref{ID-Beta-n+L2-a0-a0-LdL} has a relative factor
of $\frac{1}{^3 C_2 \times 4}$.

In figure~\ref{fig:Beta-n+L2:a0:a0} we finally process the manipulable components of 
diagram~\ref{bn:L1:a0:a0-r-R}.
\begin{center}
\begin{figure}[h]
	\[
	\begin{array}{c}
	\vspace{0.2in}
		\displaystyle
		\frac{1}{48} \sum_{r=0}^{n_+} \sum_{t=0}^r
		\dec{
			\begin{array}{c}
			\vspace{0.2in}
				\dec{
					\scd{./Beta-n+L2-a0-a0-LdL}{Vertex-n+_r-R}
				}{\bullet} 
			\\
			\vspace{0.2in}
				\displaystyle
				-3 \sum_{u=1}^t
				\left[
					\sco{\tco[-0.2]{(2(t_u-1)\beta_u + \gamma_u)
						\pder{}{\alpha}\ensuremath{\begin{array}{c}\input{pstex/Vertex-t_u.pstex_t} \end{array}}}{\ensuremath{\begin{array}{c}\input{pstex/Vertex-r_t-R.pstex_t} \end{array}}}}{\hspace{1.4in} \ensuremath{\begin{array}{c}\input{pstex/Vertex-n+_r-R.pstex_t} \end{array}}}
				\right]
			\\
				\displaystyle
				-\frac{3}{2} \left(
									\sum_{u=0}^t 
									\decB[-0.1]{
										\PEO[0.1]{5}{
											\tco[-0.3]{\ensuremath{
\left[
\begin{array}{c}\input{pstex/Dumbell-t-u.pstex_t} \end{array} \hspace{0in}
\right]_R
}}{\scd[0.1]{Vertex-r_t-R}{Vertex-n+_r-R}}
										}{bn:L2:a0:a0:a0}{fig:Beta-n+L2:a0:a0:a0:R} \hspace{-0.1in}
									}{}
									- \dec{\sco[0.1]{\ensuremath{\begin{array}{c}\input{pstex/Beta-n+L2-a0-a0-a1.pstex_t} \end{array}}}{\scd[0.1]{Vertex-r_t}{Sigma_n_r}}}{}
								\right)
			\end{array}
		}{11\Delta^2}
	\\
		\displaystyle
		\frac{1}{4} \sum_{r=0}^{n_+} \sum_{t=0}^r
		\dec{
			\left[
				\sco{\ensuremath{\begin{array}{c}\input{pstex/Dumbell-tR-r_tR-DW.pstex_t} \end{array}}}{\ensuremath{\begin{array}{c}\input{pstex/Vertex-n+_r-R.pstex_t} \end{array}}}
			\right]
			-2
			\left[
				\CEO[0.1]{4}{\sco{\ensuremath{\begin{array}{c}\input{pstex/Dumbell-tR-r_tR-S.pstex_t} \end{array}}}{\ensuremath{\begin{array}{c}\input{pstex/Vertex-n+_r-R.pstex_t} \end{array}}}}{L1:a0:a0-S}{L2-a0-a0-a0-S}
			\right]
		}{11\Delta}
	\end{array}
	\]
\caption{The result of processing the manipulable part of diagram~\ref{bn:L1:a0:a0-r-R}.}
\label{fig:Beta-n+L2:a0:a0}
\end{figure}
\end{center}

To isolate the two-point, tree level vertices and perform the subsequent attachments, we simply follow
section~\ref{sec:bn:L0:a0}. The only difference is that, rather than discarding the diagram containing
a pair of two-point, tree level vertices, we keep it since the presence of $>1$ effective propagators
means that we are not compelled to decorate at least one of these vertices with an external field.
Figure~\ref{fig:Beta-n+L2:a0:a0:a0:R} shows the result
of proceeding along these lines. 
\begin{center}
\begin{figure}[h]
	\[
	\begin{array}{c}
	\vspace{0.2in}
		\displaystyle
		-\frac{1}{32} \sum_{r=0}^{n_+} \sum_{t=0}^{r} 
		\dec{
			\sum_{u=0}^t 
			\dec{
				\LED[0.1]{}{./Beta-n+L2-a0-a0-a0-R}{bn:L2:a0:a0:a0:R} 
			}{11\Delta^2} 
			- 16 
			\dec{
				\CEO[0.1]{4}{\sco{\ensuremath{\begin{array}{c}\input{pstex/Dumbell-tR-r_tR-S.pstex_t} \end{array}}}{\ensuremath{\begin{array}{c}\input{pstex/Vertex-n+_r-R.pstex_t} \end{array}}}}{L2-a0-a0-a0-S}{L1:a0:a0-S}
			}{11\Delta}
		}{}
	\\
		\displaystyle
		+\frac{1}{2} \sum_{r=0}^{n_+} 
		\dec{
			\CED[0.1]{5}{./Beta-n+L2-a0-a0-a0-2ptx2}{bn:L2:a0:a0:a0:2ptx2}{bn:L0:a0:DW}
		}{11}
	\end{array}
	\]
\caption{Isolation of the two-point, tree level vertices in diagram~\ref{bn:L2:a0:a0:a0} and etc.}
\label{fig:Beta-n+L2:a0:a0:a0:R}
\end{figure}
\end{center}

As expected, we now find two cancellations. Since the parent diagram was formed by
the action of at least one instance of $a_0$ (two, in the current case), diagrams 
just generated cancel diagrams from both of the previous two levels.

\begin{cancel}
Diagram~\ref{L2-a0-a0-a0-S} exactly cancels diagrams~\ref{L1:a0:a0-S}.
\end{cancel}

\begin{cancel}
Diagram~\ref{bn:L2:a0:a0:a0:2ptx2} exactly cancels diagram~\ref{bn:L0:a0:DW}.
\end{cancel}

The cancellations we have seen in this section are easily generalised to diagrams containing
an arbitrary number of vertices, say $J+2$. For our current purposes, we will assume that such
diagrams are formed by the repeated action of $a_0$ only; in other words, their lineage
can be traced back to diagram~\ref{bn:L0:a0}. At each successive level of manipulation,
the term whose manipulable component we will be processing has one more vertex
and one more effective propagator compared to the analogous term from the previous level.
It is thus the case that a level-$J$ term, formed exclusively by the action of $a_0$,
possesses $J+2$ vertices. Two of these vertices are barred and form a dumbbell structure.
We take, as usual, the reduced component of these vertices.
The remaining $J$ vertices are all reduced Wilsonian effective action vertices. 
The diagram as a whole is decorated
by the usual external fields and by $J$ effective propagators.

Before moving on, it is worth discussing such diagrams in further detail. Since the parent
diagram has $J+2$ vertices and only $J+1$ objects with which to join them, (one wine
and $J$ effective propagators), it is not possible to avoid the presence of one-point and/or
two-point vertices. This is not an issue, unless such a vertex is tree level, in which
case, the diagram is to be discarded.
This puts a constraint
on the maximum level, $\crit$, at which a manipulable diagram occurs. Since the sum of the vertex arguments always adds up
to $n_+$, we will first be compelled to take a tree level vertex at level-$n$ (where there are $n+2$ vertices).
With each additional level, we will be compelled to take one further tree level vertex. Hence,
if a diagram is compelled to have $T$ tree level vertices, it must be a level
\begin{equation}
	J = n+T-1
\label{eq:level:T-TL}
\end{equation}
diagram.

Since we do not allow one-point or two-point, tree level vertices, each tree level vertex
must be decorated by three fields. To maximise the number of tree level vertices we are allowed
requires that we minimise the number of fields decorating the remaining vertices. Hence, we
suppose that these remaining vertices are each decorated by a single (necessarily internal)
field. Temporarily ignoring the external fields, this means that we have $J+2-T$ one point
vertices and $T$ three-point vertices. 

If we now include the two external fields, this allows us to decrease the number of
internal fields attached to the tree level vertices by two.

Noting that we have $2(J+1)$ internal fields,
the condition that there be no one-point or two-point, tree level vertices implies that
\[
(J+2-T)\times 1 + T\times3 -2 \leq 2(J+1) 
\]
Using equation~(\ref{eq:level:T-TL}) to substitute for $T$ we obtain:
\begin{equation}
	J \leq 2n,
\label{eq:CritLevel:s=0}
\end{equation}
telling us that $\crit = 2n$; this being last level where the candidate for manipulation is
guaranteed not to possess one-point or two-point, tree level vertices.
At the next level, however, we will be unable to avoid the appearance of such vertices.
Thus, in summary, the final diagram to be manipulated occurs at level-$2n$;
we call this the critical level.
This generates level-$2n+1$ terms but none of these are then manipulated: the 
diagrammatic procedure terminates.

There is one further property of the level-$\crit$ term whose components we
manipulate.
Since every field is required to decorate a vertex, for the diagram
to survive, this implies that the wine must be undecorated. Hence, \emph{all}
Wilsonian effective action 
contributions to the level-$\crit$ term are manipulable!

In
figure~\ref{fig:ArbitraryLine}, we show the conversion of a level-$J$ diagram,
formed by the repeated action $a_0$,
into a $\Lambda$-derivative
term and a correction.
Taking into account the factor of two yielded by joining two different vertices with an effective
propagator, the parent diagram can be generated in 
\[
2(J+1)\ ^{J+2} C_2 = (J+2)(J+1)^2
\]
different ways from the $\Lambda$-derivative term. Hence, the $\Lambda$-derivative term comes with the 
inverse of this factor, relative to the parent. The overall factor of $\norm_{J,J}$ is a normalisation factor
 that we will determine at the end.
\begin{center}
\begin{figure}[h]
	\[
	\begin{array}{l}
		 \displaystyle \norm_{J,J} \sum_{r=0}^{n_+} \sum_{t=0}^r \cdots \sum_{y=0}^{x} 
			\dec{
				\left[
					\LGD[0.1]{1}{ArbitraryLine}{ArbitraryLine}
				\right] + 
				\left[
					\LGD[0.1]{1}{ArbitraryLine-DW}{ArbitraryLine-DW}
				\right] -2 
				\left[
					\CGD[0.1]{4.5}{ArbitraryLine-S}{ArbitraryLine-S}{ArbitraryLine-a0-TP}
				\right]
			}{11\Delta^{J}}
	\\
	\vspace{-0.9in}
		\ensuremath{\begin{array}{c}\input{pstex/Arrow.pstex_t} \end{array}}
	\\
	
		\displaystyle  \frac{\norm_{J,J}}{(J+1)^2}  \sum_{r=0}^{n_+} \sum_{t=0}^r \cdots \sum_{y=0}^{x} 
		\dec{\frac{1}{J+2} 
			\dec{
				\LGD[0.1]{1}{ArbitraryLineLdL}{ArbitraryLineLdL}
			}{\bullet} 
			- 
			\left[
				\PGD[0.1]{4}{ArbitraryLineLdLC}{ArbitraryLineLdLC}{fig:ArbitraryLineM}
			\right]
		}{11\Delta^{J+1}}
	\end{array}
	\]
\caption{Processing a level-$J$ diagram formed by the repeated action of $a_0$.}
\label{fig:ArbitraryLine}
\end{figure}
\end{center}

Notice, in the parent diagram, that we have chosen to join the bottom two vertices
with the wine. This is, of course, a choice and is exactly equivalent to joining
any other pair of vertices with the wine. However, by doing this, we make it
clear that, in addition to the dumbbell structure, there can be an 
arbitrary number of additional vertices. Nonetheless, if we are to be
absolutely rigorous about the interpretation of the diagrams of figure~\ref{fig:ArbitraryLine}, 
we should really demand that there is a total of at least three vertices.

By demanding that we have at least the dumbbell structure and the vertex labelled by
$y$, the ellipsis in the sequence of sums over vertex arguments makes sense: in
the minimal case of three vertices, all we need do is identify the upper limit
on the sum over $y$ with $t$;
for terms with more vertices we simply replace the ellipsis with an appropriate number of extra sums.
If we were to take only two vertices of the dumbbell structure, then we would actually have to remove
the entire string of sums starting with the sum over $t$ and ending with the
sum over $y$. Furthermore, we would also have to remove the $t$ from the
vertex argument $r_t$.

Indeed, since in what follows we will create a term for which the effective propagator
relation can be used twice, we will be able to generate diagrams with one vertex less than the
parent. If we want to be rigorous about this diagram, we should demand that it has at least
three vertices, thereby forcing the parent to have at least four. These issues amount largely
to pedantry. We will obtain the correct result by being lax and not worrying too
much about the correctness of the notation. Nonetheless, since the cases in which the
parent diagram has a total
of either two and three vertices have been dealt with explicitly, the rigorous route
will also take us where we want to go, so long as we supplement it with the necessary,
earlier results.

In diagram~\ref{ArbitraryLineLdLC} we can allow the $\flow$ to strike any of
the vertices; we choose the bottom one. However, upon processing this term,
we rewrite the vertex arguments, to bring them into the same form as
those of the parent.
This is shown in figure~\ref{fig:ArbitraryLineM}.
\begin{center}
\begin{figure}[h]
	\[
	\frac{-\norm_{J,J}}{(J+1)^2}
	\sum_{r=0}^{n_+} \sum_{t=0}^r \cdots \sum_{y=0}^{x}
	\dec{
	\begin{array}{c}		
	\vspace{0.2in}
		\displaystyle
		\sum_{z=1}^y
		\left[
			2(y_z-1)\beta_z 
			\dec{
				\LGD[0.1]{1}{ArbitraryLine-beta}{ArbitraryLine-beta}
			}{} 
			+  \gamma_z 
			\dec{
				\LGD[0.1]{1}{ArbitraryLine-alpha}{ArbitraryLine-alpha}
			}{} 
		\right]
	\\
		\displaystyle
		+\frac{1}{2} 
		\left( 
			\sum_{z=0}^y 
			\left[
				\PGO[0.1]{4}{\sco{\ensuremath{\begin{array}{c}\input{pstex/ArbitraryLine-a0-Top.pstex_t} \end{array}}}{\ensuremath{
\left[
\begin{array}{c}\input{pstex/ArbitraryLine-a0-Bottom.pstex_t} \end{array} \hspace{0in}
\right]_R
}}}{ArbitraryLine-a0}{fig:ArbitraryLine-a0-TP}
			\right]
			- 
			\left[
				\LGD[0.1]{1.5}{ArbitraryLine-a1}{ArbitraryLine-a1}
			\right]
		\right)
	\end{array}
	}{11\Delta^{J+1}}
	\]
\caption{The result of processing diagram~\ref{ArbitraryLineLdLC}.}
\label{fig:ArbitraryLineM}
\end{figure}
\end{center}

The next step is to isolate the two-point, tree level vertices in diagram~\ref{ArbitraryLine-a0}.
Discarding any terms in which two-point, tree level vertices are decorated by
external fields, we arrive at the set of diagrams in figure~\ref{fig:ArbitraryLine-a0-TP}.
\begin{center}
\begin{figure}
	\[
	2 \norm_{J,J} \sum_{r=0}^{n_+} \sum_{t=0}^r \cdots \sum_{x=0}^{w}
	\dec{
	\begin{array}{c}
	\vspace{0.2in}
		\displaystyle
		-\frac{1}{4(J+1)^2} \sum_{y=0}^x \sum_{z=0}^y 
		\dec{
			\LGD[0.1]{1.5}{ArbitraryLine-a0-R}{ArbitraryLine-a0-R}
		}{11\Delta^{J+1}} 
	\\
		\displaystyle
		+\sum_{y=0}^x
		\dec{
			\CGD[0.1]{4.5}{ArbitraryLine-a0-TP}{ArbitraryLine-a0-TP}{ArbitraryLine-S}
		}{11\Delta^{J}} 
		+ J^2 
		\dec{
			\LGD[0.1]{1.5}{ArbitraryLine-a0-TPx2}{ArbitraryLine-a0-TPx2}
		}{11\Delta^{J-1}}
	\end{array}
	}{}
	\]
\caption[Isolation of two-point, tree level vertices in diagram~\ref{ArbitraryLine-a0}, and the results of
their subsequent decoration.]{Isolation of two-point, tree level vertices in diagram~\ref{ArbitraryLine-a0}, and the results of
their subsequent decoration. 
Diagrams possessing a two-point, tree level vertex decorated by an external field have,
together with any gauge remainders,
been discarded.}
\label{fig:ArbitraryLine-a0-TP}
\end{figure}
\end{center}

Note that if we are to allow the parent diagram to possess any number of vertices greater
than two, then
we need to be careful with diagram~\ref{ArbitraryLine-a0-TPx2}. This diagram cannot exist if the parent
diagram is a zeroth level term and so should be discarded, in this case.

As expected, we find a cancellation.

\Cancel{ArbitraryLine-a0-TP}{ArbitraryLine-S}

In turn, this cancellation implies a cancellation mechanism.

\begin{CM}
A level-$J$ diagram formed by the repeated action of $a_0$ comprises a total of $J+2$
vertices, two of which are barred and joined together by a wine. 
Stripping off all two-point, tree level vertices, we arrive at the parent, for what
follows.

 The component of this diagram for which all vertices are  Wilsonian effective
action vertices and the wine is undecorated can be further manipulated,
using the flow equations. This manipulation
is guaranteed to produce a further set of terms generated by $a_0$.
A subset of these terms  contain a dumbbell comprising a single 
two-point, tree level vertex and thus, necessarily, also a reduced
seed action vertex. If we sum over the identical diagrams
formed upon joining this two-point, tree level vertex 
to each of the loose Wilsonian effective action vertices
then the resulting diagram
will exactly cancel the seed action 
contributions to the parent diagram.
In this scenario, gauge remainders automatically vanish.

\label{CM:LineOfVertices:Seed}
\end{CM}

Our next task is to find an expression for $\norm_{J,J}$. Comparing diagrams~\ref{ArbitraryLine-a0-R}
and~\ref{ArbitraryLine}, we see that
\begin{equation}
\norm_{J+1,J+1} = -\frac{\norm_{J,J}}{2(J+1)^2}.
\label{eq:NormArbLineRecursion}
\end{equation}
Since diagram~\ref{bn:L0:a0} tells us that $\norm_{0,0} = -1/2$, it follows that
\begin{equation}
\norm_{J,J} = \frac{1}{(J!)^2} \left(-\frac{1}{2}\right)^{J+1}.
\label{eq:NormArbLine}
\end{equation}

Let us now consider iterating the diagrammatic procedure by processing
the manipulable component of diagram~\ref{ArbitraryLine-a0-R}. In making this
step, we have implicitly assumed that $J < \crit$.

The result of this manipulation is simply to reproduce 
figures~\ref{fig:ArbitraryLine}--\ref{fig:ArbitraryLine-a0-TP}
but with $J\rightarrow J+1$. The level-$J+1$ version of diagram~\ref{ArbitraryLine-a0-TP}
will just cancel the seed action components of diagram~\ref{ArbitraryLine-a0-R},
by cancellation mechanism~\ref{CM:LineOfVertices:Seed}. The level-$J+1$ version of 
diagram~\ref{ArbitraryLine-a0-TPx2} comes
with an overall factor of $2(J+1)^2\norm_{{J+1},{J+1}}$ which, by equation~(\ref{eq:NormArbLineRecursion})
is just equal to $-\norm_{J,J}$. Hence, this diagram exactly cancels 
diagram~\ref{ArbitraryLine-DW}.

The cancellation of seed action contributions is always guaranteed to occur. Even if it is the
case that we are manipulating a level-$\crit$ term, we still generate level-$\crit+1$ terms,
one of which will cancel the seed action contributions to the parent. However, the level-$\crit+1$
candidate for manipulation will vanish, apparently meaning that we will not generate the diagrams to cancel
the remaining Wilsonian effective action contribution to the level-$\crit$ parent. However,
as we have already noted, all Wilsonian effective action contributions to a
level-$\crit$ diagram are manipulable! Hence, the cancellation of all non-manipulable Wilsonian
effective action contributions is guaranteed, also.

With these points in mind, we arrive at another cancellation mechanism.

\begin{CM}
A level-$J$ diagram formed by the repeated action of $a_0$ comprises a total of $J+2$
vertices, two of which are barred and joined together by a wine. 
Stripping off all two-point, tree level vertices, we arrive at the parent, for what
follows.

Assuming that we can iterate the diagrammatic procedure twice
then, amongst the terms we generate, is a diagram with
$J+2$ reduced vertices plus an additional dumbbell structure.
Focusing on the case where the two vertices comprising the dumbbell
are two-point, tree level vertices, we now join each of
these to a different, loose vertex.
Summing over all ways of doing this,
 the resulting term will exactly cancel
the remaining Wilsonian effective action contributions to the original parent diagram.

In case that the diagrammatic procedure terminates with the generation of level-$J+1$
terms, the only Wilsonian effective action contributions to the parent diagram
are those without a decorated wine \ie\ those which have already been manipulated.

In this scenario, gauge remainders automatically vanish.

\label{CM:LineOfVertices:Wilson}
\end{CM}

We can now use cancellation mechanisms~\ref{CM:LineOfVertices:Seed} and~\ref{CM:LineOfVertices:Wilson}
to describe the ultimate fate of diagram~\ref{bn:L0:a0} and all of its children. 
When the diagrammatic procedure has been iterated until exhaustion,
only four types of term will remain: those at each level which are
analogous to diagrams~\ref{ArbitraryLineLdL}, \ref{ArbitraryLine-beta}, \ref{ArbitraryLine-alpha}
and~\ref{ArbitraryLine-a1}. All the rest are either processed, cancelled or discarded.
Those in the latter class have been discarded either at $\Op{2}$ or as a consequence
of the constraint $S^{\ 1}_m(0) = 0$. It is actually the case that certain components of
the surviving diagrams could also be discarded, for just these reasons.
However, as certain components of such diagrams do survive, 
we choose to keep the entire diagram, for the time being.

This allows us to write the result of exhaustively manipulating diagram~\ref{bn:L0:a0}
in a fantastically compact form.
We begin by introducing the vertex arguments, $V^J$, where
 upper roman indices act simply as labels. To make contact with
the current computation, $V^{J=0}=n_+$. Next,
we introduce the compact notation
\begin{eqnarray}
	\comp{J}{J_+} & \equiv & V^J - V^{J+1} \label{eq:ArbitraryVertexLabels} \\
	\CR{J}{J_+} & = & V^{J;R} - V^{J+1;R}		\label{eq:ArbitraryReducedVertexLabels}
\end{eqnarray}
where
\[
	V^{J;R} = \left(V^{J}\right)^R.
\]
Finally, we define
\begin{equation}
	\Vertex{n_+,J} \equiv \prod_{I=0}^J \sum_{V^{I_+}=0}^{V^{I}} \ensuremath{\begin{array}{c}\begin{picture}(0,0)%
\includegraphics{pstex/GV-I-I+.pstex}%
\end{picture}%
\setlength{\unitlength}{3947sp}%
\begingroup\makeatletter\ifx\SetFigFont\undefined%
\gdef\SetFigFont#1#2#3#4#5{%
  \reset@font\fontsize{#1}{#2pt}%
  \fontfamily{#3}\fontseries{#4}\fontshape{#5}%
  \selectfont}%
\fi\endgroup%
\begin{picture}(598,598)(75,75)
\put(110,360){\makebox(0,0)[lb]{\smash{\SetFigFont{8}{9.6}{\rmdefault}{\mddefault}{\updefault}{\color[rgb]{0,0,0}$\CR{I}{I_+}$}%
}}}
\end{picture}
 \end{array}}
\label{eq:n-loop:ProdOverI}
\end{equation}
where the first argument gives the value of $V^0$.

Now we can give a diagrammatic expression for the result of exhaustively manipulating
diagram~\ref{bn:L0:a0}, as shown in figure~\ref{fig:Exhaust:L0:bn:a0}.
\begin{center}
\begin{figure}[h]
	\[
		\sum_{J=0}^{2n}
		\frac{\norm_{J,J}}{(J+1)^2}
		\dec{
			\begin{array}{c}
			\vspace{0.2in}
				\displaystyle
				\frac{1}{J+2}
				\dec{
					\sco[0.1]{\ensuremath{\begin{array}{c}\input{pstex/GV-J+1.pstex_t} \end{array}}}{\Vertex{n_+,J}}
				}{\bullet}
			\\
			\vspace{0.2in}
				\displaystyle
				- \sum_{V^{J+2}=1}^{V^{J+1}} 2 \left( V^{J+1,J+2}-1 \right) \beta_{V^{\! J+2}}
				\left[
					\sco[0.1]{\ensuremath{\begin{array}{c}\input{pstex/GV-J+1-J+2.pstex_t} \end{array}}}{\Vertex{n_+,J}}
				\right]
			\\
			\vspace{0.2in}
				\displaystyle
				- \sum_{V^{J+2}=1}^{V^{J+1}} \gamma_{V^{\! J+2}}
				\left[
					\sco[0.1]{\pder{}{\alpha} \ensuremath{\begin{array}{c}\input{pstex/GV-J+1-J+2.pstex_t} \end{array}}}{\Vertex{n_+,J}}
				\right]
			\\
				\displaystyle
				-
				\left[
					\sco[0.1]{\ensuremath{\begin{array}{c}\input{pstex/GSigma_J+1-1.pstex_t} \end{array}}}{\Vertex{n_+,J}}
				\right]
			\end{array}
		}{11\Delta^{J+1}}
	\]
\caption{Final result of the exhaustive manipulation of diagram~\ref{bn:L0:a0}.}
\label{fig:Exhaust:L0:bn:a0}
\end{figure}
\end{center}

\section{Level-Three Manipulations} \label{sec:bn:L3}

As with the previous level of manipulation, we find that there is an 
excess of candidates for manipulation. In keeping with what went before,
we use the following prescription for choosing which terms to manipulate.
We begin by only considering terms formed by the exclusive action of $a_0$ and $a_1$
(as opposed to \eg\ terms formed by the $\flow$ striking a simple loop).
Of these, we further restrict our choice to those for which the final
operation was by an $a_0$ unless the diagram was formed exclusively
by the action of $a_1$.

This leaves us with three candidates for manipulation: diagrams~\ref{bn:L2:a1:a0:a0-r-R},
\ref{bn:L2:a1:a1:a1} and~\ref{bn:L2:a0:a0:a0:R}. All bar the first of these have been
implicitly
dealt with as a consequence of the
generalised manipulations of the previous section and so are not considered further.

As for diagram~\ref{bn:L2:a1:a0:a0-r-R} we find,
 as we might suspect, that its seed action contributions
are cancelled upon processing the manipulable part. This leaves over the non-manipulable,
Wilsonian effective action contribution. Again, we suspect that this will be
cancelled upon iterating the diagrammatic procedure, but we will not get to see this
explicitly. Rather, the initial manipulation of diagram~\ref{bn:L2:a1:a0:a0-r-R}
will guide us toward a set of cancellation mechanisms, one of which
guarantees the cancellation of the remaining Wilsonian effective action
contribution to diagram~\ref{bn:L2:a1:a0:a0-r-R}.

Indeed, the cancellation mechanisms obtained in the section
will complete the set of rules required to show that $\beta_{n_+}$
can, up to gauge remainders and terms which require manipulation 
at $\Op{2}$, be expressed as a set of $\Lambda$-derivative, $\alpha$
and $\beta$-terms.

\subsection{Diagram~\ref{bn:L2:a1:a0:a0-r-R}} \label{sec:bn:L2:a1:a0:a0-r-R}

We will now use the insights of section~\ref{sec:bn:L1:a0:a0-r-R}, in which we processed diagram~\ref{bn:L1:a0:a0-r-R},
to process diagram~\ref{bn:L2:a1:a0:a0-r-R}. These two diagrams are of a very similar structure. However, in 
the latter case, we have an additional effective propagator with which to decorate. 
This greatly increases the number of different topologies which can be constructed. The compensation for this is that the sum over
vertex arguments decreases by one.

Let us consider the manipulable component of diagram~\ref{bn:L2:a1:a0:a0-r-R}. Since, in this case, the wine is
undecorated, we must use one of the effective propagators to make a join between the loose vertex and either
of the other two. So far, this is just the same as with diagram~\ref{bn:L1:a0:a0-r-R}. The real 
difference occurs with the attachment of the remaining effective propagator, with which we can do one
of two things: form a simple loop or join any pair of vertices. However, despite this apparent complexity,
the conversion into a $\Lambda$-derivative term is remarkably simple. 

As we did with diagram~\ref{bn:L1:a0:a0-r-R}, let us try constructing the $\Lambda$-derivative term by
promoting the wine to a decorative effective propagator, as shown in figure~\ref{fig:Beta-n+L2:a0:a0-a0-LdL-trial}.
The currently undetermined parameter, $\sym'$, could, in principle, depend
on the structure of the diagram after all joins between vertices have been made.
\begin{center}
\begin{figure}[h]
	\[
	\displaystyle
	\frac{1}{16} \sum_{r=0}^{n} \sum_{t=0}^r
	\sym'
	\dec{
		\dec{
			\sco[0.1]{\tcd{Vertex-t-R}{Vertex-r_t-R}}{\ensuremath{\begin{array}{c}\input{pstex/Vertex-n_r-R.pstex_t} \end{array}}}
		}{\bullet}
	}{11\Delta^3}
	\]
\caption{A trial form for the $\Lambda$-derivative term arising from diagram~\ref{bn:L2:a1:a0:a0-r-R}.}
\label{fig:Beta-n+L2:a0:a0-a0-LdL-trial}
\end{figure}
\end{center}

From this diagram, the can generate the parent diagram in 
\[
\nCr{3}{2} \times 2 \times \nCr{3}{1} = 18 
\]
different ways. (We choose two vertices from three, and
one effective propagator from three, which can attach
either way round) Thus $\sym'$ just reduces to a constant factor
of $1/18$. In figure~\ref{fig:Beta-n+L3:a1:a0:a0} we process 
diagram~\ref{bn:L2:a1:a0:a0-r-R}. Due to the 
indistinguishability of the three vertices
of the $\Lambda$-derivative term, we need only compute the
flow of one of these, so long as we multiply by three.
\begin{center}
\begin{figure}
	\[
	\begin{array}{c}
	\vspace{0.2in}
		\displaystyle
		\frac{1}{288} \sum_{r=0}^n \sum_{t=0}^r
		\dec{
		\begin{array}{c}
		\vspace{0.2in}
			\decB[0.1]{
					\sco{
						\tcd[-0.2]{Vertex-t-R}{Vertex-r_t-R}
					}{\ensuremath{\begin{array}{c}\input{pstex/Vertex-n_r-R.pstex_t} \end{array}}}
			}{\bullet} 
		\\
		\vspace{0.2in}
			\displaystyle
			-3 \sum_{u=1}^t
			\left[
				2(t_u-1)\beta_u \dec{\sco{\tcd{Vertex-t_u}{Vertex-r_t-R}}{\ensuremath{\begin{array}{c}\input{pstex/Vertex-n_r-R.pstex_t} \end{array}}}}{}
				+ \gamma_u \dec{\sco{\tco[-0.2]{\pder{}{\alpha}\ensuremath{\begin{array}{c}\input{pstex/Vertex-t_u.pstex_t} \end{array}}}{\ensuremath{\begin{array}{c}\input{pstex/Vertex-r_t-R.pstex_t} \end{array}}}}{\ensuremath{\begin{array}{c}\input{pstex/Vertex-n_r-R.pstex_t} \end{array}}}}{}
			\right]
		\\
			\displaystyle
			-\frac{3}{2}
			\left(
			\sum_{u=0}^t
			\dec{
				\PEO[0.1]{}{
					\tco[-0.2]{\ensuremath{\begin{array}{c}\input{pstex/Dumbell-t-u.pstex_t} \end{array}}}{\scd[0.1]{Vertex-r_t-R}{Vertex-n_r-R}}
				}{bn:L3:a1:a0:a0-a0}{fig:Beta-n+L3:a1:a0:a0:R}
			}{}
			- 
			\dec{
					\sco[0.1]{\scd[0.1]{Vertex-t-R}{Vertex-r_t-R}}{\ensuremath{\begin{array}{c}\input{pstex/Sigma_n_r+1.pstex_t} \end{array}}}
			}{}
			\right)
		\end{array}
		}{11\Delta^3} 
	\\
		\displaystyle
		-\frac{1}{16} \sum_{r=0}^n \sum_{t=0}^r
		\dec{
			\frac{1}{2}
			\decB[-0.1]{
				\LEO[0.1]{2}{
					\sco{\ensuremath{\begin{array}{c}\input{pstex/Vertex-t-R.pstex_t} \end{array}}}{\scd{Vertex-r_t-R}{Vertex-n_r-R-Dloop}}
				}{bn:L3:a1:a0:a0-Dloop}
			}{}
			-
			\left[
				\sco{\ensuremath{\begin{array}{c}\input{pstex/Dumbell-tR-r_tR-DW.pstex_t} \end{array}}}{\ensuremath{\begin{array}{c}\input{pstex/Vertex-n_r-R.pstex_t} \end{array}}}
			\right]
			+2
			\left[
				\CEO[0.1]{5}{\sco{\ensuremath{\begin{array}{c}\input{pstex/Dumbell-tR-r_tR-S.pstex_t} \end{array}}}{\ensuremath{\begin{array}{c}\input{pstex/Vertex-n_r-R.pstex_t} \end{array}}}}{bn:L2:a1:a0:a0-S}{bn:L3:a1:a0:a0:a0-S}
			\right]
		}{11\Delta^2}
	
	\end{array}
	\]
\caption{The result of processing diagram~\ref{bn:L2:a1:a0:a0-r-R}.}
\label{fig:Beta-n+L3:a1:a0:a0}
\end{figure}
\end{center}

The correction term
formed by the $\flow$ striking a simple loop, diagram~\ref{bn:L3:a1:a0:a0-Dloop},
can be formed in nine ways as
we can choose any one of three effective 
propagators to decorate any of the three vertices. The final two diagrams
in the figure are just the un-manipulated components of the parent diagram.

In figure~\ref{fig:Beta-n+L3:a1:a0:a0:R}, we show the effect of isolating the two-point,
tree level vertices belonging to diagram~\ref{bn:L3:a1:a0:a0-a0}. These vertices have been decorated, any loose ends tied up
and the effective propagator relation has been applied.
\begin{center}
\begin{figure}
	\[
	\begin{array}{c}
	\vspace{0.2in}
		\displaystyle
		-\frac{1}{192} \sum_{r=0}^n \sum_{t=0}^r 
		\dec{
			\sum_{u=0}^t 
			\dec{	
					\tco[-0.2]{\ensuremath{\begin{array}{c}\input{pstex/Dumbell-tR-uR.pstex_t} \end{array}}}{\scd[0.1]{Vertex-r_t-R}{Vertex-n_r-R}}
			}{11\Delta^3}
			-24
			\dec{
				\left[
					\CEO[0.1]{5}{\scd[0.1]{Dumbell-hat-tR-r_tR}{Vertex-n_r-R}}{bn:L3:a1:a0:a0:a0-S}{bn:L2:a1:a0:a0-S}
				\right]
				-
				\left[
					\scd[0.1]{Dumbell-hat-tR-r_tRGR}{Vertex-n_r-R}
				\right]
			}{11\Delta^2}
		}{}
	\\
	\vspace{0.2in}
		\displaystyle
		+ \frac{1}{16} \sum_{r=0}^n \sum_{t=0}^r
		\dec{ 
			\frac{1}{6}
			\decB[-0.1]{
					\tco[-0.2]{\ensuremath{\begin{array}{c}\input{pstex/Dumbell-tR-E.pstex_t} \end{array}}}{\scd[0.1]{Vertex-r_t-R}{Vertex-n_r-R}}
			}{1 \Delta^3} 
			+
			\dec{
				\sco[0.1]{
					\ensuremath{\begin{array}{c}\input{pstex/Vertex-hat-t-R-loop.pstex_t} \end{array}}
					-\ensuremath{\begin{array}{c}\input{pstex/Vertex-hat-t-R-loopGR.pstex_t} \end{array}}
					-\ensuremath{\begin{array}{c}\input{pstex/Vertex-hat-t-R-WGR.pstex_t} \end{array}}
				}{\scd[0.1]{Vertex-r_t-R}{Vertex-n_r-R}}
			}{11\Delta^2}
		}{}
	\\
	\vspace{0.2in}
		\displaystyle
		+\frac{1}{16} \sum_{r=0}^n
		\dec{
			\dec{
				\sco{\ensuremath{\begin{array}{c}\input{pstex/Vertex-TLTP-loop.pstex_t} \end{array}}}{\scd{Vertex-r-R}{Vertex-n_r-R}}		
			}{11\Delta^2}
			+4
			\dec{
				\CED[0.1]{4}{Dumbell-rR-n_rR-RW}{bn:L3:a1:a0:a0-a0-W}{bn:L1:a1:a0-DW}
				-2\ensuremath{\begin{array}{c}\input{pstex/Dumbell-rR-n_rR-RWGR.pstex_t} \end{array}}
				+\ensuremath{\begin{array}{c}\input{pstex/Dumbell-rR-n_rR-RW-DGR.pstex_t} \end{array}}
			}{11\Delta}
		}{}
	\\
	\vspace{0.2in}
		\displaystyle
		+\frac{1}{4} \sum_{r=0}^n
		\dec{
			\sco[0.1]{\ensuremath{\begin{array}{c}\input{pstex/Vertex-r-R-loop+GRs.pstex_t} \end{array}} - 2 \left(\ensuremath{\begin{array}{c}\input{pstex/Vertex-r-R-WGRs.pstex_t} \end{array}} \right)}{\ensuremath{\begin{array}{c}\input{pstex/Vertex-n_r-R.pstex_t} \end{array}}}
		}{11\Delta}
	\\
		\displaystyle
		+\frac{1}{16} \sum_{r=0}^n
		\dec{
			-\dec{
				\sco{\ensuremath{\begin{array}{c}\input{pstex/Struc-TLTP-WGR.pstex_t} \end{array}}}{\scd{Vertex-r-R}{Vertex-n_r-R}}
			}{}
			+2 
			\dec{
				\sco{\ensuremath{\begin{array}{c}\input{pstex/Dumbell-rR-TLTP-RW.pstex_t} \end{array}} - \ensuremath{\begin{array}{c}\input{pstex/Dumbell-rR-TLTP-RW-GR.pstex_t} \end{array}}}{\ensuremath{\begin{array}{c}\input{pstex/Vertex-n_r-R.pstex_t} \end{array}}}
			}{}
		}{1\Delta^2}
	\end{array}
	\]
\caption{Isolation of two-point, tree level vertices in diagram~\ref{bn:L3:a1:a0:a0-a0} and etc.}
\label{fig:Beta-n+L3:a1:a0:a0:R}
\end{figure}
\end{center}

As expected, we find a number of cancellations between diagrams from this level
and the previous two. The first two cancellations, as we will see shortly, can be described
by the generalisation of cancellation mechanisms~\ref{CM:LineOfVertices:Seed} and~\ref{CM:LineOfVertices:Wilson}.

\begin{cancel}
Diagram~\ref{bn:L3:a1:a0:a0:a0-S} exactly cancels diagram~\ref{bn:L2:a1:a0:a0-S}.
\end{cancel}

\begin{cancel}
Diagram~\ref{bn:L3:a1:a0:a0-a0-W} exactly cancels diagram~\ref{bn:L1:a1:a0-DW}.
\end{cancel}

The remaining cancellation is similar to something we have already seen, 
but in a different context---as we will discuss shortly. First, though, recall how, in
section~\ref{sec:bn:L2}, we chose not to manipulate diagram~\ref{bn:L1:a0:a1-B}
since we claimed that it would be partially cancelled by terms arising from
the manipulation of diagram~\ref{bn:L1:a1:a0-Re}. We now see that this
diagram is, in fact, \emph{completely} cancelled by terms arising from the \emph{iterated}
manipulation of diagram~\ref{bn:L1:a1:a0-Re}.

\begin{cancel}
Diagram~\ref{bn:L3:a1:a0:a0-a0-r-R-loop} is very similar to diagram~\ref{bn:L2:a1:a0:loop-r}.
They have the same vertex arguments, the same overall factor and are decorated by the
same set of fields. However, in the former case, the explicitly drawn wine---which forms
a simple loop---must be undecorated. In the latter case, this wine must must be decorated.
Hence, we can combine the diagrams by taking either one of them and removing the
restriction on the decoration of the wine. This new diagram can then be combined with
diagram~\ref{bn:L2:a1:a0:EP-loop-r} to give the first diagram in figure~\ref{fig:CancelLoop}.

Let us analyse the topmost vertex. The restriction that this vertex be reduced excludes
the case $\Sigma_0^2 = -S_0^2$. However, this contribution is exactly provided by 
diagram~\ref{bn:L2:a1:a0:TLTP-loop}. By including this term, we are left with only the first diagram
on the second line of figure~\ref{fig:CancelLoop}. Finally, then, this diagram exactly cancels
diagram~\ref{bn:L1:a0:a1-B}.
\label{C:bn:L1:a0:a1-B}
\end{cancel}

\begin{center}
\begin{figure}[h]
	\begin{eqnarray*}
		\displaystyle
		\ref{bn:L3:a1:a0:a0-a0-r-R-loop} + \ref{bn:L2:a1:a0:loop-r}+ \ref{bn:L2:a1:a0:EP-loop-r}
	& 	= 
	&	\frac{1}{4}
		\sum_{r=0}^n
		\dec{			
			\scd[0.1]{Sigma_r-R}{Vertex-n_r-R}
		}{11\Delta}
	\\
	&&
	\\
	&	= 
	&	\frac{1}{4}
		\dec{
		\sum_{r=0}^n
			\left[
				\CEO[0.1]{5}{\scd[0.1]{Sigma_r}{Vertex-n_r-R}}{bn:L1:a0:a1-Cancel}{bn:L1:a0:a1-B} 
			\right]
			-
			\left[ 
				\CEO[0.1]{5}{\scd[0.1]{Sigma_0-2}{Vertex-n-R}}{bn:L1:a0:a1-Cancel-Correction}{bn:L2:a1:a0:TLTP-loop}
			\right]
		}{11\Delta}
	\end{eqnarray*}
\caption{Showing how diagrams~\ref{bn:L3:a1:a0:a0-a0-r-R-loop}, \ref{bn:L2:a1:a0:loop-r}, \ref{bn:L2:a1:a0:EP-loop-r}
and~\ref{bn:L2:a1:a0:TLTP-loop} combine to cancel diagram~\ref{bn:L1:a0:a1-B}.}
\label{fig:CancelLoop}
\end{figure}
\end{center}

It is apparent that cancellation~\ref{C:bn:L1:a0:a1-B} has much in common with
the combination of cancellations~\ref{C:L0:a1:S}, \ref{C:L0:a1:DW} and~\ref{C:L0:a1:TLTP}.
In each case we are cancelling components of a diagram containing a structure formed by the action of
$a_1$. In the latter case, this is all that the diagram contains and so the cancellations
arise from manipulating this structure directly. In the former case, however, the 
structure is formed after the action of an $a_0$ on the previous level and so we have
chosen not to manipulate it.

Had we chosen to manipulate it, the cancellation of the seed action contribution
and the Wilsonian effective action contributions with decorated wine would
then have proceeded in exactly the same way as 
cancellations~\ref{C:L0:a1:S}, \ref{C:L0:a1:DW} and~\ref{C:L0:a1:TLTP}.
However, there would be additional contributions arising from the $\flow$
striking other structures in the diagram. This, in turn, would then
have affected our manipulation of diagram~\ref{bn:L1:a1:a0-Re}, since
it would have removed the manipulable contributions containing simple loops.

To generalise the cancellations just seen, we will again consider a
level-$J$ diagram, but this time allow decoration by any (legal)
number of effective propagators. To ensure that the relationship
between the number of vertices, the vertex arguments and the number
of effective propagators is correct, we will start by analysing the
way in which manipulable diagrams are formed.

We assume that we have a manipulable diagram and that we have 
promoted the wine to a decorative effective propagator, thereby
converting the diagram into a $\Lambda$-derivative term. We
know that the diagram has been formed by the repeated
action of $a_1$s and $a_0$s but assume that any instances of
$a_1$s are consecutive and not preceded by any instances of $a_0$.

In terms of the $\Lambda$-derivative term that has been formed,
it is the case that each instance of $a_0$ has contributed
a vertex and an effective propagator, whilst leaving the 
sum of the vertex arguments the same. Each instance of
$a_1$ has contributed an effective propagator, has decreased
the sum of the vertex arguments by one but left the number
of vertices the same. Both $a_0$ and $a_1$ contribute a 
factor of $1/2$ each time they act. The former comes
with a relative factor of $-1$.

If we define $s$ to be the number of times $a_1$ has
acted, then
a generic level-$J$ $\Lambda$-derivative term possesses $J+2-s$
vertices, the sum of whose arguments adds up to $n_+-s$.
These vertices are decorated by the usual external fields
and by $J+1$ effective propagators.

Since we have already considered, in section~\ref{sec:bn:L1:a1:a1},
diagrams formed by the exclusive action of $a_1$, we
consider in this section those diagrams for which
$a_0$ has acted at least once.

The treatment of these terms is 
simply a generalisation of what we did in section~\ref{sec:bn:L1:a0:a0-r-R}.
The first thing to note is that the presence of more than one vertex
implies the action of at least one instance of $a_0$, which gives the
condition that
\begin{equation}
	J \geq s.
\label{eq:Jgeqs}
\end{equation}

Compared to the case in section~\ref{sec:bn:L1:a0:a0-r-R}, there are
a number of changes we must make. First, we replace $\norm_{J,J}$ by $\norm_{J,J-s}$
where it is straightforward to show that
\begin{equation}
	\norm_{J,J-s} = \frac{1}{J! (J-s)!} \left(\frac{1}{2}\right)^s \left(-\frac{1}{2} \right)^{J+1-s}.
\label{eq:norm_(J,s)}
\end{equation}
It is clear that this expression is consistent with our previous expression of
both $\norm_{J,J}$ and  $\norm_{J,0}$ (see equations~\ref{eq:NormArbLine} and~\ref{eq:norm_(J,0)}).

The second change that we must make is to the level at which the diagrammatics terminates.
Compared to the terms in section~\ref{sec:bn:L1:a0:a0-r-R}, a manipulable diagram now
has  $J+2-s$ vertices, rather than $J+2$. Thus, from equation~\ref{eq:CritLevel:s=0},
such a diagram will last appear at level
\begin{equation}
	J \leq 2n+s,
\label{eq:CritLevel}
\end{equation}
meaning that the diagrammatic procedure terminates at level $2n+s+1$.
We emphasise that the critical level now depends on $s$.

Thirdly, after casting the manipulable term as a $\Lambda$-derivative,
we must now admit the possibility of the $\flow$ striking a simple loop. 
Lastly, we can no longer discard either gauge remainders or 
two-point, tree level vertices decorated by external fields.

Figure~\ref{fig:bn:L:J...a0} shows the starting point for the manipulations: a level-$J$
diagram for which the manipulable term has been isolated. Again, to be rigorous,
the diagrams of this figure should, strictly speaking, possess at least three vertices
and to be rigorous in what follows, at least four. However, it is clear by now
that the results we obtain will be valid, so long as we are careful with how we
interpret the diagrams. We note further that the case of just two vertices has, of
course, been treated in complete generality in section~\ref{sec:bn:L1:a1:a1}.
\begin{center}
\begin{figure}[h]
	\[
	\begin{array}{c}
		\displaystyle
		\norm_{J,J-s} \sum_{r=0}^{n_{s-1}} \sum_{t=0}^r \cdots \sum_{y=0}^x
	\\[1ex]
		\times
		\dec{
			\left[
				\PGD[0.1]{}{LJ-MultipleVertices}{bn:LJ:MV-M}{fig:bn:L:J...a0-M}
			\right]
			+
			\left[ 
				\LGD[0.1]{}{LJ-MV-DW}{bn:LJ:MV-DW}
			\right]
			-2 
			\left[ 
				\CGD[0.1]{}{LJ-MV-S}{bn:LJ:MV-S}{bn:LJ+1:MV-a0-S}
			\right]
		}{11\Delta^J}
	\end{array}
	\]
\caption{Isolation of the manipulable part of a level-$J$ diagram.}
\label{fig:bn:L:J...a0}
\end{figure}
\end{center}

Figure~\ref{fig:bn:L:J...a0-M}
shows the result of performing the manipulations (including isolation of 
two-point, tree level vertices and etc.) where we have not explicitly 
drawn 
 the $\Lambda$-derivative, $\beta$ and $\gamma$-terms, gauge remainders
and diagrams which require manipulation at $\Op{2}$.

\begin{center}
\begin{figure}
	\[
	\begin{array}{l}
	\\
	\vspace{0.2in}
		\displaystyle
		\norm_{J+1,J+1-s} \sum_{r=0}^{n_{s-1}}
		\dec{
			\sum_{t=0}^r \cdots \sum_{z=0}^y
			\left[
				\LGD[0.1]{}{LJ-MV-a0}{bn:LJ:MV-a0}
			\right]
			- \sum_{r=0}^{n-s} \sum_{t=0}^r \cdots \sum_{y=0}^x
			\left[
				\LGD[0.1]{}{LJ-MV-a1}{bn:LJ:MV-a1}
			\right]
		}{11\Delta^{J+1}}
	\\	
	\vspace{0.2in}
		\displaystyle
		+ \norm_{J,J-s} \sum_{r=0}^{n_{s-1}} \sum_{t=0}^r \cdots \sum_{y=0}^x
		\dec{
			\textstyle
			2
			\left[
				\CGD[0.1]{}{LJ-MultipleVertices-S}{bn:LJ+1:MV-a0-S}{bn:LJ:MV-S}
			\right]
			-\frac{1}{J+1-s}
			\left(
			\left[
				\LGD[0.1]{}{LJ-MV-Dloop}{bn:LJ+1:MV-Dloop}
			\right]
			-
			\left[
				\LGD[0.1]{}{LJ-MV-Dloop-C}{bn:LJ+1:MV-Dloop-C}
			\right]
			\right)
		}{11\Delta^J}
	\\
		\displaystyle
		+\sum_{r=0}^{n_{s-1}}\sum_{t=0}^r \cdots \sum_{x=0}^w
		\dec{
		2J\norm_{J,J-s}
		\left[
			\LGD[0.1]{}{LJ-MV-TLTPx2-A}{bn:LJ+1:MV-TLTPx2-A}
		\right]
		- \norm_{J-1,J-1-s}
		\left[
			\LGD[0.1]{}{LJ-MV-TLTPx2-B}{bn:LJ+1:MV-TLTPx2-B}
		\right]
		}{11\Delta^{J-1}} + \cdots
	\end{array}
	\]
\caption{Selected terms arising from the manipulation of diagram~\ref{bn:LJ:MV-M}.}
\label{fig:bn:L:J...a0-M}
\end{figure}
\end{center}

\Cancel{bn:LJ+1:MV-a0-S}{bn:LJ:MV-S}

A cancellation mechanism follows trivially from
cancellation~\ref{bn:LJ+1:MV-a0-S}.
\begin{CM}
A level-$J$ diagram formed by $s$ consecutive instances of $a_1$
followed by $J-s+1$ instances of $a_0$ possesses a dumbbell
structure and an additional $J-s$ reduced vertices. The diagram as a whole is
decorated by the usual external fields and $J+1$ effective propagators.
Stripping off any remaining two-point, tree level vertices, we arrive
at the parent diagram for what follows.

The component of this diagram for which the vertices are Wilsonian effective
action vertices and the wine is undecorated can be further manipulated,
using the flow equations. This manipulation
is guaranteed to produce a further set of terms generated by $a_0$.
A subset of these terms  contain a dumbbell comprising a single 
two-point, tree level vertex and thus, necessarily, also a reduced,
seed action vertex.
If we join this two-point, tree level vertex 
to one of the loose (reduced) Wilsonian effective action vertices
then the resulting diagram
will exactly cancel the seed action 
contributions to the parent diagram, up to gauge remainders.
\label{CM:MV-a0-S}
\end{CM}

The remaining cancellation mechanisms require a little more work. First,
consider diagram~\ref{bn:LJ+1:MV-TLTPx2-B}. The level-$J+1$ version of
this diagram, assuming that it exists, 
will exactly cancel diagram~\ref{bn:LJ:MV-DW}. This
is what we expect, since we know that diagrams containing exclusively
Wilsonian effective action vertices and decorated wines always cancel
diagrams coming two levels down the line. Again, if the diagrammatic
procedure terminates at level-$J+1$, then this implies that the
only Wilsonian effective action contributions to the parent diagram are
manipulable.

\begin{CM}
A level-$J$ diagram formed by $s$ consecutive instances of $a_1$
followed by $J-s+1$ instances of $a_0$ possesses a dumbbell
structure and an additional $J-s$ reduced vertices. The diagram as a whole is
decorated by the usual external fields and $J+1$ effective propagators.
Stripping off any remaining two-point, tree level vertices, we arrive
at the parent diagram for what follows.

Assuming that we can iterate the diagrammatic procedure twice
then, amongst the terms we generate, is a diagram with
$J-s+2$ reduced vertices plus an additional dumbbell structure.
Focusing on the case where the two vertices comprising the dumbbell
are two-point, tree level vertices, we now join each of
these to a different, loose vertex. Summing over all ways of doing this,
the resulting term will exactly cancel
the remaining Wilsonian effective action contributions to the original parent diagram,
up to gauge remainders.

In case that the diagrammatic procedure terminates with the generation of level-$J+1$
terms, the only Wilsonian effective action contributions to the parent diagram
are those without a decorated wine \ie\ those which have already been manipulated.
\label{CM:MV-a0-DW}
\end{CM}

The next cancellation mechanism follows by comparing diagrams~\ref{bn:LJ:MV-a1} and~\ref{bn:LJ+1:MV-Dloop}
and~\ref{bn:LJ+1:MV-Dloop-C} and~\ref{bn:LJ+1:MV-TLTPx2-A}.
Looking at the first pair of diagrams we note that, from equation~\ref{eq:norm_(J,s)},
\[
\frac{\norm_{J+1,J-s}}{J+1-s} = -\norm_{J+1,J-s+1}.
\]
Consequently, the version of diagram~\ref{bn:LJ+1:MV-Dloop} for
which $(J,J-s) \rightarrow (J+1,J-s)$ exactly cancels diagram~\ref{bn:LJ:MV-a1}. 

Due to these diagrams occurring for different values of $J$ and $s$, we must check that
the cancellation occurs for all values of $J,s$ and not just over some range between the
boundaries. At the bottom end, we note that
diagram~\ref{bn:LJ+1:MV-Dloop} does not exist for $s=0$ (after decoration
we would be left with a loose vertex, in this case). Thus, by equation~\ref{eq:Jgeqs},
this diagram first exists for $(J,s)=(1,1)$ which cancels the $(0,0)$ instance
of diagram~\ref{bn:LJ:MV-a1}. These cancellations continue all the way up to the 
version of diagram~\ref{bn:LJ:MV-a1}, spawned from the critical level.
Recalling from equation~\ref{eq:CritLevel} that the critical level depends on $s$,
it is clear that the final instance of diagram~\ref{bn:LJ:MV-a1} is exactly
cancelled by the final instance of diagram~\ref{bn:LJ+1:MV-Dloop}.

Looking at diagrams~\ref{bn:LJ+1:MV-Dloop-C} and~\ref{bn:LJ+1:MV-TLTPx2-A} we note that, from equation~\ref{eq:norm_(J,s)},
\[
2 J \norm_{J,J-s} = -\frac{\norm_{J-1,J-s-1}}{J-s}.
\]
Consequently, the level-$J+1$ version of diagram~\ref{bn:LJ+1:MV-TLTPx2-A} exactly cancels diagram~\ref{bn:LJ+1:MV-Dloop-C}.
Again, we must check that this cancellation occurs over the complete range of values of $J$ ($s$ is the same
for both diagrams).

At the bottom end, we know that the first instance of diagram~\ref{bn:LJ+1:MV-TLTPx2-A} occurs one level after the
first instance of diagram~\ref{bn:LJ+1:MV-Dloop-C}, since the former requires the action of two instances of $a_0$, rather
than just one. Hence, the first instances of these diagrams cancel. At the top end, however, we are left
over with the version of diagram~\ref{bn:LJ+1:MV-Dloop-C} spawned from the critical level: the version
of diagram~\ref{bn:LJ+1:MV-TLTPx2-A} which should cancel it is never generated, since the diagrammatic procedure
will have terminated.

Let us consider the level-$\crit+1$ version of diagram~\ref{bn:LJ+1:MV-Dloop-C} in more detail. To maximise
the number of vertices and hence the level, the structure at the bottom of this diagram must contain a
tree level vertex, being formed by processing a tree level vertex decorated by a simple loop, and attached to
one other vertex. Hence, the structure at the bottom of the diagram is decorated by a single field.
The
constraint that both the wine and the vertex are reduced cannot be satisfied: the diagram vanishes. 
Consequently, the set of diagram~\ref{bn:LJ:MV-a1}, \ref{bn:LJ+1:MV-Dloop}, \ref{bn:LJ+1:MV-Dloop-C} and~\ref{bn:LJ+1:MV-TLTPx2-A}
are always guaranteed to disappear from the calculation.

\begin{CM}
Consider a level-$J$ diagram formed by $s$ consecutive instances of $a_1$
followed by $J-s+1$ instances of $a_0$. After stripping off any remaining
two-point, tree level vertices, we arrive at the parent diagram for what follows.

Processing the manipulable component of the parent diagram yields (amongst others)
a term formed by the action of $a_1$ and a term formed by the action of $a_0$.
In the case that $s>0$, a diagram in which the $\flow$ strikes a simple loop is
also generated. This term can be combined with terms arising from the $a_0$ term
in which the wine is made to form a simple loop (which can only be formed for $s>0$).

Up to a correction term, this combination precisely cancels the diagram formed by 
the action of $a_1$ arising from a parent diagram if we let $(J,s) \rightarrow (J+1,s+1)$.

Returning to the parent diagram, we assume that we can iterate the diagrammatic procedure twice.
We focus on the term formed by two instances of $a_0$; specifically, the component
of the resulting dumbbell comprising two two-point, tree level vertices.
If we join these two vertices together, then the resulting diagram exactly cancels
the above correction term.

In the case that the diagrammatic procedure terminates before the second iteration,
the correction term vanishes!
\label{CM:MV-a1}
\end{CM}

\section{The Critical Level} \label{sec:bn:critical}

The cancellation mechanisms~\ref{CM:a1-S}--\ref{CM:MV-a1} can now be used
to give, up to gauge remainders and terms possessing an $\Op{2}$
stub, an extremely compact expression for $\beta_{n_+}$. In fact, we
need not even use all of the cancellation mechanisms, since 
mechanisms~\ref{CM:a1...a0-S}--\ref{CM:LineOfVertices:Wilson}
are contained within mechanisms~\ref{CM:MV-a0-S} and~\ref{CM:MV-a0-DW}.

The effect of the minimal set of cancellation mechanisms is very simple:
iterating the diagrammatic procedure until exhaustion, the only terms
which survive, up to gauge remainders and terms possessing an $\Op{2}$ stub
are $\Lambda$-derivative terms, $\beta$-terms and $\gamma$-terms.
Using the methodology of section~\ref{sec:bn:L1:a0:a0-r-R}
in which we gave a diagrammatic expression for the result of
exhaustively manipulating diagram~\ref{bn:L0:a0}, we can arrive at
a similar expression for exhaustively manipulating both diagrams~\ref{bn:L0:a0}
and~\ref{bn:L0:a1}. The only change that we must make is to take
\begin{equation}
V^0 = n_+ -s.
\label{eq:General-V^0}
\end{equation}
This gives us the extremely compact diagrammatic expression for $\beta_{n_+}$,
shown in figure~\ref{fig:Beta-n+Part-I}, where we have used
the compact notation defined by~\ref{eq:n-loop:ProdOverI}.

\begin{center}
\begin{figure}
	\[
	\begin{array}{c}
	\vspace{0.2in}
		\displaystyle
		-4 \beta_{n_+} \Box_{\mu\nu}(p) = 
		-2 \sum_{s=1}^{n_+} \sum_{J=s}^{2n+s} 
		\norm_{J+1,J+1-s}
	\\
	\vspace{0.2in}
		\displaystyle
		\times
		\dec{
		\begin{array}{c}
		\vspace{0.2in}
			\displaystyle
			\frac{1}{J+2-s}
			\dec{
				\sco[0.1]{\ensuremath{\begin{array}{c}\input{pstex/GV-J+1-s.pstex_t} \end{array}}}{\ensuremath{\begin{array}{c}\input{pstex/V-n+-s,J-s.pstex_t} \end{array}}}
			}{\bullet}
		\\
			\displaystyle
			- \sum_{V^{J+2-s}=1}^{V^{J+1-s}} 
			\left[
				\sco[0.1]{
					\left[2 \left( V^{J+1-s,J+2-s}-1 \right) \beta_{V^{\! J+2-s}} + \gamma_{V^{\! J+2-s}}	\pder{}{\alpha}\right]	
					\ensuremath{\begin{array}{c}\input{pstex/GV-J+1-s-J+2-s.pstex_t} \end{array}}
				}{\ensuremath{\begin{array}{c}\input{pstex/V-n+-s,J-s.pstex_t} \end{array}}}
			\right]
		\end{array}		
		}{11\Delta^{J+1}}
	\\
	\vspace{0.2in}
		\displaystyle
		-2\sum_{J=0}^n \norm_{J+1,0}
		\dec{
			\dec{
				\ensuremath{\begin{array}{c}\input{pstex/Vertex-n_J-R.pstex_t} \end{array}}
			}{\bullet}
			- \sum_{r=0}^{n_J} 
			\left[2 \left( n_J-r-1 \right) \beta_r + \gamma_r	\pder{}{\alpha}\right]	
			\ensuremath{\begin{array}{c}\input{pstex/Vertex-n_J+r.pstex_t} \end{array}}
		}{11\Delta^{J+1}}
	\\
	+\cdots
	\end{array}
	\]
\caption[Diagrammatic expression for $\beta_{n_+}$ arising from the exhaustive manipulation
of diagrams~\ref{bn:L0:a0} and~\ref{bn:L0:a1}.]{Diagrammatic expression for $\beta_{n_+}$ arising from the exhaustive manipulation
of diagrams~\ref{bn:L0:a0} and~\ref{bn:L0:a1}. The ellipsis denotes gauge remainders and terms
with an $\Op{2}$ stub.}
\label{fig:Beta-n+Part-I}
\end{figure}
\end{center}

A word should be said about how to simplify this expression, yet further. Up until
now, we have been keeping all one-point Wilsonian effective vertices, unless they
occur at tree level. However, it is inevitable that, at some stage, we must implement the requirement
that one-point Wilsonian effective vertices are required to vanish and it is here that we must do it.
Our aim is to maximise the number of vertices each of the diagrams of figure~\ref{fig:Beta-n+Part-I}
can possess, so long as none of these vertices are one-point.

Once more, let us suppose that there are $T$ tree level vertices.
In the case of the $\Lambda$-derivative terms, each of these must be decorated
by at least three fields.
The remaining vertices must
be decorated by at least two fields.

Thus, for the $\Lambda$-derivative terms, we can obtain a constraint
on the maximum value of $J$ by equating the number of fields required
to produce a diagram with no one-point vertices with the total number of
available fields:
\[
2(J+2-s-T) + 3T  \leq 2(J+1) +2.
\]
Using equation~\ref{eq:level:T-TL} to substitute for $T$ we obtain:
\begin{equation}
	J \leq n+2s-1.
\label{eq:NewCritLevel}
\end{equation}

The new maximum value of $J$ can be used to replace the upper limit for the sum over $J$ in
figure~\ref{fig:Beta-n+Part-I}. This limit coincides with the old one when $s$
takes its maximum value, $n_+$, but is lower for all other values.

For the $\alpha$ and $\beta$-terms we might worry that there
could be a higher maximum value of $J$, since both of
these terms possess a single full vertex. Such a vertex can be
tree level, but needs only two decorative fields. However,
if this vertex is tree level, then this necessarily reduces
the sum over vertex arguments from $n_+-s$
to $n_+-s - V^{J+2-s}$, where $ V^{J+2-s} >0$. This ensures
that the maximum value of $J$ for the $\alpha$ and $\beta$-terms
is not higher than corresponding value for the $\Lambda$-derivative
terms.

At this stage, a final comment about the diagrams of figure~\ref{fig:Beta-n+Part-I}
is in order. When we come to decorate these diagrams,
terms which vanish at $\Op{2}$ will be generated. It is finally time to discard such diagrams,
just as it is now  time to discard diagrams with one-point
vertices.

%
%

\chapter{Gauge Remainders} \label{ch:bn:GRs}

In this chapter, we show how to deal with gauge remainders terms, whilst 
using the heavily compacted notation developed in  chapter~\ref{ch:bn:LambdaDerivsI}. 
The very first
thing that we must do is identify all terms containing gauge remainders.

Figure~\ref{fig:Beta-n+L3:a1:a0:a0:R} provides an example of all
the different types of terms we will encounter. On the first and
second lines, we see gauge remainders arising from a single
application of the effective propagator relation. Such terms---which we will
call Type-Ia terms---have
a seed action vertex joined to wine which ends in a gauge remainder. 
On the third and fourth lines,
we see gauge remainders arising from two applications of the
effective propagator relations. Such terms do not have any seed action vertices.
If the necessarily decorated wine ends in a gauge remainder at one end only,
then it is a gauge remainder of Type-Ib. If the wine ends in a gauge remainder
at both ends, then it is a gauge remainder of Type-II.

Finally, on the last line, we find gauge remainders formed by a single
application of the effective propagator relation but which also
possess an $\Op{2}$ stub. These terms are of Type-III.

We will find that the majority of cancellations occur between terms in the
same class. Nonetheless, as we will see, there are important cancellations
between the various classes, allowing us to incorporate the gauge remainder
terms into our compact expression for $\beta_{n_+}$.

\section{Type-Ia Gauge Remainders} \label{sec:GR-Ia}

A number of specific examples of Type-Ia gauge remainders are present,
throughout this chapter. The only ones which have been presented in
complete generality are those arising from the manipulation of
diagrams formed by the exclusive use of $a_1$, which appear in 
figure~\ref{fig:Beta-n+:LJ:a1...a1-Split}. We combine these with those formed
from the manipulation of terms generated by the action of at least one
instance of $a_0$ in figure~\ref{fig:bn:GR-Ia}.

\begin{center}
\begin{figure}
	\[
	\begin{array}{c}
	\vspace{0.2in}
		\displaystyle
		-2\sum_{s=1}^{n_+} \sum_{J=s+1}^{\JsumLim}  \norm_{J,J-s}
	\\
	\vspace{0.2in}
		\displaystyle
		\times
		\dec{
			\sco[0.3]{
				\PGD[0.1]{}{LJ+1-GR-Ia-line}{LJ+1-GR-Ia-line}{fig:bn:GR-Ia-Rest}
				+\frac{1}{J+1-s}
				\left[
					\sco{
						\PGD[0.1]{}{GV-V01-hat-R-GR}{LJ+1-GR-Ia-loop}{fig:bn:GR-Ia-Rest} 
						+ \PGD[0.1]{}{GV-V01-hat-R-WGR}{LJ+1-GR-Ia-WGR}{fig:bn:GR-Ia-Rest}
					}{\ensuremath{\begin{array}{c}\input{pstex/GV-J+1-s-R.pstex_t} \end{array}}}
				\right]
			}{\left[\VertexSum{1}{J-s} \ensuremath{\begin{array}{c}\input{pstex/GV-I-I+.pstex_t} \end{array}}\right]}
		}{11\Delta^J}
	\\
	\vspace{0.2in}
		\displaystyle
		-2\sum_{J=1}^{n_+} \sum_{r=0}^{n_{J-1}} \norm_{J,0}
		\dec{
			\PGD[0.1]{}{GR-Ia-Dumbell-n_J+r-1-S-R-rR}{LJ+1-GR-Ia-line-J=s}{fig:GR-Ia-J=s} +
			\left[
			\sco[0.1]{
				\PGD[0.1]{}{GV-n_J+r-1-S-R-GR}{LJ+1-GR-Ia-GR-J=s}{fig:bn:GR-Ia-J=s} +
				\PGD[0.1]{}{GV-n_J+r-1-S-R-WGR}{LJ+1-GR-Ia-WGR-J=s}{fig:bn:GR-Ia-J=s}
			}{\ensuremath{\begin{array}{c}\input{pstex/Vertex-r-R.pstex_t} \end{array}}}
			\right]
		}{11\Delta^J}
	\\
		\displaystyle
		+2 \sum_{J=0}^{n} \frac{\norm_{J,0}}{\vartheta_{n_J}}
		\dec{
			\PGD[0.1]{}{GV-n_J-GR}{LJ+1-1V-a0-2ptI-GR-pre}{fig:GR-Ia-1V} 
			+ \PGD[0.1]{}{GV-n_J-WGR}{LJ+1-1V-a0-2ptI-WGR-pre}{fig:GR-Ia-1V}
		}{11\Delta^J}
	\end{array}
	\]
\caption{Type-Ia gauge remainders.}
\label{fig:bn:GR-Ia}
\end{figure}
\end{center}

There are several aspects of figure~\ref{fig:bn:GR-Ia} that deserve comment. 
First, in the last two lines,  we have separated off those diagram containing two vertices
or fewer from those containing more than two. It is easy to see why: referring to the
first set of diagrams, if we were to allow the sum over $J$ to start from $s$, then
the product over $I$ does not make sense. Thus, we separate off the $J=s$ case,
which corresponds to there being two vertices. Similarly, the single vertex terms must
be treated separately. Note also that we have not used the compact
notation~(\ref{eq:n-loop:ProdOverI}) since this can only be used when the sum over the dummy
index $I$ starts from zero.

Secondly, despite the comments made under figure~\ref{fig:Beta-n+Part-I}
we use $2n+s$ as the upper limit for the sum over $J$. We know that
such diagrams can contain one-point Wilsonian effective action vertices. 
However, we wish to keep them
right up until the stage where the conversion of gauge remainder diagrams
into $\Lambda$-derivatives has been achieved.

Thirdly, we know that gauge remainders arising from terms formed by the exclusive action
$a_0$ kill the diagram. Thus we need not consider diagrams for which $s=0$: the sum over $s$ starts
from one.

We now want to generate the diagrams formed by the action of the gauge remainders.
Since these will be the very first gauge remainders performed we recall, from section~\ref{sec:GRs},
that we can collect together pushes forward and pulls back by using charge conjugation.

In keeping with what we have done already in this chapter, we
want to isolate two-point, tree level vertices.
The isolation of two-point, tree level vertices is trivial, even in the case where
the gauge remainder strikes a reduced vertex. Consider a gauge remainder
which strikes a reduced vertex, but which has not yet acted. Since the gauge
remainder kills  two-point (tree level) vertices, we can promote the reduced
vertex to a full vertex. Then, after the gauge remainder has acted, we simply split the
resulting vertex into a reduced part and a two-point, tree level part. 
 
The field that the gauge remainder strikes will be represented using 
the `socket' notation of
section~\ref{sec:socket}. When the socket decorates a two-point,
tree level vertex, we will partially decorate this vertex, not by specifying
the field which fills the socket, but by specifying the other field that
decorates the vertex.\footnote{This field carries the same momentum as the
vertex, and so we can use the effective propagator relation.}

We will start by dealing with the simplest terms: diagrams~\ref{LJ+1-1V-a0-2ptI-GR-pre} and~\ref{LJ+1-1V-a0-2ptI-WGR-pre}.
In the case of the former diagram we can drop the $\vartheta_{n_J}$ since, if the vertex is two-point and 
tree level, then it is killed by the gauge remainder anyway. In the latter case, something a little
odd is going on. Irrespective of whether or not we allow the gauge remainder to
act, the vertex of this diagram has a two-point, tree level component. The reason that this has a arisen
relates to how we have organised the calculation. Since, when we introduced the entire $\beta_{n_+}$ diagrammatics
we generated diagrams such as~\ref{LJ+1-1V-a0-2ptI-WGR-pre} with specific value of $J$, we were unable
to separate off two-point, tree level components cleanly as this would have required treating diagrams
with different values of $n$, differently. Now, with our more sophisticated treatment where $J$ is summed over,
we can happily 
separate off the two-point, tree level contribution to the vertex in diagram~\ref{LJ+1-1V-a0-2ptI-WGR-pre},
noting that this simply occurs for $J=n$. We do this in figure~\ref{fig:GR-Ia-1V} where we recall that
when we do take the two-point, tree level contribution, $\vartheta_{n_J} = 2$. We also redraw
diagram~\ref{LJ+1-1V-a0-2ptI-GR-pre}, removing the $\vartheta_{n_J}$ altogether.
\begin{center}
\begin{figure}
	\[
	\begin{array}{c}
	\vspace{0.2in}
		\displaystyle
		+2 \sum_{J=0}^{n} \norm_{J,0}
		\dec{
			\PGD[0.1]{}{GV-n_J-S-R-GR}{LJ+1-1V-a0-2ptI-GR}{fig:GR-Ia-a1s} 
			+ \PGD[0.1]{}{GV-n_J-S-R-WGR}{LJ+1-1V-a0-2ptI-WGR}{fig:GR-Ia-a1s}
		}{11\Delta^J}
	\\
		\displaystyle
		-
		\norm_{n-1,0}
		\dec{
			\PGD[0.1]{}{Struc-WGRx2}{GR-II-WGRx2}{fig:bn:GR-II}
		}{11\Delta^{n-1}}
		+
		\norm_{n,0}
		\dec{
			\PGD[0.1]{}{Struc-TLTP-WGR}{GR-III-TLTP-WGR}{fig:bn:GR-III}
		}{1\Delta^n}
		
	\end{array}
	\]
\caption{Re-expression of diagrams~\ref{LJ+1-1V-a0-2ptI-GR-pre} and~\ref{LJ+1-1V-a0-2ptI-WGR-pre}.}
\label{fig:GR-Ia-1V}
\end{figure}
\end{center}

The effects of our operations are now clear. The two diagrams~\ref{LJ+1-1V-a0-2ptI-GR}
and~\ref{LJ+1-1V-a0-2ptI-WGR} now take exactly the same form as the remaining diagrams
in figure~\ref{fig:bn:GR-Ia}, just with fewer vertices. The other two diagrams, 
\ref{GR-II-WGRx2} and~\ref{GR-III-TLTP-WGR} are more
properly classified as type-II and type-III diagrams, respectively. These will be returned
to in later sections. Note that diagram~\ref{GR-II-WGRx2} does not make sense for $n=0$
and so our prescription is simply to discard it, in this case.

We now focus on diagrams~\ref{LJ+1-1V-a0-2ptI-GR}
and~\ref{LJ+1-1V-a0-2ptI-WGR} and allow the gauge remainder to act. First, we will look at
the case where, in the former diagram, the gauge remainder bites the field to which 
the wine attaches and, in the latter diagram, the gauge remainder bites the base of
the wine. These two terms exactly cancel, via cancellation mechanism~\ref{CM:VertexJoinWine}.
In addition to this, we must of course take into account account terms formed by the gauge remainders
striking other locations. The result of processing diagrams~\ref{LJ+1-1V-a0-2ptI-GR} and~\ref{LJ+1-1V-a0-2ptI-WGR}
is shown in figure~\ref{fig:GR-Ia-a1s}.

\begin{center}
\begin{figure}
	\[
	\begin{array}{c}
	\vspace{0.2in}
		\displaystyle
		4\norm_{n,0}
		\dec{
			\CGD[0.1]{}{Struc-TLTP-WineBiteBase}{GR-Ia-a1s-TLTP-WBB}{GR-Ib-TLTP-WBB-R}
			+\CGD[0.1]{}{Struc-TLTP-WineBiteSocket}{GR-Ia-a1s-TLTP-WBS}{GR-Ib-TLTP-WBS-R}
		}{11\Delta^n}
	\\
		\displaystyle		
		+4 \sum_{J=0}^n \norm_{J,0}
		\dec{ 
			\CGD[0.1]{}{Struc-n_J-S-R-WGR-Top}{GR-Ia-n_J-S-R-WGR-T}{DGR-n_J-S-R-WGR-T}
			+ \CGD[0.1]{}{Struc-n_J-S-R-WBS}{GR-Ia-a1s-n_J-S-R-WBS}{GR-Ia-a1s-n_J-S-R-WBS-C}
			+ \CGD[0.1]{}{Struc-n_J-S-R-WGR-Out}{GR-Ia-n_J-S-R-WGR-O}{GR-Ia-n_J-S-R-WGR-O-C}
		}{11\Delta^J}
	\end{array}	
	\]
\caption{Result of processing diagrams~\ref{LJ+1-1V-a0-2ptI-GR} and~\ref{LJ+1-1V-a0-2ptI-WGR}.}
\label{fig:GR-Ia-a1s}
\end{figure}
\end{center}

The usual procedure, from here, would be to decorate all those two-point, tree level vertices
to which we can attach a field which carries the same momentum as the vertex. However, we
will see later that the candidate for decoration, diagram~\ref{GR-Ia-a1s-TLTP-WBB},
is cancelled, wholesale, and so there is no need to perform the decoration,
in this case.

Next, we process diagram~\ref{LJ+1-GR-Ia-line-J=s}, 
letting $J\rightarrow J+1$. This is 
shown in figure~\ref{fig:GR-Ia-J=s}.
\begin{center}
\begin{figure}[h]
	\[
	-4 \sum_{J=0}^{n} \norm_{J+1,0}
	\dec{
		\sum_{r=0}^{n_J} 
		\left[
			\CGD[0.1]{}{GR-Ia-Dumbell-n_J+r-S-R-r-R}{GR-Ia-Dumbell-n_J+r-S-R-r-R-Socket}{GR-Ia-Dumbell-n_J+r-S-R-r-R-Socket-C}
		\right]
		+ \PGD[0.1]{}{GR-Ia-Dumbell-n_J-S-R-TLTP}{Struc-Dumbell-n_J-S-R-TLTP}{fig:GR-Ia-J=s-Dec}
	}{11\Delta^{J+1}}
	\]
\caption{Result of processing diagram~\ref{LJ+1-GR-Ia-line-J=s}.}
\label{fig:GR-Ia-J=s}
\end{figure}
\end{center}

We now follow the usual diagrammatic procedure of decorating two-point, tree level
vertices, tying up any loose ends and then applying the effective propagator relation, as appropriate,
to give the diagrams of figure~\ref{fig:GR-Ia-J=s-Dec}. 
Notice that we have not drawn those diagrams in which the loose end of the effective
propagator has been attached to
the socket. To form such a diagram we would have to start with a three-point (tree level)
vertex decorated by a simple loop. The gauge remainder would strike the vertex along the third field.
This gauge remainder could not be fermionic since then the three-point vertex would have
an odd number of fermions. Since gauge remainders have no support in the $C$-sector, this
forces it to be in the $A$-sector; but now the diagram vanishes by charge conjugation invariance!

The validity of using the effective propagator relation in the situation where one
of the fields of the two-point, tree level vertex has been pushed forward (pulled back)
on to has been discussed in section~\ref{sec:socket}.

\begin{center}
\begin{figure}[h]
	\[
	\begin{array}{c}
	\vspace{0.2in}
		\displaystyle
		-4\sum_{J=0}^{n} \norm_{J+1,0}
		\dec{
			\LGD[0.1]{}{GR-Ia-Dumbell-n_J-S-R-TLTP-E}{GR-Ia-TLTP-E-n_J-S-R}
		}{1\Delta^{J+1}}
	\\
		\displaystyle
		-4\sum_{J=0}^{n} \norm_{J,0}
		\dec{
			\CGD[0.1]{}{Struc-n_J-S-R-WBS}{GR-Ia-a1s-n_J-S-R-WBS-C}{GR-Ia-a1s-n_J-S-R-WBS}
			- \LGD[0.1]{}{Struc-n_J-S-R-WBS-GR}{GR-Ia-WBS-GR-n_J-S-R} 
			+ \CGD[0.1]{}{Struc-n_J-S-R-WGR-Out}{GR-Ia-n_J-S-R-WGR-O-C}{GR-Ia-n_J-S-R-WGR-O}
			-\LGD[0.1]{}{Struc-n_J-S-R-WGR-Out-GR}{GR-Ia-WGR-GR-n_J-S-R}
		}{11\Delta^{J}}
	\end{array}
	\]
\caption{Partial decoration of diagram~\ref{Struc-Dumbell-n_J-S-R-TLTP}, in which all
loose ends have been tied up
and the effective propagator relation has been applied.}
\label{fig:GR-Ia-J=s-Dec}
\end{figure}
\end{center}

Immediately, we find cancellations.
\begin{cancel}
Diagram~\ref{GR-Ia-a1s-n_J-S-R-WBS-C} exactly cancels diagram~\ref{GR-Ia-a1s-n_J-S-R-WBS}.
\label{C:GR-Ia-n_J-S-R-WBS}
\end{cancel}
\begin{cancel}
Diagram~\ref{GR-Ia-n_J-S-R-WGR-O-C} exactly cancels diagram~\ref{GR-Ia-n_J-S-R-WGR-O}.
\label{C:GR-Ia-n_J-S-R-WGR-O}
\end{cancel}

These two cancellations remove all contributions arising from diagram~\ref{LJ+1-1V-a0-2ptI-GR}
in which the gauge remainder strikes a socket, up to terms in which the socket decorates
a two-point, tree level vertex. From our experiences with the calculation so far,
this is both what we expect and require if we are to be able, ultimately, to cast
the gauge remainder diagrams as $\Lambda$-derivative terms.

Currently, the only term left containing a reduced vertex decorated by a socket which
has been struck by a gauge remainder is diagram~\ref{GR-Ia-Dumbell-n_J+r-S-R-r-R-Socket}.
We can guess that this diagram will be cancelled by terms coming from processing a
term with one more vertex \ie\ the three vertex component of diagram~\ref{LJ+1-GR-Ia-line}.
However, before doing this, we process the partners of diagram~\ref{LJ+1-GR-Ia-line-J=s}:
diagrams~\ref{LJ+1-GR-Ia-GR-J=s} and~\ref{LJ+1-GR-Ia-WGR-J=s}.

The treatment of these diagrams is almost identical to the previous treatment of
diagrams~\ref{LJ+1-1V-a0-2ptI-GR} and~\ref{LJ+1-1V-a0-2ptI-WGR} and is shown in
figure~\ref{fig:bn:GR-Ia-J=s}. The only difference is that an additional vertex,
unaffected by the action of the gauge remainder, must be included.

\begin{center}
\begin{figure}
	\[
	\begin{array}{c}
	\vspace{0.2in}
		\displaystyle
		-4 \sum_{J=0}^{n} \norm_{J+1,0}
		\dec{
			\sco[0.1]{
				\left[
					\PGD[0.1]{}{Struc-TLTP-WineBiteBase}{GR-Ia-TLTP-WBB-n_J-R}{fig:bn:GR-Ia,Ib-A}
					+ \PGD[0.1]{}{Struc-TLTP-WineBiteSocket}{GR-Ia-TLTP-WBS-n_J-R}{fig:bn:GR-Ia,Ib-A}
				\right]
			}{\ensuremath{\begin{array}{c}\input{pstex/Vertex-n_J-R.pstex_t} \end{array}}}
		}{11\Delta^{J+1}}
	\\
		\displaystyle
		-4 \sum_{J=0}^{n} \sum_{r=0}^{n_J} \norm_{J+1,0}
		\dec{
			\sco[0.1]{
				\left[
					\CGD[0.1]{}{Struc-n_J+r-S-R-WBS}{Struc-n_J+r-S-R-WBS}{Struc-n_J+r-S-R-WBS-C} 
					+\CGD[0.1]{}{Struc-n_J+r-S-R-WGR-Out}{Struc-n_J+r-S-R-WGR-Out}{Struc-n_J+r-S-R-WGR-Out-C}
					+\LGD[0.1]{}{Struc-n_J+r-S-R-WGR-Top}{GR-Ia-WGR-Top-n_J+r-S-R-rR}
				\right]
			}{\ensuremath{\begin{array}{c}\input{pstex/Vertex-r-R.pstex_t} \end{array}}}
		}{11 \Delta^{J+1}}
	\end{array}
	\]
\caption{Result of processing diagrams~\ref{LJ+1-GR-Ia-GR-J=s} and~\ref{LJ+1-GR-Ia-WGR-J=s}.}
\label{fig:bn:GR-Ia-J=s}
\end{figure}
\end{center}

Now, in addition to diagram~\ref{GR-Ia-Dumbell-n_J+r-S-R-r-R-Socket},
we have two further diagrams containing two reduced vertices and a socket
which has been hit by a gauge remainder; namely diagrams~\ref{Struc-n_J+r-S-R-WBS}
and~\ref{Struc-n_J+r-S-R-WGR-Out}. To remove these three diagrams, we must complete 
our treatment of the diagrams of figure~\ref{fig:bn:GR-Ia} by
processing those containing three or more vertices. 
These are of course those terms which have been represented using a very compact notation.
It would be nice to maintain
the compact notation for all of the daughter diagrams, but this is not quite possible.
If, after the gauge remainder has acted, we take the two-point, tree level component
of the vertex with argument $V^{J+1-s}$, then this causes the argument of the final vertex 
in the product over $I$, $V^{J-s,J+1-s;R}$, to reduce to just $V^{J-s;R}$. If the parent diagram
had only three vertices, then this means that the daughter has no product over $I$. 
Hence, when taking the two-point, tree level contribution from a vertex,
only those diagrams with more than three vertices can still be represented using the
compact notation.

The first stage of the treatment of diagrams~\ref{LJ+1-GR-Ia-line}--\ref{LJ+1-GR-Ia-WGR}
is shown in figure~\ref{fig:bn:GR-Ia-Rest}. We have simplified the diagrams by
letting $J\rightarrow J+s$. The final row of diagrams correspond (after shifting) to $J=1$.
In this case, we have let $s \rightarrow s+1$ and then changed the dummy variable $s$
to $J$.
\begin{center}
\begin{figure}
	\[
	\begin{array}{c}
	\vspace{0.1in}
		\displaystyle	
		-4 \sum_{s=1}^{n_+} \sum_{J=1}^{2n-1}  \norm_{J+s,J}
	\\
	\vspace{0.1in}
		\displaystyle
		\times
		\dec{
			\sco[0.1]{
				\CGD[0.1]{}{GR-Ia-Dumbell-V01-S-R-VJ+1-R}{GR-Ia-Dumbell-V01-S-R-VJ+1-R-Socket}{GR-Ia-Dumbell-V01-S-R-VJ+1-R-Socket-C}  
				+\frac{1}{J+1} 
				\left[
					\sco[0.2]{
						\CGD[0.1]{}{Struc-V01-S-R-WBS}{Struc-V01-S-R-WBS}{Struc-V01-S-R-WBS-C} 
						+ \CGD[0.1]{}{Struc-V01-S-R-WGR-Out}{Struc-V01-S-R-WGR-Out}{Struc-V01-S-R-WGR-Out-C}
						+ \LGD[0.1]{}{Struc-V01-S-R-WGR-Top}{GR-Ia-WGR-Top-Arb}
					}{\ensuremath{\begin{array}{c}\input{pstex/Vertex-VJ+1-R.pstex_t} \end{array}}}
				\right]
			}{\left[ \ds \prod_{I=1}^J \sum_{V^{I_+}=0}^{V^I} \ensuremath{\begin{array}{c}\input{pstex/GV-I-I+.pstex_t} \end{array}} \right]}
		}{11\Delta^{J+s}}
	\\
	\vspace{0.1in}
		\displaystyle	
		-4 \sum_{s=1}^{n_+} \sum_{J=2}^{2n}  \norm_{J+s,J}
			\dec{
				\sco[0.1]{
					\PGD[0.1]{}{GR-Ia-Dumbell-V01-S-R-TLTP}{GR-Ia-Dumbell-V01-S-R-VJ+1-R-TLTP}{fig:bn:GR-Ia-Rest-B-Dec}
					+ \frac{1}{J+1} 
					\left[
						\sco[0.2]{
							\PGD[0.1]{}{Struc-TLTP-WineBiteBase}{GR-Ia-TLTP-WBB-Arb}{fig:bn:GR-Ib-Rest}
							+ \PGD[0.1]{}{Struc-TLTP-WineBiteSocket}{GR-Ia-TLTP-WBS-Arb}{fig:bn:GR-Ib-Rest}
						}{\ensuremath{\begin{array}{c}\input{pstex/Vertex-V01-R.pstex_t} \end{array}}}
					\right]
				}{\tco{\left[\ds \prod_{I=1}^J \sum_{V^{I_+}=0}^{V^I} \ensuremath{\begin{array}{c}\input{pstex/GV-I-I+.pstex_t} \end{array}} \right]}{\ensuremath{\begin{array}{c}\input{pstex/Vertex-VJ-R.pstex_t} \end{array}}}}
			}{11\Delta^{J+s}}
	\\
		\displaystyle
		-4 \sum_{J=0}^n \sum_{r=0}^{n_J} \norm_{J+2,1}
		\dec{
			\sco[0.1]{
				\PGD[0.1]{}{GR-Ia-Dumbell-n_J+r-S-R-TLTP}{GR-Ia-Dumbell-n_J+r-S-R-TLTP}{fig:bn:GR-Ia-Rest-A-Dec}
				+ \frac{1}{2} 
				\left[
						\sco[0.2]{
							\PGD[0.1]{}{Struc-TLTP-WineBiteBase}{GR-Ia-TLTP-WBB-n_J_rR-rR}{fig:bn:GR-Ib-Rest}
							+ \PGD[0.1]{}{Struc-TLTP-WineBiteSocket}{GR-Ia-TLTP-WBS-n_J_rR-rR}{fig:bn:GR-Ib-Rest}
						}{\ensuremath{\begin{array}{c}\input{pstex/Vertex-n_J+r-R.pstex_t} \end{array}}}
				\right]
			}{\ensuremath{\begin{array}{c}\input{pstex/Vertex-r-R.pstex_t} \end{array}}}
		}{11\Delta^{J+2}}
	\end{array}
	\]
\caption{Separating off the two-point, tree level vertices from diagrams~\ref{LJ+1-GR-Ia-line}--\ref{LJ+1-GR-Ia-WGR},
after the action of the gauge remainder.}
\label{fig:bn:GR-Ia-Rest}
\end{figure}
\end{center}

Note that, for diagrams~\ref{GR-Ia-Dumbell-V01-S-R-VJ+1-R-Socket}--\ref{GR-Ia-WGR-Top-Arb},
 we have reduced the upper limit of the sum over $J$ by one.
The reason is simple: when $J=2n$, the parent diagram is the highest level manipulable diagram
and hence contains only three-point, tree level vertices and one-point higher
loop vertices. If the gauge remainders strikes a vertex, then this vertex must be
one of the three-point, tree level vertices. The effect of this is to produce a two-point,
tree level vertex and so does not leave behind any reduced components. 
Since all the vertices of diagrams~\ref{GR-Ia-Dumbell-V01-S-R-VJ+1-R-Socket}--\ref{GR-Ia-WGR-Top-Arb}
are reduced, such a term cannot exist for $J=2n$.
If, instead,
the gauge remainder strikes the wine, then this means that, in processing the parent,
one of the fields attached to one of the three-point, tree level vertices has
been moved on to a wine. Again, this leaves behind a two-point, tree level vertex, but
no reduced component. 

In  figure~\ref{fig:bn:GR-Ia-Rest-A-Dec} we process diagram~\ref{GR-Ia-Dumbell-n_J+r-S-R-TLTP}
by partially decorating the two-point, tree level vertices, tying up any
loose ends and applying the effective propagator relation, as appropriate.

\begin{center}
\begin{figure}
	\[
	\begin{array}{c}
	\vspace{0.2in}
		\displaystyle
		4 \sum_{J=0}^n \sum_{r=0}^{n_J} \norm_{J+1,0}
	\\
	\vspace{0.2in}
		\left[
			\begin{array}{c}
			\vspace{0.2in}
				\displaystyle
				\frac{1}{2(J+2)}
				\dec{
					\sco[0.1]{
						\LGD[0.1]{}{GR-Ia-Dumbell-n_J+r-S-R-TLTP-E}{GR-Ia-TLTP-E-n_J+r-S-R-r-R}
					}{\ensuremath{\begin{array}{c}\input{pstex/Vertex-r-R.pstex_t} \end{array}}}
				}{1\Delta^{J+2}}
				+
				\dec{
					\CGD[0.1]{}{GR-Ia-Dumbell-n_J+r-S-R-r-R}{GR-Ia-Dumbell-n_J+r-S-R-r-R-Socket-C}{GR-Ia-Dumbell-n_J+r-S-R-r-R-Socket}
					 - \PGD[0.1]{}{GR-Ia-Dumbell-n_J+r-S-R-r-R-GR}{GR-Ia-Dumbell-n_J+r-S-R-r-R-GR}{fig:bn:GR-Ia-Nest-Ex}
				}{11\Delta^{J+1}}
			\\
			\vspace{0.2in}
				\displaystyle
				+
				\dec{
					\sco[0.1]{
						\left[
							\CGD[0.1]{}{Struc-n_J+r-S-R-WBS}{Struc-n_J+r-S-R-WBS-C}{Struc-n_J+r-S-R-WBS}
							- \LGD[0.1]{}{Struc-n_J+r-S-R-WBS-GR}{GR-Ia-WBS-GR-n_J+r-S-R-r-R} 
							+ \CGD[0.1]{}{Struc-n_J+r-S-R-WGR-Out}{Struc-n_J+r-S-R-WGR-Out-C}{Struc-n_J+r-S-R-WGR-Out}
							- \LGD[0.1]{}{Struc-n_J+r-S-R-WGR-Out-GR}{GR-Ia-WGR-GR-n_J+r-S-R-r-R} 
						\right]
			}{\ensuremath{\begin{array}{c}\input{pstex/Vertex-r-R.pstex_t} \end{array}}}
				}{11\Delta^{J+1}}
			\end{array}
		\right]
	\end{array}
	\]
\caption[Partial decoration of diagram~\ref{GR-Ia-Dumbell-n_J+r-S-R-TLTP}.]{Partial decoration of diagram~\ref{GR-Ia-Dumbell-n_J+r-S-R-TLTP}. All loose ends have been tied up
and the effective propagator relation applied.}
\label{fig:bn:GR-Ia-Rest-A-Dec}
\end{figure}
\end{center}

As we predicted earlier, by processing the three-vertex terms, we generate diagrams to cancel the
only remaining diagrams possessing less than three reduced vertices, one of which is
decorated by a socket, hit by a gauge remainder.
\begin{cancel}
Diagram~\ref{GR-Ia-Dumbell-n_J+r-S-R-r-R-Socket-C} exactly cancels diagram~\ref{GR-Ia-Dumbell-n_J+r-S-R-r-R-Socket}
\end{cancel}
\begin{cancel}
Diagram~\ref{Struc-n_J+r-S-R-WBS-C} exactly cancels diagram~\ref{Struc-n_J+r-S-R-WBS}.
\end{cancel}
\begin{cancel}
Diagram~\ref{Struc-n_J+r-S-R-WGR-Out-C} exactly cancels diagram~\ref{Struc-n_J+r-S-R-WGR-Out}.
\end{cancel}

This then leaves behind diagrams~\ref{GR-Ia-Dumbell-V01-S-R-VJ+1-R-Socket}--\ref{Struc-V01-S-R-WGR-Out}
 as the only
remaining diagrams at this stage of the calculation possessing only reduced vertices
and a socket struck by a gauge remainder. The structure of the cancellations
is now becoming clear. 
Gauge remainder diagrams of Type-Ia come in three types, as
we know already from figure~\ref{fig:bn:GR-Ia}, with the first of these being
absent when there is only a single vertex. 

Processing the one vertex terms,\footnote{Having removed 
any terms which are properly classed as gauge remainders
of type-II of III.} we find that the 
contribution in which a gauge remainder strikes the field on the
vertex to which the wine attaches cancels the contribution
in which a gauge remainder strikes the base of the wine.
Among the surviving terms, 
we are left with diagrams with reduced vertices and a socket
struck by a gauge remainder. 

Processing the two vertex terms, we find that diagrams of the second
and third topologies exactly repeat the cancellations seen with
the one vertex terms. However, we also have a diagram of the first topology
to process.
When the gauge remainder acts in diagrams of the first topology, it
generates either a two-point, tree level vertex or a reduced vertex.
Let us consider the former case and suppose that we
join the two-point, tree level vertex to some other structure,
with an effective propagator. With the term under consideration,
we must make this join 
either to the wine or to the vertex from which the wine leaves. 
We have seen how this then cancels (up to nested gauge remainders)
the remaining one vertex terms
in which the vertex is reduced and the diagram possess a socket struck by
a gauge remainder.

Processing the three vertex terms repeats the cancellations seen
with the two vertex terms but gives us something extra. Now
when we process the diagram of the first topology and generate
a two-point, tree level vertex, we can join this not only
to the wine and the vertex to which the wine attaches but also to
the loose vertex. We have seen how this then cancels (up to nested gauge remainders)
the remaining two vertex term formed from the diagram of the
first topology in which the gauge remainder acts to produce
a reduced vertex.

As we iterate the diagrammatic procedure, we expect that this pattern
of cancellations will just repeat, until the procedure terminates.
We now demonstrate that this expectation is indeed true, by processing
diagram~\ref{GR-Ia-Dumbell-V01-S-R-VJ+1-R-TLTP}---the result of which is shown in
figure~\ref{fig:bn:GR-Ia-Rest-B-Dec}.

\begin{center}
\begin{figure}
	\[
	\begin{array}{c}
	\vspace{0.2in}
		\displaystyle
		4 \sum_{s=1}^{n_+} \sum_{J=2}^{2n}  \norm_{J+s-1,J-1}		
	\\
	\vspace{0.2in}
		\left[
			\begin{array}{c}
			\vspace{0.2in}
				\displaystyle
				\frac{1}{2J(J+s)}
				\dec{
					\LGO[0.1]{}{
					\tco[-0.3]{				
						\ensuremath{\begin{array}{c}\input{pstex/GR-Ia-Dumbell-V01-S-R-TLTP-E.pstex_t} \end{array}}
					}
						{
						\sco[0.1]{\left[\ds \prod_{I=1}^{J-1} \sum_{V^{I_+}=0}^{V^I} \ensuremath{\begin{array}{c}\input{pstex/GV-I-I+.pstex_t} \end{array}}\right]}{\ensuremath{\begin{array}{c}\input{pstex/Vertex-VJ-R.pstex_t} \end{array}}}
					}}{GR-Ia-TLTP-E-Arb}
				}{1\Delta^{J+s}}
			\\
			\vspace{0.2in}
				\displaystyle
				+
				\dec{
					\begin{array}{c}
					\vspace{0.2in}
						\displaystyle
						\left[
							\sco[0.1]{
								\CGD[0.1]{}{GR-Ia-Dumbell-V01-S-R-VJ-R}{GR-Ia-Dumbell-V01-S-R-VJ+1-R-Socket-C}{GR-Ia-Dumbell-V01-S-R-VJ+1-R-Socket}
								- \LGD[0.1]{}{GR-Ia-Dumbell-V01-S-R-VJ-R-GR}{GR-Ia-Dumbell-Arb-GR}
							}{\left[ \ds \prod_{I=1}^{J-1} \sum_{V^{I_+}=0}^{V^I} \ensuremath{\begin{array}{c}\input{pstex/GV-I-I+.pstex_t} \end{array}} \right]}
						\right]
					\\
						\displaystyle
						+\frac{1}{J}
						\left[
							\sco[0.1]{
								\CGD[0.1]{}{Struc-V01-S-R-WBS}{Struc-V01-S-R-WBS-C}{Struc-V01-S-R-WBS} 
								- \LGD[0.1]{}{Struc-V01-S-R-WBS-GR}{GR-Ia-WBS-GR-Arb} 
								+ \CGD[0.1]{}{Struc-V01-S-R-WGR-Out}{Struc-V01-S-R-WGR-Out-C}{Struc-V01-S-R-WGR-Out}
								- \LGD[0.1]{}{Struc-V01-S-R-WGR-Out-GR}{GR-Ia-WGR-GR-Arb}
							}{\tco{\left[\ds \prod_{I=1}^{J-1} \sum_{V^{I_+}=0}^{V^I} \ensuremath{\begin{array}{c}\input{pstex/GV-I-I+.pstex_t} \end{array}}\right]}{\ensuremath{\begin{array}{c}\input{pstex/Vertex-VJ-R.pstex_t} \end{array}}}}
						\right]
					\end{array}
				}{11\Delta^{J+s-1}}				
			\end{array}
		\right]
	\end{array}
	\]
\caption{The result of processing diagram~\ref{GR-Ia-Dumbell-V01-S-R-VJ+1-R-TLTP}.}
\label{fig:bn:GR-Ia-Rest-B-Dec}
\end{figure}
\end{center}

\begin{cancel}
Diagram~\ref{GR-Ia-Dumbell-V01-S-R-VJ+1-R-Socket-C} cancels diagram~\ref{GR-Ia-Dumbell-V01-S-R-VJ+1-R-Socket}.
\end{cancel}
\begin{cancel}
Diagram~\ref{Struc-V01-S-R-WBS-C} cancels diagram~\ref{Struc-V01-S-R-WBS}.
\end{cancel}
\begin{cancel}
Diagram~\ref{Struc-V01-S-R-WGR-Out-C} cancels diagram~\ref{Struc-V01-S-R-WGR-Out}.
\end{cancel}

This completes the cancellation of all terms arising from Type-Ia gauge
remainders in which the gauge remainder strikes a socket and for
which all vertices are reduced. It is apparent that there is a cancellation
mechanism at work.

\begin{CM}
Consider a diagram, to be denoted by $X$, possessing two arbitrary structures, $A$ and $B$. These two 
structures can, but need not be, identified with each other. Additionally,
the diagram possesses $N_V$ reduced vertices and is decorated by the usual
external fields and by $N_P$ effective propagators.

A wine leaves the structure~$B$ and attaches, via a gauge remainder with
an arbitrary number of nestings, to the structure $A$. We are free to
place any restrictions we choose on the decoration of the wine.
Note that the attachment to $B$ need not be direct: it
can, in principle, be via a (nested) gauge remainder.
Furthermore, both of the structures~$A$ and~$B$ can be associated with the wine.
Hence, diagram~$X$ represents not only gauge remainders of type-I but also of type-II (and type-III).
An example of diagram~$X$ is shown in figure~\ref{fig:bn:Struc-Arb-W-GR-Arb} where we have taken
$N_V=0$, the attachment to $B$ to be direct, the wine to be full and no nesting in the
attachment to $A$.
\begin{center}
\begin{figure}[h]
	\[
	\decB[0.1]{
		\ensuremath{\begin{array}{c}\input{pstex/Struc-Arb-W-GR-Arb.pstex_t} \end{array}}
	}{11\Delta^{N_P}}
	\]
\label{fig:bn:Struc-Arb-W-GR-Arb}
\caption{An example of diagram $X$.}
\end{figure}
\end{center}

Suppose now that the structure $A$ is actually just one of $N_A$ identical
structures. For example, suppose that $A$ is just a reduced vertex. Then if $B$
is some other structure, we know that we could equally have hit any of the other
$N_V$ vertices with the gauge remainder. Therefore, in this case, $N_A = N_V+1$.
If, however, $B$ and $A$ are identified as being the same vertex, then $N_A=1$.
Likewise, if $A$ represents the wine, then $N_A=1$.

Finally, we allow the gauge remainder to act. We are interested in the case
where it strikes a socket on $A$. Thus, if $A$ is to be identified with the
wine, this excludes the case where the gauge remainder strikes the end of the
wine. If $A$ is identified with a vertex, we do not allow the socket to be
filled by the other end of the wine. For the purposes of
this analysis, the case where the gauge remainder generates a two-point, 
tree level vertex is ignored.

Let us now consider a second diagram~$Y$. It possesses both the structures
$A$ and $B$, but this time the wine ends in a gauge remainder which 
strikes a vertex (it makes no difference whether or not the vertex is reduced).
Any restrictions placed on the decoration of the wine of diagram~$X$
apply to the wine of diagram~$Y$.
Additionally, diagram~$Y$ possesses a further $N_V$ vertices and is decorated
by the usual external fields and by $N_P+1$ effective propagators. An 
example of diagram~$Y$ is shown in figure~\ref{fig:bn:Struc-Arb-W-GR-V-Arb}.
\begin{center}
\begin{figure}[h]
	\[
	\frac{1}{2(N_P+1)N_A}
	\decB[0.1]{
		\ensuremath{\begin{array}{c}\input{pstex/Struc-Arb-W-GR-V-Arb.pstex_t} \end{array}}
	}{11\Delta^{N_P+1}}
	\]
\label{fig:bn:Struc-Arb-W-GR-V-Arb}
\caption{An example of diagram $Y$.}
\end{figure}
\end{center}

Allowing the gauge remainder to act, this time it is the two-point,
tree level component of the struck vertex that we focus on. We now decorate
this vertex with one of the effective propagators and tie up the loose end.
Since $A$ can be identified with any of the other structures, we only need
join the loose end to $A$. Up to further nestings this will precisely
cancel the contribution we took from diagram $X$, so long as,
before any manipulation, diagram $X$
has a factor of $2(N_P+1)N_A$, relative to diagram $Y$.
\label{CM:GR-Socket}
\end{CM}

The combination of cancellation mechanisms~\ref{CM:VertexJoinWine}
and~\ref{CM:GR-Socket} are sufficient to account for all
the cancellations seen in this section.

We will now collate the terms
that have survived up to this stage of the calculation.\footnote{Generally,
these terms will be processed or cancelled but, since this occurs at a later stage
in the calculation, these actions should be ignored for the collation of current survivors.}
The terms fall very neatly into five set, the elements of which are given below:
\begin{enumerate}
\item diagrams~\ref{GR-II-WGRx2} and~\ref{GR-III-TLTP-WGR};
\label{GR-Ia-Rem-Odd}

\item diagrams~\ref{GR-Ia-a1s-TLTP-WBB} and~\ref{GR-Ia-a1s-TLTP-WBS};
diagrams~\ref{GR-Ia-TLTP-WBB-n_J-R} and~\ref{GR-Ia-TLTP-WBS-n_J-R};
diagrams~\ref{GR-Ia-TLTP-WBB-Arb} and~\ref{GR-Ia-TLTP-WBS-Arb};
and diagrams~\ref{GR-Ia-TLTP-WBB-n_J_rR-rR} and~\ref{GR-Ia-TLTP-WBS-n_J_rR-rR};
\label{GR-Ia-Rem-TLTP-WBB,WBS}

\item diagrams~\ref{GR-Ia-TLTP-E-n_J-S-R}, \ref{GR-Ia-TLTP-E-n_J+r-S-R-r-R}
and~\ref{GR-Ia-TLTP-E-Arb};
\label{GR-Ia-Op2}

\item diagrams~\ref{GR-Ia-n_J-S-R-WGR-T}, \ref{GR-Ia-WGR-Top-n_J+r-S-R-rR}
and~\ref{GR-Ia-WGR-Top-Arb};
\label{GR-Ia-WBT}

\item diagrams~\ref{GR-Ia-WBS-GR-n_J-S-R} and~\ref{GR-Ia-WGR-GR-n_J-S-R}; 
diagrams~\ref{GR-Ia-Dumbell-n_J+r-S-R-r-R-GR}, \ref{GR-Ia-WBS-GR-n_J+r-S-R-r-R} and~\ref{GR-Ia-WGR-GR-n_J+r-S-R-r-R};
and diagrams~\ref{GR-Ia-Dumbell-Arb-GR}, \ref{GR-Ia-WBS-GR-Arb}
and~\ref{GR-Ia-WGR-GR-Arb}.
\label{GR-Ia-Nested}
\end{enumerate}

The first set stands somewhat apart from the rest since, as we have already noted,
the diagrams contained therein should be reclassified as gauge remainders of
a different type.
 
The remaining four sets have a very nice structure. Recall how we arranged
the original type-Ia gauge remainders in figure~\ref{fig:bn:GR-Ia}. However,
we then noted that the final two elements should really be replaced,
in the context of bona-fide type-Ia gauge remainders, by diagrams~\ref{LJ+1-1V-a0-2ptI-GR}
and~\ref{LJ+1-1V-a0-2ptI-WGR}. We call this new set the set of proper type-Ia gauge remainders.
The set of proper type-Ia gauge remainders essentially consists of diagrams~\ref{LJ+1-1V-a0-2ptI-GR}
and~\ref{LJ+1-1V-a0-2ptI-WGR} and their multiple vertex analogues.
As we arrange
the gauge remainders in this way, so we can do likewise with the terms that
survive their manipulation.

The elements set~\ref{GR-Ia-Rem-TLTP-WBB,WBS} will be 
combined with similar diagrams formed from gauge remainders of type-Ib.

Set~\ref{GR-Ia-Op2} comprises those diagrams which require manipulation at $\Op{2}$,
the $\Op{2}$ stub of which has been formed by the action of a gauge remainder. These
terms will be commented on in chapter~\ref{ch:Op2}.

Set~\ref{GR-Ia-WBT} is of particular interest. Whereas the elements of the previous two sets contain
two-point, tree-level vertices, the elements of this set do not.
Rather, it is as if the gauge remainder has `reached the end of line',
by bitting the end of the wine. The corresponding diagrams  escape cancellation 
since there is, at this point
in the calculation, no other
way to generate them. As we have already seen in the one-loop calculation, such
terms play a crucial role, being reducible to $\Lambda$-derivative terms (\cf\ diagram~\ref{d1:1.10}
and its children).

The final set comprises nested gauge remainders which, up to the nesting, take
an identical form to the set of proper type-Ia gauge remainders.

Thus, with a bit of care, we can deduce the result of processing the nested gauge remainders. Noting that
we can not collect the nested push forward with the nested pull back, we will simply
generate the nested versions of the sets~\ref{GR-Ia-Rem-TLTP-WBB,WBS}--\ref{GR-Ia-WBT} and a doubly nested
version of set~\ref{GR-Ia-Nested}.
There is no analogue of set~\ref{GR-Ia-Rem-Odd}
since its formation would require the presence of
a full vertex in the nested versions of diagram~\ref{LJ+1-1V-a0-2ptI-WGR}. 

There is, however, one more type of diagram we can generate. To see this, consider processing
diagram~\ref{GR-Ia-Dumbell-n_J+r-S-R-r-R-GR}, as shown in figure~\ref{fig:bn:GR-Ia-Nest-Ex}.
\begin{center}
\begin{figure}[h]
	\[
	\begin{array}{c}
		\displaystyle
		-4\sum_{J=0}^n \norm_{J+1,0}
	\\
		\displaystyle
		\times
		\dec{
			\sum_{r=0}^{n_J} 
			\left[
				\ensuremath{\begin{array}{c}\input{pstex/GR-Ia-Nest1-n_J+r-S-R-r-R.pstex_t} \end{array}}
				-\ensuremath{\begin{array}{c}\input{pstex/GR-Ia-Nest1-n_J+r-S-R-r-R-PB.pstex_t} \end{array}}
			\right]
			+\PED[0.1]{}{GR-Ia-Nest1-n_J-S-R-TLTP}{GR-Ia-Nest1-n_J-S-R-TLTP}{fig:bn:GR-Ia-Nest-Dec-Ex}
			-\PED[0.1]{}{GR-Ia-Nest1-n_J-S-R-TLTP-PB}{GR-Ia-Nest1-n_J-S-R-TLTP-PB}{fig:bn:GR-Ia-Nest-Dec-Ex}
		}{11\Delta^{J+1}}
		\end{array}
	\]
\caption{Result of processing diagram~\ref{GR-Ia-Dumbell-n_J+r-S-R-r-R-GR}.}
\label{fig:bn:GR-Ia-Nest-Ex}
\end{figure}
\end{center}

So far, this is exactly what we expect, where the nested push forward has been performed independently
of the nested pull back and we have separated off the two-point, tree level component
of the vertex struck by the gauge remainder. However, when we come to decorate the two-point,
tree level vertex, we encounter something new. If we decorate this vertex with an effective
propagator, then we can tie up the loose ends in the usual ways (as before, we can discard
the case where we join the two-point, tree level vertex to itself). Additionally, though, we can attach
the loose end to the nested gauge remainder. We show this explicitly in figure~\ref{fig:bn:GR-Ia-Nest-Dec-Ex}.
\begin{center}
\begin{figure}[h]
	\[
	-4 \sum_{J=0}^n \norm_{J,0}
	\dec{
		\CGD[0.1]{}{GR-Ia-Nest1-ExtraTerm-Ex}{GR-Ia-Nest1-ExtraTerm-Ex}{DGRx2-n_J-S-R-W-GR-LR}
		-\CGD[0.1]{}{GR-Ia-Nest1-ExtraTerm-Ex-PB}{GR-Ia-Nest1-ExtraTerm-Ex-PB}{DGRx2-n_J-S-R-W-GR-RR}
	}{11\Delta^J}
	\]
\caption[An example of the new types of diagram arising from processing nested gauge remainders.]{
An example of the new types of diagram arising from processing nested gauge remainders.
These terms represent the only surviving contributions from diagrams~\ref{GR-Ia-Nest1-n_J-S-R-TLTP}
and~\ref{GR-Ia-Nest1-n_J-S-R-TLTP-PB} after all nested gauge remainders have been processed.}
\label{fig:bn:GR-Ia-Nest-Dec-Ex}
\end{figure}
\end{center}

Since these diagrams are formed by attaching the loose end to a single specific structure (as opposed to
one of many identical structures), they come with a factor of $2(J+1)$. Although we can apply the
effective propagator relation,
we cannot employ the diagrammatic identity~\ref{D-ID:PseudoEffectivePropagatorRelation},
since the gauge remainder attached to the wine bites the effective propagator at the point where it
joins to the other gauge remainder. Hence, this latter gauge remainder has a different momentum flowing
through it than the effective propagator.

Analogous diagrams at a higher level of nesting behave in a very similar manner. Now,
though, we have a choice about which of the unattached gauge remainders to join
to the two-point, tree level vertex. Each of these cases must be treated individually.

It is thus apparent that we can iterate the diagrammatic procedure, at each stage generating
ever more nested versions sets~\ref{GR-Ia-Rem-TLTP-WBB,WBS}--\ref{GR-Ia-Nested},
together with diagrams of the type just discussed. The procedure will terminate:
each successive nesting reduces the maximum number of vertices by one.

It is straightforward to compute the maximum number of nestings. We must maximise the 
number of three-point, tree level vertices; since it is these that a gauge remainder
must strike to have any hope of producing further levels of nesting.

Referring back to diagram~\ref{LJ+1-GR-Ia-line},
to maximise the number of nestings, we require two fields to decorate the struck vertex and three fields
to decorate the loose vertices. The vertex to which the wine is attached must be decorated by at least one
additional field. Since we assume that all other vertices are tree level, the argument of this vertex is just
$n_+-s$. Hence, the number of fields required to decorate this vertex, $E(s)$, depends on $s$:
if $s <n_+$, then we require only one additional field to decorate this vertex whereas, if
$s = n_+$, we require two additional fields, on account of the vertex being reduced.

Letting $J\rightarrow J+s$, the number of vertices included in the sum over $I$ is
now just $J$. It is thus apparent that, ]
of all $J+2$ vertices, $J+1$ of them could  potentially increment
the level of nesting by one. However, for each of these vertices to actually be able to
increase the level of nesting, we require that there be sufficient decorative fields.
The value of $J$ corresponding to the maximum number of nestings, $J_{MN}$, can
thus be found by equating the number of required fields
 with the number of decorative fields:
\[
2+ 3J_{MN} + E(s) = 2(J_{MN}+s) + 2.
\]
It is clear that to maximise $J_{MN}$, we should maximise $s$. We thus find that
$J_{MN} = 2n$ and so the maximum number of nestings is $2n+1$.

Since, to maximise the number of nestings requires that we maximise $s$
then, at the final level of nesting, we produce (nested) versions
of diagrams~\ref{GR-Ia-WBS-GR-n_J-S-R} and~\ref{GR-Ia-WGR-GR-n_J-S-R},
for which the vertices are tree level (\ie\ $J=n$). The wines, in both cases,
have the minimal number of decorations---being zero in the first case and one in the second.
Allowing the final gauge remainder to act will produce diagrams like those of
sets~\ref{GR-Ia-Rem-TLTP-WBB,WBS}--\ref{GR-Ia-WBT}, above.

\newpage
\section{Type-Ib Gauge Remainders} \label{sec:GR-Ib}

The treatment of type-Ib gauge remainders is very similar to those of type-Ia.
Figure~\ref{fig:bn:GR-Ib} shows all gauge remainders of type-Ib.

\begin{center}
\begin{figure}
	\[
	\begin{array}{c}
	\vspace{0.1in}
		\displaystyle
		2\sum_{s=1}^{n_+} \sum_{J=2}^{2n}  \norm_{J+s-1,J-1}
	\\
	\vspace{0.2in}
		\displaystyle
		\times
		\dec{
			\sco[0.3]{
				\PGD[0.1]{}{GR-Ib-Dumbell-V01-R-VJ-R}{GR-Ib-Dumbell-V01-R-VJ-R}{fig:bn:GR-Ib-Rest}
				+\frac{1}{J}
				\left[
					\sco{
						\PGD[0.1]{}{GV-V01-R-GR}{GR-Ib-V01-R-GR-Arb}{fig:bn:GR-Ib-Rest} 
						+ \PGD[0.1]{}{GV-V01-R-WGR}{GR-Ib-V01-R-WGR-Arb}{fig:bn:GR-Ib-Rest}	
					}{\ensuremath{\begin{array}{c}\input{pstex/GV-VJ-R.pstex_t} \end{array}}}
				\right]
			}{\left[\ds \VertexSum{1}{J-1}\ensuremath{\begin{array}{c}\input{pstex/GV-I-I+.pstex_t} \end{array}}\right]}
		}{11\Delta^{J+s-1}}
	\\
	\vspace{0.2in}
		\displaystyle
		+2 \sum_{J=1}^{n_+} \sum_{r=0}^{n_{J-1}} \norm_{J,0}
		\dec{
			\PGD[0.1]{}{GR-Ib-Dumbell-n_J+r-1-R-rR}{GR-Ib-Dumbell-n_J+r-1-R-rR}{fig:bn:GR-Ib-J=s}
			+
			\left[
			\sco[0.1]{
				\PGD[0.1]{}{GV-n_J+r-1-R-GR}{GR-Ib-n_J+r-1-R-GR}{fig:bn:GR-Ib-J=s} +
				\PGD[0.1]{}{GV-n_J+r-1-R-WGR}{GR-Ib-n_J+r-1-R-WGR}{fig:bn:GR-Ib-J=s}
			}{\ensuremath{\begin{array}{c}\input{pstex/Vertex-r-R.pstex_t} \end{array}}}
			\right]
		}{11\Delta^{J}}
	\\
		\displaystyle
		-2 \sum_{J=0}^n \norm_{J,0}
		\dec{
			\PGD[0.1]{}{GV-n_J-GR-R}{GR-Ib-GV-n_J-GR-R}{fig:GR-Ib-a1s} 
			+ \PGD[0.1]{}{GV-n_J-R-WGR}{GR-Ib-GV-n_J-WGR-R}{fig:GR-Ib-a1s}
		}{11\Delta^J}
	\end{array}
	\]
\caption{Type-Ib gauge remainders.}
\label{fig:bn:GR-Ib}
\end{figure}
\end{center}

Comparing figures~\ref{fig:bn:GR-Ia} and~\ref{fig:bn:GR-Ib} it is clear that 
the primary differences between gauge remainders of type-Ia and Ib are as
follows.
First, in the latter case, the wine leaves a Wilsonian effective action vertex,
rather than a seed action vertex. Secondly, this wine must be decorated.
Of course, this latter restriction is automatically satisfied in the case
where the gauge remainder bites the wine. It is also clear that our
set of type-Ib diagrams really are type-Ib diagrams: there are no diagrams of
types-II and III hiding, as there were for the original
set of type-Ia diagrams.

When we allow the gauge remainders to act, it is apparent what the result will be.
All diagrams possessing only reduced vertices and in which the gauge remainder
hits a socket will cancel, via cancellation mechanism~\ref{CM:GR-Socket}.
However, the surviving terms will be of a slightly different
form to the analogous type-Ia survivors, due to the reduction of the wine.
In figure~\ref{fig:GR-Ib-a1s}, we show the terms arising from diagrams~\ref{GR-Ib-GV-n_J-GR-R}
and~\ref{GR-Ib-GV-n_J-WGR-R} that are not cancelled via cancellation mechanism~\ref{CM:GR-Socket}.

\begin{center}
\begin{figure}
	\[
		\begin{array}{c}
		\vspace{0.2in}
			\displaystyle
			4\sum_{J=0}^n \norm_{J,0}
			\dec{
				\CGD[0.1]{}{Struc-n_J-R-DEPBB}{GR-Ib-n_J-R-DEPBB}{GR-Ib-n_J-R-DEPBB-C} - 
				\PGD[0.1]{}{Struc-n_J-R-WGR-Top}{GR-Ib-n_J-R-WGR-T}{fig:bn:GR-Ia,Ib-n_J-R-LdL-pre}
			}{11\Delta^J}
		\\
			\displaystyle
			- 4\norm_{n,0}
			\dec{
				\CGD[0.1]{}{Struc-TLTP-WBB-R}{GR-Ib-TLTP-WBB-R}{GR-Ia-a1s-TLTP-WBB}+
				\CGD[0.1]{}{Struc-TLTP-WBS-R}{GR-Ib-TLTP-WBS-R}{GR-Ia-a1s-TLTP-WBS}
			}{11\Delta^n}
		\end{array}
	\]
\caption{Diagrams arising from processing diagrams~\ref{GR-Ib-GV-n_J-GR-R}
and~\ref{GR-Ib-GV-n_J-WGR-R} that are not cancelled when diagram~\ref{GR-Ib-Dumbell-n_J+r-1-R-rR}
is processed.}
\label{fig:GR-Ib-a1s}
\end{figure}
\end{center}

Diagram~\ref{GR-Ib-n_J-R-DEPBB} is something we have not seen before, arising
due to the reduction of the wine in diagram~\ref{GR-Ib-GV-n_J-GR-R}. When the
gauge remainder in diagram~\ref{GR-Ib-GV-n_J-WGR-R} pulls back on to the
base of the wine, there is no contribution from diagram~\ref{GR-Ib-GV-n_J-GR-R}
to cancel the term with an undecorated wine.

The next diagram is familiar from our analysis of the type-Ia terms. Indeed,
we can combine it with diagram~\ref{GR-Ia-n_J-S-R-WGR-T} simply by replacing
the vertex argument $n_J^R$ with $\Pi_{n_J^R}$.

The final two diagrams also are of a familiar structure though the wines of
the analogous type-Ia terms, diagrams~\ref{GR-Ia-a1s-TLTP-WBB} and~\ref{GR-Ia-a1s-TLTP-WBS},
are not explicitly reduced. However, since these diagrams contain a single, two-point vertex,
we are very restricted in the decorations we can perform. Indeed, irrespective of
the value of $n$, we always have at least two fields with which to decorate (\ie\ the two
external fields). This in fact forces us to decorate wines of the type-Ia terms,
giving two cancellations.

\begin{cancel}
Diagram~\ref{GR-Ib-TLTP-WBB-R} cancels diagram~\ref{GR-Ia-a1s-TLTP-WBB},
the cancellation being forced as a consequence of the diagrammatic structure 
and the specific set of decorative fields.
\label{cancel:GR-Ib-TLTP-WBB-R}
\end{cancel}

\begin{cancel}
Diagram~\ref{GR-Ib-TLTP-WBS-R} cancels diagram~\ref{GR-Ia-a1s-TLTP-WBS},
the cancellation being forced as a consequence of the diagrammatic structure 
and the specific set of decorative fields.
\label{cancel:GR-Ib-TLTP-WBS-R}
\end{cancel}

The next stage is to process diagrams~\ref{GR-Ib-Dumbell-n_J+r-1-R-rR}--\ref{GR-Ib-n_J+r-1-R-WGR},
the result of which
which is shown in figure~\ref{fig:bn:GR-Ib-J=s}. Once again, we have retained only those terms
which are not involved in cancellations against other diagrams generated from figure~\ref{fig:bn:GR-Ib}.
\begin{center}
\begin{figure}
	\[
	\begin{array}{c}
	\vspace{0.2in}
		\displaystyle
		4 \sum_{J=0}^n \norm_{J+1,0}
		\left[
			\dec{
				\LGD[0.1]{}{GR-Ib-Dumbell-n_J-R-TLTP-E}{GR-Ib-TLTP-E-n_J-R}
			}{1\Delta^{J+1}}
			+
			\dec{
				\sco[0.2]{
					\PGD[0.1]{}{Struc-TLTP-WBB-R}{GR-Ib-TLTP-WBB-R-n_J-R}{fig:bn:GR-Ia,Ib-A} +
					\PGD[0.1]{}{Struc-TLTP-WBS-R}{GR-Ib-TLTP-WBS-R-n_J-R}{fig:bn:GR-Ia,Ib-A}
				}{\ensuremath{\begin{array}{c}\input{pstex/Vertex-n_J-R.pstex_t} \end{array}}}
			}{11\Delta^{J+1}}
		\right]		
	\\
	\vspace{0.2in}
		\displaystyle
		+4\sum_{J=0}^n \sum_{r=0}^{n_J} \norm_{J+1,0}
		\dec{
			\sco[0.2]{
				\PGD[0.1]{}{Struc-n_J+r-R-WGR-Top}{GR-Ib-n_J+r-R-WGR-Top}{fig:bn:GR-Ib-Rest}
				-\PGD[0.1]{}{Struc-n_J+r-R-RWBB}{GR-Ia-n_J+r-R-RWBB}{fig:bn:GR-Ib-Rest}
			}{\ensuremath{\begin{array}{c}\input{pstex/Vertex-r-R.pstex_t} \end{array}}}
		}{11\Delta^{J+1}}
	\\
	\vspace{0.2in}
		\displaystyle
		-4 \sum_{J=0}^n \norm_{J,0}
		\dec{
			\LGD[0.1]{}{Struc-n_J-R-RWBS-GR}{GR-Ib-n_J-R-RWBS-GR}
			+\LGD[0.1]{}{Struc-n_J-R-WGR-Out-GR}{GR-Ib-n_J-R-WGR-Out-GR}		
		}{11\Delta^J}
	\end{array}
	\]
\caption{Diagrams arising from processing diagrams~\ref{GR-Ib-Dumbell-n_J+r-1-R-rR}--\ref{GR-Ib-n_J+r-1-R-WGR}
that are not cancelled by other diagrams arising from figure~\ref{fig:bn:GR-Ib}.}
\label{fig:bn:GR-Ib-J=s}
\end{figure}
\end{center}

We now uncover a wonderful interplay between the type-Ia and type-Ib gauge remainders. First,
we note that diagrams~\ref{GR-Ib-TLTP-E-n_J-R} and~\ref{GR-Ia-TLTP-E-n_J-S-R} are very similar.
Indeed, we can decompose diagram~\ref{GR-Ia-TLTP-E-n_J-S-R} into a component with a decorated wine and
a component with an undecorated wine. The former can then be combined with
diagram~\ref{GR-Ib-TLTP-E-n_J-R} if we simply replace the argument $n_J$ with
$\Pi_{n_J}$. We will comment on these diagrams in chapter~\ref{ch:Op2}.

Secondly, we note that we can combine diagrams~\ref{GR-Ib-n_J+r-R-WGR-Top} and~\ref{GR-Ia-WGR-Top-n_J+r-S-R-rR}.
This is simply the two vertex generalisation of the combination of diagrams~\ref{GR-Ib-n_J-R-WGR-T} and~\ref{GR-Ia-n_J-S-R-WGR-T}

Lastly, diagrams~\ref{GR-Ib-TLTP-WBB-R-n_J-R} and~\ref{GR-Ib-TLTP-WBS-R-n_J-R}
can be combined, respectively, with diagrams~\ref{GR-Ia-TLTP-WBB-n_J-R} and~\ref{GR-Ia-TLTP-WBS-n_J-R}.
The former diagrams remove from the latter diagrams all contributions in which the wine is decorated.

This is exactly what we saw when we combined diagrams~\ref{GR-Ib-TLTP-WBB-R}, \ref{GR-Ib-TLTP-WBS-R} 
with diagrams~\ref{GR-Ia-a1s-TLTP-WBB}, \ref{GR-Ia-a1s-TLTP-WBS}. In this case, however,
the combination of diagrams vanished since, irrespective of the value of $n$, there was no
way for the requirement that the wine be undecorated to be reconciled with the number of
decorative fields. In the current case, this issue is circumvented by the presence of an additional
vertex.

Focusing on the remainder of diagrams~\ref{GR-Ib-TLTP-WBB-R-n_J-R} and~\ref{GR-Ia-TLTP-WBB-n_J-R},
we recognise that
joining the two-point, tree level vertex to the other vertex 
allows us to use the effective propagator relation, once more.
In the case of the remainder of diagrams~\ref{GR-Ib-TLTP-WBS-R-n_J-R} and~\ref{GR-Ia-TLTP-WBS-n_J-R}, 
the two-point, tree level vertex is already joined to a structure which carries
the same momentum (an undecorated wine) and it is clear, in this case, that we cannot use the effective propagator relation.
We \emph{can}, however, use diagrammatic identity~\ref{D-ID:VertexWineGR}.
After utilising this identity, we then decorate the socket of the
two-point, tree level vertex in all possible ways.

The result of the above operations is shown in figure~\ref{fig:bn:GR-Ia,Ib-A}. 

\begin{center}
\begin{figure}
	\[
	\begin{array}{c}
	\vspace{0.2in}
		\displaystyle
		-4\sum_{J=0}^n \norm_{J+1,0}
		\dec{
			\sco[0.2]{
				\DGD[0.1]{}{Struc-TLTP-WBB-R-E}{GR-Ib-TLTP-WBB-R-n_J-R-E} 
				-\DGD[0.1]{}{Struc-LdL-L-E}{GR-Ib-LdL-L-E}
			}{\ensuremath{\begin{array}{c}\input{pstex/Vertex-n_J-R.pstex_t} \end{array}}}
		}{1\Delta^{J+1}}
	\\
		\displaystyle
		-4\sum_{J=0}^n \norm_{J,0}
		\dec{
			\CGD[0.1]{}{Struc-n_J-R-DEPBB}{GR-Ib-n_J-R-DEPBB-C}{GR-Ib-n_J-R-DEPBB}
			- \PGD[0.1]{}{Struc-n_J-R-DEPBB-GR}{GR-Ib-n_J-R-DEPBB-GR}{fig:bn:GR-Extra}
			+ \PGD[0.1]{}{Struc-n_J-R-DWGR-Top}{GR-Ib-n_J-R-DWGR-Top}{fig:bn:GR-Ia,Ib-n_J-R-LdL-pre}
		}{11\Delta^J}
	\end{array}
	\]
\caption[The result of combining diagrams~\ref{GR-Ib-TLTP-WBB-R-n_J-R} and~\ref{GR-Ib-TLTP-WBS-R-n_J-R}
with diagrams~\ref{GR-Ia-TLTP-WBB-n_J-R} and~\ref{GR-Ia-TLTP-WBS-n_J-R}.]{The result of combining diagrams~\ref{GR-Ib-TLTP-WBB-R-n_J-R} and~\ref{GR-Ib-TLTP-WBS-R-n_J-R}
with diagrams~\ref{GR-Ia-TLTP-WBB-n_J-R} and~\ref{GR-Ia-TLTP-WBS-n_J-R}. We have used the effective propagator
relation and diagrammatic identity~\ref{D-ID:VertexWineGR}. In the final diagram we have converted, for
later convenience, a
pull back on to the top of the wine into a push forward, picking up a minus sign.}
\label{fig:bn:GR-Ia,Ib-A}
\end{figure}
\end{center}

\begin{cancel}
Diagram~\ref{GR-Ib-n_J-R-DEPBB-C} exactly cancels diagram~\ref{GR-Ib-n_J-R-DEPBB}.
\label{cancel:GR-Ib-n_J-R-DEPBB-C}
\end{cancel}

We can make further progress, since a number of diagrams can be discarded.
We focus first on diagram~\ref{GR-Ib-TLTP-WBB-R-n_J-R-E} and~\ref{GR-Ib-LdL-L-E}.
In both cases, the diagrams are disconnected. In the former case, this is
a consequence of the wine not being decorated; in the latter case it is trivial.
Hence, these diagrams can be discarded. We do,
however, make the very important observation that, at higher levels of nesting,
we must keep the analogues of these diagrams; in this circumstance, the additional
gauge remainder(s) provide a socket(s) to which we can attach the loose vertex.

Diagram~\ref{GR-Ib-n_J-R-DEPBB-GR}, too, can be discarded. The gauge remainder
striking the vertex must be bosonic, else the vertex
to which it attaches will be decorated by an odd number of fermions. 
This restriction
can only be satisfied in the $A$-sector since gauge remainders have no support
in the $C$-sector. However, this then causes the diagram to vanish
(when we sum over all independent
permutations of the fields) by charge conjugation invariance.

In this case, though, it is useful to retain the diagram.
By processing diagram~\ref{GR-Ib-n_J-R-DEPBB-GR},
we will find that the generated terms manifestly remove contributions
from diagrams later in the calculation. In some sense, this point is moot,
since we know that these contributions must vanish anyway. The real point
is that we will encounter nested versions of diagram~\ref{GR-Ib-n_J-R-DEPBB-GR}
that do not vanish and so must be processed. Processing the simplest instance
of such diagrams \ie\ diagram~\ref{GR-Ib-n_J-R-DEPBB-GR}, itself, will illuminate
the pattern of cancellations we expect for the more complicated terms.

Returning to figure~\ref{fig:bn:GR-Ia,Ib-A},
we note that diagrams~\ref{GR-Ib-n_J-R-DWGR-Top} and~\ref{GR-Ib-n_J-R-WGR-T}
have a very similar structure and can be combined
to give  a $\Lambda$-derivative
term, plus corrections. This observation will prove central to our subsequent analysis (see chapter~\ref{ch:bn:Iterate}).

We now complete
our treatment of the terms of figure~\ref{fig:bn:GR-Ib}. 
Using the results of the previous section and the cancellations seen in this
section, we can jump straight to the final set of terms. Thus, to generate these terms,
we include not only diagrams~\ref{GR-Ib-Dumbell-V01-R-VJ-R}--\ref{GR-Ib-V01-R-WGR-Arb}\footnote{Strictly 
speaking, some terms arising from the manipulation of these diagrams have already been cancelled
by the some of the terms generated by the manipulation of
diagrams~\ref{GR-Ib-Dumbell-n_J+r-1-R-rR}--\ref{GR-Ib-n_J+r-1-R-WGR}.}
but also diagrams~\ref{GR-Ia-TLTP-WBB-Arb}, \ref{GR-Ia-TLTP-WBS-Arb}, \ref{GR-Ia-TLTP-WBB-n_J_rR-rR}, 
\ref{GR-Ia-TLTP-WBS-n_J_rR-rR}, \ref{GR-Ib-n_J+r-R-WGR-Top} and~\ref{GR-Ia-n_J+r-R-RWBB}.
The result of combining these terms is shown split between figures~\ref{fig:bn:GR-Ib-Rest-Nest}
and~\ref{fig:bn:GR-Ib-Rest}. The former of these shows the nested gauge remainder terms.
\begin{center}
\begin{figure}
	\[
	\begin{array}{c}
	\vspace{0.2in}
		\displaystyle
		4\sum_{J=0}^n \sum_{r=0}^{n_J} \norm_{J+1,0}
		\dec{
			\sco[0.2]{
				\LGD[0.1]{}{GR-Ib-Dumbell-n_J+r-R-r-R-GR}{GR-Ib-Dumbell-n_J+r-R-r-R-GR}
				+\LGD[0.1]{}{Struc-n_J+r-R-RWBS-GR}{GR-Ib-n_J+r-R-RWBS-GR}
				+\LGD[0.1]{}{Struc-n_J+r-R-WGR-Out-GR}{GR-Ib-n_J+r-R-WGR-Out-GR}
				-\LGD[0.1]{}{Struc-n_J+r-R-DEPBB-GR}{GR-Ib-n_J+r-R-DEPBB-GR}
			}{\ensuremath{\begin{array}{c}\input{pstex/Vertex-r-R.pstex_t} \end{array}}}
		}{11\Delta^{J+1}}
	\\
		\displaystyle
		+ 4\sum_{s=1}^{n_+} \sum_{J=2}^{2n-1}  \norm_{J+s-1,J-1}
	\\
		\displaystyle
		\times
		\dec{
			\sco[0.1]{
				\LGD[0.1]{}{GR-Ib-Dumbell-V01-R-VJ-R-GR}{GR-Ib-Dumbell-Arb-GR}
					}{\left[ \ds \VertexSum{1}{J-1}\ensuremath{\begin{array}{c}\input{pstex/GV-I-I+.pstex_t} \end{array}} \right]}
			+\frac{1}{J}
			\left[
				\sco[0.1]{
					\LGD[0.1]{}{Struc-V01-R-WBS-GR}{GR-Ib-WBS-GR-Arb}
					\LGD[0.1]{}{Struc-V01-R-WGR-Out-GR}{GR-Ib-WGR-GR-Arb}
					-\LGD[0.1]{}{Struc-V01-R-DEPBB-GR}{GR-Ib-V01-R-DEPBB-GR}
				}{\tco{\left[ \ds \VertexSum{1}{J-1}\ensuremath{\begin{array}{c}\input{pstex/GV-I-I+.pstex_t} \end{array}} \right]}{\ensuremath{\begin{array}{c}\input{pstex/Vertex-VJ-R.pstex_t} \end{array}}}}
			\right]
		}{11\Delta^{J+s-1}}
	\end{array}
	\]
\caption{Nested gauge remainder contributions arising from the manipulation of
diagrams~\ref{GR-Ib-Dumbell-V01-R-VJ-R}--\ref{GR-Ib-V01-R-WGR-Arb}.}
\label{fig:bn:GR-Ib-Rest-Nest}
\end{figure}
\end{center}

\begin{center}
\begin{figure}
	\[
	\begin{array}{c}
	\vspace{0.2in}
		\displaystyle
		4 \sum_{J=0}^n \sum_{r=0}^{n_J} 
	\\
	\vspace{0.2in}
		\times
		\left[
			\norm_{J+2,1}
			\dec{
				\sco[0.1]{
					\LGD[0.1]{}{GR-Ib-Dumbell-n_J+r-R-TLTP-E}{GR-Ib-TLTP-E-n_J+r-R-r-R}
				}{\ensuremath{\begin{array}{c}\input{pstex/Vertex-r-R.pstex_t} \end{array}}}
			}{1\Delta^{J+2}}
			+\norm_{J+1,0}
			\dec{
				\sco[0.1]{
					\PGD[0.1]{}{Struc-n_J+r-R-W-WBT}{GR-n_J+r-R-W-WBT}{fig:bn:GR-Ia,Ib-n_J-R-LdL-pre}
					+\PGD[0.1]{}{Struc-n_J+r-R-EP-LdL-WBT}{GR-n_J+r-R-EP-LdL-WBT}{fig:bn:GR-Ia,Ib-n_J-R-LdL-pre}
				}{\ensuremath{\begin{array}{c}\input{pstex/Vertex-r-R.pstex_t} \end{array}}}
			}{11\Delta^{J+1}}
		\right]
	\\
	\vspace{0.2in}
		\displaystyle
		+4 \sum_{s=1}^{n_+}\sum_{J=2}^{2n}  \frac{\norm_{J+s-1,J-1}}{J}
		\dec{
			\sco[0.1]{
				\PGD[0.1]{}{Struc-V01-R-W-WBT}{GR-V01-R-W-WBT}{fig:bn:GR-Ia,Ib-n_J-R-LdL-pre}
				+\PGD[0.1]{}{Struc-V01-R-EP-LdL-WBT}{GR-V01-R-EP-LdL-WBT}{fig:bn:GR-Ia,Ib-n_J-R-LdL-pre}
			}{ \left[ \ds \prod_{I=1}^{J-1} \sum_{V^{I_+}=0}^{V^I} \ensuremath{\begin{array}{c}\input{pstex/GV-I-I+.pstex_t} \end{array}} \right] \ensuremath{\begin{array}{c}\input{pstex/Vertex-VJ-R.pstex_t} \end{array}}}
		}{11\Delta^{J+s-1}}
	\\
		\displaystyle
		+4 \sum_{s=1}^{n_+}\sum_{J=2}^{2n-1}  \norm_{J+s,J}
		\dec{
			\LGO[0.1]{}{
				\tco[-0.3]{				
					\ensuremath{\begin{array}{c}\input{pstex/GR-Ib-Dumbell-V01-R-TLTP-E.pstex_t} \end{array}}
				}
					{
					\sco[0.1]{\left[ \ds \prod_{I=1}^{J-1} \sum_{V^{I_+}=0}^{V^I} \ensuremath{\begin{array}{c}\input{pstex/GV-I-I+.pstex_t} \end{array}} \right]}{\ensuremath{\begin{array}{c}\input{pstex/Vertex-VJ-R.pstex_t} \end{array}}}
				}}{GR-Ib-TLTP-E-Arb}	
		}{1\Delta^{J+s}}
	\end{array}
	\]
\caption{Result of combining (the surviving contributions to) diagrams~\ref{GR-Ib-Dumbell-V01-R-VJ-R}--\ref{GR-Ib-V01-R-WGR-Arb}
with
diagrams~\ref{GR-Ia-TLTP-WBB-Arb}, \ref{GR-Ia-TLTP-WBS-Arb}, \ref{GR-Ia-TLTP-WBB-n_J_rR-rR}, 
\ref{GR-Ia-TLTP-WBS-n_J_rR-rR}, \ref{GR-Ib-n_J+r-R-WGR-Top} and~\ref{GR-Ia-n_J+r-R-RWBB}.}
\label{fig:bn:GR-Ib-Rest}
\end{figure}
\end{center}

We now collate all surviving terms---including those remaining from the type-Ia sector
of the calculation---which fall neatly into three sets:
\begin{enumerate}
\item 	Diagrams which comprise an $\Op{2}$ stub, formed by the action of 
		a gauge remainder: \ref{GR-Ia-TLTP-E-n_J-S-R}, \ref{GR-Ia-TLTP-E-n_J+r-S-R-r-R},
		\ref{GR-Ia-TLTP-E-Arb}, \ref{GR-Ib-TLTP-E-n_J-R}, \ref{GR-Ib-TLTP-E-n_J+r-R-r-R}
		and~\ref{GR-Ib-TLTP-E-Arb};

\item 	Diagrams which possess the structure $\WBT$: \ref{GR-Ia-n_J-S-R-WGR-T}, \ref{GR-Ia-WGR-Top-n_J+r-S-R-rR},
		\ref{GR-Ia-WGR-Top-Arb}, \ref{GR-Ib-n_J-R-WGR-T}, \ref{GR-Ib-n_J+r-R-WGR-Top},
		\ref{GR-Ib-n_J-R-DWGR-Top},
		\ref{GR-n_J+r-R-W-WBT}, \ref{GR-n_J+r-R-EP-LdL-WBT}, \ref{GR-V01-R-W-WBT}
		and~\ref{GR-V01-R-EP-LdL-WBT};

\item	Nested gauge remainders: \ref{GR-Ia-WBS-GR-n_J-S-R}, \ref{GR-Ia-WGR-GR-n_J-S-R};
		\ref{GR-Ia-Dumbell-n_J+r-S-R-r-R-GR}, \ref{GR-Ia-WBS-GR-n_J+r-S-R-r-R}, \ref{GR-Ia-WGR-GR-n_J+r-S-R-r-R};
		\ref{GR-Ia-Dumbell-Arb-GR}, \ref{GR-Ia-WBS-GR-Arb}, \ref{GR-Ia-WGR-GR-Arb}; 
		\ref{GR-Ib-n_J-R-RWBS-GR}, \ref{GR-Ib-n_J-R-WGR-Out-GR}, \ref{GR-Ib-n_J-R-DEPBB-GR}
		and~\ref{GR-Ib-Dumbell-n_J+r-R-r-R-GR}--\ref{GR-Ib-V01-R-DEPBB-GR}.
\end{enumerate}

We discussed, in some detail, the result of processing nested gauge remainders
at the end of the section on type-Ia gauge remainders. We will now add to this,
by including the
analysis of the nested gauge remainders of type-Ib. 

The first thing we note is that, among the set of nested gauge remainders
are a set of diagrams which have no analogue in the original set of gauge remainders:
namely diagrams~\ref{GR-Ib-n_J-R-DEPBB-GR},
\ref{GR-Ib-n_J+r-R-DEPBB-GR} and~\ref{GR-Ib-V01-R-DEPBB-GR}. We will put these to the side
for one moment.

In preparation for processing the other nested gauge remainders, 
let us recall the pattern of cancellations in the un-nested case. The bulk of terms
generated by the type-Ia and type-Ib parent diagrams cancelled among themselves,
courtesy of cancellation mechanisms~\ref{CM:VertexJoinWine} and~\ref{CM:GR-Socket}. Such
cancellations will go through in exactly the same way in the nested case. There were
then those terms that are cancelled only as a result of combining terms formed by type-Ia
gauge remainders with those formed by type-Ib gauge remainders. Examples of this are given
by cancellations~\ref{cancel:GR-Ib-TLTP-WBB-R}--\ref{cancel:GR-Ib-n_J-R-DEPBB-C}.
In the nested case, too, these cancellations will go through  just the same,
with one exception. At
the final level, the analogues of  diagrams~\ref{GR-Ia-a1s-TLTP-WBB} and~\ref{GR-Ia-a1s-TLTP-WBS}
will have, in addition to the socket on the vertex, an additional $2n+1$ locations to which
a field must be attached (\ie\ corresponding to the $2n+1$ nested gauge remainders). To fill
these slots, there are precisely the correct number of fields. Now, the type-Ib versions
of these diagrams must die at this level of nesting, since we cannot both decorate the wine
and leave no empty slots.

Note that this nicely confirms that, at any other level of nesting, the analogue of
cancellations~\ref{cancel:GR-Ib-TLTP-WBB-R} and~\ref{cancel:GR-Ib-TLTP-WBS-R} work:
other than at the final level, diagrams~\ref{GR-Ia-a1s-TLTP-WBB} and~\ref{GR-Ia-a1s-TLTP-WBS}
always have more decorative fields than empty slots, meaning that the wine must be decorated.
This guarantees cancellation against type-Ib terms which necessarily possess a reduced wine.

However, rather than performing these cancellations it will prove useful
to retain the components of diagrams~\ref{GR-Ia-a1s-TLTP-WBB} and~\ref{GR-Ia-a1s-TLTP-WBS}
which die only as a consequence of our particular choice of decorations
\ie\ those components with an un-decorated wine.
Doing so, we will find a manifest cancellation against
components of other diagrams, coming from later in the calculation.\footnote{At the
un-nested level, there is no mileage in keeping the vanishing terms.}

When dealing with nested terms, though, we must be aware that new types of diagram can
arise. There are two ways in which this can happen. First, diagrams,
or particular decorations of diagrams, that were legitimately discarded
in the un-nested case may survive in the nested case: we have commented
already that this is precisely the case with diagrams~\ref{GR-Ib-TLTP-WBB-R-n_J-R-E} 
and~\ref{GR-Ib-LdL-L-E}.

Secondly, when a nested gauge
remainder generates a two-point, tree level vertex, we now have the option of
attaching it to one of the nested gauge remainders. 
We have encountered this already
with the type-Ia gauge remainders, which yielded
diagrams of the type shown in figure~\ref{fig:bn:GR-Ia-Nest-Dec-Ex}.
Noting that we will find further examples of such terms shortly,
we show the type-Ib analogue of the aforementioned diagrams in 
figure~\ref{fig:bn:GR-Extra-A}.
\begin{center}
\begin{figure}[h]
	\[
	-4 \sum_{J=0}^n \norm_{J,0}
	\dec{
		\CGD[0.1]{}{Struc-n_J-R-RW-LL-Alg}{GR-Ib-n_J-R-RW-LL-Alg}{DGRx2-n_J-R-RW-GR-RR}
		-\CGD[0.1]{}{Struc-n_J-R-RW-LR-Alg}{GR-Ib-n_J-R-RW-LR-Alg}{DGRx2-n_J-R-RW-GR-LR}	
	}{11\Delta^{J}}
	\]
\caption{Analogues of the diagrams of figure~\ref{fig:bn:GR-Ia-Nest-Dec-Ex}
generated by type-Ib gauge remainders.}
\label{fig:bn:GR-Extra-A}
\end{figure}
\end{center}

We now return to the three diagrams that we temporarily put to one side,
diagrams~\ref{GR-Ib-n_J-R-DEPBB-GR},
\ref{GR-Ib-n_J+r-R-DEPBB-GR} and~\ref{GR-Ib-V01-R-DEPBB-GR}.
However, we note that these now have partner diagrams which are nested
with respect to the already processed gauge remainder; these additional
terms coming, of course, from having manipulated the other nested
gauge remainders.

Processing this set of diagrams is actually very easy, requiring a minor
generalisation of cancellation mechanism~\ref{CM:GR-Socket}.
Recall the set-up for cancellation mechanism~\ref{CM:GR-Socket}: a wine, whose join
to the arbitrary structure $B$ can be via nested gauge remainders, attaches to the structure
$A$ via nested gauge remainders. Allowing the gauge remainder hitting $A$ to act, we
assume that it strikes a socket. Up until now, we have demanded that this socket could not be
filled by the other end of the wine. Now, we simply relax this assumption, though we need
to do so with care.

By allowing the socket to be filled by the wine, we are implicitly identifying the structures $A$
and $B$. In the case that $A$ and $B$ are both reduced vertices, this step would be illegal for
type-Ia gauge remainders since $B$ would have to have been a seed action vertex whereas $A$
would have to have been 
a Wilsonian effective action vertex. (Note that this problem does not arise for type-Ib
gauge remainders). In the case that $A$ and $B$ are both identified with the wine, we would
discard the result in both the type-Ia and type-Ib cases, since it would correspond to a wine
biting its own tail. However, in the type-II case, we note that the attachment of the wine
to the structure $B$ is non-trivial \ie\ is via a gauge remainder. In the case that we identify
both $A$ and $B$ with the wine, this prevents the wine from biting its own tail. 

Next we must consider a similar diagram in which the gauge remainder which previously
attached to $A$ now attaches to a vertex, instead. In the case that the gauge remainder
strikes a socket and this socket is filled by the wine, we are now identifying the \emph{vertex}
with structure $B$. When we take the two-point, tree level part of this vertex, we still
go ahead and attach it to the structure $A$ with an effective propagator. We are then free
to identify $A$ as either a vertex or the wine.

With these thoughts in mind, we now supplement cancellation mechanism~\ref{CM:GR-Socket}.
\begin{CM}
Cancellation mechanism~\ref{CM:GR-Socket} holds exactly if we allow the socket 
struck by the gauge remainder to be
filled by the wine, so long as the implicit identification of structures involved
does not lead to an inconsistency.
\label{CM:GR-Socket-Supp}
\end{CM}

Thus, cancellation mechanism~\ref{CM:GR-Socket-Supp} now guarantees that 
contributions from diagrams~\ref{GR-Ib-n_J-R-DEPBB-GR},
\ref{GR-Ib-n_J+r-R-DEPBB-GR}, \ref{GR-Ib-V01-R-DEPBB-GR} and their analogues
possessing additional vertices
in which the gauge remainder bites a socket which decorates anything
other than a two-point, tree level vertex will cancel.
This equally applies to the version of these
diagrams in which the gauge remainder biting the base of the wine
is nested.

Iterating the diagrammatic
procedure once, we obtain nested versions of the parents, diagrams
possessing an $\Op{2}$ stub and diagrams in which the 
 two-point, tree level vertex has been joined to a bitten
gauge remainder. Note that the nesting is now with respect to the
gauge remainder we are currently performing. 

Iterating the diagrammatic procedure until
exhaustion, we obtain either diagrams with an $\Op{2}$ stub or diagrams
in which the two-point, tree level vertex has been joined to a bitten
gauge remainder.
In figure~\ref{fig:bn:GR-Extra} we give examples
of these latter terms.
The
diagrams that we choose to draw are the simplest that
can exist in the sense that we choose to draw the ones with the minimum
number of vertices and the minimum level of nesting. We henceforth call
such a set of diagrams a minimal set. (We no longer explicitly
draw diagrams with an $\Op{2}$ stub, since we will not be doing a complete
treatment of them, within this thesis.)
\begin{center}
\begin{figure}
	\[
	\begin{array}{c}
	\vspace{0.1in}
		\displaystyle
		+4 \norm_{n-1,0}
	\\
		\times
		\dec{
			\begin{array}{c}
			\vspace{0.2in}
				-\PGD[0.1]{}{Struc-EP-RXR}{GR-Ib-EP-RXR}{fig:bn:GR-II-Redraw}
				+\PGD[0.1]{}{Struc-EP-LXR}{GR-Ib-EP-LXR}{fig:bn:GR-II-Redraw}
			\\
			\vspace{0.2in}
				+\PGD[0.1]{}{Struc-EPBB-L-PFX}{GR-Ib-EPBB-L-PFX}{fig:bn:GR-II-N2-Wine/Combo-Redraw-B}
				-\PGD[0.1]{}{Struc-EPBB-L-PBX}{GR-Ib-EPBB-L-PBX}{fig:bn:GR-II-N2-Wine/Combo-Redraw-B}
				-\PGD[0.1]{}{Struc-EPBB-R-PFX}{GR-Ib-EPBB-R-PFX}{fig:bn:GR-II-N2-Wine/Combo-Redraw-B}
				-\PGD[0.1]{}{Struc-EPBB-R-PBX}{GR-Ib-EPBB-R-PBX}{fig:bn:GR-II-N2-Wine/Combo-Redraw-B}
			\\
			\vspace{0.2in}
				-\PGD[0.1]{}{Struc-EPGRx2-LXR-PF}{GR-Ib-EPGRx2-LXR-PF}{fig:bn:GR-II-N2-Wine/Combo-Redraw-B}
				+\PGD[0.1]{}{Struc-EPGRx2-LXR-PB}{GR-Ib-EPGRx2-LXR-PB}{fig:bn:GR-II-N2-Wine/Combo-Redraw-B}
				+\PGD[0.1]{}{Struc-EPGRx2-RXR-PF}{GR-Ib-EPGRx2-RXR-PF}{fig:bn:GR-II-N2-Wine/Combo-Redraw-B}
				-\PGD[0.1]{}{Struc-EPGRx2-RXR-PB}{GR-Ib-EPGRx2-RXR-PB}{fig:bn:GR-II-N2-Wine/Combo-Redraw-B}
			\end{array}
		}{11\Delta^{n-1}}
	\end{array}
	\]
\caption{The simplest versions of diagrams in which a two-point,
tree level vertex is attached to a bitten gauge remainder.}
\label{fig:bn:GR-Extra}
\end{figure}
\end{center}

We can now group all terms which survive upon processing the nested gauge
remainders.

\begin{enumerate}
\item 	Diagrams which comprise an $\Op{2}$ stub, formed by the action of 
		a gauge remainder;

\item	Diagrams which comprise a nested version of $\BT$, 
		including those diagrams in which one of the gauge
		remainders in this structure is hit by $\flow$;

\item	Diagrams in which a two-point, tree level vertex is attached
		to one of the nested gauge remainders; 
\end{enumerate}

To conclude this section, we analyse those diagrams in which the string of nested gauge
remainders bite each other, in a loop
\ie\ the second set
above. In the case of
such terms formed by type-Ia gauge remainders this is trivial: we just generate
nested versions of diagrams~\ref{GR-Ia-n_J-S-R-WGR-T}, \ref{GR-Ia-WGR-Top-n_J+r-S-R-rR},
and~\ref{GR-Ia-WGR-Top-Arb}. 
In the case of such terms formed by type-Ib
gauge remainders, alone, things are exactly the same \ie\
we simply generate nested versions of diagrams~\ref{GR-Ib-n_J+r-R-WGR-Top},
\ref{GR-n_J+r-R-W-WBT} and~\ref{GR-V01-R-W-WBT}.

In the case of those terms formed by a combination of gauge
remainder of type-Ia and-Ib \ie\ terms~\ref{GR-Ib-n_J-R-DWGR-Top}, \ref{GR-n_J+r-R-EP-LdL-WBT}
and~\ref{GR-V01-R-EP-LdL-WBT} we must be more careful deducing
the structure of the nested terms from the un-nested ones.
It is very easy to see where the problem comes from: in the nested case,
we are no longer compelled to join the gauge remainder hit by $\flow$
to the loose vertex; we could just as well attach an external field.\footnote{
Imagine the nested version of diagram~\ref{GR-Ib-LdL-L-E}.}

Hence, our starting point should not be diagrams~\ref{GR-Ib-n_J-R-DWGR-Top}, \ref{GR-n_J+r-R-EP-LdL-WBT}
and~\ref{GR-V01-R-EP-LdL-WBT},
but rather the nested version of the
combination of diagrams~\ref{GR-Ia-TLTP-WBS-n_J-R} and~\ref{GR-Ib-TLTP-WBS-R-n_J-R}.
We will
consider such a combination in conjunction with the nested version of diagram~\ref{GR-Ib-n_J-R-DWGR-Top},
as shown in figure~\ref{fig:bn:GR-Nested-LdL-GR}.
\begin{center}
\begin{figure}[h]
	\[
	-4 \sum_{J=0}^n 
	\left[
		\norm_{J,0}
		\dec{
			\LID[0.1]{}{Struc-GR-ring-n_J-R}{GR-ring-n_J-R-Wine}
		}{11\Delta^J}
		+\norm_{J+1,0}
		\dec{
			\sco[0.1]{
				\LID[0.1]{}{Struc-LdL-GR-ring}{LdL-GR-ring-n_J-R}
			}{\ensuremath{\begin{array}{c}\input{pstex/Vertex-n_J-R.pstex_t} \end{array}}}
		}{11\Delta^{J+1}}
	\right]
	\]
\caption{An arbitrarily nested version of diagram~\ref{GR-Ib-n_J-R-WGR-T} and a nested version of 
the combination of diagrams~\ref{GR-Ia-TLTP-WBS-n_J-R} and~\ref{GR-Ib-TLTP-WBS-R-n_J-R}.}
\label{fig:bn:GR-Nested-LdL-GR}
\end{figure}
\end{center}

The crucial feature of both diagrams is the ring, composed of
gauge remainders.
The very first gauge remainder to be performed in each of these diagrams
can be chosen to be either a push forward or a pull back, by charge conjugation
invariance. All nested gauge remainders can act in either sense, and we pick up
a minus sign for each pull back. 

Our strategy will be to try and convert this combination of
terms into a $\Lambda$-derivative term, plus corrections. If we can
do this, then this will tell us how to treat diagrams containing
nested instances of $\BT$.

If we are to be able to convert
these terms (usefully) into $\Lambda$-derivative terms, then the question
that we must answer is whether or not it makes any difference which of the arbitrary
number of gauge remainders in the ring we choose to be the first one. To answer
this question, it is useful to talk in terms of bites to the
right and bites to the left (see section~\ref{sec:GRs-Nested}), rather than pushes forward
and pulls back.
Utilising this picture, when all the gauge remainders act in the same sense, it clearly does
not make any difference which gauge remainder in the ring
we choose to be the first one. In this case, if there are $m$ gauge remainders then we could
combine the diagrams of figure~\ref{fig:bn:GR-Nested-LdL-GR} into a $\Lambda$-derivative
term, as shown in figure~\ref{fig:bn:GR-Nested-LdL-GR-B}.
\begin{center}
\begin{figure}[h]
	\[
	-\frac{2}{m} \sum_{J=0}^n 
	\left[
		\dec{
			2 \norm_{J+1,0}
			\dec{
				\sco[0.1]{
					\LID[0.1]{}{Struc-GR-ring}{GR-ring-n_J-R}
				}{\ensuremath{\begin{array}{c}\input{pstex/Vertex-n_J-R.pstex_t} \end{array}}}
			}{11\Delta^{J+1}}
		}{\bullet}
		-m\norm_{J,0}
		\dec{
			\sco[0.1]{
				\LID[0.1]{}{Struc-GR-ring-Join}{GR-ring-n_J-R-Join}	
			}{\ensuremath{\begin{array}{c}\input{pstex/Vertex-n_J-R.pstex_t} \end{array}}}
		}{11\Delta^{J}}
	\right]
	+ \cdots
	\]
\caption{Combining diagrams~\ref{GR-ring-n_J-R-Wine} and~\ref{LdL-GR-ring-n_J-R}
into a $\Lambda$-derivative term.}
\label{fig:bn:GR-Nested-LdL-GR-B}
\end{figure}
\end{center}

The explicitly  
drawn correction term is that for which the $\Lambda$-derivative
strikes an effective propagator joining two of the gauge remainders in the ring
to each other. Depending on how many gauge remainders we take the
ellipses on the ring to represent, diagram~\ref{GR-ring-n_J-R-Join}
implicitly represents all possible independent joins between 
pairs of gauge remainders.
The relative factor of this term is easy to compute. We take one end of one of
the decorative effective propagators of diagram~\ref{GR-ring-n_J-R}
and insert it into any one of $m$ equivalent locations. We then sum over
all possible insertions of the other end. The resulting set of diagrams
come with a relative factor of $mJ$. We will see in section~\ref{sec:GR-II} how such terms are
generated by gauge remainders of type-II.

We now argue that this conversion of diagrams~\ref{GR-ring-n_J-R-Wine} and~\ref{LdL-GR-ring-n_J-R}
into a $\Lambda$-derivative term (plus corrections) is valid, irrespective of the precise
arrangement of the gauge remainders.
To see this, it is simplest not to use charge conjugation to collect together
the first bite on the right with the first bite on
the left. Now we have a
total of $2^m$ independent arrangements of the gauge remainders, each of which can be represented by a string 
of $L$s and $R$s. To look at it another way, we are counting from zero to $2^{m-1}$
in binary, using $R$s and $L$s to represent ones and zeros:
\[
	\begin{array}{c}
	RR \cdots RR \\
	RR \cdots RL \\
	RR \cdots LR \\
	\vdots
	\end{array}
\]

Next, imagine that we cyclically permute each of the above strings. After a suitable
vertical
re-ordering of terms, we just reproduce the above sequence. Thus, when singling
out one of the gauge remainders of diagram~\ref{GR-ring-n_J-R}---for example
to attach a wine to or to hit with $\flowConstAl$---we can do this in $m$ equivalent ways.
Therefore, the conversion of diagrams~\ref{GR-ring-n_J-R-Wine} and~\ref{LdL-GR-ring-n_J-R}
into diagram~\ref{GR-ring-n_J-R} works, irrespective of the precise arrangement of
the gauge remainders.

\newpage
\section{Type-II Gauge Remainders} \label{sec:GR-II}

All gauge remainders of type-II have been collected together in
figure~\ref{fig:bn:GR-II}. The general pattern of terms is similar
to those for the gauge remainders of types-Ia and Ib, 
but with some marked differences.

\begin{center}
\begin{figure}
	\[
	\begin{array}{c}
	\vspace{0.1in}
		\displaystyle
		-\sum_{s=1}^{n_+} \sum_{J=2}^{2n}  \norm_{J+s-1,J-1}
	\\
	\vspace{0.2in}
		\displaystyle
		\times
		\dec{
			\sco[0.3]{
				\LGD[0.1]{}{GR-II-Dumbell-V01-R-VJ-R}{GR-II-Dumbell-V01-R-VJ-R}
				+\frac{1}{J}
				\left[
					\sco[0.1]{
						\LGD[0.1]{}{GR-II-V01-R}{GR-II-V01-R}
						+2 \LGD[0.1]{}{GR-II-V01-R-WGR}{GR-II-V01-R-WGR}
						+\frac{1}{J+1}
						\left[
							\LGO[0.1]{}{
								\scd[0.1]{Struc-WGRx2-B}{GV-V01-R}
							}{GR-II-WGRx2-Arb}
						\right]
					}{\ensuremath{\begin{array}{c}\input{pstex/GV-VJ-R.pstex_t} \end{array}}}
				\right]
			}{\left[ \ds \VertexSum{1}{J-1} \ensuremath{\begin{array}{c}\input{pstex/GV-I-I+.pstex_t} \end{array}} \right]}
		}{11\Delta^{J+s-1}}
	\\
	\vspace{0.1in}
		\displaystyle
		- \sum_{J=0}^{n} \sum_{r=0}^{n_{J}} \norm_{J+1,0}
	\\
	\vspace{0.2in}
		\displaystyle
		\times
		\dec{
			\PGD[0.1]{}{GR-II-Dumbell-n_J+r-R-rR}{GR-II-Dumbell-n_J+r-R-rR}{fig:bn:GR-II-L2-Nest}
			+
			\left[
			\sco[0.1]{
				\PGD[0.1]{}{GR-II-n_J+r-R-GR}{GR-II-n_J+r-R-GR}{fig:bn:GR-II-L2-Nest}
				+2\PGD[0.1]{}{GR-II-n_J+r-R-WGR}{GR-II-n_J+r-R-WGR}{fig:bn:GR-II-L2-Nest}
				+\frac{1}{2}
				\left[
					\LGO[0.1]{}{
						\scd[0.1]{Struc-WGRx2-B}{Vertex-n_J+r-R}
					}{GR-II-n_J+r-WGRx2}
				\right]
			}{\ensuremath{\begin{array}{c}\input{pstex/Vertex-r-R.pstex_t} \end{array}}}
			\right]
		}{11\Delta^{J+1}}
	\\
	\vspace{0.2in}
		\displaystyle
		+ \sum_{J=0}^n \norm_{J,0}
		\dec{
			\PGD[0.1]{}{Struc-n_J-R-RW-GRx2}{GR-II-n_J-R-RW-GRx2}{fig:bn:GR-II-L1}
			+2 \PGD[0.1]{}{Struc-n_J-R-WGR-GR}{GR-II-n_J-R-WGR-GR}{fig:bn:GR-II-L1}
			+\PGO[0.1]{}{
				\scd[0.1]{Struc-WGRx2-B}{Vertex-n_J-R}
			}{GR-II-WGRx2-n_J-R}{fig:bn:GR-II-L2-Nest}
		}{11\Delta^J}
	\\
	\displaystyle
		-\norm_{n-1,0}
		\dec{
			\PGD[0.1]{}{Struc-WGRx2-B}{GR-II-WGRx2-B}{fig:bn:GR-II-L1}
		}{11\Delta^{n-1}}
	\end{array}
	\]
\caption{Type-II gauge remainders.}
\label{fig:bn:GR-II}
\end{figure}
\end{center}

The most obvious difference in the pattern of terms is the appearance of diagrams~\ref{GR-II-WGRx2-Arb},
\ref{GR-II-n_J+r-WGRx2}, \ref{GR-II-WGRx2-n_J-R} and~\ref{GR-II-WGRx2-B}. However, we also
see that diagrams~\ref{GR-II-n_J-R-WGR-GR}, \ref{GR-II-n_J+r-R-WGR} and~\ref{GR-II-V01-R-WGR}
come with factor of two, relative to their partner diagrams.
This makes sense: for these diagrams, we have correctly counted separately the case where each of
the gauge remainders attaches to the wine.

We must now decide how to process these diagrams. The first thing we note is that we are guaranteed
to encounter trapped gauge remainders in diagrams~\ref{GR-II-n_J-R-RW-GRx2}, \ref{GR-II-WGRx2-B}
and their analogues possessing additional vertices. In each of these cases, we can choose to act with either of the gauge remainders
first, and one of the things it will do is bite the field on the structure to which
the tip of the other gauge remainder is attached.
If, on the other hand, the first action of one of the gauge remainders is to bite a socket
or the end of a wine, then 
we are free to perform the other gauge remainder as well. However, as we will see shortly, it is
inefficient to do this: we will be able to cancel a set of
diagrams in which an active gauge remainder
is un-processed.

Given that we are only going to process one gauge remainder, for the time being, we must decide what
to do with diagrams~\ref{GR-II-n_J-R-WGR-GR}, \ref{GR-II-n_J+r-R-WGR} and~\ref{GR-II-V01-R-WGR}.
The answer is to be democratic: we take one instance of these diagrams where one gauge remainder
acts and one instance where the other acts, dividing by two to avoid over-counting.

Once a gauge remainder has acted, we proceed in exactly the same way as we did with the
gauge remainders of types-Ia and Ib: we isolate any two-point, tree level vertices,
decorate them and apply the effective propagator relation, as appropriate. 
This procedure is trivial for diagrams in which a single vertex is hit
by both gauge remainders: if one of the gauge remainders hits a socket, then the vertex
must still be reduced, else it will be killed by the remaining gauge remainders (indeed, the vertex must
actually be three-point or greater, irrespective of loop order, though we will not make use of this).
On the other hand, if one of the gauge remainders traps the other, then a reduction of the vertex is
no longer forced. 

We now see the benefit both of performing only one of the gauge remainders and of treating
the gauge remainders in diagrams~\ref{GR-II-n_J-R-WGR-GR}, \ref{GR-II-n_J+r-R-WGR} and~\ref{GR-II-V01-R-WGR}
democratically. Starting with diagram~\ref{GR-II-WGRx2-B} let us suppose that we process one of the gauge
remainders and it strikes a socket. By cancellation mechanism~\ref{CM:GR-Socket} this will be cancelled by
a contribution arising from diagram~\ref{GR-II-n_J-R-WGR-GR}. To be precise, this contribution
will be killed if we take one half of diagram~\ref{GR-II-n_J-R-WGR-GR} and allow the bottom
gauge remainder to act, creating a two-point, tree level vertex, which we then join to the wine
with an effective propagator.

Staying with diagram~\ref{GR-II-n_J-R-WGR-GR}, the next thing to do is to find the contributions
in which one of the gauge remainders strikes a socket on anything other than a two-point, tree level
vertex. There are two ways in which this can happen. Either the bottom gauge remainder can strike
a socket on the vertex or the top gauge remainder can strike a socket on the wine. Let us now
add this collection of diagrams to those generated by the gauge remainders of diagrams~\ref{GR-II-n_J-R-RW-GRx2}
and~\ref{GR-II-WGRx2-n_J-R} striking a socket.\footnote{In the former case, if the socket decorates
a two-point, tree level vertex, then the diagram dies whereas, in the latter case, it is impossible
to generate a socket on a two-point, tree level vertex.
}
 It is not hard to check that the complete set
of these diagrams is cancelled by terms coming from diagrams~\ref{GR-II-Dumbell-n_J+r-R-rR}
and~\ref{GR-II-n_J+r-R-WGR}. This only works if we treat the gauge remainders in the
latter diagram democratically. Moving on to diagrams~\ref{GR-II-Dumbell-V01-R-VJ-R}--\ref{GR-II-WGRx2-Arb}, it
is clear that all contributions in figure~\ref{fig:bn:GR-II} for which a gauge remainder strikes
a socket are removed by cancellation mechanism~\ref{CM:GR-Socket}.

We must now address the question of the types of term that will be left over, after we allow one
of the gauge remainders to act. As usual, we will be left with terms possessing an $\Op{2}$ stub,
nested gauge remainders and terms in which a gauge remainder strikes the end of a wine. Additionally,
as mentioned previously, we will find trapped gauge remainders. 

Consider allowing one of the gauge remainders in diagram~\ref{GR-II-n_J+r-R-GR} to act,
trapping the other gauge remainder and creating a two-point tree level vertex. Now join this vertex
to either the wine or the other vertex, with one of the effective propagators. This
procedure is shown in figure~\ref{fig:bn:GR-II-Trapped-Ex}.
\begin{center}
\begin{figure}[h]
	\[
	\begin{array}{l}
	\vspace{0.2in}
		\displaystyle
		-2\sum_{J=0}^n \norm_{J+1,0}
		\dec{
			\scd[0.1]{Struc-TLTP-RWBB-Trap}{Vertex-n_J-R}
		}{11\Delta^{J+1}}
	\\
		\displaystyle
		\rightarrow
		-2\sum_{J=0}^n \norm_{J,0}
		\dec{
			\ensuremath{\begin{array}{c}\input{pstex/Struc-n_J-R-RWBB-Trap.pstex_t} \end{array}}
			\hspace{1em}
			+\scd[0.1]{Struc-WGRx2-Trap-B}{Vertex-n_J-R}
		}{11\Delta^J} + \cdots
	\end{array}
	\]
\caption{Some of the terms generated by processing diagram~\ref{GR-II-n_J+r-R-GR}.}
\label{fig:bn:GR-II-Trapped-Ex}
\end{figure}
\end{center}

The terms on the second line of figure~\ref{fig:bn:GR-II-Trapped-Ex}
will cancel against trapped gauge remainder terms generated
by diagrams~\ref{GR-II-n_J-R-RW-GRx2} and~\ref{GR-II-WGRx2-n_J-R}. These 
cancellations
are a examples of cancellation mechanism~\ref{CM:GR-Socket-Supp}.

Indeed, cancellation mechanism~\ref{CM:GR-Socket-Supp} now guarantees that all diagrams arising
from figure~\ref{fig:bn:GR-II} possessing a trapped gauge remainder are cancelled, up to
nested gauge remainders and terms with an $\Op{2}$ stub.

We are almost ready to process the diagrams of figure~\ref{fig:bn:GR-II}, but there
are a couple of  further observations we can make to simplify things. We have noted that, among
the terms left over are those in which a gauge remainder strikes the end of a wine.
Depending on which end of the wine the gauge remainder bites will determine how we treat
the corresponding term. If the gauge remainder bites the end of the wine
to which the remaining, active gauge remainder attaches, then we will go ahead and process the active
gauge remainder. If, on the other hand, the first gauge remainder bites the other end
of the wine, then we will
leave the remaining active gauge remainder unprocessed, for the time being. The reason for this
is that, later in the calculation, we will find terms that combine with the diagrams possessing
unprocessed gauge remainders; by leaving these gauge remainders unprocessed, until that
stage, we save ourselves some work.

In the case where we do allow the second gauge remainder to act,
cancellation mechanism~\ref{CM:GR-Socket} ensures that terms in which this gauge remainder
bites a socket that decorates anything other than a two-point, tree level vertex will
cancel.\footnote{If we were to allow the gauge remainder to act in the terms were we have
withheld its action, cancellation mechanism~\ref{CM:GR-Socket} would ensure the usual
cancellation of diagrams in this case, too.}
Hence, in cases
where both gauge remainders strike the wine, we are ultimately left only with those terms
in which both gauge remainders hit the ends of the wine.

Finally, then, figure~\ref{fig:bn:GR-II-L1}
shows the result of processing diagrams~\ref{GR-II-n_J-R-RW-GRx2}, \ref{GR-II-n_J-R-WGR-GR}
and~\ref{GR-II-WGRx2-B}, where all terms which would be cancelled by cancellation 
mechanisms~\ref{CM:GR-Socket} and~\ref{CM:GR-Socket-Supp} have been omitted.

\begin{center}
\begin{figure}
	\[
	\begin{array}{c}
	\vspace{0.2in}
		\displaystyle
		2 \sum_{J=0}^n \norm_{J,0}
		\dec{
			\PGD[0.1]{}{Struc-n_J-R-GR-W-DGR}{GR-II-n_J-R-GR-W-DGR}{fig:bn:GR-II/Iterate-Combo}
		}{11\Delta^J}
		-2 \norm_{n-1,0}
		\dec{
			\PGD[0.1]{}{Struc-WGRx2-Top}{GR-II-WGRx2-Top}{fig:bn:GR-II/Iterate-Combo}
		}{11\Delta^{n-1}}
	\\
	\vspace{0.1in}
		\displaystyle
		-2 \norm_{n-1,0}
	\\
	\vspace{0.2in}
		\times
		\dec{
			\begin{array}{c}
			\vspace{0.2in}
				\PGD[0.1]{}{Struc-WGRx2-Out-GR}{GR-II-WGRx2-Out-GR}{fig:bn:GR-II-N1-Survive-Class3c}
				+\PGD[0.1]{}{Struc-WGRx2-Trap-GR}{GR-II-WGRx2-Trap-GR}{fig:bn:GR-II-N1-Survive-Class2b}
				+\PGD[0.1]{}{Struc-WGR-N2-PB}{GR-II-WGR-N2-PB}{fig:bn:GR-II-N1-Survive-Class1}
				-\PGD[0.1]{}{Struc-WGR-N2-PF}{GR-II-WGR-N2-PF}{fig:bn:GR-II-N1-Survive-Class1}
			\\
				+\PGD[0.1]{}{Struc-WGR-N1-Bottom-PB}{GR-II-WGR-N1-Bottom-PB}{fig:bn:TLTP-GR-JoinedToWine}
				-\PGD[0.1]{}{Struc-WGR-N1-Bottom-PF}{GR-II-WGR-N1-Bottom-PF}{fig:bn:TLTP-GR-JoinedToWine}
				+\CGD[0.1]{}{Struc-WGR-N1-Top-PB}{GR-II-WGR-N1-Top-PB}{DGRx2-TLTP-E-RW-GR-RR}
				-\CGD[0.1]{}{Struc-WGR-N1-Top-PF}{GR-II-WGR-N1-Top-PF}{DGRx2-TLTP-E-RW-GR-LR}
			\end{array}
		}{11\Delta^{n-1}}
	\\
	\vspace{0.2in}
		\displaystyle
		+2 \norm_{n,0}
		\dec{
			\LGD[0.1]{}{Struc-TLTP-RWBB-Trap-E}{GR-II-TLTP-RWBB-Trap-E}
			+\LGD[0.1]{}{Struc-II-TLTP-WBS-WGR-E}{GR-II-TLTP-WBS-WGR-E}
			+\LGD[0.1]{}{Struc-II-TLTP-WBS-WBB-E}{GR-II-TLTP-WBS-WBB-E}
			-\LGD[0.1]{}{Struc-II-TLTP-WBS-WBB-E-PB}{GR-II-TLTP-WBS-WBB-E-PB}
		}{1\Delta^n}
	\end{array}
	\]
\caption{Result of processing diagrams~\ref{GR-II-n_J-R-RW-GRx2}, \ref{GR-II-n_J-R-WGR-GR}
and~\ref{GR-II-WGRx2-B}, up to terms which are guaranteed to cancel, 
via cancellation mechanisms~\ref{CM:GR-Socket} and~\ref{CM:GR-Socket-Supp}.}
\label{fig:bn:GR-II-L1}
\end{figure}
\end{center}

Having done all that we can with diagrams~\ref{GR-II-n_J-R-RW-GRx2}, \ref{GR-II-n_J-R-WGR-GR}
and~\ref{GR-II-WGRx2-B},
up to processing nested gauge remainders,
we now move on to diagrams~\ref{GR-II-Dumbell-n_J+r-R-rR}--\ref{GR-II-n_J+r-R-WGR} and~\ref{GR-II-WGRx2-n_J-R}.

We can immediately write down the result of processing these diagrams, by using cancellation
mechanisms~\ref{CM:GR-Socket} and~\ref{CM:GR-Socket-Supp} and by using
the preceding analysis as a template. Most straightforwardly, we obtain
a copy of figure~\ref{fig:bn:GR-II-L1}, but where each diagram
contains an extra vertex, with argument $n_J^R$, and where the factor of $-\norm_{n-1,0}$
is replaced by $\norm_{J,0}$;
$J$ being summed over in the usual manner. In addition to these terms, we obtain
a further set of nested gauge remainders, which are shown in figure~\ref{fig:bn:GR-II-L2-Nest}

\begin{center}
\begin{figure}
	\[
	2\sum_{J=0}^n \norm_{J,0}
	\dec{
		\begin{array}{c}
		\vspace{0.2in}
			\displaystyle
			\PGD[0.1]{}{Struc-II-n_J-R-N1}{GR-II-n_J-R-N1}{fig:bn:GR-II-N1-Survive-Class3c}
			+\PGD[0.1]{}{Struc-II-n_J-R-WGR-N1}{GR-II-n_J-R-WGR-N1}{fig:bn:GR-II-N1-Survive-Class3c}
			+\PGD[0.1]{}{Struc-II-n_J-R-N1-WGR}{GR-II-n_J-R-N1-WGR}{fig:bn:GR-II-N1-Survive-Class3c}
		\\
			+\PGD[0.1]{}{Struc-n_J-R-RWBB-Trap-GR}{GR-II-n_J-R-RWBB-Trap-GR}{fig:bn:GR-II-N1-Survive-Class2b}
			+\PGD[0.1]{}{Struc-II-n_J-R-N1-PF-WBB}{GR-II-n_J-R-N1-PF-WBB}{fig:bn:GR-II-N1-Survive-Class1}
			-\PGD[0.1]{}{Struc-II-n_J-R-N1-PB-WBB}{GR-II-n_J-R-N1-PB-WBB}{fig:bn:GR-II-N1-Survive-Class1}
		\end{array}
	}{11\Delta^J}
	\]
\caption{Nested gauge remainder terms generated from diagrams~\ref{GR-II-Dumbell-n_J+r-R-rR}--\ref{GR-II-n_J+r-R-WGR} and~\ref{GR-II-WGRx2-n_J-R}.}
\label{fig:bn:GR-II-L2-Nest}
\end{figure}
\end{center}

Clearly, the result of processing the remaining diagrams of figure~\ref{fig:bn:GR-II} is just to
create versions of figures~\ref{fig:bn:GR-II-L1} and~\ref{fig:bn:GR-II-L2-Nest} with greater number
of vertices. In the new versions of the latter figure
we will, of course,  also have a nested version of diagram~\ref{GR-II-Dumbell-n_J+r-R-rR}.

Our task now is to understand what happens when we allow the gauge remainders
in these diagrams to act, recalling that we choose to
hold back with diagrams~\ref{GR-II-n_J-R-GR-W-DGR} and~\ref{GR-II-WGRx2-Top}.
To do this, it is useful to split the diagrams into
different classes, listed below. Each class constitutes a minimal set of diagrams, which
we give explicitly, together with copies of the diagrams of this minimal set 
supplemented by additional
vertices.

\begin{enumerate}
\item	Diagrams possessing a single, unprocessed gauge remainder, which is active.
		The minimal set of such term constitutes
		diagrams~\ref{GR-II-WGR-N2-PB}, \ref{GR-II-WGR-N2-PF},
		\ref{GR-II-n_J-R-N1-PF-WBB} and~\ref{GR-II-n_J-R-N1-PB-WBB};

\item	Diagrams possessing two unprocessed gauge remainders, one of
		which is active and the other of which is inactive. The minimal
		set of such terms constitutes diagrams~\ref{GR-II-WGRx2-Trap-GR} 
		and~\ref{GR-II-n_J-R-RWBB-Trap-GR};

\item	Diagrams possessing two unprocessed, active gauge remainders. The
		minimal set such terms constitutes diagrams~\ref{GR-II-WGRx2-Out-GR},
		\ref{GR-II-n_J-R-N1}--\ref{GR-II-n_J-R-N1-WGR} and the nested 
		version of diagram~\ref{GR-II-Dumbell-n_J+r-R-rR}.
\end{enumerate}

Allowing the gauge remainders to act, we will find that all terms in which
the gauge remainder strikes a socket will cancel by cancellation
mechanisms~\ref{CM:GR-Socket} and~\ref{CM:GR-Socket-Supp}, unless the socket
decorates a two-point, tree level vertex. Furthermore, these cancellations
occur within the above classes.

For the diagrams of the final class---in each of which there are two active gauge
remainders, one of which is nested---we might suspect that, once more we should
be diplomatic and take contributions in which each of the gauge remainders
act. However, it is unnecessary to do this.

We can understand the need for our original diplomacy as being down to the indistinguishability
of the two gauge remainders structures \ie\ each of these structures constitutes
an un-nested, single gauge remainder. In the current case, however, the two structures
are distinguishable, since one is nested whilst one is not. If we consistently allow just one of
these structures to act, then cancellation mechanisms~\ref{CM:GR-Socket} and~\ref{CM:GR-Socket-Supp}
will guarantee that
all contributions in which the gauge remainder
bites a socket (unless the socket is on a two-point, tree level vertex), including the case where the
socket is filled by the wine, cancel among themselves. The question is with which gauge
remainder
structure should we act? Of course, either choice is equivalent though we can hope that
a judicious choice will simplify the rest of the calculation.

We choose to allow the nested structure to act. As we will see, this does generate a particularly
appealing set of diagrams. Moreover, had we allowed the un-nested gauge remainder to act then, among
the diagrams of the next level of nesting, we would have found terms possessing two identical (nested)
gauge remainder structures. In this case, we would once again have had to turn to diplomacy;
by always choosing to process the nested contribution, we avoid ever having to do this again.

We now describe the terms which survive the action of the gauge remainders, up to 
those diagram with an $\Op{2}$ stub which is generated by the action of a gauge remainder. 
Let us begin
with those coming from the first class, above, which are shown 
in figure~\ref{fig:bn:GR-II-N1-Survive-Class1}.

\begin{center}
\begin{figure}
	\[
	\begin{array}{c}
	\vspace{0.2in}
		\displaystyle
		2 \sum_{J=0}^n \norm_{J,0}
		\dec{
			\LGD[0.1]{}{Struc-II-n_J-R-N2-PF-PF-WBB}{GR-II-n_J-R-N2-PF-PF-WBB}
			-\LGD[0.1]{}{Struc-II-n_J-R-N2-PF-PB-WBB}{GR-II-n_J-R-N2-PF-PB-WBB}
			-\LGD[0.1]{}{Struc-II-n_J-R-N2-PB-PF-WBB}{GR-II-n_J-R-N2-PB-PF-WBB}
			+\LGD[0.1]{}{Struc-II-n_J-R-N2-PB-PB-WBB}{GR-II-n_J-R-N2-PB-PB-WBB}
		}{11\Delta^J}
	\\
	\vspace{0.1in}
		\displaystyle
		-2 \norm_{n-1,0}
	\\
		\times
		\dec{
			\begin{array}{c}
			\vspace{0.2in}
				\LGD[0.1]{}{Struc-WGR-N2-PF-PF}{GR-WGR-N2-PF-PF}
				-\LGD[0.1]{}{Struc-WGR-N2-PF-PB}{GR-WGR-N2-PF-PB}
				-\LGD[0.1]{}{Struc-WGR-N2-PB-PF}{GR-WGR-N2-PB-PF}
				+\LGD[0.1]{}{Struc-WGR-N2-PB-PB}{GR-WGR-N2-PB-PB}
			\\
			\vspace{0.2in}
				-\PGD[0.1]{}{Struc-II-N1-WBB-PF-PF-TLTP-Alg}{GR-II-N1-WBB-PF-PF-TLTP-Alg}{fig:bn:GR-II-N2-Wine/Combo-Redraw}
				+\PGD[0.1]{}{Struc-II-N1-WBB-PF-PB-TLTP-Alg}{GR-II-N1-WBB-PF-PB-TLTP-Alg}{fig:bn:GR-II-N2-Wine/Combo-Redraw}
				+\PGD[0.1]{}{Struc-II-N1-WBB-PB-PF-TLTP-Alg}{GR-II-N1-WBB-PB-PF-TLTP-Alg}{fig:bn:GR-II-N2-Wine/Combo-Redraw}
				-\PGD[0.1]{}{Struc-II-N1-WBB-PB-PB-TLTP-Alg}{GR-II-N1-WBB-PB-PB-TLTP-Alg}{fig:bn:GR-II-N2-Wine/Combo-Redraw}
			\\
			\vspace{0.2in}
				+\PGD[0.1]{}{Struc-WGR-N2-B-PB-PB}{GR-WGR-N2-B-PB-PB}{fig:bn:GR-II-N2-Wine/Combo}
				-\PGD[0.1]{}{Struc-WGR-N2-B-PB-PF}{GR-WGR-N2-B-PB-PF}{fig:bn:GR-II-N2-Wine/Combo}
				+\LGD[0.1]{}{Struc-WGR-N2-T-PB-PB}{GR-WGR-N2-T-PB-PB}
				-\LGD[0.1]{}{Struc-WGR-N2-T-PB-PF}{GR-WGR-N2-T-PB-PF}
			\\
				-\PGD[0.1]{}{Struc-WGR-N2-B-PF-PB}{GR-WGR-N2-B-PF-PB}{fig:bn:GR-II-N2-Wine/Combo}
				+\PGD[0.1]{}{Struc-WGR-N2-B-PF-PF}{GR-WGR-N2-B-PF-PF}{fig:bn:GR-II-N2-Wine/Combo}
				-\LGD[0.1]{}{Struc-WGR-N2-T-PF-PB}{GR-WGR-N2-T-PF-PB}
				+\LGD[0.1]{}{Struc-WGR-N2-T-PF-PF}{GR-WGR-N2-T-PF-PF}
			\end{array}
		}{11\Delta^{n-1}}
	\end{array}
	\]
\caption{The minimal set of
terms arising from processing diagrams~\ref{GR-II-WGR-N2-PB}, \ref{GR-II-WGR-N2-PF},
\ref{GR-II-n_J-R-N1-PF-WBB} and~\ref{GR-II-n_J-R-N1-PB-WBB}
and their analogues with additional vertices,
up to terms with an $\Op{2}$ stub.}
\label{fig:bn:GR-II-N1-Survive-Class1}
\end{figure}
\end{center}

Despite the apparent slew of terms, what we have is actually very simple.
Diagrams~\ref{GR-II-n_J-R-N2-PF-PF-WBB}--\ref{GR-WGR-N2-PB-PB}
are just nested versions of the parent diagrams; 
diagrams~\ref{GR-II-N1-WBB-PF-PF-TLTP-Alg}--\ref{GR-II-N1-WBB-PB-PB-TLTP-Alg} 
are further cases of the by-now-familiar diagrams in which a two-point,
tree level vertex is joined to a bitten gauge remainder
and  
diagrams~\ref{GR-WGR-N2-B-PB-PB}--\ref{GR-WGR-N2-T-PF-PF} are just nested versions
of diagrams~\ref{GR-II-WGR-N1-Bottom-PB}--\ref{GR-II-WGR-N1-Top-PF}.

The sense in which we can consider certain diagrams of figure~\ref{fig:bn:GR-II-N1-Survive-Class1}
to be nestings of old diagrams is very precise. To see this let us consider first 
diagrams~\ref{GR-II-n_J-R-N2-PF-PF-WBB}-\ref{GR-II-n_J-R-N2-PB-PB-WBB} as nestings of 
diagrams~\ref{GR-II-n_J-R-N1-PF-WBB} and~\ref{GR-II-n_J-R-N1-PB-WBB}.
In this case the rule to take us from the old diagram to the new diagram is simple: we 
insert an additional gauge remainder between the active gauge remainder and 
the preceding processed gauge remainder, picking up a sign if the insertion
corresponds to a pull back. In the parent diagrams there is, however, a second
processed gauge remainder and we might well wonder whether this can become nested as well.
The answer is yes: we will see how this comes about shortly.

Similarly, all other diagrams which we have said can be thought of as nestings of
earlier diagrams can, at this stage of the calculation, only be thought of
a nestings with respect to a particular processed gauge remainder. With our experiences
of the calculation thus far, it is not surprising that we will find that this
restriction can be lifted, upon continuing with the diagrammatic procedure.

Next, we move on to consider the surviving terms coming from the second class of
diagrams, above, collecting together all such terms in
figure~\ref{fig:bn:GR-II-N1-Survive-Class2b}, up to diagrams with an $\Op{2}$ stub.
\begin{center}
\begin{figure}[h]
	\[
	\begin{array}{c}
	\vspace{0.2in}
	\displaystyle
		2 \sum_{J=0}^n \norm_{J,0}
		\dec{
			\LGD[0.1]{}{Struc-n_J-R-RWBB-Trap-GR-N1-PF}{GR-n_J-R-RWBB-Trap-GR-N1-PF}
			-\LGD[0.1]{}{Struc-n_J-R-RWBB-Trap-GR-N1-PB}{GR-n_J-R-RWBB-Trap-GR-N1-PB}
		}{11\Delta^J}
		+2\norm_{n-1,0}
		\dec{
			\LGD[0.1]{}{Struc-WGR-N2-PB-trap}{GR-II-WGR-N2-PB-trap}
			-\LGD[0.1]{}{Struc-WGR-N2-PF-trap}{GR-II-WGR-N2-PF-trap}	
		}{11\Delta^{n-1}}
	\\
	\vspace{0.1in}
		\displaystyle
		+2 \norm_{n-1,0}
	\\
		\times
		\dec{
			+\PGD[0.1]{}{Struc-WGR-N1-Bottom-PB-GR-Trap}{GR-WGR-N1-Bottom-PB-GR-Trap}{fig:bn:TLTP-GR-JoinedToWine}
			-\PGD[0.1]{}{Struc-WGR-N1-Bottom-PF-GR-Trap}{GR-WGR-N1-Bottom-PF-GR-Trap}{fig:bn:TLTP-GR-JoinedToWine}
			+\PGD[0.1]{}{Struc-WGR-N1-Top-PB-LHGR}{GR-WGR-N1-Top-PB-LHGR}{fig:bn:GR-II-Redraw}
			-\PGD[0.1]{}{Struc-WGR-N1-Top-PF-RHGR}{GR-WGR-N1-Top-PF-RHGR}{fig:bn:GR-II-Redraw}
		}{11\Delta^{n-1}}
	\end{array}
	\]
\caption{Surviving contributions from diagrams~\ref{GR-II-WGRx2-Trap-GR} 
and~\ref{GR-II-n_J-R-RWBB-Trap-GR}, modulo terms with an $\Op{2}$ stub.}
\label{fig:bn:GR-II-N1-Survive-Class2b}
\end{figure}
\end{center}

Once again, we have generated nested versions of the parent diagrams but, 
once again, we have generated only one of the possible types of nesting.
Referring back to the parents, diagrams~\ref{GR-II-WGRx2-Trap-GR} 
and~\ref{GR-II-n_J-R-RWBB-Trap-GR}, we note that each has one active
gauge remainder, one performed gauge remainder and one trapped
gauge remainder. Diagrams~\ref{GR-n_J-R-RWBB-Trap-GR-N1-PF}--\ref{GR-II-WGR-N2-PF-trap}
are nested with respect to the active gauge remainder. We will shortly generate
the nestings with respect to the performed gauge remainder; trapped gauge remainders will never
become nested since, by definition, they are unable to act.

The final four diagrams of figure~\ref{fig:bn:GR-II-N1-Survive-Class2b} will
prove to be of particular interest. Notice that diagram~\ref{GR-WGR-N1-Bottom-PF-GR-Trap}
(\ref{GR-WGR-N1-Bottom-PB-GR-Trap}) naturally combines with diagram~\ref{GR-II-WGR-N1-Bottom-PF}
(\ref{GR-II-WGR-N1-Bottom-PB}); we return to this at the end of the section.
Having combined terms in this manner, we will find in section~\ref{sec:D-ID-New}
that they can
be redrawn via diagrammatic identity~\ref{D-ID-Alg-2}. The
resultant diagram will then take the same from as diagram~\ref{GR-WGR-N1-Top-PB-LHGR}
(\ref{GR-WGR-N1-Top-PF-RHGR}) but with the gauge remainder on the other end
of the wine.\footnote{The two diagrams with a gauge remainders on
either end of the wine can actually be shown to be equivalent to each other, by a trivial
redrawing.} We will return to this observation in section~\ref{sec:D-ID-App}.

Finally, then, we can treat the surviving terms coming from the
third class of diagrams, above, for which we
recall that the minimal set comprises diagrams~\ref{GR-II-WGRx2-Out-GR},
\ref{GR-II-n_J-R-N1}--\ref{GR-II-n_J-R-N1-WGR} and the nested 
version of diagram~\ref{GR-II-Dumbell-n_J+r-R-rR}.
Remembering to perform,  undemocratically, the nested gauge remainder first,
we split the resultant terms into three sets. 

First, we have those diagrams with an $\Op{2}$ stub, which we comment on
in chapter~\ref{ch:Op2}. Secondly, we have diagrams with a further level
of nesting. As usual, we produce nested versions of the parent diagrams,
where we note that the nesting is with respect to the already nested gauge
remainder. There are, however, additional nested diagrams, as we should expect.
Returning to the original set of type-II gauge remainders in figure~\ref{fig:bn:GR-II},
we know that among the nested gauge remainders generated are diagrams with a different
structure to the parents: namely diagrams~\ref{GR-II-WGRx2-Trap-GR}--\ref{GR-II-WGR-N2-PF}
and~\ref{GR-II-n_J-R-RWBB-Trap-GR}--\ref{GR-II-n_J-R-N1-PB-WBB}.
Thus, we expect nested versions of these to be produced when we process nested versions
of the original gauge remainders.

Now we can start to see how everything meshes together. We have already seen that by processing
diagrams~\ref{GR-II-WGRx2-Trap-GR}--\ref{GR-II-WGR-N2-PF}
and~\ref{GR-II-n_J-R-RWBB-Trap-GR}--\ref{GR-II-n_J-R-N1-PB-WBB}, we generate their nested
counterparts, but that this nesting is with respect to the active gauge remainder. By processing
the diagrams of the third class, above, we now generate nested versions of 
diagrams~\ref{GR-II-WGRx2-Trap-GR}--\ref{GR-II-WGR-N2-PF}
and~\ref{GR-II-n_J-R-RWBB-Trap-GR}--\ref{GR-II-n_J-R-N1-PB-WBB}
where the nesting
is with respect to the gauge remainder which bites the wine.

The third set of terms---which constitutes all those diagrams 
not in the first or second sets---is shown in figure~\ref{fig:bn:GR-II-N1-Survive-Class3c}.
 As with that which was done previously,
if the first gauge remainder bites the opposite end of the wine to the one which the other gauge
remainder attaches, then we delay further manipulation.
We show the resultant diagrams in figure~\ref{fig:bn:GR-II-N1-Survive-Class3c}.
\begin{center}
\begin{figure}
	\[
	\begin{array}{c}
	\vspace{0.1in}
		\displaystyle
		-2\sum_{J=0}^n \norm_{J,0}
	\\
		\times
			\dec{
				\PGD[0.1]{}{Struc-II-n_J-R-PF-PF-TLTP-EP-GR}{GR-II-n_J-R-PF-PF-TLTP-EP-GR}{fig:bn:DGRx2-GR-II/Combo}
				-\PGD[0.1]{}{Struc-II-n_J-R-PF-PB-TLTP-EP-GR}{GR-II-n_J-R-PF-PB-TLTP-EP-GR}{fig:bn:DGRx2-GR-II/Combo}
				+\PGD[0.1]{}{Struc-n_J-R-GR-W-RR}{GR-II-n_J-R-GR-W-RR}{fig:bn:DGRx2-GR-II/Combo}
				-\PGD[0.1]{}{Struc-n_J-R-GR-W-LR}{GR-II-n_J-R-GR-W-LR}{fig:bn:DGRx2-GR-II/Combo}
			}{11\Delta^{J}}
	\\
	\vspace{0.1in}	
		\displaystyle
		-2 \norm_{n-1,0}
	\\
		\times
		\dec{
			\begin{array}{c}
				-\PGD[0.1]{}{Struc-WGR-RR}{GR-II-WGR-RR}{fig:bn:DGRx2-GR-II/Combo}
				+\PGD[0.1]{}{Struc-WGR-LR}{GR-II-WGR-LR}{fig:bn:DGRx2-GR-II/Combo}
			\\
				+\PGD[0.1]{}{Struc-WGR-N2-B-PB-PB2}{GR-WGR-N2-B-PB-PB2}{fig:bn:GR-II-N2-Wine/Combo}
				-\PGD[0.1]{}{Struc-WGR-N2-B-PB-PF2}{GR-WGR-N2-B-PB-PF2}{fig:bn:GR-II-N2-Wine/Combo}
				+\LGD[0.1]{}{Struc-WGR-N2-T-PB-PB2}{GR-WGR-N2-T-PB-PB2}
				-\LGD[0.1]{}{Struc-WGR-N2-T-PB-PF2}{GR-WGR-N2-T-PB-PF2}
			\\
				-\PGD[0.1]{}{Struc-WGR-N2-B-PF-PB2}{GR-WGR-N2-B-PF-PB2}{fig:bn:GR-II-N2-Wine/Combo}
				+\PGD[0.1]{}{Struc-WGR-N2-B-PF-PF2}{GR-WGR-N2-B-PF-PF2}{fig:bn:GR-II-N2-Wine/Combo}
				-\LGD[0.1]{}{Struc-WGR-N2-T-PF-PB2}{GR-WGR-N2-T-PF-PB2}
				+\LGD[0.1]{}{Struc-WGR-N2-T-PF-PF2}{GR-WGR-N2-T-PF-PF2}
			\\
				+\CGD[0.1]{}{Struc-WGRx2-Trap-Alg-A}{GR-II-WGRx2-Trap-Alg-A}{DGRx2-WGR-N1-Top-PB-GRx2}
				-\CGD[0.1]{}{Struc-WGRx2-Trap-Alg-B}{GR-II-WGRx2-Trap-Alg-B}{DGRx2-WGR-N1-Top-PF-GRx2}
				-\PGD[0.1]{}{Struc-WGRx2-Active-Alg-A}{Struc-WGRx2-Active-Alg-A}{fig:bn:DGRx2-GR-II/Combo}
				+\PGD[0.1]{}{Struc-WGRx2-Active-Alg-B}{Struc-WGRx2-Active-Alg-B}{fig:bn:DGRx2-GR-II/Combo}
			\\
				-\PGD[0.1]{}{Struc-WGRx2-B-N2-Alg-A}{Struc-WGRx2-B-N2-Alg-A}{fig:bn:TLTP-GR-JoinedToWine}
				+\PGD[0.1]{}{Struc-WGRx2-B-N2-Alg-B}{Struc-WGRx2-B-N2-Alg-B}{fig:bn:TLTP-GR-JoinedToWine}
				+\PGD[0.1]{}{Struc-WGRx2-B-N2-Alg-C}{Struc-WGRx2-B-N2-Alg-C}{fig:bn:TLTP-GR-JoinedToWine}
				-\PGD[0.1]{}{Struc-WGRx2-B-N2-Alg-D}{Struc-WGRx2-B-N2-Alg-D}{fig:bn:TLTP-GR-JoinedToWine}
			\end{array}
		}{11\Delta^{n-1}}
	\end{array}
		\]
\caption{Surviving terms from diagrams~\ref{GR-II-WGRx2-Out-GR},
\ref{GR-II-n_J-R-N1}--\ref{GR-II-n_J-R-N1-WGR} and the nested 
version of diagram~\ref{GR-II-Dumbell-n_J+r-R-rR}, up to additionally nested diagrams
and terms with an $\Op{2}$ stub.}
\label{fig:bn:GR-II-N1-Survive-Class3c}
\end{figure}
\end{center}

There are a number of comments worth making.
First, note how diagrams~\ref{GR-II-n_J-R-GR-W-RR}--\ref{GR-II-WGR-LR}
are nested versions of diagrams~\ref{GR-II-n_J-R-GR-W-DGR} and~\ref{GR-II-WGRx2-Top},
respectively, where the nesting is with respect to the performed gauge
remainder. These diagrams, like diagrams~\ref{GR-II-n_J-R-PF-PF-TLTP-EP-GR}, \ref{GR-II-n_J-R-PF-PB-TLTP-EP-GR},
\ref{Struc-WGRx2-Active-Alg-A} and~\ref{Struc-WGRx2-Active-Alg-B},
still possess an active gauge remainder, which we choose not to
process, for the time being. 

Secondly, diagrams~\ref{GR-WGR-N2-B-PB-PB2}--\ref{GR-WGR-N2-T-PF-PF2}
naturally combine with diagrams~\ref{GR-WGR-N2-B-PB-PB}--\ref{GR-WGR-N2-T-PF-PF}.
Together, we can think of these as nested versions of 
diagrams~\ref{GR-II-WGR-N1-Bottom-PB}--\ref{GR-II-WGR-N1-Top-PF} where now we
perform the nesting in all possible ways.  Note that these diagrams---or, at any
rate, their analogues possessing additional vertices---are precisely the type of
diagrams we alluded to at the end of section~\ref{sec:GR-Ib}. Taking the undecorated component
of the wines, we see that we produce the diagrams necessary to exactly cancel a special case of
diagram~\ref{GR-ring-n_J-R-Join}\footnote{Recalling that, in diagram~\ref{GR-ring-n_J-R-Join}
we must pick up a minus sign for each pull back and that we sum over all realisations of
the un-drawn gauge remainders.} \ie\ the case where $m=3$.

We are now in a position to deduce the diagrams that remain once
we have iterated the diagrammatic procedure, until exhaustion, which
we list below.
\begin{enumerate}
\item	terms with an $\Op{2}$ stub, which has
		been formed by the action of a gauge remainder;

\item	Diagrams possessing a wine which has been bitten by
		two gauge remainders. To generate the terms this set comprises,
		we start with just
		diagrams~\ref{GR-II-WGR-N1-Bottom-PB}--\ref{GR-II-WGR-N1-Top-PF}
		and~\ref{GR-WGR-N1-Bottom-PB-GR-Trap}--\ref{GR-WGR-N1-Top-PF-RHGR}.
		For each of these terms, we now include nested versions where the nesting
		is respect to either of the performed gauge remainders. Note, of course,
		that we have already explicitly encountered some of these diagrams.
		For every diagram now in the set, we include copies possessing
		additional vertices.

\item	Diagrams possessing a wine bitten by a single gauge remainder at one
		end and having an active gauge remainder at the other. To generate the terms this set comprises,
		we start with just diagrams~\ref{GR-II-n_J-R-GR-W-DGR} and~\ref{GR-II-WGRx2-Top}. For both
		of these terms, we  now include nested versions where the nesting
		is respect to either of the performed gauge remainder and etcetera;

\item	Diagrams possessing a two-point, tree level
		vertex joined to a bitten gauge remainder. The examples of this
		we have seen so far are: diagrams~\ref{GR-II-N1-WBB-PF-PF-TLTP-Alg}--\ref{GR-II-N1-WBB-PB-PB-TLTP-Alg},
		\ref{GR-II-n_J-R-PF-PF-TLTP-EP-GR}, \ref{GR-II-n_J-R-PF-PB-TLTP-EP-GR} and
		\ref{GR-II-WGRx2-Trap-Alg-A}--\ref{Struc-WGRx2-B-N2-Alg-D}.
\end{enumerate}

Whereas we have stated how to generate the complete set of diagrams for the second
and third sets above, we have not done so with the fourth set. The reason is that,
whilst we certainly expect the fourth set of terms to contain nestings of the diagrams
already listed, there are additional terms that we have not yet encountered.

The diagrams we are considering occur whenever a diagram possesses the following:
a two-point, tree level vertex, formed by the action of a gauge remainder, and a nested gauge remainder,
which is not attached to anything.
This tells us that, indeed, there is a diagram of the form we are interested in that
we are yet to encounter. To see this, return to the diagrams of figure~\ref{fig:bn:GR-II-L2-Nest}.
With the exception of diagram~\ref{GR-II-n_J-R-RWBB-Trap-GR} all these
diagrams possess a nested gauge remainder, which is not attached to anything. Hence, after processing
the gauge remainder(s) of these diagrams, we will generate two-point, tree level vertices,
which can give us diagrams of the type we are currently studying. However, in the case of 
diagram~\ref{GR-II-n_J-R-RWBB-Trap-GR} we must wait until the next level of nesting until
we are able to generate a two-point, tree level vertex which is joined to a bitten
gauge remainder. The additionally nested versions of this diagram have actually
been generated---see diagrams~\ref{GR-n_J-R-RWBB-Trap-GR-N1-PF} and~\ref{GR-n_J-R-RWBB-Trap-GR-N1-PB}---but
are yet to be processed. When we do process them, among the terms generated will
be those in figure~\ref{fig:bn:GR-II-N2-Algebra-Supp}.
\begin{center}
\begin{figure}[h]
	\[
	2 \norm_{n-1,0}
	\dec{
		\LGD[0.1]{}{Struc-n_J-R-RWBB-Trap-GR-PF-PF-TLTP-EPGR}{GR-n_J-R-RWBB-Trap-GR-PF-PF-TLTP-EPGR}
		-\LGD[0.1]{}{Struc-n_J-R-RWBB-Trap-GR-PF-PB-TLTP-EPGR}{GR-n_J-R-RWBB-Trap-GR-PF-PB-TLTP-EPGR}
		-\LGD[0.1]{}{Struc-n_J-R-RWBB-Trap-GR-PB-PF-TLTP-EPGR}{GR-n_J-R-RWBB-Trap-GR-PB-PF-TLTP-EPGR}
		+\LGD[0.1]{}{Struc-n_J-R-RWBB-Trap-GR-PB-PB-TLTP-EPGR}{GR-n_J-R-RWBB-Trap-GR-PB-PB-TLTP-EPGR}
	}{11\Delta^{n-1}}
	\]
\caption{Additional diagrams comprising a two-point, tree level vertex joined to
a bitten gauge remainder. They are 
spawned by diagrams~\ref{GR-n_J-R-RWBB-Trap-GR-N1-PF} and~\ref{GR-n_J-R-RWBB-Trap-GR-N1-PB}.}
\label{fig:bn:GR-II-N2-Algebra-Supp}
\end{figure}
\end{center}

We are not quite done. Let us return to 
diagrams~\ref{GR-n_J-R-RWBB-Trap-GR-N1-PF} and~\ref{GR-n_J-R-RWBB-Trap-GR-N1-PB}, the
parent diagrams
of the terms in the first row of figure~\ref{fig:bn:GR-II-N2-Algebra-Supp}. If we iterate the diagrammatic
procedure then we will just generate nested versions of 
diagrams~\ref{GR-n_J-R-RWBB-Trap-GR-PF-PF-TLTP-EPGR}--\ref{GR-n_J-R-RWBB-Trap-GR-PB-PB-TLTP-EPGR},
where this nesting is with respect to the the string of gauge remainders, the first of which
is the gauge remainder contracted into the trapped gauge remainder.
The point is that we also expect nestings with respect to the gauge remainder that
\emph{traps} the full gauge remainder. So long as, in such diagrams, the two-point tree level
vertex still attaches to the same string of gauge remainders as in figure~\ref{fig:bn:GR-II-N2-Algebra-Supp},
 then these new terms can
just be derived from diagrams~\ref{GR-n_J-R-RWBB-Trap-GR-PF-PF-TLTP-EPGR}--\ref{GR-n_J-R-RWBB-Trap-GR-PB-PB-TLTP-EPGR}
by demanding that we sum over all possible nestings. The new diagrams occur if the two-point,
tree level vertex attaches to the newly nested gauge remainder. These 
diagrams---shown in figure~\ref{fig:bn:GR-II-N2-Algebra-Supp2}--- can have their 
lineage traced back to diagram~\ref{GR-II-n_J-R-N1}.
\begin{center}
\begin{figure}[h]
	\[
	2 \norm_{n-1,0}
	\dec{
		\LGD[0.1]{}{Struc-n_J-R-RWBB-PF-Trap-GR-PF-TLTP-EPGR}{GR-n_J-R-RWBB-PF-Trap-GR-PF-TLTP-EPGR}
		-\LGD[0.1]{}{Struc-n_J-R-RWBB-PF-Trap-GR-PB-TLTP-EPGR}{GR-n_J-R-RWBB-PF-Trap-GR-PB-TLTP-EPGR}
		-\LGD[0.1]{}{Struc-n_J-R-RWBB-PB-Trap-GR-PF-TLTP-EPGR}{GR-n_J-R-RWBB-PB-Trap-GR-PF-TLTP-EPGR}
		+\LGD[0.1]{}{Struc-n_J-R-RWBB-PB-Trap-GR-PB-TLTP-EPGR}{GR-n_J-R-RWBB-PB-Trap-GR-PB-TLTP-EPGR}
	}{11\Delta^{n-1}}
	\]
\caption{Additional diagrams comprising a two-point, tree level vertex joined to
a bitten gauge remainder, whose lineage can be traced back to diagram~\ref{GR-II-n_J-R-N1}.}
\label{fig:bn:GR-II-N2-Algebra-Supp2}
\end{figure}
\end{center}

Now we have all the diagrams we need from which we generate the complete set of
diagrams comprising a two-point, tree level vertex attached to a bitten gauge
remainder.
Starting from diagrams~\ref{GR-II-N1-WBB-PF-PF-TLTP-Alg}--\ref{GR-II-N1-WBB-PB-PB-TLTP-Alg},
\ref{GR-II-n_J-R-PF-PF-TLTP-EP-GR}, \ref{GR-II-n_J-R-PF-PB-TLTP-EP-GR},
\ref{GR-II-WGRx2-Trap-Alg-A}--\ref{Struc-WGRx2-B-N2-Alg-D} 
and~\ref{GR-n_J-R-RWBB-Trap-GR-PF-PF-TLTP-EPGR}--\ref{GR-n_J-R-RWBB-PB-Trap-GR-PB-TLTP-EPGR}
we now include all possible nestings of each of these terms. 
Finally, for every diagram now in the set, we include copies possessing
additional vertices.

We might worry that the considerations that led us
to the diagrams of figures~\ref{fig:bn:GR-II-N2-Algebra-Supp} and~\ref{fig:bn:GR-II-N2-Algebra-Supp2}
imply that we are missing contributions that could arise from \eg\ diagram~\ref{GR-II-n_J-R-WGR-N1}.
If we allow the nested gauge remainder to act first and it bites an end of the wine,
then we know we must allow the un-nested gauge remainder to act. 
If this second gauge remainder were nested, then when we generate a two-point tree level
vertex, we would have a new option of where to attach it. Since we have not explicitly
gone to this level of nesting in the calculation, we might worry that we are missing such terms.
They are, however, taken care of by diagrams~\ref{GR-II-n_J-R-N1-PF-WBB} and~\ref{GR-II-n_J-R-N1-PB-WBB}.
Upon processing these terms we generate, among others, diagrams~\ref{GR-II-N1-WBB-PF-PF-TLTP-Alg}--\ref{GR-II-N1-WBB-PB-PB-TLTP-Alg}.
Provided that we consider all possible nestings of these diagrams, then we will generate the
terms we were worrying about.

At this stage, it perhaps seems that the set of diagrams comprising a two-point,
tree level vertex joined to a bitten gauge remainder form a disparate
set of terms. There is, however, a deep and illuminating relationship
between them which we now begin to explore.

First, we look at diagrams~\ref{GR-II-WGR-RR}, \ref{GR-II-WGR-LR}
\ref{Struc-WGRx2-Active-Alg-A} and~\ref{Struc-WGRx2-Active-Alg-B}.
Suppose that we were to allow the active gauge remainder in
these diagrams to act: then the diagrams in which the gauge remainder
strikes a socket decorating anything other than a two-point,
tree level vertex would cancel, courtesy of cancellation mechanism~\ref{CM:GR-Socket}.
We return to this in chapter~\ref{ch:bn:Iterate}.

Secondly, we look at diagrams~\ref{GR-II-N1-WBB-PF-PF-TLTP-Alg}--\ref{GR-II-N1-WBB-PB-PB-TLTP-Alg}
and diagrams~\ref{Struc-WGRx2-B-N2-Alg-A}--\ref{Struc-WGRx2-B-N2-Alg-D}. To this set of
eight terms, we now add diagrams~\ref{GR-II-WGR-N1-Bottom-PB}, \ref{GR-II-WGR-N1-Bottom-PF},
\ref{GR-WGR-N1-Bottom-PB-GR-Trap} and~\ref{GR-WGR-N1-Bottom-PF-GR-Trap}. We have already recognised
that these latter four diagrams can be combined, in pairs. If we do this,
then we generate a two-point, tree level vertex joined to a bitten gauge
remainder \ie\ we generate two terms of the same type as the other eight!
We collect together all ten diagrams in figure~\ref{fig:bn:TLTP-GR-JoinedToWine}.
\begin{center}
\begin{figure}
	\[
	\begin{array}{c}
	\vspace{0.1in}
		\displaystyle
		-2 \norm_{n-1,0}
	\\
		\times
		\dec{
			\begin{array}{c}
			\vspace{0.2in}
				-\PGD[0.1]{}{Struc-W-RXR}{GR-II-W-RXR}{fig:bn:GR-II-Redraw}
				+\PGD[0.1]{}{Struc-W-LXR}{GR-II-W-LXR}{fig:bn:GR-II-Redraw}	
			\\
			\vspace{0.2in}
				+\PGD[0.1]{}{Struc-WBB-L-PFX}{GR-Ib-WBB-L-PFX}{fig:bn:GR-II-N2-Wine/Combo-Redraw}
				-\PGD[0.1]{}{Struc-WBB-L-PBX}{GR-Ib-WBB-L-PBX}{fig:bn:GR-II-N2-Wine/Combo-Redraw}
				-\PGD[0.1]{}{Struc-WBB-R-PFX}{GR-Ib-WBB-R-PFX}{fig:bn:GR-II-N2-Wine/Combo-Redraw}
				-\PGD[0.1]{}{Struc-WBB-R-PBX}{GR-Ib-WBB-R-PBX}{fig:bn:GR-II-N2-Wine/Combo-Redraw}
			\\
			\vspace{0.2in}
				-\PGD[0.1]{}{Struc-WGRx2-B-N2-Alg-A}{Struc-WGRx2-B-N2-Alg-A2}{fig:bn:GR-II-N2-Wine/Combo-Redraw}
				+\PGD[0.1]{}{Struc-WGRx2-B-N2-Alg-B}{Struc-WGRx2-B-N2-Alg-B2}{fig:bn:GR-II-N2-Wine/Combo-Redraw}
				+\PGD[0.1]{}{Struc-WGRx2-B-N2-Alg-C}{Struc-WGRx2-B-N2-Alg-C2}{fig:bn:GR-II-N2-Wine/Combo-Redraw}
				-\PGD[0.1]{}{Struc-WGRx2-B-N2-Alg-D}{Struc-WGRx2-B-N2-Alg-D2}{fig:bn:GR-II-N2-Wine/Combo-Redraw}
			\end{array}
		}{11\Delta^{n-1}}
	\end{array}
	\]
\caption{Collecting together diagrams~\ref{GR-II-WGR-N1-Bottom-PB}, \ref{GR-II-WGR-N1-Bottom-PF},
\ref{GR-WGR-N1-Bottom-PB-GR-Trap} and~\ref{GR-WGR-N1-Bottom-PF-GR-Trap}, which have 
been combined in pairs, and diagrams~\ref{GR-II-N1-WBB-PF-PF-TLTP-Alg}--\ref{GR-II-N1-WBB-PB-PB-TLTP-Alg},
and~\ref{Struc-WGRx2-B-N2-Alg-A}--\ref{Struc-WGRx2-B-N2-Alg-D}.}
\label{fig:bn:TLTP-GR-JoinedToWine}
\end{figure}
\end{center}

Note that, in the second row of diagrams, we have used \CC\
to replace a bite on one side of the wine with a bite on the
other side.

The most striking feature of figure~\ref{fig:bn:TLTP-GR-JoinedToWine} is its similarity
to figure~\ref{fig:bn:GR-Extra}. Indeed, there is a one-to-one correspondence between the
diagrams of these to figures: to generate the diagrams of the latter figure from
those of the former, we simply remove the restriction that the wines be un-decorated and
multiply by $-1/2$.

To understand the interrelationship between the diagrams of figure~\ref{fig:bn:TLTP-GR-JoinedToWine},
let us start by taking diagrams~\ref{GR-II-W-RXR} and~\ref{GR-II-W-LXR} and 
perform the following operations: we
strip
off the effective propagator and detach the end of the wine attached to the vertex.
This gives us the diagrams of figure~\ref{fig:bn:TLTP-GR-SplitFromWine}.
\begin{center}
\begin{figure}[h]
	\[
	-2 \norm_{n,0}
	\dec{
		-\LID[0.1]{}{Struc-W-RXR-detach}{GR-II-W-RXR-detach}
		+\LID[0.1]{}{Struc-W-LXR-detach}{GR-II-W-LXR-detach}
	}{11\Delta^{n}}
	\]
\caption{Resulting of promoting the effective propagator of diagrams~\ref{GR-II-W-RXR} and~\ref{GR-II-W-LXR}
to a decoration and detaching the wine from the vertex.}
\label{fig:bn:TLTP-GR-SplitFromWine}
\end{figure}
\end{center}

We now consider adding an additional level of nesting to diagrams~\ref{GR-II-W-RXR-detach}
and~\ref{GR-II-W-LXR-detach}. Then we reattach the wine---but let it join to
any available socket. If it attaches to the vertex, then we just generate nested versions
of diagrams~\ref{GR-II-W-RXR} and~\ref{GR-II-W-LXR}. If, instead, it attaches to one
of the nested gauge remainders then, upon joining the other gauge remainder to the
un-drawn socket of the vertex, we simply generate diagrams~\ref{GR-Ib-WBB-L-PFX}--\ref{Struc-WGRx2-B-N2-Alg-D2}.
This deep relationship between the diagrams of figure~\ref{fig:bn:TLTP-GR-JoinedToWine}
will be exploited in section~\ref{sec:D-ID-App} and chapter~\ref{ch:bn:Iterate}.

To conclude this section, we comment on the final set
of diagrams comprising a two-point,
tree level vertex joined to a bitten gauge remainder.
This set constitutes diagrams~\ref{GR-II-WGRx2-Trap-Alg-A}, \ref{GR-II-WGRx2-Trap-Alg-B}
and~\ref{GR-n_J-R-RWBB-Trap-GR-PF-PF-TLTP-EPGR}--\ref{GR-n_J-R-RWBB-PB-Trap-GR-PB-TLTP-EPGR}.
We can arrange these diagrams as we arranged those in figure~\ref{fig:bn:TLTP-GR-JoinedToWine},
providing we make three minor changes: first, change the overall sign; secondly, include
a trapped gauge remainder; thirdly reduce the wine.

\newpage
\section{Type-III Gauge Remainders} \label{sec:GR-III}

All gauge remainders of type-III have been collected together in
figure~\ref{fig:bn:GR-III}.
\begin{center}
\begin{figure}
	\[
	\begin{array}{c}
	\vspace{0.2in}
		\displaystyle
		-2 \sum_{s=1}^{n_+} \sum_{J=2}^{2n}  \norm_{J+s,J}
		\dec{
			\sco[0.1]{
				\LGD[0.1]{}{Struc-III-V01-R-TLTP}{GR-III-V01-R-TLTP}
				+\frac{1}{J+1}
				\left[
					\sco[0.1]{
						\LGD[0.1]{}{Struc-TLTP-WGR}{GR-III-TLTP-WGR-Arb}
					}{\ensuremath{\begin{array}{c}\input{pstex/Vertex-V01-R.pstex_t} \end{array}}}
				\right]
			}{\left[ \ds \VertexSum{1}{J-1} \ensuremath{\begin{array}{c}\input{pstex/GV-I-I+.pstex_t} \end{array}}\right]  \ensuremath{\begin{array}{c}\input{pstex/GV-VJ-R.pstex_t} \end{array}} }
		}{1\Delta^{J+s}}
	\\
	\vspace{0.2in}
		\displaystyle
		+2 \sum_{J=0}^n \sum_{r=0}^{n_J} \norm_{J+2,0}
		\dec{
			\sco[0.1]{
				\LGD[0.1]{}{Struc-III-n_J+r-R-TLTP}{GR-III-n_J+r-R-TLTP}
				+ \frac{1}{2}
				\left[
					\sco[0.1]{
						\LGD[0.1]{}{Struc-TLTP-WGR}{GR-III-TLTP-WGR-n_J+r-R}
					}{\ensuremath{\begin{array}{c}\input{pstex/Vertex-n_J+r-R.pstex_t} \end{array}}}
				\right]
			}{\ensuremath{\begin{array}{c}\input{pstex/Vertex-r-R.pstex_t} \end{array}}}
		}{1\Delta^{J+2}}
	\\
		\displaystyle
		-2 \sum_{J=0}^n \norm_{J+1,0} 
		\dec{
			\LGD[0.1]{}{Struc-III-n_J-R-TLTP}{GR-III-n_J-R-TLTP}
			\left[
			-\sco[0.1]{
				\LGD[0.1]{}{Struc-TLTP-WGR}{GR-III-TLTP-WGR-n_J-R}
			}{\ensuremath{\begin{array}{c}\input{pstex/Vertex-n_J-R.pstex_t} \end{array}}}
		\right]
		}{1\Delta^{J+1}}
		\displaystyle
		+2 \norm_{n,0}
		\dec{
			\LGD[0.1]{}{Struc-TLTP-WGR}{GR-III-TLTP-WGR-B}
		}{1\Delta^n}
	\end{array}
	\]
\caption{Type-III gauge remainders.}
\label{fig:bn:GR-III}
\end{figure}
\end{center}

Given what we have done already in this chapter, we can immediately deduce
the diagrams that will remain, after we allow the gauge remainders to act.
Diagrams in which the gauge remainder strikes a socket which
decorates anything other than a two-point, tree level vertex
will be cancelled, via cancellation mechanism~\ref{CM:GR-Socket}. The surviving
diagrams will either be nested versions of the parents or diagrams in which
the gauge remainder strikes an end of the wine. Note that when we create a
two-point, tree level vertex, we cannot decorate it with the remaining external
field, since then the diagram as a whole will vanish at $\Op{2}$.

Iterating the diagrammatic procedure, we will ultimately be left with only two
types of diagram:
\begin{enumerate}
	\item	diagrams in which an arbitrarily nested gauge remainder bites
			either end of the wine;
	
	\item	diagrams in which a two-point, tree level  vertex (formed by
			the action of a gauge remainder) is joined to a bitten gauge
			remainder.
\end{enumerate}

We give the minimal set of these terms in figure~\ref{fig:bn:GR-III-Terminate}.
We recall that, to obtain the complete set we do the following. For each diagram,
include versions in which the level of nesting is arbitrarily higher and nest in
all possible ways. Finally, for every diagram now in the set, we include copies possessing
additional vertices.
\begin{center}
\begin{figure}
	\[
	\begin{array}{c}
	\vspace{0.1in}
		\displaystyle
		4 \norm_{n,0}
	\\
		\dec{
			\LGD[0.1]{}{Struc-III-WBB}{GR-III-WBB}
			+\CGD[0.1]{}{Struc-TLTP-E-W-R}{GR-III-WBT}{DGR-TLTP-WGR-T-E}
			-\CGD[0.1]{}{Struc-III-N1-PF-TLTP-EPGR}{GR-III-N1-PF-TLTP-EPGR}{DGRx2-TLTP-E-RW-GR-LR}
			+\CGD[0.1]{}{Struc-III-N1-PB-TLTP-EPGR}{GR-III-N1-PB-TLTP-EPGR}{DGRx2-TLTP-E-RW-GR-RR}
		}{1\Delta^n}
	\end{array}
	\]
\caption{The minimal set of terms produced by applying the diagrammatic procedure,
until exhaustion, to diagrams~\ref{GR-III-V01-R-TLTP}--\ref{GR-III-TLTP-WGR-B}.}
\label{fig:bn:GR-III-Terminate}
\end{figure}
\end{center}

\chapter{Further Diagrammatic Identities} \label{ch:FurtherD-ID}

\section{The New Identities} \label{sec:D-ID-New}

In order to make further progress, one of the things that
we must do is introduce a new set of diagrammatic identities.
These will enable us to treat diagrams like the ones in
figure~\ref{fig:bn:GR-Ia-Nest-Dec-Ex}. Rather than simply
stating the diagrammatic identities, it will prove more illuminating
to provide some motivation, by sketching how the terms involved
arise.

We know already that we plan to convert certain diagrams possessing
processed gauge remainders into $\Lambda$-derivative terms; we have,
of course, seen an explicit example of this in the
one-loop calculation. Whilst we are yet to demonstrate that
we can, in the general case, make this conversion,
let us suppose that we are able to do so. An example of
the types of term that we expect to encounter is shown in 
figure~\ref{fig:bn:GR-LdL-Ex}.
\begin{center}
\begin{figure}[h]
	\[	
	-\sum_{J=0}^n \norm_{J+1,0}
	\dec{
		\dec{
			\sco[0.1]{
				\LID[0.1]{}{Struc-4GR}{Struc-4GR-n_J-R-LdL}
			}{\ensuremath{\begin{array}{c}\input{pstex/Vertex-n_J-R.pstex_t} \end{array}}}
		}{\bullet}
		-
		\left[
			\sco[0.1]{
				\LID[0.1]{}{Struc-4GR}{Struc-4GR-LdL-n_J-R}
			}{\ensuremath{\begin{array}{c}\input{pstex/Vertex-n_J-R-LdL.pstex_t} \end{array}}}
		\right]
		-\cdots
	}{11\Delta^{J+1}}
	\]
\caption{An example of the type of $\Lambda$-derivative term
we expect to arise from the gauge remainder sector of the
calculation.}
\label{fig:bn:GR-LdL-Ex}
\end{figure}
\end{center}

We now iterate the usual diagrammatic procedure but with one
crucial difference: every time we generate a two-point,
tree level vertex---whether it be via the flow equations, 
the action of a gauge remainder or manipulations 
at $\Op{2}$---we have the new option of
attaching it to the gauge remainder structure depicted in the
diagrams of figure~\ref{fig:bn:GR-LdL-Ex}. 
In figure~\ref{fig:bn:GR-LdL-NewTerms-Ex} we schematically
represent examples of the types
of diagram we expect to arise in the former two cases.
\begin{center}
\begin{figure}
	\[
	\left[
		\LID[0.1]{}{Struc-V-S-W-4GR}{Ex-V-S-W-4GR}
		- \LID[0.1]{}{Struc-V-S-W-GR-4GR}{Ex-V-S-W-GR-4GR}
	\right],
	\left[
		\LID[0.1]{}{Struc-GR-4GR}{Ex-GR-4GR}
		-\LID[0.1]{}{Struc-GR-GR-4GR}{Ex-GR-GR-4GR}
	\right]
	\]
\caption{Examples of some some of the new types of term
we expect to generate when we process diagrams like~\ref{Struc-4GR-LdL-n_J-R}.}
\label{fig:bn:GR-LdL-NewTerms-Ex}
\end{figure}
\end{center}

These examples are not exhaustive. For instance, there are diagrams
similar to~\ref{Ex-V-S-W-4GR} and~\ref{Ex-V-S-W-GR-4GR} in which
the vertex is a Wilsonian effective action vertex, and the wine
is reduced. Similarly, we have implicitly assumed that diagrams~\ref{Ex-GR-4GR}
and~\ref{Ex-GR-GR-4GR} have been created by an un-nested
type-Ia gauge remainder. Nonetheless, these example diagrams
capture all the features
that we require. The first diagrammatic identity follows from
considering a version of sub-diagram~\ref{Ex-GR-4GR} in
which the gauge remainder structure at the top is not nested.
For the sake of clarity, we will be more explicit about where
this sub-diagram comes from.

To this end, consider decoration of two-point, tree level vertices by $\WBT$. 
We will not
specify the number of vertices in the diagram and will suppose that the
decorations include the usual external fields and $M\geq1$ effective propagators.
This is illustrated in figure~\ref{fig:bn:C-sec:TLTP-WBT}, where we use the effective
propagator relation and note that the gauge remainder vanishes, since it must be in the $C$-sector.
\begin{center}
\begin{figure}[h]
	\[
	\dec{
		\ensuremath{\begin{array}{c}\input{pstex/Struc-GR-TLTP-WBT.pstex_t} \end{array}}
	}{11\Delta^M}
	=
	\dec{
		\ensuremath{\begin{array}{c}\input{pstex/Struc-GR-hook.pstex_t} \end{array}}
	}{11\Delta^M}
	\]
\caption{A two-point, tree level vertex, generated by the action of a gauge remainder
and decorated by $\protect \WBT $.}
\label{fig:bn:C-sec:TLTP-WBT}
\end{figure}
\end{center}

Further progress can be made with the final diagram. The socket must be decorated by one of
the effective propagators,\footnote{We suppose
that the wine is attached to one of the un-drawn vertices, rather than 
the socket, but this restriction is unnecessary
for the argument that follows.} since the gauge remainder does not 
carry a Lorentz index. We perform this decoration and assume, for simplicity, that the loose
ends are both joined to the same, lone vertex (whose argument is irrelevant) and that the only remaining
decorations are the external fields.
This is shown in figure~\ref{fig:bn:C-sec:TLTP-WBT-B} where the second diagram has just been 
obtained by charge conjugation of the first.
\begin{center}
\begin{figure}[h]
	\[
	\dec{
		\ensuremath{\begin{array}{c}\input{pstex/Struc-GR-hook-dec.pstex_t} \end{array}}
	}{11}
	=
	\dec{
		\ensuremath{\begin{array}{c}\input{pstex/Struc-GR-hook-dec-CC.pstex_t} \end{array}}
	}{11}
	\]
\caption{Further decoration of the previous diagram.}
\label{fig:bn:C-sec:TLTP-WBT-B}
\end{figure}
\end{center}

We now consider specific decorations of these diagram and
order the fields. We decorate the first diagram
by placing both external fields on the `outside' of the vertex,
so that the ordering of the fields on the vertex is $11F\bar{F}$.
We decorate the second diagram by placing 
both external fields on the `inside' of the vertex. This is shown in
figure~\ref{fig:bn:C-sec:TLTP-WBT-C}, where we also indicate the flavour
of the empty loops.
\begin{center}
\begin{figure}[h]
	\[
	\ensuremath{\begin{array}{c}\input{pstex/Struc-GR-hook-dec-FO.pstex_t} \end{array}} + \ensuremath{\begin{array}{c}\input{pstex/Struc-GR-hook-dec-CC-FO.pstex_t} \end{array}} \mathrm{FAS}
	\]
\caption{A possible pair of field ordered diagrams formed from a term
in which the structure $\protect \WBT $ is attached to a two-point, tree level
vertex which has been formed by the action of a gauge remainder.}
\label{fig:bn:C-sec:TLTP-WBT-C}
\end{figure}
\end{center}

The ordering
of the fields is the same in both cases. Indeed,
the bottom parts of the diagrams are the same in each case:
the (field ordered) vertices are identical, as is the wine,
the gauge remainder it goes into and the effective propagator. Moreover,
in both cases there is any empty loop giving a factor of $-N$.
The difference occurs in the top part of the diagram.

We focus first on the push forward / pull back on to the top of the effective
propagator. The two cases correspond to pushing forward / pulling back an ${F}$ on to
an $\bar{F}$. Referring to table~\ref{tab:GR:Flavour}, we see that both cases come with the same
sign, but generate a different field. However, this field has been removed by the effective
propagator relation!
Finally, then, all that remains is the hook at the top of the diagram, around which we
assume flows momentum $l$. Up to group theory, they both reduce to $\int_l g_l$,
as we know from equations~(\ref{eq:hook-R-1})--(\ref{eq:hookCC}).
 However, in one case the hook encloses
a loop in the two-sector, whereas in the other case, it enclose a loop in the one-sector.
These come with opposite signs, causing the two terms to cancel.

It should be clear that the various restrictions we have imposed, such as there only begin on
vertex, do not affect the final conclusion: whenever a diagram possesses a component like that
of figure~\ref{fig:bn:C-sec:TLTP-WBT}, it is guaranteed to vanish. This provides us with a 
new diagrammatic identity.
\begin{D-ID}
Consider a vertex attached to an effective propagator, which ends in a hook. Suppose that a gauge remainder
strikes this vertex. Focusing on the terms produced when the gauge remainder
pushes forward / pulls back on to a socket, we suppose that a two-point, tree level vertex is generated.
Applying the effective propagator relation, the gauge remainder dies, leaving us with the
diagrammatic identity shown in figure~\ref{fig:D-ID-Alg-1}.
\begin{center}
\begin{figure}[h]
	\[
	\ensuremath{\begin{array}{c}\input{pstex/Struc-AR0-A.pstex_t} \end{array}} - \ensuremath{\begin{array}{c}\input{pstex/Struc-AR0-B.pstex_t} \end{array}} = 0
	\]
\caption{Diagrammatic identity~\ref{D-ID-Alg-1}.}
\label{fig:D-ID-Alg-1}
\end{figure}
\end{center}

\paragraph{Corollary} 

The diagrammatic identity immediately implies that $\ensuremath{\begin{array}{c}\begin{picture}(0,0)%
\includegraphics{pstex/Struc-AR0-A.pstex}%
\end{picture}%
\setlength{\unitlength}{3947sp}%
\begingroup\makeatletter\ifx\SetFigFont\undefined%
\gdef\SetFigFont#1#2#3#4#5{%
  \reset@font\fontsize{#1}{#2pt}%
  \fontfamily{#3}\fontseries{#4}\fontshape{#5}%
  \selectfont}%
\fi\endgroup%
\begin{picture}(230,285)(197,457)
\end{picture}
 \end{array}}$, alone,
vanishes; this follows because we can re-express this sub-diagram as:
\[
1/2 \left[\ensuremath{\begin{array}{c}\begin{picture}(0,0)%
\includegraphics{pstex/Struc-AR0-A.pstex}%
\end{picture}%
\setlength{\unitlength}{3947sp}%
\begingroup\makeatletter\ifx\SetFigFont\undefined%
\gdef\SetFigFont#1#2#3#4#5{%
  \reset@font\fontsize{#1}{#2pt}%
  \fontfamily{#3}\fontseries{#4}\fontshape{#5}%
  \selectfont}%
\fi\endgroup%
\begin{picture}(230,285)(197,457)
\end{picture}
 \end{array}} - \ensuremath{\begin{array}{c}\begin{picture}(0,0)%
\includegraphics{pstex/Struc-AR0-B.pstex}%
\end{picture}%
\setlength{\unitlength}{3947sp}%
\begingroup\makeatletter\ifx\SetFigFont\undefined%
\gdef\SetFigFont#1#2#3#4#5{%
  \reset@font\fontsize{#1}{#2pt}%
  \fontfamily{#3}\fontseries{#4}\fontshape{#5}%
  \selectfont}%
\fi\endgroup%
\begin{picture}(132,273)(193,469)
\end{picture}
 \end{array}}\right]
\]
(see section~\ref{sec:GR-Wines}).

\label{D-ID-Alg-1}
\end{D-ID}

The next diagrammatic identity is more interesting, finally resolving the issue of
what to do with the diagrams of figure~\ref{fig:bn:GR-Ia-Nest-Dec-Ex}. We do not
consider these diagrams on their own, but rather combine them with two diagrams
which are special cases of diagram~\ref{Ex-V-S-W-GR-4GR}. These latter two 
diagrams are shown in figure~\ref{fig:bn:Nestx2-Trapped-GR-Ex}.
\begin{center}
\begin{figure}[h]
	\[
	4 \sum_{J=0}^n \norm_{J,0}
	\dec{
		\LID[0.1]{}{Struc-n_J-S-R-WGRx2-Top-PB-GR}{C-sec-n_J-S-R-WGRx2-T-PB-GR}
		-\LID[0.1]{}{Struc-n_J-S-R-WGRx2-Top-GR}{C-sec-n_J-S-R-WGRx2-T-GR}
	}{11\Delta^J}
	\]
\caption{Two diagrams which are special cases of diagram~\ref{Ex-V-S-W-GR-4GR}.}
\label{fig:bn:Nestx2-Trapped-GR-Ex}
\end{figure}
\end{center}

For the purposes of the
imminent diagrammatic identity, we 
strip off the common structures from diagrams~\ref{C-sec-n_J-S-R-WGRx2-T-PB-GR}, \ref{C-sec-n_J-S-R-WGRx2-T-GR},
\ref{GR-Ia-Nest1-ExtraTerm-Ex}
and~\ref{GR-Ia-Nest1-ExtraTerm-Ex-PB}.

The resulting sub-diagrams
corresponding to diagrams~\ref{GR-Ia-Nest1-ExtraTerm-Ex-PB} 
and~\ref{C-sec-n_J-S-R-WGRx2-T-GR}
are collected together in figure~\ref{fig:bn:GR-II-ACM-2}.\footnote{The remaining sub-diagrams will be treated later.}
\begin{center}
\begin{figure}[h]
	\[
	\LID[0.1]{}{Struc-AR1-A}{Struc-AR1-A}  \hspace{1em} 
	- \LID[0.1]{}{Struc-AR1-B}{Struc-AR1-B} \hspace{1em} 
	- \LID[0.1]{}{Struc-AR1-C}{Struc-AR1-C}
	\]
\caption{The first two sub-diagrams show the algebraic part of diagram~\ref{GR-Ia-Nest1-ExtraTerm-Ex-PB},
where we have used the effective propagator relation.
The final sub-diagram shows the algebraic part of diagram~\ref{C-sec-n_J-S-R-WGRx2-T-GR}.}
\label{fig:bn:GR-II-ACM-2}
\end{figure}
\end{center}

The letters $U$--$Z$ will be used to denote field flavour, and the subscripts
$S$ and $R$ represent indices. Each of these sub-diagrams is really part of
a complete diagram and so the loose ends are somehow tied up. This immediately
tells us that $W$ and $X$ must both be either bosonic or fermionic. In the former
case, they must both be in the $A$-sector: gauge remainders have no support in the
$C$-sector, and the mixed case---where $W$ is in the $A$-sector and $X$ is in the
$C$-sector---will vanish, due to charge conjugation invariance. Depending
on how we tie the legs up determines how many closed loops each diagram
possesses, which in turn affects the group theory factors. This is only
important when both the fields $W$ and $X$ are fermionic, as we will see below.
In table~\ref{tab:ACM-1}, the possible sets of field flavours a listed.

\begin{center}
\begin{table}
	\[
	\begin{array}{cc|cc|cc}
		W_S 		& X_R 		& Y 		& Z 		& U 		& V 		\\ \hline \hline
		A^1_\sigma 	& A^1_\rho	& A^1 		& A^1 		& A^1 		& A^1 		\\
					&			& F 		& B			& \bar{B}	& \bar{B} 	\\ \hline
		A^2_\sigma 	& A^2_\rho	& A^2 		& A^2 		& A^2 		& A^2 		\\
					&			& \bar{F}	& \bar{B}	& B			& B			\\ \hline
		\bar{F}_S	& F_R		& A^2		& F			& A^2		& \bar{F}	\\
					&			& \bar{B}	& A^1		& F			& A^1		\\ \hline
		F_S			& \bar{F}_R	& A^1		& \bar{F}	& A^1		& F			\\
					&			& B			& A^2		& \bar{F}	& A^2	
 	\end{array}
	\]
\caption{Possible flavours of the fields of the diagrams of figure~\ref{fig:bn:GR-II-ACM-2}.}
\label{tab:ACM-1}
\end{table}
\end{center}

We now evaluate the three diagrams of figure~\ref{fig:bn:GR-II-ACM-2}, algebraically, in the different sectors.
The easiest case to do is when all of the fields are in one of the $A$-sectors, say $A^1$. Strictly
speaking, we should not look at this case on its own, as it is UV divergent; rather, we should look 
at it together with its partner, in which the fields $U,V,Y,Z$ are fermionic.

Treating these contributions together, we can express the sub-diagrams thus:
\begin{equation}
	N \frac{k_\sigma}{k^2} \int_l
	\left\{
	\begin{array}{l}
	\frac{l_\rho}{l^2}
		\left[
			1 -  \frac{l \cdot(l+k)}{(l+k)^2} - \frac{ k \cdot(l+k)}{(l+k)^2}
		\right]
	\\
	- \frac{f_l l_\rho}{\Lambda^2}
	\left[
		1 - \frac{l \cdot(l+k)}{\Lambda^2} -2g_{l+k} - \frac{k \cdot(l+k)}{\Lambda^2}
	\right]
	\end{array}
	\right\} = 0,
\end{equation}
where the group theory factors of $\pm N$ arise from the explicitly drawn loop and
we have used equation~(\ref{eq:app-xf+2g}).

It is trivial to show that we obtain the same overall result when we take the fields $W$ and $X$
to be in the $A^2$ sector. The remaining sectors are more subtle and must, in fact, all
be treated together. At first, this looks a little strange, since we cannot obviously
combine the case where $W,X = \bar{F}, F$ with $W,X = F, \bar{F}$: when we tie up all
the loose ends, it looks as though we are trying to equate diagrams possessing arbitrary structure
that are clearly different. However, as we will now demonstrate, we can take these
structures to be the same by utilising charge-conjugation invariance.

In figure~\ref{fig:bn:GR-II-ACM-2-CC}, we have taken 
sub-diagram~\ref{Struc-AR1-C}, with fermionic external fields
and have redrawn it. The first step is done
by using charge-conjugation invariance. The second step is
just a trivial redrawing.

\begin{center}
\begin{figure}[h]
	\[
	\ensuremath{\begin{array}{c}\input{pstex/Struc-AR1-Ci.pstex_t} \end{array}} \stackrel{\mathrm{CC}}{\longrightarrow} \ensuremath{\begin{array}{c}\input{pstex/Struc-AR1-Cii.pstex_t} \end{array}} \equiv \ensuremath{\begin{array}{c}\input{pstex/Struc-AR1-Ciii.pstex_t} \end{array}}
	\]
\caption{Sub-diagram~\ref{fig:bn:GR-II-ACM-2-CC} with specific field content.}
\label{fig:bn:GR-II-ACM-2-CC}
\end{figure}
\end{center}

The point is that we should now consider together the final diagram of figure~\ref{fig:bn:GR-II-ACM-2-CC}
with a copy of 
the first diagram for which we take the fields $W,X$ to be not  $\bar{F},F$ but $F, \bar{F}$.
To this pair of diagrams, we can now attach exactly the same arbitrary structure. 

There is one final subtlety. From table~\ref{tab:ACM-1} it is apparent that, for a given
fermionic choice of $W$ and $X$ we can, in each of the sub-diagrams~\ref{Struc-AR1-A}--\ref{Struc-AR1-C}, 
choose one of the internal fields
to be either an $A^1$ or an $A^2$. These two choice come with different group theory factors.
Once all loose ends are tied up, we are interested in the flavours of the loops
carrying momenta $l$ and $k$. There are three such loops and we denote their
flavours by an ordered triplet.\footnote{There can, of course, be additional loops, but they will contribute a group theory
factor common to both cases.}
In the case that we choose one of the fields to be an $A^1$, the loop flavours are $\{1,1,2\}$
whereas, if we choose one of the field to be an $A^2$, then the loop flavours are $\{2,2,1\}$.
If all loops are empty, then the group theory factors are $-N^3$ and $N^3$, respectively.
If one of the loops is decorated by the external fields, then this loop contributes
$\str A^1 A^1$ and so the group theory factors are just $-N^2$ and $N^2$, respectively.
Either way, the two choices of the flavour of the bosonic fields give rise to opposite signs.

With these points in mind, we can evaluate diagrams~\ref{Struc-AR1-A}--\ref{Struc-AR1-C}
when the external fields are fermionic: it is straightforward to show that they sum to zero.
Combining this with the similar result from the $A$-sector, we have a diagrammatic identity,
since the diagrams of figure~\ref{fig:bn:GR-II-ACM-2} sum to zero.

However, we are not quite done. 
Referring back to figure~\ref{fig:bn:GR-II-ACM-2},
we should also analyse the case in which the nested gauge remainder is either a push forward,
rather than a pull back or vice-versa, as appropriate. 
The corresponding diagrams are shown in figure~\ref{fig:bn:GR-II-ACM-2-B}
\begin{center}
\begin{figure}[h]
	\[
	\ensuremath{\begin{array}{c}\input{pstex/Struc-AR1-A+B-2.pstex_t} \end{array}} 
	-\ensuremath{\begin{array}{c}\input{pstex/Struc-AR1-C-bare-2.pstex_t} \end{array}}
	\]
\caption{Partner diagrams to those of figure~\ref{fig:bn:GR-II-ACM-2}, in which the nested gauge
remainder has changed character.}
\label{fig:bn:GR-II-ACM-2-B}
\end{figure}
\end{center}

These sub-diagrams are even easier to analyse than the previous set: we know from 
trivial extension of the arguments presented in section~\ref{sec:GRs-Nested}
that the external fields cannot be fermionic and so we need take them only in the $A$-sector. Note, however,
that the internal fields can be fermionic. In this case, one of the external fields must be in the $A^1$
sector, whilst the other must be in the $A^2$-sector. Taking this into account, it is easy to show
that the two diagrams of figure~\ref{fig:bn:GR-II-ACM-2-B} sum to zero. This completes our analysis
and yields the required diagrammatic identity.

\begin{D-ID} 

Consider a two-point, tree level vertex, attached to an effective propagator
and then to a gauge remainder.
The gauge remainder bites the remaining socket of the vertex in either sense. A second gauge
remainder bites the first gauge remainder. These sub-diagrams 
can be redrawn as shown in figure~\ref{fig:D-ID-Alg-2}.
\begin{center}
\begin{figure}[h]
	\[
	\ensuremath{\begin{array}{c}\input{pstex/Struc-AR1-A+B.pstex_t} \end{array}} \equiv \ensuremath{\begin{array}{c}\input{pstex/Struc-AR1-C-bare.pstex_t} \end{array}}, \hspace{1em} \ensuremath{\begin{array}{c}\input{pstex/Struc-AR1-A+B-2.pstex_t} \end{array}} \equiv \ensuremath{\begin{array}{c}\input{pstex/Struc-AR1-C-bare-2.pstex_t} \end{array}}
	\]
\caption{Diagrammatic identity~\ref{D-ID-Alg-2}.}
\label{fig:D-ID-Alg-2}
\end{figure}
\end{center}

\label{D-ID-Alg-2}
\end{D-ID}

The next logical step is to generalise this relationship to one that is suitable
for an arbitrary level of nesting. To begin with, we will increase the level
of nesting by one. Before stating the new diagrammatic identity,
we will indicate where the terms involved come from.

Let us start by considering nested versions of diagrams~\ref{GR-Ia-Nest1-ExtraTerm-Ex}
and~\ref{GR-Ia-Nest1-ExtraTerm-Ex-PB}. The extra gauge remainder can either go
inside or outside the loop, as shown 
in figure~\ref{fig:bn:GR-Ia-Nest2-Dec-Ex}.
\begin{center}
\begin{figure}[h]
	\[
	4 \sum_{J=0}^n \norm_{J,0}
	\dec{
	\begin{array}{c}
	\vspace{0.2in}
		-\LID[0.1]{}{GR-Ia-Nest2-ETa2}{GR-Ia-Nest2-ETa2-n_J-S-R}
		+\LID[0.1]{}{GR-Ia-Nest2-ETd2}{GR-Ia-Nest2-ETd2-n_J-S-R}
		-\LID[0.1]{}{GR-Ia-Nest2-ETb2}{GR-Ia-Nest2-ETb2-n_J-S-R}
		+\LID[0.1]{}{GR-Ia-Nest2-ETc2}{GR-Ia-Nest2-ETc2-n_J-S-R}
	\\
		\LID[0.1]{}{GR-Ia-Nest2-ETa}{GR-Ia-Nest2-ETa-n_J-S-R}
		-\LID[0.1]{}{GR-Ia-Nest2-ETd}{GR-Ia-Nest2-ETd-n_J-S-R}
		+\LID[0.1]{}{GR-Ia-Nest2-ETb}{GR-Ia-Nest2-ETb-n_J-S-R}
		-\LID[0.1]{}{GR-Ia-Nest2-ETc}{GR-Ia-Nest2-ETc-n_J-S-R}
	\end{array}
	}{11\Delta^J}
	\]
\caption{The nested versions of diagrams~\ref{GR-Ia-Nest1-ExtraTerm-Ex} and~\ref{GR-Ia-Nest1-ExtraTerm-Ex-PB}.}
\label{fig:bn:GR-Ia-Nest2-Dec-Ex}
\end{figure}
\end{center}

The diagrams of figure~\ref{fig:bn:GR-Ia-Nest2-Dec-Ex} leave us with two problems.
On the one had, whilst
diagrams~\ref{GR-Ia-Nest2-ETa2-n_J-S-R}, \ref{GR-Ia-Nest2-ETd2-n_J-S-R}, \ref{GR-Ia-Nest2-ETa-n_J-S-R} and~\ref{GR-Ia-Nest2-ETd-n_J-S-R}
can be redrawn using diagrammatic
identity~\ref{D-ID-Alg-2}, we need to find the terms against which they cancel.
On the other hand,
we suspect that
diagrams~\ref{GR-Ia-Nest2-ETb2-n_J-S-R}, \ref{GR-Ia-Nest2-ETc2-n_J-S-R}, \ref{GR-Ia-Nest2-ETb-n_J-S-R} 
and~\ref{GR-Ia-Nest2-ETc-n_J-S-R} will 
somehow cancel against nested versions of diagrams~\ref{C-sec-n_J-S-R-WGRx2-T-PB-GR}
and~\ref{C-sec-n_J-S-R-WGRx2-T-GR}. As it stands, this cancellation does not quite work:
we need further diagrams. 

The solutions to both of these two problems are intimately related.
The terms we require to make
both sets of cancellations work are generated by the same set of diagrams.
Imagine converting a diagram containing a singly nested version of $\BT$
into a $\Lambda$-derivative term. Among the corrections generated will
by the (processed) type-Ia gauge remainders of figure~\ref{fig:bn:NestGRManip-GR-Ia-Ex}.

\begin{center}
\begin{figure}[h]
	\[
	4 \sum_{J=0}^n \norm_{J,0}
	\dec{
		\LID[0.1]{}{Struc-GRx2-GR-Ia-Ex-1}{Struc-GRx2-GR-Ia-Ex-1}
		-\LID[0.1]{}{Struc-GRx2-GR-Ia-Ex-2}{Struc-GRx2-GR-Ia-Ex-2}
		-\LID[0.1]{}{Struc-GRx2-GR-Ia-Ex-3}{Struc-GRx2-GR-Ia-Ex-3}
		+\LID[0.1]{}{Struc-GRx2-GR-Ia-Ex-4}{Struc-GRx2-GR-Ia-Ex-4}
	}{11\Delta^J} 
	\]
\caption{Particular type-Ia gauge remainders arising from the 
conversion of diagrams possessing a nested version of \protect \BT
into a $\Lambda$-derivative term.}
\label{fig:bn:NestGRManip-GR-Ia-Ex}
\end{figure}
\end{center}

Utilising the effective propagator relation, we generate eight terms.
The two not involving further gauge remainders are precisely those
that we need for the next diagrammatic identity. What of the
other terms? These cancel diagrams~\ref{GR-Ia-Nest2-ETa2-n_J-S-R}, 
\ref{GR-Ia-Nest2-ETd2-n_J-S-R}, \ref{GR-Ia-Nest2-ETa-n_J-S-R} and~\ref{GR-Ia-Nest2-ETd-n_J-S-R}
via diagrammatic identity~\ref{D-ID-Alg-2}!

Now we are in a position to uncover the new diagrammatic identity.
Consider the set of
diagrams shown in figure~\ref{fig:bn:GR-II-ACM-3}.
\begin{center}
\begin{figure}[h]
	\[
	\LID[0.1]{}{Struc-AR3-A-1}{Struc-AR3-A-1}  \hspace{1em}
	-\LID[0.1]{}{Struc-AR3-B-1}{Struc-AR3-B-1} \hspace{1em}
	+\LID[0.1]{}{Struc-AR3-C-1}{Struc-AR3-C-1} \hspace{1em}
	-\LID[0.1]{}{Struc-AR3-D-1}{Struc-AR3-D-1}
	\]
\caption{A set of diagrams, possessing three performed gauge remainders, which sum to zero.}
\label{fig:bn:GR-II-ACM-3}
\end{figure}
\end{center}

The first sub-diagram has been stripped off from diagram~\ref{GR-Ia-Nest2-ETc-n_J-S-R};
the second sub-diagram has been stripped off from a nested version of diagram~\ref{C-sec-n_J-S-R-WGRx2-T-PB-GR}
and the final two sub-diagrams have been stripped off from diagrams~\ref{Struc-GRx2-GR-Ia-Ex-3}
and~\ref{Struc-GRx2-GR-Ia-Ex-4}.

It intuitively makes sense that these four diagrams sum to zero: suppose that we remove a nested gauge
remainder from each diagram.\footnote{In the last two diagrams, the top-most gauge remainder is
independent of the other two; it is certainly not nested in the usual sense and it is not this
that we remove.} The  set of
four diagrams would then decompose: the first two would cancel via diagrammatic identity~\ref{D-ID-Alg-2},
whereas the last two would cancel via diagrammatic identity~\ref{D-ID-Alg-1}!

Let us now examine the sub-diagrams of figure~\ref{fig:bn:GR-II-ACM-3} in more detail.
Ultimately, these sub-diagrams must form full diagrams and so all empty
sockets or loose ends will be tied up. Cancellations will occur between terms
for which the field ordered structure which completes the diagram is the same.
It is clear that if we attach a given field ordered structure to these
sub-diagrams, then the group theory factors will be the same for each of the
complete diagrams. This is, of course, crucial.

We will now compute these diagrams, taking  all fields to be in the $A$-sector
(it does not matter whether they are $A^1$s or $A^2$s). Up to the common structure
and a common group theory factor, we have:
\begin{equation}
	\frac{k_\rho}{k^2}  \int_l \frac{l_\sigma}{l^2}
	\left[
		\begin{array}{l}
			\frac{(l+m+k)_\tau}{(l+m+k)^2} - \frac{(l+m)_\tau}{(l+m)^2} \frac{(l+m)\cdot(l+m+k)}{(l+m+k)^2}
		\\
			- \frac{(l+m)_\tau}{(l+m)^2} \frac{k \cdot (l+m+k)}{(l+m+k)^2}
		\\
			+ \frac{(l+m)_\tau}{(l+m)^2} - \frac{(l+m+k)_\tau}{(l+m+k)^2}
		\end{array}
	\right] = 0,
\label{eq:Alg-2-A-sec}
\end{equation}
where we have implicitly assumed that the individual terms are regulated by diagrams
in which the internal fields of the sub-diagrams are fermionic. This is straightforward
to show; these regulating diagrams also sum to zero. Finally, we must consider diagrams
in which two of the external fields are fermionic. As with diagrammatic identity~\ref{D-ID-Alg-2},
we  must use charge conjugation to identify all those diagrams with the same common structure.
Though the pattern of cancellations is different from that of equation~(\ref{eq:Alg-2-A-sec}),
we have verified that the diagrams sum to zero.

This completes the analysis of the diagrams of figure~\ref{fig:bn:GR-II-ACM-3}
and so now we should consider the partner diagrams which have different
patterns of pushes forward and pulls back. This is subtle. The salient points
are most easily introduced by focusing on diagram~\ref{Struc-AR3-B-1}. As drawn,
each of the three gauge remainders in the loop is bitten on the right
and it is implicitly assumed that this diagram has been combined with its
charge conjugate, in which each of the gauge remainders is bitten on 
the left.

By changing pushes forward into pulls back---or, equivalently, bites on the right (left) to
bites on the left (right)---we can generate three independent new
diagrams. We can characterise this by the sense in which each of the gauge
remainders is bitten. We have  $-RRL$, $-RLR$, $+RLL$, where we have picked
up a sign for each change from right to left. Using charge conjugation
to rewrite the final term as $-LRR$, we see that set of three terms are related
to each other by cyclic permutations. This leads us to consider the three diagrams
of figure~\ref{fig:bn:GR-II-ACM-3-Cross}, together.
\begin{center}
\begin{figure}[h]
	\[
	\LID[0.1]{}{Struc-AR3-B-X1}{Struc-AR3-B-X1} 
	+\LID[0.1]{}{Struc-AR3-B-X2}{Struc-AR3-B-X2}  
	+\LID[0.1]{}{Struc-AR3-B-X3}{Struc-AR3-B-X3}
	\]
\caption{The three partner diagrams to diagram~\ref{Struc-AR3-B-1}.}
\label{fig:bn:GR-II-ACM-3-Cross}
\end{figure}
\end{center}

Now we encounter a potential problem. Whilst diagram~\ref{Struc-AR3-A-1}, too,
has three partner diagrams, diagrams~\ref{Struc-AR3-C-1} and~\ref{Struc-AR3-D-1}
have only one, each. It is thus clear that we will not be able to
show that the sum of these diagrams vanishes via a direct analogue of equation~\ref{eq:Alg-2-A-sec}.

The key to understanding how these diagrams do cancel
revolves around understanding both the 
group theory factors of the diagrams and the
flavours of the external
legs. We examine the former first.

Returning to diagram~\ref{Struc-AR3-B-X1}, 
let us suppose that all three legs are attached to the same field-ordered structure.
Attaching the legs in the order shown, \ie\ without crossing any of them over, the order
on the structure is $RST$, in the counter-clockwise sense. 
Now, however, consider crossing the leg labelled $S$ with the
leg labelled $R$. Na\"ively, one might expect this to change the group theory but this is
not the case. The diagram can be untied, as illustrated in figure~\ref{fig:bn:GR-II-ACM-3-Cross-B}.
\begin{center}
\begin{figure}[h]
	\[
	\LID[0.1]{}{Struc-AR3-B-X1-B}{Struc-AR3-B-X1-B} 
	\]
\caption{A trivial redrawing of sub-diagram~\ref{Struc-AR3-B-X1}.}
\label{fig:bn:GR-II-ACM-3-Cross-B}
\end{figure}
\end{center}

Similar considerations apply to the partners of diagrams~\ref{Struc-AR3-A-1},
\ref{Struc-AR3-C-1} and~\ref{Struc-AR3-D-1}:
crossing over legs which attach to the same structure 
does not, counter intuitively, change the group theory factor.

Next, we examine the flavours of the external legs. First, we note
that the external legs cannot be fermionic (in just the same way that
the external legs of the analogous diagrams involved in diagrammatic
identity~\ref{D-ID-Alg-2} could not be fermionic). Hence, the external
legs must be either $A$s or $C$s. 

If an odd number of the external legs are
$A$s then 
forming a complete
diagram  will give us something which vanishes by \CC\
when we sum over field ordered structures. This follows because,
even if the three external legs attach to the same structure,
we can interchange two of them without changing the group theory factors
(and we will not get any flavour changes, since the fields are bosonic).

Now, due to gauge remainders having no support in the $C$-sector, it must
be the case that at least one of the external fields is an $A$; for
complete diagrams not to vanish by \CC, it follows that there must be precisely two $A$s.
But, if any of the legs are $C$s, then a pair of the internal lines must be fermionic,
since gauge remainders have no support in the $C$-sector. This then means that whilst
two of the bosonic external legs must be in the 1(2) sector, the remaining one must
be in the 2(1) sector. However, the single field in the 2(1) sector must be on a different
supertrace from the other two; if this field is an $A$, then once more the diagram vanishes
by \CC. This thus leaves only those sub-diagrams with two external $A^{1,2}$s and one
external $C^{2,1}$. We see how this applies to the partners of diagrams~\ref{Struc-AR3-C-1}
and~\ref{Struc-AR3-D-1} in figure~\ref{fig:bn-Alg2-PartnersC,D}.
\begin{center}
\begin{figure}[h]
	\[
	-\LID[0.1]{}{Struc-AR3-C-1b}{Struc-AR3-C-1b} \hspace{1em}
	+\LID[0.1]{}{Struc-AR3-C-2b}{Struc-AR3-C-2b}
	\]
\caption{The two partners of sub-diagrams~\ref{Struc-AR3-C-1} and~\ref{Struc-AR3-D-1}.
Any other combination of external fields will cause the diagrams to vanish.}
\label{fig:bn-Alg2-PartnersC,D}
\end{figure}
\end{center}

The crucial point is that sub-diagrams~\ref{Struc-AR3-C-1b} and~\ref{Struc-AR3-C-2b}
simply cancel: the un-nested pull back of the first diagram cancels
the corresponding  push forward of the second. Now let us examine diagrams~\ref{Struc-AR3-B-X1}--\ref{Struc-AR3-B-X3}.
The final one of these vanishes; the trapped gauge remainder cannot be in 
the $C$-sector, but since the field that attaches to this must be on a different
supertrace from the other two, the diagram vanishes by \CC. This then leaves diagrams~\ref{Struc-AR3-B-X1} 
and~\ref{Struc-AR3-B-X2}. Similarly, of the partners to diagram~\ref{Struc-AR3-A-1},
only two survive. Evaluating these four surviving diagrams, algebraically, yields
zero.


We have demonstrated that the set of diagrams in figure~\ref{fig:bn:GR-II-ACM-3}
sums to zero and that this still holds if we change pushes forward into pulls back (or \emph{vice-versa})
in all independent ways. Having made these changes, whilst it is true that the complete
set of diagrams sum to zero, it is also true that there are more stringent relationships.
However, we are not interested in these relationships, as such. For the purposes
of this thesis, we require only the weaker constraint.
With this in mind, we obtain a further diagrammatic identity.

\begin{D-ID}
Consider the diagrams of figure~\ref{fig:bn:D-ID-Alg-3}.
This equality holds, if we change nested gauge remainders from
being struck on the left (right) to being struck on the right (left),
in all independent ways.
\begin{center}
\begin{figure}[h]
	\[
		\ensuremath{\begin{array}{c}\input{pstex/Struc-AR3-A.pstex_t} \end{array}} - \ensuremath{\begin{array}{c}\input{pstex/Struc-AR3-B.pstex_t} \end{array}} + \ensuremath{\begin{array}{c}\input{pstex/Struc-AR3-C.pstex_t} \end{array}} - \ensuremath{\begin{array}{c}\input{pstex/Struc-AR3-D.pstex_t} \end{array}} = 0
	\]
\caption{Diagrammatic identity~\ref{D-ID-Alg-3}.}
\label{fig:bn:D-ID-Alg-3}
\end{figure}
\end{center}
\label{D-ID-Alg-3}
\end{D-ID}

We conclude our discussion of diagrammatic identities, for the time being,
by going some way towards generalising diagrammatic identity~\ref{D-ID-Alg-3}.
Our aim is to show that an analogous relation holds for sub-diagrams
with arbitrarily nested gauge remainders. Of course, if this nesting is with respect
to the gauge remainder that we have been taking to carry index $R$,
then we can still use diagrammatic identity~\ref{D-ID-Alg-3}. Rather, we are interested
in the case where the nesting
is within the loops formed out of gauge remainders.

We will not, however, rigorously
demonstrate the resulting identity to be true. Rather, what we will do is demonstrate it to be
true for the following case: first, we will take 
all gauge remainders in a loop to bite each other in the same sense (\ie\ all 
to the right (left)); second, we will demand that all fields are in the $A$-sector.
Having done this, we will then assert, without proof, that if it is true in this
case, then it is generally true.
The specific case we will consider is depicted in figure~\ref{fig:bn:D-ID-Alg-G}.

\begin{center}
\begin{figure}[h]
	\[
	\ensuremath{\begin{array}{c}\input{pstex/Struc-Gen-Alg-A.pstex_t} \end{array}} 
		\hspace{1em} - \hspace{1em} 
	\ensuremath{\begin{array}{c}\input{pstex/Struc-Gen-Alg-B.pstex_t} \end{array}} 
		\hspace{1em} + \hspace{1em}
	\ensuremath{\begin{array}{c}\input{pstex/Struc-Gen-Alg-C.pstex_t} \end{array}}
		\hspace{1em} - \hspace{1em}
	\ensuremath{\begin{array}{c}\input{pstex/Struc-Gen-Alg-D.pstex_t} \end{array}}
	\]
\caption{Four diagrams which sum to zero when all fields are in the $A$-sector.}
\label{fig:bn:D-ID-Alg-G}
\end{figure}
\end{center}

\begin{assert}
If the diagrams of figure~\ref{fig:bn:D-ID-Alg-G} can be shown to sum to zero in 
the $A$-sector then we assert the following, without proof:
\begin{enumerate}
	\item 	these diagrams will sum to zero in all sectors;
	\item 	the sets of partner diagrams in which we exchange pushes forward
			for pulls back (and \emph{vice-versa}) also sum to zero.
\end{enumerate}
\label{assert:spec-gen}
\end{assert}

It is worth emphasising that there is no particular barrier to proving this
assertion. Indeed, intuitively it makes sense and a future task is to demonstrate
it explicitly.

We now take all fields in the diagrams of figure~\ref{fig:bn:D-ID-Alg-G} to be
in the $A$-sector.  In this sector, the roman indices reduce to Latin indices \emph{viz.}
$S_j \rightarrow \sigma_j$.
Assuming that these sub-diagrams attach to the same common structure, we can combine them,
algebraically: 
\begin{equation}
\frac{k_\rho}{k^2} \int_l \frac{l_{\sigma_1}}{l^2} \prod_{j=2}^r \frac{(l+\sum_{i=2}^j m_j)_{\sigma_j}}{(l+\sum_{i=2}^j m_j)^2}
\left[
	\begin{array}{l}
		\frac{(l-m_1)_\tau}{(l-m_1)^2} - \frac{(l-m_1-k)_\tau}{(l-m_1-k)^2} \frac{(l-m_1-k) \cdot (l-m_1)}{(l-m_1)^2}
	\\
		- \frac{(l-m_1-k)_\tau}{(l-m_1-k)^2}  \frac{k \cdot (l-m_1)}{(l-m_1)^2}
	\\
		+ \frac{(l-m_1-k)_\tau}{(l-m_1-k)^2}
	\\
		- \frac{(l-m_1)_\tau}{(l-m_1)^2}
	\end{array}
\right] = 0.
\label{eq:Alg-G-A-sec}
\end{equation}

Combining this result with assertion~\ref{assert:spec-gen} gives us a diagrammatic identity.

\begin{D-ID}
The set of sub-diagrams of figure~\ref{fig:bn:D-ID-Alg-G} sum to zero.
The complimentary sets of diagrams in which arbitrary numbers of pushes forward (pulls back)
are consistently changed to pulls back (pushes forward) also sum to zero.
\label{D-ID-Alg-G}
\end{D-ID}

\section{Application} \label{sec:D-ID-App}

We now apply the new diagrammatic identities to the
types of diagram that arise in our computation of $\beta_{n_+}$.
This is actually extremely easy! In figure~\ref{fig:D-ID-App}
we show how to re-express a typical series of
sub-diagrams. By enclosing each sub-diagram by $\AGRO{\hspace{1em}}$,
we mean that we should sum over Every Sense ($ES$) in which the gauge
remainders can push forward or pull back on to the designated
location. 

For example, sub-diagram~\ref{D-ID-App-LXL} which has two
gauge remainders---one of which is a push forward and the other of
 which is a pull back---is taken to 
represent four diagrams. The other three can be characterised
by $PF/PF$, $PB/PF$ and $PB/PB$. Note that we do not combine the
very first push forward with the corresponding pull back.
This is because we are dealing with sub-diagrams, and it is quite
possible that we have already `used up' this luxury, elsewhere in
the diagram.

\begin{center}
\begin{figure}
	\[
	\begin{array}{ll}
	\vspace{0.2in}
		&
		\AGRO{
			\LID[0.1]{}{Struc-LXL}{D-ID-App-LXL} \hspace{0.225em}
			+
			\left[
				\begin{array}{llll}
					& \LID[0.1]{}{Struc-LLXL}{D-ID-App-LLXL} & + & \cdots
				\\
					+ & \LID[0.1]{}{Struc-LXLL}{D-ID-App-LXLL} & + & \cdots
				\\ 
					& & & \vdots
				\end{array}
			\right]
		}
	\\
		= &
		\AGRO{
			\begin{array}{lll}
			\vspace{0.2in}
				\LID[0.1]{}{Struc-GR-LL}{D-ID-App-GR-LL} & + & \hspace{2.3em} \LID[0.1]{}{Struc-GR-LLL}{D-ID-App-GR-LLL} \hspace{1.6em} +\cdots
			\\
				& + &
				\left[
					\begin{array}{lll}
						\hspace{1em} \LID[0.1]{}{Struc-PB-TLTP-EP-LL}{D-ID-App-PB-TLTP-EP-LL} 	
						& \hspace{0.5em} +  \LID[0.1]{}{Struc-PB-TLTP-EP-LLL}{D-ID-App-PB-TLTP-EP-LLL}  
						& +\cdots 
					\\
					 																
						& 
						\hspace{0.5em} + \LID[0.1]{}{Struc-L-PB-TLTP-EP-LL}{D-ID-App-L-PB-TLTP-EP-LL}
						& +\cdots
					\end{array}
				\right]
			\end{array}
		}
	\end{array}
	\]
\caption{A re-expression of a series of diagrams, using diagrammatic 
identities~\ref{D-ID-Alg-2}--\ref{D-ID-Alg-G}.}
\label{fig:D-ID-App}
\end{figure}
\end{center}

It is worth clarifying the diagrams represented by the various ellipses.
The ellipsis following diagram~\ref{D-ID-App-LLXL} denotes additionally
nested diagrams, where the two-point, tree level vertex is still
joined to the innermost gauge remainder. The ellipsis after
diagram~\ref{D-ID-App-LXLL} denotes additionally nested diagrams,
where the two-point, tree level vertex joins to the innermost
but one gauge  remainder. The vertical dots on the next line represent additionally
nested diagrams in which the two-point, tree level vertex joins to
gauge remainders successively further away from the innermost one.
Note, though, that we never make the join to the outermost gauge remainder.

The meaning of the remaining ellipses should be clear. We draw attention
to the fact that $\AGRO{\hspace{1em}}$ refers to \emph{all} gauge remainders. Thus, for example, in
diagram~\ref{D-ID-App-PB-TLTP-EP-LL} we must sum over not just the different senses
in which the two gauge remainders at the top of the diagram can bite each other,
but also the sense in which the other gauge remainder strikes the vertex.
Diagrams related to each other by the diagrammatic identities are those which
possess the same total number of processed gauge remainders.

The relationship shown in figure~\ref{fig:D-ID-App}
can now be directly applied to specific diagrams,
coming from the computation of $\beta_{n_+}$. There are two cases
to deal with. Focusing on the sub-diagrams enclosed by the first $\AGRO{\hspace{1em}}$,
we can either attach the bottommost gauge remainder to some independent
structure or to one of the empty sockets in the sub-diagrams.

To obtain the set of terms in the first class, we start with
diagrams~\ref{GR-Ia-Nest1-ExtraTerm-Ex}, \ref{GR-Ia-Nest1-ExtraTerm-Ex-PB},
\ref{GR-Ib-EP-RXR} and~\ref{GR-Ib-EP-LXR}. We then add their
analogues with a higher level of nesting. Finally, for each diagram
now in the set, we include their analogues possessing additional vertices.

To obtain the set of terms in the second class
we start with: diagrams~\ref{GR-Ib-EP-RXR}--\ref{GR-Ib-EPGRx2-RXR-PB},
diagrams~\ref{GR-II-W-RXR}--\ref{Struc-WGRx2-B-N2-Alg-D2}
and the version of the final set possessing a trapped gauge remainder.
To generate the complete set of terms, we now employ the usual
procedure of including additional levels of nesting and extra vertices,
as appropriate.

Redrawing the terms of the first class is trivial. The second
class is not much harder but, having applied the diagrammatic
identities, some of the resulting terms can be combined both
with other redrawn terms and also terms from elsewhere in the calculation.
Consequently, it is the second set of terms we will focus on.

We start by looking at diagrams~\ref{GR-II-W-RXR} and~\ref{GR-II-W-LXR}.
Redrawing these terms 
 using diagrammatic identity~\ref{D-ID-Alg-2} leads them to take a very 
similar form to diagrams~\ref{GR-WGR-N1-Top-PB-LHGR}
and~\ref{GR-WGR-N1-Top-PF-RHGR}---they look identical but for the fact that the
trapped gauge remainder is on the other side of the wine. In fact, 
it makes no difference which side of the wine the trapped gauge remainder lies; the two
cases are identical as can be demonstrated by a trivial redrawing.

To the four diagrams comprising the two we have redrawn and diagrams~\ref{GR-WGR-N1-Top-PB-LHGR}
and~\ref{GR-WGR-N1-Top-PF-RHGR} we now add diagrams~\ref{GR-Ib-EP-RXR} and~\ref{GR-Ib-EP-LXR}.
This has the effect of removing the undecorated component of the wine.\footnote{At this
level of nesting, the wine must be decorated, irrespective of the value
of $n$. This is an example of how retaining terms which can be argued to vanish,
obviates the need to make such an argument.} The result is shown in
figure~\ref{fig:bn:GR-II-Redraw}.
\begin{center}
\begin{figure}
	\[
	\begin{array}{c}
		\displaystyle
		2\norm_{n-1,0}
	\\
		\times
		\dec{
			\CGD[0.1]{}{Struc-RW-GR-RR}{GR-WGR-N1-Top-PB-LHGR-B}{DGRx2-WGR-N1-Top-PB-LHGR}
			-\CGD[0.1]{}{Struc-GR-RW-LR}{GR-WGR-N1-Top-PF-RHGR-B}{DGRx2-WGR-N1-Top-PF-RHGR}
			+\CGD[0.1]{}{Struc-GR-RW-RR}{GR-WGR-N1-Top-PB-RHGR-B}{DGRx2-WGR-N1-Top-PB-RHGR}
			-\CGD[0.1]{}{Struc-RW-GR-LR}{GR-WGR-N1-Top-PF-LHGR-B}{DGRx2-WGR-N1-Top-PF-LHGR}
		}{11\Delta^{n-1}}
	\end{array}
	\]
\caption{Combining diagrams~\ref{GR-WGR-N1-Top-PB-LHGR}, \ref{GR-WGR-N1-Top-PF-RHGR},
\ref{GR-II-W-RXR} and~\ref{GR-II-W-LXR}
with diagrams~\ref{GR-Ib-EP-RXR} and~\ref{GR-Ib-EP-LXR}.}
\label{fig:bn:GR-II-Redraw}
\end{figure}
\end{center}

Now we want to move on to see what happens in the additionally nested case. 
Our first task is to generate the nested versions of 
diagrams~\ref{GR-II-W-RXR} and~\ref{GR-II-W-LXR}. This is done
by combining the diagram~\ref{GR-WGR-N2-B-PB-PB}, \ref{GR-WGR-N2-B-PB-PF}, \ref{GR-WGR-N2-B-PF-PB}
and~\ref{GR-WGR-N2-B-PF-PF}; \ref{GR-WGR-N2-B-PB-PB2}, \ref{GR-WGR-N2-B-PB-PF2},
\ref{GR-WGR-N2-B-PF-PB2} and~\ref{GR-WGR-N2-B-PF-PF2} and the nested versions
of diagrams~\ref{GR-WGR-N1-Bottom-PB-GR-Trap} and~\ref{GR-WGR-N1-Bottom-PF-GR-Trap}.
Combining terms with trapped gauge remainders with the corresponding terms
that do not gives us the eight diagrams of figure~\ref{fig:bn:GR-II-N2-Wine/Combo}.
\begin{center}
\begin{figure}[h]
	\[
	\begin{array}{c}
	\vspace{0.1in}
		\displaystyle
		2 \norm_{n-1,0}
	\\
		\times
		\dec{
			\begin{array}{c}
			\vspace{0.2in}
				-\PGD[0.1]{}{Struc-W-RLR-TLTP-EP}{GR-II-W-RLR-TLTP-EP}{fig:bn:GR-II-N2-Wine/Combo-Redraw}
				+\PGD[0.1]{}{Struc-W-RLL-TLTP-EP}{GR-II-W-RLL-TLTP-EP}{fig:bn:GR-II-N2-Wine/Combo-Redraw}
				+\PGD[0.1]{}{Struc-W-RRR-TLTP-EP}{GR-II-W-RRR-TLTP-EP}{fig:bn:GR-II-N2-Wine/Combo-Redraw}
				-\PGD[0.1]{}{Struc-W-RRL-TLTP-EP}{GR-II-W-RRL-TLTP-EP}{fig:bn:GR-II-N2-Wine/Combo-Redraw}
			\\
				-\PGD[0.1]{}{Struc-W-RLR-TLTP-EP-B}{GR-II-W-RLR-TLTP-EP-B}{fig:bn:GR-II-N2-Wine/Combo-Redraw}
				+\PGD[0.1]{}{Struc-W-RLL-TLTP-EP-B}{GR-II-W-RLL-TLTP-EP-B}{fig:bn:GR-II-N2-Wine/Combo-Redraw}
				+\PGD[0.1]{}{Struc-W-RRR-TLTP-EP-B}{GR-II-W-RRR-TLTP-EP-B}{fig:bn:GR-II-N2-Wine/Combo-Redraw}
				-\PGD[0.1]{}{Struc-W-RRL-TLTP-EP-B}{GR-II-W-RRL-TLTP-EP-B}{fig:bn:GR-II-N2-Wine/Combo-Redraw}
			\end{array}
		}{11\Delta^{n-1}}
	\end{array}
	\]
\caption{Result of combining the nested versions of diagrams~\ref{GR-II-WGR-N1-Bottom-PB}
and~\ref{GR-II-WGR-N1-Bottom-PF} with the nested versions
of diagrams~\ref{GR-WGR-N1-Bottom-PB-GR-Trap} and~\ref{GR-WGR-N1-Bottom-PF-GR-Trap}.}
\label{fig:bn:GR-II-N2-Wine/Combo}
\end{figure}
\end{center}

The eight diagrams of figure~\ref{fig:bn:GR-II-N2-Wine/Combo}
naturally combine with diagrams~\ref{GR-Ib-WBB-L-PFX}--\ref{Struc-WGRx2-B-N2-Alg-D2}.
We now apply diagrammatic
identity~\ref{D-ID-Alg-3} to diagrams~\ref{GR-II-W-RLR-TLTP-EP}--\ref{GR-II-W-RRL-TLTP-EP}
and~\ref{Struc-WGRx2-B-N2-Alg-A2}--\ref{Struc-WGRx2-B-N2-Alg-D2}
and diagrammatic
identity~\ref{D-ID-Alg-2} to diagrams~\ref{GR-II-W-RLR-TLTP-EP-B}--\ref{GR-II-W-RRL-TLTP-EP-B}
and~\ref{GR-Ib-WBB-L-PFX}--\ref{GR-Ib-WBB-R-PBX}.
The result of these redrawing is shown in figure~\ref{fig:bn:GR-II-N2-Wine/Combo-Redraw}
where, in particular cases, we have collected terms together to recreate a two-point, tree level
vertex attached to an effective propagator.

\begin{center}
\begin{figure}
	\[
	\begin{array}{c}
	\vspace{0.1in}
		\displaystyle
		2 \norm_{n-1,0}
	\\
		\times
		\dec{
			\begin{array}{c}
			\vspace{0.2in}
				\PGD[0.1]{}{Struc-W-GR-LLXL}{GR-II-W-GR-LLXL}{fig:bn:GR-II-N2-Wine/Combo-Redraw-B}
				+\PGD[0.1]{}{Struc-W-GR-LLLX}{GR-II-W-GR-LLLX}{fig:bn:GR-II-N2-Wine/Combo-Redraw-B}
				-\PGD[0.1]{}{Struc-W-GR-LRLX}{GR-II-W-GR-LRLX}{fig:bn:GR-II-N2-Wine/Combo-Redraw-B}
				-\PGD[0.1]{}{Struc-W-GR-LRXL}{GR-II-W-GR-LRXL}{fig:bn:GR-II-N2-Wine/Combo-Redraw-B}
			\\
			\vspace{0.2in}
				-\PGD[0.1]{}{Struc-W-GR-LLXR}{GR-W-GR-LLXR}{fig:bn:GR-II-N2-Wine/Combo-Redraw-B}
				-\PGD[0.1]{}{Struc-W-GR-LLRX}{GR-W-GR-LLRX}{fig:bn:GR-II-N2-Wine/Combo-Redraw-B}
				-\PGD[0.1]{}{Struc-W-GR-RLLX}{GR-W-GR-RLLX}{fig:bn:GR-II-N2-Wine/Combo-Redraw-B}
				-\PGD[0.1]{}{Struc-W-GR-RLXL}{GR-W-GR-RLXL}{fig:bn:GR-II-N2-Wine/Combo-Redraw-B}
			\\
			\vspace{0.2in}
				-\PGD[0.1]{}{Struc-W-PF-TLTP-EP-RR}{GR-II-W-PF-TLTP-EP-RR}{fig:bn:GR-II-N2-Wine/Combo-Redraw-B}
				+\PGD[0.1]{}{Struc-W-PB-TLTP-EP-RR}{GR-II-W-PB-TLTP-EP-RR}{fig:bn:GR-II-N2-Wine/Combo-Redraw-B}
				+\PGD[0.1]{}{Struc-W-PF-TLTP-EP-LR}{GR-II-W-PF-TLTP-EP-LR}{fig:bn:GR-II-N2-Wine/Combo-Redraw-B}
				-\PGD[0.1]{}{Struc-W-PB-TLTP-EP-LR}{GR-II-W-PB-TLTP-EP-LR}{fig:bn:GR-II-N2-Wine/Combo-Redraw-B}
			\\
			\vspace{0.2in}
				-\CGD[0.1]{}{Struc-WGRx2-N1-T-T-PF-PF2}{GR-II-WGRx2-N1-T-T-PF-PF2}{DGRx2-WGRx2-N1-T-T-PF-PF2}
				+\CGD[0.1]{}{Struc-WGRx2-N1-T-T-PB-PF2}{GR-II-WGRx2-N1-T-T-PB-PF2}{DGRx2-WGRx2-N1-T-T-PB-PF2}
				+\CGD[0.1]{}{Struc-WGRx2-N1-T-T-PF-PB}{GR-II-WGRx2-N1-T-T-PF-PB}{DGRx2-WGRx2-N1-T-T-PF-PB}
				-\CGD[0.1]{}{Struc-WGRx2-N1-T-T-PB-PB}{GR-II-WGRx2-N1-T-T-PB-PB}{DGRx2-WGRx2-N1-T-T-PB-PB}
			\\
				+\CGD[0.1]{}{Struc-WBT-GR-LR}{GR-II-N1-WBB-PB-PB-TLTP-Alg-RD}{DGRx2-WBT-GR-LR}
				-\CGD[0.1]{}{Struc-WBT-PB-GR-LR}{GR-II-N1-WBB-PF-PF-TLTP-Alg-RD}{DGRx2-WBT-PB-GR-LR}
				-\CGD[0.1]{}{Struc-WBT-GR-RR}{GR-II-N1-WBB-PB-PF-TLTP-Alg-RD}{DGRx2-WBT-GR-RR}
				+\CGD[0.1]{}{Struc-WBT-PB-GR-RR}{GR-II-N1-WBB-PF-PB-TLTP-Alg-RD}{DGRx2-WBT-PB-GR-RR}
			\end{array}
		}{11\Delta^{n-1}}
	\end{array}
	\]
\caption{Result of applying diagrammatic identities~\ref{D-ID-Alg-2} and~\ref{D-ID-Alg-3},
as appropriate, to  diagrams~\ref{GR-II-W-RLR-TLTP-EP}--\ref{GR-II-W-RRL-TLTP-EP-B}
and~\ref{GR-Ib-WBB-L-PFX}--\ref{Struc-WGRx2-B-N2-Alg-D2}. }
\label{fig:bn:GR-II-N2-Wine/Combo-Redraw}
\end{figure}
\end{center}

We are nearly at the nested version of figure~\ref{fig:bn:GR-II-Redraw}. To
get there, there are two more things we must do. First, we need to include
the nested versions of diagrams~\ref{GR-WGR-N1-Top-PB-LHGR} and~\ref{GR-WGR-N1-Top-PF-RHGR}.
This will simply have the effect of doubling the contribution
coming from diagrams~\ref{GR-II-W-GR-LLXL}--\ref{GR-W-GR-RLXL}.
Secondly, we should add diagrams~\ref{GR-Ib-EPBB-L-PFX}--\ref{GR-Ib-EPGRx2-RXR-PB}
and the nested version of diagrams~\ref{GR-Ib-EP-RXR} and~\ref{GR-Ib-EP-LXR},
having redrawn the lot via diagrammatic identity~\ref{D-ID-Alg-2}.
The combined result of both of these procedures is shown in figure~\ref{fig:bn:GR-II-N2-Wine/Combo-Redraw-B}.
\begin{center}
\begin{figure}
	\[
	\begin{array}{c}
	\vspace{0.1in}
		\displaystyle
		2 \norm_{n-1,0}
	\\
		\times
		\dec{
			\begin{array}{c}
			\vspace{0.2in}
				2
				\left[
					\begin{array}{c}
					\vspace{0.2in}
						\LGD[0.1]{}{Struc-RW-GR-LLXL}{GR-II-W-RGR-LLXL}
						+\LGD[0.1]{}{Struc-RW-GR-LLLX}{GR-II-W-RGR-LLLX}
						-\LGD[0.1]{}{Struc-RW-GR-LRLX}{GR-II-W-RGR-LRLX}
						-\LGD[0.1]{}{Struc-RW-GR-LRXL}{GR-II-W-RGR-LRXL}
					\\
					\vspace{0.2in}
						-\LGD[0.1]{}{Struc-RW-GR-LLXR}{GR-W-RGR-LLXR}
						-\LGD[0.1]{}{Struc-RW-GR-LLRX}{GR-W-RGR-LLRX}
						-\LGD[0.1]{}{Struc-RW-GR-RLLX}{GR-W-RGR-RLLX}
						-\LGD[0.1]{}{Struc-RW-GR-RLXL}{GR-W-RGR-RLXL}
					\\
						+\LGD[0.1]{}{Struc-DEPBB-PB-TLTP-EP-LL}{GR-DEPBB-PB-TLTP-EP-LL}
						-\LGD[0.1]{}{Struc-DEPBB-PB-TLTP-EP-RL}{GR-DEPBB-PB-TLTP-EP-RL}
						-\LGD[0.1]{}{Struc-DEPBB-PF-TLTP-EP-LL}{GR-DEPBB-PF-TLTP-EP-LL}
						+\LGD[0.1]{}{Struc-DEPBB-PF-TLTP-EP-RL}{GR-DEPBB-PF-TLTP-EP-RL}
					\end{array}
				\right]
			\\
			\vspace{0.2in}
				+\CGD[0.1]{}{Struc-RXR-DEP-PF-TLTP-EP}{GRs-RXR-DEP-PF-TLTP-EP}{DGRx2-RXR-DEP-PF-TLTP-EP}
				-\CGD[0.1]{}{Struc-RXR-DEP-PB-TLTP-EP}{GRs-RXR-DEP-PB-TLTP-EP}{DGRx2-RXR-DEP-PB-TLTP-EP}
				-\CGD[0.1]{}{Struc-LXR-DEP-PF-TLTP-EP}{GRs-LXR-DEP-PF-TLTP-EP}{DGRx2-LXR-DEP-PF-TLTP-EP}
				+\CGD[0.1]{}{Struc-LXR-DEP-PB-TLTP-EP}{GRs-LXR-DEP-PB-TLTP-EP}{DGRx2-LXR-DEP-PB-TLTP-EP}
			\\
			\vspace{0.2in}
				-\CGD[0.1]{}{Struc-RXR-DW-PF-TLTP-EP}{GRs-RXR-DW-PF-TLTP-EP}{DGRx2-RXR-DW-PF-TLTP-EP}
				+\CGD[0.1]{}{Struc-RXR-DW-PB-TLTP-EP}{GRs-RXR-DW-PB-TLTP-EP}{DGRx2-RXR-DW-PB-TLTP-EP}
				+\CGD[0.1]{}{Struc-LXR-DW-PF-TLTP-EP}{GRs-LXR-DW-PF-TLTP-EP}{DGRx2-LXR-DW-PF-TLTP-EP}
				-\CGD[0.1]{}{Struc-LXR-DW-PB-TLTP-EP}{GRs-LXR-DW-PB-TLTP-EP}{DGRx2-LXR-DW-PB-TLTP-EP}
			\end{array}
		}{11\Delta^{n-1}}
	\end{array}
	\]
\caption[Result of combining diagrams~\ref{GR-II-W-GR-LLXL}--\ref{GR-II-W-PB-TLTP-EP-LR},
 nested versions of diagrams~\ref{GR-WGR-N1-Top-PB-LHGR} and~\ref{GR-WGR-N1-Top-PF-RHGR},
diagrams~\ref{GR-Ib-EPBB-L-PFX}--\ref{GR-Ib-EPGRx2-RXR-PB} and
the nested version of diagrams~\ref{GR-Ib-EP-RXR} and~\ref{GR-Ib-EP-LXR}.]{
Result of combining diagrams~\ref{GR-II-W-GR-LLXL}--\ref{GR-II-W-PB-TLTP-EP-LR},
 nested versions of diagrams~\ref{GR-WGR-N1-Top-PB-LHGR} and~\ref{GR-WGR-N1-Top-PF-RHGR},
diagrams~\ref{GR-Ib-EPBB-L-PFX}--\ref{GR-Ib-EPGRx2-RXR-PB} and
the nested version of diagrams~\ref{GR-Ib-EP-RXR} and~\ref{GR-Ib-EP-LXR}. We have concluded that
there is only one thing more tedious than drawing a thousand diagrams by hand;
and that is drawing them on a computer.}
\label{fig:bn:GR-II-N2-Wine/Combo-Redraw-B}
\end{figure}
\end{center}

It is now straightforward to predict the result of going
to arbitrarily higher levels of nesting: we generate
ever more nested versions of 
diagrams~\ref{GR-II-WGRx2-N1-T-T-PF-PF2}--\ref{GRs-LXR-DW-PB-TLTP-EP}.
This, in turn, tells us what to expect from those diagrams
possessing both a two-point, tree level vertex attached
to a bitten gauge remainder and a trapped gauge remainder
(\eg\ diagrams~\ref{GR-II-WGRx2-Trap-Alg-A}, \ref{GR-II-WGRx2-Trap-Alg-B}
and~\ref{GR-n_J-R-RWBB-Trap-GR-PF-PF-TLTP-EPGR}--\ref{GR-n_J-R-RWBB-PB-Trap-GR-PB-TLTP-EPGR}).
At the lowest level of nesting, we will generate versions of diagrams~\ref{GR-WGR-N1-Top-PB-LHGR-B}
and~\ref{GR-WGR-N1-Top-PF-RHGR-B} with an additional trapped gauge remainder
and the opposite sign. At the next level of nesting, we simply generate
a version of figure~\ref{fig:bn:GR-II-N2-Wine/Combo-Redraw}
where we make the following changes: first, change the overall sign; secondly, include
a trapped gauge remainder; thirdly reduce the wine. For successive levels of
nesting, we take the set of diagrams just generated and add further nested gauge
remainders.

\chapter{Iterating the Diagrammatic Procedure} \label{ch:bn:Iterate}

The next phase of the calculation is to convert sets of surviving gauge
remainder diagrams into $\Lambda$-derivative terms, plus corrections. This
is essentially an iteration of what we have done already: we expect the
first $\Lambda$-derivative terms to spawn further $\Lambda$-derivative
terms, together with gauge remainders and etc. This part of the 
calculation will just draw on the techniques given in 
chapter~\ref{ch:bn:LambdaDerivsI}.  Then we can
process these new gauge remainders, using the techniques of chapter~\ref{ch:bn:GRs},
 ultimately arriving at a set of terms
we can once more convert into $\Lambda$-derivatives, and so forth.

The advantage of having left this phase until now is that the diagrammatic
identities of the chapter~\ref{ch:FurtherD-ID} will enable us to immediately identify
cancellations which would otherwise have remained hidden. We begin
by collecting terms spawned by gauge remainders of types-Ia, Ib and II,
components of which naturally combine to give $\Lambda$-derivative terms. 

The first case we will examine is the very simplest, corresponding to diagrams in which
the gauge remainder structure
is un-nested. Starting from diagrams~\ref{GR-Ib-n_J-R-WGR-T},  \ref{GR-Ib-n_J-R-DWGR-Top}
and~\ref{GR-n_J+r-R-W-WBT}--\ref{GR-V01-R-EP-LdL-WBT}
we re-express them as shown in figure~\ref{fig:bn:GR-Ia,Ib-n_J-R-LdL-pre}.

\begin{center}
\begin{figure}[h]
	\[
	\begin{array}{c}
	\vspace{0.2in}
		\displaystyle
		-4 \sum_{J=0}^n \norm_{J,0}
		\dec{
			\PGD[0.1]{}{Struc-n_J-R-LdL-WBT}{GR-n_J-R-LdL-WBT}{fig:bn:GR-Ia,Ib-n_J-R-LdL-pre-B}
			+\CGD[0.1]{}{Struc-n_J-R-RW-WBT}{GR-n_J-R-RW-WBT}{DGR-n_J-R-RW-WBT}
		}{11\Delta^J}	
	\\
	\vspace{0.2in}
		\displaystyle
		4 \sum_{J=0}^n \sum_{r=0}^{n_J} \norm_{J+1,0}
		\dec{
			\sco[0.1]{
				\CGD[0.1]{}{Struc-n_J+r-R-LdL-WBT}{GR-n_J+r-R-LdL-WBT}{DGR-n_J+r-R-LdL-WBT}
				+\LGD[0.1]{}{Struc-n_J+r-R-RW-WBT}{GR-n_J+r-R-RW-WBT}
			}{\ensuremath{\begin{array}{c}\input{pstex/Vertex-r-R.pstex_t} \end{array}}}
		}{11\Delta^{J+1}}
	\\
		\displaystyle
		+4 \sum_{s=1}^{n_+}\sum_{J=2}^{2n}  \frac{\norm_{J+s-1,J-1}}{J}
		\dec{
			\sco[0.1]{
				\LGD[0.1]{}{Struc-V01-R-LdL-WBT}{GR-V01-R-LdL-WBT}
				+\LGD[0.1]{}{Struc-V01-R-RW-WBT}{GR-V01-R-RW-WBT}
			}{\left[ \ds\prod_{I=1}^{J-1} \sum_{V^{I_+}=0}^{V^I} \ensuremath{\begin{array}{c}\input{pstex/GV-I-I+.pstex_t} \end{array}} \right] \ensuremath{\begin{array}{c}\input{pstex/Vertex-VJ-R.pstex_t} \end{array}} }
		}{11\Delta^{J+s-1}}
	\end{array}
	\]
\caption{Combination of diagrams~\ref{GR-Ib-n_J-R-WGR-T},  \ref{GR-Ib-n_J-R-DWGR-Top}
and~\ref{GR-n_J+r-R-W-WBT}--\ref{GR-V01-R-EP-LdL-WBT}.}
\label{fig:bn:GR-Ia,Ib-n_J-R-LdL-pre}
\end{figure}
\end{center}

We note in passing that we can remove the restriction that the vertex of
diagram~\ref{GR-n_J-R-LdL-WBT}  be reduced, as the decorations
force it to be at least three-point, irrespective of the value of $n$. 
However, we will not perform this step as
we will then have to undo it, when we come to deal with diagrams in which the
gauge remainder structure is nested. As we iterate the diagrammatic procedure,
though, we will encounter diagrams for which such restrictions can be removed,
irrespective of the level of nesting of the gauge remainder structure.

Diagram~\ref{GR-n_J-R-LdL-WBT} is converted
into a $\Lambda$-derivative term, as shown in figure~\ref{fig:bn:GR-Ia,Ib-n_J-R-LdL-pre-B}.
\begin{center}
\begin{figure}
	\[
	-4 \sum_{J=0}^n \norm_{J,0}
	\left[
		\begin{array}{c}
		\vspace{0.2in}
			\dec{
				\dec{
					\PGD[0.1]{}{Struc-n_J-R-WGR-Top-LdL}{GR-n_J-R-WGR-Top-LdL}{fig:bn:GR-Ia,Ib-n_J-R-LdL}
				}{\bullet}
				-\PGD[0.1]{}{Struc-LdL-n_J-R-WGR-Top}{GR-LdL-n_J-R-WGR-Top}{fig:bn:GR-Ia,Ib-n_J-R-LdL}
			}{11\Delta^J}
		\\ 
			-J
			\dec{
				\PGD[0.1]{}{Struc-n_J-R-WGR-Top-Dloop}{GR-n_J-R-WGR-Top-Dloop}{fig:bn:GR-Ia,Ib-n_J-R-LdL}
			}{11\Delta^{J-1}}
		\end{array}
	\right]
	\]
\caption{Converting diagram~\ref{GR-n_J-R-LdL-WBT} into a $\Lambda$-derivative term.}
\label{fig:bn:GR-Ia,Ib-n_J-R-LdL-pre-B}
\end{figure}
\end{center}

To prepare for the treatment of the diagrams of figure~\ref{fig:bn:GR-Ia,Ib-n_J-R-LdL-pre-B}
and, in the process, for the treatment of terms containing nested gauge remainders,
we can guess from our earlier work that it will be desirable to promote $\WBT$
into a decoration. 

Having done this, we will detach the gauge remainder from the effective propagator.
Since it can now attach to either end of any of the effective propagators, we must compensate for
this step with a factor of $1/2(J+1)$.
The rule for its attachment
is that it must either bite another gauge remainder (the nested case) or the end of a wine.
Furthermore, it can attach either as a
push-forward or a pull-back. Since, for the term under consideration these add, we must
compensate by a further factor of two.

Figure~\ref{fig:bn:GR-Ia,Ib-n_J-R-LdL} shows the re-expression of the diagrams of 
figure~\ref{fig:bn:GR-Ia,Ib-n_J-R-LdL-pre-B} using the new notation, where
the vertex struck by the $\Lambda$-derivative has been processed.
\begin{center}
\begin{figure}
	\[
	\begin{array}{c}
	\vspace{0.2in}
		\displaystyle
		-2 \sum_{J=0}^n \norm_{J+1,0}
		\dec{
			\begin{array}{c}
			\vspace{0.2in}
				\displaystyle
				\dec{
					\LGD[0.1]{}{Vertex-n_J-R}{Vertex-n_J-R-GR-LdL}
				}{\bullet}
				-\sum_{r=1}^{n_J}
				\left[
					2(n_{J+r}-1)\beta_r + \gamma_r \pder{}{\alpha}
				\right]
				\ensuremath{\begin{array}{c}\input{pstex/Vertex-n_J+r-R.pstex_t} \end{array}}
			\\
				\displaystyle
				-\frac{1}{2}\sum_{r=0}^{n_J}
				\left[
					\PGD[0.1]{}{Dumbell-n_J+r-r}{GR-1-LdL-a0}{fig:bn:DGR:TLTP}
				\right]_R
				+\frac{1}{2}
				\ensuremath{\begin{array}{c}\input{pstex/Sigma_n-J+1.pstex_t} \end{array}}
			\end{array}
		}{11\Delta^{J+1}\wedge}
	\\
		\displaystyle
		+ \sum_{J=1}^n \norm_{J,0}
		\dec{
			\ensuremath{\begin{array}{c}\input{pstex/Sigma_n_J+1-Dloop.pstex_t} \end{array}}
		}{11\Delta^J\wedge}
	\end{array}
	\]
\caption{Re-expression of the diagrams of figure~\ref{fig:bn:GR-Ia,Ib-n_J-R-LdL-pre-B}.}
\label{fig:bn:GR-Ia,Ib-n_J-R-LdL}
\end{figure}
\end{center}

In the final diagram, we start the sum over $J$ from one, rather than from zero,
since we must have at least one effective propagator available with which to attach the
gauge remainder to some other structure.
Furthermore, we have removed the restriction that the vertex be reduced.
We do this because the vertex must be at least three-point, irrespective of both how many 
gauge remainders decorate the
diagram and the value of $n$.

This is looking extremely familiar. What we have here is essentially
a version of what we were doing in chapter~\ref{ch:bn:LambdaDerivsI}. The crucial 
difference is the presence of the decorative structure $\WBT$. 
Its behaviour is somewhere between that of the usual (internal) effective
propagators and the usual external fields. Like the former, it can be struck
by $\flowConstAl$ but, like the latter, it cannot be used to join different structures.

It is not hard to predict what our strategy will be: we will process diagram~\ref{GR-1-LdL-a0}
using the usual diagrammatic procedure. However, compared to the analogous manipulations performed
at the beginning of this chapter, we will have  extra terms, arising from the presence of the
structure $\WBT$.

First, we isolate the two-point, tree level vertices in diagram~\ref{GR-1-LdL-a0}. As
usual, we can decorate such vertices with external fields or use an effective propagator
to join them to some other structure. Now, however, we have the additional choice of
decorating with the structure $\WBT$. This is where all the subtleties in converting
terms containing gauge remainders into $\Lambda$-derivative terms arise: our entire
diagrammatic procedure revolves around isolating two-point, tree level vertices and
attaching them to other structures. Now, every time we create such a vertex, we can
join it to the decorative gauge remainder, which was never previously an option.

In this
particular case, though, there is very little complication, since an effective propagator
which attaches to the (un-nested) gauge remainder must be in the $C$-sector. Hence, if this effective
propagator attaches at the other end to a two-point, tree level vertex then, after
application of the effective propagator relation, there will be no additional gauge remainders.
When we move on to diagrams where we can join two-point, tree level vertices to
nested gauge remainders, however, the full effective propagator relation must be used.

Returning to the  case in hand, we iterate the usual diagrammatic 
procedure. The intermediate step of stripping off the two-point, tree
level vertices from diagram~\ref{GR-1-LdL-a0} is shown in figure~\ref{fig:bn:DGR:TLTP}.
\begin{center}
\begin{figure}[h]
	\[
	\begin{array}{c}
	\vspace{0.2in}
		\displaystyle
		\sum_{J=0}^n \norm_{J+1,0}
		\dec{
			\sum_{r=0}^{n_J} \LGD[0.1]{}{Dumbell-n_J+rR-rR}{DGR-Dumbell-n_J+rR-rR}
			-2 \PGD[0.1]{}{Dumbell-n_J-S-R-TLTP}{DGR-Dumbell-n_J-S-R-TLTP}{fig:bn:DGR:Dec}
		}{11\Delta^{J+1}\wedge}
	\\
		\displaystyle	
		-\norm_{n+1,0}
		\dec{
			\PGD[0.1]{}{Dumbell-TLTPx2}{DGR-Dumbell-TLTPx2}{fig:bn:DGR:Dec}
		}{11\Delta^{n+1}\wedge}
	\end{array}
	\]
\caption{Isolation of the two-point, tree level vertices of diagram~\ref{GR-1-LdL-a0}.}
\label{fig:bn:DGR:TLTP}
\end{figure}
\end{center}

The partial decoration of diagrams~\ref{DGR-Dumbell-n_J-S-R-TLTP} and~\ref{DGR-Dumbell-TLTPx2},
followed by the tying together of any loose ends and the 
application of the effective propagator relation, as appropriate,
is shown in figure~\ref{fig:bn:DGR:Dec}. Note that
the attachment of $\WBT$ to a two-point, tree level
vertex will come with a factor of $2\times 2(J+1)$,
corresponding to the number of different ways in which we can create $\WBT$. 
\begin{center}
\begin{figure}
	\[
	\begin{array}{c}
	\vspace{0.2in}
		\displaystyle
		-2\sum_{J=0}^n 
		\left[
			\norm_{J+1,0}
			\dec{
				\LGD[0.1]{}{Dumbell-n_J-S-R-TLTP-E}{DGR-n_J-S-R-TLTP-E}
			}{1\Delta^{J+1}\wedge}
			+2 \norm_{J,0}
			\dec{
				\CGD[0.1]{}{Struc-n_J-S-R-WGR-Top}{DGR-n_J-S-R-WGR-T}{GR-Ia-n_J-S-R-WGR-T}
			}{11\Delta^J}
		\right]
	\\
	\vspace{0.2in}
		\displaystyle
		-2\sum_{J=1}^n \norm_{J,0}
		\dec{
			\frac{1}{\vartheta_{n_J}} \LGD[0.1]{}{Sigma_n-J-S}{DGR-Sigma_n-J-S}
			-\LGD[0.1]{}{GV-n_J-S-R-GR}{DGR-n_J-S-R-GR}
			-\LGD[0.1]{}{GV-n_J-S-R-WGR}{DGR-n_J-S-R-WGR}
		}{11\Delta^J \wedge}
	\\
	\vspace{0.2in}
		\displaystyle
		2\norm_{n,0}
		\left[
			\dec{
				\LGD[0.1]{}{Struc-TLTP-WGR}{DGR-TLTP-WGR}
			}{1\Delta^n \wedge}
			-2
			\dec{
				\CGD[0.1]{}{Struc-TLTP-WGR-T-E}{DGR-TLTP-WGR-T-E}{GR-III-WBT}
			}{1\Delta^n}
		\right]
	\\
		\displaystyle
		+\norm_{n-1,0}
		\left[
			4
			\dec{
				\PGD[0.1]{}{Struc-WGRx2-Top}{DGR-WGRx2-Top}{fig:bn:GR-II/Iterate-Combo}
			}{11\Delta^{n-1}}
		-
			\dec{
				\LGD[0.1]{}{Struc-WGRx2}{DGR-WGRx2}
			}{11\Delta^{n-1} \wedge}
		\right]
	\end{array}
	\]
\caption{The result of processing diagrams~\ref{DGR-Dumbell-n_J-S-R-TLTP} and~\ref{DGR-Dumbell-TLTPx2}.}
\label{fig:bn:DGR:Dec}
\end{figure}
\end{center}

Diagrams~\ref{DGR-n_J-S-R-TLTP-E}, \ref{DGR-Sigma_n-J-S}--\ref{DGR-n_J-S-R-WGR}
\ref{DGR-TLTP-WGR} and~\ref{DGR-WGRx2}
are the types of diagram we were obtaining in chapter~\ref{ch:bn:LambdaDerivsI} 
with the difference that,
amongst the decorative fields, we have $\wedge$. Again, the 
constraint that there
must be an effective propagator with which to attach the gauge remainder
to some other structure
means that for 
diagrams~\ref{DGR-Sigma_n-J-S}--\ref{DGR-n_J-S-R-WGR}
we start the sum over $J$ from one, rather than zero. Encouragingly,
we find cancellations against terms generated by the original
set of gauge remainders.
\begin{cancel}
Diagram~\ref{DGR-n_J-S-R-WGR-T} exactly cancels diagram~\ref{GR-Ia-n_J-S-R-WGR-T}.
\label{cancel:DGR-n_J-S-R-WGR-T}
\end{cancel}

\begin{cancel}
Diagram~\ref{DGR-TLTP-WGR-T-E} cancels diagram~\ref{GR-III-WBT}
by virtue of the fact that the decorations force the wine of the latter
diagram to be decorated, for all values of $n$.
\label{cancel:DGR-TLTP-WGR-T-E}
\end{cancel}

We now consider continuing with the usual diagrammatic procedure, by processing the manipulable part of
diagram~\ref{DGR-Dumbell-n_J+rR-rR}. It is clear that this will just give us a version of 
figure~\ref{fig:Beta-n+:LJ:a1...a1-a0-M}, so long as we include the gauge remainder in the decorations
and add one extra term, corresponding to the $\Lambda$-derivative striking $\WBT$.

Let us compute the factor of this extra term, compared to the parent diagram.
First of all, being a correction term, it will come with a relative minus sign.
Promoting the wine to a decorative effective propagator will have yielded a factor of $1/2(J+2)$.
This $J$ dependence is compensated for by the $4(J+2)$ different ways in which we can create $\WBT$. 
We then have a choice of two vertices
to which to attach the structure $\dec{\WBT}{\bullet}$. Hence, the overall
factor, relative to the parent diagram, is $-4$. The corresponding term is the first
diagram in figure~\ref{fig:bn:DGR-L2-NewTerms}.

At the next stage of the diagrammatic procedure, too, we will have new diagrams. These arise
from attaching $\WBT$ to two-point, tree level vertices. Two such diagrams we have essentially
encountered already, as these are just versions of diagrams~\ref{DGR-n_J-S-R-WGR-T} 
and~\ref{DGR-WGRx2-Top} possessing an additional vertex. There are two more that we have
not yet encountered: taking the usual dumbbell structure with a two-point, tree level vertex
at either end, these correspond to attaching $\WBT$ to one of the vertices and joining
the other vertex to a reduced Wilsonian effective action vertex. The resulting terms
correspond to the final two diagrams of figure~\ref{fig:bn:DGR-L2-NewTerms}.
\begin{center}
\begin{figure}
	\[
	\begin{array}{c}
	\vspace{0.2in}
		\displaystyle
		-4 \sum_{J=0}^n \sum_{r=0}^{n_J} \norm_{J+1,0}
		\dec{
			\sco[0.1]{
				\CGD[0.1]{}{Struc-n_J+r-R-LdL-WBT}{DGR-n_J+r-R-LdL-WBT}{GR-n_J+r-R-LdL-WBT}
			}{\ensuremath{\begin{array}{c}\input{pstex/Vertex-r-R.pstex_t} \end{array}}}
		}{11\Delta^{J+1}}
	\\
		\displaystyle
		+4\sum_{J=0}^n \norm_{J,0}
		\dec{
			\CGD[0.1]{}{Struc-n_J-R-RW-WBT}{DGR-n_J-R-RW-WBT}{GR-n_J-R-RW-WBT}
			-\PGD[0.1]{}{Struc-n_J-R-GR-RW-WBT}{DGR-n_J-R-GR-RW-WBT}{fig:bn:GR-II/Iterate-Combo}
		}{11\Delta^J}	
	\end{array}
	\]
\caption{New types of diagram arising from iterating the diagrammatic procedure.}
\label{fig:bn:DGR-L2-NewTerms}
\end{figure}
\end{center}

\begin{cancel}
Diagram~\ref{DGR-n_J+r-R-LdL-WBT} exactly cancels diagram~\ref{GR-n_J+r-R-LdL-WBT}.
\label{cancel:DGR-n_J+r-R-LdL-WBT}
\end{cancel}

\begin{cancel}
Diagram~\ref{DGR-n_J-R-RW-WBT} exactly cancels diagram~\ref{GR-n_J-R-RW-WBT}.
\label{cancel:DGR-n_J-R-RW-WBT}
\end{cancel}

Iterating the diagrammatic procedure, it is apparent that 
the four cancellations~\ref{cancel:DGR-n_J-S-R-WGR-T}--\ref{cancel:DGR-n_J-R-RW-WBT} will go through just the same
at each level of manipulation.
Given the cancellation of these terms, cancellation 
mechanisms~\ref{CM:a1-S}--\ref{CM:MV-a1} now guarantee that we can reduce this sector of the
calculation to $\Lambda$-derivative, $\beta$ and $\gamma$ terms, up to further gauge remainders
and terms that require manipulation at $\Op{2}$.

Processing the new gauge remainders is easy! We have copies of all of 
the old gauge remainder diagrams, with the only difference that we must include
$\wedge$ as a decoration. There is also an
additional set of type-Ib gauge remainders,
corresponding to diagrams~\ref{DGR-WGRx2-Top} and~\ref{DGR-n_J-R-GR-RW-WBT} and their
analogues possessing additional vertices. 

Before doing anything with these terms, we note that we can combine them with two
diagrams coming from the \emph{original} set of type-II gauge remainders;
namely, diagrams~\ref{GR-II-n_J-R-GR-W-DGR} and~\ref{GR-II-WGRx2-Top}. The result of
this is shown in figure~\ref{fig:bn:GR-II/Iterate-Combo}.
\begin{center}
\begin{figure}
	\[
	\begin{array}{c}
	\vspace{0.2in}
		\ds
		2\sum_{J=0}^n \norm_{J,0}
		\dec{
			\PGD[0.1]{}{Struc-n_J-R-GR-DEP-R}{DGR-n_J-R-GR-DEP-R}{fig:bn:DGR-GR-Extra}
			-\PGD[0.1]{}{Struc-n_J-R-GR-RW-WBT}{DGR-n_J-R-GR-RW}{fig:bn:DGR-GR-Extra}
		}{11\Delta^J}
	\\
		\ds
		+2\norm_{n-1,0}
		\dec{
			\PGD[0.1]{}{Struc-WGRx2-Top}{DGR-WGRx2-Top-B}{fig:bn:DGR-GR-Extra}
		}{11\Delta^{n-1}}
	\end{array}
	\]
\caption{Result of combining diagrams~\ref{DGR-WGRx2-Top} and~\ref{DGR-n_J-R-GR-RW-WBT} with
diagrams~\ref{GR-II-n_J-R-GR-W-DGR} and~\ref{GR-II-WGRx2-Top}.}
\label{fig:bn:GR-II/Iterate-Combo}
\end{figure}
\end{center}

We can discard diagram~\ref{DGR-n_J-R-GR-DEP-R} since the wine, not being
decorated, is forced to be in the $C$-sector: it is clear that the 
unprocessed gauge remainder has no support. However, it is illuminating
to keep this diagram since it will serve to illustrate the structure
of cancellations in the case where the processed gauge remainder is
nested.

Allowing the gauge remainder to act in
diagrams~\ref{DGR-n_J-R-GR-DEP-R}--\ref{DGR-WGRx2-Top-B},
it is apparent that all generated terms in which the gauge remainder strikes
a socket which decorates anything other than a two-point, tree level vertex will 
simply cancel, via cancellation mechanism~\ref{CM:GR-Socket}.

Remembering to include terms that survive from 
the manipulation of diagrams~\ref{DGR-n_J-R-GR-DEP-R}--\ref{DGR-WGRx2-Top-B},
our analysis of the complete set of gauge remainders
goes through almost exactly as it did before. 
 We might worry that there are new terms
which arise when we decorate a two-point, tree level vertex generated by a gauge
remainder with $\WBT$. Upon utilising the effective propagator relation, 
however, these terms vanish, courtesy of diagrammatic identity~\ref{D-ID-Alg-1}.
The set of surviving terms is then just given by the survivors from 
sections~\ref{sec:GR-Ia}--\ref{sec:GR-III}---where we include $\wedge$ in the
decorations---and by the diagrams of figure~\ref{fig:bn:DGR-GR-Extra} and their
analogues containing greater number of vertices.
\begin{center}
\begin{figure}
	\[
	\begin{array}{c}
	\vspace{0.2in}
		\displaystyle
		4 \norm_{n-1,0}
		\dec{
			\DGD[0.1]{}{Struc-WGRx2-Top-Top}{DGR-WGRx2-Top-Top-PC}
			+\CGD[0.1]{}{Struc-WGRx2-Top-Bottom}{DGR-WGRx2-Top-Bottom-PC}{DGRx2-R-R}
			+\PGD[0.1]{}{Struc-WGRx2-WBS-Top-GR}{DGR-II-WGRx2-WBS-Top-GR-B}{fig:bn:DGR-GR-P}
		}{11\Delta^{n-1}}
	\\
		\displaystyle
		4 \sum_{J=0}^n \norm_{J,0}
		\dec{
			\PGD[0.1]{}{Struc-DGR-DEP-PF-n_J-R}{DGR-DEP-PF-n_J-R}{fig:bn:DGR-GR-P}
			-\PGD[0.1]{}{Struc-DGR-RW-PF-n_J-R}{DGR-RW-PF-n_J-R}{fig:bn:DGR-GR-P}
		}{11\Delta^J}
	\end{array}
	\]
\caption{The minimal set of new types of diagram arising from processing the gauge remainders spawned
by terms with $\wedge$ among the decorations.}
\label{fig:bn:DGR-GR-Extra}
\end{figure}
\end{center}


As we will see later, diagram~\ref{DGR-WGRx2-Top-Bottom-PC} will be cancelled, completely.
What, however, are we to do with diagram~\ref{DGR-WGRx2-Top-Top-PC}? The answer is that this
diagram vanishes via diagrammatic identity~\ref{D-ID-Alg-1}. 

Now we process the gauge remainders in diagrams~\ref{DGR-II-WGRx2-WBS-Top-GR-B}--\ref{DGR-RW-PF-n_J-R}
and their analogues containing additional vertices. All contributions in which a gauge remainder strikes 
a socket which decorates anything other than a two-point, tree level vertex will 
cancel, via cancellation mechanism~\ref{CM:GR-Socket}.
 Iterating the diagrammatic
procedure once will give us:
\begin{enumerate}
\item	versions of the parent diagrams, nested with respect
		to the already nested gauge remainder;

\item 	terms with an $\Op{2}$ stub;
 
\item	the diagrams of figure~\ref{fig:bn:DGR-GR-P} and their analogues, containing additional vertices.
\end{enumerate}

\begin{center}
\begin{figure}
	\[
	\begin{array}{c}
	\vspace{0.2in}
		\ds
		4 \norm_{n-1,0}
	\\
		\times
		\dec{
			\begin{array}{c}
			\vspace{0.2in}
				\DGD[0.1]{}{Struc-WGRx2-N1-T-T-PF-PF}{GR-WGRx2-N1-T-T-PF-PF}
				-\DGD[0.1]{}{Struc-WGRx2-N1-T-T-PB-PF}{GR-WGRx2-N1-T-T-PB-PF}
					+\PGD[0.1]{}{Struc-WGRx2-N1-T-B-PF-PF}{GR-WGRx2-N1-T-B-PF-PF}{fig:bn:DGRx2-GR-Extra/Combo}
				-\PGD[0.1]{}{Struc-WGRx2-N1-T-B-PB-PF}{GR-WGRx2-N1-T-B-PB-PF}{fig:bn:DGRx2-GR-Extra/Combo}
			\\
				+\PGD[0.1]{}{Struc-DGR-DEP-L-PF-Alg}{DGR-DEP-L-PF-Alg}{fig:bn:DGRx2-GR-Extra/Combo}
				-\PGD[0.1]{}{Struc-DGR-DEP-L-PB-Alg}{DGR-DEP-L-PB-Alg}{fig:bn:DGRx2-GR-Extra/Combo}
				-\PGD[0.1]{}{Struc-R-DEP-PF-PF-TLTP-Alg}{DGR-R-DEP-PF-PF-TLTP-Alg}{fig:bn:DGRx2-GR-Extra/Combo}
				+\PGD[0.1]{}{Struc-R-DEP-PF-PB-TLTP-Alg}{DGR-R-DEP-Pf-PF-TLTP-Alg}{fig:bn:DGRx2-GR-Extra/Combo}
		\end{array}
		}{11\Delta^{n-1}}
	\end{array}
	\]
\caption{The minimal set of terms spawned by diagrams~\ref{DGR-II-WGRx2-WBS-Top-GR-B}--\ref{DGR-RW-PF-n_J-R},
up to nested versions of the parent diagrams and diagrams with an $\Op{2}$ stub.}
\label{fig:bn:DGR-GR-P}
\end{figure}
\end{center}

Diagrams~\ref{GR-WGRx2-N1-T-T-PF-PF} and~\ref{GR-WGRx2-N1-T-T-PB-PF}
vanish by virtue of diagrammatic identity~\ref{D-ID-Alg-1}.

Iterating the diagrammatic procedure until exhaustion will give us,
up to terms with an $\Op{2}$ stub, increasingly nested versions
of the diagrams of figure~\ref{fig:bn:DGR-GR-P}, together with 
versions of these diagrams accompanied by additional vertices,
as appropriate. The nesting is with respect to the already
nested gauge remainder and so we need not bother considering
the nested versions of diagrams~\ref{GR-WGRx2-N1-T-T-PF-PF} and~\ref{GR-WGRx2-N1-T-T-PB-PF}
in this sense, since they will still vanish.

Having processed the gauge remainders, we arrive inevitably (for generic $n$)  at yet 
another set of terms which can be converted into $\Lambda$-derivatives: we obtain a
copy of figure~\ref{fig:bn:GR-Ia,Ib-n_J-R-LdL-pre} but where we have a $\wedge$ among the
decorations. The analogue of diagram~\ref{GR-n_J-R-LdL-WBT} is shown in 
figure~\ref{fig:bn:DGRx2-Nested-LdL-pre}.
\begin{center}
\begin{figure}
	\[
	-4 \sum_{J=0}^n \norm_{J,0}
	\dec{
		\ensuremath{\begin{array}{c}\input{pstex/Struc-n_J-R-LdL-WBT.pstex_t} \end{array}}
	}{11\Delta^{J}\wedge}
	=
	-8 \sum_{J=1}^n \norm_{J-1,0}
	\dec{
		\ensuremath{\begin{array}{c}\input{pstex/Struc-n_J-R-LdL-WBT-WBT.pstex_t} \end{array}}
	}{11\Delta^{J-1}}
	\]
\caption{An example of a candidate for conversion into a $\Lambda$-derivative term,
possessing the structure $\protect \WBT $ both as a decoration and as the component
of the diagram hit by $-\flow$.}
\label{fig:bn:DGRx2-Nested-LdL-pre}
\end{figure}
\end{center}

We see that we must take some care converting such a term into a $\Lambda$-derivative
term, since the structure $\WBT$ is repeated. The new rule is simple to guess:
to convert the first diagram into a $\Lambda$-derivative term, promote $\WBT$
to a decoration and divide not only by a factor of $4(J+1)$, but also by an
additional factor of two. This extra factor recognises that we now have two
gauge remainders with which to decorate the effective propagators.
The resulting $\Lambda$-derivative term is shown in figure~\ref{fig:bn:DGRx2-Nested-LdL}.
\begin{center}
\begin{figure}
	\[
	-\sum_{J=0}^n \norm_{J+1,0}
	\dec{
		\dec{
			\LGD[0.1]{}{Vertex-n_J-R}{DGRx2-Vertex-n_J-R}
		}{\bullet}
	}{11\Delta^{J+1}\wedge^2}
	+ \cdots
	\]
\caption{A $\Lambda$-derivative term for a diagram decorated by two gauge remainders.}
\label{fig:bn:DGRx2-Nested-LdL}
\end{figure}
\end{center}

To check that we can correctly reproduce the parent diagram, we will try
to recreate the rightmost diagram of figure~\ref{fig:bn:DGRx2-Nested-LdL-pre}.
To generate this term,
we must choose two
effective propagators from $(J+1)$ and two gauge remainders from two. We then use the
gauge remainders to decorate the ends of the effective propagators in all possible ways.
This gives a factor of
\[
\frac{(J+1)(J)}{2} \times 8.
\] 
Adding the push-forward to the pull-back---which, we recall, is legal for both gauge remainders
in this case---gives a further factor of four.
Combining with the factor in front of the $\Lambda$-derivative term and including
an extra factor of two since the $\Lambda$-derivative can hit either of the
$\WBT$ structures, it is clear that the parent diagram is reproduced.

There is, needless to say, a subtlety in what we have done that has not been
explored. Whilst the notation we have employed for the $\Lambda$-derivative
term most certainly includes decoration by two instances of the structure $\WBT$, 
it is natural to allow it to include also decoration with a structure in which
one of the effective propagators is bitten by a nested gauge remainder.
This would correspond to the diagrams of figure~\ref{fig:bn:DGRx2-Nested-LdL-missing}.
\begin{center}
\begin{figure}[h]
	\[
	2 \sum_{J=0}^n \norm_{J+1,0}
	\dec{
	\dec{
		\sco[0.1]{
			\LID[0.1]{}{Struc-GRx2}{Struc-GRx2-n_J-R} 
			- \LID[0.1]{}{Struc-GRx2-b}{Struc-GRx2-b-n_J-R} 
		}{\ensuremath{\begin{array}{c}\input{pstex/Vertex-n_J-R.pstex_t} \end{array}}}
	}{\bullet}
	}{11\Delta^{J+1}}
	\]
\caption{Further diagrams which should be amongst those represented by
diagram~\ref{DGRx2-Vertex-n_J-R}.}
\label{fig:bn:DGRx2-Nested-LdL-missing}
\end{figure}
\end{center}

Note the overall factor of diagrams~\ref{Struc-GRx2-n_J-R} and~\ref{Struc-GRx2-b-n_J-R}:
given that we collect together the first push-forward and pull-back, 
yielding a factor of two, there is only one independent 
way to generate each of these diagrams. 

Wonderfully, we see that such $\Lambda$-derivative terms arise, naturally,
from the nested version of diagram~\ref{GR-n_J-R-LdL-WBT}. (Note that the nested
version of diagram~\ref{GR-n_J-R-LdL-WBT} are derived from diagrams~\ref{GR-Ib-n_J-R-WGR-Out-GR},
\ref{GR-Ib-n_J+r-R-RWBS-GR} and~\ref{GR-Ia-WBS-GR-n_J+r-S-R-r-R}.)
This is, of course, no coincidence. Before we process diagram~\ref{DGRx2-Vertex-n_J-R},
we would first like to show that an analogous term correctly represents all
$\Lambda$-derivative terms possessing an arbitrary number of decorative gauge remainders.

We begin by considering a diagram with $m$ decorative gauge remainders.
The case where, upon decoration, these gauge remainders form a single structure 
has appeared already---see 
figures~\ref{fig:bn:GR-Nested-LdL-GR} and~\ref{fig:bn:GR-Nested-LdL-GR-B}. We now
wish to promote the gauge remainders comprising the `ring' of diagram~\ref{GR-ring-n_J-R}
to decorations. 

To find the associated factor that this yields we will work backwards, recreating the ring
from the decorations. We have already done this in the case where there are two decorative 
gauge remainders. For consistency with this, we require that the rule to create a ring
from decorations is simply that it is done in all independent ways. 
We start by picking out
one of the decorative gauge remainders, which can be done in $m$ ways.
 This can then
bite any of the remaining $m-1$ gauge remainders, which in turn can bite any of the
$m-2$ remaining gauge remainders and etc. We also pick up a factor of two from combining
the first push-forward and pull-back. However, the overall factor of $2m!$ over-counts 
the number of independent arrangements by a factor of $m$ as our $m$ objects
are arranged in a ring, rather than in a line. Consequently, when promoting
the ring of gauge remainders to a decoration, we must compensate by a factor
of $1/2(m-1)!$.

Referring back to diagram~\ref{GR-ring-n_J-R} 
(see figure~\ref{fig:bn:GR-Nested-LdL-GR-B}), and particularly noting 
the overall factor of $1/m$, it is apparent that a $\Lambda$-derivative term
comprising a single vertex and $m$-decorative gauge remainders, derived
from a diagram in which the gauge remainders form a single structure, can be
drawn as shown in figure~\ref{fig:bn:LdL-mGR}.
\begin{center}
\begin{figure}[h]
	\[
	-\frac{2}{m!} \sum_{J=0}^n \norm_{J+1,0}
	\dec{
		\dec{
			\LID[0.1]{}{Vertex-n_J-R}{Vertex-n_J-R-mGR-LdL}
		}{\bullet}
	}{11\Delta^{J+1} \GRkp^m}
	\]
\caption{The candidate for a $\Lambda$-derivative term with $m$
decorative gauge remainders.}
\label{fig:bn:LdL-mGR}
\end{figure}
\end{center}

Our task now is to show that diagram~\ref{Vertex-n_J-R-mGR-LdL} is the correct representation
for a $\Lambda$-derivative term comprising a single vertex and $m$ decorative gauge remainders,
however we choose to arrange the gauge remainders. In actual fact, our aim at this stage is
not quite this grand. In $\Lambda$-derivative terms where effective propagators join gauge remainders,
we know that parent diagrams of the form of diagram~\ref{GR-ring-n_J-R-Join}	should exist.
We are yet to show this even for the case where the gauge remainders form a single structure
and for the time being just assume that these terms will come with the correct
factors.

To begin our analysis of diagram~\ref{Vertex-n_J-R-mGR-LdL}, we must first ask what
constitutes a valid arrangement of gauge remainders. A valid arrangement possesses
an arbitrary number of clusters of gauge remainders. In each cluster, every gauge remainder
is bitten by another gauge remainder, from the same cluster. (In the case of \BT, we
consider the gauge remainder to be biting itself.) This means that diagrams such as
$\ensuremath{\begin{array}{c}\begin{picture}(0,0)%
\includegraphics{pstex/Struc-W-GR-BT-hook.pstex}%
\end{picture}%
\setlength{\unitlength}{3947sp}%
\begingroup\makeatletter\ifx\SetFigFont\undefined%
\gdef\SetFigFont#1#2#3#4#5{%
  \reset@font\fontsize{#1}{#2pt}%
  \fontfamily{#3}\fontseries{#4}\fontshape{#5}%
  \selectfont}%
\fi\endgroup%
\begin{picture}(659,235)(320,22)
\end{picture}
 \end{array}}$ are excluded, since the rightmost gauge remainder does
not bite another gauge remainder (see also section~\ref{sec:Diagrammatics:DoubleGRs}).

We can deduce the maximum value of $m$ (note that this will change for diagrams
with additional vertices). For every instance of $\GRkp$, there must be a
field to which it can be attached. The maximum number of available fields occurs when $J=n$.
In this case, we require a minimum of three fields to decorate the vertex\footnote{
If $J<n$, we need one less field to decorate the vertex, but the number of total 
available fields decreases by at least two.}. Therefore,
\[
	m_{\mathrm{max}} = 2n+1.
\]

The first arrangement of gauge remainders we will look at is one in which we suppose that each of
the gauge remainders forms an independent structure \BT. 
We will now suppose that diagram~\ref{Vertex-n_J-R-mGR-LdL} correctly represents such
terms for some value of $m<m_{\mathrm{max}}$, say $m'$. Following through the formation of diagram~\ref{DGRx2-Vertex-n_J-R},
it is apparent the corrections generated by the formation of diagram~\ref{Vertex-n_J-R-mGR-LdL}
yield, amongst others, the term shown in figure~\ref{fig:bn:LdL-m'GR-hook}
\begin{center}
\begin{figure}[h]
	\[
	-\frac{4}{m'!} \sum_{J=0}^n \norm_{J,0}
	\dec{
		\LID[0.1]{}{Struc-n_J-R-LdL-WBT}{LdL-m'GR-hook}
	}{11\Delta^{J} \GRkp^{m'}}
	\]
\caption{A term generated by the corrections to diagram~\ref{Vertex-n_J-R-mGR-LdL}
for $m=m'<m_{\mathrm{max}}$.}
\label{fig:bn:LdL-m'GR-hook}
\end{figure}
\end{center}

Converting diagram~\ref{LdL-m'GR-hook} into a $\Lambda$-derivative term, and
promoting $\WBT$ to a decoration yields a version of diagram~\ref{Vertex-n_J-R-mGR-LdL}
in which $m=m'+1$. Given that diagram~\ref{Vertex-n_J-R-mGR-LdL} correctly represents
a term in which all decorative gauge remainders form a \BT\ for $m=1$, it therefore
follows by
induction that  diagram~\ref{Vertex-n_J-R-mGR-LdL} correctly represents such terms for all $m$.

We now generalise this argument to confirm that diagram~\ref{Vertex-n_J-R-mGR-LdL}
works, however we arrange the gauge remainders. To this end, let us suppose
that we form $r$ clusters, each containing $q_i$ gauge remainders. The first thing
we note is that there are 
\[
	\frac{m!}{\sum_{i=1}^r q_i!}
\]
ways of partitioning $m$ gauge remainders into these $r$ clusters. Assuming that we do
not collect together any pushes-forward with pulls-back, the actual formation of each
cluster yields a factor of $(q_i-1)!$. Supposing now that for one and only one of the clusters
we combine the first push-forward with the corresponding pull-back, it is clear
that we can redraw diagram~\ref{Vertex-n_J-R-mGR-LdL}, as shown in figure~\ref{fig:bn:LdL-mGR-dec}.
\begin{center}
\begin{figure}
	\[
	-\frac{2}{\sum_{i=1}^r q_i} \sum_{J=0}^n \norm_{J+1,0}
	\dec{
	\dec{
		\begin{array}{c}
		\vspace{0.1in}
			\LabIllOb{Vertex-n_J-R-mGR-dec-LdL}
		\\
		\vspace{0.2in}
			\underbrace{\ensuremath{\begin{array}{c}\input{pstex/Struc-GR-ring.pstex_t} \end{array}}}_{q_1} \underbrace{\ensuremath{\begin{array}{c}\input{pstex/Struc-GR-ring.pstex_t} \end{array}}}_{q_2} \cdots
		\\
			\ensuremath{\begin{array}{c}\input{pstex/Vertex-n_J-R.pstex_t} \end{array}}
		\end{array}
	}{\bullet}
	}{11\Delta^{J+1}}
	\]
\caption[A generic form of diagram~\ref{Vertex-n_J-R-mGR-LdL}, in which the
decorative $\GRkp$s have been explicitly arranged.]{
A generic form of diagram~\ref{Vertex-n_J-R-mGR-LdL}, in which the
decorative $\GRkp$s have been explicitly arranged. One push-forward
has been implicitly combined with the corresponding pull-back, via \CC.}
\label{fig:bn:LdL-mGR-dec}
\end{figure}
\end{center}

It may of course be possible to simplify diagram~\ref{Vertex-n_J-R-mGR-dec-LdL}
in cases where  further pushes-forward and pulls-back can be combined;
for our current purposes, however, there is no need to do this.

We now assume that diagram~\ref{Vertex-n_J-R-mGR-dec-LdL} has the correct
factor for some value of $m < m_\mathrm{max}$, say $m'$. There are two types
of diagram possessing $m'+1$ gauge remainders, which we must consider.
The first is where the extra gauge remainder forms \BT. The second is where this
gauge remainder supplements one of the existing gauge remainder clusters.

In the former case, we need to find the factor for the diagrams in which:
\begin{enumerate}
	\item	the new gauge remainder, which forms $\BT$, is hit by $\flowConstAl$;

	\item	the effective propagator leaving (the new) $\BT$
			is struck by $\flowConstAl$;

	\item	one of the other gauge remainder structures is
			hit by $\flowConstAl$;

	\item	an effective propagator leaving one of the gauge remainders
			of one of the original structures is hit by $\flowConstAl$.
\end{enumerate}

This task is made somewhat easier by the fact that, for the time being, we will not
worry about the factor of diagrams in which differentiated effective propagators
join two gauge remainders. Our hope is that the sum of the four diagrams above
is consistent with a version of diagram~\ref{Vertex-n_J-R-mGR-LdL} with $m= m'+1$.

Following our earlier analysis, it is trivial to show that the first two
diagrams come with the correct factor.  Our strategy for the remaining two
diagrams is as follows. Returning to diagram~\ref{Vertex-n_J-R-mGR-dec-LdL},
consider extracting a gauge remainder from the $i$th structure---which we assume
to possess  $q_i = q'_i > 1$ gauge remainders---and using
it to form $\BT$.
Such a diagram still possesses $m'$ gauge remainders
and so, under our initial assumption, still comes with the correct factor.

In this scenario, consider the parents for a diagram in which either
the structure containing $q'_i-1$ gauge remainders itself, or an effective
propagator joining this structure to the vertex is hit by $\flowConstAl$.
These parents are shown in figure~\ref{fig:bn:LdL-mGR-ith}.
\begin{center}
\begin{figure}[h]
	\[
	\begin{array}{c}
	\vspace{0.2in}
		\ds
		-\frac{2}{(m'-(q'_i-1))!} \sum_{J=0}^n
	\\
		\times
		\left[
			\norm_{J,0}
			\dec{
				\LID[0.1]{}{Struc-GR-ring-n_J-R}{GR-ring-W-n_J-R}
			}{11\Delta^{J} \GRkp^{m'-(q'_i-1)}}
			+ \norm_{J+1,0}
			\dec{
				\sco[0.1]{
					\LID[0.1]{}{Struc-GR-ring-SingleLdL}{GR-ring-SingleLdL-n_J-R}
				}{\ensuremath{\begin{array}{c}\input{pstex/Vertex-n_J-R.pstex_t} \end{array}}}
			}{11\Delta^{J+1} \GRkp^{m'-(q'_i-1)}}
		\right]
	\end{array}
	\]
\caption{The diagrams which spawned the $i$th cluster of diagram~\ref{Vertex-n_J-R-mGR-dec-LdL}
with $q_i = q'_i -1$.}
\label{fig:bn:LdL-mGR-ith}
\end{figure}
\end{center}

Referring back to our work on gauge remainders of types-Ia and Ib, we know that additionally
nested versions of diagrams~\ref{GR-ring-W-n_J-R} and~\ref{GR-ring-SingleLdL-n_J-R}
come with the same overall factor, up to a possible implicit minus sign coming from the extra
gauge remainder being a pull-back. 
Taking the explicitly drawn gauge remainder structures
in these diagrams to contain $q_{i}$, rather than $q_i-1$, gauge remainders what we want to do 
now is convert  into a $\Lambda$-derivative term. 

Focusing first on diagram~\ref{GR-ring-SingleLdL-n_J-R}, such a $\Lambda$-derivative
term---with all gauge remainders promoted to decorations---would 
have to come with a relative factor of
\[
	\frac{1}{q'_i!} \left( \nCr{m'+1}{q'_i} \right)^{-1}.
\]
Similarly for diagram~\ref{GR-ring-W-n_J-R}, but now with an additional
factor of $1/2(J+1)$. These factors are precisely those we need to make
these $\Lambda$-derivative terms consistent with a version of
diagram~\ref{Vertex-n_J-R-mGR-LdL} with $m=m'+1$. We now apply this argument
to each of the $r$ gauge remainder structures of diagram~\ref{Vertex-n_J-R-mGR-dec-LdL}.
We might wonder what happens if one of the $q_i$ is unity. Removing
a gauge remainder from the $i$th structure to create \BT\ and then adding a gauge remainder
to the $i$th structure leads us to analyse  the factor of diagrams corresponding to the
first two items above, which we have done already. If all of the $q_i$ are unity, 
then we are just back to the case of $m$ instances of $\BT$; again, which has already
been analysed.

Our penultimate task is to look at the case where 
one of the existing structures of
diagram~\ref{Vertex-n_J-R-mGR-dec-LdL}---for the case that this diagram
possesses $m'$ gauge remainders---is supplemented by an additional gauge
remainder. The procedure is very similar to what we have just done. We suppose
that it is the $i$th structure that we supplement. First, we show that the
supplemented $i$th structure, struck by $\flowConstAl$, comes with the correct factor.
Next, we return to the diagram with just $m'$ gauge remainders
and remove a gauge remainder from the $j$th structure, adding it to the
$i$th structure. Then we increase the number of gauge remainders in the $j$th
structure by one and suppose that it is this structure that is struck by 
$\flowConstAl$. It is straightforward to show that the resulting terms can be combined
into a $\Lambda$-derivative term, consistent with a version of
diagram~\ref{Vertex-n_J-R-mGR-LdL} for which $m=m'+1$.

To complete the proof by induction that diagram~\ref{Vertex-n_J-R-mGR-LdL}
is correct for all $m$, we must explicitly show
that it
is actually correct for $m=3$ and where there are two gauge remainder
structures. This is very easy to do and will not be presented here;
suffice to say that it has been checked.

We now return to diagram~\ref{DGRx2-Vertex-n_J-R} and 
analyse what happens when it is processed. First of all,
we obtain a version of figure~\ref{fig:bn:GR-Ia,Ib-n_J-R-LdL} where the diagrams come
with a relative factor
of $1/2$ and have an extra decorative $\wedge$. Similarly, with figure~\ref{fig:bn:DGR:TLTP}.
The next step is to decorate the two-point, tree level vertices of the diagrams in the new
version of this figure. When we decorate the diagram like~\ref{DGR-Dumbell-n_J-S-R-TLTP},
there are four things we can do:
\begin{enumerate}
\item	decorate the two-point, tree level vertex with an external field;

\item 	attach the two-point, tree level vertex to either the wine or a vertex, 
		using an effective propagator;

\item	decorate the two-point, tree level vertex with $\WBT$;

\item	decorate the two-point, tree level vertex with the nested version of $\WBT$.
\end{enumerate}

The first option will reproduce diagram~\ref{DGR-n_J-S-R-TLTP-E} but with a relative
factor of $1/2$ and an additional decorative $\wedge$. The second option will do likewise
with diagrams~\ref{DGR-Sigma_n-J-S}--\ref{DGR-n_J-S-R-WGR}. The third
option will reproduce diagram~\ref{DGR-n_J-S-R-WGR-T} with the same factor but with an additional
decorative $\wedge$. The reason that the factor is the same is because the relative factor of $1/2$
we were expecting is cancelled out by the choice of two $\wedge$s with which we can form $\WBT$.
The fourth option will yield a version of diagram~\ref{DGR-n_J-S-R-WGR-T}, with the same factor
and the same decorations, but in which the $\WBT$ has been replaced by the two independent, singly nested
versions.
These two terms are depicted by the first two diagrams of figure~\ref{fig:bn:DGRx2:Dec}.
Additionally, these diagrams will come with gauge remainders contributions, where the gauge remainder
is trapped by the nested structure, as depicted by the third and fourth diagrams of figure~\ref{fig:bn:DGRx2:Dec}.
\begin{center}
\begin{figure}[h]
	\[
	4 \sum_{J=0}^n
		\dec{
			\LGD[0.1]{}{Struc-n_J-S-R-W-RR}{DGRx2-n_J-S-R-W-RR}
			-\LGD[0.1]{}{Struc-n_J-S-R-W-LR}{DGRx2-n_J-S-R-W-LR}
			-\CGD[0.1]{}{Struc-n_J-S-R-W-GR-RR}{DGRx2-n_J-S-R-W-GR-RR}{GR-Ia-Nest1-ExtraTerm-Ex-PB}
			+\CGD[0.1]{}{Struc-n_J-S-R-W-GR-LR}{DGRx2-n_J-S-R-W-GR-LR}{GR-Ia-Nest1-ExtraTerm-Ex}
		}{11\Delta^J}
	\]
\caption{New types of diagram arising from the decoration of version of
diagram~\ref{DGR-Dumbell-n_J-S-R-TLTP}.}
\label{fig:bn:DGRx2:Dec}
\end{figure}
\end{center}

Having considered the decoration of the diagram like~\ref{DGR-Dumbell-n_J-S-R-TLTP}, we move on
to consider the decoration of the diagram like~\ref{DGR-Dumbell-TLTPx2}. From our
previous arguments it is clear that we will produce:
\begin{enumerate}
\item	versions of diagrams~\ref{DGR-TLTP-WGR}
		and~\ref{DGR-WGRx2} with a relative factor of $1/2$ and an additional decorative $\wedge$;

\item	versions of diagrams~\ref{DGR-TLTP-WGR-T-E} and~\ref{DGR-WGRx2-Top} with the same factor
		and a decorative $\wedge$;

\item	versions of diagrams~\ref{DGR-TLTP-WGR-T-E} and~\ref{DGR-WGRx2-Top} with the same factor,
		the same decorations but in which the $\WBT$  has been replaced by the two independent,
		singly  nested versions.
		These contributions have partners possessing gauge remainders;

\item	an entirely new type of term in which we attach an instance of $\WBT$ to both of the
		two-point, tree level vertices;

\item	an entirely new type of term in which we attach each of the two-point, tree level
		vertices to one of the gauge remainders which constitute the nested versions of $\BT$.
\end{enumerate}

The diagrams described in the last three items are shown  in figure~\ref{fig:bn:DGRx2:Dec-B}.
\begin{center}
\begin{figure}
	\[
	\begin{array}{c}
		\displaystyle
		4 \norm_{n,0}
	\\
	\vspace{0.2in}
		\times
		\dec{
			\LGD[0.1]{}{Struc-TLTP-E-RW-RR}{DGRx2-TLTP-E-RW-RR}
			-\LGD[0.1]{}{Struc-TLTP-E-RW-LR}{DGRx2-TLTP-E-RW-LR}
			-\CGD[0.1]{}{Struc-TLTP-E-RW-GR-RR}{DGRx2-TLTP-E-RW-GR-RR}{GR-III-N1-PB-TLTP-EPGR}
			+\CGD[0.1]{}{Struc-TLTP-E-RW-GR-LR}{DGRx2-TLTP-E-RW-GR-LR}{GR-III-N1-PF-TLTP-EPGR}
			+\CGD[0.1]{}{Struc-R-RW-R}{DGRx2-R-R}{DGR-WGRx2-Top-Bottom-PC}
		}{1\Delta^n}
	\\
		\displaystyle
		-4 \norm_{n-1,0}
	\\
	\vspace{0.2in}
		\times
		\dec{
			\PGD[0.1]{}{Struc-WGR-RR}{DGRx2-WGR-RR}{fig:bn:DGRx2-GR-II/Combo}
			-\PGD[0.1]{}{Struc-WGR-LR}{DGRx2-WGR-LR}{fig:bn:DGRx2-GR-II/Combo}
			-\PGD[0.1]{}{Struc-WGR-GR-RR}{DGRx2-WGR-GR-RR}{fig:bn:DGRx2-GR-II/Combo}
			+\PGD[0.1]{}{Struc-WGR-GR-LR}{DGRx2-WGR-GR-LR}{fig:bn:DGRx2-GR-II/Combo}
		}{11\Delta^{n-1}}
		\displaystyle
	\\
		\displaystyle
		+2 \norm_{n-1,0}
	\\
	\vspace{0.2in}
		\times
		\dec{
			\begin{array}{c}
				\CGD[0.1]{}{Struc-RW-RR}{DGRx2-WGR-N1-Top-PB}{GR-II-WGR-N1-Top-PB}
				-\CGD[0.1]{}{Struc-RW-LR}{DGRx2-WGR-N1-Top-PF}{GR-II-WGR-N1-Top-PF}
				-\CGD[0.1]{}{Struc-RW-GR-RR}{DGRx2-WGR-N1-Top-PB-LHGR}{GR-WGR-N1-Top-PB-LHGR-B}
				+\CGD[0.1]{}{Struc-GR-RW-LR}{DGRx2-WGR-N1-Top-PF-RHGR}{GR-WGR-N1-Top-PF-RHGR-B}	
			\\
				-\CGD[0.1]{}{Struc-GR-RW-RR}{DGRx2-WGR-N1-Top-PB-RHGR}{GR-WGR-N1-Top-PB-RHGR-B}
				+\CGD[0.1]{}{Struc-RW-GR-LR}{DGRx2-WGR-N1-Top-PF-LHGR}{GR-WGR-N1-Top-PF-LHGR-B}	
				+\CGD[0.1]{}{Struc-GR-RW-GR-RR}{DGRx2-WGR-N1-Top-PB-GRx2}{GR-II-WGRx2-Trap-Alg-A}
				-\CGD[0.1]{}{Struc-GR-RW-GR-LR}{DGRx2-WGR-N1-Top-PF-GRx2}{GR-II-WGRx2-Trap-Alg-B}
			\end{array}
		}{11\Delta^{n-1}}
	\end{array}
	\]
\caption{New types of diagram arising from the decoration of version of
diagram~\ref{DGR-Dumbell-TLTPx2}.}
\label{fig:bn:DGRx2:Dec-B}
\end{figure}
\end{center}

We now find a number of cancellations, some implicit and others explicit.
Turning first to the former, we see that there will be a repetition of
cancellation~\ref{cancel:DGR-n_J-S-R-WGR-T}.  To be precise, diagrams~\ref{DGRx2-n_J-S-R-W-RR}
and~\ref{DGRx2-n_J-S-R-W-LR} will cancel against the nested versions of
diagram~\ref{GR-Ia-n_J-S-R-WGR-T}. Similarly, the nested version of cancellations~\ref{cancel:DGR-n_J+r-R-LdL-WBT}
and~\ref{cancel:DGR-n_J-R-RW-WBT} go through just the same. 
We must be more careful with the nested version of
cancellation~\ref{cancel:DGR-TLTP-WGR-T-E}, since it occurs
only as a consequence of the particular diagrammatic structure and
set of decorative fields. The nested version of this
cancellation will fail in the following circumstances: at the final
level of nesting and at any level of nesting beyond the current one
(but before the final level) when there are additional vertices.

 From 
cancellations~\ref{cancel:DGR-n_J-S-R-WGR-T} and~\ref{cancel:DGR-n_J-R-RW-WBT}
we will derive another cancellation mechanism. (Cancellation~\ref{cancel:DGR-TLTP-WGR-T-E} 
involves diagrams possessing an $\Op{2}$ stub; since these
will not be fully treated in this thesis, we do not worry about cancellation mechanisms
for such terms.)

\begin{CM}

Consider a manipulable diagram comprising  
gauge remainders structures and a single vertex:
this is the parent for what follows. Iterating
the diagrammatic procedure,
amongst the correction terms generated at each stage
are those diagrams
possessing a dumbbell structure with at 
least one two-point, tree level vertex. 

In the case where there is a single two-point, tree level
vertex, we will attach it to a gauge remainder structure. In
the case that there are two such vertices, we will attach one
to a gauge remainder structure and the other to
a vertex. For the purposes of this cancellation
mechanism, we utilise the effective propagator relation, 
wherever possible, but retain only the Kronecker delta part.
The resulting diagrams are guaranteed to exactly
cancel against identical
terms generated earlier in the calculation,
together with the parent diagram.

\label{CM:GR-LdL-A}
\end{CM}

A further cancellation mechanism follows from cancellation~\ref{cancel:DGR-n_J+r-R-LdL-WBT}
\begin{CM}
Consider the conversion of diagrams possessing
gauge remainder structures and at least one vertex
into $\Lambda$-derivative
terms. For what follows, we take the parent diagrams
to have only manipulable components; \ie\ a decomposition
like that of figure~\ref{fig:bn:GR-Ia,Ib-n_J-R-LdL-pre}
has been performed.

For a given set of gauge remainder
structures, there is a tower of manipulable terms; each successive
term possessing an additional vertex.

Starting from the single vertex term, we
apply the diagrammatic procedure. This generates
a diagram which cancels the two vertex term coming from the
tower. Successive iterations generate diagrams which cancel successive 
terms from the tower.

\label{CM:GR-LdL-B}
\end{CM}

The explicit cancellations are more exciting as a selection  of them involve the first use of
one of the diagrammatic identities.
\begin{cancel}
Diagram~\ref{DGRx2-n_J-S-R-W-GR-RR} exactly cancels diagram~\ref{GR-Ia-Nest1-ExtraTerm-Ex-PB}
via diagrammatic identity~\ref{D-ID-Alg-2}.
\label{cancel:DGRx2-n_J-S-R-W-GR-RR}
\end{cancel}

\begin{cancel}
Diagram~\ref{DGRx2-n_J-S-R-W-GR-LR} exactly cancels diagram~\ref{GR-Ia-Nest1-ExtraTerm-Ex}
via diagrammatic identity~\ref{D-ID-Alg-2}.
\label{cancel:DGRx2-n_J-S-R-W-GR-LR}
\end{cancel}

\begin{cancel}
Diagram~\ref{DGRx2-TLTP-E-RW-GR-RR} exactly cancels diagram~\ref{GR-III-N1-PB-TLTP-EPGR}
via diagrammatic identity~\ref{D-ID-Alg-2}.
\label{cancel:DGRx2-TLTP-E-RW-GR-RR}
\end{cancel}

\begin{cancel}
Diagram~\ref{DGRx2-TLTP-E-RW-GR-LR} exactly cancels diagram~\ref{GR-III-N1-PF-TLTP-EPGR}
via diagrammatic identity~\ref{D-ID-Alg-2}.
\label{cancel:DGRx2-TLTP-E-RW-GR-LR}
\end{cancel}

\CancelCom{DGRx2-R-R}{DGR-WGRx2-Top-Bottom-PC}{by virtue of the fact 
that the decorations force the wine of the latter
diagram to be decorated, for all values of $n$.}

\begin{cancel}
Diagram~\ref{DGRx2-WGR-N1-Top-PB} cancels diagram~\ref{GR-II-WGR-N1-Top-PB},
by virtue of the fact that the decorations force the wine of the latter
diagram to be decorated, for all values of $n$.
\label{cancel:DGRx2-WGR-N1-Top-PB}
\end{cancel}

\begin{cancel}
Diagram~\ref{DGRx2-WGR-N1-Top-PF} cancels diagram~\ref{GR-II-WGR-N1-Top-PF},
by virtue of the fact that the decorations force the wine of the latter
diagram to be decorated, for all values of $n$.
\label{cancel:DGRx2-WGR-N1-Top-PF}
\end{cancel}

\begin{cancel}
Diagrams~\ref{DGRx2-WGR-N1-Top-PB-LHGR}--\ref{DGRx2-WGR-N1-Top-PF-LHGR}
exactly cancel diagrams~\ref{GR-WGR-N1-Top-PB-LHGR-B}--\ref{GR-WGR-N1-Top-PF-LHGR-B}.
\label{cancel:GR-WGR-N1-Top-PB-LHGR-B}
\end{cancel}

\begin{cancel}
Diagram~\ref{DGRx2-WGR-N1-Top-PB-GRx2} exactly cancels diagram~\ref{GR-II-WGRx2-Trap-Alg-A}
via diagrammatic identity~\ref{D-ID-Alg-2}.
\label{cancel:DGRx2-WGR-N1-Top-PB-GRx2}
\end{cancel}

\begin{cancel}
Diagram~\ref{DGRx2-WGR-N1-Top-PF-GRx2} exactly cancels diagram~\ref{GR-II-WGRx2-Trap-Alg-B}
via diagrammatic identity~\ref{D-ID-Alg-2}.
\label{cancel:DGRx2-WGR-N1-Top-PF-GRx2}
\end{cancel}

Taking into account cancellation mechanisms~\ref{CM:a1-S}--\ref{CM:MV-a1}, \ref{CM:GR-LdL-A} and~\ref{CM:GR-LdL-B},
we can deduce the terms we will be left with after
iterating the procedure of converting terms into $\Lambda$-derivatives.
The above cancellations are repeated, at every stage of
the calculation. This includes cancellations~\ref{cancel:DGRx2-WGR-N1-Top-PB} 
and~\ref{cancel:DGRx2-WGR-N1-Top-PF}: even in the versions of diagrams~\ref{GR-II-WGR-N1-Top-PB}
and~\ref{GR-II-WGR-N1-Top-PF} with additional vertices, we are still forced to 
decorate the wine, as this is the only way to ensure that there are no disconnected structures.
Note, though, that at higher levels of nesting, decoration of the wine is not necessarily
forced, as sockets are available for the decorative fields.

Thus, we will reduce this sector of the cancellation to $\Lambda$-derivative, 
$\beta$ and $\gamma$ terms, up to further gauge remainders
and terms that require manipulation at $\Op{2}$. The gauge remainder terms are as follows.
First, we have a copy of the original set of gauge remainders, which 
come with a relative factor of $1/2$ and two decorative $\wedge$s. Secondly,
we have copies of diagrams~\ref{DGR-n_J-R-GR-DEP-R}--\ref{DGR-WGRx2-Top-B} with an additional
decorative gauge remainder\footnote{Obtaining these terms involves partial cancellations
against diagrams generated by manipulating gauge remainders of type-II,
possessing a single, decorative gauge remainder.}. Lastly, we have an entirely new set of terms, shown in 
figure~\ref{fig:bn:DGRx2-L2-NewTerms}.
\begin{center}
\begin{figure}
	\[
	\begin{array}{c}
	\vspace{0.1in}
		\ds
		4 \sum_{J=0}^n \norm_{J,0}
	\\
		\times
		\dec{
			\begin{array}{c}
			\vspace{0.2in}
				\CGD[0.1]{}{Struc-n_J-R-RW-GR-RR}{DGRx2-n_J-R-RW-GR-RR}{GR-Ib-n_J-R-RW-LL-Alg}
				-\CGD[0.1]{}{Struc-n_J-R-RW-GR-LR}{DGRx2-n_J-R-RW-GR-LR}{GR-Ib-n_J-R-RW-LR-Alg}
			\\
				\PGD[0.1]{}{Struc-n_J-R-GR-RW-RR}{DGRx2-n_J-R-GR-RW-RR}{fig:bn:DGRx2-GR-II/Combo}
				-\PGD[0.1]{}{Struc-n_J-R-GR-RW-LR}{DGRx2-n_J-R-GR-RW-LR}{fig:bn:DGRx2-GR-II/Combo}
				-\PGD[0.1]{}{Struc-n_J-R-GR-RW-GR-RR}{DGRx2-n_J-R-GR-RW-GR-RR}{fig:bn:DGRx2-GR-II/Combo}
				+\PGD[0.1]{}{Struc-n_J-R-GR-RW-GR-LR}{DGRx2-n_J-R-GR-RW-GR-LR}{fig:bn:DGRx2-GR-II/Combo}
			\end{array}
		}{11\Delta^J}
	\end{array}
	\]
\caption{New types of gauge remainder diagrams arising from iterating the diagrammatic procedure.}
\label{fig:bn:DGRx2-L2-NewTerms}
\end{figure}
\end{center}

\begin{cancel}
Diagram~\ref{DGRx2-n_J-R-RW-GR-RR} exactly cancels diagram~\ref{GR-Ib-n_J-R-RW-LL-Alg} via diagrammatic identity~\ref{D-ID-Alg-2}.
\label{cancel:DGRx2-n_J-R-RW-GR-RR}
\end{cancel}

\begin{cancel}
Diagram~\ref{DGRx2-n_J-R-RW-GR-LR} exactly cancels diagram~\ref{GR-Ib-n_J-R-RW-LR-Alg} via diagrammatic identity~\ref{D-ID-Alg-2}.
\label{cancel:DGRx2-n_J-R-RW-GR-LR}
\end{cancel}

Notice that diagrams~\ref{DGRx2-n_J-R-GR-RW-RR} and~\ref{DGRx2-n_J-R-GR-RW-LR}
are just nested versions of diagram~\ref{DGR-n_J-R-GR-RW-WBT} 
 and that diagrams~\ref{DGRx2-WGR-RR} and~\ref{DGRx2-WGR-LR}
are just nested versions of diagram~\ref{DGR-WGRx2-Top}.
Just as the un-nested diagrams combined with diagrams generated by the original type-II
gauge remainders, so too in the nested case. 

We can do more than this: precisely half of the contributions to diagrams~\ref{DGRx2-WGR-GR-RR},
\ref{DGRx2-WGR-GR-LR}, \ref{DGRx2-n_J-R-GR-RW-GR-RR} and~\ref{DGRx2-n_J-R-GR-RW-GR-LR}
are cancelled, respectively, by diagrams~\ref{Struc-WGRx2-Active-Alg-B}, \ref{Struc-WGRx2-Active-Alg-A},
\ref{GR-II-n_J-R-PF-PB-TLTP-EP-GR} and~\ref{GR-II-n_J-R-PF-PF-TLTP-EP-GR}. This is courtesy
of diagrammatic identity~\ref{D-ID-Alg-2}.

Figure~\ref{fig:bn:DGRx2-GR-II/Combo} shows the results of combining terms, in the
manner discussed, above.
\begin{center}
\begin{figure}
	\[
	\begin{array}{c}
		\displaystyle
		-2\sum_{J=0}^n \norm_{J,0}
	\\
	\vspace{0.2in}
		\times
		\dec{
			\begin{array}{c}
				\PGD[0.1]{}{Struc-n_J-R-GR-DEP-RR}{DGRx2-n_J-R-GR-DEP-RR}{fig:bn:DGRx2-GR-Extra-B}
				-\PGD[0.1]{}{Struc-n_J-R-GR-DEP-LR}{DGRx2-n_J-R-GR-DEP-LR}{fig:bn:DGRx2-GR-Extra-B}
				-\PGD[0.1]{}{Struc-n_J-R-GR-RW-RR}{DGRx2-n_J-R-GR-RW-RR-B}{fig:bn:DGRx2-GR-Extra-B}
			\\
				+\PGD[0.1]{}{Struc-n_J-R-GR-RW-LR}{DGRx2-n_J-R-GR-RW-LR-B}{fig:bn:DGRx2-GR-Extra-B}
				+\PGD[0.1]{}{Struc-n_J-R-GR-RW-GR-RR}{DGRx2-n_J-R-GR-RW-GR-RR-B}{fig:bn:DGRx2-GR-Extra-B}
				-\PGD[0.1]{}{Struc-n_J-R-GR-RW-GR-LR}{DGRx2-n_J-R-GR-RW-GR-LR-B}{fig:bn:DGRx2-GR-Extra-B}
			\end{array}
		}{11\Delta^J}
	\\
		\displaystyle
		-2\norm_{n-1,0}
	\\
		\times
		\dec{
			\PGD[0.1]{}{Struc-WGR-RR}{DGRx2-WGR-RR-B}{fig:bn:DGRx2-GR-Extra}
			-\PGD[0.1]{}{Struc-WGR-LR}{DGRx2-WGR-LR-B}{fig:bn:DGRx2-GR-Extra}
			-\PGD[0.1]{}{Struc-WGR-GR-RR}{DGRx2-WGR-GR-RR-B}{fig:bn:DGRx2-GR-Extra}
			+\PGD[0.1]{}{Struc-WGR-GR-LR}{DGRx2-WGR-GR-LR-B}{fig:bn:DGRx2-GR-Extra}
		}{11\Delta^{n-1}}
	\end{array}
	\]
\caption{The result of combining diagrams~\ref{DGRx2-WGR-RR}--\ref{DGRx2-WGR-GR-LR}
and~\ref{DGRx2-n_J-R-GR-RW-RR}--\ref{DGRx2-n_J-R-GR-RW-GR-LR} with
diagrams~\ref{GR-II-n_J-R-GR-W-RR}--\ref{GR-II-WGR-LR}, \ref{GR-II-n_J-R-PF-PF-TLTP-EP-GR},
\ref{GR-II-n_J-R-PF-PB-TLTP-EP-GR},
\ref{Struc-WGRx2-Active-Alg-A}
and~\ref{Struc-WGRx2-Active-Alg-B}. }
\label{fig:bn:DGRx2-GR-II/Combo}
\end{figure}
\end{center}

This time around, processing the gauge remainders requires a little more thought. When we
process the copy of the original set of gauge remainders, we must now take seriously
those terms in which we attach a nested gauge remainder to a two-point, tree level
vertex; unlike the un-nested case, such diagrams do not vanish courtesy of
diagrammatic identity~\ref{D-ID-Alg-2}. 

To begin, let us consider such diagrams formed by gauge remainders of type-Ia.
An example of a term of this type is shown in figure~\ref{fig:bn:GR-Ia-TLTP-Ep-GRx2}.
\begin{center}
\begin{figure}[h]
	\[
	4 \sum_{J=0}^n \norm_{J,0}
	\dec{
		\LID[0.1]{}{Struc-n_J-S-R-W-GR-TLTP-EP-RR}{n_J-S-R-W-GR-TLTP-EP-RR}
	}{11\Delta^J}
	\]
\caption{Example of a new type of diagram formed when a type-Ia gauge remainder
generates a two-point, tree level vertex, to which a nested gauge remainder is
attached, via an effective propagator.}
\label{fig:bn:GR-Ia-TLTP-Ep-GRx2}
\end{figure}
\end{center}

The fate of such a diagram is easy to deduce. 
Consider cancellations~\ref{cancel:DGRx2-n_J-S-R-W-GR-RR}
and~\ref{cancel:DGRx2-n_J-S-R-W-GR-LR}. At a higher level
of nesting, we know from our work in chapter~\ref{ch:FurtherD-ID}
that these cancellations will not quite go through; they require
a supplementary term. This term is precisely provided by diagrams
of the form~\ref{n_J-S-R-W-GR-TLTP-EP-RR}!

Similarly, diagrams in which a nested gauge remainder structure is
attached to a two-point, tree level vertex formed by the action of
a type-Ib gauge remainder ensure that cancellations~\ref{cancel:DGRx2-n_J-R-RW-GR-RR}
and~\ref{cancel:DGRx2-n_J-R-RW-GR-LR} go through at any level of nesting.

There is a pattern here, which leads to another cancellation mechanism.

\begin{CM}

Consider some cancellation that occurs due to diagrammatic identity~\ref{D-ID-Alg-2}.
There are two ingredients to a cancellation of this type. The first is a diagram
in which an active gauge remainder is trapped by a (nested) gauge remainder
structure. Such a term is formed when a two-point, tree level vertex is generated
as a correction to a $\Lambda$-derivative term and subsequently attached
to a nested gauge remainder structure.
The second ingredient is a diagram in which a two-point, tree level vertex, formed by the action
of a singly nested gauge remainder, is attached to this nested gauge remainder.

At the next level of nesting, the corresponding terms do not quite cancel. The
term required to  complete the cancellation is closely related to the second
ingredient, above. Given that a diagram like the second ingredient exists,
it then follows that a similar diagram must exist, in which the two-point,
tree level vertex has been formed by the action of an un-nested gauge remainder
(and is yet to be attached to anything). Now consider the version of this
diagram with two decorative gauge remainders (this will exist if the additionally
nested version of the first and second ingredients exist). Joining these gauge remainders
to the two-point, tree level vertex, we create the term we require for
our cancellation to go ahead.  

Additionally nested versions of this supplementary
diagram now ensure, via diagrammatic identities~\ref{D-ID-Alg-3}
and~\ref{D-ID-Alg-G} that the original cancellation is guaranteed to occur, at all levels
of nesting.

\label{CM:NewD-IDs}
\end{CM}

Consequently, not only are analogues of cancellations~\ref{cancel:DGRx2-n_J-S-R-W-GR-RR},
\ref{cancel:DGRx2-n_J-S-R-W-GR-LR}, \ref{cancel:DGRx2-n_J-R-RW-GR-RR}
and~\ref{cancel:DGRx2-n_J-R-RW-GR-LR} guaranteed to go ahead at all levels
of nesting but so to are analogues of cancellations~\ref{cancel:DGRx2-TLTP-E-RW-GR-RR}, 
\ref{cancel:DGRx2-TLTP-E-RW-GR-LR}, \ref{cancel:DGRx2-WGR-N1-Top-PB-GRx2} 
and~\ref{cancel:DGRx2-WGR-N1-Top-PF-GRx2}.

Moreover, diagrammatic identity~\ref{D-ID-Alg-2} was applied 
in the generation of the diagrams of figure~\ref{fig:bn:DGRx2-GR-II/Combo}.
Thus, the additionally nested version of this figure is generated also,
so long as we incorporate the the correct set of terms in which a two-point,
tree level vertex generated by a type-II gauge remainder is attached to 
a nested gauge remainder structure.

With this in mind, let us now analyse the diagrams of figure~\ref{fig:bn:DGRx2-GR-II/Combo}.
This is easy, because there are no remaining
decorative gauge remainders, in this case.
Allowing the gauge remainder to act in these
diagrams, it is apparent that all generated terms in which the gauge remainder strikes
a socket which decorates anything other than a two-point, tree level vertex will 
simply cancel, via cancellation mechanism~\ref{CM:GR-Socket}. Iterating the diagrammatic
procedure until exhaustion, we focus first on the surviving terms coming from
diagrams~\ref{DGRx2-WGR-RR-B}--\ref{DGRx2-WGR-GR-LR-B}. The minimal set
of such terms with the lowest level of nesting, is shown in figure~\ref{fig:bn:DGRx2-GR-Extra} 
up to diagrams with an $\Op{2}$ stub. We note that the additionally nested diagrams
are nested with respect to the currently un-nested, processed gauge remainder.

\begin{center}
\begin{figure}
	\[
	\begin{array}{c}
	\vspace{0.1in}
		\ds
		2 \norm_{n-1,0}
	\\
		\times
		\dec{
			\begin{array}{c}
			\vspace{0.2in}
				\CGD[0.1]{}{Struc-WGRx2-N1-T-T-PF-PF2}{DGRx2-WGRx2-N1-T-T-PF-PF2}{GR-II-WGRx2-N1-T-T-PF-PF2}
				-\CGD[0.1]{}{Struc-WGRx2-N1-T-T-PB-PF2}{DGRx2-WGRx2-N1-T-T-PB-PF2}{GR-II-WGRx2-N1-T-T-PB-PF2}
				-\CGD[0.1]{}{Struc-WGRx2-N1-T-T-PF-PB}{DGRx2-WGRx2-N1-T-T-PF-PB}{GR-II-WGRx2-N1-T-T-PF-PB}
				
			\\
			\vspace{0.2in}
				+\CGD[0.1]{}{Struc-WGRx2-N1-T-T-PB-PB}{DGRx2-WGRx2-N1-T-T-PB-PB}{GR-II-WGRx2-N1-T-T-PB-PB}
				-\CGD[0.1]{}{Struc-WBT-GR-LR}{DGRx2-WBT-GR-LR}{GR-II-N1-WBB-PB-PB-TLTP-Alg-RD}
				+\CGD[0.1]{}{Struc-WBT-PB-GR-LR}{DGRx2-WBT-PB-GR-LR}{GR-II-N1-WBB-PF-PF-TLTP-Alg-RD}
			\\
			\vspace{0.2in}
				
				+\CGD[0.1]{}{Struc-WBT-GR-RR}{DGRx2-WBT-GR-RR}{GR-II-N1-WBB-PB-PF-TLTP-Alg-RD}
				-\CGD[0.1]{}{Struc-WBT-PB-GR-RR}{DGRx2-WBT-PB-GR-RR}{GR-II-N1-WBB-PF-PB-TLTP-Alg-RD}
				+2 \PGD[0.1]{}{Struc-WGRx2-N1-T-B-PF-PF}{DGRx2-WGRx2-N1-T-B-PF-PF2}{fig:bn:DGRx2-GR-Extra/Combo}
			\\
				+2
				\left[
					
					-\PGD[0.1]{}{Struc-WGRx2-N1-T-B-PB-PF}{DGRx2-WGRx2-N1-T-B-PB-PF2}{fig:bn:DGRx2-GR-Extra/Combo}
					-\PGD[0.1]{}{Struc-R-W-GR-LR}{DGRx2-R-W-GR-LR}{fig:bn:DGRx2-GR-Extra/Combo}
					+\PGD[0.1]{}{Struc-R-W-GR-RR}{DGRx2-R-W-GR-RR}{fig:bn:DGRx2-GR-Extra/Combo}
				\right]
			\end{array}
		}{11\Delta^{n-1}}
	\end{array}
	\]
\caption{The minimal set of diagrams surviving diagrams 
arising from diagrams~\ref{DGRx2-WGR-RR-B}--\ref{DGRx2-WGR-GR-LR-B}.} 
\label{fig:bn:DGRx2-GR-Extra}
\end{figure}
\end{center}

Directly, we find a very encouraging set of cancellations.
\CancelRange{DGRx2-WGRx2-N1-T-T-PF-PF2}{DGRx2-WGRx2-N1-T-T-PB-PB}{GR-II-WGRx2-N1-T-T-PF-PF2}{GR-II-WGRx2-N1-T-T-PB-PB}

\CancelRange{DGRx2-WBT-GR-LR}{DGRx2-WBT-PB-GR-RR}{GR-II-N1-WBB-PB-PB-TLTP-Alg-RD}{GR-II-N1-WBB-PF-PB-TLTP-Alg-RD}

Focusing now on the terms that survive, we can combine them with diagrams coming from 
earlier in the calculation. Diagrams~\ref{DGRx2-WGRx2-N1-T-B-PF-PF2} and~\ref{DGRx2-WGRx2-N1-T-B-PB-PF2}
are identical to diagrams~\ref{GR-WGRx2-N1-T-B-PF-PF} and~\ref{GR-WGRx2-N1-T-B-PB-PF} and can be combined directly.
All four diagrams~\ref{DGR-DEP-L-PF-Alg}--\ref{DGR-R-DEP-Pf-PF-TLTP-Alg} can be redrawn,
using diagrammatic identity~\ref{D-ID-Alg-2}. The first two now manifestly remove from
diagrams~\ref{DGRx2-R-W-GR-LR} and~\ref{DGRx2-R-W-GR-RR} the components in which the
wine is undecorated.\footnote{As we noted earlier, this contribution 
vanishes anyway, since gauge remainders have no support in the $C$-sector. However,
this illustrates the kind of cancellations which are necessary at higher levels
of nesting.} The resultant diagrams take exactly the same form as
the final two redrawn diagrams and so can be combined. The end product of these
procedures is shown in figure~\ref{fig:bn:DGRx2-GR-Extra/Combo}
\begin{center}
\begin{figure}[h]
	\[
	8 \norm_{n-1,0}
	\dec{
		\LGD[0.1]{}{Struc-WGRx2-N1-T-B-PF-PF}{DGRx2-WGRx2-N1-T-B-PF-PF2-B}
		-\LGD[0.1]{}{Struc-WGRx2-N1-T-B-PB-PF}{DGRx2-WGRx2-N1-T-B-PB-PF2-B}
		-\LGD[0.1]{}{Struc-R-RW-GR-LR}{DGRx2-R-RW-GR-LR}
		+\LGD[0.1]{}{Struc-R-RW-GR-RR}{DGRx2-R-RW-GR-RR}
	}{11\Delta^{n-1}}
	\]
\caption{Combination of diagrams~\ref{DGRx2-WGRx2-N1-T-B-PF-PF2}--\ref{DGRx2-R-W-GR-RR}
and~\ref{GR-WGRx2-N1-T-B-PF-PF}--\ref{DGR-R-DEP-Pf-PF-TLTP-Alg}.} 
\label{fig:bn:DGRx2-GR-Extra/Combo}
\end{figure}
\end{center}

This kind of pattern of terms is precisely what we anticipate needing. Consider
the kind of terms we expect to arise when we manipulate 
diagrams with three decorative gauge remainders. Amongst these terms
will be those formed by decorating a version of diagram~\ref{DGR-Dumbell-TLTPx2}.
This version of diagram~\ref{DGR-Dumbell-TLTPx2}
will come with an overall factor of $1/6$ and two additional decorative gauge
remainders. 

Now, if we attach a $\WBT$ to one of the two-point, tree level vertices
and the nested version of $\BT$ to the other, then we will generate
terms very similar to those in figure~\ref{fig:bn:DGRx2-GR-Extra/Combo}.
Indeed, the diagram with the trapped gauge remainder, \ref{DGRx2-R-RW-GR-LR}
and~\ref{DGRx2-R-RW-GR-RR} will be removed completely, as we would hope.
The components of the other two diagrams, in which the wine is un-decorated
will, however survive. But this is exactly what we want: these are precisely
the terms we need in order to construct $\Lambda$-derivative terms
in which gauge remainder structures are joined by effective propagators!

Next, we consider the result of processing 
diagrams~\ref{DGRx2-n_J-R-GR-DEP-RR}--\ref{DGRx2-n_J-R-GR-RW-GR-LR-B}.
We need not explicitly give all of the resultant terms,
since many of them are just nestings of earlier diagrams.
In particular, diagrams~\ref{DGRx2-n_J-R-GR-DEP-RR}--\ref{DGRx2-n_J-R-GR-RW-LR-B}
are just the nested versions of diagrams~\ref{DGR-n_J-R-GR-DEP-R}--\ref{DGR-n_J-R-GR-RW}.
Thus, we can follow through the manipulation of the latter diagrams to deduce
the result of manipulating the former. 

First, we will produce nested versions
of diagrams~\ref{DGR-DEP-PF-n_J-R} and~\ref{DGR-RW-PF-n_J-R}. Iterating
the diagrammatic procedure until exhaustion, the minimal set of terms will
be given by a version of diagrams~\ref{DGR-DEP-L-PF-Alg}--\ref{DGR-R-DEP-Pf-PF-TLTP-Alg},
where the nesting is with respect to $\WBT$. We must be careful with the overall
factor: the factor of four in front of 
diagrams~\ref{DGR-DEP-L-PF-Alg}--\ref{DGR-R-DEP-Pf-PF-TLTP-Alg} arises because we
have been able to collect pushes-forward and pulls-back twice. With the new diagrams,
we can do it only once. Hence, the overall factor is just two, but each of the original
diagrams will now have four nested versions.

Going beyond the minimal set, we include diagrams possessing additional vertices and
diagrams additionally nested, with respect to the top-most gauge remainder structure.

These points will be illuminated when we consider diagrams~\ref{DGRx2-n_J-R-GR-RW-GR-RR-B}
and~\ref{DGRx2-n_J-R-GR-RW-GR-LR-B}. Although these diagrams have a virtually
identical structure to the ones just discussed, they do not have analogues with a
lower degree of nesting, earlier in the calculation, since such terms die. Hence, we should
include the result of processing these diagrams, explicitly.

Together with this, we will include the other terms spawned by 
diagrams~\ref{DGRx2-n_J-R-GR-DEP-RR}--\ref{DGRx2-n_J-R-GR-RW-GR-LR-B}
which have no analogue earlier in this section of the calculation. These are those
for which the active gauge remainder generates a two-point, tree
level vertex which we then join to the nested gauge remainder at the top of
the diagram. The resulting collection of terms is shown in 
figure~\ref{fig:bn:DGRx2-GR-Extra-B}.
\begin{center}
\begin{figure}
	\[
	\begin{array}{c}
	\vspace{0.1in}
		\ds
		-2 \norm_{n-1,0}
	\\
		\times
		\dec{
			\begin{array}{c}
			\vspace{0.2in}
				\CGD[0.1]{}{Struc-RXR-DEP-PF-TLTP-EP}{DGRx2-RXR-DEP-PF-TLTP-EP}{GRs-RXR-DEP-PF-TLTP-EP}
				-\CGD[0.1]{}{Struc-RXR-DEP-PB-TLTP-EP}{DGRx2-RXR-DEP-PB-TLTP-EP}{GRs-RXR-DEP-PB-TLTP-EP}
				-\CGD[0.1]{}{Struc-LXR-DEP-PF-TLTP-EP}{DGRx2-LXR-DEP-PF-TLTP-EP}{GRs-LXR-DEP-PF-TLTP-EP}
				+\CGD[0.1]{}{Struc-LXR-DEP-PB-TLTP-EP}{DGRx2-LXR-DEP-PB-TLTP-EP}{GRs-LXR-DEP-PB-TLTP-EP}
			\\
			\vspace{0.2in}
				-\CGD[0.1]{}{Struc-RXR-DW-PF-TLTP-EP}{DGRx2-RXR-DW-PF-TLTP-EP}{GRs-RXR-DW-PF-TLTP-EP}
				+\CGD[0.1]{}{Struc-RXR-DW-PB-TLTP-EP}{DGRx2-RXR-DW-PB-TLTP-EP}{GRs-RXR-DW-PB-TLTP-EP}
				+\CGD[0.1]{}{Struc-LXR-DW-PF-TLTP-EP}{DGRx2-LXR-DW-PF-TLTP-EP}{GRs-LXR-DW-PF-TLTP-EP}
				-\CGD[0.1]{}{Struc-LXR-DW-PB-TLTP-EP}{DGRx2-LXR-DW-PB-TLTP-EP}{GRs-LXR-DW-PB-TLTP-EP}
			\\
			\vspace{0.2in}
				+\LGD[0.1]{}{Struc-RXR-GR-DW-PF-TLTP-EP}{DGRx2-RXR-GR-DW-PF-TLTP-EP}
					-\LGD[0.1]{}{Struc-RXR-GR-DW-PB-TLTP-EP}{DGRx2-RXR-GR-DW-PB-TLTP-EP}
				-\LGD[0.1]{}{Struc-LXR-GR-DW-PF-TLTP-EP}{DGRx2-LXR-GR-DW-PF-TLTP-EP}
				+\LGD[0.1]{}{Struc-LXR-GR-DW-PB-TLTP-EP}{DGRx2-LXR-GR-DW-PB-TLTP-EP}
			\\
			\vspace{0.2in}
				+\LGD[0.1]{}{Struc-RR-GR-DW-LPFX-TLTP}{DGRx2-RR-GR-DW-LPFX-TLTP}
				-\LGD[0.1]{}{Struc-RR-GR-DW-LPBX-TLTP}{DGRx2-RR-GR-DW-LPBX-TLTP}
				-\LGD[0.1]{}{Struc-RR-GR-DW-RPFX-TLTP}{DGRx2-RR-GR-DW-RPFX-TLTP}
				+\LGD[0.1]{}{Struc-RR-GR-DW-RPBX-TLTP}{DGRx2-RR-GR-DW-RPBX-TLTP}
			\\
				+\LGD[0.1]{}{Struc-LR-GR-DW-LPFX-TLTP}{DGRx2-LR-GR-DW-LPFX-TLTP}
				-\LGD[0.1]{}{Struc-LR-GR-DW-LPBX-TLTP}{DGRx2-LR-GR-DW-LPBX-TLTP}
				-\LGD[0.1]{}{Struc-LR-GR-DW-RPFX-TLTP}{DGRx2-LR-GR-DW-RPFX-TLTP}
				+\LGD[0.1]{}{Struc-LR-GR-DW-RPBX-TLTP}{DGRx2-LR-GR-DW-RPBX-TLTP}
			\end{array}
		}{11\Delta^{n-1}}
	\end{array}
	\]
\caption{The minimal set of diagrams surviving diagrams 
arising from diagrams~\ref{DGRx2-n_J-R-GR-DEP-RR}--\ref{DGRx2-n_J-R-GR-RW-GR-LR-B},
which cannot be written as nestings of terms generated
earlier in this section, up to diagrams with an $\Op{2}$ stub.} 
\label{fig:bn:DGRx2-GR-Extra-B}
\end{figure}
\end{center}

\CancelRange{DGRx2-RXR-DEP-PF-TLTP-EP}{DGRx2-LXR-DEP-PB-TLTP-EP}{GRs-RXR-DEP-PF-TLTP-EP}{GRs-LXR-DEP-PB-TLTP-EP}

\CancelRange{DGRx2-RXR-DW-PF-TLTP-EP}{DGRx2-LXR-DW-PB-TLTP-EP}{GRs-RXR-DW-PF-TLTP-EP}{GRs-LXR-DW-PB-TLTP-EP}

These are the final cancellations we will explicitly need to see to complete the
treatment of the gauge remainders! The pattern of cancellations is now clear.
Each gauge remainder that strikes a vertex can generate a two-point, tree
level vertex. We already know that if we attach this vertex to another vertex or a
wine, the resulting diagrams are guaranteed to cancel, via cancellation 
mechanisms~\ref{CM:GR-Socket} and~\ref{CM:GR-Socket-Supp}.
There are now only two things we can do with the two-point, tree level
vertex (up to generating terms with an $\Op{2}$ stub): either we can attach it to
one of the gauge remainders which 
explicitly strikes some
structure, or we can use the decorative gauge remainders to create a structure
to which to attach it.

In the former case, we can then re-express the diagrams using diagrammatic
identities~\ref{D-ID-Alg-1}--\ref{D-ID-Alg-G}. From figure~\ref{fig:D-ID-App},
we know that amongst the redrawn terms are those necessary to
cancel the diagrams of the latter case, up to terms
possessing sub-diagrams like~\ref{D-ID-App-GR-LL},
\ref{D-ID-App-GR-LLL} and etcetera. But these are cancelled also. Here, the cancellation
is not against terms formed by active gauge remainders. Rather, it occurs when one or more of
the two-point, tree level components of 
dumbbell
structures, generated by the flow equation are joined to decorative gauge remainders.
The trapped gauge remainder term generated is the term we require for
the cancellation to work. 

Let us illustrate this further by showing how some diagrams which we have not explicitly
cancelled will disappear. Diagrams~\ref{GR-DEPBB-PB-TLTP-EP-LL}--\ref{GR-DEPBB-PF-TLTP-EP-RL}
will vanish once we come to process the versions of the type-Ia and Ib gauge
remainder terms with two decorative gauge remainders.

Next, consider diagrams~\ref{DGRx2-RXR-GR-DW-PF-TLTP-EP}--\ref{DGRx2-LXR-GR-DW-PB-TLTP-EP}.
These will be cancelled by 
diagrams~\ref{GR-n_J-R-RWBB-Trap-GR-PF-PF-TLTP-EPGR}--\ref{GR-n_J-R-RWBB-Trap-GR-PB-PB-TLTP-EPGR},
\ref{GR-n_J-R-RWBB-PF-Trap-GR-PF-TLTP-EPGR}--\ref{GR-n_J-R-RWBB-PB-Trap-GR-PB-TLTP-EPGR}
and the nested versions of diagrams~\ref{GR-WGR-N2-B-PB-PB2} and~\ref{GR-WGR-N2-B-PB-PF2}.
From this we will be left over with the nested version of diagrams~\ref{GR-II-WGRx2-Trap-Alg-A}
and~\ref{GR-II-WGRx2-Trap-Alg-B}---ready for cancellation when we process
the $\Lambda$-derivative terms for diagrams with three decorative gauge remainders!

Finally, diagrams~\ref{DGRx2-RR-GR-DW-LPFX-TLTP}--\ref{DGRx2-LR-GR-DW-RPBX-TLTP}
will cancel terms in which a singly nested version of \BT\ is attached
to each of the two-point, tree level vertices of a dumbbell structure
formed by a $\Lambda$-derivative term with four decorative gauge remainders.

With these myriad cancellations, the question of course arises as to what are
we actually left with. There are a number of cancellations in this section
which are not exact, but dependent on the precise diagrammatic structure
and the decorative fields; namely, 
cancellations~\ref{cancel:DGRx2-R-R}--\ref{cancel:DGRx2-WGR-N1-Top-PF} and~\ref{cancel:DGR-TLTP-WGR-T-E}.
These cancellations fail in the following circumstances: at the final
level of nesting and at any level of nesting beyond the current one
(but before the final level) when there are additional vertices.
In this case, the surviving terms comprise an un-decorated, differentiated wine.
The terms surviving from the failure of the first three cancellations
are exactly the terms we require in order to be able to perform the conversion
into $\Lambda$-derivative terms, as anticipated in figure~\ref{fig:bn:GR-Nested-LdL-GR-B}.
The terms surviving from the failure of the cancellation~\ref{cancel:DGR-TLTP-WGR-T-E} possess an
$\Op{2}$ stub. Since they also possess an undecorated wine, attached to this
stub, it is apparent that they are higher loop manifestations of the $\BPZ$
terms that we encountered in the one-loop calculation.
These will be commented on in chapter~\ref{ch:Op2}.

 Indeed, we can now write down the final
result of processing the gauge remainders, up to terms with an $\Op{2}$ stub. This is shown
in figure~\ref{fig:Beta-n+W/GR} where, once again, we have used the compact notation 
defined by~(\ref{eq:n-loop:ProdOverI}).
\begin{center}
\begin{figure}
	\begin{equation}
	\begin{array}{c}
	\vspace{0.2in}
		\displaystyle
		-4 \beta_{n_+} \Box_{\mu\nu}(p) = 
		-2 \sum_{s=1}^{n_+} \sum_{m=0}^{2(s-1)}  \sum_{J=0}^{2n-m} 
		\frac{\norm_{J+s+1,J+1}}{m!}
	\\
	\vspace{0.2in}
		\displaystyle
		\times
		\dec{
		\begin{array}{c}
		\vspace{0.2in}
			\displaystyle
			\frac{1}{J+2}
			\dec{
				\sco[0.1]{\ensuremath{\begin{array}{c}\input{pstex/GV-J+1.pstex_t} \end{array}}}{\Vertex{\scriptstyle n_+-s, J}}
			}{\bullet}
		\\
			\displaystyle
			- \sum_{V^{J+2}=1}^{V^{J+1}} 
			\left[
				\sco[0.1]{
					\left[2 \left( V^{J+1,J+2}-1 \right) \beta_{V^{\! J+2}} + \gamma_{V^{\! J+2}}	\pder{}{\alpha}\right]	
					\ensuremath{\begin{array}{c}\input{pstex/GV-J+1-J+2.pstex_t} \end{array}}
				}{\Vertex{\scriptstyle n_+-s, J}}
			\right]
		\end{array}		
		}{11\Delta^{J+s+1}\wedge^m}
	\\
	\vspace{0.1in}
		\displaystyle
		-2\sum_{J=0}^n \sum_{m=0}^{2J+1} \frac{\norm_{J+1,0}}{m!}
	\\
	\vspace{0.2in}
		\displaystyle
		\times
		\dec{
			\dec{
				\ensuremath{\begin{array}{c}\input{pstex/Vertex-n_J-R.pstex_t} \end{array}}
			}{\bullet}
			- \sum_{r=1}^{n_J} 
			\left[2 \left( n_J-r-1 \right) \beta_r + \gamma_r	\pder{}{\alpha}\right]	
			\ensuremath{\begin{array}{c}\input{pstex/Vertex-n_J+r.pstex_t} \end{array}}
		}{11\Delta^{J+1} \wedge^m}
	\\
	\vspace{0.2in}
		\displaystyle
		+\frac{2\norm_{n,0}}{(2n+2)!} 
		\dec{
			\dec{
				\hspace{1em}
			}{\bullet}
		}{11\Delta^n \wedge^{2n+2}}
	\\
	+\cdots
	\end{array}
	\label{eq:beta-n+}
	\end{equation}
\caption[Diagrammatic expression for $\beta_{n_+}$ including
gauge remainder terms.]{Diagrammatic expression for $\beta_{n_+}$ including
gauge remainder terms. The ellipsis denotes terms
with an $\Op{2}$ stub.}
\label{fig:Beta-n+W/GR}
\end{figure}
\end{center}

Comparing this expression to that of figure~\ref{fig:Beta-n+Part-I},
the most striking thing is how similar they are. Whilst 
processing the gauge remainders took a huge amount of time and effort,
the final result can be represented in an incredibly concise fashion.
This is a compelling indication of the potential
of this formalism.

We now conclude this chapter by re-deriving the expression for $\beta_1$,
in terms of $\OLDs$, up to $\Op{2}$ terms. In equation~(\ref{eq:beta-n+}),
we set $n=0$ to yield:
\[
	-4 \beta_1 \Box_{\mu \nu}(p) = 
	\dec{
		\begin{array}{c}
			-\norm_{2,1}
			\dec{
				\scd[0.1]{Vertex-0-R}{Vertex-0-R}
			}{11\Delta^2}
			-2 \sum_{m=0}^1 \frac{\norm_{1,0}}{m!}
			\dec{
				\ensuremath{\begin{array}{c}\begin{picture}(0,0)%
\includegraphics{pstex/Vertex-0-R.pstex}%
\end{picture}%
\setlength{\unitlength}{3947sp}%
\begingroup\makeatletter\ifx\SetFigFont\undefined%
\gdef\SetFigFont#1#2#3#4#5{%
  \reset@font\fontsize{#1}{#2pt}%
  \fontfamily{#3}\fontseries{#4}\fontshape{#5}%
  \selectfont}%
\fi\endgroup%
\begin{picture}(264,262)(672,1303)
\put(722,1415){\makebox(0,0)[lb]{\smash{\SetFigFont{8}{9.6}{\rmdefault}{\mddefault}{\updefault}{\color[rgb]{0,0,0}$0^R$}%
}}}
\end{picture}
 \end{array}}
			}{11\Delta \GRkp^m}
		\\[3ex]
			+\norm_{0,0} 
			\dec{
			}{11\GRkp^2} 
		\end{array}
	}{\bullet}+ \cdots
\]

From equation~(\ref{eq:norm_(J,s)}) we have:
\[
	\norm_{2,1} = 1/16, \ \norm_{1,0} = -1/4, \ \norm_{0,0} = -1/2
\]

Let us start by analysing the pure gauge remainder term. Given that
the two external fields must be on the same supertrace, we can only
form a gauge remainder structure in which both gauge remainders are
bitten in the same sense (both on the left or both on the right). 
Since there are an even number of gauge
remainders we will pick up a minus sign (see section~\ref{sec:GRs-Nested}), 
whichever sense we choose.

Combining the two different senses via \CC\ yields a factor
of two. We can now choose to attach either of the external fields
to either of the gauge remainders; this yields a further factor of
two. Thus, including the $\norm_{0,0}$,
the overall factor of the diagram is $+2$, as it should be.

Next we look at the single vertex term. First, suppose that $m=0$.
Attaching all fields to the vertex comes with a factor of unity;
including the $-2 \norm_{1,0}$ the overall factor of the diagram is
$+1/2$. Secondly, suppose that $m=1$. We must attach the decorative
gauge remainder to the effective propagator. There are four ways to do this,
corresponding to the two ends of the effective propagator and the
equality of the push-forward and pull-back. Hence, the overall
factor of this diagram is $+2$.

Finally, we look at the double vertex term. We discussed, in detail,
in section~\ref{sec:bn:L1:a1:a0-Re} that the two possible ways of
performing the decorations (\ie\ with or without a simple loop)
both come with a factor of eight. Thus, including the factor
of $-\norm_{2,1}$, the overall factor of both fully
decorated diagrams is  $-1/2$.

Summing up all terms, it is clear that we have correctly
reproduced the
diagrammatic expression for $\beta_1$ in terms of $\OLDs$,
up to the $\Op{2}$ terms.

\chapter{Terms with an $\Op{2}$ Stub} \label{ch:Op2}

It is, unfortunately, beyond the scope of this thesis
to give a complete treatment of diagrams possessing
an $\Op{2}$ stub. This is due to constraints of time
and space, rather than any particular conceptual
difficulty; indeed, certain terms possessing an $\Op{2}$
stub---the type-III gauge remainders---have already been processed.
Moreover, as we will sketch below, the full
treatment of such terms just reuses many of the
techniques that we have already introduced.

To begin, it is useful to split the terms with an $\Op{2}$
stub into two groups, one of which further sub-divides. To make
the initial distinction between terms, we focus on the wine / effective propagator
leaving the $\Op{2}$ stub. If the field attached to the stub carries the
same momentum as the stub---\ie\ it has not been bitten
by a gauge remainder---then we call the corresponding diagrams
`$\Op{2}$ terms of type-I'. If, on the other hand, the field attached to the stub
has been hit by a gauge remainder then we call the 
 corresponding diagrams
`$\Op{2}$ terms of type-II'.

Let us now focus on the terms of type-I. If the wine is
undecorated, then we will call the corresponding diagrams
terms of type-Ia. If the wine is decorated, then they will
be called terms of type-Ib.

\section{\Op{2}\ Terms of Type Ia}

The terms of type-Ia are the higher loop analogues of
the \BPZ\ terms that we encountered in the
one-loop calculation. We recall here the reason why
the one-loop  \BPZ\ terms vanish (see section~\ref{sec:beta1:A'(0)}).

The final
form of $\beta_1$ is, up to a factor, the $\Lambda$-derivative of $\OLDs$.
For what
follows, we will define $\POLDs$ to be $\OLDs$, modulo
the terms with an $\Op{2}$ stub. Let us now go back a step
in the computation of $\beta_1$ and reinstate the \BPZ\
terms. These comprise an $\Op{2}$ stub attached to an undecorated
wine, the other end of which attaches to something very similar
to $\POLDs$; to be specific, it attaches to a version of $\POLDs$
where, in each diagram, one of the vertices decorated by an 
external field is replaced by a seed action version of that vertex.
However, at $\Op{2}$, we can replace this seed action vertex with a Wilsonian
effective action vertex, thereby forming a instance of $\POLDs$. Now,
the transversality of $\POLDs$ guarantees that the \BPZ\ terms
vanish at $\Op{2}$.

At higher loops, we expect something similar to happen. 
Indeed, every time we form
a $\Lambda$-derivative term, we will generate an $\BPZ$ term amongst the corrections. 
Whilst we no longer expect to be able to always replace seed action vertices by Wilsonian
effective action vertices at \Op{2}, we do expect the \BPZ\ sub-diagrams
to which the undecorated wine attaches
to be transverse,
nonetheless. Though it will not be as direct as in the one-loop case, we predict
that this transversality to be related to the transversality of the $\Lambda$-derivative 
terms. This has been explicitly checked at two-loop
order; we intend to prove that the \BPZ\ terms are guaranteed to vanish at any loop-order
in~\cite{GeneralMethods}.

\section{\Op{2}\ Terms of Types Ib and II}

Whilst we can argue that terms of type-Ia must vanish we have no choice but to
manipulate terms of types-Ib and II.
The
basic idea is that, since the diagram possess an $\Op{2}$ stub
and we are working to $\Op{2}$, we hope be able to Taylor expand the rest of
the diagram in $p$. There are, of course, a number of very important caveats.
The first is that, just because a diagram possess an $\Op{2}$ stub,
does not necessarily mean that it is at least $\Op{2}$. 
For example, consider the
diagram in figure~\ref{fig:bn:Op2-II-Ex}.
\begin{center}
\begin{figure}
	\[
		-8 \sum_{J=0}^n \norm_{J,0}
		\dec{
			\LID[0.1]{}{Struc-III-WBB-EP-V}{Op2-III-WBB-EP-V}
		}{\Delta^J}
	\]
\caption{An example of a diagram which, whilst possessing an $\Op{2}$
stub has $\Op{0}$ components.}
\label{fig:bn:Op2-II-Ex}
\end{figure}
\end{center}

The key point about this diagram is that, if the decorative
effective propagators form simple loops, it possesses an effective propagator
carrying the external momentum. This, of course, goes as $1/p^2$,
nullifying the effect of the $\Op{2}$ stub. What, then, are we to do with
such diagrams? 
We note that, if the decorative effective propagators do not decorate the top vertex,
then this vertex must be at least $\Op{2}$, washing out the effect of the
$1/p^2$. Of course, if the effective propagators do form simple loops on the vertex,
then the vertex is no longer manifestly transverse in external momentum, and
so we are back to the original problem.  The solution is to combine sets of diagrams. 
Referring to diagram~\ref{Op2-III-WBB-EP-V} we want to keep
the bottom sub-diagram exactly the same and sum over all the different things
we can attach to the top end of the effective propagator.
We expect this sum to be transverse in external momentum---something which is borne out
at the two-loop level. 

The reason for this expectation is that, by summing over all legal sub-diagrams we can attach
to the top end of the effective propagator, we expect to reproduce sets of diagrams
which contribute to the final expression for $\beta$-function coefficients. The transversality
of these diagrams will then guarantee that the effect of the $1/p^2$ is compensated for.
Once more, a complete proof of this will have to wait until~\cite{GeneralMethods}.

The second caveat is as to the validity of Taylor expanding a diagram in external
momentum. Whilst we can always Taylor expand vertices~\cite{YM-1-loop} we have seen
already in 
section~\ref{sec:TotalDerivativeMethods:Op2-NTE}
that, at the two-loop level, we encounter diagrams which, as a whole, cannot
be Taylor expanded in $p$. The solution here is to construct subtractions,
an example of which was given in the aforementioned section. It is interesting
to note that in the calculation of $\beta_2$, it is only a tiny fraction of
terms with an \Op{2}\ stub for which it is necessary to construct subtractions;
the vast majority can just be blindly Taylor expanded.

Irrespective of whether we Taylor expand directly, or construct subtractions,
the strategy then is to try and combine diagrams
either into total momentum derivatives or into $\Lambda$-derivatives.
We have seen an example of how the latter works in the one-loop
calculation. To get a feel for how the former works,
consider the three two-loop diagrams shown in the first
row of figure~\ref{fig:bn:TotMomDerivEx}.
\begin{center}
\begin{figure}
	\[
	\begin{array}{c}
	\vspace{0.2in}
		-\LID[0.1]{}{Op2-OneLoopVertexA}{Op2-OneLoopVertexA}
		-\LID[0.1]{}{Op2-OneLoopVertexB}{Op2-OneLoopVertexB}
		+\LID[0.1]{}{Op2-OneLoopVertexC}{Op2-OneLoopVertexC}
	\\
		\rightarrow
		-2
		\left[
			\LID[0.1]{}{Op2-OneLoopVertexA-M}{Op2-OneLoopVertexA-M}
			+\LID[0.1]{}{Op2-OneLoopVertexB-M}{Op2-OneLoopVertexB-M}
			-\PID[0.1]{}{Op2-OneLoopVertexC-M}{Op2-OneLoopVertexC-M}{fig:bn:TotMomDerivEx-M}
		\right]
	\end{array}
	\]
\caption{Example of three two-loop diagrams, components of which
can be combined into total momentum derivatives, at $\Op{2}$.}
\label{fig:bn:TotMomDerivEx}
\end{figure}
\end{center}

To go from the first row to the second row, we have
set $p=0$ everywhere except in the $\Op{2}$ stub, which is a perfectly
valid step, in these cases. We have then 
diagrammatically expanded to $\Op{0}$, using the techniques of
section~\ref{sec:MomentumExpansions}. The next step is
to apply diagrammatic identity~\ref{D-ID:TwoWines} to diagram~\ref{Op2-OneLoopVertexC-M},
as shown in figure~\ref{fig:bn:TotMomDerivEx-M}, where we
note that any gauge remainder which hits the two-point, one-loop
vertex will kill it. 
\begin{center}
\begin{figure}
	\[
	-
	\left[	
		2\LID[0.1]{}{Op2-OneLoopVertexC-MA}{Op2-OneLoopVertexC-MA}
		+\LID[0.1]{}{Op2-OneLoopVertexC-MB}{Op2-OneLoopVertexC-MB}
		+\LID[0.1]{}{Op2-OneLoopVertexC-MC}{Op2-OneLoopVertexC-MC}
	\right]
	\]
\caption{Result of processing diagram~\ref{Op2-OneLoopVertexC-M},
using diagrammatic identity~\ref{D-ID:TwoWines}.}
\label{fig:bn:TotMomDerivEx-M}
\end{figure}
\end{center}

Now we see that the combination of diagrams~\ref{Op2-OneLoopVertexA-M},
\ref{Op2-OneLoopVertexB-M} and~\ref{Op2-OneLoopVertexC-MA} can be rewritten
as a total momentum derivative. We have commented already how total
momentum derivatives can be discarded, since all integrals
are pre-regularised using dimensional regularisation.
This then leaves the gauge remainder terms, diagrams~\ref{Op2-OneLoopVertexC-MB}
and~\ref{Op2-OneLoopVertexC-MC}. At first sight, it looks as though we are stuck,
since the gauge remainders are differentiated \wrt\ momentum. However,
trivially redrawing
\[
	\ensuremath{\begin{array}{c}\begin{picture}(0,0)%
\includegraphics{pstex/DiffGR.pstex}%
\end{picture}%
\setlength{\unitlength}{3947sp}%
\begingroup\makeatletter\ifx\SetFigFont\undefined%
\gdef\SetFigFont#1#2#3#4#5{%
  \reset@font\fontsize{#1}{#2pt}%
  \fontfamily{#3}\fontseries{#4}\fontshape{#5}%
  \selectfont}%
\fi\endgroup%
\begin{picture}(240,398)(266,-72)
\end{picture}
 \end{array}} = \ensuremath{\begin{array}{c}\begin{picture}(0,0)%
\includegraphics{pstex/DiffGRk.pstex}%
\end{picture}%
\setlength{\unitlength}{3947sp}%
\begingroup\makeatletter\ifx\SetFigFont\undefined%
\gdef\SetFigFont#1#2#3#4#5{%
  \reset@font\fontsize{#1}{#2pt}%
  \fontfamily{#3}\fontseries{#4}\fontshape{#5}%
  \selectfont}%
\fi\endgroup%
\begin{picture}(324,398)(266,-72)
\end{picture}
 \end{array}} + \ensuremath{\begin{array}{c}\begin{picture}(0,0)%
\includegraphics{pstex/DiffGRkpr.pstex}%
\end{picture}%
\setlength{\unitlength}{3947sp}%
\begingroup\makeatletter\ifx\SetFigFont\undefined%
\gdef\SetFigFont#1#2#3#4#5{%
  \reset@font\fontsize{#1}{#2pt}%
  \fontfamily{#3}\fontseries{#4}\fontshape{#5}%
  \selectfont}%
\fi\endgroup%
\begin{picture}(286,398)(360,-76)
\end{picture}
 \end{array}}
\]
it is clear that progress can be made. In the case that the derivative hits
the $\GRkp$, the $\GRk$ is just a perfectly normal, active gauge remainder
which can be treated in the usual manner. In the case that the derivative
hits $\GRk$, we then use diagrammatic identity~\ref{D-ID:PseudoEffectivePropagatorRelation} to 
redraw
\[
	\ensuremath{\begin{array}{c}\begin{picture}(0,0)%
\includegraphics{pstex/DiffGRkEP.pstex}%
\end{picture}%
\setlength{\unitlength}{3947sp}%
\begingroup\makeatletter\ifx\SetFigFont\undefined%
\gdef\SetFigFont#1#2#3#4#5{%
  \reset@font\fontsize{#1}{#2pt}%
  \fontfamily{#3}\fontseries{#4}\fontshape{#5}%
  \selectfont}%
\fi\endgroup%
\begin{picture}(594,398)(266,-72)
\end{picture}
 \end{array}} \equiv \ensuremath{\begin{array}{c}\begin{picture}(0,0)%
\includegraphics{pstex/DiffGRkpw.pstex}%
\end{picture}%
\setlength{\unitlength}{3947sp}%
\begingroup\makeatletter\ifx\SetFigFont\undefined%
\gdef\SetFigFont#1#2#3#4#5{%
  \reset@font\fontsize{#1}{#2pt}%
  \fontfamily{#3}\fontseries{#4}\fontshape{#5}%
  \selectfont}%
\fi\endgroup%
\begin{picture}(651,398)(266,-72)
\end{picture}
 \end{array}}.
\]
Now in this case, too, we have an active gauge remainder which, again,
can be processed in the usual way.

When we do process these gauge remainders we expect one of three things to
occur. First, the usual gauge remainder cancellation mechanisms
occur via cancellation mechanisms~\ref{CM:GR-Socket} and~\ref{CM:GR-Socket-Supp}.
Secondly, processed gauge remainder terms can, themselves, be combined into total momentum
derivatives. Thirdly, processed gauge remainder terms can be converted into
$\Lambda$-derivative terms. It is instructive to see this latter operation
in action, particularly as it gives a flavour for how $\Op{2}$ terms of types-Ib and
II interact.

Our starting point is diagram~\ref{Op2-OneLoopVertexC-MB} which, of course,
is derived from an $\Op{2}$ term of type-Ib. Focusing on the case in which the momentum
derivative hits $\GRkp$\footnote{Indeed, if the momentum derivative hits $\GRk$,
the we can allow $\GRkp$ to act backwards, through the wine, courtesy of
diagrammatic identity~\ref{D-ID:PseudoEffectivePropagatorRelation}. 
The diagram then dies, since $\GRk$ strikes a two-point Wilsonian effective
action vertex.},
 we will further restrict ourselves to considering
the diagram formed when the gauge remainder pulls-back along the wine, to
meet the two-point tree level vertex.\footnote{We do not consider the partner push-forward
on to the two-point vertex,
since this corresponds to the two external fields being on different supertraces.}
We now combine the resultant diagram with two $\Op{2}$ diagrams of type-II,
as shown in figure~\ref{fig:bn:Op2-LdL-Ex-pre}.
\begin{center}
\begin{figure}[h]
	\[
	2
	\left[
		\PID[0.1]{}{Op2-Ex-LdLCompt-A}{Op2-Ex-LdLCompt-A}{fig:bn:Op2-LdL-Ex}
		-\PID[0.1]{}{Op2-Ex-LdLCompt-B}{Op2-Ex-LdLCompt-B}{fig:bn:Op2-LdL-Ex}
		-\PID[0.1]{}{Op2-Ex-LdLCompt-C}{Op2-Ex-LdLCompt-C}{fig:bn:Op2-LdL-Ex}
	\right]
	\]
\caption{Combination of a diagram derived from~\ref{Op2-OneLoopVertexC-MB}
and two $\Op{2}$ diagrams of type-II.}
\label{fig:bn:Op2-LdL-Ex-pre}
\end{figure}
\end{center}

Diagrammatically Taylor expanding the final two diagrams, 
we can then combine diagrams into a $\Lambda$-derivative.
The easiest way to see this is to note that
\begin{eqnarray*}
	\ensuremath{\begin{array}{c}\begin{picture}(0,0)%
\includegraphics{pstex/GR-W-dTLTP-EP.pstex}%
\end{picture}%
\setlength{\unitlength}{3947sp}%
\begingroup\makeatletter\ifx\SetFigFont\undefined%
\gdef\SetFigFont#1#2#3#4#5{%
  \reset@font\fontsize{#1}{#2pt}%
  \fontfamily{#3}\fontseries{#4}\fontshape{#5}%
  \selectfont}%
\fi\endgroup%
\begin{picture}(869,371)(1219,5)
\put(1615,194){\makebox(0,0)[lb]{\smash{\SetFigFont{11}{13.2}{\rmdefault}{\mddefault}{\updefault}{\color[rgb]{0,0,0}0}%
}}}
\end{picture}
 \end{array}} 	& = & \ensuremath{\begin{array}{c}\begin{picture}(0,0)%
\includegraphics{pstex/GR-W-TLTP-d-EP.pstex}%
\end{picture}%
\setlength{\unitlength}{3947sp}%
\begingroup\makeatletter\ifx\SetFigFont\undefined%
\gdef\SetFigFont#1#2#3#4#5{%
  \reset@font\fontsize{#1}{#2pt}%
  \fontfamily{#3}\fontseries{#4}\fontshape{#5}%
  \selectfont}%
\fi\endgroup%
\begin{picture}(1187,369)(901,86)
\put(1615,194){\makebox(0,0)[lb]{\smash{\SetFigFont{11}{13.2}{\rmdefault}{\mddefault}{\updefault}{\color[rgb]{0,0,0}0}%
}}}
\end{picture}
 \end{array}} - \ensuremath{\begin{array}{c}\begin{picture}(0,0)%
\includegraphics{pstex/GR-dW-TLTP-EP.pstex}%
\end{picture}%
\setlength{\unitlength}{3947sp}%
\begingroup\makeatletter\ifx\SetFigFont\undefined%
\gdef\SetFigFont#1#2#3#4#5{%
  \reset@font\fontsize{#1}{#2pt}%
  \fontfamily{#3}\fontseries{#4}\fontshape{#5}%
  \selectfont}%
\fi\endgroup%
\begin{picture}(869,386)(1219,-10)
\put(1615,194){\makebox(0,0)[lb]{\smash{\SetFigFont{11}{13.2}{\rmdefault}{\mddefault}{\updefault}{\color[rgb]{0,0,0}0}%
}}}
\end{picture}
 \end{array}} -\ensuremath{\begin{array}{c}\begin{picture}(0,0)%
\includegraphics{pstex/dGR-W-TLTP-EP.pstex}%
\end{picture}%
\setlength{\unitlength}{3947sp}%
\begingroup\makeatletter\ifx\SetFigFont\undefined%
\gdef\SetFigFont#1#2#3#4#5{%
  \reset@font\fontsize{#1}{#2pt}%
  \fontfamily{#3}\fontseries{#4}\fontshape{#5}%
  \selectfont}%
\fi\endgroup%
\begin{picture}(919,431)(1169,-55)
\put(1615,194){\makebox(0,0)[lb]{\smash{\SetFigFont{11}{13.2}{\rmdefault}{\mddefault}{\updefault}{\color[rgb]{0,0,0}0}%
}}}
\end{picture}
 \end{array}} \\
						& = & - \ensuremath{\begin{array}{c}\begin{picture}(0,0)%
\includegraphics{pstex/LdLdGR-EP.pstex}%
\end{picture}%
\setlength{\unitlength}{3947sp}%
\begingroup\makeatletter\ifx\SetFigFont\undefined%
\gdef\SetFigFont#1#2#3#4#5{%
  \reset@font\fontsize{#1}{#2pt}%
  \fontfamily{#3}\fontseries{#4}\fontshape{#5}%
  \selectfont}%
\fi\endgroup%
\begin{picture}(456,467)(1174,-48)
\end{picture}
 \end{array}} -\ensuremath{\begin{array}{c}\begin{picture}(0,0)%
\includegraphics{pstex/GR-dW.pstex}%
\end{picture}%
\setlength{\unitlength}{3947sp}%
\begingroup\makeatletter\ifx\SetFigFont\undefined%
\gdef\SetFigFont#1#2#3#4#5{%
  \reset@font\fontsize{#1}{#2pt}%
  \fontfamily{#3}\fontseries{#4}\fontshape{#5}%
  \selectfont}%
\fi\endgroup%
\begin{picture}(404,370)(1219,-8)
\end{picture}
 \end{array}} + \ensuremath{\begin{array}{c}\begin{picture}(0,0)%
\includegraphics{pstex/GR-dW-GR.pstex}%
\end{picture}%
\setlength{\unitlength}{3947sp}%
\begingroup\makeatletter\ifx\SetFigFont\undefined%
\gdef\SetFigFont#1#2#3#4#5{%
  \reset@font\fontsize{#1}{#2pt}%
  \fontfamily{#3}\fontseries{#4}\fontshape{#5}%
  \selectfont}%
\fi\endgroup%
\begin{picture}(467,370)(1219,-8)
\end{picture}
 \end{array}} -\ensuremath{\begin{array}{c}\begin{picture}(0,0)%
\includegraphics{pstex/dGR-W.pstex}%
\end{picture}%
\setlength{\unitlength}{3947sp}%
\begingroup\makeatletter\ifx\SetFigFont\undefined%
\gdef\SetFigFont#1#2#3#4#5{%
  \reset@font\fontsize{#1}{#2pt}%
  \fontfamily{#3}\fontseries{#4}\fontshape{#5}%
  \selectfont}%
\fi\endgroup%
\begin{picture}(456,396)(1174,-48)
\end{picture}
 \end{array}} + \ensuremath{\begin{array}{c}\begin{picture}(0,0)%
\includegraphics{pstex/dGR-W-GR.pstex}%
\end{picture}%
\setlength{\unitlength}{3947sp}%
\begingroup\makeatletter\ifx\SetFigFont\undefined%
\gdef\SetFigFont#1#2#3#4#5{%
  \reset@font\fontsize{#1}{#2pt}%
  \fontfamily{#3}\fontseries{#4}\fontshape{#5}%
  \selectfont}%
\fi\endgroup%
\begin{picture}(522,396)(1174,-48)
\end{picture}
 \end{array}}.
\end{eqnarray*}

The first diagram on the second line is obtained from the diagram above
by means of diagrammatic identity~\ref{D-ID:VertexWineGR}. The other
diagrams on the second line are simply obtained by using the effective propagator
relation. Applying the above relation to diagram~\ref{Op2-Ex-LdLCompt-C} we will 
generate five diagrams. The third and fifth involve an active gauge remainder
striking the two-point, one-loop vertex and so die. The second diagram exactly
cancels diagram~\ref{Op2-Ex-LdLCompt-B}. The remaining two diagrams
combine with diagram~\ref{Op2-Ex-LdLCompt-A} to form a $\Lambda$-derivative term,
as shown in figure~\ref{fig:bn:Op2-LdL-Ex}.
\begin{center}
\begin{figure}[h]
	\[
	2
	\dec{
		\LID[0.1]{}{Diagram326.0}{Diagram326.0}
	}{\bullet}
	-2\LID[0.1]{}{Diagram326.0-C}{Diagram326.0-C}
	\]
\caption{Rewriting diagrams~\ref{Op2-Ex-LdLCompt-A}--\ref{Op2-Ex-LdLCompt-C}
as a $\Lambda$-derivative term.}
\label{fig:bn:Op2-LdL-Ex}
\end{figure}
\end{center}

To conclude this section, we give a final diagrammatic identity, which the two-loop
calculation taught us is necessary for the treatment of terms with an $\Op{2}$ stub.
\begin{D-ID}
It is true in all sectors for which the gauge remainder
is not null that
\[
	\ensuremath{\begin{array}{c}\input{pstex/dGRk.pstex_t} \end{array}} = \delta_{\alpha\nu}.
\]
It therefore follows that
\[
	\ensuremath{\begin{array}{c}\begin{picture}(0,0)%
\includegraphics{pstex/dGRk-Arb.pstex}%
\end{picture}%
\setlength{\unitlength}{3947sp}%
\begingroup\makeatletter\ifx\SetFigFont\undefined%
\gdef\SetFigFont#1#2#3#4#5{%
  \reset@font\fontsize{#1}{#2pt}%
  \fontfamily{#3}\fontseries{#4}\fontshape{#5}%
  \selectfont}%
\fi\endgroup%
\begin{picture}(249,608)(-110,336)
\put(-67,367){\makebox(0,0)[lb]{\smash{\SetFigFont{8}{9.6}{\rmdefault}{\mddefault}{\updefault}{\color[rgb]{0,0,0}$k$}%
}}}
\end{picture}
 \end{array}} = \ensuremath{\begin{array}{c}\begin{picture}(0,0)%
\includegraphics{pstex/TE-Ext-Arb.pstex}%
\end{picture}%
\setlength{\unitlength}{3947sp}%
\begingroup\makeatletter\ifx\SetFigFont\undefined%
\gdef\SetFigFont#1#2#3#4#5{%
  \reset@font\fontsize{#1}{#2pt}%
  \fontfamily{#3}\fontseries{#4}\fontshape{#5}%
  \selectfont}%
\fi\endgroup%
\begin{picture}(273,375)(-134,450)
\put(-79,741){\makebox(0,0)[lb]{\smash{\SetFigFont{8}{9.6}{\rmdefault}{\mddefault}{\updefault}{\color[rgb]{0,0,0}$k$}%
}}}
\end{picture}
 \end{array}}.
\]
\label{D-ID-Deriv-GR}
\end{D-ID}

\part{Computation of $\beta_2$ and Conclusions}

\chapter{Two Loop Diagrammatics} \label{ch:bulk}

The principle result of the last chapter is
the reduction of $\beta_{n_+}$ to
$\Lambda$-derivative, $\alpha$ and $\beta$-terms,
up to diagrams with an $\Op{2}$ stub. In this chapter,
we specialise equation~(\ref{eq:beta-n+}) to help us generate a manifestly gauge
invariant, diagrammatic expression for $\beta_2$, from which we
can readily extract the numerical coefficient.

Of course, due to the incompleteness of equation~(\ref{eq:beta-n+}),
we will have to do some additional work. Though the details will not be
presented here, we have completed the diagrammatic procedure for
the $\Op{2}$ terms for the specific case of $\beta_2$. Rather than
using the highly efficient techniques of the previous chapter, this
has been done exactly along the lines of our computation of $\beta_1$
(see also chapter~\ref{ch:Op2}).
Indeed, it should be noted that the entire $\beta_2$ calculation 
was originally done in this manner. The arduous nature of this 
approach cannot be over emphasised. 
Whereas the 1-loop calculation generates
less than 100 diagrams, of which only seven survive, 
the two loop calculation generates 
of $\mathcal{O}(10^4)$ diagrams, of which $\mathcal{O}(100)$ survive.

In addition to the extra work required to treat the $\Op{2}$ terms,
the $\alpha$ and $\beta$-terms require further processing. As we have
noted previously, diagrams of these types possess a full vertex which,
necessarily, has a two-point, tree level component. Hence, we can
play diagrammatic games with these terms, reducing them to a simpler
form.

The first section will give the $\Lambda$-derivative diagrams
obtained by specialising 
equation~(\ref{eq:beta-n+}) to the form corresponding to $\beta_2$. We note that, at the two-loop
level, assertion~\ref{assert:spec-gen} is unnecessary: we never employ
diagrammatic identity~\ref{D-ID-Alg-G}. Hence, the expression we obtain
is not reliant on any assumptions.
In section~\ref{sec:b2:Op2} the $\Lambda$-derivative terms obtained by
processing the terms with an $\Op{2}$ stub are given.

Accompanying the diagrams of section~\ref{sec:b2:bulk} and~\ref{sec:b2:Op2}
are $\alpha$ and $\beta$-terms. To perform
manipulations on the $\beta$-terms does not require any new methodology
and so, in section~\ref{sec:b2:beta},  we jump straight to the final set of
diagrams. To manipulate the $\alpha$-terms, on the other hand, requires some
new diagrammatic identities. These identities are presented in section~\ref{sec:b2:alpha},
together with both the original and recast $\alpha$-terms.

We conclude this chapter by utilising the diagrammatic expression for $\beta_1$
to simplify the diagrammatic expression for $\beta_2$. This will allow
us to cast $\beta_2$ as a set of $\Lambda$-derivative terms, for every one of which the
whole diagram---as opposed to just a factorisable sub-diagram---is hit by \flowConstAl, and $\alpha$-terms.

\section{Specialising Equation~(\ref{eq:beta-n+})} \label{sec:b2:bulk}

To extract the numerical coefficient for $\beta_2$, we need to go from 
equation~(\ref{eq:beta-n+}) to a set of fully decorated diagrams. Since
we must sum over all independent decorations, this generates a large
number of terms. There are, however, several simplifications---which 
fall into three classes. First, there are those 
diagrams which vanish, for some reason. 
Secondly, there are pairs of diagrams which exactly cancel.
Lastly, we can compact our
 notation to reduce the
number of diagrams we must draw.

\subsection{Diagrams which can be Discarded}

For what follows, we will take a diagram to be fully decorated, a consequence
of which is that the sense in which any gauge remainders act is completely
determined. With this in mind, we are able to discard any diagrams for which:
\begin{enumerate}
	\item	there is manifestly no contribution at $\Op{2}$;

	\item	the two external $A^1$s are forced to be on different
			supertraces.
\end{enumerate}

The first characteristic is straightforward, and will not be discussed further.
The second characteristic has been discussed already in the context of gauge 
remainders (see section~\ref{sec:GRs-Nested}): certain patterns of pushes
forward and pulls back can force the external $A^1$s to be on 
different supertraces, if at
least one nested gauge remainder is decorated by an external field.
We choose only to discard complete diagrams for this reason,
as opposed to explicitly discarding components of diagrams for
which certain choices of field ordering cause the $A^1$s to be on different
supertraces.

\subsection{Cancellations Between Diagrams}

The cancellations in this section are between pushes forward and
pulls back. If we can find a pair of diagrams, identical in every way,
up to the sense in which one of the gauge remainders acts, then these will
cancel. Such diagrams are non-planar. It is only in this case that
we have any hope of finding the desired cancellations; changing the sense in
which any of the gauge remainders act in a planar diagram generates either a non-planar
diagram or a planar diagram in which the external
fields are on different supertraces or a planar diagram, the non-vanishing components
of which have a changed field content. As an example,
consider the diagrams of figure~\ref{fig:b2:Cancel-Ex}.
\begin{center}
\begin{figure}[h]
	\[
	4
	\dec{
		\begin{array}{ccccccc}
		\vspace{0.1in}
			\LabIllOb{Ill:FR27.6.1}	&	&\LabIllOb{Ill:FR27.6.4}&	& \LabIllOb{Ill:FR27.6.2} 	&	& \LabIllOb{Ill:FR27.6.3}\\
			\ensuremath{\begin{array}{c}\input{pstex/DiagramFR27.6.pstex_t} \end{array}}		& -	&\ensuremath{\begin{array}{c}\input{pstex/DiagramFR27.6.4.pstex_t} \end{array}}	& - & \ensuremath{\begin{array}{c}\input{pstex/DiagramFR27.6.2.pstex_t} \end{array}}		& +	& \ensuremath{\begin{array}{c}\input{pstex/DiagramFR27.6.3.pstex_t} \end{array}}
		\end{array}
	}{\bullet}
	\]
\caption{Four diagrams, two of which cancel and two of which do not.}
\label{fig:b2:Cancel-Ex}
\end{figure}
\end{center}

Diagrams~\ref{Ill:FR27.6.1} and~\ref{Ill:FR27.6.4} have the same pattern of pushes forward
and pulls back, up to the gauge remainder which bites the vertex; similarly
with diagrams~\ref{Ill:FR27.6.2} and~\ref{Ill:FR27.6.3}. In the latter case, there is
only one supertrace: the pair of diagrams cancel. In the former case, the external fields
are on separate supertraces: diagram~\ref{Ill:FR27.6.4} vanishes and diagram~\ref{Ill:FR27.6.1}
survives.

\subsection{Compact Notation}

There are two ways in which we can compact our notation. First, having identified
pairs of gauge remainder diagrams which cancel, we now recombine sets of those
which survive, to avoid having to explicitly draw all patterns of pushes forward and pulls back.
Let us consider a diagram possessing several, independent, nested 
gauge remainder structures. Now consider also those partner diagrams in which 
the nestings are performed in different senses. 
For each of these structures, we will draw only
one possible arrangement of the gauge remainders but will take this to stand for
all possible arrangements, with one proviso: we will use \CC\ to collect
together the first push forward from any \emph{one} of the structures with
the corresponding pull back. This yields a factor of two and so
we must be careful not to double count if we ever want to expand the diagram
out to explicitly show all possible arrangements of the gauge remainders. In diagrams
possessing instances of the un-nested structure \BT\, we will always combine
the push forward and pull back for each and every instance.
To distinguish diagrams where we employ these conventions from
diagrams in which we have actually made a specific choice about
the pattern of pushes forward and pulls back, we will tag them
with $\AGROB{\hspace{1em}}$ (\cf\ $\AGRO{\hspace{1em}}$ defined
in section~\ref{sec:D-ID-App}.).

Secondly, we can use equation~(\ref{eq:beta-n+}) to show that certain sub-diagrams
always occur together. In fact, we will use only the simplest manifestation of
this; fully exploiting the dependence of $\beta_{n_+}$ on the diagrammatic
expressions for all $\beta_{m \leq n}$ will be saved for the future~\cite{GeneralMethods}.
Let us focus on a diagram possessing $J'+1$ vertices, 
for $J'>1$. We now suppose that 
one of the vertices
is a three-point, tree level vertex, decorated by a simple loop.
Next, we focus on a diagram with $J'$ vertices, to one of which
we attach $\WBT$. It is easy to check that the two diagrams thus described
always come with a relative factor of $-4$. We represent the combination
of these two diagrams, as shown in figure~\ref{fig:b2:PatternedVertex}.
\begin{center}
\begin{figure}[h]
	\[
		\ensuremath{\begin{array}{c}\input{pstex/PatternedVertex.pstex_t} \end{array}} \equiv \ensuremath{\begin{array}{c}\input{pstex/TLThP-SimpleLoop.pstex_t} \end{array}} -4 \ensuremath{\begin{array}{c}\input{pstex/WBT-B.pstex_t} \end{array}}
	\]
\caption{Shorthand for two diagrams always guaranteed to occur
together.}
\label{fig:b2:PatternedVertex}
\end{figure}
\end{center}

\subsection{Generation of Terms} \label{sec:b2-diags}

We now generate the explicitly decorated terms contributing to $\beta_2$,
derived from equation~(\ref{eq:beta-n+}). Since there are so many diagrams,
we will split them into sets. These sets will be designed so that
either each element is transverse in $p$ on it own, but shares some common feature
with the other elements, or only the sum over elements is transverse. 

All $\Lambda$-derivative terms which vanish in the $\epsilon \rightarrow 0$ limit
and which are not involved in  the simplifications of section~\ref{sec:bulk:simplify} are tagged $\rightarrow 0$.
The reasons for which they vanish are dealt with in section~\ref{sec:TLD:vanishing}.
We could, of course, immediately discard all diagrams which vanish in the 
$\epsilon \rightarrow 0$ limit. However, for possible future work, we want
the expression 
obtained after the simplifications to be valid in $D=4-2\epsilon$.

The first set of terms comprises those that contain a one-loop vertex
but are not transverse in $p$, alone. These are shown in figure~\ref{fig:b2:1LV-NT}.
\begin{center}
\begin{figure}[h]
	\[
	\frac{1}{2}
	\dec{
		\begin{array}{cccccccl}
		\vspace{0.1in}
			\LabOLObLB{diag:2.1.0} & & \LabOLOb{diag:6.1} & & \LabOLOb{diag:8.1} & & \LabOLObRB{diag:7.14} & \Discard
		\\
			\ensuremath{\begin{array}{c}\input{pstex/Diagram2.1.0.pstex_t} \end{array}} & - & \ensuremath{\begin{array}{c}\input{pstex/Diagram6.1.pstex_t} \end{array}} & -2 & \ensuremath{\begin{array}{c}\input{pstex/Diagram8.1.pstex_t} \end{array}} & +2 & \ensuremath{\begin{array}{c}\input{pstex/Diagram7.14.pstex_t} \end{array}} & 
		\end{array}
	}{\bullet}
	\]
\caption{The set of diagrams possessing one-loop vertices, the elements
of which are not individually transverse in $p$.}
\label{fig:b2:1LV-NT}
\end{figure}
\end{center}

To demonstrate that the sum of diagrams~\ref{diag:2.1.0}--\ref{diag:7.14} is transverse
in $p$, we use a modified version of the technique we employed to demonstrate the
transversality of the standard set. To understand why a modification is necessary, it
is instructive to compare diagram~\ref{diag:8.1} with the second element of the standard
set. Now consider contracting these diagrams with the momentum of one the external fields,
and processing them using the techniques of section~\ref{sec:GRs}. Due to the symmetry of
the second element
of the standard set, it is manifestly clear that it does not matter which of the
external fields we choose. For diagram~\ref{diag:8.1}, on the other hand, separately
contracting momenta with each of the external fields does not manifestly yield
the same set of terms.\footnote{Though Lorentz invariance or Bose symmetry tells us that the sum of the two sets
of terms must be equivalent.} The solution is to be democratic: we take one
instance of diagram~\ref{diag:8.1} where we contract with the momentum of the external
field decorating the one loop vertex and one instance where we 
contract with the momentum of the external
field decorating the tree level vertex, dividing by two to avoid overcounting. Proceeding in
this manner, it is straightforward to demonstrate the transversality of diagrams~\ref{diag:2.1.0}--\ref{diag:7.14}.

We now move to the second set of terms derived from equation~(\ref{eq:beta-n+}).
These are  shown in figure~\ref{fig:b2:b1-sq-A} and comprise two one-loop sub-diagrams,
both of which are present in the diagrammatic expression
for $\beta_1$, modulo the little set. The nested gauge remainders are particularly simple in 
this case: the only ones that survive are those drawn, as all others vanish by group theory
considerations.
\begin{center}
\begin{figure}
	\[
	-\frac{1}{4}
	\dec{
		\begin{array}{cccccccl}
		\vspace{0.1in}		
			\LabOLObLB{diag:22.1} && \LabOLOb{diag:29.1} && \LabOLOb{diag:58e.12} && \LabOLOb{diag:35.1} &
		\\
			\ensuremath{\begin{array}{c}\input{pstex/Diagram22.1.pstex_t} \end{array}} &-2 &\ensuremath{\begin{array}{c}\input{pstex/Diagram29.1.pstex_t} \end{array}} &+8  & \ensuremath{\begin{array}{c}\input{pstex/Diagram58e.12.pstex_t} \end{array}} & -2 &\ensuremath{\begin{array}{c}\input{pstex/Diagram35.1.pstex_t} \end{array}} &
		\\
		\vspace{0.1in}
								&& \LabOLOb{diag:44.1} && \LabOLOb{diag:58g.1}&&  \LabOLOb{diag:50.1}&
		\\
							&+ & \ensuremath{\begin{array}{c}\input{pstex/Diagram44.1.pstex_t} \end{array}} &-8& \ensuremath{\begin{array}{c}\input{pstex/Diagram58g.1.pstex_t} \end{array}} &+2& \ensuremath{\begin{array}{c}\input{pstex/Diagram50.1.pstex_t} \end{array}} &
		\\
		\vspace{0.1in}
								&&	&&	\LabOLOb{diag:58j.1}	&& \LabOLOb{diag:58f.26,58i.2b} &
		\\
								&&	&+16 & \ensuremath{\begin{array}{c}\input{pstex/Diagram58j.1.pstex_t} \end{array}}	&-8&   \ensuremath{\begin{array}{c}\input{pstex/Diagram58f.26-58i.2b.pstex_t} \end{array}}&
		\\
		\vspace{0.1in}
								&&	&&	&& \LabOLObRB{diag:52.1} & \Process{fig:b2:Diagrams223b}
		\\
								&&	&&	&& \ensuremath{\begin{array}{c}\input{pstex/Diagram52.1.pstex_t} \end{array}} &
		\end{array}
	}{\bullet}
	\]
\caption{Diagrams for which both one-loop sub-diagrams are
found in the diagrammatic expression for $\beta_1$.}
\label{fig:b2:b1-sq-A}
\end{figure}
\end{center}

Notice the suggestive way in which these terms have been arranged. We could
fill in the blank entires in the `matrix' of figure~\ref{fig:b2:b1-sq-A}
by halving the values of the off-diagonal elements and
then reflecting about the diagonal. This highlights that the terms of figure~\ref{fig:b2:b1-sq-A}
can be obtained by `squaring' the diagrammatic expression for $\beta_1$, modulo the little set,
and joining pairs of diagrams together with an effective propagator.

Let us be more precise with what we mean by `squaring'. Consider extracting the diagrams
which gives us $\beta_1 \Box_{\mu \alpha}(p)$, modulo the little set, from under the $\Lambda$-derivative.
Now join these, via $\Delta^{1 \, 1}_{\alpha \beta}(p)$, to the analogous
set of diagrams extracted from $\beta_1 \Box_{\nu \beta}(p)$.
Enclosing the resultant diagrams under a $\Lambda$-derivative, we simply obtain the diagrams
of figure~\ref{fig:b2:b1-sq-A}, up to a factor of $-16$.
Trivially, this makes it immediately apparent that the sum over
diagrams of figure~\ref{fig:b2:b1-sq-A} is transverse in $p$.

The third set of terms derived from equation~(\ref{eq:beta-n+}), shown in figure~\ref{fig:b2:AAC}, 
comprises all diagrams with an $A^1_\mu A^1_\nu C$ vertex.
Each of these terms is automatically transverse.

\begin{center}
\begin{figure}
	\[
	\frac{1}{4}
	\dec{
		\begin{array}{l}
			\begin{array}{cccccccc}
			\vspace{0.1in}		
					& \LabOLObLB{diag:9a.1}	&				& \LabOLOb{diag:9b.1}	&		& \LabOLOb{diag:7.1}		&  & \LabOLOb{diag:9c.1}
			\\
				-2 	& \ensuremath{\begin{array}{c}\input{pstex/Diagram9a.1.pstex_t} \end{array}} 		& -2			& \ensuremath{\begin{array}{c}\input{pstex/Diagram9b.1.pstex_t} \end{array}} 		& +2	& \ensuremath{\begin{array}{c}\input{pstex/Diagram7.1.pstex_t} \end{array}} 			&+2& \ensuremath{\begin{array}{c}\input{pstex/Diagram9c.1.pstex_t} \end{array}} 	
			\end{array}	
		\\
			\begin{array}{cccccc}
			\vspace{0.1in}
					& \LabOLOb{diag:15.1}	&				& \LabOLOb{diag:26.1}	&		& \LabOLOb{diag:397.5}
			\\	
				-\hf& \ensuremath{\begin{array}{c}\input{pstex/Diagram15.1.pstex_t} \end{array}}		&+\frac{2}{3}	&\ensuremath{\begin{array}{c}\input{pstex/Diagram26.1.pstex_t} \end{array}} 		& +8	& \AGROB{\ensuremath{\begin{array}{c}\input{pstex/Diagram397.5.pstex_t} \end{array}}}
			\end{array}
		\\
			\begin{array}{cccccccc}
			\vspace{0.1in}	
					& \LabOLOb{diag:25.1}	&				& \LabOLOb{diag:38.1}	&		& \LabOLOb{diag:381-E}		& 			& \LabOLOb{diag:39.1}
			\\
				+	& \ensuremath{\begin{array}{c}\input{pstex/Diagram25.1.pstex_t} \end{array}}		& - 			& \ensuremath{\begin{array}{c}\input{pstex/Diagram38.1.pstex_t} \end{array}}		& +4 	& \AGROB{\ensuremath{\begin{array}{c}\input{pstex/Diagram381-E.pstex_t} \end{array}}}	&-\frac{1}{2}& \ensuremath{\begin{array}{c}\input{pstex/Diagram39.1.pstex_t} \end{array}}
			\end{array}
		\\
			\begin{array}{ccccccl}
			\vspace{0.1in}
					& \LabOLOb{diag:27.1}	&				& \LabOLOb{diag:40.1}	&		& \LabOLObRB{diag:377.17}		& \Discard
			\\
				+	& \ensuremath{\begin{array}{c}\input{pstex/Diagram27.1.pstex_t} \end{array}}		& -				& \ensuremath{\begin{array}{c}\input{pstex/Diagram40.1.pstex_t} \end{array}}		& +4	& \AGROB{\ensuremath{\begin{array}{c}\input{pstex/Diagram377.17.pstex_t} \end{array}}}&
			\end{array}
		\end{array}
	}{\bullet}
	\]
\caption{Diagrams with an $A^1_\mu A^1_\nu C$ vertex.}
\label{fig:b2:AAC}
\end{figure}
\end{center}

The fourth set of terms, split between figures~\ref{fig:b2:Trans-A} 
and~\ref{fig:b2:Trans-B}, is the final set of $\Lambda$-derivatives which we obtain directly from
equation~(\ref{eq:beta-n+}). 
\begin{center}
\begin{figure}
	\[
	\frac{1}{4}
	\dec{
		\begin{array}{l}
			\begin{array}{ccccccl}
			\vspace{0.1in}
					& \LabOLObLB{diag:21.1} &	& \LabOLOb{diag:36.1}	&	& \LabOLObRB{diag:375.1} 		& \hspace{-2em}\Process{fig:TLD:VanishingSet} \hspace{2em}
			\\
				- 	& \ensuremath{\begin{array}{c}\input{pstex/Diagram21.1.pstex_t} \end{array}}		&+	& \ensuremath{\begin{array}{c}\input{pstex/Diagram36.1.pstex_t} \end{array}}		& -4& \AGROB{\ensuremath{\begin{array}{c}\input{pstex/Diagram375.1.pstex_t} \end{array}}} 	&
			\end{array}
		\\
			\begin{array}{cccc}
			\vspace{0.1in}
					&\OLdiscard{diag:34.1}&	&\Dprocess{diag:51.1}{fig:TLD:C-stub-Diagrams}
			\\	
				+	& \ensuremath{\begin{array}{c}\input{pstex/Diagram34.1.pstex_t} \end{array}}	&-2	&\ensuremath{\begin{array}{c}\input{pstex/Diagram51.1.pstex_t} \end{array}}
			\end{array}
		\\
			\begin{array}{cccc}
			\vspace{0.1in}
					& \LabOLObLB{diag:30.1}	&	& \LabOLOb{diag:41.1}
			\\
				+2	& \ensuremath{\begin{array}{c}\input{pstex/Diagram30.1.pstex_t} \end{array}}		&-2 & \ensuremath{\begin{array}{c}\input{pstex/Diagram41.1.pstex_t} \end{array}}
			\end{array}
		\\
			\begin{array}{ccl}
			\vspace{0.1in}
					& \LabOLObRB{diag:376.1}	& \hspace{-4em} \Process{fig:TLD:Diagrams224}	
			\\			
				+8	& \AGROB{\ensuremath{\begin{array}{c}\input{pstex/Diagram376.1.pstex_t} \end{array}}}	&
			\end{array}
		\end{array}
	\hspace{-2em}
	}{\bullet}
	\]
\caption{The final set of $\Lambda$-derivatives which can be obtained
directly from equation~(\ref{eq:beta-n+}): part~I.}
\label{fig:b2:Trans-A}
\end{figure}
\end{center}

\begin{center}
\begin{figure}
	\[
	\frac{1}{4}
	\dec{
		\begin{array}{l}
			\begin{array}{ccccccccl}
			\vspace{0.1in}
					& \LabOLObLB{diag:10.10}&				& \LabOLOb{diag:19.1}	&	& \LabOLOb{diag:18.1}	&				& \LabOLObRB{diag:33.1}&\Discard
			\\
					& \ensuremath{\begin{array}{c}\input{pstex/Diagram10.10.pstex_t} \end{array}}		&-\frac{2}{3}	& \ensuremath{\begin{array}{c}\input{pstex/Diagram19.1.pstex_t} \end{array}}		&-	& \ensuremath{\begin{array}{c}\input{pstex/Diagram18.1.pstex_t} \end{array}}		&+\frac{1}{2}	& \ensuremath{\begin{array}{c}\input{pstex/Diagram33.1.pstex_t} \end{array}} &
			\end{array}
		\\
			\begin{array}{cccccc}
			\vspace{0.1in}
							&	\Dprocess{diag:23.1}{fig:TLD:d23.1}			&	&\Dprocess{diag:17.1}{fig:TLD:d:217}&   
							&\Dprocess{diag:28.1}{fig:TLD:Diagram28.1WithSubtractions}
			\\
			-\frac{2}{3}	&	\ensuremath{\begin{array}{c}\input{pstex/Diagram23.1.pstex_t} \end{array}}							&-2	& \ensuremath{\begin{array}{c}\input{pstex/Diagram17.1.pstex_t} \end{array}}					& 
							+4&\ensuremath{\begin{array}{c}\input{pstex/Diagram28.1.pstex_t} \end{array}}	
			\end{array}
		\\
			\begin{array}{cccccc}
			\vspace{0.1in}
					&\Dprocess{diag:31.1}{fig:TLD:C-stub-Diagrams}	&	&\Dprocess{diag:32.1}{fig:TLD:Diagram32.1WithSubtractions}&	&\Dprocess{diag:42.1}{fig:TLD:Diagram42.1WithSubtractions}
			\\
				+2& \ensuremath{\begin{array}{c}\input{pstex/Diagram31.1.pstex_t} \end{array}}	& + &\ensuremath{\begin{array}{c}\input{pstex/Diagram32.1.pstex_t} \end{array}}	&-2	& \ensuremath{\begin{array}{c}\input{pstex/Diagram42.1.pstex_t} \end{array}}
			\end{array}
		\\
			\begin{array}{cccc}
			\vspace{0.1in}
					& \Dprocess{diag:392.4}{fig:TLD:Diagram392.4WithSubtractions}	&	&\OLdiscard{diag:395.4}
			\\
				+16	& \ensuremath{\begin{array}{c}\input{pstex/Diagram392.4.pstex_t} \end{array}}												&-8 &\AGROB{\ensuremath{\begin{array}{c}\input{pstex/Diagram395.4.pstex_t} \end{array}}}
			\end{array}
		\\
			\begin{array}{cccccc}
			\vspace{0.1in}
					&\Dprocess{diag:FR26.1}{fig:TLD:DFR26.1,6:WS}	&	& \Dprocess{diag:FR26.2}{fig:TLD:DFR26.1,6:WS}	&	
					& \Dprocess{diag:FR26.11}{fig:TLD:DFR26.11:WS}
			\\
				+16	&\ensuremath{\begin{array}{c}\input{pstex/DiagramFR26.1.pstex_t} \end{array}}								&+16& \ensuremath{\begin{array}{c}\input{pstex/DiagramFR26.2.pstex_t} \end{array}}							&+8
					&\ensuremath{\begin{array}{c}\input{pstex/DiagramFR26.1+6.pstex_t} \end{array}}
			\end{array}
		\end{array}
	}{\bullet}
	\]
\caption{The final set of $\Lambda$-derivatives which can be obtained
directly from equation~(\ref{eq:beta-n+}): part~II.}
\label{fig:b2:Trans-B}
\end{figure}
\end{center}

To demonstrate that the set of diagrams~\ref{diag:21.1}--\ref{diag:FR26.11} is transverse is not
too difficult. For diagrams where there is no symmetry ensuring that contraction with the momentum
of one of the external fields gives a set of terms manifestly the same as those obtained by contracting
with the momentum of the other external field, we should once more be democratic. To complete
the demonstration of transversality, it is necessary to utilise diagrammatic 
identities~\ref{D-ID-Alg-1}--\ref{D-ID-Alg-3}.

Note that whilst we sum over different nestings in diagram~\ref{diag:395.4}
some, but not all, of these will cancel in pairs. Specifically, if we take
the gauge remainder attached to the effective propagator to bite the external
field on the other side, then this will ensure that the external fields
are on the same supertrace, irrespective of they sense in which their
associated gauge remainders act. Pairs of such diagrams will cancel.

\section{Terms with an $\Op{2}$ Stub} \label{sec:b2:Op2}

As anticipated in chapter~\ref{ch:Op2}, the result of processing all diagrams with
an $\Op{2}$ stub ultimately yields terms of four types: the two-loop analogue of
the \BPZ\ terms and $\Lambda$-derivative, $\alpha$ and $\beta$-terms. The set of \BPZ\
terms simply comprises an $\Op{2}$ stub joined, via a zero-point wine, to the complete
set of diagrams coming from the previous section. We have commented already that this set of
diagrams is transverse in $p$; in turn, this guarantees that the \BPZ\ terms can
be discarded at $\Op{2}$.

We will deal with all $\alpha$ and $\beta$-terms---including those generated
directly from equation~(\ref{eq:beta-n+})---together; hence, in this section, we will
present only those $\Op{2}$ terms which can be cast as $\Lambda$-derivatives. These
terms break down into four sets.

The first set of terms is shown in figure~\ref{fig:b2:Op2-A}. This set
comprises those diagrams for which the $\Op{2}$ stub is formed by the action
of an un-nested gauge remainder and for which the field attached to this gauge
remainder does not join directly to the free socket on the $\Op{2}$ stub. Note also that
the aforementioned gauge remainder is differentiated \wrt\ its momentum.
\begin{center}
\begin{figure}[h]
	\[
	2
	\dec{
		\begin{array}{l}
			\begin{array}{cccccc}
			\vspace{0.1in}
					& \OLdiscard{diag:326.1,325.9}	&	& \Dprocess{diag:328.23}{fig:TLD:C-stub-Diagrams}	& 	&\LabOLObLB{diag:327.37}
			\\
				2	& \ensuremath{\begin{array}{c}\input{pstex/Diagram326.1+325.9.pstex_t} \end{array}}		& -	& \ensuremath{\begin{array}{c}\input{pstex/Diagram328.23.pstex_t} \end{array}}									&+ 	& \ensuremath{\begin{array}{c}\input{pstex/Diagram327.37.pstex_t} \end{array}}	
			\end{array}
		\\
			\begin{array}{ccccccl}
			\vspace{0.1in}
						&\LabOLOb{diag:329.29}	&	&\LabOLOb{diag:331b.9}	&	& \LabOLObRB{diag:331b.10}		& \Process{fig:TLD:Diagrams224}
			\\
					- 	& \ensuremath{\begin{array}{c}\input{pstex/Diagram329.29.pstex_t} \end{array}}	&+2	& \ensuremath{\begin{array}{c}\input{pstex/Diagram331b.9.pstex_t} \end{array}}	& +2& \ensuremath{\begin{array}{c}\input{pstex/Diagram331b.10.pstex_t} \end{array}}			&	
			\end{array}
		\end{array}
	}{\bullet}
	\]
\caption{The first set of $\Lambda$-derivative terms with an $\Op{2}$ stub.}
\label{fig:b2:Op2-A}
\end{figure}
\end{center}

There is a clear pattern to the terms of figure~\ref{fig:b2:Op2-A}, with each possessing a
sub-diagram familiar from our calculation of $\beta_1$. Upon examination of diagram~\ref{diag:326.1,325.9},
this comes as no surprise. Diagram~\ref{diag:326.1,325.9} contains a two-point, one loop vertex; given
that $\beta_1$ is obtained by computing the flow of just such a vertex (but with the fields in the $A^1$-sector)
this explains how the remaining terms are generated.

There are, however, some differences between the sub-diagrams formed by computing the flow
of the two-point, one-loop vertex of diagram~\ref{diag:326.1,325.9} and the diagrams which
contribute to $\beta_1$. These differences arise because, in the current case, the two-point,
one-loop vertex is decorated by internal, as opposed to external fields. This immediately
explains why there are no diagrams possessing little set-like sub-diagrams: the parents
of the little set possess two-point, tree level vertices; if this vertex is decorated by an
internal field then, not only are we no longer required to Taylor expand the rest of
the diagram, but we can also simply utilise the effective propagator relation.

Furthermore, notice that there are two diagrams, \ref{diag:331b.9} and~\ref{diag:331b.10},
containing sub-diagrams like the third element of the standard set.
If the external
 fields of these sub-diagrams
were in the $A$-sector, then we could simply combine 
diagrams~\ref{diag:331b.9} and~\ref{diag:331b.10}. This would then give the third element
of the standard set with the correct relative factor, compared to diagrams~\ref{diag:327.37}
and~\ref{diag:329.29}.

Returning to diagram~\ref{diag:326.1,325.9}, and taking all fields to be in 
the $A$-sector, the renormalisation condition forces
the two-point, one loop vertex to go at least as the fourth power of momentum. This 
ensures that the Taylor expansion in $p$ which was performed to derive this diagram
is valid. Since diagrams~\ref{diag:328.23}--\ref{diag:331b.10} are spawned by 
diagram~\ref{diag:326.1,325.9} it immediately follows that $p$ is set to zero
everywhere other than the $\Op{2}$ stub, in these diagrams.

The second set of $\Lambda$-derivative terms with an $\Op{2}$ stub is shown in figure~\ref{fig:b2:Op2-b1}.
As with the previous set, the $\Op{2}$ stub is formed by the action of an un-nested gauge remainder.
This time, however, the effective propagator attached to the gauge remainder loops round
to decorate the free socket of the stub.

\begin{center}
\begin{figure}
	\[
	\begin{array}{c}
	\vspace{0.1in}
		-
		\dec{
			2 \ensuremath{\begin{array}{c}\input{pstex/LittleSet-A.pstex_t} \end{array}} + \ensuremath{\begin{array}{c}\input{pstex/LittleSet-B.pstex_t} \end{array}}
		}{\bullet}
		\Delta^{1 \, 1}_{\alpha \beta} (p)
	\\
		\times
		\left[
			\begin{array}{c}
				\begin{array}{ccccccccl}
					&\LabOLObLB{diag:383.2a}&	&\LabOLOb{diag:383.2b}	&	&\LabOLOb{diag:382.2c}	&	& \LabOLObRB{diag:383.2d} & \Process{fig:b2:Diagrams223b}
				\\
					& \ensuremath{\begin{array}{c}\input{pstex/Beta1-A-al.pstex_t} \end{array}} 		& - & \ensuremath{\begin{array}{c}\input{pstex/Beta1-B-al.pstex_t} \end{array}} 		& +4&\ensuremath{\begin{array}{c}\input{pstex/Beta1-C-al.pstex_t} \end{array}}		& - & \ensuremath{\begin{array}{c}\input{pstex/Beta1-D-al.pstex_t} \end{array}}		&
				\end{array}
			\\
				\begin{array}{ccccc}
						& \LabOLObLB{diag:FR28.15a,b}	&	&\LabOLObRB{diag:FR28.15c,d} & \Process{fig:b2:beta-Op2}
				\\
					+4	& \ensuremath{\begin{array}{c}\input{pstex/LittleSet-A-al.pstex_t} \end{array}} 			& +2& \ensuremath{\begin{array}{c}\input{pstex/LittleSet-B-al.pstex_t} \end{array}} 		&
				\end{array}
			\end{array}
		\right]
	\end{array}
	\]
\caption{The second set of $\Lambda$-derivative terms with an $\Op{2}$ stub.}
\label{fig:b2:Op2-b1}
\end{figure}
\end{center}

There are a number of things to note about the diagrams of figure~\ref{fig:b2:Op2-b1},
which we list below.
\begin{enumerate}

	\item	Though we have chosen to label the sub-diagrams carrying index $\alpha$, these
			labels refer to complete diagrams; thus each label refers to two diagrams.

	\item 	The set of sub-diagrams carrying index $\alpha$ 
			is \emph{not} the same as the set obtained in our computation
			of $\beta_1$: the final two diagrams come with a relative factor of $1/2$. 

	\item	Upon attaching the effective propagator, $\Delta^{1 \, 1}_{\alpha \beta} (p)$, 
			in diagrams~\ref{diag:FR28.15a,b} and~\ref{diag:FR28.15c,d},
			we can use the effective propagator relation but are free to discard the gauge remainder. This follows
			as a consequence of Lorentz invariance and the transversality of the $\Op{2}$ stub.
			To see this, start by looking at the diagrams carrying index $\beta$, and suppose that
			the contracted index of the $\Op{2}$ stub is just $\gamma$. Then, by Lorentz invariance, the loop
			integral must yield the index structure $ \delta_{\beta \gamma}$. Now, the gauge remainder generated by
			the application of the effective propagator relation is $p_\alpha p_\beta /p^2$; the $ \delta_{\beta \gamma}$
			ensures that the $p_\beta$ is contracted into the $\Op{2}$ stub, killing the diagram.

	\item 	Diagrams~\ref{diag:FR28.15a,b} and~\ref{diag:FR28.15c,d} can be cast, at $\Op{2}$,
			such that \flowConstAl\ strikes a set of whole diagrams (rather than just factorisable sub-diagrams).
			This follows because
			the combination of the two sub-diagrams with index $\alpha$ take 
			exactly the same form
			as the sub-diagrams struck by \flowConstAl.

	\item	The Taylor expansion in $p$ used to derive the two diagrams carrying index $\beta$
			is valid. The steps are  just a repetition of some of the one loop diagrammatics
			and we know this to be legal. 
\end{enumerate}

The third set of $\Lambda$-derivative terms with an $\Op{2}$ stub is shown in figure~\ref{fig:b2:Op2-B}.
These are distinguished by the fact the the $\Op{2}$ stub is formed by the action of a nested gauge
remainder.

\begin{center}
\begin{figure}
	\[
	4
	\dec{
		\begin{array}{l}	
			\begin{array}{ccccccccl}
			\vspace{0.1in}
					&\LabOLObLB{diag:FR22.2}& 	&\LabOLOb{diag:FR22.1}	& 	&\LabOLOb{diag:FR24.6}	& 	&\LabOLObRB{diag:FR24.1}& \Process{fig:TLD:DFR22.1...:WS}
			\\
				-	&\ensuremath{\begin{array}{c}\input{pstex/DiagramFR22.2.pstex_t} \end{array}}		& -	&\ensuremath{\begin{array}{c}\input{pstex/DiagramFR22.1.pstex_t} \end{array}}		& +	& \ensuremath{\begin{array}{c}\input{pstex/DiagramFR24.6.pstex_t} \end{array}}	& +	&\ensuremath{\begin{array}{c}\input{pstex/DiagramFR24.1.pstex_t} \end{array}}		&
			\end{array}
		\\
			\begin{array}{ccccccccl}
			\vspace{0.1in}
					&\LabOLObLB{diag:FR36.1}&	&\LabOLOb{diag:FR22.3}	& 	&\LabOLOb{diag:FR24.11}	&	&\LabOLObRB{diag:FR27.1}	&\Process{fig:TLD:FR36.1etc-U,NU}
			\\
				+	&\ensuremath{\begin{array}{c}\input{pstex/DiagramFR36.1.pstex_t} \end{array}}		&+	&\ensuremath{\begin{array}{c}\input{pstex/DiagramFR22.3.pstex_t} \end{array}}		& - &\ensuremath{\begin{array}{c}\input{pstex/DiagramFR24.11.pstex_t} \end{array}}		&+	&\ensuremath{\begin{array}{c}\input{pstex/DiagramFR27.1.pstex_t} \end{array}}		&
			\end{array}
		\\
			\begin{array}{ccccccl}
			\vspace{0.1in}
					&\PO{diag:FR27.6}{fig:TLD:FR36.1etc-U,NU}	&		&\LabOLObLB{diag:382.1}			&		&\LabOLObRB{diag:351.5}	& \hspace{-2em} \Discard
			\\
				-\hf&\ensuremath{\begin{array}{c}\input{pstex/DiagramFR27.6.pstex_t} \end{array}}							&-\hf	&\AGROB{\ensuremath{\begin{array}{c}\input{pstex/Diagram382.1.pstex_t} \end{array}}}		&+\hf	&\AGROB{\ensuremath{\begin{array}{c}\input{pstex/Diagram351.5.pstex_t} \end{array}}}		&
			\end{array}
		\end{array}
	}{\bullet}
	\]
\caption{The third set of $\Lambda$-derivative terms with an $\Op{2}$ stub.}
\label{fig:b2:Op2-B}
\end{figure}
\end{center}

A subset of the parent diagrams for these terms are not Taylor expandable to the desired order in $p$.
As detailed in section~\ref{sec:TotalDerivativeMethods:Op2-NTE} it is necessary to construct subtractions,
in this case. Manipulation of the additions helps generate the diagrams of figure~\ref{fig:b2:Op2-B};
the parents and subtractions combine to give the diagrams of figure~\ref{fig:b2:Op2-NTE}.
\begin{center}
\begin{figure}
	\[
	-4
	\dec{
		\left[
			\begin{array}{ccc}
				\ensuremath{\begin{array}{c}\input{pstex/Diagram180.7-be.pstex_t} \end{array}}	&+	&\ensuremath{\begin{array}{c}\input{pstex/Diagram180.9-be.pstex_t} \end{array}}
			\end{array}
		\right]_{\Pep}
		\times \Delta^{1 \,1}_{\alpha \beta}(p)
		\left[
			\begin{array}{ccccl}
			\vspace{0.1in}
					&\LOLB{NTE-A}			&	&\LORB{NTE-B}		&\Process{fig:b2:beta-Op2-B}
			\\
				2	&\ensuremath{\begin{array}{c}\input{pstex/LittleSet-A-al.pstex_t} \end{array}}&+	&\ensuremath{\begin{array}{c}\input{pstex/LittleSet-B-al.pstex_t} \end{array}}	&
			\end{array}
		\right]
	}{\bullet}
	\]
\caption{Terms with an $\Op{2}$ stub which are not Taylor expandable to the desired order in $p$.}
\label{fig:b2:Op2-NTE}
\end{figure}
\end{center}

\section{$\beta$-Terms} \label{sec:b2:beta}

In this section, we collect together all  $\beta$-terms, as shown in figure~\ref{fig:b2:beta}.

\begin{center}
\begin{figure}
	\[
	\begin{array}{c}
		\ds
		-2\beta_1(\alpha)
	\\[2ex]
		\left[
			\begin{array}{c}
				\begin{array}{cccccccl}
				\vspace{0.1in}
					\LabOLObLB{diag:55.15}	&	&\LabOLOb{diag:56.17}	&	&\LabOLOb{diag:58e.3}	&	&\LabOLObRB{diag:56.4b,58.1}&\Process{fig:b2:Diagrams223b}
				\\
					\ensuremath{\begin{array}{c}\input{pstex/Diagram55.15.pstex_t} \end{array}}		&-	&\ensuremath{\begin{array}{c}\input{pstex/Diagram56.17.pstex_t} \end{array}}		&+4	&\ensuremath{\begin{array}{c}\input{pstex/Diagram58e.3.pstex_t} \end{array}}		&-	&\ensuremath{\begin{array}{c}\input{pstex/Diagram56.4b,58.1.pstex_t} \end{array}}		&
				\end{array}
			\\
				\begin{array}{ccccccccl}
				\vspace{0.1in}
						&\LabOLObLB{diag:180.7}	&	&\LabOLOb{diag:180.9}	&	&\LabOLOb{diag:326.3}	&	&\LabOLObRB{diag:326.4}	&\Process{fig:b2:beta-Op2}
				\\
					+2	&\ensuremath{\begin{array}{c}\input{pstex/Diagram180.7.pstex_t} \end{array}}		&+2	&\ensuremath{\begin{array}{c}\input{pstex/Diagram180.9.pstex_t} \end{array}}		&-4	&\ensuremath{\begin{array}{c}\input{pstex/Diagram326.3.pstex_t} \end{array}}		&+4	&\ensuremath{\begin{array}{c}\input{pstex/Diagram326.4.pstex_t} \end{array}}		&
				\end{array}
			\end{array}
		\right]
	\end{array}
	\]
\caption{All $\beta$-terms.}
\label{fig:b2:beta}
\end{figure}
\end{center}

There is a great temptation to try and manipulate diagrams~\ref{diag:180.7} and~\ref{diag:180.9}
at $\Op{2}$. However, we must be extremely careful: when all fields are in the $A$-sector,
the loop integral has components which are not Taylor expandable to the desired order in $p$. Thus, whilst we can Taylor expand the three point
vertex of diagram~\ref{diag:180.7}, we cannot set $\Delta^{11}(l-p)$ to $\Delta^{11}(l)$.
Nonetheless, we can make progress with these terms, as we will see in section~\ref{sec:bulk:simplify}.

\section{$\alpha$-Terms} \label{sec:b2:alpha}

The $\alpha$-terms split naturally into two sets, shown in figures~\ref{fig:b2:alpha-1}
and~\ref{fig:b2:alpha-2}. The first set are those for which the $\alpha$-derivative strikes
an $\Op{2}$ stub. The second set are those for which the $\alpha$-derivative strikes any
other vertex.

\begin{center}
\begin{figure}
	\[
	\gamma_1(\alpha)
	\left[
		\begin{array}{cccccccl}
		\vspace{0.1in}
			\LabOLObLB{diag:55.16}	&	&\LabOLOb{diag:56.18}	&	&\LabOLOb{diag:58e.4}	&	&\LabOLObRB{diag:56.4c,58.2}& \Discard
		\\
			\ensuremath{\begin{array}{c}\input{pstex/Diagram55.16.pstex_t} \end{array}}		&-	&\ensuremath{\begin{array}{c}\input{pstex/Diagram56.18.pstex_t} \end{array}}		&+4	&\ensuremath{\begin{array}{c}\input{pstex/Diagram58e.4.pstex_t} \end{array}}		&-	&\ensuremath{\begin{array}{c}\input{pstex/Diagram56.4c,58.2.pstex_t} \end{array}}		&
		\end{array}
	\right]
	\]
\caption{$\alpha$-terms of the first type.}
\label{fig:b2:alpha-1}
\end{figure}
\end{center}

It is straightforward to argue that terms of the first type vanish, at $\Op{2}$.
The $\Op{2}$ stub of diagrams~\ref{diag:55.16}--\ref{diag:56.4c,58.2} is attached, via
an effective propagator, to something we know to be transverse in $p$. Hence, the $1/p^2$
coming from the effective propagator is ameliorated. Given that the $\Op{2}$
part of the vertex $S^{\ 1\, 1}_{0 \mu\alpha}(p)$ is independent of $\alpha$,
the diagrams can be discarded.

\begin{center}
\begin{figure}
	\[
	\frac{1}{2} \gamma_1(\alpha)
	\left[
		\begin{array}{l}
			\begin{array}{cccccc}
			\vspace{0.1in}
					&\LabOLObLB{diag:2.1.2}	&	&\LabOLOb{diag:6.4}	&	&\LabOLOb{diag:9.10}	
			\\
				-	&\ensuremath{\begin{array}{c}\input{pstex/Diagram2.1.2.pstex_t} \end{array}}		&+	&\ensuremath{\begin{array}{c}\input{pstex/Diagram6.4.pstex_t} \end{array}}		&+2	&\ensuremath{\begin{array}{c}\input{pstex/Diagram9.10.pstex_t} \end{array}}
			\end{array}
		\\
			\begin{array}{cccccc}
			\vspace{0.1in}
					&\LabOLOb{diag:9a.7}&	& \LabOLOb{diag:9b.3}	&	&\LabOLOb{diag:7.4}
			\\
				+	&\ensuremath{\begin{array}{c}\input{pstex/Diagram9a.7.pstex_t} \end{array}}	&+	& \ensuremath{\begin{array}{c}\input{pstex/Diagram9b.3.pstex_t} \end{array}}		&-	&\ensuremath{\begin{array}{c}\input{pstex/Diagram7.4.pstex_t} \end{array}}
			\end{array}
		\\
			\begin{array}{ccccccl}
			\vspace{0.1in}
					&\LabOLOb{diag:9c.8}&	&\LabOLOb{diag:7.16}&	&\LabOLObRB{diag:326.5}	& \Process{fig:bulk:alpha-terms}
			\\
				-	&\ensuremath{\begin{array}{c}\input{pstex/Diagram9c.8.pstex_t} \end{array}}	&-2	&\ensuremath{\begin{array}{c}\input{pstex/Diagram7.16.pstex_t} \end{array}}	&-8	&\ensuremath{\begin{array}{c}\input{pstex/Diagram326.5.pstex_t} \end{array}}		&
			\end{array}
		\end{array}
	\right]
	\]
\caption{$\alpha$-terms of the second type.}
\label{fig:b2:alpha-2}
\end{figure}
\end{center}

We conclude this section by developing---and then applying---a set of diagrammatic identities
to enable us to process the remaining $\alpha$-terms. The first key
ingredient is the observation that $\GRk$ is independent of $\alpha$ and so is killed
by $\alpha$-derivatives. Recall that we can use \Dal\ as shorthand for differentiation
\wrt\ $\alpha$.

\begin{D-ID}
A sub-diagram comprising a two-point, tree level vertex, differentiated \wrt\ $\alpha$ and attached to an effective propagator can, using the effective
propagator relation,
 be redrawn, in the following form:
\[
	\ensuremath{\begin{array}{c}\input{pstex/D-ID-V-al-EP-A.pstex_t} \end{array}} \equiv - \ensuremath{\begin{array}{c}\input{pstex/D-ID-V-al-EP-B.pstex_t} \end{array}} - \ensuremath{\begin{array}{c}\begin{picture}(0,0)%
\includegraphics{pstex/D-ID-V-al-EP-C.pstex}%
\end{picture}%
\setlength{\unitlength}{3947sp}%
\begingroup\makeatletter\ifx\SetFigFont\undefined%
\gdef\SetFigFont#1#2#3#4#5{%
  \reset@font\fontsize{#1}{#2pt}%
  \fontfamily{#3}\fontseries{#4}\fontshape{#5}%
  \selectfont}%
\fi\endgroup%
\begin{picture}(319,361)(1593,-466)
\put(1629,-201){\makebox(0,0)[lb]{\smash{\SetFigFont{8}{9.6}{\rmdefault}{\mddefault}{\updefault}{\color[rgb]{0,0,0}$\alpha$}%
}}}
\end{picture}
 \end{array}}.
\]
\label{D-ID:V-al-EP}
\end{D-ID}

\begin{D-ID}
A sub-diagram comprising a two-point, tree level  vertex, differentiated \wrt\ $\alpha$ and attached to two effective
propagators can, using diagrammatic 
identity~\ref{D-ID:V-al-EP} and the effective
propagator relation, be redrawn in the following form:
\[
	\ensuremath{\begin{array}{c}\input{pstex/D-ID-V-al-EPx2-A.pstex_t} \end{array}} 
	\equiv 
	- \ensuremath{\begin{array}{c}\begin{picture}(0,0)%
\includegraphics{pstex/D-ID-V-al-EPx2-B.pstex}%
\end{picture}%
\setlength{\unitlength}{3947sp}%
\begingroup\makeatletter\ifx\SetFigFont\undefined%
\gdef\SetFigFont#1#2#3#4#5{%
  \reset@font\fontsize{#1}{#2pt}%
  \fontfamily{#3}\fontseries{#4}\fontshape{#5}%
  \selectfont}%
\fi\endgroup%
\begin{picture}(328,204)(514,-10)
\put(635, 98){\makebox(0,0)[lb]{\smash{\SetFigFont{8}{9.6}{\rmdefault}{\mddefault}{\updefault}{\color[rgb]{0,0,0}$\alpha$}%
}}}
\end{picture}
 \end{array}} 
	+\frac{1}{2}
	\left[
		\begin{array}{ccc}
			\ensuremath{\begin{array}{c}\begin{picture}(0,0)%
\includegraphics{pstex/D-ID-V-al-EPx2-C.pstex}%
\end{picture}%
\setlength{\unitlength}{3947sp}%
\begingroup\makeatletter\ifx\SetFigFont\undefined%
\gdef\SetFigFont#1#2#3#4#5{%
  \reset@font\fontsize{#1}{#2pt}%
  \fontfamily{#3}\fontseries{#4}\fontshape{#5}%
  \selectfont}%
\fi\endgroup%
\begin{picture}(356,250)(182,555)
\put(352,709){\makebox(0,0)[lb]{\smash{\SetFigFont{8}{9.6}{\rmdefault}{\mddefault}{\updefault}{\color[rgb]{0,0,0}$\alpha$}%
}}}
\end{picture}
 \end{array}}	& - & \ensuremath{\begin{array}{c}\begin{picture}(0,0)%
\includegraphics{pstex/D-ID-V-al-EPx2-D.pstex}%
\end{picture}%
\setlength{\unitlength}{3947sp}%
\begingroup\makeatletter\ifx\SetFigFont\undefined%
\gdef\SetFigFont#1#2#3#4#5{%
  \reset@font\fontsize{#1}{#2pt}%
  \fontfamily{#3}\fontseries{#4}\fontshape{#5}%
  \selectfont}%
\fi\endgroup%
\begin{picture}(627,344)(355,494)
\put(711,742){\makebox(0,0)[lb]{\smash{\SetFigFont{8}{9.6}{\rmdefault}{\mddefault}{\updefault}{\color[rgb]{0,0,0}$\alpha$}%
}}}
\end{picture}
 \end{array}} \\
			\ensuremath{\begin{array}{c}\begin{picture}(0,0)%
\includegraphics{pstex/D-ID-V-al-EPx2-E.pstex}%
\end{picture}%
\setlength{\unitlength}{3947sp}%
\begingroup\makeatletter\ifx\SetFigFont\undefined%
\gdef\SetFigFont#1#2#3#4#5{%
  \reset@font\fontsize{#1}{#2pt}%
  \fontfamily{#3}\fontseries{#4}\fontshape{#5}%
  \selectfont}%
\fi\endgroup%
\begin{picture}(356,250)(223,555)
\put(352,709){\makebox(0,0)[lb]{\smash{\SetFigFont{8}{9.6}{\rmdefault}{\mddefault}{\updefault}{\color[rgb]{0,0,0}$\alpha$}%
}}}
\end{picture}
 \end{array}}	& - & \ensuremath{\begin{array}{c}\begin{picture}(0,0)%
\includegraphics{pstex/D-ID-V-al-EPx2-F.pstex}%
\end{picture}%
\setlength{\unitlength}{3947sp}%
\begingroup\makeatletter\ifx\SetFigFont\undefined%
\gdef\SetFigFont#1#2#3#4#5{%
  \reset@font\fontsize{#1}{#2pt}%
  \fontfamily{#3}\fontseries{#4}\fontshape{#5}%
  \selectfont}%
\fi\endgroup%
\begin{picture}(627,337)(514,501)
\put(711,742){\makebox(0,0)[lb]{\smash{\SetFigFont{8}{9.6}{\rmdefault}{\mddefault}{\updefault}{\color[rgb]{0,0,0}$\alpha$}%
}}}
\end{picture}
 \end{array}}
		\end{array}
	\right].
\]
We could re-express this in a less symmetric form by taking either of the two rows in the square brackets and removing the
factor of half.
\label{D-ID:V-al-Epx2}
\end{D-ID}

\begin{D-ID}
A sub-diagram comprising a two-point, tree level  vertex, differentiated \wrt\ $\alpha$ and attached to
an effective propagator which terminates in a $\GRk'$ vanishes:
\[
	\ensuremath{\begin{array}{c}\input{pstex/D-ID-V-al-EP-GR.pstex_t} \end{array}} \equiv \ensuremath{\begin{array}{c}\input{pstex/D-ID-V-al-EP-GR-B.pstex_t} \end{array}} = 0,
\]
as follows from diagrammatic identities~\ref{D-ID:PseudoEffectivePropagatorRelation} and~\ref{D-ID:TLTP-GR}.
\label{D-ID:V-al-EP-GR}
\end{D-ID}

\begin{D-ID}
A sub-diagram comprising a two-point, tree level  vertex attached to an
effective propagator, differentiated \wrt\
$\alpha$, which terminates in a $\wedge$ can be redrawn, using diagrammatic
identity~\ref{D-ID:V-al-EP-GR}, the effective propagator relation diagrammatic
identity~\ref{D-ID:GaugeRemainderRelation}:
\[
	\ensuremath{\begin{array}{c}\input{pstex/D-ID-V-EP-al-GR.pstex_t} \end{array}} \equiv - \ensuremath{\begin{array}{c}\begin{picture}(0,0)%
\includegraphics{pstex/D-ID-V-EP-al-GR-B.pstex}%
\end{picture}%
\setlength{\unitlength}{3947sp}%
\begingroup\makeatletter\ifx\SetFigFont\undefined%
\gdef\SetFigFont#1#2#3#4#5{%
  \reset@font\fontsize{#1}{#2pt}%
  \fontfamily{#3}\fontseries{#4}\fontshape{#5}%
  \selectfont}%
\fi\endgroup%
\begin{picture}(165,369)(151,367)
\put(203,640){\makebox(0,0)[lb]{\smash{\SetFigFont{8}{9.6}{\rmdefault}{\mddefault}{\updefault}{\color[rgb]{0,0,0}$\alpha$}%
}}}
\end{picture}
 \end{array}}.
\]
\label{D-ID:V-al-EPx2-GR}
\end{D-ID}

Applying the new diagrammatic identities to diagrams~\ref{diag:6.4}--\ref{diag:326.5} and their
daughters, as appropriate, we obtain the terms of figure~\ref{fig:bulk:alpha-terms}.
\begin{center}
\begin{figure}[h]
\[
-\frac{1}{2}\gamma_1 \pder{}{\alpha}
\left[
	\begin{array}{c}
		\begin{array}{ccccccc}
		\vspace{0.1in}
			\LabOLOb{b2-alA}&	&\LabOLOb{b2-alB}	&	&\LabOLOb{b2-alC}	&	&\LabOLOb{b2-alDE}
		\\
			\ensuremath{\begin{array}{c}\input{pstex/Beta1-LdL-A.pstex_t} \end{array}}& -	&\ensuremath{\begin{array}{c}\input{pstex/Beta1-LdL-B.pstex_t} \end{array}} 	& +4& \ensuremath{\begin{array}{c}\input{pstex/Beta1-LdL-C.pstex_t} \end{array}} 	& -	& \ensuremath{\begin{array}{c}\input{pstex/Beta1-LdL-DE.pstex_t} \end{array}}
		\end{array}
	\\
		\begin{array}{cccc}
		\vspace{0.1in}
				& \LabOLOb{b2-alF}	&	&\LabOLOb{b2-alG}
		\\
			+8	& \ensuremath{\begin{array}{c}\input{pstex/Beta1-LdL-F.pstex_t} \end{array}}	&+4	& \ensuremath{\begin{array}{c}\input{pstex/Beta1-LdL-G.pstex_t} \end{array}}
		\end{array}
	\end{array}
\right]
\]
\caption{All remaining $\alpha$-terms.}
\label{fig:bulk:alpha-terms}
\end{figure}
\end{center}

It is thus apparent that the $\alpha$-terms arising from the computation of $\beta_2$
can be reduced to the $\alpha$-derivative of the set of terms which we differentiate
\wrt\ \LConstAl\ to give
$\beta_1$!

\section{Simplifications} \label{sec:bulk:simplify}

In this section we simplify the diagrammatic expression 
for $\beta_2$ by utilising our diagrammatic expression for
$\beta_1$. We begin by examining 
diagrams~\ref{diag:55.15}--\ref{diag:56.4b,58.1} which,
having re-expressed, we will combine with the
diagrams of figure~\ref{fig:b2:b1-sq-A} and 
diagrams~\ref{diag:383.2a}--\ref{diag:383.2d}

Referring to diagrams~\ref{diag:55.15}--\ref{diag:56.4b,58.1},
the first thing to note is that we can use the effective propagator relation
to get rid of the $\Op{2}$ stub and the effective propagator to which
it attaches. Having done this, we can discard the resulting gauge remainder,
as it strikes a set of diagrams transverse in $p$. The next step is to expand the
overall factor of $\beta_1$, using our earlier diagrammatic expression.
To do this, we perform two intermediate steps. First, change the index $\nu$ to $\alpha$ and multiply
by $\delta_{\alpha \nu}$.
Secondly, we use the  fact that
diagrams~\ref{diag:55.15}--\ref{diag:56.4b,58.1} are transverse in $p$ to recognise that, at $\Op{2}$,
we can replace $\delta_{\alpha \nu}$
by $\Box_{\beta \nu}(p) 2 \Delta^{1 \, 1}_{\alpha \beta}(p)$. 
We now insert our diagrammatic expression for $\beta_1 \Box_{\beta \nu}(p)$ and, having done this, join
together one-loop
sub-diagrams with the effective propagator, $\Delta^{1 \, 1}_{\alpha \beta}(p)$. Now we are ready to identify cancellations.

At $\Op{2}$, the diagrams of figure~\ref{fig:b2:b1-sq-A} are completely cancelled.
The remaining terms are shown in figure~\ref{fig:b2:Diagrams223b}; they are the same as diagrams~\ref{diag:383.2a}--\ref{diag:383.2d},
but with opposite sign.
\begin{center}
\begin{figure}[h]
	\[
	\begin{array}{c}
	\vspace{0.1in}
		\dec{
			2 \ensuremath{\begin{array}{c}\input{pstex/LittleSet-A.pstex_t} \end{array}} + \ensuremath{\begin{array}{c}\input{pstex/LittleSet-B.pstex_t} \end{array}}
		}{\bullet}
		\Delta^{1 \, 1}_{\alpha \beta} (p)
	\\
		\times
		\left[
			\begin{array}{ccccccccl}
				&\LabOLObLB{diag:383.2a-B}&	&\LabOLOb{diag:383.2b-B}	&	&\LabOLOb{diag:382.2c-B}	&	& \LabOLObRB{diag:383.2d-B} & \Process{fig:b2:beta-Op2}
			\\
				& \ensuremath{\begin{array}{c}\input{pstex/Beta1-A-al.pstex_t} \end{array}} 		& - & \ensuremath{\begin{array}{c}\input{pstex/Beta1-B-al.pstex_t} \end{array}} 		& +4&\ensuremath{\begin{array}{c}\input{pstex/Beta1-C-al.pstex_t} \end{array}}		& - & \ensuremath{\begin{array}{c}\input{pstex/Beta1-D-al.pstex_t} \end{array}}		&
			\end{array}
		\right]
	\end{array}
	\]
\caption{Result of combining the
diagrams of figure~\ref{fig:b2:b1-sq-A}, 
diagrams~\ref{diag:383.2a}--\ref{diag:383.2d} 
and~\ref{diag:55.15}--\ref{diag:56.4b,58.1}.}
\label{fig:b2:Diagrams223b}
\end{figure}
\end{center}

The next step is to examine diagrams~\ref{diag:180.7}--\ref{diag:326.4}.
Replacing $\beta_1$ proceeds exactly as before: first  we change the
index $\nu$ to $\alpha$ and multiply by 
$2 \Box_{\beta \nu}(p) \Delta^{1\, 1}_{\alpha \beta}(p)$;
secondly, we substitute for $\beta_1 \Box_{\beta \nu}(p)$.

Noting that the diagrams contributing to $\beta_1$ are hit by
\flowConstAl\ but the one-loop sub-diagrams to which they are joined
by the effective propagator, $\Delta^{1\, 1}_{\alpha \beta}(p)$, are not,
we now move the \flowConstAl. At $\Op{2}$, this will generate a set of terms
where whole two-loop diagrams are hit by \flowConstAl\
minus a correction in which the explicitly drawn sub-diagrams
of diagrams~\ref{diag:180.7}--\ref{diag:326.4} are hit by $\flowConstAl$.
This is precisely what we want: with these sub-diagrams now under a $\Lambda$-derivative,
they are Taylor expandable to the desired order in $p$. Performing this Taylor expansion, 
we can reduce the four sub-diagrams to an instance of the little set,
struck by \flowConstAl. We now find cancellations. First, those terms formed
from $\beta_1$ modulo the little set
exactly cancel diagrams~\ref{diag:383.2a-B}--\ref{diag:383.2d-B}. Secondly,
those diagrams from the little set contribution to $\beta_1$ can
be combined with diagrams~\ref{diag:FR28.15a,b}--\ref{diag:FR28.15c,d}.

The set of  surviving terms are shown in figure~\ref{fig:b2:beta-Op2}.
The labels for the first set of terms each refer to four diagrams. Upon attachment of
the effective propagator to a two-point, tree level vertex, the resulting gauge remainder
can be thrown away.
\begin{center}
\begin{figure}
	\[
	\begin{array}{c}
	\vspace{0.2in}
		\dec{
			\begin{array}{l}
			\vspace{0.2in}
				\left[
					\begin{array}{ccccccc}
						\ensuremath{\begin{array}{c}\input{pstex/Diagram180.7-be.pstex_t} \end{array}}	&+	&\ensuremath{\begin{array}{c}\input{pstex/Diagram180.9-be.pstex_t} \end{array}}	&-2	&\ensuremath{\begin{array}{c}\input{pstex/Diagram326.3-be.pstex_t} \end{array}}	&+2	&\ensuremath{\begin{array}{c}\input{pstex/Diagram326.4-be.pstex_t} \end{array}}
					\end{array}
				\right]	
			\\		
				\times \Delta^{1 \,1}_{\alpha \beta}(p)
				\left[
					\begin{array}{c}
						\begin{array}{cccccl}
						\vspace{0.1in}
							\LabOLObLB{b2-b1A}&	&\LabOLOb{b2-b1B}	&	&\LORB{b2-b1C}		&	\Process{fig:TLD:Diagrams224}
						\\
							\ensuremath{\begin{array}{c}\input{pstex/Beta1-A-al.pstex_t} \end{array}}& -	&\ensuremath{\begin{array}{c}\input{pstex/Beta1-B-al.pstex_t} \end{array}} 	& +4& \ensuremath{\begin{array}{c}\input{pstex/Beta1-C-al.pstex_t} \end{array}} 	& 
						\end{array}
					\\
						\begin{array}{ccccccl}
						\vspace{0.1in}
								&\PO{b2-b1DE}{fig:TLD:C-stub-Diagrams}	& 	&\LOLB{b2-b1F}			&	&\LabOLObRB{b2-b1G}		& \Process{fig:b2:beta-Op2-B}
						\\
							-	& \ensuremath{\begin{array}{c}\input{pstex/Beta1-D-al.pstex_t} \end{array}}						&+8	& \ensuremath{\begin{array}{c}\input{pstex/LittleSet-A-al.pstex_t} \end{array}}	&+4	& \ensuremath{\begin{array}{c}\input{pstex/LittleSet-B-al.pstex_t} \end{array}}	&
						\end{array}
					\end{array}
				\right]
			\end{array}
		}{\bullet}
	\\
		-3
		\dec{
			\begin{array}{ccccccl}
			\vspace{0.1in}
					&\LabOLObLB{LittleSetSq-A}	&	&\LabOLOb{LittleSetSq-B}&	&\LabOLObRB{LittleSetSq-C}	&\Process{fig:b2:beta-Op2-B}
			\\
				4	&\ensuremath{\begin{array}{c}\input{pstex/LittleSetSq-A.pstex_t} \end{array}}			&+4	&\ensuremath{\begin{array}{c}\input{pstex/LittleSetSq-B.pstex_t} \end{array}}		& +	&\ensuremath{\begin{array}{c}\input{pstex/LittleSetSq-C.pstex_t} \end{array}}			&
			\end{array}
		}{\bullet}
	\end{array}
	\]
\caption{Result of combining diagrams~}
\label{fig:b2:beta-Op2}
\end{figure}
\end{center}

There is now one final step to perform. We focus on diagrams~\ref{b2-b1F} and~\ref{b2-b1G}.
We know that the first two diagrams which they multiply (carrying index $\beta$)
have components which are not Taylor expandable to the desired order in $p$. These components
are exactly removed by diagrams~\ref{NTE-A}--\ref{NTE-B}. With these components removed,
we now Taylor expand these diagrams in $p$ and perform the usual diagrammatic manipulations,
reducing the four diagrams carrying index $\beta$ to an instance of the standard set. The
resulting diagrams can then be combined with diagrams~\ref{LittleSetSq-A}--\ref{LittleSetSq-C},
to yield the diagrams of figure~\ref{fig:b2:beta-Op2-B}.
\begin{center}
\begin{figure}
	\[
	\dec{
		\begin{array}{ccccccl}
		\vspace{0.1in}
				&\LabOLObLB{LSSq-A}	&	&\LabOLOb{LSSq-B}	&	&\LabOLObRB{LSSq-C}	&\Process{fig:TLD:LittleSetSq:U}
		\\
			4	&\ensuremath{\begin{array}{c}\input{pstex/LittleSetSq-A.pstex_t} \end{array}}	&+4	&\ensuremath{\begin{array}{c}\input{pstex/LittleSetSq-B.pstex_t} \end{array}}	& +	&\ensuremath{\begin{array}{c}\input{pstex/LittleSetSq-C.pstex_t} \end{array}}	&
		\end{array}
	}{\bullet}
	\]
\caption{Result of combining diagrams~\ref{b2-b1F} and~\ref{b2-b1G} with~\ref{NTE-A}--\ref{NTE-B}
and then combining the result with diagrams~\ref{LittleSetSq-A}--\ref{LittleSetSq-C}.}
\label{fig:b2:beta-Op2-B}
\end{figure}
\end{center}

We have thus demonstrated that $\beta_2$ can be reduced to a set of $\Lambda$-derivative
terms, where the entire diagram is hit by \flowConstAl, and $\alpha$-terms.
For ease of future reference, we present the simplified diagrammatic expression
for $\beta_2$ in appendix~\ref{app:beta_2}.

\chapter{Numerical Evaluation of $\beta_2$}	\label{ch:TotalLambdaDerivatives:Application}

In this chapter we apply the methodology of chapter~\ref{ch:LambdaDerivatives:Methodology} 
to the computation of $\beta_2$. The starting point is
the set of diagrams left over from chapter~\ref{ch:bulk}, which we will call the original set
(see also appendix~\ref{app:beta_2}). 

There are three things we must do. First, ignoring the $\alpha$-terms, we must demonstrate
that all non-computable contributions cancel, between diagrams. 
Secondly, we must show that the $\alpha$-terms
vanish in the limit that $\alpha \rightarrow 0$. Lastly, we must evaluate $\beta_2$.

We begin our first task
in section~\ref{sec:TLD:vanishing} with the simplest diagrams to treat---those which 
vanish in the $\epsilon \rightarrow 0$ limit---and build up in complexity from there. In section~\ref{sec:TLD:finite}
we treat a sub-set of those diagrams which are manifestly finite.\footnote{
There are some diagrams which, although they appear to 
contain IR divergences, turn out not to, for purely numerical reasons. It is not useful to
treat such diagrams in this section as they are more naturally dealt with in conjunction with those that do contain IR divergences.
}
There are two types of such terms that we will encounter. First, we treat a non-factorisable diagram
which yields a computable contribution to $\beta_2$. Secondly, we treat a set of factorisable diagrams
which, up to \Oep\ corrections, give a computable coefficient multiplying the fourth and fifth diagrams
of $\OLDs$. This computable coefficient turns out to vanish, as a consequence of the
transversality of the non-universal one-loop sub-diagrams it multiplies. 

We can guess from the pattern of terms that cancel in this way that we expect similar
cancellations involving the members of the standard set.
This is one of the things we 
investigate in section~\ref{sec:TLD:SSasSubD}. As we see in subsection~\ref{sec:TLD:SS-Mirror},
the cancellations  of section~\ref{sec:TLD:finite} are indeed mirrored, though things are
much more complex. This complexity arises, on the one hand, because the standard set comprises not only Taylor
expandable, finite parts but also both Taylor and non-Taylor expandable IR divergent
coefficients. On the other hand, the complete set of terms we seek in order to mirror
the earlier cancellations is not manifestly present in the original set. The philosophy we
take is to add zero to the calculation by constructing the terms we are missing,
together with identical diagrams with opposite sign. Our next task is to try and cancel
these latter diagrams.

This is trivial for those diagrams involving the first element of
the standard set, which is not surprising, since such diagrams are manifestly
finite. We choose not to treat these terms in section~\ref{sec:TLD:finite},
with the other manifestly finite diagrams, but treat them in 
subsection~\ref{sec:TLD:SS-Mirror-ComplementaryTerms},
in order to keep the standard set together.

Also in this section, we generate the required diagrams
involving the second and third members of the standard set. 
This is harder: we must construct subtractions 
for a subset of the original set; upon the manipulation
of the additions, we generate the terms we are after, and many
more, besides. We conclude section~\ref{sec:TLD:SSasSubD}
by manipulating the semi-computable partners of the additions.

In section~\ref{sec:TLD:FurtherSubtractions} we construct a further
set of subtractions. First, we do this, as appropriate, for the un-treated terms
in the original set. Secondly, we construct subtractions for a subset of
the diagrams
resulting from the manipulation of the additions generated
in section~\ref{sec:TLD:SSasSubD}. In section~\ref{sec:TLD:FurtherManipulations}
we perform a final set of manipulations, which remove all remaining non-computable
contributions.

In section~\ref{sec:TLD:alpha} we treat the $\alpha$-terms;
in section~\ref{sec:TLD:Compute} we give the result of our numerical evaluation of
$\beta_2$ and in section~\ref{sec:TLD:conc}, we conclude this chapter.

\section{Vanishing Diagrams} \label{sec:TLD:vanishing}

There are a number of diagrams in the original set which are IR safe, 
even before differentiation \wrt\ $\LConstAl$.
After differentiation, such terms will vanish in the 
$\epsilon \rightarrow 0$ limit. In almost all cases, diagrams vanish
on an individual level and, in this case, have been tagged 
$\rightarrow 0$ in chapter~\ref{ch:bulk}.
Whilst there are many diagrams which are trivially IR safe, 
there are cases in which it is necessary
to use some property of the diagram in question to 
demonstrate this, as we will discuss in 
section~\ref{sec:TLD:VanishAlone}.

In section~\ref{sec:TLD:VanishingSet} we encounter diagrams which 
can only be shown to vanish in the \eptoz\ limit when combined into a set.
In section~\ref{sec:TL:VansishingCompts} we recognise that
for certain diagrams possessing gauge remainders, the $ES'$
tag can be removed.

\subsection{Diagrams which Vanish Individually} \label{sec:TLD:VanishAlone}

To check that individual diagrams tagged $\rightarrow 0$ do indeed vanish, 
the first thing we do is put all fields carrying loop momenta\footnote{This excludes
fields
forced to carry zero momentum which must be $C$s.} in the $A$-sector;
if a diagram is IR safe in this sector then it will be IR safe in all sectors.

There are now three observations  
which are necessary to show that certain $\Lambda$-derivative terms vanish
in the \eptoz\ limit:
\begin{enumerate}
	\item	the renormalisation condition~(\ref{eq:Intro:RenormCondition}),
			together with the form for $S_{0 \alpha \beta}^{\ 1 \, 1}(q)$
			(equation~(\ref{eq:app:TLTP-11}))
			demands that $S_{1 \alpha \beta}^{\ 1 \, 1}(q) \sim \mathcal{O}(q^4)$,
			which is sufficient to guarantee that all diagrams containing
			one-loop vertices vanish;

	\item 	gauge invariance forces $S^{AAC}_{\alpha \, \beta}(q) \sim \mathcal{O}(q^2)$ 
			which is sufficient to guarantee that diagrams such
			as~\ref{diag:34.1} vanish;

	\item	sub-diagrams which are divergent as a consequence of all fields
			being in the $A$-sector but that attach to a $C$-sector effective 
			propagator no longer diverge because, since $A$-fields do
			not carry a fifth index, the sub-diagram has no support
			in the divergent sector (\eg\ diagram~\ref{diag:395.4}).
\end{enumerate}

\subsection{Diagrams which Vanish as a Set} \label{sec:TLD:VanishingSet}

Diagrams~\ref{diag:21.1}--\ref{diag:375.1} have been reproduced in figure~\ref{fig:TLD:VanishingSet}. 
\begin{center}
\begin{figure}[h]
\[
	-\frac{1}{4}
	\dec{
		\begin{array}{cccccl}
		\vspace{0.1in}
				\LabOLObLB{diag:21.1-B}	&	& \LabOLOb{diag:36.1-B}	&	& \LabOLObRB{diag:375.1-B} 		& \Discard
			\\
				\ensuremath{\begin{array}{c}\input{pstex/Diagram21.1.pstex_t} \end{array}}		&+	& \ensuremath{\begin{array}{c}\input{pstex/Diagram36.1.pstex_t} \end{array}}		& -4& \AGROB{\ensuremath{\begin{array}{c}\input{pstex/Diagram375.1.pstex_t} \end{array}}} 	&
		\end{array}
	}{\bullet}
\]
\caption{Three diagrams which do not separately vanish as $\eptoz$, but vanish when combined into a set related by gauge invariance.}
\label{fig:TLD:VanishingSet}
\end{figure}
\end{center}

For the individual
diagrams to have any chance of surviving differentiation \wrt\ \flowConstAl, 
the internal legs leaving the common four-point vertex must be in the A-sector (and we must take only the
$\Op{2}$ part of this vertex). However, if these legs are in the $A$-sector then, summing over the three diagrams, it is clear
that the common four-point vertex attaches to the standard set.
Since the standard set is transverse in momentum, the IR behaviour of the diagrams is improved---to the extent that the set of
diagrams clearly vanishes in the $\epsilon \rightarrow 0$ limit. Note that diagrams~\ref{diag:27.1}--\ref{diag:377.17}
combine into a set in exactly the same way. However, these diagrams actually vanish on an individual level, anyway, 
by virtue of the second point
above.

\subsection{Diagrams which Possess Vanishing Components}
\label{sec:TL:VansishingCompts}

Particular patterns of pushes forward
and pulls back in diagrams possessing nested gauge remainder structures
can be discarded, up to $\Oep$ corrections. To illustrate this,
consider the diagrams shown in figure~\ref{fig:b2:Discard-Ex}.
\begin{center}
\begin{figure}[h]
	\[
	2
	\left[
		\ensuremath{\begin{array}{c}\input{pstex/Diagram376.1.pstex_t} \end{array}}
	\right]_{ES'}^{\bullet}
	=
	2
	\dec{
		\begin{array}{ccc}
		\vspace{0.1in}
			\LabIllOb{Ill:376.1.1}	&	& \LabIllOb{Ill:376.1.2} \\
			\ensuremath{\begin{array}{c}\input{pstex/Diagram376.1.pstex_t} \end{array}} 		& -	& \ensuremath{\begin{array}{c}\input{pstex/Diagram376.1.2.pstex_t} \end{array}}
		\end{array}
	}{\bullet}
	\]
\caption{Two components of a diagram, one of which can and one of which cannot
be discarded.}
\label{fig:b2:Discard-Ex}
\end{figure}
\end{center}

Taking diagrams~\ref{Ill:376.1.1} and~\ref{Ill:376.1.2}
to have their fields as shown (FAS), let us now suppose
that we move the external field on, say, the left hand vertex to the
`inside' of each diagram. The key point is that the only
components of these diagrams which survive in the $\eptoz$ limit are
those for which all fields leaving the three-point vertices
are in the $A^1$-sector; for what follows, we discard
all other contributions. Thus, moving one field
from the `outside' to the `inside', 
whilst temporarily ignoring the group theory
factors,  produces a relative minus sign and a possible change to the
flavours of the gauge remainders.

Now consider the components of diagram~\ref{Ill:376.1.1} where there
are attachment corrections. The 
supertrace structure of the diagram as a whole is \strAA,  
irrespective of the location of the external fields. Hence, in
this case, components with the left-hand external field `inside' are exactly cancelled
by the corresponding components with the left-hand field `outside'.
However, for the components of
diagram~\ref{Ill:376.1.1} for which both all attachments are direct, 
moving the left-hand external field `inside' the diagram
will cause it to be on a different supertrace from the other external field.
In this case, there can be no cancellation.

Turning now to diagram~\ref{Ill:376.1.2}, things are different for the case
where all attachments are
direct:
the diagram has
no empty supertraces; the external fields are always on the same supertrace, 
irrespective of location. Note that this further implies that all internal 
fields must be bosonic:
fermionic fields separate a portion of supertrace in the 1-sector from
a piece in the 2-sector, but the latter does not exist, in this scenario! Thus, all
components with the left-hand external field `inside' are exactly cancelled
by the corresponding components with the left-hand field `outside'.
Hence, whilst we keep diagram~\ref{Ill:376.1.1}, mindful that certain components
cancel but some survive, we discard diagram~\ref{Ill:376.1.2}. In other
words, we can drop the $ES'$ of the parent.

\section{Finite Diagrams}	\label{sec:TLD:finite}

In this section, we deal with some, but not all, of the manifestly finite
diagrams. First, we will encounter a single diagram which yields a computable
contribution to $\beta_2$. Next, we will examine a set of diagrams which
result in a computable  coefficient times a non-universal, one loop diagram; the computable coefficient
will be shown to vanish. The one-loop diagram is none other than the usual combination of the
two diagrams possessing an $A^1_\mu A^1_\nu C$ vertex that contribute to $\OLDs$.

With this in mind, we expect a find a set of diagrams in which the stub is the standard set. Though
some of the resulting contributions are manifestly finite, we choose not to treat them in this section,
so that we can keep the standard set intact.

\subsection{Universal Diagrams}	\label{sec:TLD:Universal}

Diagram~\ref{diag:23.1} is the sole $\Lambda$-derivative term which yields simply a finite,
universal contribution to $\beta_2$.\footnote{Since the leading order contribution to this
diagram is finite, it is not merely computable but actually universal: it
is independent of the way in which we compute it (see~\cite{scalar2loop} for a different way of
evaluating this diagram).} We reproduce this diagram, having chosen a
particular momentum routing, in figure~\ref{fig:TLD:d23.1}.
\begin{center}
\begin{figure}[h]
	\[
	-\frac{1}{6}
	\dec{
		\LOLD[0.1]{Diagram23.1-Lab}{diag:23.1-B} \hspace{1em}
	}{\bullet}
	\]
\caption{Reproduction of diagram~\ref{diag:23.1}.}
\label{fig:TLD:d23.1}
\end{figure}
\end{center}

The requirement that we take contributions which survive the \eptoz\ limit
places useful constraints on the diagram. First, all fields must be in the $A$-sector;
given this, we are compelled to take $\Omom{0}$ from each of the vertices. In turn,
this forces both vertices to comprise a single supertrace: 
it is forbidden to have a single gauge field on a
supertrace; if we take two supertraces, each with two gauge fields, 
then gauge invariance demands that we cannot take 
$\mathcal{O}(\mathrm{mom}^0)$
from such a vertex.

Now that we know that both vertices have only a single supertrace,
all fields are forced to be in the $A^1$ sector. Temporarily ignoring
attachment corrections, the group theory factor
of the diagram must be either $N^2$ or unity. However, we can show that contributions
of the latter type cancel. To see this, recall from equation~(\ref{eq:MomExp:4pt}) that
\begin{eqnarray*}
S^{1\, 1\, 1\, 1}_{\mu\alpha\beta\gamma}(\underline{0}) & = & -2(2\delta_{\mu \beta}\delta_{\alpha \gamma} - \delta_{\alpha \beta} \delta_{\mu \gamma}
- \delta_{\mu \alpha} \delta_{\beta \gamma}) \\
S^{1\, 1\, 1\, 1}_{\nu\gamma\beta\alpha}(\underline{0}) & = & -2(2\delta_{\nu \beta}\delta_{\alpha \gamma} - \delta_{\gamma \beta} \delta_{\nu \alpha}
- \delta_{\nu \gamma} \delta_{\beta \alpha}).
\end{eqnarray*}

Focusing on the component of diagram~\ref{diag:23.1-B} with a group theory
factor of unity, the locations of the external fields are independent,  
since they are always guaranteed to be on the
same supertrace. Summing over all independent locations of the external fields
yields something proportional to
\begin{equation}
	S^{1\, 1\, 1\, 1}_{\mu\alpha\beta\gamma}(\underline{0})
	+S^{1\, 1\, 1\, 1}_{\alpha\mu\beta\gamma}(\underline{0})
	+S^{1\, 1\, 1\, 1}_{\alpha\beta\mu\gamma}(\underline{0}) = 0.
\label{eq:TLD:4ptSum}
\end{equation}

Similarly, all attachment corrections can be ignored. If we suppose
that one of the effective propagators attaches via a correction
(see figure~\ref{fig:NFE:Attachment-A1s,A2s}) then the supertrace structure
of the diagram is left invariant under independently placing the ends of this
effective propagator in all independent locations. Hence the diagram vanishes
courtesy of~(\ref{eq:TLD:4ptSum}).
Increasing the number of effective propagators which attach via a
correction clearly does not change this result.

Returning to the case of direct attachment,
if the group theory goes as $N^2$,
then the locations of the external gauge fields are dependent, since it must be ensured that they are on the
same supertrace. Up to insertions of $A^1_{\mu,\nu}$, we can use charge conjugation invariance to fix the order of the three internal 
fields so long as we multiply by two. Now, there are three identical pairs of locations that we can place the
pair of fields $A^1_{\mu,\nu}$. 
 Including the diagram's overall factor of $-1/6$ we have:
\begin{equation}
-N^2\times S^{1\, 1\, 1\, 1}_{\mu\alpha\beta\gamma}(\underline{0}) S^{1\, 1\, 1 \, 1}_{\nu\gamma\beta\alpha}(\underline{0}) = -72N^2\delta_{\mu \nu} + \Oep.
\label{eq:diag23.1-GT}
\end{equation}

To obtain the contribution to $\beta_2$ coming from diagram~\ref{diag:23.1-B}, we must multiply
the above factor by the number obtained from the loop integral. Since the integral yields a finite
contribution, we simply Taylor expand the effective propagator $\Delta^{11}(l-p)$
to $\Op{2}$. Remembering to evaluate the cutoff
functions at zero momentum---which yields a factor of $1/2$ for each of the effective propagators---we have:
\[
	\frac{1}{8} \left[\int_{l,k} \frac{1}{k^2 (l-k)^2 l^4} \left(p.p - \frac{4(l.p)^2}{l^2} \right)\right]^\bullet,
\]

Looking at this expression, we might worry that the presence of $l^4$ in the denominator
means that the integral is actually IR divergent, even after differentiation
\wrt\ \flowConstAl.
However,
due to the form of the $\Op{2}$ contributions, averaging over angles in the $l$-integral will produce a factor of 
\[
	1 - \frac{4}{D} \sim \epsilon,
\]
in addition to the power of $\epsilon$
coming from the $\Lambda$-derivative. This renders the contribution from diagram~\ref{diag:23.1-B} finite.

To evaluate the integral, we use the 
techniques of chapter~\ref{ch:LambdaDerivatives:Methodology}.
Specifically, we perform the $l$ integral first, with unrestricted
range of integration, and then perform the $k$-integral with
the radial integral cutoff at $\Lambda$.
After differentiation \wrt\ \flowConstAl, the integral gives \PowAngVol{D}{2}$p^2/32$.
Combining this with the factor coming from equation~(\ref{eq:diag23.1-GT}) yields:
\[
	\begin{array}{ccccc}
	\vspace{0.1in}
												&	& \LO{diag:23.1-Ans}							&	&
	\\
		\mathrm{diagram} \ \ref{diag:23.1-B} 	& = & \ds -\frac{9N^2}{(4\pi)^4 }p^2\delta_{\mu \nu}& + &\Oep.
	\end{array}
\]

Before moving on, it is worth commenting further on the
fact that all attachment corrections in diagram~\ref{diag:23.1}
effectively vanish.
When we finally come to evaluate the numerical value of
$\beta_2$, we will be dealing with diagrams for which all
fields are in the $A^1$ sector. The highest point vertex that
we will encounter is four-point: we have already seen how
attachment corrections to such a vertex vanish. Three-point
vertices are even easier to treat. Suppose that an effective
propagator attaches via a correction to a three-point vertex,
decorated exclusively by $A$s. We can sum over the two
locations to which the effective propagator can attach,
but these two contributions cancel, by \CC.

If nested gauge remainders are in the $A$-sector, we know
from the analysis of 
section~\ref{sec:GR:CompleteDiagrams} that we can ignore attachment
corrections. 

Thus, when we come to extract numerical contributions to
$\beta_2$, we will neglect attachment corrections. 
Similarly, for direct attachments, we need focus only on the cases where
the group theory goes as $N^2$.

\subsection{Non-Universal Diagrams} \label{sec:TLD:C-sector}

In this section, we treat all diagrams which possess the one-loop diagram
shown in figure~\ref{fig:TLD:C-stub}. We use the fact that gauge invariance forces the diagram to be transverse in its
external momentum, to extract the momentum dependence: the bottom structure of the diagram
on the \rhs\ of the figure represents the $\mathcal{O}(q^2)$ coefficient of the bottom vertex
of the diagram on the \lhs
\begin{center}
\begin{figure}[h]
\[
	\begin{array}{cccl}
		\ensuremath{\begin{array}{c}\input{pstex/Cstub-A.pstex_t} \end{array}} & = & \ensuremath{\begin{array}{c}\input{pstex/Cstub-B.pstex_t} \end{array}} & \hspace{-1em} \times \Box_{\alpha \beta}(q) + \Oq{4}
	\end{array}
\]
\caption{Extracting the dependence of  a transverse one-loop diagram on its external momentum.}
\label{fig:TLD:C-stub}
\end{figure}
\end{center}

The diagram of figure~\ref{fig:TLD:C-stub} will turn up in one of two ways:
either with its external momentum being $p$ or with its external momentum being a loop momentum. The key point is that, in both
cases, we can discard contributions higher order in momentum. In the former case, this is because we are working at $\Op{2}$ whereas, in the 
latter case, it is because additional powers of loop momentum will kill the divergence that keeps the diagram alive. Stripping off this 
$\mathcal{O}(\mathrm{mom}^2)$ part leaves behind the same coefficient function.

The set of diagrams containing the diagram of figure~\ref{fig:TLD:C-stub} as a common one-loop sub-diagram is shown in figure~\ref{fig:TLD:C-stub-Diagrams}. Whilst
it is not immediately apparent that all the diagrams of this set possess the common one-loop sub-diagram,
we will find that they can be shown to do so in the $\epsilon \rightarrow 0$ limit.
\begin{center}
\begin{figure}
\[
	\begin{array}{l}
	\vspace{0.2in}
		\dec{
			\begin{array}{c}
			\vspace{0.2in}
				\begin{array}{cccccc}
				\vspace{0.1in}
							& \LOLB{d:31.1-B}	&			& \LO{d:51.1-B}		&	& \LO{d:328.23-B}
				\\	
					\ds \hf & \ensuremath{\begin{array}{c}\input{pstex/Diagram31.1.pstex_t} \end{array}} 	& \ds - \hf & \ensuremath{\begin{array}{c}\input{pstex/Diagram51.1.pstex_t} \end{array}} 	& -2 & \ensuremath{\begin{array}{c}\input{pstex/Diagram328.23.pstex_t} \end{array}}
				\end{array}
			\\
				- 
				\left[
					\begin{array}{cccccccl}
					\vspace{0.1in}
						\LO{d:CS-E-A}			&	&\LO{d:CS-E-B}			&	&\LO{d:CS-E-C}			&	&\LORB{d:CS-E-D}		& =
					\\
						\ensuremath{\begin{array}{c}\input{pstex/Diagram180.7-be.pstex_t} \end{array}}	&+	&\ensuremath{\begin{array}{c}\input{pstex/Diagram180.9-be.pstex_t} \end{array}}	&-2	&\ensuremath{\begin{array}{c}\input{pstex/Diagram326.3-be.pstex_t} \end{array}}	&+2	&\ensuremath{\begin{array}{c}\input{pstex/Diagram326.4-be.pstex_t} \end{array}}	&
					\end{array}
				\right]
			\\
				\times \Delta^{1 \,1}_{\alpha \beta}(p)
				\ensuremath{\begin{array}{c}\input{pstex/Beta1-D-al.pstex_t} \end{array}}
			\end{array}
		}{\bullet}
	\\
		\displaystyle
		= 
		\dec{
			\begin{array}{c}
			\vspace{0.1in}
				\LO{d:CS}
			\\
				\ds
				\frac{N\Lambda^{-2 \epsilon}}{(4 \pi)^{D/2}} \left(u + v \frac{p^{-2\epsilon}}{\Lambda^{-2 \epsilon}}\right)\frac{1}{\epsilon} 
				\times \hspace{-1em} \ensuremath{\begin{array}{c}\input{pstex/Cstub-B.pstex_t} \end{array}} \Box_{\mu \nu}(p)
			\end{array}
		}{\bullet} + \Oep
	\end{array}
\]
\caption{The complete set of diagrams which contain the one-loop diagram shown in figure~\ref{fig:TLD:C-stub}.}
\label{fig:TLD:C-stub-Diagrams}
\end{figure}
\end{center}

A number of comments are in order. First is that  diagrams~\ref{d:CS-E-A}--\ref{d:CS-E-D}
manifestly contain the
desired one-loop sub-diagram. 
Upon restricting the internal fields carrying a loop
momentum to the $A$-sector, it is clear that diagrams~\ref{d:51.1-B} and~\ref{d:328.23-B}
also possess the one-loop sub-diagram. What about the first diagram?
Let us denote the loop momentum by $k$. Given that the fields carrying $k$ are in the $A$-sector,
we must take a single power of momentum from the four-point vertex. This means that we can
Taylor expand this vertex to zeroth order in either $k$ or $p$. The result will be a momentum
derivative of an $AAC$ vertex, carrying either $p$ or $k$, respectively; we show this diagrammatically
in figure~\ref{fig:TLD:C-stub-TE}.
\begin{center}
\begin{figure}[h]
\[
	\dec{
		\LID[0.1]{}{Diagram31.1-TE-k}{Diagram31.1-TE-k} + \LID[0.1]{}{Diagram31.1-TE-p}{Diagram31.1-TE-p}
	}{\bullet} + \Oep
\]
\caption{Taylor expanding the four-point vertex of diagram~\ref{d:31.1-B}.}
\label{fig:TLD:C-stub-TE}
\end{figure}
\end{center}

Diagrams~\ref{Diagram31.1-TE-p} and~\ref{Diagram31.1-TE-k} now contain the desired one-loop
sub-diagram, albeit differentiated \wrt\ its external momentum.
However, since we know
the momentum dependence of the differentiated vertex, we can strip it off, all the same.

Having stripped the momentum dependence off the common one-loop sub-diagram, we find that the sum of diagrams
in figure~\ref{fig:TLD:C-stub-Diagrams}, given by contribution~\ref{d:CS}, is transverse in $p$. 
Hence, we could, if we wanted,  reabsorb the $\Box_{\mu \nu}(p)$ of this term into the 
diagrammatic part. 

The coefficients $u$ and $v$ are to be determined. There are two ways
to do this. First, it is straightforward, using the techniques of 
chapter~\ref{ch:LambdaDerivatives:Methodology}  to show that
\[
	u=0, \ v=0.
\]
It is important that $v=0$: hidden inside the common one-loop diagram is a factor of $\Lambda^{-2 \epsilon}$;
hence, the $p^{-2\epsilon}$ terms survive for generic $v$.

However, it is also possible to demonstrate that $u$ and $v$ vanish, diagrammatically!
First, recall that we take all internal fields in the diagrams of figure~\ref{fig:TLD:C-stub-Diagrams}
to be in the $A$-sector. Secondly, note that  when our common stub attaches to an internal field, we can
use the fact that it is transverse in momentum to apply the effective propagator relation,
so long as we compensate with a factor of $1/\NewCO_k$.

In this way, diagram~\ref{d:328.23-B} reduces to diagrams~\ref{d:CS-E-C} and~\ref{d:CS-E-D},
up to a relative factor of $-\NewCO(k)/\NewCO(p)$. However, at \Op{2}, we can evaluate $\NewCO(p)$ at zero
momentum. Up to \Oep\ correction, we can do likewise with $\NewCO(k)$. Thus, at leading order, 
diagram~\ref{d:328.23-B} cancels diagrams~\ref{d:CS-E-C} and~\ref{d:CS-E-D}.

Now let us turn to diagram~\ref{d:51.1-B}. Applying the effective propagator
relation, remembering to compensate with $1/\NewCO(k)$, the Kronecker delta contribution
gives a diagram which cancels diagram~\ref{d:31.1-B}, at leading order. Processing the
gauge remainder we get three terms. The first cancels diagram~\ref{d:CS-E-A},
at leading order. The second and third involve the iterated use of the effective
propagator relation. The Kronecker delta contribution can be processed by noting that
the processed gauge remainder can act backwards, through the effective propagator, by means
of diagrammatic identity~\ref{D-ID:PseudoEffectivePropagatorRelation}. This gives two contributions, which
cancel each other at \Op{2}. The gauge remainder contribution gives one diagram which cancels~\ref{d:CS-E-B},
at leading order and a second which vanishes via diagrammatic identity~\ref{D-ID:TLTP-GR}.

We conclude this section with an observation that will prove very useful, in the next section. 

\begin{com}
The cancellations in figure~\ref{fig:TLD:C-stub-Diagrams} are 
guaranteed to occur at, leading order in $\epsilon$, for any choice of
common one-loop sub-diagram, so long as the sub-diagram is 
both transverse and Taylor expandable in its external momentum. This latter requirement
can be understood as follows. Suppose that the 
external momentum of the one-loop sub-diagram corresponds to a loop
momentum of the complete diagram, $k$. If the one-loop sub-diagram is not
Taylor expandable, then powers of $k^{-2 \epsilon}$ will affect the
loop integral over $k$, spoiling the cancellations.
\label{com:TLD:cancellations}
\end{com}

\section{The Standard Set as a Sub-Diagram} \label{sec:TLD:SSasSubD}

Inspired by comment~\ref{com:TLD:cancellations}, we try and 
repeat the cancellations of the previous section but with the one-loop
diagram of figure~\ref{fig:TLD:C-stub} replaced by the standard set.
Immediately, we know that life is going to get harder:
the standard set has contributions which are not Taylor expandable
in momentum; moreover, the standard set is IR divergent, so even if
we do manage to repeat the cancellations of the previous
section, the sub-leading contributions will not necessarily
vanish as \eptoz. Nonetheless, the pattern of cancellations
we have just observed will be mirrored; the difference is that
there will be terms that survive.

The first challenge is to identify the analogues of the diagrams
of figure~\ref{fig:TLD:C-stub-Diagrams}. This is easy for
all but the first diagram. Putting this diagram temporarily
to one side, the diagrams we will need are: 
\ref{diag:30.1}--\ref{diag:376.1},
\ref{diag:327.37}--\ref{diag:331b.10}
and~\ref{b2-b1A}--\ref{b2-b1C}.\footnote{Diagrams~\ref{diag:30.1}--\ref{diag:376.1}
appear to be highly IR divergent. However, the degree of divergence is lessened
by the transversality of the standard set.}

As for the first diagram, we need to think what it is that we really want.
The cancellations of the last section involved us partially Taylor expanding
this diagram, to produce diagrams~\ref{Diagram31.1-TE-p} and~\ref{Diagram31.1-TE-k}.
Now, the key point is that the momentum derivative acting in these diagrams
hits a structure which is transverse. Given that the standard set is transverse,
it is thus apparent that we want
to find a set of diagrams in which  the whole of the standard set is  hit by
a momentum derivative; it is no use if the momentum derivative strikes just
the vertices of the standard set but not the effective propagators or
gauge remainders as well.

It is immediately clear that the set of such diagrams do not exist. Particularly, there
are no diagrams which, after Taylor expansion, directly yield momentum derivatives
striking effective propagators or gauge remainders. In section~\ref{sec:TLD:SS-Mirror}, 
we will simply 
construct the diagrams that we need, both adding and subtracting\footnote{These  subtractions
are distinct from those discussed in  section~\ref{sec:TLD:Subtractions}: we are not
starting from some parent diagram and constructing a set of terms directly derived from the
parent to cancel all Taylor expandable components. Rather, we are using an empirical observation
to find a set of diagrams for which certain Taylor expandable components cancel.} them
from the calculation.
One set of terms will be involved in cancellations similar to those of the previous
section; the complementary set with opposite sign will cancel against terms generated
in section~\ref{sec:TLD:SS-Mirror-ComplementaryTerms}.

\subsection{Mirroring the Cancellations of Section~\ref{sec:TLD:C-sector}} \label{sec:TLD:SS-Mirror}

The terms we need to construct to mirror the cancellations of
section~\ref{sec:TLD:C-sector} are shown in figure~\ref{fig:TLD:Diagrams224-A}. Since each diagram
is both added to and subtracted from the calculation, each diagram comes with two
labels.
\begin{center}
\begin{figure}
\[
	\mp
	\dec{
		\begin{array}{ccccl}
		\vspace{0.1in}
			&\OLLabLBC[-2]{d:224b.4}{d:224b.10}{d:217.4} 	&	&\OLLabC[-2]{d:224b.5}{d:224b.11}{d:258.1}
		\\
			&\ensuremath{\begin{array}{c}\input{pstex/Diagram224b.4.pstex_t} \end{array}}								& -	& \ensuremath{\begin{array}{c}\input{pstex/Diagram224b.5.pstex_t} \end{array}}
		\\
		\vspace{0.1in}
				&\OLLabC[-2]{d:224b.6}{d:224b.12}{d:258.2}	&	& \OLLabC[-2]{d:224b.6b}{d:224b.12b}{d:258.6}
		\\				
			+2	&\ensuremath{\begin{array}{c}\input{pstex/Diagram224b.12.pstex_t} \end{array}}						&+2	& \ensuremath{\begin{array}{c}\input{pstex/Diagram224b.12b.pstex_t} \end{array}}
		\\
		\vspace{0.1in}
				&\OLLabC[-2]{d:248.1,4}{d:248.7}{d:217.2} 	&	&\OLLabC[-2]{d:248.2,5}{d:248.8}{d:257.6}
		\\
			+	&\ensuremath{\begin{array}{c}\input{pstex/Diagram248.7.pstex_t} \end{array}}							& -	& \ensuremath{\begin{array}{c}\input{pstex/Diagram248.8.pstex_t} \end{array}}
		\\
		\vspace{0.1in}
				&\OLLabC[-2]{d:248.3a,6a}{d:248.9a}{d:257.7a} 	&	&\OLLabRBC[-2]{d:248.3b,6b}{d:248.9b}{d:257.7b} & \hspace{-4em} \ProBlank[-2]{fig:TLD:Diagrams224}
		\\
			+2	&\ensuremath{\begin{array}{c}\input{pstex/Diagram248.9a.pstex_t} \end{array}}								& +2	& \ensuremath{\begin{array}{c}\input{pstex/Diagram248.9b.pstex_t} \end{array}}
		\end{array}
	}{\bullet}
\]
\caption{Terms we construct to allow us to mirror the cancellations of the previous section.}
\label{fig:TLD:Diagrams224-A}
\end{figure}
\end{center}

We can now ask what happens when we combine the diagrams of figure~\ref{fig:TLD:Diagrams224-A}
with the overall minus sign with diagrams~\ref{diag:30.1}--\ref{diag:376.1},
\ref{diag:327.37}--\ref{diag:331b.10}
and~\ref{b2-b1A}--\ref{b2-b1C}.

Our starting point is equation~(\ref{eq:TLD:StandardSetGeneralForm}),
the algebraic form of the standard set. 
Given comment~\ref{com:TLD:cancellations}, we know that the coefficients 
$a_i$---which correspond to the parts of the standard set, Taylor expandable in 
external momentum---will be involved in cancellations. Specifically, the leading 
order contributions involving the $a_i$ will cancel exactly. Hence, terms involving
$a_1$ will disappear completely, in the \eptoz\ limit. Terms involving $a_0$, on
the other hand, will leave behind \Oone, as opposed to \Oep\ contributions,
which will survive.

Let us examine these surviving terms in more detail. 
We begin by noting that, in the case that the
external momentum of the standard set corresponds to a loop momentum,
the standard set must attach to something computable, as all
other contributions will die in the \eptoz\ limit. Surviving sub-leading corrections
will arise from the expansion of functions of $D$ in $\epsilon$.\footnote{
If the standard set has external momentum $l$, we might worry that
we could pick up momentum derivative contributions in the $l$-integral from
expanding, say, $\NewCO(l-p)$ in $p$. However, we could always ensure that
the $l$-integral is done first in which case, as we know
from section~\ref{sec:TLD:Subtractions:Gauge}, such contributions will vanish.}

When the standard set has external momentum $p$, we are free to pick
up any sub-leading contributions from the one-loop sub-diagrams to which
it attaches; we are not restricted to computable contributions only.

We must now move on to consider the $b_i$ contributions to the
standard set. Once again, if the external momentum of
the standard set is a loop momentum, then the standard
set must attach to something computable. 
Next, let us suppose that the external momentum of the standard set is
$p$. Since all the $b_i$ are computable, if we take the most divergent part
of the sub-diagram to which the standard set is attached, then we have
computable contributions. If we take the sub-leading part of the diagram
to which the standard set attaches, we will be left with a mixture of
computable and non-computable contributions.

We represent all the surviving terms using a mixture of algebra
and diagrams in figure~\ref{fig:TLD:Diagrams224}. Contributions containing non-computable parts are
represented diagrammatically; these are combined with 
computable contributions 
which take the same diagrammatic form.
 The remaining computable
contributions are represented algebraically, in terms of
the undetermined coefficients $U_{i}^{a,b}$ and $V_i^{a,b}$.

Since non-computable contributions occur only at sub-leading order,
we must take a sub-leading contribution from the sub-diagrams to which
the standard set attaches. This is denoted by the tag \Oepz. 
\begin{center}
\begin{figure}
\[
	\dec{
		\begin{array}{l}
		\vspace{0.2in}
			\begin{array}{cc}
			\vspace{0.1in}
					&\OLdiscard{d:248.1,4:NU}	
			\\
				-	&\ensuremath{\begin{array}{c}\input{pstex/Diagram248.1,4Oep.pstex_t} \end{array}}
			\end{array}
		\\
		\vspace{0.2in}
			\begin{array}{ccccccl}
			\vspace{0.1in}
					& \LOLB{d:248.2,5:NU}		&	&\LO{d:248.3a,6a:NU}		&	&\LORB{d:248.3b,6b:NU}		&\Process{fig:TLD:Diagrams269B}
			\\	
				+	&\ensuremath{\begin{array}{c}\input{pstex/Diagram248.2,5Oep.pstex_t} \end{array}}		&-2	&\ensuremath{\begin{array}{c}\input{pstex/Diagram248.3a,6aOep.pstex_t} \end{array}}	&-2	&\ensuremath{\begin{array}{c}\input{pstex/Diagram248.3b,6bOep.pstex_t} \end{array}}	&
			\end{array}
		\\
		\vspace{0.2in}
			+
			\left[
				\begin{array}{l}
				\vspace{0.2in}
					\left[
						\begin{array}{ccccccc}
							\ensuremath{\begin{array}{c}\input{pstex/Diagram180.7-be.pstex_t} \end{array}}	&+	&\ensuremath{\begin{array}{c}\input{pstex/Diagram180.9-be.pstex_t} \end{array}}	&-2	&\ensuremath{\begin{array}{c}\input{pstex/Diagram326.3-be.pstex_t} \end{array}}	&+2	&\ensuremath{\begin{array}{c}\input{pstex/Diagram326.4-be.pstex_t} \end{array}}
						\end{array}
					\right]_{\Oepz}	
				\\
					\times \Delta^{1 \,1}_{\alpha \beta}(p)
					\left[
						\begin{array}{cccccl}
							\vspace{0.1in}
							\LabOLObLB{diag:223b-E1}&	&\LabOLOb{diag:223b-E2}	&	&\LabOLObRB{diag:223b-E3}	& \Process{fig:TLD:SS-rems-A}
						\\		
							\ensuremath{\begin{array}{c}\input{pstex/Beta1-A-al.pstex_t} \end{array}}			& -	&\ensuremath{\begin{array}{c}\input{pstex/Beta1-B-al.pstex_t} \end{array}} 		& +4& \ensuremath{\begin{array}{c}\input{pstex/Beta1-C-al.pstex_t} \end{array}}			&
						\end{array}
					\right]
				\end{array}
			\right]
		\\
			\begin{array}{c}
				\hspace{8em} \LabOLOb{Diagrams224-Alg}
			\\	
				\displaystyle
				+\frac{N^2\Lambda^{-4\epsilon}}{(4\pi)^D} \sum_{i=0}^1 \frac{1}{\epsilon^{i-2}}	
				\left[
					\left(U_i^a + V_i^a  \PoLep \right) p^2 \delta_{\mu \nu} +
					\left(U_i^b + V_i^b  \PoLep \right) p_\mu p_\nu
				
				\right]
			\end{array}
		\end{array}
	}{\bullet}
\]
\caption{The terms surviving from the analogue of figure~\ref{fig:TLD:C-stub-Diagrams}, 
in which the common
one-loop sub-diagram is the standard set.}
\label{fig:TLD:Diagrams224}
\end{figure}
\end{center}

The are several things to note. First, is that
 in all diagrams possessing the tag \Oepz, we must take
the most divergent part of the instance of the standard
set to which they attach. Although some of the contributions
corresponding to the \OepPow{-1}\ parts of the tagged sub-diagrams
are included in~\ref{Diagrams224-Alg}, we cannot include them in the diagrams
tagged by \Oepz\ by simply 
removing the tag: this would cause $a_1$ terms
contributing to the standard set to spuriously reappear.

Secondly, the coefficients $U_0^{a,b}$ and $V_0^{a,b}$ both
vanish, by construction.

Lastly, the coefficients $U_1^{a,b}$ are easy to compute,
since each of the diagrams contributing involves the computable part of the standard set,
for which we know the algebraic form. Moreover, to extract these coefficients we can
Taylor expand the sub-diagram to which the standard set attaches,
in $p$ (we can not do this for $V_1^{a,b}$, which would make life
harder if we were to ever evaluate these coefficients).

Using our knowledge of the standard set coefficients $a_0$, $b_0$ and $b_1$ (see
equations~(\ref{eq:TLD:StandardSetGeneralForm}), (\ref{eq:SS:b_0}) and~(\ref{eq:SS:b_1})) we find:
\begin{eqnarray}
	U_1^a	& = &	\frac{571}{36}  -\frac{80\EM}{3}	\label{eq:b2-contr-U_1^a} \\
	U_1^b	& = &	-\frac{451}{36}	+\frac{80\EM}{3}	\label{eq:b2-contr-U_1^b}.
\end{eqnarray}

It may seem a little odd that $\EM$ appears here; this is an artifact
of the way in which we have chosen to split up the computable contributions in 
figure~\ref{fig:TLD:Diagrams224} between diagrams and algebra. Thus, the diagrams tagged \Oepz,
contain $\EM$ in such a way as to exactly cancel the contributions contained in the
algebraic coefficients.

\subsection{Generating the Constructed Diagrams} \label{sec:TLD:SS-Mirror-ComplementaryTerms}

Referring back to the constructed diagrams of the
previous section, we show how to cancel the unprocessed
diagrams of figure~\ref{fig:TLD:Diagrams224-A};
in other words, we demonstrate how the constructed diagrams
can be generated, naturally.

This generation of terms requires two steps. We begin by processing a
set of $\Lambda$-derivative diagrams.  The simplest of these---the IR finite
diagram~\ref{diag:17.1}---can be dealt with by diagrammatic Taylor expansion.
Having been manipulated, we immediately find two of the desired cancellations.
 
However, the remaining diagrams of this section are not manifestly
finite and so we must construct subtractions (see section~\ref{sec:TLD:Subtractions}). 
The corresponding additions
are not of the right form to cancel the rest of the unprocessed terms
of the previous section, and so we must move on to the second step of
our procedure. This involves the diagrammatic manipulation of a subset of the
additions. This completes the generation of the terms we are looking for.

We conclude this section by manipulating the semi-computable terms
formed by combining parent diagrams with their subtractions. We choose
to do this here, as these manipulations are very similar to those
which we perform on the additions.

\subsubsection{Manipulation of $\Lambda$-derivative Terms}

The first diagram aside, the treatment of terms in this section
requires the construction of subtractions.
As we will see, for certain diagrams there is no unique way in which to 
do this.
However, since we know the set of terms we are ultimately trying to
generate, we can use this as a guide. Of course, the ultimate effect
of constructing any particular set of subtractions and their corresponding
additions must be the same: all we are really doing is adding zero
to the calculation in a convenient form. 
The point is that by a cunning choice of subtraction,
we can generate a set of terms which are manifestly of the form we
desire.

\paragraph{Diagram~\ref{diag:17.1}}

To manipulate diagram~\ref{diag:17.1}, we note that it
is manifestly finite, after differentiation \wrt\ \LConstAl. Taking the simple
loop of this diagram to carry momentum $k$ and the other loop to carry $l$,
we are forced to take a single power of either $l$ or $p$ from both
the three-point and five-point vertices. This
is shown in figure~\ref{fig:TLD:d:217}.

\begin{center}
\begin{figure}
	\[
	-\frac{1}{2}
	\dec{
		\begin{array}{ccccc}
		\vspace{0.1in}
								&				&\CancelOLObject{d:217.1}{d:221b.1}	&	&\CancelOLObject{d:217.2}{d:248.7}
		\\
			\ensuremath{\begin{array}{c}\input{pstex/Diagram17.1-Lab.pstex_t} \end{array}}&\rightarrow -2	&\ensuremath{\begin{array}{c}\input{pstex/Diagram217.1.pstex_t} \end{array}}					&+2	& \ensuremath{\begin{array}{c}\input{pstex/Diagram217.2.pstex_t} \end{array}}
		\\\vspace{0.1in}
								&				&\CancelOLObject{d:217.3}{d:225.8}	&	&\CancelOLObject{d:217.4}{d:224b.10}
		\\
								&+2				&\ensuremath{\begin{array}{c}\input{pstex/Diagram217.3.pstex_t} \end{array}}					&+2	& \ensuremath{\begin{array}{c}\input{pstex/Diagram217.4.pstex_t} \end{array}}
		\end{array}
	}{\bullet}
	\]
\caption{Manipulation of diagram~\ref{diag:17.1} under $\Lambda \partial_\Lambda|_\alpha$.}
\label{fig:TLD:d:217}
\end{figure}
\end{center}

Diagrams~\ref{d:217.1}--\ref{d:217.4} constitute a momentum derivative of 
a vertex which belongs to
an instance of the first element of the standard set. 
Diagrams~\ref{d:217.2} and~\ref{d:217.4} are the analogues of
diagrams~\ref{Diagram31.1-TE-p} and~\ref{Diagram31.1-TE-k}
and, as we might hope, are involved in cancellations
against some of the terms that we constructed in the last section.
We note that a key feature of the first element of the standard set
is that the effective propagator forming the loop does not carry
the external momentum of the diagram. Hence, the momentum derivatives
in diagrams~\ref{d:217.2} and~\ref{d:217.4} can be taken to strike
the whole of the sub-diagram of which they currently strike only part.

\Cancel{d:217.2}{d:248.7}
\Cancel{d:217.4}{d:224b.10}

Diagrams~\ref{d:217.1} and~\ref{d:217.3},
on the other hand, have no analogue that we have encountered already:
the momentum derivative in 
these diagrams is not \wrt\ the external momentum of the one-loop
sub-diagram which they strike.

\paragraph{Diagram~\ref{diag:28.1}}

Diagram~\ref{diag:28.1} is the first term for which we construct subtractions, and it is one for
which we can choose different ways in which to do this. Rather than considering diagram~\ref{diag:28.1}
as is, we will---for reasons which will become 
apparent---explicitly consider two different momentum routings. 
Whilst these diagrams are, of course, equivalent by momentum rerouting invariance,
they guide us to different subtractions,
as shown in figure~\ref{fig:TLD:Diagram28.1WithSubtractions}.

\begin{center}
\begin{figure}
	\[
	\frac{1}{2}
	\dec{
		\begin{array}{ccccccl}
		\vspace{0.1in}
			&\OLLabBlankLB[-2]{d:28.1a}	&		&\OLLabPro[-2]{d:28.1a:sub1}{d:28.1a:add1}{fig:TLD:Diagram258.1}&
				&\OLLabPro[-2]{d:28.1a:sub2}{d:28.1a:add2}{fig:TLD:Diagram258.1}	&
		\\
			&\ensuremath{\begin{array}{c}\input{pstex/Diagram28.1a.pstex_t} \end{array}}			&\pm 2	&\ensuremath{\begin{array}{c}\input{pstex/Diagram28.1a-s1.pstex_t} \end{array}}											&
			\mp4&\ensuremath{\begin{array}{c}\input{pstex/Diagram28.1a-s2.pstex_t} \end{array}}												&
		\\
		\vspace{0.1in}
			&							&		&\OLLabRBPro[-2]{d:28.1a:sub3}{d:28.1a:add3}{fig:TLD:Additions-Ia}&  \hspace{-6.5em} \DiscardBlank&&
		\\
			&							&\mp 2	&\ensuremath{\begin{array}{c}\input{pstex/Diagram28.1a-s3.pstex_t} \end{array}}&&&
		\\
		\vspace{0.1in}
				&\OLLabBlankLB[-2]{d:28.1b}	&		&\OLLabLab[-2]{d:28.1b:sub1}{d:28.1b:add1}	&		&\OLLabLab[-2]{d:28.1b:sub2}{d:28.1b:add2}	&
		\\
			+	&\ensuremath{\begin{array}{c}\input{pstex/Diagram28.1b.pstex_t} \end{array}}			&\mp 2	&\ensuremath{\begin{array}{c}\input{pstex/Diagram28.1b-s1.pstex_t} \end{array}}						&\pm 4	&\ensuremath{\begin{array}{c}\input{pstex/Diagram28.1b-s2.pstex_t} \end{array}}						&
		\\
		\vspace{0.1in}
				&							&		&\OLLabLab[-2]{d:28.1b:sub3}{d:28.1b:add3}	&		&\OLLabRBLabRB[-2]{d:28.1b:sub4}{d:28.1b:add4} 
				&\hspace{-2em}\ProPro{fig:TLD:D28.1bU}{fig:TLD:Diagram258.1}
		\\
				&							&\pm 2	&\ensuremath{\begin{array}{c}\input{pstex/Diagram28.1b-s3.pstex_t} \end{array}} 						&\mp 4	&\ensuremath{\begin{array}{c}\input{pstex/Diagram28.1b-s4.pstex_t} \end{array}}
				&
		\end{array}
	}{\bullet}
	\]
\caption[Diagram~\ref{diag:28.1} and its subtractions.]{
Diagram~\ref{diag:28.1} and its subtractions. The two diagrams on the LHS are copies of diagram~\ref{diag:28.1} 
with differing momentum routings (hence the overall factor of $1/2$).}
\label{fig:TLD:Diagram28.1WithSubtractions}
\end{figure}
\end{center}

The reason for constructing the subtractions in this manner is now hopefully
apparent: diagrams~\ref{d:28.1a:add1}, \ref{d:28.1a:add2}, \ref{d:28.1b:add1}
and~\ref{d:28.1b:add2} can be very nearly combined into a momentum derivative
of the second element of the standard set (we are currently missing the
diagram in which the momentum derivative strikes the effective propagators).
Had we only constructed subtractions~\ref{d:28.1a:sub1}--\ref{d:28.1a:sub3},
this would not have been manifestly obvious.

To begin the analysis of diagram~\ref{d:28.1a} and its subtractions, we focus on the 
three-legged sub-diagram carrying loop momentum $k$. For the diagram as a whole to have any chance of
surviving in the $\epsilon \rightarrow 0$ limit, the (internal) legs leaving the sub-diagram must be in
the $A$-sector. Thus, by Lorentz invariance, the sub-diagram carrying loop momentum $k$ must go as odd powers of momentum
(up to additional non-Taylor expandable functions of $l$).
Noting that, in $D=4$, the  sub-diagram carrying loop momentum $k$ goes as, at worst, $(\ln l)\times \mathcal{O}(\mathrm{mom}, \ldots)$ in the IR,
it is clear that we must take only the $\mathcal{O}$(mom) part of the sub-diagram.

The effect of diagrams~\ref{d:28.1a:sub1} and~\ref{d:28.1a:sub2} is now immediately clear: they completely remove from
diagram~\ref{d:28.1a} all contributions in which the  sub-diagram carrying loop momentum $k$ goes as $l$.

Let us now suppose that we take $\mathcal{O}(p)$ from the sub-diagram carrying loop momentum $k$. We start by noting that the only
place for this power of $p$ to come from is the four-point vertex. Now, if all the fields leaving the four-point vertex are in the $A$-sector,
then Lorentz invariance forces us to take an additional power of momentum from the four-point vertex. Recalling that we should not take any 
further powers of $l$ or $p$, we see that we must take (at least) one power of $k$ from the four-point vertex (and a  further 
power of $k$ from the 3pt).
The $k$-integral is now Taylor expandable in $l$. In the case that the $k$-dependent fields leaving the four-point vertex are not in 
the $A$-sector, the $k$-integral is trivially Taylor expandable in $l$.

The remaining subtraction, diagram~\ref{d:28.1a:sub3}, removes all these contributions; hence diagram~\ref{d:28.1a}
turns out to be completely cancelled by its subtractions.

The same cannot, however, be said for diagram~\ref{d:28.1b}.
It is clear that whilst diagrams~\ref{d:28.1b:sub1} and~\ref{d:28.1b:sub2} remove from~\ref{d:28.1b}
all contributions from the $k$-integral that are Taylor expandable in $p$,
these diagrams possess components which are not Taylor expandable in $p$ and so will survive.
Diagrams~\ref{d:28.1b:sub3} and~\ref{d:28.1b:sub4} remove
all contributions that are Taylor expandable in $l$ (as always, this statement is correct only up to contributions that
vanish anyway in the $\epsilon \rightarrow 0$ limit). Non-computable contributions from the $k$-integral are precisely those
which are Taylor expandable in $l$ and $p$ and so cancel between the parent diagram and its subtractions. 

Figure~\ref{fig:TLD:D28.1bU} shows the surviving 
contributions to diagrams~\ref{d:28.1b}, \ref{d:28.1b:sub1}, \ref{d:28.1b:sub2}, \ref{d:28.1b:sub3} and~\ref{d:28.1b:sub4}.
\begin{center}
\begin{figure}[h]
	\[
	\frac{1}{2}
	\dec{
		\begin{array}{cccccl}
			\vspace{0.1in}
				\LabOLOb{d:28.1bU}		&		&\LabOLOb{d:28.1b:sub1U}	&		&\LabOLOb{d:28.1b:sub2U}	&
		\\
				\Uni{\ensuremath{\begin{array}{c}\input{pstex/Diagram28.1b.pstex_t} \end{array}}}	&- 2	&\Uni{\ensuremath{\begin{array}{c}\input{pstex/Diagram28.1b-s1.pstex_t} \end{array}}}	&+4		&\Uni{\ensuremath{\begin{array}{c}\input{pstex/Diagram28.1b-s1.pstex_t} \end{array}}}	&
		\\
		\vspace{0.1in}
											&		&\LabOLObLB{d:268.3}	&		&\LabOLObRB{d:268.4}	& \hspace{-2em}\Process{fig:TLD:Add-I:semiU}
		\\
											&+2		&\ensuremath{\begin{array}{c}\input{pstex/Diagram28.1b-s3SU.pstex_t} \end{array}} &-4		&\ensuremath{\begin{array}{c}\input{pstex/Diagram28.1b-s4SU.pstex_t} \end{array}}	&
		\end{array}
	}{\bullet}
	\]
\caption[The cancellation of non-computable contributions between diagram~\ref{d:28.1b} and its subtractions.]{
The cancellation of non-computable contributions between diagram~\ref{d:28.1b} and its subtractions.
The $\COMP$ symbol has been expanded to cover the whole of the first three diagrams, which is valid up to corrections
which vanish in the $\epsilon \rightarrow 0$ limit.}
\label{fig:TLD:D28.1bU}
\end{figure}
\end{center}

We conclude the treatment of diagram~\ref{diag:28.1} by
discussing whether or not it is necessary to consider derivatives of cutoff functions (or other functions),
since we found in section~\ref{sec:TLD:Subtractions:Gauge} that such objects can, in principle, contribute. Any 
such derivatives arising from the sub-diagram carrying loop momentum $k$ will cancel between parent and subtractions, since the 
accompanying powers of momentum render the $k$ integral Taylor expandable in $l$ and $p$. 

Next, we must consider derivatives arising in the $l$-integral. 
If we Taylor expand the $p$ dependent cutoff function to
yield $p.l \NewCO'(l)$, then the $k$ integral becomes Taylor expandable in $p$, to the order in $p$ to which we are
working. Hence, such contributions will have been cancelled.

These points lead us directly to consider some of the subtleties involved in precisely what we mean by $\COMP$. As we have just
discussed, there is no need for us to consider derivatives of cutoff functions. Now, if the $k$-integral is the outer integral,
then it is regulated by $\NewCO(k-p)\NewCO(l-k)$ which we can simply  Taylor expand to zeroth order in $p$ and $l$.
Doing this, of course, causes the diagram to lose invariance under momentum routing. Indeed, to maintain this invariance---at
least up to $\mathcal{O}(\epsilon^0)$---it would be necessary for us to keep the terms $l.k\NewCO'(k)$ and $k.p\NewCO'(k)$
(which will provide a computable
contribution at $\mathcal{O}(\epsilon^0$)) from the Taylor expansion of the cutoff functions.\footnote{Recall that taking more than a
single power of either $l$ or $p$ from the sub-diagram carrying loop momentum $k$ will kill the diagram as a whole.}

Knowing that the contributions from these terms will be removed by the subtractions, we can drop them as long as we
ensure that the form we choose for the four-point vertex is consistent between parent diagram and subtractions. 
This last point is
important. We know from section~\ref{sec:MomExp:CompleteDiagrams} that the differences between the alternative forms of the
four-point vertex disappear as a consequence of the freedom to reroute momenta.
Here, then, we must take real care. 

There are several ways to proceed. On the one hand, we could ensure that the choice of 
four point vertex must be consistent between parent and \emph{both}
the factorisable and non-factorisable subtractions which involves making sure that all dummy indices match up, in all
diagrams.
On the other hand, we can make sure that we keep the first order terms in the Taylor expansion
of both $\NewCO(k-p)$ and $\NewCO(l-k)$ so that then we can forget about the consistency of the four-point vertices, as we have
restored momentum rerouting invariance up to the required order in $\epsilon$.
Alternatively, we could Taylor expand (say) $\NewCO(l-k)$ to zeroth order only and choose the four-point
vertex to be consistent between parent and non-factorisable subtraction. Then we would only need to keep $k.p\NewCO'(k)$.
Either way, these points demonstrate the subtleties of 
interpreting $\COMP$ and the danger of trying to apply $\COMP$ to a single diagram.

\paragraph{Diagram~\ref{diag:32.1}}

Diagram~\ref{diag:32.1} is shown, together with its subtractions, in figure~\ref{fig:TLD:Diagram32.1WithSubtractions}.
\begin{center}
\begin{figure}[h]
	\[
	\frac{1}{4}
	\dec{
		\begin{array}{cccccccl}
		\vspace{0.1in}
			\OLLabBlankLB[-2]{d:32.1}	&		&\OLLabPro[-2]{d:32.1:sub1}{d:32.1:add1}{fig:TLD:Additions-Ia}	
			&		&\OLLabLabLB[-2]{d:32.1:sub2}{d:32.1:add2}	&		&\OLLabRBLabRB[-2]{d:32.1:sub3}{d:32.1:add3}& \hspace{-1em} \ProPro{fig:TLD:D32.1U}{fig:TLD:Diagram258.1}
		\\
			\ensuremath{\begin{array}{c}\input{pstex/Diagram32.1.pstex_t} \end{array}}			&\mp 4	&\ensuremath{\begin{array}{c}\input{pstex/Diagram32.1-s1.pstex_t} \end{array}}						
			&\mp 4	&\ensuremath{\begin{array}{c}\input{pstex/Diagram32.1-s2.pstex_t} \end{array}}						&\pm 8	&\ensuremath{\begin{array}{c}\input{pstex/Diagram32.1-s3.pstex_t} \end{array}}						&
		\end{array}
	}{\bullet}
	\]
\caption{Diagram~\ref{diag:32.1} and its subtractions.}
\label{fig:TLD:Diagram32.1WithSubtractions}
\end{figure}
\end{center}

To construct the subtractions, we must take into account the symmetry of the parent diagram.
For example, the first subtraction, diagram~\ref{d:32.1:sub1}, is designed to remove non-computable contributions arising when one of the
sub-diagrams is Taylor expandable in $p$. Since it could be either of the sub-diagrams which is Taylor expandable in $p$, this
gives rise to factor of two in addition to the usual factor coming from charge conjugation symmetry.
The effect of the subtractions is shown in figure~\ref{fig:TLD:D32.1U}.

\begin{center}
\begin{figure}
	\[
	\frac{1}{4}
	\dec{
		\begin{array}{cccccl}
		\vspace{0.1in}
			\LabOLOb{d:32.1U}		&	&\LabOLObLB{d:268.5}	&	&\LabOLObRB{d:268.6}	&\Process{fig:TLD:Add-I:semiU}
		\\
			\Uni{\ensuremath{\begin{array}{c}\input{pstex/Diagram32.1.pstex_t} \end{array}}}	&-4	&\ensuremath{\begin{array}{c}\input{pstex/Diagram32.1-s1SU.pstex_t} \end{array}}	&+8	&\ensuremath{\begin{array}{c}\input{pstex/Diagram32.1-s2SU.pstex_t} \end{array}}	&
		\end{array}
	}{\bullet}
	\]
\caption{The cancellation of non-computable contributions between diagram~\ref{d:32.1} and its subtractions.}
\label{fig:TLD:D32.1U}
\end{figure}
\end{center}

\paragraph{Diagram~\ref{diag:42.1}}

As in the previous case, we must take into account the symmetry of the diagram, when constructing the subtractions.
Figures~\ref{fig:TLD:Diagram42.1WithSubtractions} and~\ref{fig:TLD:D42.1U} show the construction of the subtractions
and their effects on the parent diagram. Note that we have extended the influence of $\COMP$
to cover the whole of diagrams~\ref{d:42.1U} and~\ref{d:42.1sub1U}, which is valid up to \Oep\ corrections.

\begin{center}
\begin{figure}
	\[
	-\frac{1}{2}
	\dec{
		\begin{array}{cccccl}
			\OLLabBlankLB[-2]{d:42.1}	&		&\OLLabLabLB[-2]{d:42.1:sub1}{d:42.1:add1}	&		&\OLLabLabRB[-2]{d:42.1:sub2}{d:42.1:add2}	
			&\ProPro{fig:TLD:D42.1U}{fig:TLD:Additions-Ia}
		\\
			\ensuremath{\begin{array}{c}\input{pstex/Diagram42.1.pstex_t} \end{array}}			&\mp4	&\ensuremath{\begin{array}{c}\input{pstex/Diagram42.1-s1.pstex_t} \end{array}}						&\mp4	&\ensuremath{\begin{array}{c}\input{pstex/Diagram42.1-s2.pstex_t} \end{array}}					
			&
		\end{array}
	}{\bullet}
	\]
\caption{Diagram~\ref{diag:42.1} and its subtractions.}
\label{fig:TLD:Diagram42.1WithSubtractions}
\end{figure}
\end{center}

\begin{center}
\begin{figure}
	\[
	-\frac{1}{2}
	\dec{
		\begin{array}{ccccc}
		\vspace{0.1in}
			\LabOLOb{d:42.1U}		&	&\LabOLOb{d:42.1sub1U}		&	&\Dprocess{d:268.7}{fig:TLD:Add-I:semiU}	
		\\
			\Uni{\ensuremath{\begin{array}{c}\input{pstex/Diagram42.1.pstex_t} \end{array}}}	&-4	&\Uni{\ensuremath{\begin{array}{c}\input{pstex/Diagram42.1-s1.pstex_t} \end{array}}}	&-4	&\ensuremath{\begin{array}{c}\input{pstex/Diagram42.1-s2SU.pstex_t} \end{array}}
		\end{array}
	}{\bullet}
	\]
\caption{The cancellation of non-computable contributions between diagram~\ref{d:42.1} and its subtractions.}
\label{fig:TLD:D42.1U}
\end{figure}
\end{center}

\paragraph{Diagram~\ref{diag:392.4}}

Diagram~\ref{diag:392.4} is another example of a case where we must be somewhat sneaky in the choice our subtractions.
As with diagram~\ref{diag:28.1}, we split the parent diagram into two copies, with differing momentum routings, and
construct subtractions for each. This is shown in figure~\ref{fig:TLD:Diagram392.4WithSubtractions}.
Note that, although the diagrammatics is unambiguous, we have explicitly indicated
the momentum routing for the subtractions.
\begin{center}
\begin{figure}[h]
	\[
	2
	\dec{
		\begin{array}{ccccccccl}
		\vspace{0.1in}
			&\OLLabBlankLB[-2]{d:392.4a}&	&\OLLabC[-2]{d:392.4a:sub1}{d:392.4a:add1}{d:255.25}&	&\OLLabC[-2]{d:392.4a:sub2}{d:392.4a:add2}{d:256.27,30}
			&	&\OLLabPro[-2]{d:392.4a:sub3}{d:392.4a:add3}{fig:TLD:Subs392.4a3-P}
		\\
			&\ensuremath{\begin{array}{c}\input{pstex/Diagram392.4a.pstex_t} \end{array}}			&\mp&\ensuremath{\begin{array}{c}\input{pstex/Diagram392.4a-s1.pstex_t} \end{array}}								&\mp&\ensuremath{\begin{array}{c}\input{pstex/Diagram392.4a-s2.pstex_t} \end{array}}
			&\mp&\ensuremath{\begin{array}{c}\input{pstex/Diagram392.4a-s3.pstex_t} \end{array}}
		\\
		\vspace{0.1in}
				&\LOBl[-2]{d:392.4b}&	&\OLLabC[-2]{d:392.4b:sub1}{d:392.4b:add1}{d:255.26}&	&\OLLabC[-2]{d:392.4b:sub2}{d:392.4b:add2}{d:256.25,33}
			&	&\OLLabRBPro[-2]{d:392.4b:sub3}{d:392.4b:add3}{fig:TLD:Subs392.4a3-P}	&\hspace{-4em} \ProBlank{fig:TLD:D392.4U}
		\\
			-	&\ensuremath{\begin{array}{c}\input{pstex/Diagram392.4b.pstex_t} \end{array}}	&\pm&\ensuremath{\begin{array}{c}\input{pstex/Diagram392.4b-s1.pstex_t} \end{array}}								&\pm&\ensuremath{\begin{array}{c}\input{pstex/Diagram392.4b-s2.pstex_t} \end{array}}
			&\pm&\ensuremath{\begin{array}{c}\input{pstex/Diagram392.4b-s3.pstex_t} \end{array}}
		\end{array}
	}{\bullet}
	\]
\caption{Diagram~\ref{diag:392.4} and its subtractions.}
\label{fig:TLD:Diagram392.4WithSubtractions}
\end{figure}
\end{center}

The pattern of subtractions demands comment. To understand them, it suffices to focus
on the three-legged sub-diagram, carrying loop momentum $k$, of diagram~\ref{d:392.4a}. The components
of this diagram that are Taylor expandable in $p$ or $k$ are subtracted off by 
diagrams~\ref{d:392.4a:sub1} and~\ref{d:392.4a:sub2}, respectively. The surviving non-Taylor
expandable components fall into two classes. First, there are those which survive only at small
$l$---in this case, the entire diagram gives a computable contribution, up to \Oep\ corrections.
However, there are also contributions which survive at large $l$. These non-computable
contributions are removed by diagram~\ref{d:392.4a:sub3}.

The structure of subtractions perhaps seems different from the previous two cases. However,
for diagrams~\ref{diag:32.1} and~\ref{diag:42.1} we were able to exploit symmetry
to reduce the number of subtractions; in other words, the analogue of diagram~\ref{d:392.4a:sub3}
exists in these cases, but can simply be combined with the analogue of 
diagrams~\ref{d:392.4a:sub1} and~\ref{d:392.4a:sub2}.

Figure~\ref{fig:TLD:D392.4U} shows the cancellation of non-computable contributions between the parents
and subtractions of figure~\ref{fig:TLD:Diagram392.4WithSubtractions}. 
Notice that, in all bar diagrams~\ref{d:268.8} and~\ref{d:268.9}
we are able to extend the influence of $\COMP$ to
cover the entire diagram, up to terms which vanish as $\epsilon \rightarrow 0$.
\begin{center}
\begin{figure}[h]
	\[
	2
	\dec{
		\begin{array}{cccccccc}
		\vspace{0.1in}
			&\LabOLOb{d:392.4aU}		&	&\LabOLOb{d:392.4a:sub1U}		&	&\Dprocess{d:268.8}{fig:TLD:Add-I:semiU}&	&\CancelOLObject{d:392.4a:sub3U}{d:277c.3}
		\\
			&\Uni{\ensuremath{\begin{array}{c}\input{pstex/Diagram392.4aU.pstex_t} \end{array}}}	&-	&\Uni{\ensuremath{\begin{array}{c}\input{pstex/Diagram392.4a-s1U.pstex_t} \end{array}}}	&-	&\ensuremath{\begin{array}{c}\input{pstex/Diagram392.4a-s2U.pstex_t} \end{array}}					&-	&\Uni{\ensuremath{\begin{array}{c}\input{pstex/Diagram392.4a-s3U.pstex_t} \end{array}}}
		\\
		\vspace{0.1in}
				&\LabOLOb{d:392.4bU}		&	&\LabOLOb{d:392.4b:sub1U}		&	&\Dprocess{d:268.9}{fig:TLD:Add-I:semiU}&	&\CancelOLObject{d:392.4b:sub3U}{d:277c.4}
		\\
			-	&\Uni{\ensuremath{\begin{array}{c}\input{pstex/Diagram392.4bU.pstex_t} \end{array}}}	&+	&\Uni{\ensuremath{\begin{array}{c}\input{pstex/Diagram392.4b-s1U.pstex_t} \end{array}}}	&+	&\ensuremath{\begin{array}{c}\input{pstex/Diagram392.4b-s2U.pstex_t} \end{array}}					&+	&\Uni{\ensuremath{\begin{array}{c}\input{pstex/Diagram392.4b-s3U.pstex_t} \end{array}}}
		\end{array}
	}{\bullet}
	\]
\caption{The terms remaining after 
cancellation of non-computable contributions to diagrams~\ref{d:392.4a} and diagram~\ref{d:392.4b} by their subtractions.}
\label{fig:TLD:D392.4U}
\end{figure}
\end{center}

\subsubsection{Manipulation of Additions}

Having completed the construction of subtractions for this section,
we note that diagrams~\ref{d:28.1a:add3}, 
\ref{d:32.1:add1}, \ref{d:42.1:add1} and~\ref{d:42.1:add2} 
can be manipulated,
using diagrammatic identity~\ref{D-ID:TwoWines}.\footnote{
It is also true that diagrams~\ref{d:392.4a:add3} and~\ref{d:392.4b:add3} could be manipulated in this way. However, the resulting
gauge remainders are trapped. It is more efficient to construct subtractions for these diagrams first, and then manipulate them, rather
than doing things the other way around.}
However, in preparation for this, we choose first to combine some
of the non-manipulable additions.

We can, up to a missing term and a discarded total momentum derivative, combine 
diagrams~\ref{d:28.1a:add1}, \ref{d:28.1a:add2}, \ref{d:28.1b:add1} and~\ref{d:28.1b:add2}
such that they cancel diagram~\ref{d:224b.11}. In turn, this missing term will 
be generated when we process the manipulable additions. Likewise, 
we can combine diagrams~\ref{d:28.1b:add3}, \ref{d:28.1b:add4}, \ref{d:32.1:add2} and~\ref{d:32.1:add3}
such that they cancel diagram~\ref{d:248.8}, up to a missing term which will shortly be generated
and a discarded total momentum derivative.
This recasting of terms is shown in figure~\ref{fig:TLD:Diagram258.1}. 
\begin{center}
\begin{figure}[h]
\[
	\dec{
		\begin{array}{c}
			\begin{array}{ccc}	
			\vspace{0.1in}
				\CancelOLObject{d:258.1}{d:224b.11}	&	&\CancelOLObject{d:258.1b}{d:253.6}	
			\\
				\ensuremath{\begin{array}{c}\input{pstex/Diagram224b.5.pstex_t} \end{array}}					&-2	&\ensuremath{\begin{array}{c}\input{pstex/Diagram253.6.pstex_t} \end{array}}			
			\end{array}
		\\
			\begin{array}{cccc}
			\vspace{0.1in}
					&\CancelOLObject{d:257.6}{d:248.8}	&	&\CancelOLObject{d:248.2,5b}{d:256.1}
			\\
				+	&\ensuremath{\begin{array}{c}\input{pstex/Diagram248.8.pstex_t} \end{array}}					&-2	&\ensuremath{\begin{array}{c}\input{pstex/Diagram256.1.pstex_t} \end{array}}
			\end{array}
		\end{array}
	}{\bullet}
\]
\caption{The first two diagrams are a 
re-expression of diagrams~\ref{d:28.1a:add1}, \ref{d:28.1a:add2}, \ref{d:28.1b:add1} and~\ref{d:28.1b:add2}
and the final two are a re-expression of 
diagrams~\ref{d:28.1b:add3}, \ref{d:28.1b:add4}, \ref{d:32.1:add2} and~\ref{d:32.1:add3}.}
\label{fig:TLD:Diagram258.1}
\end{figure}
\end{center}

As hoped, we find two cancellations.

\Cancel{d:258.1}{d:224b.11}
\Cancel{d:257.6}{d:248.8}

Now we turn to the diagrammatic manipulation of the aforementioned
additions. The final result of this procedure---where we have processed
all gauge remainders---is split between figures~\ref{fig:TLD:Additions-Ia}
and~\ref{fig:TLD:Additions-Ia-B}.

\begin{center}
\begin{figure}
	\[
	2
	\dec{
		\begin{array}{l}
			\begin{array}{cccccccc}
			\vspace{0.1in}
							&\CO{d:221b.1}{d:217.1}	&			&\CO{d:225.8}{d:217.3}	&	&\CO{d:253.6}{d:258.1b}	&	&\CO{d:256.1}{d:248.2,5b}
			\\
				-\ds \hf	&\ensuremath{\begin{array}{c}\input{pstex/Diagram221b.1.pstex_t} \end{array}}		& -	\ds \hf	&\ensuremath{\begin{array}{c}\input{pstex/Diagram225.8.pstex_t} \end{array}}		&+	&\ensuremath{\begin{array}{c}\input{pstex/Diagram253.6.pstex_t} \end{array}}		&+	&\ensuremath{\begin{array}{c}\input{pstex/Diagram256.1.pstex_t} \end{array}}
			\end{array}
		\\
			\begin{array}{ccccccccl}
			\vspace{0.1in}
					&\LOLB{d:225.11,14}		&			&\LO{d:225.15}		&	&\LO{d:221d.2,4}		&			&\LORB{d:221d.8}	&\hspace{-2em}\Process{fig:TLD:SemiNonUniversal}
			\\
				+	&\ensuremath{\begin{array}{c}\input{pstex/Diagram225.11,14.pstex_t} \end{array}}	&+\ds \hf	&\ensuremath{\begin{array}{c}\input{pstex/Diagram225.15.pstex_t} \end{array}}	&-	&\ensuremath{\begin{array}{c}\input{pstex/Diagram221d.2,4.pstex_t} \end{array}}	&-\ds \hf	&\ensuremath{\begin{array}{c}\input{pstex/Diagram221d.8.pstex_t} \end{array}}	&
			\end{array}
		\\
			\begin{array}{cccccccc}
			\vspace{0.1in}
					&\CO{d:258.6}{d:224b.12b}	&	&\CO{d:258.2}{d:224b.12}&	&\CO{d:255.25}{d:392.4a:add1}	&	&\CO{d:255.26}{d:392.4b:add1}
			\\
				-	&\ensuremath{\begin{array}{c}\input{pstex/Diagram258.6.pstex_t} \end{array}}			&-	&\ensuremath{\begin{array}{c}\input{pstex/Diagram258.2.pstex_t} \end{array}}		&+	&\ensuremath{\begin{array}{c}\input{pstex/Diagram255.25.pstex_t} \end{array}}				&+	&\ensuremath{\begin{array}{c}\input{pstex/Diagram255.26.pstex_t} \end{array}}
			\end{array}
		\\
			\begin{array}{cccccccc}
			\vspace{0.1in}
					&\CO{d:257.7a}{d:248.9a}&	&\CO{d:257.7b}{d:248.9b}&	&\CO{d:256.25,33}{d:392.4a:add2}&	&\CO{d:256.27,30}{d:392.4b:add2}
			\\
				-	&\ensuremath{\begin{array}{c}\input{pstex/Diagram257.7a.pstex_t} \end{array}}		&-	&\ensuremath{\begin{array}{c}\input{pstex/Diagram257.7b.pstex_t} \end{array}}		&+	&\ensuremath{\begin{array}{c}\input{pstex/Diagram256.25,33.pstex_t} \end{array}}			&+	& \ensuremath{\begin{array}{c}\input{pstex/Diagram256.27,30.pstex_t} \end{array}}
			\end{array}
		\end{array}
	}{\bullet}
	\]
\caption{The result of the diagrammatic manipulation of diagrams~\ref{d:28.1a:add3}, 
\ref{d:32.1:add1}, \ref{d:42.1:add1} and~\ref{d:42.1:add2}---part~I.}
\label{fig:TLD:Additions-Ia}
\end{figure}
\end{center}

\begin{center}
\begin{figure}
	\[
	2
	\dec{
		\begin{array}{l}
			\begin{array}{cccccccc}
			\vspace{0.1in}
					&\LOLB{d:255b.1}	&	&\LO{d:255c.7}		&	&\LO{d:255c.8}		&	&\LO{d:255d.7}
			\\
				+	&\ensuremath{\begin{array}{c}\input{pstex/Diagram255b.1.pstex_t} \end{array}}	&+	&\ensuremath{\begin{array}{c}\input{pstex/Diagram255c.7.pstex_t} \end{array}}	&-	&\ensuremath{\begin{array}{c}\input{pstex/Diagram255c.8.pstex_t} \end{array}}	&+	&\ensuremath{\begin{array}{c}\input{pstex/Diagram255d.7.pstex_t} \end{array}}
			\end{array}
		\\
			\begin{array}{cccccccc}
			\vspace{0.1in}
					&\LO{d:255c.16}					&	&\LO{d:255c.17}		&	&\LO{d:255d.1}		&	&\LO{d:255d.2}
			\\
				+	&\ensuremath{\begin{array}{c}\input{pstex/Diagram255c.16.pstex_t} \end{array}}			&-	&\ensuremath{\begin{array}{c}\input{pstex/Diagram255c.17.pstex_t} \end{array}}&-	&\ensuremath{\begin{array}{c}\input{pstex/Diagram255d.1.pstex_t} \end{array}}	&+	&\ensuremath{\begin{array}{c}\input{pstex/Diagram255d.2.pstex_t} \end{array}}
			\end{array}
		\\
			\begin{array}{ccccccl}
			\vspace{0.1in}
					&\LO{d:255b.11}		&	&\LO{d:255d.8}		&	&\LORB{d:255c.1}		&\hspace{-1.5em}\Process{fig:TLD:Universal-IIa}
			\\
				-	&\ensuremath{\begin{array}{c}\input{pstex/Diagram255b.11.pstex_t} \end{array}}&+	&\ensuremath{\begin{array}{c}\input{pstex/Diagram255d.8.pstex_t} \end{array}}	&-	&\ensuremath{\begin{array}{c}\input{pstex/Diagram255c.1.pstex_t} \end{array}}	&
			\end{array}
		\\
			\begin{array}{ccccccccl}
			\vspace{0.1in}
					&\LOLB{d:255b.16}		&	&\LO{d:255b-Corr}		&	&\LO{d:255d.4}		&	&\LORB{d:255d.5} 	& \Process{fig:TLD:255b.16etc-U,NU}
			\\
				-	&\ensuremath{\begin{array}{c}\input{pstex/Diagram255b.16.pstex_t} \end{array}}&-	&\ensuremath{\begin{array}{c}\input{pstex/Diagram255b-Corr.pstex_t} \end{array}}	&-	&\ensuremath{\begin{array}{c}\input{pstex/Diagram255d.4.pstex_t} \end{array}}	&+	&\ensuremath{\begin{array}{c}\input{pstex/Diagram255d.5.pstex_t} \end{array}}		&
			\end{array}
		\end{array}
	}{\bullet}
	\]
\caption{The result of the diagrammatic manipulation of diagrams~\ref{d:28.1a:add3}, 
\ref{d:32.1:add1}, \ref{d:42.1:add1} and~\ref{d:42.1:add2}---part~II.}
\label{fig:TLD:Additions-Ia-B}
\end{figure}
\end{center}

It looks as though we may have missed some simplifications. 
In particular,
diagrams~\ref{d:255b.16}--\ref{d:255d.5} possess an \Op{2}\ stub,
which we might hope will allow us to remove the discontinuity in momentum arguments.
This discontinuity occurs where an effective 
propagator carrying momentum, say
$k-p$, attaches to structures which have been Taylor expanded in $p$. However, we cannot
set $p=0$ in the denominator of the effective propagator, since $p$ provides IR
regularisation for the $k$-integral. Hence, we cannot simplify these diagrams.

As promised, we cancel the remaining unprocessed diagrams from figure~\ref{fig:TLD:Diagrams224-A}.

\CancelTmD{d:258.6}{d:224b.12b}
\CancelTmD{d:258.2}{d:224b.12}
\CancelTmD{d:257.7a}{d:248.9a}
\CancelTmD{d:257.7b}{d:248.9b}

A comment about these cancellations is in order. Consider
the
momentum derivative in  diagrams~\ref{d:258.6}, \ref{d:258.2},
\ref{d:257.7a} and~\ref{d:257.7b}. As drawn, each momentum derivative
strikes part of the sub-diagram. However, we can take this momentum derivative
to strike the whole of the sub-diagram, where this derivative
is \wrt\ the sub-diagram's external momentum. This step is
correct up to a total momentum derivative \wrt\ the sub-diagram's
loop momentum (see section~\ref{sec:MomExp:CompleteDiagrams}). However,
we know that such total momentum derivative terms can be discarded~\cite{SU(N|N)}.

The two surviving diagrams from figure~\ref{fig:TLD:Diagram258.1} are removed.

\Cancel{d:253.6}{d:258.1b}
\Cancel{d:256.1}{d:248.2,5b}

Finally, we get six bonus cancellations.

\CancelTmD{d:221b.1}{d:217.1}
\CancelTmD{d:225.8}{d:217.3}
\CancelCom{d:255.25}{d:392.4a:add1}{courtesy of diagrammatic identity~\ref{D-ID-Deriv-GR}.}
\CancelCom{d:255.26}{d:392.4b:add1}{courtesy of diagrammatic identity~\ref{D-ID-Deriv-GR}.}
\CancelCom{d:256.25,33}{d:392.4a:add2}{courtesy of diagrammatic identity~\ref{D-ID-Deriv-GR}.}
\CancelCom{d:256.27,30}{d:392.4b:add2}{courtesy of diagrammatic identity~\ref{D-ID-Deriv-GR}.}

\subsubsection{Manipulation of Semi-Computable Terms}

We conclude this section by manipulating the semi-computable
partners of the additions which were manipulated in the
previous section. Note, though, that not all of the additions have semi-computable
partners: in the case where the loop integrals of a semi-computable
diagram do not factorise, the $\COMP$ can be extended to cover the entire
diagram. We choose not manipulate these terms as we can just compute them, directly;
we comment on this further in section~\ref{sec:TLD:FinalRemainders}.

For the terms whose loop integrals factorise, we 
focus on the computable sub-diagrams.
Ideally, we would like to manipulate these. However, by doing so we 
generate diagrams where we would like to have kept 
momentum derivatives of cutoff functions (see section~\ref{sec:TLD:Subtractions:Gauge}).
Hence, what we choose to do is
split the computable part of the semi-computable terms into \OepPow{-1}\ and \Oepz\ components. We can then
manipulate the former, keeping only the \OepPow{-1}\ part of the resultant terms. We leave the
\Oepz\ components un-manipulated. Again, this will be commented on  in section~\ref{sec:TLD:FinalRemainders}.

Figure~\ref{fig:TLD:Add-I:semiU} shows the result of this procedure.

\begin{center}
\begin{figure}
\[
	2
	\dec{
		\begin{array}{l}
			\begin{array}{ccccccl}
			\vspace{0.1in}
							&\LOLB{d:269.1}			&	&\LO{d:269.2a}				&	&\LORB{d:269.2b}	&\hspace{-1em}\Process{fig:TLD:Diagrams269B}
			\\
				-\ds \hf	&\ensuremath{\begin{array}{c}\input{pstex/Diagram269.1.pstex_t} \end{array}}\emol	&+	&\ensuremath{\begin{array}{c}\input{pstex/Diagram269.2a.pstex_t} \end{array}}	\emol	&+	&\ensuremath{\begin{array}{c}\input{pstex/Diagram269.2b.pstex_t} \end{array}}	&
			\end{array}
		\\
			\begin{array}{ccccl}
			\vspace{0.1in}
					&\LOLB{d:269.11b}			&	&\LORB{d:269.12b}	&\hspace{-1em}\Process{fig:TLD:Universal-III}
			\\
				2	&\ensuremath{\begin{array}{c}\input{pstex/Diagram269.11b.pstex_t} \end{array}}\emol	&+	&\ensuremath{\begin{array}{c}\input{pstex/Diagram269.12b.pstex_t} \end{array}}&
			\end{array}
		\\
			\begin{array}{cccccc}
			\vspace{0.1in}
						&\LO{d:269.4}			&	&\LO{d:269.5}			&			&\LO{d:269.6}
			\\
				+\ds \hf&\ensuremath{\begin{array}{c}\input{pstex/Diagram269.4.pstex_t} \end{array}}\etozl&-	&\ensuremath{\begin{array}{c}\input{pstex/Diagram269.5.pstex_t} \end{array}}\etozl&-\ds \hf	&\ensuremath{\begin{array}{c}\input{pstex/Diagram269.6.pstex_t} \end{array}}
			\end{array}
		\\
			\begin{array}{cccccccc}
			\vspace{0.1in}
					&\LO{d:269.7}			&	&\LO{d:269.8}			&	&\LO{d:269.9}			&	&\LO{d:269.10}
			\\
				+	&\ensuremath{\begin{array}{c}\input{pstex/Diagram269.7.pstex_t} \end{array}}\etozl&+	&\ensuremath{\begin{array}{c}\input{pstex/Diagram269.8.pstex_t} \end{array}}\etozl&-	&\ensuremath{\begin{array}{c}\input{pstex/Diagram269.9.pstex_t} \end{array}}\etozl&+	&\ensuremath{\begin{array}{c}\input{pstex/Diagram269.10.pstex_t} \end{array}}	\etozl
			\end{array}
		\end{array}
	}{\bullet}
\]
\caption{The result of processing diagrams~\ref{d:268.3}, \ref{d:268.4}, \ref{d:268.5}, \ref{d:268.6}, 
\ref{d:268.7},   \ref{d:268.8} and~\ref{d:268.9}. }
\label{fig:TLD:Add-I:semiU}
\end{figure}
\end{center}

Since, in diagrams~\ref{d:269.4}--\ref{d:269.10} we take the \Oepz\ part of the topmost sub-diagram,
we are compelled to take the most divergent contribution from the bottom diagram, up to terms
which vanish as \eptoz. Consequently, these seven diagrams are computable.

\section{Further Subtractions}	\label{sec:TLD:FurtherSubtractions}

In this section, we construct two further sets of
subtractions. First, we do this for the remaining
$\Lambda$-derivative terms generated in chapter~\ref{ch:bulk}.
Secondly, we note that we can construct `second level'
subtractions, for a number of the terms
generated in section~\ref{sec:TLD:SSasSubD}.

\subsection{Subtractions for Terms Generated in Chapter~\ref{ch:bulk}}

\paragraph{Diagrams~\ref{diag:FR22.2}--\ref{diag:FR24.1}}

To generate the subtractions for the
four diagrams analysed in this section will require 
a minor development of the current formalism.

In the subsequent analysis, we need look only at the first two diagrams, since all conclusions drawn will apply to
the other pair, as well. 
We will suppose that the effective propagator which terminates in a processed gauge remainder
which is differentiated \wrt\ momentum carries loop momentum $l$. We will take the other loop
momentum to be $k$. (See also figure~\ref{fig:TLD:DFR22.1...:WS}.)
In order for the diagrams to have any chance of surviving, the effective propagator carrying momentum $l$ must be
in the $A$-sector. If this is the case, then diagrams~\ref{diag:FR22.2} and~\ref{diag:FR22.1} become algebraically equivalent;
likewise for diagrams~\ref{diag:FR24.6} and~\ref{diag:FR24.1}.
Consequently, for the following analysis, we will replace diagram~\ref{diag:FR22.1} (\ref{diag:FR24.1}) with twice 
diagram~\ref{diag:FR22.2} (\ref{diag:FR24.6}). This step is valid, up to $\Oep$ corrections.

With the diagrams partially in the $A$-sector---as just described---it is clear that, by Lorentz invariance, the $k$-integral must
go as odd powers of $l$ (up to additional factors of $l^{-2\epsilon}$). For the diagram to survive, we take just the single power of $l$.
Given that we wish to cancel off terms for which the $k$-integral is Taylor expandable in $l$, we construct our subtraction such that
it contains
\begin{equation}
	l \cdot \partial^l F(l-k) |_{l=0}. \label{eq:TLD:OrderMomentum}
\end{equation}
where $F$ is the structure which depends on $l-k$; precisely which structure this is will depend on the momentum routing. 
This expression can be rewritten as
\[
	-l \cdot \partial^k F(-k).
\]

Given that we want to be able to construct the same subtraction,
irrespective of the momentum routing of the parent, we
must use the prescription that, in the $k$-integral, \emph{we keep
derivatives of cutoff functions, if they appear}.
Hence, we construct the subtractions as shown in figure~\ref{fig:TLD:DFR22.1...:WS};
we are free to move the $\partial^k$ from 
gauge-remainder to effective propagator at the expense of a minus sign.
\begin{center}
\begin{figure}[h]
	\[
	-8
	\dec{
		\begin{array}{ccccl}
		\vspace{0.1in}
				&\OLLabBlankLB[-2]{d:FR22.1,2}		&	&\OLLabLabLB[-2]{d:FR22.1,2:sub1}{d:FR22.1,2:add1} 	&
		\\
				&\ensuremath{\begin{array}{c}\input{pstex/DiagramFR22.1,2.pstex_t} \end{array}}				&\mp&\ensuremath{\begin{array}{c}\input{pstex/DiagramFR22.1,2-s1.pstex_t} \end{array}}							&
		\\
		\vspace{0.1in}
				&\LO{d:FR24.1,6}					&	&\OLLabRBLabRB[-2]{d:FR24.1,6:sub1}{d:FR24.1,6:add1}&\ProPro[-2]{fig:TLD:FR22.1...U}{fig:TLD:FR22.1...U}
		\\
			-	&\ensuremath{\begin{array}{c}\input{pstex/DiagramFR24.1,6.pstex_t} \end{array}}				&\pm&\ensuremath{\begin{array}{c}\input{pstex/DiagramFR24.1,6-s1.pstex_t} \end{array}}							&
		\end{array}
	}{\bullet}
	\]
\caption[Diagrams~\ref{diag:FR22.1}, \ref{diag:FR22.2}, \ref{diag:FR24.1} and~\ref{diag:FR24.6} and their subtractions.]{
Diagrams~\ref{diag:FR22.1}, \ref{diag:FR22.2}, \ref{diag:FR24.1} and~\ref{diag:FR24.6} and their subtractions.
The first pair of diagrams have been combined into a single term, with an additional factor of two. Likewise, for the
second pair.}
\label{fig:TLD:DFR22.1...:WS}
\end{figure}
\end{center}

Note that in the subtractions, the dummy field leaving the vertex
is implicitly in the $A$-sector. This is because the only fields
which can attach to circles denoting momentum derivatives
are in the $A$-sector.

 Figure~\ref{fig:TLD:FR22.1...U} shows both
the cancellation of non-computable components between parent and subtraction and the subsequent cancellation between the computable part
of the subtraction and the addition. This is different from the way in which we treated
additions and subtractions in the previous section, where we chose to keep them apart.
Note that the $\NUn$ of diagrams~\ref{d:274b.5} and~\ref{d:274b.6}
refers just to the `innermost' sub-diagram.
\begin{center}
\begin{figure}[h]
	\[
	-8
	\dec{
		\begin{array}{cccc}
		\vspace{0.1in}
				&\LO{d:FR22.1,2U}			&	&\PO{d:274b.5}{fig:TLD:LittleSetSq:NU-I}
		\\
				&\Uni{\ensuremath{\begin{array}{c}\input{pstex/DiagramFR22.1,2.pstex_t} \end{array}}}	&+	&\ensuremath{\begin{array}{c}\input{pstex/Diagram274b.5.pstex_t} \end{array}} \NUl
		\\
		\vspace{0.1in}
				&\LO{d:FR24.1,6U}		&	&\PO{d:274b.6}{fig:TLD:LittleSetSq:NU-I}
		\\
			-	&\Uni{\ensuremath{\begin{array}{c}\input{pstex/DiagramFR24.1,6.pstex_t} \end{array}}}	&-	&\ensuremath{\begin{array}{c}\input{pstex/Diagram274b.6.pstex_t} \end{array}} \NUl
		\end{array}
	}{\bullet}
	\]
\caption[The cancellation of contributions between the diagrams of figure~\ref{fig:TLD:DFR22.1...:WS}.]{
The cancellation of contributions between the diagrams of figure~\ref{fig:TLD:DFR22.1...:WS}. 
Two steps have been performed on the RHS: first, non-computable contributions have been cancelled between
parent and subtraction; second, diagrams~\ref{d:FR22.1,2:add1} and~\ref{d:FR24.1,6:add1} have been added.}
\label{fig:TLD:FR22.1...U}
\end{figure}
\end{center}

\paragraph{Diagrams~\ref{diag:FR26.1} and~\ref{diag:FR26.2}}

The analysis of diagrams~\ref{diag:FR26.1} and~\ref{diag:FR26.2} requires the techniques of the previous section. Similarly,
we can choose to keep just one of the diagrams (\ref{diag:FR26.1}) and multiply by two. This is because
the two gauge remainders attached to external fields must be in the $A^1$-sector for the diagrams not to vanish. 
Now let us examine the consequences of this for the remaining gauge remainders, both of which are attached to an internal field.
The one which bites a gauge remainder attached to an external field must also be in the $A^1$ sector. The final gauge
remainder can either be in the $A^1$ or $B$-sector.

Once again, it is
necessary for the subtraction to contain a term of the form given by equation~(\ref{eq:TLD:OrderMomentum}). We can simplify the diagrammatics using a
trick most easily seen with the momentum routing such that the  gauge remainder on the `inside' of diagram~\ref{diag:FR26.1}
carries $l-k$ (see also figure~\ref{fig:TLD:DFR26.1,6:WS}). This gauge remainder, which can
be in any sector, is contracted into the gauge remainder $l_\alpha/l^2$, at the other end of the effective propagator. Now,
\[
l \cdot \partial^l [(l-k)'_{\alpha}]_{l=0} \frac{l_\alpha}{l^2} = \partial^k \cdot (k').
\]
Rather than constructing the subtraction such that the $k$-integral contains no indices whereas the $l$-integral possesses both
$\mu$ and $\nu$, we use Lorentz invariance to draw the diagram in an equivalent, but more intuitive form. 
Figure~\ref{fig:TLD:DFR26.1,6:WS} comprises two separate parts. The first part
shows the construction of the subtraction. The second part shows both
the cancellation of non-computable components between parent and subtraction and the subsequent cancellation between the computable part
of the subtraction and the addition.
\begin{center}
\begin{figure}[h]
\[
	8
	\dec{
		\begin{array}{ccc}
			\ensuremath{\begin{array}{c}\input{pstex/DiagramFR26.1,2-Lab.pstex_t} \end{array}} &\mp& \ensuremath{\begin{array}{c}\input{pstex/DiagramFR26.1,2-s.pstex_t} \end{array}}
		\end{array}
	}{\bullet}
	\rightarrow
	8
	\dec{
		\begin{array}{ccc}
		\vspace{0.1in}
			\LO{d:FR26.1,2U}			&	&\PO{d:FR26.1,2:sub1NU}{fig:TLD:SemiNonUniversal}
		\\
			\Uni{\ensuremath{\begin{array}{c}\input{pstex/DiagramFR26.1,2.pstex_t} \end{array}}}	&+	&\ensuremath{\begin{array}{c}\input{pstex/DiagramFR26.1,2NU.pstex_t} \end{array}}
		\end{array}
	}{\bullet}
\]
\caption{The LHS of the figure shows diagrams~\ref{diag:FR26.1}, \ref{diag:FR26.2} and their subtractions.
The parent diagrams have been combined into a single term, with an additional factor of two.
The right hand side shows the cancellation of contributions between the three diagrams of the RHS.
}
\label{fig:TLD:DFR26.1,6:WS}
\end{figure}
\end{center}

There is a curious feature to the parent and its subtraction that deserves comment. In the parent, the gauge remainders carrying $l$ or
$l-p$ can always be chosen to be in the $A$-sector. UV regularisation is provided by the $l-k$ gauge-remainder being in the F-sector.
However, in the subtraction term, we cannot throw away the F-sector part of the $l$-dependent (factorisable) sub-diagram. Such details
concerning the regularisation will not, of course affect the universal quantity which we ultimately compute.

\paragraph{Diagram~\ref{diag:FR26.11}}

Due to the symmetry of
diagram~\ref{diag:FR26.11}, the subtraction comes with a relative factor of two.
Figure~\ref{fig:TLD:DFR26.11:WS} shows both the construction of the subtraction and the cancellation of terms between
parent, subtraction and addition.
\begin{center}
\begin{figure}
	\[
	\begin{array}{r}
	\vspace{0.2in}
		2
		\dec{
			\begin{array}{ccc}
				\ensuremath{\begin{array}{c}\input{pstex/DiagramFR26.1+6.pstex_t} \end{array}} & \mp 2&\ensuremath{\begin{array}{c}\input{pstex/DiagramFR26.1+6-s.pstex_t} \end{array}}
			\end{array}
		}{\bullet}
	\\
		\rightarrow
		2
		\dec{
		\begin{array}{ccc}
		\vspace{0.1in}
			\LO{d:FR26.11U}				&	&\CO{d:284.4}{d:284.1}
			\\
		\Uni{\ensuremath{\begin{array}{c}\input{pstex/DiagramFR26.1+6.pstex_t} \end{array}}}	&+2	&\ensuremath{\begin{array}{c}\input{pstex/DiagramFR26.1+6NU.pstex_t} \end{array}} \NUl
		\end{array}
		}{\bullet}
	\end{array}
	\]
\caption{The top row of the figure shows diagram~\ref{diag:FR26.11} and its subtraction.
The bottom shows the cancellation of contributions between all three diagrams of the top row.}
\label{fig:TLD:DFR26.11:WS}
\end{figure}
\end{center}

\paragraph{Diagrams~\ref{diag:FR36.1}--\ref{diag:FR27.6}}

The subtractions for diagrams~\ref{diag:FR36.1}--\ref{diag:FR27.6}
are shown in figure~\ref{fig:TLD:FR36.1etc-S}.

\begin{center}
\begin{figure}
	\[
	\mp 4
	\dec{
		\begin{array}{l}
		\vspace{0.2in}
			\begin{array}{ccccc}
			\vspace{0.1in}
				\LOLBLOLB{d:FR36.1-s}{d:FR36.1-a}	&	&\LObLOb{d:FR22.3-s}{d:FR22.3-a}&	&\LObLOb{d:FR24.11-s}{d:FR24.11-a}
			\\
				\ensuremath{\begin{array}{c}\input{pstex/DiagramFR36.1-s.pstex_t} \end{array}}			&+	&\ensuremath{\begin{array}{c}\input{pstex/DiagramFR22.3-s.pstex_t} \end{array}}			&-	&\ensuremath{\begin{array}{c}\input{pstex/DiagramFR24.11-s.pstex_t} \end{array}}
			\end{array}
		\\
			\begin{array}{ccccl}
			\vspace{0.1in}
					&\LObLOb{d:FR27.1-s}{d:FR27.1-a}&			&\LORBLORB{d:FR27.6-s}{d:FR27.6-a} 	& \ProPro{fig:TLD:FR36.1etc-U,NU}{fig:TLD:FR36.1etc-U,NU}
			\\
				+	&\ensuremath{\begin{array}{c}\input{pstex/DiagramFR27.1-s.pstex_t} \end{array}}			&-\ds \hf	&\ensuremath{\begin{array}{c}\input{pstex/DiagramFR27.6-s.pstex_t} \end{array}}				&
			\end{array}
		\end{array}
	}{\bullet}
	\]
\caption{Subtractions (and additions) for diagrams~\ref{diag:FR36.1}--\ref{diag:FR27.6}.}
\label{fig:TLD:FR36.1etc-S}
\end{figure}
\end{center}

In figure~\ref{fig:TLD:FR36.1etc-U,NU} we combine both the subtractions and additions
with their corresponding parent diagrams.
\begin{center}
\begin{figure}
	\[
	4
	\dec{
		\begin{array}{l}
		\vspace{0.2in}
			\begin{array}{ccccc}
			\vspace{0.1in}
				\LO{d:FR36.1-U}			&	&\LO{d:FR22.3-U}			&	&\LO{d:FR24.11-U}
			\\
				\Uni{\ensuremath{\begin{array}{c}\input{pstex/DiagramFR36.1.pstex_t} \end{array}}}&+	&\Uni{\ensuremath{\begin{array}{c}\input{pstex/DiagramFR22.3.pstex_t} \end{array}}}	&-	&\Uni{\ensuremath{\begin{array}{c}\input{pstex/DiagramFR24.11.pstex_t} \end{array}}}
			\end{array}
		\\
		\vspace{0.2in}
			\begin{array}{cccc}
			\vspace{0.1in}
					&\LO{d:FR27.1-U}			&			&\LO{d:FR27.6-U}
			\\
				+	&\Uni{\ensuremath{\begin{array}{c}\input{pstex/DiagramFR27.1.pstex_t} \end{array}}}	&-\ds \hf	&\Uni{\ensuremath{\begin{array}{c}\input{pstex/DiagramFR27.6.pstex_t} \end{array}}}
			\end{array}
		\\
		\vspace{0.2in}
			\begin{array}{ccccccl}
			\vspace{0.1in}
					&\LOLB{d:FR36.1-NU}			&	&\LO{d:FR22.3-NU}			&	&\LORB{d:FR24.11-NU}	& \Process{fig:TLD:LS-rems-B}
			\\
				+	&\ensuremath{\begin{array}{c}\input{pstex/DiagramFR36.1-NU.pstex_t} \end{array}}\NUl	&+	&\ensuremath{\begin{array}{c}\input{pstex/DiagramFR22.3-NU.pstex_t} \end{array}}\NUl	&-	&\ensuremath{\begin{array}{c}\input{pstex/DiagramFR24.11-NU.pstex_t} \end{array}}	&
			\end{array}
		\\
			\begin{array}{ccccl}
			\vspace{0.1in}
					&\PO{d:FR27.1-NU}{fig:TLD:LS-rems-B}		&			&\CO{d:FR27.6-NU}{d:255b-Corr-B-NU}	& 
			\\
				+	&\ensuremath{\begin{array}{c}\input{pstex/DiagramFR27.1-NU.pstex_t} \end{array}}						&-\ds \hf	&\ensuremath{\begin{array}{c}\input{pstex/DiagramFR27.6-NU.pstex_t} \end{array}}				&
			\end{array}
		\end{array}
	}{\bullet}
	\]
\caption{Result of combining diagrams~\ref{diag:FR36.1}--\ref{diag:FR27.6}
with both their subtractions and additions.}
\label{fig:TLD:FR36.1etc-U,NU}
\end{figure}
\end{center}

\subsection{Subtractions for Terms Generated in Section~\ref{sec:TLD:SSasSubD}}

\paragraph{Diagrams~\ref{d:392.4a:add3} and~\ref{d:392.4b:add3}}

Figure~\ref{fig:TLD:Subs392.4a3} shows the subtractions for
diagrams~\ref{d:392.4a:add3} and~\ref{d:392.4b:add3}.
\begin{center}
\begin{figure}
	\[
	\mp2
	\dec{
		\begin{array}{cccl}
		\vspace{0.1in}
			\LOLBLOLB{d:277.7}{d:277.9}	&	&\LORBLORB{d:277.8}{d:277.10}	&\hspace{-1.5em}\ProPro{fig:TLD:Subs392.4a3-P}{fig:TLD:Subs392.4a3-P}
		\\
			\ensuremath{\begin{array}{c}\input{pstex/Diagram277.7.pstex_t} \end{array}}			&-	&\ensuremath{\begin{array}{c}\input{pstex/Diagram277.8.pstex_t} \end{array}}				&
		\end{array}
	}{\bullet}
	\]
\caption{Subtractions for diagrams~\ref{d:392.4a:add3} and~\ref{d:392.4b:add3}.}
\label{fig:TLD:Subs392.4a3}
\end{figure}
\end{center}

Figure~\ref{fig:TLD:Subs392.4a3-P} shows the result of adding the subtractions of
figure~\ref{fig:TLD:Subs392.4a3} to their parents.
\begin{center}
\begin{figure}[h]
	\[
	2
	\dec{
		\begin{array}{ccccccc}
		\vspace{0.1in}
			\CO{d:277c.3}{d:392.4a:sub3U}	&	&\PO{d:277c.5a}{fig:TLD:SemiNonUniversal}		&	&\CO{d:277c.4}{d:392.4b:sub3U}	&	&\PO{d:277c.5b}{fig:TLD:SemiNonUniversal}
		\\
			\Uni{\ensuremath{\begin{array}{c}\input{pstex/Diagram392.4a-s3U.pstex_t} \end{array}}}	&+	&\ensuremath{\begin{array}{c}\input{pstex/Diagram277c.5a.pstex_t} \end{array}}							&-	&\Uni{\ensuremath{\begin{array}{c}\input{pstex/Diagram392.4b-s3U.pstex_t} \end{array}}}	&-	&\ensuremath{\begin{array}{c}\input{pstex/Diagram277c.5b.pstex_t} \end{array}}
		\end{array}
	}{\bullet}
	\]
\caption{The result of combining diagrams~\ref{d:392.4a:add3} and~\ref{d:392.4b:add3} with their
subtractions.}
\label{fig:TLD:Subs392.4a3-P}
\end{figure}
\end{center}

At this stage, we find two cancellations. Whilst cancellations are always welcome, these
turn out to be between diagrams under the influence of $\COMP$; hence, they do not
help us to demonstrate the computability of $\beta_2$.

\Cancel{d:277c.3}{d:392.4a:sub3U}
\Cancel{d:277c.4}{d:392.4b:sub3U}

\paragraph{Diagrams~\ref{d:255b.1}--\ref{d:255c.1}}

Figure~\ref{fig:TLD:Subtractions-II-A} shows the construction of subtractions
for diagrams~\ref{d:255b.1}--\ref{d:255c.1}.
\begin{center}
\begin{figure}
	\[
	\mp 2
	\dec{
		\begin{array}{c}
			\begin{array}{ccccccc}
			\vspace{0.1in}
				\LOLBLOLB{d:255b.1:sub1}{d:255b.1:add1}	&	&\LObLOb{d:255c.7:sub1}{d:255c.7:add1}
				&	&\LObLOb{d:255c.8:sub1}{d:255c.8:add1}	&	&\LObLOb{d:255d.7:sub1}{d:255d.7:add1}
			\\
				\ensuremath{\begin{array}{c}\input{pstex/Diagram255b.1-s.pstex_t} \end{array}}					&+	&\ensuremath{\begin{array}{c}\input{pstex/Diagram255c.7-s1.pstex_t} \end{array}}					
				&-	&\ensuremath{\begin{array}{c}\input{pstex/Diagram255c.8-s.pstex_t} \end{array}}					&+	&\ensuremath{\begin{array}{c}\input{pstex/Diagram255d.7-s.pstex_t} \end{array}}
			\end{array}
		\\
			\begin{array}{cccc}
			\vspace{0.1in}
					&\LObLOb{d:255c.16:sub1}{d:255c.16:add1}	&	&\LObLOb{d:255c.17:sub1}{d:255c.17:add1}		
			\\
				+	&\ensuremath{\begin{array}{c}\input{pstex/Diagram255c.16-s.pstex_t} \end{array}}						&-	&\ensuremath{\begin{array}{c}\input{pstex/Diagram255c.17-s.pstex_t} \end{array}}									
			\end{array}
		\\
			\begin{array}{cccc}
			\vspace{0.1in}
					&\LObLOb{d:255d.1:sub1}{d:255d.1:add1}	&	&\LObLOb{d:255d.2:sub1}{d:255d.2:add1}
			\\
				-	&\ensuremath{\begin{array}{c}\input{pstex/Diagram255d.1-s.pstex_t} \end{array}}					&+	&\ensuremath{\begin{array}{c}\input{pstex/Diagram255d.2-s.pstex_t} \end{array}}
			\end{array}
		\\
			\begin{array}{cccccccl}
			\vspace{0.1in}
					&\LObLOb{d:255b.11:sub1}{d:255b.11:add1}	&	&\LObLOb{d:255d.8:sub1}{d:255d.8:add1}
					&	&\LORBLORB{d:255c.1:sub1}{d:255c.1:add1}& \hspace{-2em} \ProPro{fig:TLD:Universal-IIa}{fig:TLD:Universal-IIa}
			\\
				-	&\ensuremath{\begin{array}{c}\input{pstex/Diagram255b.11-s.pstex_t} \end{array}}											&+	&\ensuremath{\begin{array}{c}\input{pstex/Diagram255d.8-s.pstex_t} \end{array}}
					&-	&\ensuremath{\begin{array}{c}\input{pstex/Diagram255c.1-s.pstex_t} \end{array}}							&
			\end{array}
		\end{array}
	}{\bullet}
	\]
\caption{Subtractions for diagrams~\ref{d:255b.1}--\ref{d:255c.1}.}
\label{fig:TLD:Subtractions-II-A}
\end{figure}
\end{center}

Figures~\ref{fig:TLD:Universal-IIa} and~\ref{fig:TLD:Universal-IIB} show the result of
combining both the subtractions and additions of figure~\ref{fig:TLD:Subtractions-II-A} with their
parents.
\begin{center}
\begin{figure}
	\[
	2
	\dec{
		\begin{array}{l}
			\begin{array}{cccccccc}
			\vspace{0.1in}
					&\LO{d:259b.1}		&	&\LO{d:261b.1}		&	&\LO{d:261b.3}		&	&\LO{d:272b.1}
			\\
				+	&\ensuremath{\begin{array}{c}\input{pstex/Diagram255b.1.pstex_t} \end{array}}	&+	&\ensuremath{\begin{array}{c}\input{pstex/Diagram255c.7.pstex_t} \end{array}}	&-	&\ensuremath{\begin{array}{c}\input{pstex/Diagram255c.8.pstex_t} \end{array}}	&+	&\ensuremath{\begin{array}{c}\input{pstex/Diagram255d.7.pstex_t} \end{array}}
			\end{array}
		\\
			\begin{array}{cccccccc}
			\vspace{0.1in}
					&\LO{d:259c.3}					&	&\LO{d:259c.5}		&	&\LO{d:261b.9}		&	&\LO{d:261b.11}
			\\
				+	&\ensuremath{\begin{array}{c}\input{pstex/Diagram255c.16.pstex_t} \end{array}}			&-	&\ensuremath{\begin{array}{c}\input{pstex/Diagram255c.17.pstex_t} \end{array}}&-	&\ensuremath{\begin{array}{c}\input{pstex/Diagram255d.1.pstex_t} \end{array}}	&+	&\ensuremath{\begin{array}{c}\input{pstex/Diagram255d.2.pstex_t} \end{array}}
			\end{array}
		\\
			\begin{array}{cccccc}
			\vspace{0.1in}
					&\LO{d:259c.1}		&	&\LO{d:272b.5}		&	&\LO{d:260b.3}		
			\\
				-	&\ensuremath{\begin{array}{c}\input{pstex/Diagram255b.11.pstex_t} \end{array}}&+	&\ensuremath{\begin{array}{c}\input{pstex/Diagram255d.8.pstex_t} \end{array}}	&-	&\ensuremath{\begin{array}{c}\input{pstex/Diagram255c.1.pstex_t} \end{array}}
			\end{array}
		\end{array}
	}{\bullet}_{\mbox{\normalsize $\COMP$}}
	\]
\caption{Result of the cancellation of components between diagrams~\ref{d:255b.1}--\ref{d:255c.1}
and both the subtractions and additions of figure~\ref{fig:TLD:Subtractions-II-A}---part~I.}
\label{fig:TLD:Universal-IIa}
\end{figure}
\end{center}

\begin{center}
\begin{figure}
	\[
	2
	\dec{
		\begin{array}{c}
			\begin{array}{cccccccl}
			\vspace{0.1in}
				\LOLB{d:259b.2}		&	&\LO{d:261b.2}		&	&\LO{d:261b.4}		&	&\LORB{d:272b.2}	& \hspace{-1.5em} \Process{fig:TLD:Additions-IIa}
			\\
				\ensuremath{\begin{array}{c}\input{pstex/Diagram259b.2.pstex_t} \end{array}}	&+	&\ensuremath{\begin{array}{c}\input{pstex/Diagram261b.2.pstex_t} \end{array}}	&-	&\ensuremath{\begin{array}{c}\input{pstex/Diagram261b.4.pstex_t} \end{array}}	&+	&\ensuremath{\begin{array}{c}\input{pstex/Diagram272b.2.pstex_t} \end{array}}	&
			\end{array}
		\\
			\begin{array}{cccc}
			\vspace{0.1in}
					&\CO{d:259c.4}{d:282.3}	&	&\CO{d:259c.6}{d:272b.12}
			\\
				+	&\ensuremath{\begin{array}{c}\input{pstex/Diagram259c.4.pstex_t} \end{array}}		&-	&\ensuremath{\begin{array}{c}\input{pstex/Diagram259c.6.pstex_t} \end{array}}
			\end{array}
		\\
			\begin{array}{cccc}
			\vspace{0.1in}
					&\CO{d:261b.10}{d:282.5}&	&\CO{d:261b.12}{d:283.13b}
			\\
				-	&\ensuremath{\begin{array}{c}\input{pstex/Diagram261b.10.pstex_t} \end{array}}	&+	&\ensuremath{\begin{array}{c}\input{pstex/Diagram261b.12.pstex_t} \end{array}}
			\end{array}
		\\
			\begin{array}{cccccc}
			\vspace{0.1in}
					&\PO{d:259c.2}{fig:TLD:Additions-IIa}	&	&\CO{d:272b.6}{d:283.13a}	&	&\CO{d:260b.4}{d:283.12}
			\\
				-	&\ensuremath{\begin{array}{c}\input{pstex/Diagram259c.2.pstex_t} \end{array}}						&+	&\ensuremath{\begin{array}{c}\input{pstex/Diagram272b.6.pstex_t} \end{array}}			&-	&\ensuremath{\begin{array}{c}\input{pstex/Diagram260b.4.pstex_t} \end{array}}
			\end{array}
		\end{array}
	}{\bullet}
	\]
\caption{Result of the cancellation of components between diagrams~\ref{d:255b.1}--\ref{d:255c.1}
and both the subtractions and additions of figure~\ref{fig:TLD:Subtractions-II-A}---part~II.}
\label{fig:TLD:Universal-IIB}
\end{figure}
\end{center}

\paragraph{Diagrams~\ref{d:255b.16}--\ref{d:255d.5}}

The construction of subtractions for diagrams~\ref{d:255b.16}--\ref{d:255d.5}
is straightforward, since it almost exactly mirrors the construction
of subtractions for diagrams~\ref{diag:FR36.1}--\ref{diag:FR27.6}, which we have done already. 
The primary
difference between the diagrams we treat here and those we have treated already
is the presence of a discontinuity in momentum arguments. Specifically,
diagrams~\ref{d:255b.16}--\ref{d:255d.5} each possess a bar, indicating that
we have not set $p=0$ in the denominator of the ($A$-sector) effective propagator
joined to the two-point, tree level vertex.

However, we can ignore the bar, for the purposes of constructing
subtractions. The point is that $p$ acts as an IR
regulator only if we take computable contributions. For non-computable contributions, 
we can just set $p=0$ in the effective propagator and hence remove the bar. Since it
is these contributions that our subtractions are designed to remove, we
can just construct our subtractions without any discontinuity in momentum arguments.
Note that if we were to remove the bar from diagram~\ref{d:255b-Corr}, then we could
apply the effective propagator relation. We will exploit this, shortly.

Rather than explicitly constructing the subtractions and additions, we use our experience
to jump straight to the set of terms arising from combining these diagrams with
their parents. This is shown in figure~\ref{fig:TLD:255b.16etc-U,NU}.

\begin{center}
\begin{figure}
	\[
	-2
	\dec{
		\begin{array}{c}
		\vspace{0.2in}
			\begin{array}{cccccccl}
			\vspace{0.1in}
				\LO{d:255b.16-U}			&	&\LO{d:255b-Corr-U}				&	&\LO{d:255d.4-U}			&	&\LO{d:255d.5-U} 			& 
			\\
				\Uni{\ensuremath{\begin{array}{c}\input{pstex/Diagram255b.16.pstex_t} \end{array}}}	&+	&\Uni{\ensuremath{\begin{array}{c}\input{pstex/Diagram255b-Corr.pstex_t} \end{array}}}	&+	&\Uni{\ensuremath{\begin{array}{c}\input{pstex/Diagram255d.4.pstex_t} \end{array}}}	&-	&\Uni{\ensuremath{\begin{array}{c}\input{pstex/Diagram255d.5.pstex_t} \end{array}}}	&
			\end{array}
		\\
		\vspace{0.2in}
			\begin{array}{ccccccl}
			\vspace{0.1in}
					&\LOLB{d:255b.16-NU}		&	&\LO{d:255d.4-NU}			&	&\LORB{d:255d.5-NU}		& \Process{fig:TLD:LS-rems-B}
			\\
				+	&\ensuremath{\begin{array}{c}\input{pstex/DiagramFR36.1-NU.pstex_t} \end{array}}\NUl	&+	&\ensuremath{\begin{array}{c}\input{pstex/DiagramFR22.3-NU.pstex_t} \end{array}}\NUl	&-	&\ensuremath{\begin{array}{c}\input{pstex/DiagramFR24.11-NU.pstex_t} \end{array}}	&
			\end{array}
		\\
			\begin{array}{ccccl}
			\vspace{0.1in}
					&\PO{d:255b-Corr-A-NU}{fig:TLD:LS-rems-B}	&	&\CO{d:255b-Corr-B-NU}{d:FR27.6-NU}
			\\
				+	&\ensuremath{\begin{array}{c}\input{pstex/DiagramFR27.1-NU.pstex_t} \end{array}}						&-	&\ensuremath{\begin{array}{c}\input{pstex/DiagramFR27.6-NU.pstex_t} \end{array}}
			\end{array}
		\end{array}
	}{\bullet}
	\]
\caption{Result of isolating computable and non-computable contributions to 
diagrams~\ref{d:255b.16}--\ref{d:255d.5}, achieved via an implicit intermediate
step involving the construction of subtractions.}
\label{fig:TLD:255b.16etc-U,NU}
\end{figure}
\end{center}

\Cancel{d:255b-Corr-B-NU}{d:FR27.6-NU}

\section{Further Manipulations}	\label{sec:TLD:FurtherManipulations}

In this section, we complete the demonstration that all non-computable
contributions to the original set (ignoring the $\alpha$-terms)
cancel. To do this, we organise the calculation in a specific way. First,
we recognise that by manipulating certain diagrams, we can completely remove
all non-computable contributions arising from instances of
the standard set. Secondly,
we remove all remaining non-computable contributions arising from
instances of the little set. Finally, we demonstrate that the small
number of remaining non-computable contributions to $\beta_2$ cancel,
amongst themselves.

\subsection{Removal of Non-Computable Contributions to the Standard Set}

In this section, we process a subset of the
terms of figure~\ref{fig:TLD:Universal-IIB},
which we note are under the influence of \NUn.
To make progress, we  can interpret
\NUn\ as the identity minus $\COMP$; thus, if we are happy
manipulating diagrams under $\COMP$, then we are happy 
manipulating them under \NUn.

For the diagrams we wish to manipulate, it turns
out that the manipulations do not generate diagrams
for which we would, according to our current prescription, 
have to reinstate derivatives of cutoff
functions; hence we proceed.

Diagram~\ref{d:259b.2}	 can be processed using the effective
propagator relation, since the rightmost two-point, tree level vertex
is attached to an effective propagator carrying the same momentum.
If we take the Kronecker delta arising from the effective propagator
relation, then we note that we can combine this term with
diagram~\ref{d:259c.2}.
If we take the gauge remainder part, then the \GRk\ strikes a two-point, tree level
vertex differentiated \wrt\ its momentum. We can move this derivative
from the vertex to the \GRk, by means of diagrammatic identity~\ref{D-ID:TLTP-GR}.
This then allows us to use the effective propagator relation, once more.

Both diagrams~\ref{d:261b.2} and~\ref{d:261b.4}
possess, amongst other structures, an undifferentiated \GRkp\
which attaches to a wine carrying the same momentum. Such structures
can be redrawn using diagrammatic identity~\ref{D-ID:PseudoEffectivePropagatorRelation}
to give a pseudo effective propagator ending in a \GRk. This \GRk\
strikes the three-point vertex, and this can just be processed in the
usual way.

Diagram~\ref{d:272b.2} can be processed using diagrammatic identity~\ref{D-ID:OneWine}. 
It is not obvious that we should be performing this manipulation, since one of the resulting diagrams has the 
same complexity as its parent. The reason that we choose to do this is to bring diagram~\ref{d:272b.2} into the same form as 
diagram~\ref{d:259c.2}---and, by doing so, we will find a number of cancellations. This illustrates a key tenet of the diagrammatics,
in general: if it is possible to bring the common sub-diagram of two full diagrams into the same form, it is always right to do so!

As usual, a subset of
the diagrams produced by the manipulations cancel between themselves. The remaining terms are shown in figure~\ref{fig:TLD:Additions-IIa}.

\begin{center}
\begin{figure}
	\[
	2
	\dec{
		\begin{array}{l}
			\begin{array}{cccccc}
			\vspace{0.1in}
								&\PO{d:259.4,13}{fig:TLD:SemiNonUniversal}	&	&\PO{d:272.4}{fig:TLD:SemiNonUniversal}	&	&\CO{d:282.3}{d:259c.4}	
			\\
				-2\hspace{-3em}	&\ensuremath{\begin{array}{c}\input{pstex/Diagram259c.2.pstex_t} \end{array}}						&-	&\ensuremath{\begin{array}{c}\input{pstex/Diagram272.4.pstex_t} \end{array}}							&-	&\ensuremath{\begin{array}{c}\input{pstex/Diagram259c.4.pstex_t} \end{array}}
			\end{array}
		\\
			\begin{array}{cccccc}
			\vspace{0.1in}
					&\CO{d:272b.12}{d:259c.6}	&	&\PO{d:259.8-CORR}{fig:TLD:LS-rems-C}	&	&\PO{d:272.7-CORR}{fig:TLD:LS-rems-C}
			\\
				+	&\ensuremath{\begin{array}{c}\input{pstex/Diagram259c.6.pstex_t} \end{array}}			& +	&\ensuremath{\begin{array}{c}\input{pstex/Diagram259.8-CORR.pstex_t} \end{array}}	\NUl			&+	&\ensuremath{\begin{array}{c}\input{pstex/Diagram272.7-CORR.pstex_t} \end{array}}	\NUl
			\end{array}
		\\
			\begin{array}{cccccccc}
			\vspace{0.1in}
						& \CO{d:282.5}{d:261b.10}	&	&\CO{d:283.13b}{d:261b.12}	&	&\PO{d:261.7-CORR}{fig:TLD:LS-rems-C}
			\\
					+	&\ensuremath{\begin{array}{c}\input{pstex/Diagram261b.10.pstex_t} \end{array}}		&-	&\ensuremath{\begin{array}{c}\input{pstex/Diagram261b.12.pstex_t} \end{array}}		& -	&\ensuremath{\begin{array}{c}\input{pstex/Diagram261.7-CORR.pstex_t} \end{array}}	
			\end{array}
		\\
			\begin{array}{cccccccc}
			\vspace{0.1in}
						& \PO{d:261.8-CORR}{fig:TLD:LS-rems-C}	&	&\CO{d:283.13a}{d:272b.6}	&	&\CO{d:283.12}{d:260b.4}
			\\
					+	&\ensuremath{\begin{array}{c}\input{pstex/Diagram261.8-CORR.pstex_t} \end{array}}					&-	&\ensuremath{\begin{array}{c}\input{pstex/Diagram272b.6.pstex_t} \end{array}}			&+	&\ensuremath{\begin{array}{c}\input{pstex/Diagram260b.4.pstex_t} \end{array}}
			\end{array}
		\end{array}
	}{\bullet}
	\]
\caption[The result of manipulating diagrams~\ref{d:259b.2}--\ref{d:272b.2}.]{The result of manipulating diagrams~\ref{d:259b.2}--\ref{d:272b.2}. 
The factor of two in front of
the first diagram arises from combining one of the generated terms with diagram~\ref{d:259c.2}.}
\label{fig:TLD:Additions-IIa}
\end{figure}
\end{center}

As expected, many of the diagrams cancel against un-manipulated terms from figure~\ref{fig:TLD:Universal-IIB}.
\Cancel{d:282.3}{d:259c.4}
\Cancel{d:272b.12}{d:259c.6}
\Cancel{d:282.5}{d:261b.10}
\Cancel{d:283.13b}{d:261b.12}
\Cancel{d:283.13a}{d:272b.6}
\Cancel{d:283.12}{d:260b.4}

The next step is to collect together a set of
terms, possessing non-universal sub-diagrams
which are related by gauge invariance. Up to
\Oep\ corrections, we will be able to recast
this set as a computable term plus a set of 
diagrams comprising the standard set
atop the little set. The set of diagrams is
shown in figure~\ref{fig:TLD:SemiNonUniversal}.
For all sub-diagrams possessing the tag \NUn,
we are using the prescription that we keep derivatives
of cutoff functions; hence we are free to move the
momentum derivative around this sub-diagram at will,
discarding any total momentum derivatives.
\begin{center}
\begin{figure}
	\[
	2
	\dec{
		\begin{array}{c}
			\begin{array}{cccccc}
			\vspace{0.1in}
					&\PO{d:278.2}{fig:TLD:SS+LS:SU}	&	&\PO{d:278.4}{fig:TLD:SS+LS:SU}	&	&\CO{d:284.1}{d:284.4}
			\\
				-4	&\ensuremath{\begin{array}{c}\input{pstex/Diagram278.2.pstex_t} \end{array}}				&-2	&\ensuremath{\begin{array}{c}\input{pstex/Diagram278.4.pstex_t} \end{array}}				&+2	&\ensuremath{\begin{array}{c}\input{pstex/Diagram284.1.pstex_t} \end{array}}
			\end{array}
		\\
			\begin{array}{cccccc}
			\vspace{0.1in}
					&\PO{d:273.1}{fig:TLD:SS+LS:SU}	&	&\PO{d:273.2}{fig:TLD:SS+LS:SU}	&	&\LO{d:282b.1}
			\\
				+2	&\ensuremath{\begin{array}{c}\input{pstex/Diagram273.1.pstex_t} \end{array}}				&+	&\ensuremath{\begin{array}{c}\input{pstex/Diagram273.2.pstex_t} \end{array}}				&+2	&\Uni{\ensuremath{\begin{array}{c}\input{pstex/Diagram282b.1.pstex_t} \end{array}}}
			\end{array}
		\\
			\begin{array}{cccccc}
			\vspace{0.1in}
					&\LO{d:282b.2}				&	&\PO{d:269.13}{fig:TLD:SS+LS:SU}&	&\PO{d:269.14}{fig:TLD:SS+LS:SU}
			\\
				+	&\Uni{\ensuremath{\begin{array}{c}\input{pstex/Diagram282b.2.pstex_t} \end{array}}}	&+2	&\ensuremath{\begin{array}{c}\input{pstex/Diagram269.13.pstex_t} \end{array}}				&+	&\ensuremath{\begin{array}{c}\input{pstex/Diagram269.14.pstex_t} \end{array}}
			\end{array}
		\\
			\begin{array}{cccc}
			\vspace{0.1in}
					&\PO{d:268.1:NU}{fig:TLD:Universal-III}	&	&\PO{d:268.2:NU}{fig:TLD:Universal-III}
			\\
				-2	&\ensuremath{\begin{array}{c}\input{pstex/Diagram268.1-NU.pstex_t} \end{array}}					&-	&\ensuremath{\begin{array}{c}\input{pstex/Diagram268.2-NU.pstex_t} \end{array}}
			\end{array}
		\end{array}
	}{\bullet}
	\]
\caption{The rearrangement of diagrams~\ref{d:FR26.1,2:sub1NU}, \ref{d:221d.2,4} and~\ref{d:221d.8}; the cancellation of
components between~\ref{d:225.11,14}, \ref{d:225.15} and \ref{d:259.4,13}, \ref{d:272.4}; the manipulation of diagrams~\ref{d:277c.5a} 
and~\ref{d:277c.5b}.}
\label{fig:TLD:SemiNonUniversal}
\end{figure}
\end{center}

Diagram~\ref{d:278.2} is (up to a discarded total momentum derivative) simply diagram~\ref{d:FR26.1,2:sub1NU}, 
which we have moved here to make the subsequent steps clearer.
The next two diagrams are obtained by manipulating diagrams~\ref{d:277c.5a} and~\ref{d:277c.5b}.

\CancelCom{d:284.1}{d:284.4}{, though this is not immediately obvious. First, redraw the topmost
sub-diagram of diagram~\ref{d:284.4} using diagrammatic identity~\ref{D-ID-Deriv-GR} and use \CC\
to reflect the bottom sub-diagram. This yields
the the first diagram shown in figure~\ref{fig:TLD:CancellationIllustration}.
\begin{center}
\begin{figure}[h]
	\[
	\begin{array}{cccccc}
		4&\ensuremath{\begin{array}{c}\input{pstex/Diagram284.1-B.pstex_t} \end{array}} \NUl &\equiv -4 & \ensuremath{\begin{array}{c}\input{pstex/Diagram284.1-C.pstex_t} \end{array}}\NUl & \equiv -4 &\ensuremath{\begin{array}{c}\input{pstex/Diagram284.1-D.pstex_t} \end{array}}
	\end{array}
	\]
\caption{Showing how we can redraw diagram~\ref{d:284.4}.}
\label{fig:TLD:CancellationIllustration}
\end{figure}
\end{center}

In turn, we redraw this once using diagrammatic identity~\ref{D-ID:GaugeRemainderRelation} and
then again using diagrammatic identity~\ref{D-ID:PseudoEffectivePropagatorRelation}. Finally, the cancellation
works courtesy of diagrammatic identity~\ref{D-ID-GR-W-dGR}.
}

Diagrams~\ref{d:273.1}--\ref{d:282b.2} come from combining diagrams~\ref{d:225.11,14} and \ref{d:225.15} with diagrams~\ref{d:259.4,13}
and~\ref{d:272.4}, where we have discarded terms which vanish in
the \eptoz\ limit.

Diagrams~\ref{d:269.13}--\ref{d:268.2:NU} come from re-expressing diagrams~\ref{d:221d.2,4} and~\ref{d:221d.8}. Although it may not look like it,
diagram~\ref{d:269.13} (\ref{d:269.14}) is actually the same as diagram~\ref{d:268.1:NU} (\ref{d:268.2:NU}). This
follows because Lorentz invariance demands that the little set goes as $\delta_{\alpha \beta}$, allowing us to redraw the diagrams in the
manner shown. Summing diagram~\ref{d:269.13} (\ref{d:268.1:NU}) and~\ref{d:268.1:NU} (\ref{d:268.2:NU}) simply sums $\NUn + \COMP$ and so 
reproduces the parent diagram.

Having described the origin of the diagrams in figure~\ref{fig:TLD:SemiNonUniversal} we move on to why  they have been collected in
this manner. Start by focusing on the first column of diagrams. It is apparent that the \NUn\ part is common between all three. 
Furthermore, if we are to take the \NUn\ part as shown, then for the contribution not to vanish in the $\epsilon \rightarrow 0$ limit,
we must take the divergent part of the bottom sub-diagram. However, when we sum over all three diagrams, the divergent part
of the bottom sub-diagrams is none other than the divergent part of the standard set! 

Since the standard set is transverse, and the little set goes as $\delta_{\alpha \beta}$, we can 
redraw this set of diagrams to take form shown in figure~\ref{fig:TLD:SS+LS:SU}.
At \Oep, we must take the divergent---and hence universal---part of the standard set.
\begin{center}
\begin{figure}[h]
\[
	\dec{
		\begin{array}{c}
		\vspace{0.2in}
			\left[
				\begin{array}{cccc}
					-2	& \ensuremath{\begin{array}{c}\input{pstex/LittleSet-A-be.pstex_t} \end{array}}	&-	& \ensuremath{\begin{array}{c}\input{pstex/LittleSet-B-be.pstex_t} \end{array}}
				\end{array}
			\right]_{\NUn}
		\\
			\times	\Delta^{1 \,1}_{\alpha \beta}(p)
			\left[
				\begin{array}{cccccl}
				\vspace{0.1in}
					\LOLB{d:278.10,11:NU}	&	&\LO{d:278.12,13:NU}	&	&\LORB{d:278.14:NU,15}	&\Process{fig:TLD:SS-rems-A}
				\\
					\ensuremath{\begin{array}{c}\input{pstex/Beta1-A-al.pstex_t} \end{array}}			& -	&\ensuremath{\begin{array}{c}\input{pstex/Beta1-B-al.pstex_t} \end{array}} 		& +4& \ensuremath{\begin{array}{c}\input{pstex/Beta1-C-al.pstex_t} \end{array}}		&
				\end{array}
			\right]
		\end{array}
	}{\bullet}
\]
\caption{A redrawing of the first two columns of 
figure~\ref{fig:TLD:SemiNonUniversal} (up to $\Oep$ corrections). }
\label{fig:TLD:SS+LS:SU}
\end{figure}
\end{center}

We conclude this section by completing the removal of
all non-computable contributions which come from the standard set.
In figure~\ref{fig:TLD:Diagrams269B}, we combine 
diagrams~\ref{d:269.1}--\ref{d:269.2b}
with diagrams~\ref{d:248.2,5:NU}--\ref{d:248.3b,6b:NU}.

\begin{center}
\begin{figure}
\[
	-
	\dec{
		\left[
			\begin{array}{ccccc}
			\vspace{0.1in}
				\LO{d:248.2,5:U}	&	&\LO{d:248.3a,6a:U}	&	&\LO{d:248.3b,6b-U}
			\\
				\ensuremath{\begin{array}{c}\input{pstex/Diagram248.8.pstex_t} \end{array}}	&-2	&\ensuremath{\begin{array}{c}\input{pstex/Diagram248.9a.pstex_t} \end{array}}	&-2	&\ensuremath{\begin{array}{c}\input{pstex/Diagram248.9b.pstex_t} \end{array}}
			\end{array}
		\right]_{\OepPow{-2}}
	}{\bullet}
\]
\caption{Result of combining diagrams~\ref{d:269.1}--\ref{d:269.2b}
with diagrams~\ref{d:248.2,5:NU}--\ref{d:248.3b,6b:NU}.}
\label{fig:TLD:Diagrams269B}
\end{figure}
\end{center}

\subsection{Removal of Non-Computable Contributions to the Little Set}

Diagrams~\ref{d:278.10,11:NU}--\ref{d:278.14:NU,15}
cancel components of diagrams~\ref{diag:223b-E1}--\ref{diag:223b-E3},
to leave behind a computable contribution. To be specific,
let us focus on the sub-diagrams multiplying the standard set,
for diagrams~\ref{diag:223b-E1}--\ref{diag:223b-E3}. The non-computable
parts of these diagrams are completely cancelled, to leave behind
the components shown in figure~\ref{fig:TLD:SS-rems-A}.
\begin{center}
\begin{figure}[h]
\[
	\dec{
		\begin{array}{l}
		\vspace{0.2in}
			\left[
				\begin{array}{ccccccc}
					\ensuremath{\begin{array}{c}\input{pstex/Diagram180.7-be.pstex_t} \end{array}}	&+	&\ensuremath{\begin{array}{c}\input{pstex/Diagram180.9-be.pstex_t} \end{array}}	&-2	&\ensuremath{\begin{array}{c}\input{pstex/Diagram326.3-be.pstex_t} \end{array}}	&+2	&\ensuremath{\begin{array}{c}\input{pstex/Diagram326.4-be.pstex_t} \end{array}}					
				\end{array}
			\right]^{\mathcal{O}\left(\epsilon^0 \PoLep\right)}_{\COMP}	
		\\
			\times \Delta^{1 \,1}_{\alpha \beta}(p)
			\left[
				\begin{array}{ccccc}
					\vspace{0.1in}
					\LO{d:SS-rem-A1}	&	&\LO{d:SS-rem-A2}	&	&\LO{d:SS-rem-A3}
				\\		
					\ensuremath{\begin{array}{c}\input{pstex/Beta1-A-al.pstex_t} \end{array}}		& -	&\ensuremath{\begin{array}{c}\input{pstex/Beta1-B-al.pstex_t} \end{array}} 	& +4& \ensuremath{\begin{array}{c}\input{pstex/Beta1-C-al.pstex_t} \end{array}}
				\end{array}
			\right]_{\COMP}
		\end{array}
	}{\bullet}
\]
\caption{The result of combining diagrams~\ref{d:278.10,11:NU}--\ref{d:278.14:NU,15} with
diagrams~\ref{diag:223b-E1}--\ref{diag:223b-E3}.}
\label{fig:TLD:SS-rems-A}
\end{figure}
\end{center}

Note that the final two sub-diagrams on the first row of figure~\ref{fig:TLD:SS-rems-A}
can be discarded,
as they do not possess an $\Op{-2 \epsilon}$ component

To complete the removal of the non-computable contributions to
the little set, there are a number of intermediate steps we must
perform. First, we can simplify the remaining terms by
recognising that diagrams~\ref{d:255b.16-NU}--\ref{d:255b-Corr-A-NU}
exactly halve the overall factor of 
diagrams~\ref{d:FR36.1-NU}--\ref{d:FR27.1-NU}.
The resulting partial cancellations yield the diagrams of figure~\ref{fig:TLD:LS-rems-B}.
\begin{center}
\begin{figure}
\[
	2
	\dec{
		\begin{array}{l}
			\begin{array}{ccc}
			\vspace{0.1in}
				\PO{d:FR36.1-NU-B}{fig:TLD:LS-rems-B-supp}	&	&\CO{d:FR22.3-NU-B}{d:261.7-CORR-rd-B}	
			\\
				\ensuremath{\begin{array}{c}\input{pstex/DiagramFR36.1-NU.pstex_t} \end{array}}\NUl					&+	&\ensuremath{\begin{array}{c}\input{pstex/DiagramFR22.3-NU.pstex_t} \end{array}}\NUl				
			\end{array}
		\\
			\begin{array}{cccc}
			\vspace{0.1in}
					&\CO{d:FR24.11-NU-B}{d:261.8-CORR-rd-B}		& 	&\PO{d:FR27.1-NU-B}{fig:TLD:LittleSetSq:NU-I}
			\\
				-	&\ensuremath{\begin{array}{c}\input{pstex/DiagramFR24.11-NU.pstex_t} \end{array}}\NUl					&+	&\ensuremath{\begin{array}{c}\input{pstex/DiagramFR27.1-NU.pstex_t} \end{array}}	\NUl
			\end{array}
		\end{array}
	}{\bullet}
\]
\caption{The result of combining diagrams~\ref{d:FR36.1-NU}--\ref{d:FR27.1-NU} with
diagrams~\ref{d:255b.16-NU}--\ref{d:255b-Corr-A-NU}.}
\label{fig:TLD:LS-rems-B}
\end{figure}
\end{center}

Immediately, we will process diagram~\ref{d:FR36.1-NU-B}. The simplest way to do this
is to use diagrammatic identity~\ref{D-ID:OneWine} to redraw the differentiated two-point,
tree level vertex and the effective propagator to which it attaches on the \emph{right}.
The term in which the momentum derivative is moved from the vertex to the effective propagator
just dies: the \GRkp\ now acts backwards along the other effective propagator to kill the
vertex. Hence, we are left only with the term in which the momentum derivative strikes
a (trapped) gauge remainder. We draw the diagrams in which this derivative strikes \GRk\
and \GRkp\ separately, in figure~\ref{fig:TLD:LS-rems-B-supp}.

Note that these manipulations are valid under \NUn, without the need to reinstate derivatives
of cutoff functions.
\begin{center}
\begin{figure}[h]
\[
	-2
	\dec{
		\begin{array}{ccc}
		\vspace{0.1in}
			\CO{d:259.8-CORR-rd-B}{d:259.8-CORR-rd}			&	& \PO{d:259.8-CORR-rd-B2}{fig:TLD:LittleSetSq:NU-I}
		\\
			\ensuremath{\begin{array}{c}\input{pstex/Diagram259.8-CORR-rd-B.pstex_t} \end{array}}\NUl					&+	& \ensuremath{\begin{array}{c}\input{pstex/Diagram259.8-CORR-rd-B2.pstex_t} \end{array}}
		\end{array}
	}{\bullet}
\]
\caption{Result of manipulating diagram~\ref{d:FR36.1-NU-B}.}
\label{fig:TLD:LS-rems-B-supp}
\end{figure}
\end{center}

The next step is to recognise that
there are four diagrams whose momentum discontinuity can be removed. 
First, look at diagram~\ref{d:259.8-CORR}.
Having removed the momentum discontinuity, we will recast this diagram. In the sub-diagram
under the influence of \NUn, we move the momentum derivative which strikes \GRk\ to the \GRkp\ with
which it is contracted. (See diagrammatic identity~\ref{D-ID:GaugeRemainderRelation}.) Next, we use diagrammatic
identity~\ref{D-ID:PseudoEffectivePropagatorRelation}.

The second diagram we treat is~\ref{d:272.7-CORR} which, having removed the momentum discontinuity,
we leave as is. The final two diagrams, \ref{d:261.7-CORR} and~\ref{d:261.8-CORR} are related,
since the sub-diagram under the influence of \NUn\ is common to both. Removing the momentum discontinuity
allows us to apply the effective propagator relation. The resulting Kronecker delta term is
redrawn using diagrammatic identity~\ref{D-ID-Deriv-GR}; the gauge remainder term is redrawn
using diagrammatic identities~\ref{D-ID-GR-W-dGR} and~\ref{D-ID:PseudoEffectivePropagatorRelation}.
Figure~\ref{fig:TLD:LS-rems-C} shows the result of these manipulations.
\begin{center}
\begin{figure}[h]
\[
	-2
	\dec{
		\begin{array}{l}
		\vspace{0.2in}
			\begin{array}{ccc}
			\vspace{0.1in}
				\CO{d:259.8-CORR-rd}{d:259.8-CORR-rd-B}	&	& \PO{d:272.7-CORR-rd}{fig:TLD:LittleSetSq:NU-I}
			\\
				\ensuremath{\begin{array}{c}\input{pstex/Diagram259.8-CORR-rd.pstex_t} \end{array}}\NUl			&-	&\ensuremath{\begin{array}{c}\input{pstex/Diagram272.7-CORR-rd.pstex_t} \end{array}}
			\end{array}
		\\
		\vspace{0.2in}
			\begin{array}{cccc}
			\vspace{0.1in}
					&\PO{d:261.7-CORR-rd-A}{fig:TLD:LittleSetSq:NU-I}	&	&\CO{d:261.7-CORR-rd-B}{d:FR22.3-NU-B}		
			\\
				+	&\ensuremath{\begin{array}{c}\input{pstex/Diagram261.7-CORR-rd-A.pstex_t} \end{array}}						&+	&\ensuremath{\begin{array}{c}\input{pstex/DiagramFR22.3-NU.pstex_t} \end{array}}\NUl				
			\end{array}
		\\
			\begin{array}{cccc}
			\vspace{0.1in}
					&\PO{d:261.8-CORR-rd-A}{fig:TLD:LittleSetSq:NU-I}		&	&\CO{d:261.8-CORR-rd-B}{d:FR24.11-NU-B}
			\\
				-	&\ensuremath{\begin{array}{c}\input{pstex/Diagram261.8-CORR-rd-A.pstex_t} \end{array}}							&-	&\ensuremath{\begin{array}{c}\input{pstex/DiagramFR24.11-NU.pstex_t} \end{array}}\NUl
			\end{array}
		\end{array}
	}{\bullet}
\]
\caption{Result of redrawing diagrams~\ref{d:259.8-CORR}, \ref{d:272.7-CORR}, \ref{d:261.7-CORR} and~\ref{d:261.8-CORR}.}
\label{fig:TLD:LS-rems-C}
\end{figure}
\end{center}

\CancelCom{d:259.8-CORR-rd}{d:259.8-CORR-rd-B}{courtesy of diagrammatic identity~\ref{D-ID-GR-W-dGR}.}
\Cancel{d:261.7-CORR-rd-B}{d:FR22.3-NU-B}
\Cancel{d:261.8-CORR-rd-B}{d:FR24.11-NU-B}

We are now ready to perform the penultimate step: we collect together all diagrams which possess
an element of the little set, under the influence of \NUn. There are five such diagrams
coming from this section: \ref{d:FR27.1-NU-B}, \ref{d:259.8-CORR-rd-B2}, \ref{d:272.7-CORR-rd}, \ref{d:261.7-CORR-rd-A} 
and~\ref{d:261.8-CORR-rd-A}. In addition, there are two more coming from earlier in the calculation:
diagrams~\ref{d:274b.5} and~\ref{d:274b.6}. 

We can redraw these latter diagrams.
If we take the indices of the sub-diagram under the influence of \NUn\ to be $\alpha$ and $\beta$,
then
Lorentz invariance forces this sub-diagram to go like $\delta_{\alpha \beta}$. This then contracts
together a \GRk\ with a \GRkp\, giving unity. Thus, these diagrams look like the little
set, under the influence of \NUn, joined to a set of sub-diagrams best described as a `stretched out'
version of the little set; these diagrams are actually just elements of the little
set restricted to the $A^1$-sector. 
Noting that we must take the divergent part of
these stretched out sub-diagram, we can just redraw them to take the form of the
little set, up to \Oep\ corrections.

We now combine the redrawn versions of diagrams~\ref{d:274b.5} and~\ref{d:274b.6}
with the five diagrams coming from this section. We always take momentum derivatives
to strike (pseudo) effective propagators, rather than gauge remainders. 
This means that will need to include derivative of cutoff functions, as appropriate.
The resulting diagrams are shown in figure~\ref{fig:TLD:LittleSetSq:NU-I}.
\begin{center}
\begin{figure}
\[
	-2
	\dec{
		\begin{array}{ccccccccl}
		\vspace{0.1in}
				&\LOLB{d:274b.7}	&	&\LO{d:274b.8B}		&	&\LO{d:279.6}		&	&\LORB{d:279.7} 	& \Process{fig:TLD:LittleSetSq:U}
		\\
			4	&\ensuremath{\begin{array}{c}\input{pstex/Diagram274b.7.pstex_t} \end{array}}	&+2	&\ensuremath{\begin{array}{c}\input{pstex/Diagram274b.8B.pstex_t} \end{array}}&+2	&\ensuremath{\begin{array}{c}\input{pstex/Diagram279.6.pstex_t} \end{array}}	&+	&\ensuremath{\begin{array}{c}\input{pstex/Diagram279.7.pstex_t} \end{array}} 	&
		\end{array}
	}{\bullet}
\]
\caption{Combining the remaining terms which possess an element of the little set explicitly
under the influence of \NUn.}
\label{fig:TLD:LittleSetSq:NU-I}
\end{figure}
\end{center}

Diagrams~\ref{d:274b.7}--\ref{d:279.7} are precisely those we need to remove the non-computable
components from diagrams~\ref{LSSq-A}--\ref{LSSq-C}	(up to \Oep\ corrections).
The remaining computable contributions are shown in figure~\ref{fig:TLD:LittleSetSq:U}.
\begin{center}
\begin{figure}
\[
	\left[
		\begin{array}{cccccc}
		\vspace{0.1in}
				&\LO{d:279.8}		&	&\LO{d:279.9.10}	&	&\LO{d:279.11}
		\\
			4	&\ensuremath{\begin{array}{c}\input{pstex/LittleSetSq-A.pstex_t} \end{array}}	&+4	&\ensuremath{\begin{array}{c}\input{pstex/LittleSetSq-B.pstex_t} \end{array}}	& +	&\ensuremath{\begin{array}{c}\input{pstex/LittleSetSq-C.pstex_t} \end{array}}
		\end{array}
	\right]^\bullet_\COMP
\]
\caption{Cancellation of non-computable contributions between diagrams~\ref{d:274b.7}--\ref{d:279.7}
and~\ref{LSSq-A}--\ref{LSSq-C}.}
\label{fig:TLD:LittleSetSq:U}
\end{figure}
\end{center}

\subsection{Final Remainders}	\label{sec:TLD:FinalRemainders}

Diagrams~\ref{d:268.1:NU} and~\ref{d:268.2:NU} combine with diagrams~\ref{d:269.11b} and~\ref{d:269.12b}
to give the diagrams of figure~\ref{fig:TLD:Universal-III}
\begin{center}
\begin{figure}
\[
	2
	\dec{
		\begin{array}{cccc}
		\vspace{0.1in}
				&\LO{d:269.11}				&	&\LO{d:269.12}
		\\		
			2	&\ensuremath{\begin{array}{c}\input{pstex/Diagram269.11.pstex_t} \end{array}}\etozl	&-	&\ensuremath{\begin{array}{c}\input{pstex/Diagram269.12.pstex_t} \end{array}} \etozl
		\end{array}
	}{\bullet}
\]
\caption{Showing diagrams~\ref{d:268.1:NU} and~\ref{d:268.2:NU} 
combine with diagrams~\ref{d:269.11b} and~\ref{d:269.12b}.}
\label{fig:TLD:Universal-III}
\end{figure}
\end{center}

This completes the demonstration that, up to the $\alpha$-terms,
all non-computable contributions to the original set cancel.
The surviving diagrams are
either explicitly under the influence of $\COMP$ or contain
a factorisable sub-diagram from which we take the \Oepz,
computable coefficient. In the latter case
we 
are forced to take the (computable) divergent part of the 
other sub-diagram, as all other contributions die in the \eptoz\
limit.

Although we are left with a large number of computable
contributions to $\beta_2$, it is possible to simplify
our expression. For example, we can manipulate diagrams
such as~\ref{d:42.1sub1U}. This would reproduce
many of the cancellations that we saw when we manipulated
the addition, corresponding to this diagram. However, not all
of the cancellations would go through. Specifically, those which involved
throwing away total momentum derivative contributions will survive,
since we are working under the influence of $\COMP$ (see section~\ref{sec:TLD:Subtractions:Gauge}). Consequently, we
choose not to perform these manipulations, for the time being. However,
we return to this issue in the conclusions of this chapter and envisage a thorough treatment
in~\cite{FurtherWork}.

\section{The $\alpha$-terms}	\label{sec:TLD:alpha}

\subsection{The Problem}

Before computing the numerical value of $\beta_2$, 
we must analyse the $\alpha$-terms. Our aim is to show that,
as anticipated in section~\ref{sec:NFE:TheNeed}, this set
of terms vanishes in the $\altoz$ limit.

Our starting point is the simplified set of $\alpha$-terms,
collected together in figure~\ref{fig:bulk:alpha-terms}.
As recognised already, we can write the 
contribution to $\beta_2 \Box_{\mu \nu}(p)$ coming from the
$\alpha$-terms
in the following form:
\begin{equation}
	\frac{1}{8} \gamma_1 \pder{\OLDs}{\alpha},
\label{eq:SimplifiedAlphaTerms}
\end{equation}
where, of course,
\begin{equation}
	4 \beta_1 \Box_{\mu \nu}(p) = -\frac{1}{2}  \dec{\OLDs}{\bullet}.
\label{eq:beta_1-D_1}
\end{equation}

Since we know from equation~(\ref{eq:gamma_1}) that $\gamma_1 \sim \alpha$,
for small $\alpha$, it is clear that the $\alpha$-terms will not vanish
as required, if $\partial\OLDs / \partial\alpha  \sim 1/\alpha$, or
worse.
The task, then, is to analyse the $\alpha$-dependence of $\OLDs$.

Whilst we have an explicit algebraic form for
three of the diagrams  (the final member of
the standard set and the two members of the little set), the rest
contain either three-point or four-point vertices, for which we
do not have an explicit an algebraic realisation. 
It is not our aim to choose
specific forms for these structures (\ie\ through
a choice of seed action etc.); rather we wish to focus on
general properties that will tell us all we need to know
about the $\alpha$-dependence.

We note that the issue is complicated by the fact that in
deducing the $\alpha$-dependence of $\OLDs$, we must
perform a loop integral. This integral must be done
before we take the $\altoz$ limit, as the two procedures do
not commute. However, whilst we cannot set $\altoz$ too soon,
we can work at small $\alpha$, and we will do so henceforth.

From our work in chapter~\ref{ch:LambdaDerivatives:Methodology},
we know how to parameterise the $\alpha$-dependence of
$\OLDs$. For our purposes, this is most easily done in $D=4$:
\begin{equation}
	\OLDs = \left[ 
				4 \beta_1 \ln 
				\left(
					\frac{\left(\mbox{IR scale}\right)^2}{\Lambda^2} 
				\right) + H(\alpha) 
			\right] \Box_{\mu \nu}(p).
\label{eq:OLDs-charateristic}
\end{equation}

The non-universal function $H(\alpha)$ is independent of $\Lambda$.
We now choose to recast this equation. When constructing the (dimensionless)
argument of the logarithm, we divide the IR scale by
the only other scale available, $\Lambda$.

Let us examine this in the context of actually performing the loop integrals
to obtain~(\ref{eq:OLDs-charateristic}). The appearance of the IR scale has been discussed,
in depth, in chapter~\ref{ch:LambdaDerivatives:Methodology}. The scale $\Lambda$ has
a natural interpretation as the scale at which the loop integrals are effectively cutoff.
However, we should not preclude the possibility that the loop integrals are actually cutoff
at some scale $h(\alpha) \Lambda$, where $h(\alpha)$ is a dimensionless function, independent of
$\Lambda$. Of course, this has no effect on the value of $\beta_1$ obtained by differentiating
$\OLDs$ \wrt\ \LConstAl. With this in mind, we rewrite equation~(\ref{eq:OLDs-charateristic})
as follows:
\begin{equation}
	\OLDs = \left[ 
				4 \beta_1 \ln 
				\left(
					\frac{\left(\mbox{IR scale}\right)^2}{\Lambda^2 h(\alpha)} 
				\right) + \tilde{H}(\alpha) 
			\right] \Box_{\mu \nu}(p).
\label{eq:OLDs-charateristic-B}
\end{equation}

This recasting now allows us to break the problem of the $\alpha$-terms into two
parts. On the one hand, we have potential $\alpha$-dependence coming
from any non-trivial $\alpha$-dependence of the effective cutoff scale, parameterised by
$h(\alpha)$.
On the other hand, we have $\alpha$-dependence
coming from the region of the loop integral with support, parameterised by $\tilde{H}(\alpha)$.
We deal with these cases in turn.

\subsection{Behaviour of $h(\alpha)$}

The treatment of this problem is slightly easier than one might expect.
The crucial point is that the logarithm term in equation~(\ref{eq:OLDs-charateristic-B})
comes only from (UV regulated) terms with non-trivial IR behaviour.
There are four diagrams with non-trivial IR behaviour: the final two elements
of the standard set and both elements of the little set. Our strategy is
to examine these diagrams and, through a choice of cutoff functions,
ensure that the momentum integrals are cutoff at $\Lambda$; equivalently
that $h(\alpha)$ is independent of $\alpha$.

Sufficient UV regularisation can be provided by cutoff
regularisation alone---\eg\ for the little set---or entirely by
the regulating sector---\eg\ for the final element of the standard set.
In the latter case, we are interested simply in the scale at which 
the $B$-sector diagrams regulate the $A^1$-sector diagrams. 
In the former case we are interested
not only in this, but also the scale at which the $A^1$ sector 
diagrams die off on their own.

For all that follows, we assume that the momentum of
the cutoff functions
$c(k^2/\Lambda^2)$ and $\tilde{c}(k^2/\Lambda^2)$ crosses over from large to small
for $x \equiv k^2 / \Lambda^2 \sim \order{1}$. This amounts to an implicit choice of
the non-universal details of the set-up. In this section,
we will demonstrate that, given this choice, we can
consistently arrange for all momentum integrals to 
be cutoff at $x \sim \order{1}$.

We start by looking at the
final element of the standard set, 
which we recall
has the algebraic form
\begin{equation}
	4N \int_k \left(\frac{(k-p)_\nu}{(k-p)^2} \frac{k_\mu}{k^2} - \frac{f_{k-p} (k-p)_\nu}{\Lambda^2} \frac{f_k k_\mu}{\Lambda^2}  \right),
\label{eq:SS-C-Alg}
\end{equation}
where
\begin{equation}
	f_k = \f{x}.
\label{eq:f_k-B}
\end{equation}
At large $x$, where $f_k = 1/x$ (see equation~(\ref{eq:CutoffConstraints})), 
we recover unbroken $SU(N|N)$, as we must,
for the theory to be regularised. 
Given our assumption
that $\tilde{c}_x$ crosses over at $x \sim \order{1}$,
we need
only ensure that the denominator of $f_k$ 
crosses over at $x \sim \order{1}$.
The crossover of the denominator
occurs at 
\[
	x \tilde{c}_x \sim 4 \alpha c_x.
\]
Now, since we are working at small $\alpha$ (and $x \tilde{c}_x \gg c_x$ for large
$x$), the crossover must happen for small values of $x$. Taylor expanding,
we therefore find that the crossover occurs at
\[
	x \tilde{c}_0 \sim \alpha c_0,
\]
(in the limit of small $\alpha$). Now, we know that $c_0$ is
fixed to be unity; however, there is no such constraint on $\tilde{c}_0$.
In turn, this implies that the momentum integral for
the third element of the standard set is cutoff 
at $x \sim \mathcal{O}(\alpha/ \tilde{c}_0)$. Note that if
we set $\tilde{c}_0 \sim 1$, then we would indeed find the
problem that $\partial \OLDs / \partial \alpha \sim 1/ \alpha$. 
Demanding that the cutoff
scale occurs at $x \sim \Oone$ then forces us to choose
\begin{equation}
	\tilde{c}_0 \sim \Oal,
\label{eq:c-tilde(0)}
\end{equation}
(which is perfectly compatible with $\tilde{c}_x$ crossing 
over at $x \sim \order{1}$).

Let us now turn to the remaining three diagrams with non-trivial
IR behaviour. The treatment of these is somewhat different
from what we have just done, as 
we have regularisation provided not only by the $B$-sector, but
also by cutoff function regularisation. The crossover scale 
in the $B$-sector follows trivially from the 
observation that
\[
	 \DB{k} = \frac{\alpha c_x f_x}{(1+\alpha) \Lambda^2}.
\]
Hence, we immediately know that, given the choice $\tilde{c}_0 \sim \Oal$, 
the crossover occurs at $x \sim \order{1}$,
in this sector.

However, we now need to show that the scale at which the
cutoff regularisation kicks in in the $A^1$-sector also 
occurs at this scale. If a diagram is sufficiently
regulated by cutoff regularisation alone, then the $B$-sector
becomes effectively redundant. If the $B$-sector is required,
in addition to cutoff regularisation, then there can be two scales
in the problem: the first is where the momentum in the $A^1$-sector
can be considered large and the second is where the momentum in the $B$-sector
can be considered large. The $A^1$ and $B$-sectors cancel
each other at the highest of these scales.

Turning now to the $A^1$-sector, we recall that
\[
	 \DAone{k} = \EPAone{k}{x},
\]
which goes as $\alpha c_x/ x$ for large $x$ and as $1/2x$ for small $x$.
The crossover occurs, for small $\alpha$, at
\[
	\alpha(c_x + 1) \sim 1-c_x.
\]
(Note that the \lhs\ dominates at sufficiently small $x$, whereas the \rhs\ dominates
at large $x$.) Again, due to the smallness of $\alpha$, we can Taylor expand in $x$
to find the crossover point, which 
occurs at
\[
	\frac{x \mod{c'_0}}{\alpha} \sim \Oone.
\]
This implies that, for the effective cutoff to be at $x \sim \Oone$,
we must choose $\mod{c'_0} \sim \Oal$.

This actually completes the analysis necessary to show that
$h(\alpha)$ can always be arranged to be independent
of $\alpha$. However, for the purposes of the next section,
it is useful to show that we can, in fact, ensure
that all momentum integrals are cutoff at $x \sim \order{1}$.
The reason that this is useful is because we expect
a one dimensional integral with an integrand of $\Oone$ but 
with support only over a range $\Delta x$ to go like $\Delta x$.
Hence, it is desirable for this range to be $\Oone$ as opposed to,
\eg, $\OalPow{-1}$.

In the $A^2$ sector, recalling that
\[
	\Delta^{22}(k) = \frac{1}{k^2} \Atwo{x},
\]
we see that the large momentum behaviour is $\alpha c_x / x$,
whereas the small momentum behaviour is $\alpha / 2x$. Since
the $\alpha$-dependence is the same for both, the crossover
scale is clearly set by $c$, which is assumed to 
crossover at $x \sim \order{1}$.

In the $D$ sector, given that
\[
	\DD{k} = \frac{\tilde{c}_x f_x}{\Lambda^4},
\]
our choice that $\tilde{c}_0 \sim \Oal$
ensures that the crossover occurs at
$x \sim \order{1}$, assuming that $\tilde{c}_x$
crosses over at $x \sim \Oone$.

In the $C$-sector, we recall that
\[
	\DCs{k} = \frac{1}{\Lambda^4} \frac{\tilde{c}_x}{x + 2 \lambda \tilde{c}_x},
\]
from which it is clear that the crossover scale
is controlled by $\tilde{c}$ and $\lambda$; we
use this freedom to ensure that the crossover occurs at
$x \sim \order{1}$.

We have thus demonstrated that, by suitable choices of
the behaviours of our cutoff functions, we can guarantee that
all momentum integrals are cutoff at the scale $x \sim \order{1}$
(working at small $\alpha$); one consequence of this is that
$h(\alpha)$ is independent of $\alpha$ and so does not
generate a contribution to $\beta_2$ in the $\alpha \rightarrow 0$
limit.

We conclude this section with an interesting comment on universality.
It is clear from our analysis so far that the freedom to choose the non-universal
parts of our cutoff functions enables us to choose
$h(\alpha)$. Returning to equation~(\ref{eq:OLDs-charateristic-B}), it thus looks
like we could generate a universal contribution to $\beta_2$ by choosing \eg\
$h(\alpha) = \alpha^m$, for some $m \neq 0$. However, the universal appearance 
of this contribution is accidental,
as can be appreciated from the fact that it arises from a particular choice of
a non-universal function. Indeed, the universal $\beta_2$ will only
be obtained by arranging things so that all contributions from the
running of $\alpha$ can be removed in the $\alpha \rightarrow 0$
limit.

\subsection{Behaviour of $\tilde{H}(\alpha)$}

To start our analysis of $\tilde{H}(\alpha)$,
we begin by returning to the third element
of the standard set. We know that the integrand
of equation~(\ref{eq:SS-C-Alg})
effectively has support only over the region
$0 \leq x < \Oone$. Moreover, any non-trivial 
$\alpha$-dependence of
$\tilde{H}(\alpha)$ must come from the $B$-sector,
as $A$-sector (processed) gauge remainders are 
independent of $\alpha$.

From our algebraic choice for $f$ (see equation~(\ref{eq:f_k-B})),
the most obvious  possible source of problematic $\alpha$-behaviour comes when
$x$ is small. However, this is ameliorated by our 
previous choice of $\tilde{c}_0$.
To complete our analysis of this diagram, we must now perform the loop integral
but we already know that, 
since the effective cutoff of this integral in $\Oone$ as opposed to,
say, $\OalPow{-1}$, we do not expect the loop integral to 
generate any bad $\alpha$-dependence (\ie\ dependence which diverges as $\alpha \rightarrow 0$).

Looking now at the little set, the situation is similar:
in the $B$-sector, our choice of $\tilde{c}_0$ cures
any bad $\alpha$-dependence; in the $A^1$-sector there
is not even a potential problem. In both cases, the effective cutoff
for the loop integral is $\Oone$.

This exhausts the analysis of the diagrams for which 
we have an explicit algebraic form and so now we turn to 
the diagrams possessing three and four-point
vertices. 

First, we will look at the diagrams with an $A^1_\mu A^1_\nu C$
vertex. Neither this vertex nor the effective propagator to
which it attaches carries the loop momentum of the diagram
and so we analyse them separately. The effective propagator---which
carries zero momentum---goes as
\[
	\frac{1}{2 \lambda \Lambda^4}.
\]
We see that the $\alpha$-dependence of this can always be
controlled by a suitable choice of $\lambda$. (It can always be
ensured that this choice of $\lambda$ is compatible with the choice
which guarantees that the crossover for the $C$-sector effective propagator
occurs at $x \sim \order{1}$.)

What about the $A^1_\mu A^1_\nu C$ vertex? We have encountered this already in
section~\ref{sec:TLD:Couplings}: at $\Op{2}$ it is a dimensionless coupling
and so we tune its flow to zero. This means that the flow equations tell us
nothing about its $\alpha$-dependence: this dependence is a boundary condition. 
The solution is simply to choose the boundary condition to have sufficiently
good $\alpha$-dependence, which is something we are always at liberty to do.

Now let us examine the remaining part of these diagrams. Attached to the other end
of the $C$-sector effective propagator is either a hook or a three-point
vertex, decorated by a simple loop. Since the hook simply goes
as
\[
	N \int_k g_k,
\]
where $g_x= (1-xf)/2$, it is clear that the loop integral does not produce
any troublesome $\alpha$-dependence.

The case where the top part of the diagram constitutes a three-point vertex
is almost as easy to treat. Let us consider the flow of this vertex. If we
take the dimensionless part, then we know that the flow has already been tuned
to zero; in this case, we choose the boundary condition, appropriately. 
Taking the flow of the dimensionful part of the vertex we once again tune the
seed action. This time, though, we do so to ensure sufficiently good $\alpha$-dependence.
Note that we need not worry about constraints coming from gauge invariance. If the top
vertex contains two fields in the $A$-sector (it cannot contain a single one) then
gauge invariance simply tells us that the vertex is transverse; it is not related to
lower point vertices. Performing the loop integral does not generate
any bad $\alpha$-dependence.

The penultimate diagram to deal with is the second element of the
standard set, which comprises two three-point vertices, joined together
by two effective propagators. Once again, our aim is to choose
the seed action such that the $\alpha$-dependence of the three-point 
vertices is sufficiently
good. This time, however, we must worry about gauge invariance.

The first effect of gauge invariance is to relate the longitudinal
part of each of these vertices to two-point, tree level 
vertices. Referring to our list of two-point, tree level vertices
(equations~(\ref{eq:app:TLTP-11})--(\ref{eq:app:TLTP-C2,C2})), it is clear that
our three-point vertices have buried in them, necessarily,  components which go
as $\OalPow{-1}$.

To examine the transverse components of these vertices,
we must use the flow equations. Figure~\ref{fig:flow-3pt-E+2I}
shows the flow of a three-point vertex comprising one
external field and two identical internal fields (\ie\ this
vertex should be viewed as part of a whole diagram).
\begin{center}
\begin{figure}[h]
	\[
		- \ensuremath{\begin{array}{c}\input{pstex/Vertex-TLThP-E-2I.pstex_t} \end{array}} = \ensuremath{\begin{array}{c}\input{pstex/Dumbell-TLTP-E-TLThP.pstex_t} \end{array}} + 2\ensuremath{\begin{array}{c}\input{pstex/Dumbell-TLTP-I-TLThP.pstex_t} \end{array}} +\ensuremath{\begin{array}{c}\input{pstex/Dumbell-TLTP-DW-E-TLTP.pstex_t} \end{array}} +2 \ensuremath{\begin{array}{c}\input{pstex/Dumbell-TLTP-DW-I-TLTP.pstex_t} \end{array}}
	\]
\caption{Flow of a three-point, tree level 
vertex viewed as part of a whole diagram.}
\label{fig:flow-3pt-E+2I}
\end{figure}
\end{center}

We expect the critical case to occur when all three fields are in 
the $A$-sector, since gauge invariance will then force us to
take $\Omom{2}$ from each two-point, tree level vertex. In particular,
this means that we will be unable to tune the three-point seed action
vertices if we take the $\Omom{3}$ part of the vertex whose flow
we are computing (the $\Omom{1}$ part is, of course, universal and independent of $\alpha$).

Given our choice that $c'_0 \sim \Oal$, it is straightforward to show  that
\[
	- \left. \ensuremath{\begin{array}{c}\input{pstex/Vertex-TLThP-E-2I.pstex_t} \end{array}} \right|_{\Omom{3}} \sim  \frac{\OalPow{0} \Omom{3}}{\Lambda^2}.
\]

Thus it is clear that we can always tune the seed action to ensure that
worst behaviour of the three-point, tree level vertices is the $\OalPow{-1}$
dependence forced by gauge invariance. This leading $\alpha$-dependence now cancels
between the vertices and effective propagators of the second element of the standard
set. The loop integral does not generate any bad $\alpha$-dependence

The treatment of the first element of the standard set follows, similarly:
by tuning the seed action (and choosing suitable boundary conditions
for the flow) we can ensure that the worst $\alpha$-dependence is
that forced on us by gauge invariance. This dependence is then
cancelled by the effective propagator.
The loop integral does not generate any
bad $\alpha$-dependence.

We have thus demonstrated that, by suitable choice of seed action,
the non-universal details of the cutoff functions and the
boundary conditions for our flow, we can ensure 
that $\tilde{H}(\alpha) \sim \OalPow{0}$. It therefore follows
that we can always arrange for the $\alpha$-terms to vanish in the
$\altoz$ limit.

\section{Computation of $\beta_2$} \label{sec:TLD:Compute}

We are now in a position to compute $\beta_2$. Recalling the form for a generic two loop integral, given by equation~(\ref{eq:TLD:TwoLoopGenericForm}),
we expect to be able to write:
\begin{equation}
-4 \beta_2 \Box_{\mu \nu}(p) = 
\left[\frac{\Lambda^{-4 \epsilon}}{4} \sum_{i=0}^{1} \left(C_i + D_i \frac{p^{-2 \epsilon}}{\Lambda^{-2 \epsilon}} \right) \frac{1}{\epsilon^{i-2}}
\right]^\bullet \Box_{\mu \nu}(p)
+ \Oep.
\label{eq:b2-penpenultimate}
\end{equation}
As we have argued already, it must be the case that $C_0$ and $D_i$ vanish,
else the \rhs\ of equation~(\ref{eq:b2-penpenultimate}) is not 
compatible with the \lhsB. Since we have shown that the
non-computable contributions to $C_1$ cancel, we could just compute the
surviving, computable contributions to $C_1$ and hence obtain $\beta_2$.
However, as a check on the consistency of the formalism, we will do more
than this. We have, in fact, already shown that the non-computable contributions to
$D_1$ cancel. We will now also check that $C_0$ vanishes. (We could demonstrate
that $D_0$ and the computable part of $D_1$ vanish, too.)

To collate the contributions to $C_0$ and $C_1$, it is easier to
introduce a slightly different form for $\beta_2$:
\begin{equation}
-4 \beta_2 \Box_{\mu \nu}(p) = \frac{N^2}{(4 \pi)^D} 
								\left[
									\left(\frac{1}{\epsilon} C_0^a + C_1^a \right) p^2 \delta_{\mu \nu}
									+
									\left(\frac{1}{\epsilon} C_0^b + C_1^b \right) p_\mu p_\nu
								\right]+\Oep
\label{eq:b2-penultimate}
\end{equation}

We now split up the contributions to $C_0^{a,b}$ and $C_1^{a,b}$ into five sets, $A$--$E$,
as follows:
\begin{enumerate}
	\item[$A$]	contribution~\ref{diag:23.1-Ans};

	\item[$B$]	contribution~\ref{Diagrams224-Alg};

	\item[$C$]	diagrams~\ref{d:269.4}--\ref{d:269.10}, \ref{d:269.11} and~\ref{d:269.12};
	
	\item[$D$]	diagrams~\ref{d:248.2,5:U}--\ref{d:248.3b,6b-U};

	\item[$E$]	diagrams~\ref{d:28.1bU}--\ref{d:28.1b:sub2U}; \ref{d:32.1U}; \ref{d:42.1U} and~\ref{d:42.1sub1U}; 
				\ref{d:392.4aU}, \ref{d:392.4a:sub1U}, \ref{d:392.4bU} and~\ref{d:392.4b:sub1U}; 
				\ref{d:FR22.1,2U} and~\ref{d:FR24.1,6U}; \ref{d:FR26.1,2U}; \ref{d:FR26.11U}; \ref{d:FR36.1-U}--\ref{d:FR27.6-U};
				\ref{d:259b.1}--\ref{d:260b.3}; \ref{d:255b.16-U}--\ref{d:255d.5-U}; \ref{d:279.8}--\ref{d:279.11}.
\end{enumerate}

The result of numerically evaluating the contribution to $C_0^{a,b}$ and $C_1^{a,b}$
is summarised in table~\ref{tab:beta_2-contributions}.

\begin{center}
\begin{table}[h]
	\[
	\begin{array}{c|c|c|c|c}
			& C_0^a	& C_0^b & C_1^a	& C_1^b	
	\\  \hline
		A	&	0				&	0				& -9		& 0
	\\ [2ex]
		B	&	0				&	0				&\ds \frac{571}{9}	-\frac{80\EM}{3}	&\ds -\frac{451}{9} +\frac{80\EM}{3}
	\\ [2ex]
		C	&	\mbox{---}		&	\mbox{---}		&\ds -\frac{111}{2}	+ 60 \EM			& \ds \frac{111}{2} -60\EM
	\\ [2ex]
		D	& \ds-\frac{50}{3}	& \ds\frac{50}{3}	&	\mbox{---}							& \mbox{---}
	\\ [2ex]
		E	& \ds \frac{50}{3}	& \ds-\frac{50}{3}	&\ds \frac{835}{18}	-\frac{100\EM}{3}	&\ds -\frac{913}{18} + \frac{100\EM}{3}
	\\ [2ex] \hline \hline
		&&&&	
	\\ [-3ex]
			
			&	0				& 0					&\ds\frac{136}{3}						& \ds-\frac{136}{3}
	\end{array}
	\]
\caption{Contributions to $\beta_2$.}
\label{tab:beta_2-contributions}
\end{table}
\end{center}

The null entries correspond to terms which are restricted to a specific
order in $\epsilon$; this is distinct from an entry of zero. We note that
the evaluation of the elements of sets $A$, $B$ and $D$ is independent of
the prescription we choose for treating derivatives of cutoff functions,
total momentum derivatives and etc. However, the remaining two sets are,
individually, susceptible to this prescription.

Inserting the net contributions to $C_0^{a,b}$ and $C_1^{a,b}$ into equation~(\ref{eq:b2-penultimate})
we find that, in the \eptoz\ limit,
\begin{equation}
	\beta_2 = -\frac{34}{3} \frac{N^2}{(4\pi)^4},
\label{eq:beta_2}
\end{equation}
thereby recovering the standard result~\cite{I&Z}.

\section{Conclusions}	\label{sec:TLD:conc}

The correct numerical evaluation of $\beta_2$ represents
the final result of this thesis---and perhaps the most
important. More than anything else, this demonstrates
beyond all reasonable doubt the consistency of the
formalism, thereby fulfilling the fundamental
aim of this piece of work.

With this in mind, it is worth now looking at the calculation---as we did
in the one-loop case---as a piece of work in its own right, particularly
in regard to its complexity.
Our starting point will be the starting point for this chapter:
the set of diagrams in appendix~\ref{app:beta_2}, which we denote by $\TLDs$. By beginning 
our evaluation of the complexity here,
we are essentially
assuming that this set of diagrams can be trivially generated.\footnote{
The actual generation of decorated diagrams should be natural
to implement on a computer, removing the need to do it by hand.
} We know that
this is true for those not involving $\Op{2}$ stubs; it is certainly reasonable to
expect that our $n_+$-loop formula~(\ref{eq:beta-n+}) can be extended to 
include the missing diagrams. 

Even given our starting point, it still seems that we had to work very hard
to actually extract the numerical value of $\beta_2$. However, as I will
argue here, I believe this to be an artifact of being the very first
time a two-loop calculation has been performed, within this framework. Let us analyse
the extraction of $\beta_2$, in more detail.

The set $\TLDs$ contains 83 terms.\footnote{We are counting diagrams in which we must
sum over $ES'$ only once, but this makes no difference to our argument as all
these terms vanish, anyway.} Including the $\alpha$-terms, 35 of these can
be discarded. Of those that remain, there are 19 factorisable diagrams which explicitly
contain elements of $\OLDs$. There are  10
non-factorisable diagrams which also explicitly contain elements of $\OLDs$, 
once  we realise that survival in the \eptoz\ limit forces 
certain internal fields to be in the $A$-sector. The remaining 19 diagrams are
what we will refer to as the irreducible two-loop diagrams.

Much of the work in this chapter went into constructing subtractions
for the irreducible diagrams and then performing manipulations on either
the additions or the \NUn\ diagrams resulting from combining parent,
subtraction and addition. However there is a very clear pattern to 
both the construction of the subtractions, their subsequent manipulation
and the diagrams we are left with.

Indeed, recall that when we reached the end of the one-loop calculation, we
concluded that we had been essentially repeating the same steps, over
and over again; this led us to the vastly more powerful formalism of part III.
I believe the situation here to be somewhat analogous: to me, the structure
of the computation of $\beta_2$ is highly suggestive of a greatly
simplified, underlying formalism.

Further support for this is given by the set of diagrams from which we 
ultimately evaluate the computable parts. We will denote this set of terms by $\TLDU$.
To generate this set, we can start
by simply taking each of the irreducible two-loop diagrams and placing it,
wholesale, under the influence of $\COMP$. We then do likewise with the three factorisable
diagrams whose two sub-diagrams are each elements of the little set. 
As we have noted already, many of the remaining
diagrams of $\TLDU$ 
can be cancelled if we are willing to perform manipulations under the influence
of $\COMP$. Up until now, we have been hesitant to do this, but it is clear that
this is the natural way to proceed. This would enable us to replace 20 of the diagrams
under the influence of $\COMP$ with a small set of total momentum derivative
terms (which, of course, generically survive in this scenario).

Indeed, if we were always to manipulate diagrams under the influence of $\COMP$, then 
this suggests a different way to organise the calculation. Given that these manipulations
just mirror those performed on the corresponding additions, we should \emph{always}
combine parent, subtraction \emph{and} addition to yield the parent under the
influence of $\COMP$ and a term under the influence of \NUn. This will give the simplifications
just discussed; it will also affect our treatment of the remaining elements of $\TLDU$.

The remaining elements are basically those which we recognised could be
combined into finite sets (see sections~\ref{sec:TLD:C-sector} 
and~\ref{sec:TLD:SS-Mirror}). Now, however, the combination of terms is slightly different:
all the cancellations we saw will still occur. However, the set of surviving terms will
be smaller, with some of them having been involved in the aforementioned simplifications.

Ultimately, we are left with $\mathcal{O}(30)$ contributions to $\beta_2$, all of
which are very easy to compute.
Thus, if  the procedure of going from
$\TLDs$ to $\TLDU$ can be greatly simplified---as I believe to be the case---then
the extraction of $\beta_2$ within this framework could be reduced to a
remarkably straightforward calculation. These ideas will be further explored 
in~\cite{FurtherWork}.

\chapter{Conclusion} \label{ch:Conclusion}

\section{Summary}

The basis of all new work done in this
thesis is the new flow 
equation,~\ref{eq:Flow:General-New}--\ref{eq:flow:new:a_1}.
Like the flow equation which it replaces 
(equations~\ref{eq:Flow:General}--\ref{eq:flow:old:a_1}),
the new flow equation is manifestly gauge invariant.
Where it differs is in its ability to distinguish the
two fields $A^1$ and $A^2$. In turn, this enables us
to isolate the effects of
the unphysical field $A^2$ and thus to
compute useful quantities beyond one loop.

The computational method draws its inspiration from the
diagrammatic techniques of~\cite{YM-1-loop}
and~\cite{scalar2loop}. The former paper presents a 
computation of the one-loop Yang-Mills $\beta$-function
using the (old) \MGIERG\
whereas
the latter paper presents  a 
computation of the scalar two-loop
$\beta$-function. (Henceforth, we will
use $\beta_n$ to refer to Yang-Mills $\beta$-function
coefficients.)
In both papers, it was recognised
that the \ERG\ provides a natural framework for isolating
universal contributions to universal objects. By leaving
all non-universal quantities unspecified, the calculational
procedure becomes so constrained that computation can
be done diagrammatically.

Given that the form of the flow equation has changed
from that of~\cite{YM-1-loop}, the first challenge
was to see whether diagrammatic techniques can still
be employed. As discussed in section~\ref{sec:NFE:NewDiags-I},
the diagrammatics of~\cite{YM-1-loop} can indeed be
extended, at the expense of additional complication.
The essence of this complication is that wines---which
had previously only ever comprised single supertrace components---are
effectively generalised to multi-supertrace objects.

Whilst the initial reaction was that this
had made life harder, such multi-supertrace objects
can, in fact, be naturally absorbed into the diagrammatics.
Indeed, in section~\ref{sec:NFE:NewDiags-II} we developed
an entirely new form of diagrammatics, in which single
and multi-supertrace objects are treated in parallel,
having been absorbed into a composite object.

In retrospect, that such a simplification is possible
is hardly surprising. Although all ingredients
in the treatment of~\cite{YM-1-loop}
were restricted to single supertrace terms, the
structure of the diagrammatic cancellations strongly
suggested that multi-supertrace terms, if included, would cancel
in the same way. Indeed, since all non-universal
contributions must cancel anyway, it is natural that
sets of them can be packaged up together and thus
removed in one go.

Moreover, our scheme now amounts to using standard
Feynman diagrammatic expansions, except that the
Feynman rules are novel and, embedded within the
diagrams, there is a prescription for automatically
evaluating the group theory factors.

An immediate consequence of these new diagrammatics
is that the calculation of~\cite{YM-1-loop} can be
essentially repeated, line for line. However, in 
the new way of doing things, multi-supertrace terms
come along for the ride, without really adding any
complication.\footnote{Effectively, the only additional terms
arise from the existence of $AAC$ vertices, which could be discarded
in the original treatment, as a consequence of the action
being restricted to single supertrace terms.}

Now, although much of the calculation of $\beta_1$
in~\cite{YM-1-loop} was done diagrammatically, these
techniques were not pushed to their limit, since 
gauge remainders and $\Op{2}$ terms were treated
algebraically.

In section~\ref{sec:GRs} we showed how
the gauge remainders, too, can be treated diagrammatically.
The $\Op{2}$ terms are those possessing a manifestly
$\Op{2}$ stub. Since we work at $\Op{2}$, other structures
in the diagram carrying momentum $p$ can be Taylor
expanded to zeroth order in $p$, assuming that this step does
not generate any IR divergences. In section~\ref{sec:MomentumExpansions}
we showed how to perform these Taylor expansions diagrammatically.

With the techniques for treating gauge remainders and Taylor
expansions, diagrammatically, under our belts, we illustrated their use
by applying them to the calculation of $\beta_1$. In chapter~\ref{ch:beta1}
we showed, for the first time, that $\beta_1$ can be reduced to a
set of $\Lambda$-derivative terms. This represents a radical
improvement over the approach in~\cite{YM-1-loop} and proves
crucial for performing calculations beyond one-loop.

Given the reduction of $\beta_1$ to $\Lambda$-derivative terms,
in chapter~\ref{ch:LambdaDerivatives:Methodology} we discussed
how to numerically evaluate them. This is easy for
those diagrams contributing to $\beta_1$, and serves as
a warm-up exercise for a more general treatment of $\Lambda$-derivative
terms, suitable for two-loop calculations.

At the two-loop level, there are a number of novelties. The
first is that individual diagrams may possess genuine
IR divergences (as opposed to \emph{pseudo} IR divergences,
which arise as a consequence of the way in which we evaluate
$\Lambda$-derivative terms). The second is that individual
diagrams generically possess non-universal contributions
which do not die in the limit that we take $D\rightarrow 4$.

To demonstrate that these non-universal contributions cancel
amongst themselves, we developed the subtraction techniques
of section~\ref{sec:TLD:Subtractions}. Wonderfully, even 
these have a diagrammatic interpretation, which 
can be applied not only to $\Lambda$-derivative terms
but also to $\Op{2}$ terms which are not Taylor expandable
in $p$.

However, despite the appeal and elegance of the diagrammatic
techniques developed in chapters~\ref{ch:NewFlowEquation}--\ref{ch:LambdaDerivatives:Methodology}
there is a problem when it comes to actually using them
to compute $\beta$-function coefficients. Although
the final expression for $\beta_1$ is both small
and extremely easy to extract the answer from, it
requires a considerable amount of work to actually
derive it. Part~III is devoted to finding the solution
to this problem.

The principle behind the methodology of part-III is
an extension of the idea that non-universal contributions
cancel in parallel. In the development of the new
diagrammatics of section~\ref{sec:NFE:NewDiags-II}
it was recognised that the flow of a vertex 
decorated by some set of fields $\{f\}$ can be represented
by leaving the elements of $\{f\}$ as implicit (or
unrealised) decorations. In the computation of $\beta_1$,
this was not exploited, and we explicitly performed all
these decorations. This, of course, generated many terms,
since the rule for performing explicit decorations is that one
does so in all possible, independent ways. However, when applying
the diagrammatic procedure to these myriad diagrams, it
became apparent that exactly the same steps were being performed on
sets of diagrams. Thus, it made sense to try and perform
these steps in parallel.

The key to doing this, as first discussed in chapter~\ref{ch:bn:LambdaDerivsI},
is to make use of the fact that we can leave decorations unrealised.
Let us review the diagrammatics for $\beta_{n_+}$. We start be
computing the flow of $S_{n_+ \mu \nu}^{\ \ \; 1 \, 1}(p)$, at $\Op{2}$.
However, rather than explicitly decorating our diagrams with the two
external fields $A^1_\mu(p)$ and $A^1_\nu(p)$, we leave them as
unrealised decorations. The diagrammatic procedure now proceeds as follows:
\begin{enumerate}
	\item	isolate any two-point, tree level vertices and fully decorate
			them, applying the effective propagator relationship, if
			possible, and identify any cancellations;
	
	\item	of the terms that remain, isolate the manipulable components
			\ie\ those comprising only Wilsonian effective action vertices,
			effective propagators and an undecorated wine;	
	
	\item	convert the manipulable diagrams into $\Lambda$-derivative
			terms, plus corrections, having promoted the wine of the parent
			into an unrealised decoration;

	\item	repeat the procedure.
\end{enumerate}

Thus, as we iterate the procedure, the only decorations we perform (of two-point,
tree level vertices) all facilitate further manipulation---either because we can
apply the effective propagator relation or because we can perform manipulations
at $\Op{2}$. For the $\Lambda$-derivative terms, we go the other way, promoting
explicit decorations to unrealised decorations. Since this is done at each
stage of the calculation, the $\Lambda$-derivative terms in fact possess only
unrealised decorations.

Let us now consider the partner non-manipulable diagrams to a $\Lambda$-derivative
term. Since these were spawned by a $\Lambda$-derivative term from the previous
level of manipulation, they each possess only one explicit decoration: the
wine formed by the action of $a_0$ or $a_1$. At each stage of the calculation,
these non-manipulable terms cancel without the need for any further
decoration, which represents a massive simplification
of the calculational procedure.

By iterating the diagrammatic procedure until exhaustion we were
able to show that an arbitrary $\beta_{n_+}$ can be reduced to $\Lambda$-derivative,
$\beta$ and $\alpha$-terms, up to gauge remainders and $\Op{2}$ terms 
(figure~\ref{fig:Beta-n+Part-I}).

Chapter~\ref{ch:bn:GRs} is dedicated to the initial processing
of the gauge remainders. Again, this procedure can be performed
without the need to fully decorate diagrams. As with our initial
strategy, we decorate only two-point, tree level vertices. This
procedure
generates further $\Op{2}$ terms, diagrams which cancel
non-manipulable terms and nested gauge remainders. These latter
terms are then processed.

By iterating this procedure until exhaustion, entire sets of
gauge remainder terms are shown to cancel, to all orders in 
perturbation theory. In preparation for the treatment of those
diagrams which remain, we introduced a new set of diagrammatic
identities in chapter~\ref{ch:FurtherD-ID}. We note that the last
of these, which is applicable to calculations at the three-loop
level and beyond remains to be proven. However, that it is plausible
has been demonstrated by proving its validity in a special case
and so we made the unproven assertion that it is true in all cases.

Armed with these diagrammatic identities, we showed in 
chapter~\ref{ch:bn:Iterate} how the remaining gauge remainder
terms can be converted into $\Lambda$-derivative, $\alpha$
and $\beta$-terms, thereby supplementing the 
expression of figure~\ref{fig:Beta-n+Part-I}.
In this way, we arrived at figure~\ref{fig:Beta-n+W/GR}
which gives an incredibly compact form for any $\beta_{n_+}$,
given our earlier assertion, up to $\Op{2}$ terms.

The full treatment of the $\Op{2}$ terms, though not conceptually
of any greater difficulty than the preceding analysis is
beyond the scope of this thesis. A discussion of the general
principles is given in chapter~\ref{ch:Op2}.

The final part of the thesis is devoted to the computation
of $\beta_2$. Up to $\Op{2}$ terms, we jumped straight to
the $\Lambda$-derivative, $\beta$ and $\alpha$-terms by
specialising the expression of figure~\ref{fig:Beta-n+W/GR}
to $n=1$. The explicit treatment of the $\Op{2}$ terms is
not presented in this thesis but it turns out that, by
following chapter~\ref{ch:Op2}, such terms can be
reduced to $\Lambda$-derivative, $\beta$ and $\alpha$-terms, also.

Having done this, we then used our earlier
diagrammatic expression for $\beta_1$ to simplify the
expression for $\beta_2$. The set of diagrams from 
which we extract the numerical coefficient comprises
just $\Lambda$-derivative and $\alpha$-terms and is
given in appendix~\ref{app:beta_2}.

In chapter~\ref{ch:TotalLambdaDerivatives:Application} we used
the techniques of chapter~\ref{ch:LambdaDerivatives:Methodology}
to numerically evaluate $\beta_2$. After demonstrating that the $\alpha$-terms
vanish in the $\alpha \rightarrow 0$, subject to constraints on the $\alpha$
dependence of the cutoff functions and seed action,
we obtained 
\[
	\beta_2 = -\frac{34}{3} \frac{N^2}{(4\pi)^4},
\]
thereby reproducing the standard result.

\section{Discussion}	\label{sec:Conclusion:Discussion}

In terms of the original aims of the thesis, the computation
of $\beta_2$ is the most important result of this work. That this highly
non-trivial calculation produced the correct answer surely
confirms the consistency of the approach, beyond reasonable
doubt. Moreover, it should be emphasised that this is the
very first analytical, continuum calculation of a two-loop
quantity to be performed without fixing the gauge.

However, we have achieved very much more than this. As
mentioned in chapter~\ref{ch:bulk}, this two-loop
calculation was initially performed in a very arduous fashion,
since it was first done before the discovery of the methods
of part~III. Consequently, every single decoration was
performed explicitly, generating $\mathcal{O}(10^4)$
diagrams, almost all of which cancelled, to
leave the diagrams of chapter~\ref{ch:bulk}.

At this stage, the status of continuum, manifestly gauge invariant
calculations was in a peculiar state. On the one hand,
leaving the gauge unfixed had guided us to a set of powerful
and elegant diagrammatic techniques. On the other hand,
the sheer number of diagrams produced was almost overwhelming.
Certainly, for perturbative computation, these methods
could not be considered anything other than a curiosity.

However, the discoveries of part~III have changed everything.
Although the work started in part~III is yet to be completed
(assertion~1 must be proven and the $\Op{2}$ terms must be
treated), it is quite clear there is a deep structure underlying
our original computations of $\beta_1$ and $\beta_2$.

It is certainly my belief that it should be straightforward
to complete the work of part~III and thus to arrive
at an expression for $\beta_{n_+}$ in terms of only $\Lambda$-derivative,
$\beta$ and $\alpha$-terms. Indeed, if the computation of $\beta_2$
is anything to go by, then it may very well be that such an 
expression can be further simplified. Recall how, at the two-loop
level, we were able to remove the $\beta$-terms and also
those $\Lambda$-derivative diagrams which 
were compelled (in dimensional regularisation) to go as $\Op{2-2\epsilon}$.

Irrespective of whether these final simplifications take place or not,
being able to reduce $\beta_{n_+}$ to $\Lambda$-derivative ($\beta$)
and $\alpha$-terms would have profound implications. First, at a stroke,
almost all of the difficulty in extracting $\beta$-function
coefficients within the \MGIERG\ would be removed.
(That said, the numerical extraction of $\beta_2$ was hardly easy,
as chapter~\ref{ch:bulk} demonstrated; we will return to this
point shortly.)

Perhaps more interesting is that such an expression for $\beta_{n_+}$
would contain only Wilsonian effective action vertices and effective 
propagators. 
A very important consequence of this is
that, in such an expression for $\beta_{n_+}$, there can be no diagrams
in which a wine bites its own tail: this is trivially the case, since there
are no wines! Now, claiming that we have solved the problem associated
with wines biting their tails is specious, at this stage, since they have
been excluded from our calculations, since the very beginning.
However, the analysis of part-III
makes it abundantly clear that, had we kept them, they would all have cancelled
anyway.

This is a very deep statement. That the flow equation generates
wine biting their tail diagrams means that it is not
manifestly regulated. By excluding these terms, we cure this
problem, but at the expense of a temporary loss of gauge 
invariance~\cite{TRM-Faro, GI-ERG-I}. Now we see this problem in
a different light. Certainly, when it comes to perturbatively
computing $\beta$-function coefficients, it turns out that the flow
equation is effectively regulated (so long as the analysis of part~III can indeed 
be completed, as intended), since one can demonstrate that
all diagrams in which a wine bites its own tail cancel, amongst
themselves.

In some sense, it seems that the real physics of the flow equation
is better expressed in the compact expression of figure~\ref{fig:Beta-n+W/GR}
than in the flow equation itself (at least for calculations of
$\beta$-function coefficients). There are further indications
in this direction. When constructing the new flow equation, we
commented that this flow equation is just one of an infinite number
of manifestly gauge invariant flow equations that can distinguish
between $A^1$ and $A^2$, whilst respecting no-$\SF^0$ symmetry. 
Each of these flow equations will
have different, old-style diagrammatic rules. However, it now seems
clear that these differences should not matter. We should
be able to package all details into the new diagrammatic rules and
since we do not expect such details to affect universal quantities,
they should cancel out when we iterate the diagrammatic procedure.
In other words, so long as we have a valid flow equation that
can distinguish between $A^1$ and $A^2$, we should always
be able to generate figure~\ref{fig:Beta-n+W/GR}.

Furthermore, if we continue to suppose that the $\Op{2}$ terms can be
reduced to $\Lambda$-derivative, $\alpha$ and $\beta$-terms,
then it would seem that there is a diagrammatic prescription
to implement the pre-regularisation.  Recall that the pre-regularisation
demands we throw away all surface terms. In our computation of $\beta_2$,
the only place that these arose was as total momentum derivative terms,
which have a unique diagrammatic form---see section~\ref{sec:GR:CompleteDiagrams}.
By adopting the prescription that we always throw away these diagrams,
we can implement the pre-regularisation in an entirely diagrammatic fashion,
without the need for explicitly using \eg\ dimensional regularisation.\footnote{
The use of dimensional regularisation to implement
the pre-regularisation is entirely independent of its use in
extracting the numerical coefficient, $\beta_2$.}

Let us now return to the issue of the actual difficulty
in extracting the numerical value of $\beta_2$. As discussed in
section~\ref{sec:TLD:conc}
the calculation actually has a very clear
structure. The construction of subtractions is done in a 
systematic way and the final set of computable terms can
 be considerably
simplified.

Given our experiences of refining the diagrammatic techniques of 
parts~I and~II in part~III, this is highly suggestive. It is my
belief that the entire procedure of constructing subtractions
and re-expressing the $\Lambda$-derivative terms as the sum of computable
terms can be radically simplified by working with partially
decorated diagrams, rather than fully decorated diagrams. 

If this is possible, then it opens up some
very intriguing questions. Assuming that we can
extend the analysis of part~III to include $\Op{2}$ terms,
then this will demonstrate that arbitrary $\beta_{n_+}$ has
no explicit dependence on the seed action and choice of
covariantisation. If the process of constructing subtractions
can be simplified in the way envisioned, then perhaps it is possible
to also demonstrate that arbitrary $\beta_{n_+}$ has
no implicit dependence on the seed action and choice
of covariantisation. The end of this line of reasoning
is to ask whether, within the \MGIERG,
it is possible to arrange for all $\beta_{n_+}$ to depend only
on universal details.

One very encouraging indication in this direction is that
the analysis of section~\ref{sec:TLD:Couplings} tells us
that all running couplings, other than $g$ and $\alpha$,
can be tuned to zero. This removes an entire source
of non-universality of $\beta$-function coefficients,
to all orders in perturbation theory.

We note, though, that even if it is the case that $\beta_{n_+}$ 
can be arranged to depend only on universal details, within
the \MGIERG, we are not claiming that all $\beta_n$ are universal.
We would expect our values of $\beta_{n>2}$ to disagree with
those obtained in other schemes. However, our claim would
be that, within the \MGIERG\ scheme, there are an infinite
number of sub-schemes which yield the same $\beta_{n}$ (and
also an infinite number that do not) and that this $\beta_n$
can be arranged to depend only on universal details of the \MGIERG.

If it proves to be possible to arrange such $\beta_n$,
then this would represent some kind of `factorisation of universality'.
In the computation of a physical observable, non-universal
ingredients such as $\beta$-function coefficients have to
combine with other non-universal objects, such that the final
answer is universal. However, if the $\beta_n$ can be arranged
to depend only on universal details of the \MGIERG, then the objects with
which they must be combined would have to be universal in 
this sense, too.

\section{Future Work} \label{sec:Conclusion:FutureWork}

There are a number of very exciting avenues of research to
pursue, as a result of the work in this thesis. Most obviously,
there is an immediate need to complete the analysis of
part~III and thus hopefully demonstrate that an arbitrary $\beta_{n_+}$
can, indeed, be reduced to $\Lambda$-derivative, ($\beta$) and $\alpha$-terms.

If this can be achieved, then the next logical step is to 
perform a full investigation of whether or not $\beta_{>2}$
can be arranged to depend only on universal quantities.
This is certainly an interesting question but it must be borne
in mind that the real purpose of constructing the \MGIERG\ is
to do physics. Simply examining the properties of $\beta$-function
coefficients in a scheme which currently has no mapping to
existing schemes, in which calculations of amplitudes are done, is not
a big enough step in this direction. One solution is to attempt
to perturbatively match our scheme to (say) $\overline{MS}$.

However, we intend to be very much more ambitious than this. It is
clear from this thesis that the \MGIERG\ has far more than just
novelty value: it is possible to harness the
benefits of manifest gauge invariance whilst entirely
removing
the associated difficulties. Thus, an obvious project is
to begin the examination of perturbative Yang-Mills amplitudes.

Nevertheless, we should remember that the possibility of
these perturbative investigations
came about entirely
as a by-product of demonstrating the consistency of the
formalism. The original aim, which should not be
lost sight of, it to apply the formalism to the
non-perturbative domain.

If it is the case that progress can be made in any of these directions,
then it will be desirable to move from Yang-Mills to full QCD.
We note, though, that the insertion of quarks is not trivial. The
most obvious way of including physical quark fields, $\psi$, is to embed
them, together with the unphysical bosonic spinor $\varphi$,
in a field, $\Psi$, which transforms under
the fundamental representation of $SU(N|N)$
\[
	\Psi =
	\left(
		\begin{array}{c}
			\psi
		\\
			\varphi
		\end{array}
	\right).
\]

Now, the interaction of the field $\Psi$ with the supergauge field $\SF$
is just
\[
	\sim \bar{\Psi} \gamma^\mu \SF_\mu  \Psi.
\]
Immediately, we see that we have a problem since this term has
a component involving $\SF^0$. 
Indeed, $\SF^0$ now acts as a Lagrange multiplier enforcing the
constraint
\[
	\bar{\varphi} \gamma_\mu \varphi + \bar{\psi} \gamma_\mu \psi = 0,
\]
which seems to prevent the implementation of quarks
in this manner. To
make progress, one must perhaps look at embedding
the quarks in fields which transform under higher dimensional
representations of $SU(N|N)$, with a view to constructing
a supergauge invariant interaction term which is independent
of $\SF^0$~\cite{FurtherWork}.

\section{Outlook}

We have demonstrated the consistency
of a formalism which allows continuum calculations
to be done in Yang-Mills without fixing the
gauge.  The initial difficulties have been overcome
and the benefits of manifest gauge invariance
are being revealed. The stage is
now set for application of this formalism to 
real physical problems, upon which I have utter
faith that the \MGIERG\ will prove itself
to be an indispensable tool in gauge theory.

\appendix

\chapter{Ingredients of the Weak Coupling Flow Equations} 

\section{Two-Point, Tree Level Vertices} \label{app:TLTPs}

There follows a list of our choices for all independent
(\ie\ not related by no-$\SF^0$ symmetry) seed action, two-point,
tree level vertices.

\begin{eqnarray}
	\hat{S}^{\ \, 1 1}_{0 \mu \nu}(p) 		& = &  	\NCOAlg{p} \Box_{\mu \nu}(p),
\label{eq:app:TLTP-11}
\\
	\hat{S}^{\ \, 2 2}_{0 \mu \nu}(p) 		& = &  	\NCTAlg{p}	\Box_{\mu \nu}(p),
\label{eq:app:TLTP-22}
\\
	\hat{S}^{\ B\bar{B}}_{0 \mu \, \nu}(p) 	& = &	\frac{\alpha + 1}{\alpha c_p}\Box_{\mu \nu}(p) 
													+ \frac{4 \Lambda^2}{\tilde{c}_p} \delta_{\mu \nu},
\label{eq:app:TLTP-BB}
\\
	\hat{S}^{\ D\bar{B}}_{0 \ \  \mu} (p)	& = & 	\frac{2 \Lambda^2 p_\mu}{\tilde{c}_p},
\label{eq:app:TLTP-BD}
\\
	\hat{S}^{\ D\bar{D}}_0(p) 				& = & 	\frac{\Lambda^2 p^2}{\tilde{c}_p},
\label{eq:app:TLTP-DD}
\\
	\hat{S}^{\ C^1C^1}_0(p)					& = &	\frac{\Lambda^2 p^2}{\tilde{c}_p} + 2 \lambda \Lambda^4,
\label{eq:app:TLTP-C1C1}
\\
	\hat{S}^{\ C^2C^2}_0(p) 				& = &	\frac{\Lambda^2 p^2}{\tilde{c}_p} + 2 \lambda \Lambda^4,
\label{eq:app:TLTP-C2C2}
\\
	\hat{S}^{\ C^1, C^1}_0(p)				& = & 0,
\label{eq:app:TLTP-C1,C1}
\\
	\hat{S}^{\ C^1, C^2}_0(p)				& = & 0,
\label{eq:app:TLTP-C1,C2}
\\
	\hat{S}^{\ C^2, C^2}_0(p)				& = & 0.
\label{eq:app:TLTP-C2,C2}
\end{eqnarray}

Equations~\ref{eq:app:TLTP-C1,C1}--\ref{eq:app:TLTP-C2,C2} allow us
to set the corresponding Wilsonian effective action vertices to zero.
All other two-point, tree level Wilsonian effective action vertices are
taken to be equal to the corresponding seed action vertices.

\section{The Zero-Point Kernels and Effective Propagators} \label{app:Kernels}

Using the definitions
\begin{eqnarray}
	f_p & = & \fB{p}{x},
\label{eq:app:NFE-f}
\\[1ex]
	g_p	& = & \gB{p}{x},
\label{eq:app:NFE-g}
\end{eqnarray}
we give the zero-point kernels
\begin{eqnarray}
	\DEP{1}{1}(p)		& = &	\frac{2}{\Lambda^2}	\left[ \Aone{p} \right]',
\label{eq:app:NFE-DEP-11}
\\[1ex]
	\DEP{2}{2}(p)		& = &	\frac{2}{\Lambda^2}	\left[ \Atwo{p} \right]',
\label{eq:app:NFE-DEP-22}
\\[1ex]
	\DEP{B}{\bar{B}}(p)	& = &	\frac{1}{\Lambda^2} \left[ x \tilde{c}_p g_p \right]',
\label{eq:app:NFE-DEP-BB}
\\[1ex]
	\DEP{D}{\bar{D}}(p) & = &	\frac{2}{\Lambda^4}\frac{1}{x} \left[ x^2 \tilde{c}_p f_p \right]',
\label{eq:app:NFE-DEP-DD}
\\[1ex]
	\DEP{C^1}{C^1}(p)	& = &	\frac{1}{\Lambda^4}\frac{1}{x} \left[ \frac{2x^2 \tilde{c}_p}{x + 2\lambda \tilde{c}_p} \right]',
\label{eq:app:NFE-DEP-C1C1}
\\[1ex]
	\DEP{C^2}{C^2}(p)	& = &	\frac{1}{\Lambda^4}\frac{1}{x} \left[ \frac{2x^2 \tilde{c}_p}{x + 2\lambda \tilde{c}_p} \right]',
\label{eq:app:NFE-DEP-C2C2}
\end{eqnarray}
and the effective propagators
\begin{eqnarray}
	\EP{11}{}(p)		& = & \frac{1}{p^2} \Aone{p},
\label{eq:app:NFE:EP-11}
\\[1ex]
	\EP{22}{}(p)		& = & \frac{1}{p^2} \Atwo{p},
\label{eq:app:NFE:EP-22}
\\[1ex]
	\EP{B\bar{B}}{}(p)	& = & \frac{1}{2 \Lambda^2} \tilde{c}_p g_p,
\label{eq:app:NFE:EP-BB}
\\[1ex]
	\EP{D\bar{D}}{}(p)	& = & \frac{1}{\Lambda^4} \tilde{c}_p f_p,
\label{eq:app:NFE:EP-DD}
\\[1ex]
	\EP{C^1C^1}{}(p)	& = & \frac{1}{\Lambda^4}	\frac{\tilde{c}_p}{x+ 2\lambda \tilde{c}_p},
\label{eq:app:NFE:EP-C1C1}
\\[1ex]
	\EP{C^2C^2}{}(p)	& = & \frac{1}{\Lambda^4}	\frac{\tilde{c}_p}{x+ 2\lambda \tilde{c}_p},
\label{eq:app:NFE:EP-C2C2}
\end{eqnarray}
where we note that
\begin{equation}
	xf_p + 2g_p = 1.
\label{eq:app-xf+2g}
\end{equation}

\section{The Gauge Remainders}	\label{app:GRs}

The gauge remainders are defined via the equation
\[
		S_{0 MS}^{\ \, X \, Y}(p) \EP{Y Y}{SN}(p) = \delta_{MN} - p'_M p_N,
\]
where we sum over all realisations of the field $Y$ and the components of
the \rhs\ are identified in table~\ref{tab:app:k,k'}.
\begin{center}
\begin{table}[h]
	\[
	\begin{array}{c|ccc}
				& \delta_{MN}		& p_M'							& p_N
	\\ \hline
		F		& \delta_{MN}		& (f_k p_\mu / \Lambda^2, g)	& (p_\nu, 2)
	\\
		A		& \delta_{\mu \nu}	& p_\mu / p^2					& p_\nu
	\\
		C		& \one				& \mbox{---}					& \mbox{---}
	\end{array}
	\]
\caption{Summary of the gauge remainders in each sector }
\label{tab:app:k,k'}
\end{table}
\end{center}

In both the $A$ and $F$-sectors,
\[
	p_M p'_M = 1.
\]

\chapter{Diagrammatic Identities}	\label{app:D-IDs}

\setcounter{D-ID}{0}

\begin{D-ID}
Consider a two point, tree level (single supertrace) vertex, 
attached to an effective propagator which,
only ever appearing as an internal line,
attaches to an arbitrary structure. This is shown
diagrammatically in figure~\ref{fig:NFE:EffProp}, where the effective propagator
is
denoted by a long, solid line.
\begin{center}
\begin{figure}[h]
	\[
		\ensuremath{\begin{array}{c}\input{pstex/EffPropReln.pstex_t} \end{array}}	\equiv \ensuremath{\begin{array}{c}\input{pstex/K-Delta.pstex_t} \end{array}} - \ensuremath{\begin{array}{c}\input{pstex/FullGaugeRemainder.pstex_t} \end{array}} 
							\equiv \ensuremath{\begin{array}{c}\input{pstex/K-Delta.pstex_t} \end{array}} - \ensuremath{\begin{array}{c}\input{pstex/DecomposedGR.pstex_t} \end{array}}
	\]
\caption{The effective propagator relation.}
\end{figure}
\end{center}

The object $\ensuremath{\begin{array}{c}\begin{picture}(0,0)%
\includegraphics{pstex/FullGaugeRemainder-B.pstex}%
\end{picture}%
\setlength{\unitlength}{3947sp}%
\begingroup\makeatletter\ifx\SetFigFont\undefined%
\gdef\SetFigFont#1#2#3#4#5{%
  \reset@font\fontsize{#1}{#2pt}%
  \fontfamily{#3}\fontseries{#4}\fontshape{#5}%
  \selectfont}%
\fi\endgroup%
\begin{picture}(174,174)(2239,-823)
\end{picture}
 \end{array}}$ decomposes into two constituents:
\begin{eqnarray}
	\ensuremath{\begin{array}{c}\input{pstex/GaugeRemainder-ppr.pstex_t} \end{array}} & \equiv & p'_M,
\label{eq:app:Diag:p'}
\\[1ex]
	\ensuremath{\begin{array}{c}\input{pstex/GaugeRemainder-p.pstex_t} \end{array}} & \equiv & p_N.
\label{eq:app:Diag:p}
\end{eqnarray}
\end{D-ID}

\begin{D-ID}
Using the definitions~\ref{eq:app:Diag:p'} and~\ref{eq:app:Diag:p},
together with the information of table~\ref{tab:app:k,k'} and equation~\ref{eq:app-xf+2g}
gives us the diagrammatic identity below.
\begin{equation}
	\ensuremath{\begin{array}{c}\begin{picture}(0,0)%
\includegraphics{pstex/GR-relation.pstex}%
\end{picture}%
\setlength{\unitlength}{3947sp}%
\begingroup\makeatletter\ifx\SetFigFont\undefined%
\gdef\SetFigFont#1#2#3#4#5{%
  \reset@font\fontsize{#1}{#2pt}%
  \fontfamily{#3}\fontseries{#4}\fontshape{#5}%
  \selectfont}%
\fi\endgroup%
\begin{picture}(314,174)(2239,-821)
\end{picture}
 \end{array}} = 1
\label{eq:app:GR-relation}
\end{equation}
The index at the apex of $\GRk$ is contracted into the index carried
by $\GRkp$.
Note that this relationship exists only in the $A^{1,2}$ and $F$-sectors.

A number of corollaries follow from equation~\ref{eq:app:GR-relation} and the
fact that $p_N$ is independent of both $\Lambda$ and $\alpha$,
in all sectors.

\paragraph{Corollaries}
\begin{eqnarray*}
	\dec{\ensuremath{\begin{array}{c}\begin{picture}(0,0)%
\includegraphics{pstex/GR-relation.pstex}%
\end{picture}%
\setlength{\unitlength}{3947sp}%
\begingroup\makeatletter\ifx\SetFigFont\undefined%
\gdef\SetFigFont#1#2#3#4#5{%
  \reset@font\fontsize{#1}{#2pt}%
  \fontfamily{#3}\fontseries{#4}\fontshape{#5}%
  \selectfont}%
\fi\endgroup%
\begin{picture}(314,174)(2239,-821)
\end{picture}
 \end{array}}}{\bullet}					& = 			& 0	
\\
													& \Rightarrow 	& \ensuremath{\begin{array}{c}\begin{picture}(0,0)%
\includegraphics{pstex/kkpr-LdL.pstex}%
\end{picture}%
\setlength{\unitlength}{3947sp}%
\begingroup\makeatletter\ifx\SetFigFont\undefined%
\gdef\SetFigFont#1#2#3#4#5{%
  \reset@font\fontsize{#1}{#2pt}%
  \fontfamily{#3}\fontseries{#4}\fontshape{#5}%
  \selectfont}%
\fi\endgroup%
\begin{picture}(314,293)(2239,-821)
\put(2433,-612){\makebox(0,0)[lb]{\smash{\SetFigFont{8}{9.6}{\rmdefault}{\mddefault}{\updefault}{\color[rgb]{0,0,0}$\bullet$}%
}}}
\end{picture}
 \end{array}} = 0	
\\
	\pder{}{\alpha} \left[ \ensuremath{\begin{array}{c}\begin{picture}(0,0)%
\includegraphics{pstex/GR-relation.pstex}%
\end{picture}%
\setlength{\unitlength}{3947sp}%
\begingroup\makeatletter\ifx\SetFigFont\undefined%
\gdef\SetFigFont#1#2#3#4#5{%
  \reset@font\fontsize{#1}{#2pt}%
  \fontfamily{#3}\fontseries{#4}\fontshape{#5}%
  \selectfont}%
\fi\endgroup%
\begin{picture}(314,174)(2239,-821)
\end{picture}
 \end{array}} \right] & = 			& 0
\\
													& \Rightarrow	& \ensuremath{\begin{array}{c}\begin{picture}(0,0)%
\includegraphics{pstex/k.pstex}%
\end{picture}%
\setlength{\unitlength}{3947sp}%
\begingroup\makeatletter\ifx\SetFigFont\undefined%
\gdef\SetFigFont#1#2#3#4#5{%
  \reset@font\fontsize{#1}{#2pt}%
  \fontfamily{#3}\fontseries{#4}\fontshape{#5}%
  \selectfont}%
\fi\endgroup%
\begin{picture}(174,174)(2239,-821)
\end{picture}
 \end{array}} \pder{}{\alpha} \ensuremath{\begin{array}{c}\begin{picture}(0,0)%
\includegraphics{pstex/kpr.pstex}%
\end{picture}%
\setlength{\unitlength}{3947sp}%
\begingroup\makeatletter\ifx\SetFigFont\undefined%
\gdef\SetFigFont#1#2#3#4#5{%
  \reset@font\fontsize{#1}{#2pt}%
  \fontfamily{#3}\fontseries{#4}\fontshape{#5}%
  \selectfont}%
\fi\endgroup%
\begin{picture}(174,174)(2379,-821)
\end{picture}
 \end{array}} \equiv \ensuremath{\begin{array}{c}\begin{picture}(0,0)%
\includegraphics{pstex/GRk-dal-GRkpr.pstex}%
\end{picture}%
\setlength{\unitlength}{3947sp}%
\begingroup\makeatletter\ifx\SetFigFont\undefined%
\gdef\SetFigFont#1#2#3#4#5{%
  \reset@font\fontsize{#1}{#2pt}%
  \fontfamily{#3}\fontseries{#4}\fontshape{#5}%
  \selectfont}%
\fi\endgroup%
\begin{picture}(314,357)(2239,-821)
\put(2445,-560){\makebox(0,0)[lb]{\smash{\SetFigFont{8}{9.6}{\rmdefault}{\mddefault}{\updefault}{\color[rgb]{0,0,0}$\alpha$}%
}}}
\end{picture}
 \end{array}} = 0
\end{eqnarray*}
We have introduced \Dal\ to represent $\partial / \partial \alpha$.
\end{D-ID}

\begin{D-ID}
Consider an effective propagator attached, at one end
to the $p'_M$ part of a gauge remainder. This can be
redrawn as follow:
\[
	\ensuremath{\begin{array}{c}\begin{picture}(0,0)%
\includegraphics{pstex/EP-GRpr.pstex}%
\end{picture}%
\setlength{\unitlength}{3947sp}%
\begingroup\makeatletter\ifx\SetFigFont\undefined%
\gdef\SetFigFont#1#2#3#4#5{%
  \reset@font\fontsize{#1}{#2pt}%
  \fontfamily{#3}\fontseries{#4}\fontshape{#5}%
  \selectfont}%
\fi\endgroup%
\begin{picture}(628,174)(1926,-821)
\end{picture}
 \end{array}} = \ensuremath{\begin{array}{c}\begin{picture}(0,0)%
\includegraphics{pstex/GR-PEP.pstex}%
\end{picture}%
\setlength{\unitlength}{3947sp}%
\begingroup\makeatletter\ifx\SetFigFont\undefined%
\gdef\SetFigFont#1#2#3#4#5{%
  \reset@font\fontsize{#1}{#2pt}%
  \fontfamily{#3}\fontseries{#4}\fontshape{#5}%
  \selectfont}%
\fi\endgroup%
\begin{picture}(786,174)(2089,-821)
\end{picture}
 \end{array}},
\]
where the dash-dot line represents the pseudo effective propagators.

\upshape
\paragraph{Corollary}
\[
	\ensuremath{\begin{array}{c}\begin{picture}(0,0)%
\includegraphics{pstex/EP-FullGR.pstex}%
\end{picture}%
\setlength{\unitlength}{3947sp}%
\begingroup\makeatletter\ifx\SetFigFont\undefined%
\gdef\SetFigFont#1#2#3#4#5{%
  \reset@font\fontsize{#1}{#2pt}%
  \fontfamily{#3}\fontseries{#4}\fontshape{#5}%
  \selectfont}%
\fi\endgroup%
\begin{picture}(778,174)(1926,-826)
\end{picture}
 \end{array}} = \ensuremath{\begin{array}{c}\begin{picture}(0,0)%
\includegraphics{pstex/GR-PEP-GR.pstex}%
\end{picture}%
\setlength{\unitlength}{3947sp}%
\begingroup\makeatletter\ifx\SetFigFont\undefined%
\gdef\SetFigFont#1#2#3#4#5{%
  \reset@font\fontsize{#1}{#2pt}%
  \fontfamily{#3}\fontseries{#4}\fontshape{#5}%
  \selectfont}%
\fi\endgroup%
\begin{picture}(929,174)(2089,-824)
\end{picture}
 \end{array}} = \ensuremath{\begin{array}{c}\begin{picture}(0,0)%
\includegraphics{pstex/FullGR-EP.pstex}%
\end{picture}%
\setlength{\unitlength}{3947sp}%
\begingroup\makeatletter\ifx\SetFigFont\undefined%
\gdef\SetFigFont#1#2#3#4#5{%
  \reset@font\fontsize{#1}{#2pt}%
  \fontfamily{#3}\fontseries{#4}\fontshape{#5}%
  \selectfont}%
\fi\endgroup%
\begin{picture}(778,174)(1965,-825)
\end{picture}
 \end{array}}
\]
\end{D-ID}

\begin{D-ID}
A sub-diagram comprising a two-point, tree level vertex, differentiated \wrt\ momentum and attached to an effective propagator can, using the effective
propagator relation,
 be redrawn, in the following way:
\[
\ensuremath{\begin{array}{c}\input{pstex/D-ID-OneWine.pstex_t} \end{array}}.
\]
The final diagram is interpreted as the derivative \wrt\ the momentum entering the encircled structure from the left. Had the
arrow pointed the other way, then the derivative would be \wrt\ the momentum entering the structure from the right.
\end{D-ID}

\begin{D-ID}
A sub-diagram comprising a two-point, tree level  vertex, differentiated \wrt\ momentum and attached to two effective
propagators can, using diagrammatic 
identity~\ref{D-ID:OneWine} and the effective
propagator relation, be redrawn in the following way:
\[
\ensuremath{\begin{array}{c}\input{pstex/D-ID-TwoWines-A.pstex_t} \end{array}} 
+ \frac{1}{2}
\left[
\ensuremath{\begin{array}{c}\input{pstex/D-ID-TwoWines-B.pstex_t} \end{array}} 
\right].
\]
We could re-express this in a less symmetric form by taking either of the two rows in the square brackets and removing the
factor of half.
\end{D-ID}

\begin{D-ID}
A sub-diagram comprising a two-point, tree level vertex, struck by $\GRk$
vanishes:
\[
	\ensuremath{\begin{array}{c}\begin{picture}(0,0)%
\includegraphics{pstex/TLTP-GR.pstex}%
\end{picture}%
\setlength{\unitlength}{3947sp}%
\begingroup\makeatletter\ifx\SetFigFont\undefined%
\gdef\SetFigFont#1#2#3#4#5{%
  \reset@font\fontsize{#1}{#2pt}%
  \fontfamily{#3}\fontseries{#4}\fontshape{#5}%
  \selectfont}%
\fi\endgroup%
\begin{picture}(454,238)(1438,-114)
\put(1648,-57){\makebox(0,0)[lb]{\smash{\SetFigFont{11}{13.2}{\rmdefault}{\mddefault}{\updefault}{\color[rgb]{0,0,0}0}%
}}}
\end{picture}
 \end{array}} = 0.
\]
\paragraph{Corollaries}
\begin{eqnarray*}
	\ensuremath{\begin{array}{c}\begin{picture}(0,0)%
\includegraphics{pstex/dTLTP-GR.pstex}%
\end{picture}%
\setlength{\unitlength}{3947sp}%
\begingroup\makeatletter\ifx\SetFigFont\undefined%
\gdef\SetFigFont#1#2#3#4#5{%
  \reset@font\fontsize{#1}{#2pt}%
  \fontfamily{#3}\fontseries{#4}\fontshape{#5}%
  \selectfont}%
\fi\endgroup%
\begin{picture}(457,396)(1432,-114)
\put(1648,-57){\makebox(0,0)[lb]{\smash{\SetFigFont{11}{13.2}{\rmdefault}{\mddefault}{\updefault}{\color[rgb]{0,0,0}0}%
}}}
\end{picture}
 \end{array}} 		& \equiv & - \ensuremath{\begin{array}{c}\begin{picture}(0,0)%
\includegraphics{pstex/TLTP-dGR.pstex}%
\end{picture}%
\setlength{\unitlength}{3947sp}%
\begingroup\makeatletter\ifx\SetFigFont\undefined%
\gdef\SetFigFont#1#2#3#4#5{%
  \reset@font\fontsize{#1}{#2pt}%
  \fontfamily{#3}\fontseries{#4}\fontshape{#5}%
  \selectfont}%
\fi\endgroup%
\begin{picture}(550,431)(1438,-114)
\put(1648,-57){\makebox(0,0)[lb]{\smash{\SetFigFont{11}{13.2}{\rmdefault}{\mddefault}{\updefault}{\color[rgb]{0,0,0}0}%
}}}
\end{picture}
 \end{array}}
\\
	\ensuremath{\begin{array}{c}\input{pstex/LdL-TLTP-GR.pstex_t} \end{array}} 	& \equiv & -\ensuremath{\begin{array}{c}\input{pstex/TLTP-LdL-GR.pstex_t} \end{array}} = 0
\\
	\ensuremath{\begin{array}{c}\input{pstex/dal-TLTP-GR.pstex_t} \end{array}}	& \equiv & -\ensuremath{\begin{array}{c}\input{pstex/TLTP-dal-GR.pstex_t} \end{array}} = 0
\end{eqnarray*}

Recall that \Dal\ represents differentiation \wrt\ $\alpha$. We have used the
independence of $\GRk$ on both $\Lambda$ and $\alpha$ (in all sectors).
\end{D-ID}

\begin{D-ID}
A sub-diagram comprising a two-point, tree level vertex, differentiated \wrt\ momentum and attached to two effective propagators, 
one of which terminates with a $\GRk'$ can,
using diagrammatic 
identities~\ref{D-ID:PseudoEffectivePropagatorRelation} and~\ref{D-ID:OneWine}, be redrawn in the following form:
\[
\ensuremath{\begin{array}{c}\input{pstex/./D-ID-TwoWinesGR.pstex_t} \end{array}}.
\]
\end{D-ID}

\begin{D-ID}
A sub-diagram comprising a two-point, tree level vertex, attached to an un-decorated wine
which terminates in a $\GRk'$ can be redrawn, using diagrammatic identities~\ref{D-ID:PseudoEffectivePropagatorRelation} 
and~\ref{D-ID:GaugeRemainderRelation},
the effective propagator relation and the tree level flow equation:
\begin{eqnarray*}
	\ensuremath{\begin{array}{c}\input{pstex/D-ID-VWG-A.pstex_t} \end{array}} & \equiv &
		\left[
			\ensuremath{\begin{array}{c}\begin{picture}(0,0)%
\includegraphics{pstex/D-ID-VWG-B.pstex}%
\end{picture}%
\setlength{\unitlength}{3947sp}%
\begingroup\makeatletter\ifx\SetFigFont\undefined%
\gdef\SetFigFont#1#2#3#4#5{%
  \reset@font\fontsize{#1}{#2pt}%
  \fontfamily{#3}\fontseries{#4}\fontshape{#5}%
  \selectfont}%
\fi\endgroup%
\begin{picture}(656,238)(376,-654)
\put(537,-596){\makebox(0,0)[lb]{\smash{\SetFigFont{11}{13.2}{\rmdefault}{\mddefault}{\updefault}{\color[rgb]{0,0,0}0}%
}}}
\end{picture}
 \end{array}} 
		\right]^\bullet -
		\ensuremath{\begin{array}{c}\input{pstex/D-ID-VWG-C.pstex_t} \end{array}} -
		\ensuremath{\begin{array}{c}\input{pstex/D-ID-VWG-D.pstex_t} \end{array}} 
\\
	& \equiv & \ensuremath{\begin{array}{c}\begin{picture}(0,0)%
\includegraphics{pstex/D-ID-VWG-E.pstex}%
\end{picture}%
\setlength{\unitlength}{3947sp}%
\begingroup\makeatletter\ifx\SetFigFont\undefined%
\gdef\SetFigFont#1#2#3#4#5{%
  \reset@font\fontsize{#1}{#2pt}%
  \fontfamily{#3}\fontseries{#4}\fontshape{#5}%
  \selectfont}%
\fi\endgroup%
\begin{picture}(155,227)(878,-583)
\put(935,-440){\makebox(0,0)[lb]{\smash{\SetFigFont{8}{9.6}{\rmdefault}{\mddefault}{\updefault}{\color[rgb]{0,0,0}$\bullet$}%
}}}
\end{picture}
 \end{array}} 
\end{eqnarray*}
\end{D-ID}

\begin{D-ID}
Consider the diagram shown on the LHS of figure~\ref{fig:app:D-ID-GR-W-dGR-A}.
We can re-draw this as shown on the RHS since,
if the newly drawn $\GRk$ hits the external field, then the diagram
vanishes.
\begin{center}
\begin{figure}[h]
	\[
	\ensuremath{\begin{array}{c}\input{pstex/Beta1-2.16-A.pstex_t} \end{array}} \equiv \ensuremath{\begin{array}{c}\input{pstex/Beta1-2.16-B.pstex_t} \end{array}}
	\]
\caption{The first step in re-drawing a diagram to yield
a diagrammatic identity.}
\label{fig:app:D-ID-GR-W-dGR-A}
\end{figure}
\end{center}

We now allow the original $\GRk$ to act. It must trap
the remaining $\GRk$. Taking the \CC\ of the resultant
diagram, we reflect it but pick up no net sign, since we
have both one processed gauge remainder and one momentum
derivative. Having done this, we then reverse the direction
of the arrow on the derivative, which does yield a minus sign.
This gives us the diagrammatic identity, shown in figure~\ref{fig:app:D-ID-GR-W-dGR}.
\begin{center}
\begin{figure}[h]
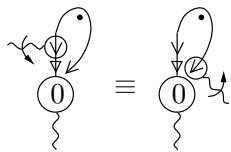

	\[
	\ensuremath{\begin{array}{c}\input{pstex/Beta1-2.16-A.pstex_t} \end{array}} \equiv \ensuremath{\begin{array}{c}\input{pstex/Beta1-2.16-C.pstex_t} \end{array}}
	\]
\caption{A diagrammatic identity.}
\label{fig:app:D-ID-GR-W-dGR-pre}
\end{figure}
\end{center}

We can, however, generalise this diagrammatic identity by stripping
off the two-point, tree level vertices. We are justified in doing this:
the integral over loop momentum will, by Lorentz invariance, yield
a Kronecker $\delta$ for both diagrams of figure~\ref{fig:app:D-ID-GR-W-dGR}.
In other words, we need not worry that stripping off the two-point,
tree level vertices is valid only up to some term which kills the vertex.
\begin{center}
\begin{figure}[h]
	\[
	\ensuremath{\begin{array}{c}\input{pstex/Beta1-2.16-A-NoV.pstex_t} \end{array}} \equiv \ensuremath{\begin{array}{c}\input{pstex/Beta1-2.16-C-NoV.pstex_t} \end{array}}
	\]
\caption{Diagrammatic identity~\ref{D-ID-GR-W-dGR}.}
\label{fig:app:D-ID-GR-W-dGR}
\end{figure}
\end{center}

Note that this identity holds if we replace the wine by an effective propagator.
\end{D-ID}

\begin{D-ID}
Consider a vertex attached to an effective propagator, which ends in a hook. Suppose that a gauge remainder
strikes this vertex. Focusing on the terms produced when the gauge remainder
pushes forward / pulls back onto a socket, we suppose that a two-point, tree level vertex is generated.
Applying the effective propagator relation, the gauge remainder dies, leaving us with the
diagrammatic identity shown in figure~\ref{fig:app:D-ID-Alg-1}.
\begin{center}
\begin{figure}[h]
	\[
	\ensuremath{\begin{array}{c}\input{pstex/Struc-AR0-A.pstex_t} \end{array}} - \ensuremath{\begin{array}{c}\input{pstex/Struc-AR0-B.pstex_t} \end{array}} = 0
	\]
\caption{Diagrammatic identity~\ref{D-ID-Alg-1}.}
\label{fig:app:D-ID-Alg-1}
\end{figure}
\end{center}

\paragraph{Corollary} 

The diagrammatic identity immediately implies that $\ensuremath{\begin{array}{c}\begin{picture}(0,0)%
\includegraphics{pstex/Struc-AR0-A.pstex}%
\end{picture}%
\setlength{\unitlength}{3947sp}%
\begingroup\makeatletter\ifx\SetFigFont\undefined%
\gdef\SetFigFont#1#2#3#4#5{%
  \reset@font\fontsize{#1}{#2pt}%
  \fontfamily{#3}\fontseries{#4}\fontshape{#5}%
  \selectfont}%
\fi\endgroup%
\begin{picture}(230,285)(197,457)
\end{picture}
 \end{array}}$, alone,
vanishes; this follows because we can re-express this sub-diagram as:
\[
1/2 \left[\ensuremath{\begin{array}{c}\begin{picture}(0,0)%
\includegraphics{pstex/Struc-AR0-A.pstex}%
\end{picture}%
\setlength{\unitlength}{3947sp}%
\begingroup\makeatletter\ifx\SetFigFont\undefined%
\gdef\SetFigFont#1#2#3#4#5{%
  \reset@font\fontsize{#1}{#2pt}%
  \fontfamily{#3}\fontseries{#4}\fontshape{#5}%
  \selectfont}%
\fi\endgroup%
\begin{picture}(230,285)(197,457)
\end{picture}
 \end{array}} - \ensuremath{\begin{array}{c}\begin{picture}(0,0)%
\includegraphics{pstex/Struc-AR0-B.pstex}%
\end{picture}%
\setlength{\unitlength}{3947sp}%
\begingroup\makeatletter\ifx\SetFigFont\undefined%
\gdef\SetFigFont#1#2#3#4#5{%
  \reset@font\fontsize{#1}{#2pt}%
  \fontfamily{#3}\fontseries{#4}\fontshape{#5}%
  \selectfont}%
\fi\endgroup%
\begin{picture}(132,273)(193,469)
\end{picture}
 \end{array}}\right]
\]
(see section~\ref{sec:GR-Wines}).
\end{D-ID}

\begin{D-ID} 

Consider a two-point, tree level vertex, attached to an effective propagator
and then to a gauge remainder.
The gauge remainder bites the remaining socket of the vertex in either sense. A second gauge
remainder bites the first gauge remainder. These sub-diagrams 
can be redrawn as shown in figure~\ref{fig:app:D-ID-Alg-2}.
\begin{center}
\begin{figure}[h]
	\[
	\ensuremath{\begin{array}{c}\input{pstex/Struc-AR1-A+B.pstex_t} \end{array}} \equiv \ensuremath{\begin{array}{c}\input{pstex/Struc-AR1-C-bare.pstex_t} \end{array}}, \hspace{1em} 
	\ensuremath{\begin{array}{c}\input{pstex/Struc-AR1-A+B-2.pstex_t} \end{array}} \equiv \ensuremath{\begin{array}{c}\input{pstex/Struc-AR1-C-bare-2.pstex_t} \end{array}}
	\]
\caption{Diagrammatic identity~\ref{D-ID-Alg-2}.}
\label{fig:app:D-ID-Alg-2}
\end{figure}
\end{center}
\end{D-ID}

\begin{D-ID}
Consider the diagrams of figure~\ref{fig:app:D-ID-Alg-3}.
This equality holds, if we change nested gauge remainders from
being struck on the left (right) to being struck on the right (left),
in all independent ways.
\begin{center}
\begin{figure}[h]
	\[
		\ensuremath{\begin{array}{c}\input{pstex/Struc-AR3-A.pstex_t} \end{array}} - \ensuremath{\begin{array}{c}\input{pstex/Struc-AR3-B.pstex_t} \end{array}} + \ensuremath{\begin{array}{c}\input{pstex/Struc-AR3-C.pstex_t} \end{array}} - \ensuremath{\begin{array}{c}\input{pstex/Struc-AR3-D.pstex_t} \end{array}} = 0
	\]
\caption{Diagrammatic identity~\ref{D-ID-Alg-3}}
\label{fig:app:D-ID-Alg-3}
\end{figure}
\end{center}
\end{D-ID}

\setcounter{assert}{0}

\begin{center}
\begin{figure}[h]
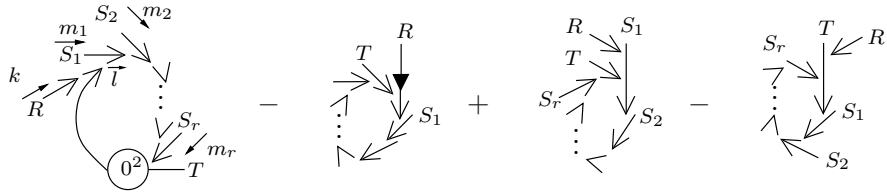

	\[
	\ensuremath{\begin{array}{c}\input{pstex/Struc-Gen-Alg-A.pstex_t} \end{array}} 
		\hspace{1em} - \hspace{1em} 
	\ensuremath{\begin{array}{c}\input{pstex/Struc-Gen-Alg-B.pstex_t} \end{array}} 
		\hspace{1em} + \hspace{1em}
	\ensuremath{\begin{array}{c}\input{pstex/Struc-Gen-Alg-C.pstex_t} \end{array}}
		\hspace{1em} - \hspace{1em}
	\ensuremath{\begin{array}{c}\input{pstex/Struc-Gen-Alg-D.pstex_t} \end{array}}
	\]
\caption{Four diagrams for which it has been demonstrated that they
sum to zero when all fields are in the $A$-sector.}
\label{fig:app:D-ID-Alg-G}
\end{figure}
\end{center}

\begin{assert}
If the diagrams of figure~\ref{fig:app:D-ID-Alg-G} can be shown to sum to zero in 
the $A$-sector then we assert the following, without proof:
\begin{enumerate}
	\item 	these diagrams will sum to zero in all sectors;
	\item 	the sets of partner diagrams in which we exchange pushes forward
			for pulls back (and \emph{vice-versa}) also sum to zero.
\end{enumerate}
\end{assert}

\begin{D-ID}
The set of sub-diagrams of figure~\ref{fig:app:D-ID-Alg-G} sum to zero.
The complimentary sets of diagrams in which arbitrary numbers of pushes forward (pulls back)
are consistently changed to pulls back (pushes forward) also sum to zero.
\end{D-ID}

\begin{D-ID}
It is true in all sectors for which the gauge remainder
is not null that
\[
	\ensuremath{\begin{array}{c}\input{pstex/dGRk.pstex_t} \end{array}} = \delta_{\alpha\nu}.
\]
It therefore follows that
\[
	\ensuremath{\begin{array}{c}\begin{picture}(0,0)%
\includegraphics{pstex/dGRk-Arb.pstex}%
\end{picture}%
\setlength{\unitlength}{3947sp}%
\begingroup\makeatletter\ifx\SetFigFont\undefined%
\gdef\SetFigFont#1#2#3#4#5{%
  \reset@font\fontsize{#1}{#2pt}%
  \fontfamily{#3}\fontseries{#4}\fontshape{#5}%
  \selectfont}%
\fi\endgroup%
\begin{picture}(249,608)(-110,336)
\put(-67,367){\makebox(0,0)[lb]{\smash{\SetFigFont{8}{9.6}{\rmdefault}{\mddefault}{\updefault}{\color[rgb]{0,0,0}$k$}%
}}}
\end{picture}
 \end{array}} = \ensuremath{\begin{array}{c}\begin{picture}(0,0)%
\includegraphics{pstex/TE-Ext-Arb.pstex}%
\end{picture}%
\setlength{\unitlength}{3947sp}%
\begingroup\makeatletter\ifx\SetFigFont\undefined%
\gdef\SetFigFont#1#2#3#4#5{%
  \reset@font\fontsize{#1}{#2pt}%
  \fontfamily{#3}\fontseries{#4}\fontshape{#5}%
  \selectfont}%
\fi\endgroup%
\begin{picture}(273,375)(-134,450)
\put(-79,741){\makebox(0,0)[lb]{\smash{\SetFigFont{8}{9.6}{\rmdefault}{\mddefault}{\updefault}{\color[rgb]{0,0,0}$k$}%
}}}
\end{picture}
 \end{array}}.
\]
\end{D-ID}

\begin{D-ID}
A sub-diagram comprising a two-point, tree level vertex, differentiated \wrt\ $\alpha$ and attached to an effective propagator can, using the effective
propagator relation,
 be redrawn, in the following form:
\[
	\ensuremath{\begin{array}{c}\input{pstex/D-ID-V-al-EP-A.pstex_t} \end{array}} \equiv - \ensuremath{\begin{array}{c}\input{pstex/D-ID-V-al-EP-B.pstex_t} \end{array}} - \ensuremath{\begin{array}{c}\begin{picture}(0,0)%
\includegraphics{pstex/D-ID-V-al-EP-C.pstex}%
\end{picture}%
\setlength{\unitlength}{3947sp}%
\begingroup\makeatletter\ifx\SetFigFont\undefined%
\gdef\SetFigFont#1#2#3#4#5{%
  \reset@font\fontsize{#1}{#2pt}%
  \fontfamily{#3}\fontseries{#4}\fontshape{#5}%
  \selectfont}%
\fi\endgroup%
\begin{picture}(319,361)(1593,-466)
\put(1629,-201){\makebox(0,0)[lb]{\smash{\SetFigFont{8}{9.6}{\rmdefault}{\mddefault}{\updefault}{\color[rgb]{0,0,0}$\alpha$}%
}}}
\end{picture}
 \end{array}}.
\]
\end{D-ID}

\begin{D-ID}
A sub-diagram comprising a two-point, tree level  vertex, differentiated \wrt\ $\alpha$ and attached to two effective
propagators can, using diagrammatic 
identity~\ref{D-ID:V-al-EP} and the effective
propagator relation, be redrawn in the following form:
\[
	\ensuremath{\begin{array}{c}\input{pstex/D-ID-V-al-EPx2-A.pstex_t} \end{array}} 
	\equiv 
	- \ensuremath{\begin{array}{c}\begin{picture}(0,0)%
\includegraphics{pstex/D-ID-V-al-EPx2-B.pstex}%
\end{picture}%
\setlength{\unitlength}{3947sp}%
\begingroup\makeatletter\ifx\SetFigFont\undefined%
\gdef\SetFigFont#1#2#3#4#5{%
  \reset@font\fontsize{#1}{#2pt}%
  \fontfamily{#3}\fontseries{#4}\fontshape{#5}%
  \selectfont}%
\fi\endgroup%
\begin{picture}(328,204)(514,-10)
\put(635, 98){\makebox(0,0)[lb]{\smash{\SetFigFont{8}{9.6}{\rmdefault}{\mddefault}{\updefault}{\color[rgb]{0,0,0}$\alpha$}%
}}}
\end{picture}
 \end{array}} 
	+\frac{1}{2}
	\left[
		\begin{array}{ccc}
			\ensuremath{\begin{array}{c}\begin{picture}(0,0)%
\includegraphics{pstex/D-ID-V-al-EPx2-C.pstex}%
\end{picture}%
\setlength{\unitlength}{3947sp}%
\begingroup\makeatletter\ifx\SetFigFont\undefined%
\gdef\SetFigFont#1#2#3#4#5{%
  \reset@font\fontsize{#1}{#2pt}%
  \fontfamily{#3}\fontseries{#4}\fontshape{#5}%
  \selectfont}%
\fi\endgroup%
\begin{picture}(356,250)(182,555)
\put(352,709){\makebox(0,0)[lb]{\smash{\SetFigFont{8}{9.6}{\rmdefault}{\mddefault}{\updefault}{\color[rgb]{0,0,0}$\alpha$}%
}}}
\end{picture}
 \end{array}}	& - & \ensuremath{\begin{array}{c}\begin{picture}(0,0)%
\includegraphics{pstex/D-ID-V-al-EPx2-D.pstex}%
\end{picture}%
\setlength{\unitlength}{3947sp}%
\begingroup\makeatletter\ifx\SetFigFont\undefined%
\gdef\SetFigFont#1#2#3#4#5{%
  \reset@font\fontsize{#1}{#2pt}%
  \fontfamily{#3}\fontseries{#4}\fontshape{#5}%
  \selectfont}%
\fi\endgroup%
\begin{picture}(627,344)(355,494)
\put(711,742){\makebox(0,0)[lb]{\smash{\SetFigFont{8}{9.6}{\rmdefault}{\mddefault}{\updefault}{\color[rgb]{0,0,0}$\alpha$}%
}}}
\end{picture}
 \end{array}} \\
			\ensuremath{\begin{array}{c}\begin{picture}(0,0)%
\includegraphics{pstex/D-ID-V-al-EPx2-E.pstex}%
\end{picture}%
\setlength{\unitlength}{3947sp}%
\begingroup\makeatletter\ifx\SetFigFont\undefined%
\gdef\SetFigFont#1#2#3#4#5{%
  \reset@font\fontsize{#1}{#2pt}%
  \fontfamily{#3}\fontseries{#4}\fontshape{#5}%
  \selectfont}%
\fi\endgroup%
\begin{picture}(356,250)(223,555)
\put(352,709){\makebox(0,0)[lb]{\smash{\SetFigFont{8}{9.6}{\rmdefault}{\mddefault}{\updefault}{\color[rgb]{0,0,0}$\alpha$}%
}}}
\end{picture}
 \end{array}}	& - & \ensuremath{\begin{array}{c}\begin{picture}(0,0)%
\includegraphics{pstex/D-ID-V-al-EPx2-F.pstex}%
\end{picture}%
\setlength{\unitlength}{3947sp}%
\begingroup\makeatletter\ifx\SetFigFont\undefined%
\gdef\SetFigFont#1#2#3#4#5{%
  \reset@font\fontsize{#1}{#2pt}%
  \fontfamily{#3}\fontseries{#4}\fontshape{#5}%
  \selectfont}%
\fi\endgroup%
\begin{picture}(627,337)(514,501)
\put(711,742){\makebox(0,0)[lb]{\smash{\SetFigFont{8}{9.6}{\rmdefault}{\mddefault}{\updefault}{\color[rgb]{0,0,0}$\alpha$}%
}}}
\end{picture}
 \end{array}}
		\end{array}
	\right].
\]
We could re-express this in a less symmetric form by taking either of the two rows in the square brackets and removing the
factor of half.
\end{D-ID}

\begin{D-ID}
A sub-diagram comprising a two-point, tree level  vertex, differentiated \wrt\ $\alpha$ and attached to
an effective propagator which terminates in a $\GRk'$ vanishes:
\[
	\ensuremath{\begin{array}{c}\input{pstex/D-ID-V-al-EP-GR.pstex_t} \end{array}} \equiv \ensuremath{\begin{array}{c}\input{pstex/D-ID-V-al-EP-GR-B.pstex_t} \end{array}} = 0,
\]
as follows from diagrammatic identities~\ref{D-ID:PseudoEffectivePropagatorRelation} and~\ref{D-ID:TLTP-GR}.
\end{D-ID}

\begin{D-ID}
A sub-diagram comprising a two-point, tree level  vertex attached to an
effective propagator, differentiated \wrt\
$\alpha$, which terminates in a $\wedge$ can be re-drawn, using 
diagrammatic
identity~\ref{D-ID:V-al-EP-GR}, the effective propagator relation diagrammatic
identity~\ref{D-ID:GaugeRemainderRelation}:
\[
	\ensuremath{\begin{array}{c}\input{pstex/D-ID-V-EP-al-GR.pstex_t} \end{array}} \equiv - \ensuremath{\begin{array}{c}\begin{picture}(0,0)%
\includegraphics{pstex/D-ID-V-EP-al-GR-B.pstex}%
\end{picture}%
\setlength{\unitlength}{3947sp}%
\begingroup\makeatletter\ifx\SetFigFont\undefined%
\gdef\SetFigFont#1#2#3#4#5{%
  \reset@font\fontsize{#1}{#2pt}%
  \fontfamily{#3}\fontseries{#4}\fontshape{#5}%
  \selectfont}%
\fi\endgroup%
\begin{picture}(165,369)(151,367)
\put(203,640){\makebox(0,0)[lb]{\smash{\SetFigFont{8}{9.6}{\rmdefault}{\mddefault}{\updefault}{\color[rgb]{0,0,0}$\alpha$}%
}}}
\end{picture}
 \end{array}}.
\]
\end{D-ID}

\chapter{Examples of Classical Flows}	\label{app:Ex-ClassicalFlows}

The first vertex whose flow we will need is a three-point,
tree level vertex,
decorated by three wildcard fields labelled $R$--$T$. This
is shown in figure~\ref{fig:App:TLThP-WildCardx3}.

\begin{center}
\begin{figure}[h]
	\[
	-\ensuremath{\begin{array}{c}\input{pstex/V0-RST-LdL.pstex_t} \end{array}}= \ensuremath{\begin{array}{c}\input{pstex/V-RSX-DEP-V-XT.pstex_t} \end{array}} + \ensuremath{\begin{array}{c}\input{pstex/V-STX-DEP-V-XR.pstex_t} \end{array}} + \ensuremath{\begin{array}{c}\input{pstex/V-TRX-DEP-V-XS.pstex_t} \end{array}} + \ensuremath{\begin{array}{c}\input{pstex/V-RX-W-S-V-TX.pstex_t} \end{array}} \hspace{1ex}
		+ \ensuremath{\begin{array}{c}\input{pstex/V-SX-W-T-V-RX.pstex_t} \end{array}} \hspace{1ex} + \ensuremath{\begin{array}{c}\input{pstex/V-TX-W-R-V-SX.pstex_t} \end{array}}
	\]
\caption{The flow of a three-point, tree level vertex decorated by three wildcard fields.}
\label{fig:App:TLThP-WildCardx3}
\end{figure}
\end{center}

\begin{center}
\begin{figure}[h]
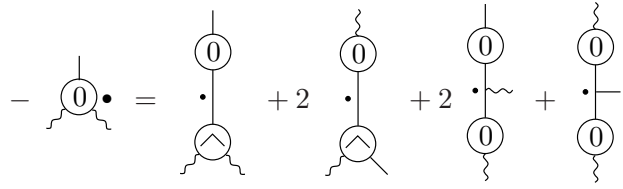

	\[
	-\ensuremath{\begin{array}{c}\input{pstex/V0-A1A1X-LdL.pstex_t} \end{array}}= \ensuremath{\begin{array}{c}\input{pstex/V-A1A1X-DEP-V-XX.pstex_t} \end{array}} + 2 \ensuremath{\begin{array}{c}\input{pstex/V-A1XX-DEP-V-A1X.pstex_t} \end{array}} +2 \ensuremath{\begin{array}{c}\input{pstex/V-A1X-W-A1-XX.pstex_t} \end{array}} + \ensuremath{\begin{array}{c}\input{pstex/V-A1X-W-X-V-A1X.pstex_t} \end{array}} 
	\]
\caption[Flow of a three-point, tree level vertex decorated by $A^1_\mu(p)$, $A^1_\nu(-p)$
and a dummy field.]{Flow of a three-point, tree level vertex decorated by $A^1_\mu(p)$, $A^1_\nu(-p)$
and a dummy field. Lorentz indices, sub-sector labels and momentum arguments are suppressed.}
\label{fig:App-TLThP-A1A1X}
\end{figure}
\end{center}

We now specialise the previous example to give the flow of a three-point, tree
level vertex decorated by $A^1_\mu(p)$, $A^1_\nu(-p)$ and a wildcard field,
which we note carries zero momentum. This is shown in figure~\ref{fig:App-TLThP-A1A1X}
where we have suppressed all labels.

The third diagram vanishes. First, we note that the wine must be bosonic. Now,
it cannot be in the $C$-sector, because $AC$
vertices do not exist. If the wine is in the $A$-sector then,
since the wildcard field carries zero momentum, this
would require a two-point $A$-vertex carrying zero
momentum, which is forbidden by gauge invariance.

The final diagram vanishes at $\Op{2}$. The wine must be in
the $A$-sector and so each of the vertices
contributes at least $\Op{2}$, as a consequence of gauge invariance.

Note that, if the wildcard field is in the $C$-sector, then the second
diagram also vanishes at $\Op{2}$. The top vertex contributes at least $\Op{2}$.
The bottom vertex must also contribute $\Op{2}$, by gauge invariance, since
$AC$ vertices do not exist.

The last example is of the flow of a four-point, tree level vertex decorated by 
$A^1_\mu(p)$, $A^1_\nu(-p)$ and two dummy fields, as shown 
in figure~\ref{fig:App:TLFP-A1A1XX}.
Summing over the flavours of the dummy fields 
and noting
that, in the current example, the dummy fields carry equal and opposite momenta,
we can treat these fields as identical.
\begin{center}
\begin{figure}
	\[
	-\ensuremath{\begin{array}{c}\input{pstex/V-A1A1XX.pstex_t} \end{array}}= 
	\begin{array}[t]{cccccccc}
		 	& 2\ensuremath{\begin{array}{c}\input{pstex/V-A1A1XX-DEP-XX.pstex_t} \end{array}}		& + & 4\ensuremath{\begin{array}{c}\input{pstex/V-A1XX-W-A1-V-XX.pstex_t} \end{array}}	& +	& \ensuremath{\begin{array}{c}\input{pstex/V-XX-W-A1A1-V-XX.pstex_t} \end{array}}		& -	& 2\ensuremath{\begin{array}{c}\input{pstex/V0-A1XX-DEP-V0-A1XX.pstex_t} \end{array}}	
	\\
		+	& 4\ensuremath{\begin{array}{c}\input{pstex/V0-A1XX-DEP-V-A1XX.pstex_t} \end{array}}	&+	& 2\ensuremath{\begin{array}{c}\input{pstex/V-A1A1X-W-X-V-XX.pstex_t} \end{array}}	& +	& 4\ensuremath{\begin{array}{c}\input{pstex/V-A1X-W-A1X-V-XX.pstex_t} \end{array}}	& +	& 4\ensuremath{\begin{array}{c}\input{pstex/V-A1X-W-X-V-A1XX.pstex_t} \end{array}}
	\\
		+	& 2\ensuremath{\begin{array}{c}\input{pstex/V-A1X-V-A1A1XX.pstex_t} \end{array}}		& -	& \ensuremath{\begin{array}{c}\input{pstex/V0-A1A1X-DEP-V0-XXX.pstex_t} \end{array}}	& +	& \ensuremath{\begin{array}{c}\input{pstex/V0-A1A1X-DEP-V-XXX.pstex_t} \end{array}}	& +& \ensuremath{\begin{array}{c}\input{pstex/V-A1A1X-DEP-V0-XXX.pstex_t} \end{array}}
	\\
		+	& 4\ensuremath{\begin{array}{c}\input{pstex/V-A1X-W-A1-VXXX.pstex_t} \end{array}}		& + & \ensuremath{\begin{array}{c}\input{pstex/VA1X-W-XX-VA1X.pstex_t} \end{array}}
	\end{array}
	\]
\caption[Flow of a four-point, tree level vertex decorated by $A^1_\mu(p)$, $A^1_\nu(-p)$
and two dummy fields.]{Flow of a four-point, tree level vertex decorated by $A^1_\mu(p)$, $A^1_\nu(-p)$
and two dummy fields. Lorentz indices and momentum arguments are suppressed.}
\label{fig:App:TLFP-A1A1XX}
\end{figure}
\end{center}

The penultimate diagram vanishes at $\Op{2}$. The wine must be in the $A$-sector, but
then the diagram possesses an $\Op{2}$ stub. However, the wine, which carries
three Lorentz indices, cannot have an $\Op{0}$ contribution by Lorentz invariance.
The final diagram, which possesses two $\Op{2}$ stubs, clearly vanishes at $\Op{2}$.

\chapter{Simplified Expression for $\beta_2$} \label{app:beta_2}

There follows a reproduction of the simplified expression for $-4\beta_2 \Box_{\mu \nu}(p)$.

\begin{center}
\begin{figure}
	\[
	\frac{1}{4}
	\dec{
		\begin{array}{l}
			\begin{array}{cccccccc}
				2&\ensuremath{\begin{array}{c}\input{pstex/Diagram2.1.0.pstex_t} \end{array}} & -2 & \ensuremath{\begin{array}{c}\input{pstex/Diagram6.1.pstex_t} \end{array}} & -4 & \ensuremath{\begin{array}{c}\input{pstex/Diagram8.1.pstex_t} \end{array}} & +4 & \ensuremath{\begin{array}{c}\input{pstex/Diagram7.14.pstex_t} \end{array}} 
			\end{array}
		\\
			\begin{array}{cccccccc}
				-2 	& \ensuremath{\begin{array}{c}\input{pstex/Diagram9a.1.pstex_t} \end{array}} 		& -2			& \ensuremath{\begin{array}{c}\input{pstex/Diagram9b.1.pstex_t} \end{array}} 		& +2	& \ensuremath{\begin{array}{c}\input{pstex/Diagram7.1.pstex_t} \end{array}} 			&+2& \ensuremath{\begin{array}{c}\input{pstex/Diagram9c.1.pstex_t} \end{array}}
			\end{array}
		\\
			\begin{array}{cccccccc}
				-\hf& \ensuremath{\begin{array}{c}\input{pstex/Diagram15.1.pstex_t} \end{array}}		&+\frac{2}{3}	&\ensuremath{\begin{array}{c}\input{pstex/Diagram26.1.pstex_t} \end{array}} 		& +8	& \AGROB{\ensuremath{\begin{array}{c}\input{pstex/Diagram397.5.pstex_t} \end{array}}}	& &
			\end{array}
		\\
			\begin{array}{cccccccc}
				+	& \ensuremath{\begin{array}{c}\input{pstex/Diagram25.1.pstex_t} \end{array}}		& - 			& \ensuremath{\begin{array}{c}\input{pstex/Diagram38.1.pstex_t} \end{array}}		& +4 	& \AGROB{\ensuremath{\begin{array}{c}\input{pstex/Diagram381-E.pstex_t} \end{array}}}	&-\frac{1}{2}& \ensuremath{\begin{array}{c}\input{pstex/Diagram39.1.pstex_t} \end{array}}		
			\end{array}
		\\
			\begin{array}{cccccccc}
				+	& \ensuremath{\begin{array}{c}\input{pstex/Diagram27.1.pstex_t} \end{array}}		& -				& \ensuremath{\begin{array}{c}\input{pstex/Diagram40.1.pstex_t} \end{array}}		& +4	& \AGROB{\ensuremath{\begin{array}{c}\input{pstex/Diagram377.17.pstex_t} \end{array}}}&&
			\end{array}	
		\end{array}
	}{\bullet}
	\]
\end{figure}
\end{center}

\begin{center}
\begin{figure}
	\[
	\frac{1}{4}
	\dec{
		\begin{array}{l}
			\begin{array}{cccccccc}
				- 	& \ensuremath{\begin{array}{c}\input{pstex/Diagram21.1.pstex_t} \end{array}}		&+	& \ensuremath{\begin{array}{c}\input{pstex/Diagram36.1.pstex_t} \end{array}}		& -4& \AGROB{\ensuremath{\begin{array}{c}\input{pstex/Diagram375.1.pstex_t} \end{array}}} & + & \ensuremath{\begin{array}{c}\input{pstex/Diagram34.1.pstex_t} \end{array}}
			\end{array}
		\\
			\begin{array}{ccccc}
				+2	& \ensuremath{\begin{array}{c}\input{pstex/Diagram30.1.pstex_t} \end{array}}			&-2 & \ensuremath{\begin{array}{c}\input{pstex/Diagram41.1.pstex_t} \end{array}}
			\\
				+8	& \AGROB{\ensuremath{\begin{array}{c}\input{pstex/Diagram376.1.pstex_t} \end{array}}}	&-2	& \ensuremath{\begin{array}{c}\input{pstex/Diagram51.1.pstex_t} \end{array}}
			\end{array}
		\\
			\begin{array}{cccccccc}
				+ 	& \ensuremath{\begin{array}{c}\input{pstex/Diagram10.10.pstex_t} \end{array}}		&-\frac{2}{3}	& \ensuremath{\begin{array}{c}\input{pstex/Diagram19.1.pstex_t} \end{array}}		&-	& \ensuremath{\begin{array}{c}\input{pstex/Diagram18.1.pstex_t} \end{array}}		&+\frac{1}{2}	& \ensuremath{\begin{array}{c}\input{pstex/Diagram33.1.pstex_t} \end{array}}
			\end{array}
		\\
			\begin{array}{cccccccc}
			-\frac{2}{3}	&	\ensuremath{\begin{array}{c}\input{pstex/Diagram23.1.pstex_t} \end{array}}							&-2	& \ensuremath{\begin{array}{c}\input{pstex/Diagram17.1.pstex_t} \end{array}}					& 
							+4&\ensuremath{\begin{array}{c}\input{pstex/Diagram28.1.pstex_t} \end{array}}											&+2& \ensuremath{\begin{array}{c}\input{pstex/Diagram31.1.pstex_t} \end{array}}
			\end{array}
		\\
			\begin{array}{cccccccc}
				+ 	&\ensuremath{\begin{array}{c}\input{pstex/Diagram32.1.pstex_t} \end{array}}	&-2	& \ensuremath{\begin{array}{c}\input{pstex/Diagram42.1.pstex_t} \end{array}}	& +16
					& \ensuremath{\begin{array}{c}\input{pstex/Diagram392.4.pstex_t} \end{array}}	&-8 &\AGROB{\ensuremath{\begin{array}{c}\input{pstex/Diagram395.4.pstex_t} \end{array}}}
			\end{array}
		\\
			\begin{array}{cccccc}
			
			+16	&\ensuremath{\begin{array}{c}\input{pstex/DiagramFR26.1.pstex_t} \end{array}}								&+16& \ensuremath{\begin{array}{c}\input{pstex/DiagramFR26.2.pstex_t} \end{array}}							&+8
				&\ensuremath{\begin{array}{c}\input{pstex/DiagramFR26.1+6.pstex_t} \end{array}}
			\end{array}
		\end{array}
	}{\bullet}
	\]
\end{figure}
\end{center}

\begin{center}
\begin{figure}
	\[
	2
	\dec{
		\begin{array}{l}
			\begin{array}{cccc}
				2	& \ensuremath{\begin{array}{c}\input{pstex/Diagram326.1+325.9.pstex_t} \end{array}}		& -& \ensuremath{\begin{array}{c}\input{pstex/Diagram328.23.pstex_t} \end{array}}
			\end{array}
		\\
			\begin{array}{ccccccccccl}
				+ 	& \ensuremath{\begin{array}{c}\input{pstex/Diagram327.37.pstex_t} \end{array}}		& - & \ensuremath{\begin{array}{c}\input{pstex/Diagram329.29.pstex_t} \end{array}}	&+2	& \ensuremath{\begin{array}{c}\input{pstex/Diagram331b.9.pstex_t} \end{array}}	& +2& \ensuremath{\begin{array}{c}\input{pstex/Diagram331b.10.pstex_t} \end{array}}		&&&
			\end{array}
		\\
			\begin{array}{ccccccccl}
				-2	&\ensuremath{\begin{array}{c}\input{pstex/DiagramFR22.2.pstex_t} \end{array}}		& -2	&\ensuremath{\begin{array}{c}\input{pstex/DiagramFR22.1.pstex_t} \end{array}}		& +	2& \ensuremath{\begin{array}{c}\input{pstex/DiagramFR24.6.pstex_t} \end{array}}	& +	2&\ensuremath{\begin{array}{c}\input{pstex/DiagramFR24.1.pstex_t} \end{array}}		&
			\end{array}
		\\
			\begin{array}{ccccccccl}
				+2	&\ensuremath{\begin{array}{c}\input{pstex/DiagramFR36.1.pstex_t} \end{array}}		&+2	&\ensuremath{\begin{array}{c}\input{pstex/DiagramFR22.3.pstex_t} \end{array}}	& -2 &\ensuremath{\begin{array}{c}\input{pstex/DiagramFR24.11.pstex_t} \end{array}}	&+2	&\ensuremath{\begin{array}{c}\input{pstex/DiagramFR27.1.pstex_t} \end{array}}	&
			\end{array}
		\\
			
			\begin{array}{ccccccl}
				-&\ensuremath{\begin{array}{c}\input{pstex/DiagramFR27.6.pstex_t} \end{array}}							&-	&\AGROB{\ensuremath{\begin{array}{c}\input{pstex/Diagram382.1.pstex_t} \end{array}}}		&+	&\AGROB{\ensuremath{\begin{array}{c}\input{pstex/Diagram351.5.pstex_t} \end{array}}}		&
			\end{array}
		\end{array}
	}{\bullet}
	\]
\end{figure}
\end{center}

\begin{center}
\begin{figure}[h]
	\[
	\begin{array}{c}
	\vspace{0.2in}
		\dec{
			\begin{array}{c}
			\vspace{0.2in}
				\left[
					\begin{array}{ccccccc}
						\ensuremath{\begin{array}{c}\input{pstex/Diagram180.7-be.pstex_t} \end{array}}	&+	&\ensuremath{\begin{array}{c}\input{pstex/Diagram180.9-be.pstex_t} \end{array}}	&-2	&\ensuremath{\begin{array}{c}\input{pstex/Diagram326.3-be.pstex_t} \end{array}}	&+2	&\ensuremath{\begin{array}{c}\input{pstex/Diagram326.4-be.pstex_t} \end{array}}
					\end{array}
				\right]	
			\\		
				\times \Delta^{1 \,1}_{\alpha \beta}(p)
				\left[
					\begin{array}{ccccccc}
						\ensuremath{\begin{array}{c}\input{pstex/Beta1-A-al.pstex_t} \end{array}}& -	&\ensuremath{\begin{array}{c}\input{pstex/Beta1-B-al.pstex_t} \end{array}} 	& +4& \ensuremath{\begin{array}{c}\input{pstex/Beta1-C-al.pstex_t} \end{array}} 	& -	& \ensuremath{\begin{array}{c}\input{pstex/Beta1-D-al.pstex_t} \end{array}}
					\end{array}
				\right]
			\end{array}
		}{\bullet}
	\\
		+
		\dec{
			\begin{array}{ccccccl}
				4	&\ensuremath{\begin{array}{c}\input{pstex/LittleSetSq-A.pstex_t} \end{array}}			&+4	&\ensuremath{\begin{array}{c}\input{pstex/LittleSetSq-B.pstex_t} \end{array}}		& +	&\ensuremath{\begin{array}{c}\input{pstex/LittleSetSq-C.pstex_t} \end{array}}			&
			\end{array}
		}{\bullet}
	\end{array}
	\]
\end{figure}
\end{center}

\begin{center}
\begin{figure}[h]
\[
-\frac{1}{2}\gamma_1 \pder{}{\alpha}
\left[
	\begin{array}{c}
		\begin{array}{ccccccc}
			\ensuremath{\begin{array}{c}\input{pstex/Beta1-LdL-A.pstex_t} \end{array}}& -	&\ensuremath{\begin{array}{c}\input{pstex/Beta1-LdL-B.pstex_t} \end{array}} 	& +4& \ensuremath{\begin{array}{c}\input{pstex/Beta1-LdL-C.pstex_t} \end{array}} 	& -	& \ensuremath{\begin{array}{c}\input{pstex/Beta1-LdL-DE.pstex_t} \end{array}}
		\end{array}
	\\
		\begin{array}{cccc}
			+8	& \ensuremath{\begin{array}{c}\input{pstex/Beta1-LdL-F.pstex_t} \end{array}}	&+4	& \ensuremath{\begin{array}{c}\input{pstex/Beta1-LdL-G.pstex_t} \end{array}}
		\end{array}
	\end{array}
\right]
\]
\end{figure}
\end{center}


\begin{thebibliography}{105}

\addcontentsline{toc}{chapter}{Bibliography}

\bibitem{Wilson}			K.~Wilson and J.~Kogut, \PhysRep{12}{C}{1974}{75}.

\bibitem{W&H}				F.~J.~Wegner and A.~Houghton, \PhysRev{A}{8}{1973}{401}.

\bibitem{Pol}				J.~Polchinski, \NuclPhys{B}{231}{1984}{269}.

\bibitem{pre-SU(N|N)-A} 	S.~Arnone, Y.~A.~Kubyshin and T.~R.~Morris, J.~F.~Tighe, in
							\emph{Proc. XVth Workshop on High Energy Physics and Quantum Field Theory},
							eds. M.~N.~Dubinin and V.~I.~Savrin (Moscow University Press, 2001), p.~297, \hepth{0102011}.

\bibitem{pre-SU(N|N)-B} 	S.~Arnone, Y.~A.~Kubyshin and T.~R.~Morris, J.~F.~Tighe,
							\IntJModPhys{A}{16}{2001}{1989}.

\bibitem{SU(N|N)} 			S.~Arnone, Y.~A.~Kubyshin and T.~R.~Morris, J.~F.~Tighe,
							\IntJModPhys{A}{17}{2002}{2283}.


\bibitem{Wetterich:1993:Scalar}	C.~Wetterich, \PhysLett{B}{301}{1993}{90}.

\bibitem{Bonini:1993}			M.~Bonini \etal, \NuclPhys{B}{409}{1993}{441}.


\bibitem{Bonini} 				M.~Bonini, G.~Marchesini and M.~Simionato, \NuclPhys{B}{483}{1997}{475}.

\bibitem{Papenbrock:1995-b2}	T.~Papenbrock and C.~Wetterich, \ZPhys{C}{65}{1995}{519}.

\bibitem{Pernici:1998-b2}		M.~Pernici and M.~Raciti, \NuclPhys{B}{531}{1998}{560}.

\bibitem{Kopietz:2001-b2}		P.~Kopietz, \NuclPhys{B}{595}{2001}{493}.

\bibitem{Zappala:2002-b2}		D.~Zappala, \PhysRev{D}{66}{2002}{105020}.

\bibitem{Hasenfratz^2:1986}		A.~Hasenfratz and P.~Hasenfratz \NuclPhys{B}{270}{1986}{687}.

\bibitem{Sumi:2000-Rep}			J.-I.~Sumi, W.~Soumma, K.-I.~Aoki, H.~Terao and K.~Morikawa, \hepth{0002231}.

\bibitem{Bagnuls:2001-Rep}		C.~Bagnuls and C.~Bervillier, \PhysRept{348}{2001}{91}.

\bibitem{Berges:2002-Rep}		J.~Berges, N.~Tetradis, C.~Wetterich, \PhysRept{363}{2002}{233}.

\bibitem{Polonyi:2004-Rep}		J.~Polonyi, \CEurJPhys{1}{2004}{1}.

\bibitem{TRM-ERG+ApproxSolns} 	T.~R.~Morris, \IntJModPhys{A}{9}{1994}{2411}.

\bibitem{TRM-elements} 			T.~R.~Morris, \emph{Prog.\ Theor.\ Phys.\ Suppl.\ } 131 (1998) 395.

\bibitem{TRM-MassiveScalar} 	T.~R.~Morris, \NuclPhys{B}{495}{1997}{477}.

\bibitem{TRM-NewDevs}		 	T.~R.~Morris, in \emph{New Developments in Quantum Field Theory}, NATO ASI series
								366, (Plenum Press, 1998),
								\IntJModPhys{B}{12}{1998}{1343}.

\bibitem{TRM-1999-b2}			T.~R.~Morris and J.~F.~Tighe, \jhep{08}{1999}{7}.

\bibitem{Scalar-1-loop} 		S.~Arnone, A.~Gatti and T.~R.~Morris \jhep{0205}{2002}{059}.

\bibitem{scalar2loop} 			S.~Arnone, A.~Gatti,  T.~R.~Morris and O.~J.~Rosten, \PhysRev{D}{69}{2004}{065009}.





\bibitem{Bonini:1993kt}			M.~Bonini, M.~D'Attanasio and G.~Marchesini,
								\NuclPhys{B}{418}{1994}{81}.

\bibitem{Bonini:1993sj}			M.~Bonini, M.~D'Attanasio and G.~Marchesini,
								\NuclPhys{B}{421}{1994}{429}.

\bibitem{Bonini:1994kp}			M.~Bonini, M.~D'Attanasio and G.~Marchesini,
								\NuclPhys{B}{437}{1995}{163}.


\bibitem{Bonini:2001}			M.~Bonini and E.~Tricarico, \NuclPhys{B}{606}{2001}{231}.

\bibitem{Becchi:1993}			C.~Becchi, in Elementary Partilces, Field Theory and Statistical Mechanics (Parma, 1993), \hepth{9607188}.

\bibitem{Aoki:1997}				K.-I.~Aoki \etal, \ProgTheorPhys{97}{1997}{479}.

\bibitem{Aoki:1999}				K.-I.~Aoki \etal, \PhysRev{D}{61}{2000}{045008}.

\bibitem{Kubota:1999}			K.-I.~Kubota and H.~Terao, \ProgTheorPhys{102}{1999}{1163}.


\bibitem{Pernici:1998}			M.~Pernici \etal, \NuclPhys{B}{520}{1998}{469}.

\bibitem{Simionato:2000}		M.~Simionato, \IntJModPhys{A}{15}{2000}{2121};
								\IntJModPhys{A}{15}{2000}{2153}.
								
\bibitem{Simionato:2000-B}		M.~Simionato \IntJModPhys{A}{15}{2000}{4811}.

\bibitem{Simionato:2001}		M.~Simionato, \IntJModPhys{A}{16}{2001}{2125}.
								
								

\bibitem{Panza:2000}			A.~Panza and R.~Soldati \PhysLett{B}{493}{2000}{197}.


\bibitem{Reuter:1994sg}			M.~Reuter and C.~Wetterich,
								\NuclPhys{B}{427}{1994}{291}.

\bibitem{Reuter:1993kw}			M.~Reuter and C.~Wetterich,
								\NuclPhys{B}{417}{1994}{181}.

\bibitem{Freire:2000bq}			F.~Freire, D.~F.~Litim and J.~M.~Pawlowski,
								\PhysLett{B}{495}{256}{2000}.

\bibitem{Litim:2002hj}			D.~F.~Litim and J.~M.~Pawlowski,
								\PhysLett{B}{546}{2002}{279}.



\bibitem{Ellwanger:1994iz}		U.~Ellwanger,
								\PhysLett{B}{335}{1994}{364}.



\bibitem{D'Attanasio:1996jd}	M.~D'Attanasio and T.~R.~Morris,
								\PhysLett{B}{378}{1996}{213}.

\bibitem{Litim:1998qi}			D.~F.~Litim and J.~M.~Pawlowski,
								\PhysLett{B}{435}{1998}{181}.

\bibitem{Litim:1998nf}
								D.~F.~Litim and J.~M.~Pawlowski,
								\hepth{9901063}.

\bibitem{Pawlowski:2002eb}		J.~M.~Pawlowski,
								\Acta{52}{2002}{475}.

\bibitem{Litim:2002ce}			D.~F.~Litim and J.~M.~Pawlowski,
								\jhep{0209}{2002}{049}.








\bibitem{Veneziano}				V.~Branchina, K.~A.~Meissner and G.~Veneziano,
								Phys.\ Lett.\ B {\bf 574} (2003) 319.

\bibitem{Pawlowski-VDW}			J.~M.~Pawlowski,
								\hepth{0310018}.




\bibitem{TRM-Faro} 			T.~R.~Morris, in: The Exact Renormalization Group, Eds. Kraznitz et al.\
							(World Sci, Singapore, 1999) p.1, hep-th/9810104.

\bibitem{GI-ERG-I} 			T.~R.~Morris, \NuclPhys{B}{573}{2000}{97}.

\bibitem{GI-ERG-II} 		T.~R.~Morris, \jhep{0012}{2000}{012}.

\bibitem{TRM-Rome} 			T.~R.~Morris, Int.\ J.\ Mod.\ Phys.\ \textbf{A 16} (2001) 1899.

\bibitem{Antonio'sThesis}	A.~Gatti, `A Gauge Invariant Flow Equation', Ph.D.\ Thesis \hepth{0301201}.

\bibitem{TRM-Acta}			S.~Arnone, A.~Gatti and T.~R.~Morris, \Acta{52}{2002}{621}.


\bibitem{YM-1-loop} 		S.~Arnone, A.~Gatti and T.~R.~Morris, \PhysRev{D}{67}{2003}{085003}.

\bibitem{Quarks2004}		O.~J.~Rosten, T.~R.~Morris and S.~Arnone, The Gauge Invariant ERG,
							Proceedings of Quarks 2004, Pushkinskie Gory, Russia, 24-30 May 2004,
							http://quarks.inr.ac.ru,
							\hepth{0409042}.

\bibitem{GeneralMethods} 	O.~J.~Rosten, A  Manifestly Gauge Invariant and
							Universal Calculus in Yang-Mills Theory, in preparation.

\bibitem{FurtherWork} 		Work in progress.





\bibitem{TRM+JL-0}			J.I.~Latorre and T.~R.~Morris, \jhep{0011}{2000}{004}.

\bibitem{TRM+JL} 			J.I.~Latorre and T.~R.~Morris, \IntJModPhys{A}{16}{2001}{2071}.




\bibitem{Bergerhoff:1995zm}			
									B.~Bergerhoff, D.~Litim, S.~Lola and C.~Wetterich,
									\IntJModPhys{A}{11}{1996}{4273}.

\bibitem{Bergerhoff:1995zq}			B.~Bergerhoff, F.~Freire, D.~Litim, S.~Lola and C.~Wetterich,
									\PhysRev{B}{53}{1996}{5734}.
\bibitem{Aoki:1996fh}				K.-I.~Aoki, K.~Morikawa, J.-I.~Sumi, H.~Terao and M.~Tomoyose,
									\ProgTheorPhys{\bf}{1997}{479}.

\bibitem{Reuter:1997gx}				M.~Reuter and C.~Wetterich,
									\PhysRev{D}{56}{1997}{7893}.

\bibitem{Bergerhoff:1997cv}			B.~Bergerhoff and C.~Wetterich,
									\PhysRev{D}{57}{1998}{1591}.

\bibitem{Gies:2002af}				H.~Gies,
									\PhysRev{D}{66}{2002}{025006}.

\bibitem{Ellwanger:1995qf}			U.~Ellwanger, M.~Hirsch and A.~Weber,
									\ZPhys{C}{69}{1996}{687}.

\bibitem{Ellwanger:1996wy}			U.~Ellwanger, M.~Hirsch and A.~Weber,
									\EurPhysJ{C}{1}{1998}{563}.

\bibitem{Pawlowski:2003hq}			J.~M.~Pawlowski, D.~F.~Litim, S.~Nedelko and L.~von Smekal,
									\PhysRevLett{93}{2004}{152002}.

\bibitem{Fischer:2004uk}			C.~S.~Fischer and H.~Gies,
									\jhep{0410}{2004}{048}.

\bibitem{Ellwanger:1997wv}			U.~Ellwanger,
									\NuclPhys{B}{531}{1998}{593}.

\bibitem{Ellwanger:1999vc}			U.~Ellwanger,
									\NuclPhys{B}{560}{1999}{587}.

\bibitem{Ellwanger:2002xa}			U.~Ellwanger and N.~Wschebor,
									\EurPhysJ{C}{28}{2003}{415}.

\bibitem{Pawlowski:1996ch}			J.~M.~Pawlowski,
									\PhysRev{D}{58}{1998}{045011}.


\bibitem{Gies:2004:chiSB}			H.~Gies and C.~Wetterich,
									\PhysRev{D}{69}{2004}{025001}.

\bibitem{Ellwanger:1999IR-fp}		U.~Ellwanger,
									\EurPhysJ{C}{7}{1999}{673}.

\bibitem{Pawlowski:2004IR-fp}		J.~M.~Pawlowski, D.~F.~Litim, S.~Nedelko and L.~von~Smekal,
									\PhysRevLett{93}{2004}{152002}.

\bibitem{Aoki:2000chiSB}			K.-I.~Aoki, K.~Morikawa, J.-I.~Sumi, H.~Terao and M.Tomoyose,
									\PhysRev{D}{61}{2000}{045008};
									K.~I.~Aoki, K.~Morikawa, J.-I.~Sumi, H.~Terao and M.Tomoyose,
									\ProgTheorPhys{102}{1999}{1151}.

\bibitem{Arnone:2001:SUSY}			S.~Arnone, S.~Chiantese and K.~Yoshida, \IntJModPhys{A}{16}{2001}{1811}.
					
\bibitem{Arnone:2004:SUSY}			S.~Arnone, and K.~Yoshida,	\IntJModPhys{B}{18}{2004}{469}.





\bibitem{Ellwanger:1997tp}			U.~Ellwanger,
									\ZPhys{C}{76}{1997}{721}.





\bibitem{Litim:1998yn}				D.~F.~Litim,
									\hepph{9811272}.


\bibitem{D'Attanasio:1996fy}		M.~D'Attanasio and M.~Pietroni,
									\NuclPhys{B}{498}{1997}{443}.




\bibitem{Bonini:1997yv}				M.~Bonini and F.~Vian,
									\NuclPhys{B}{511}{1998}{479}.


\bibitem{Falkenberg:1998bg}			S.~Falkenberg and B.~Geyer,
									\PhysRev{D}{58}{1998}{085004}.

\bibitem{Bonini:1998ec}				M.~Bonini and F.~Vian,
									\NuclPhys{B}{532}{1998}{473}.



\bibitem{Reuter:1996cp}				M.~Reuter,
									\PhysRev{D}{57}{971}{1998}.


\bibitem{Lauscher:2001rz}			O.~Lauscher and M.~Reuter,
									\ClassQuantGrav{19}{2002}{483}.

\bibitem{Litim:2003vp}				D.~F.~Litim,
									\PhysRevLett{92}{2004}{201301}.


\bibitem{SelfSimilarFlow} 	D.~V.~Shirkov, \TheorMathPhys{60}{1985}{778}.

\bibitem{PerfectActions}	P.~Hasenfratz and F.~Niedermayer, \NuclPhys{B}{414}{1994}{785}.



\bibitem{Vilkovisky:1984st}		G.~A.~Vilkovisky, in {\it Quantum Theory of Gravity}, ed. S.~M.~Christensen 
								(Adam Hilger, 1984), 169-209;
								\NuclPhys{B}{234}{1984}{125}.


\bibitem{DeWitt} 				B.~DeWitt, {\it The Effective Action} in 
								{\it Quantum Field Theory and Quantum Statistics}, eds. I.~A.~Batalin. 
								C.~J.~Isham and G.~A.~Vilkovisky (Adam Hilger: Bristol 1987). 


\bibitem{Slavnov'sReg-A} 	A.~A.~Slavnov, \TheorMathPhys{33}{1977}{977}; 
							[Teor.\ Mat.\ Fiz.\  {\bf 33}, 210 (1977)].
							L.~D.~Faddeev and A.~A.~Slavnov,
							\emph{Gauge Fields, Introduction to Quantum Field Theory}
							2nd ed. (Benjamin/Cummings, Reading, Massachusetts, 1980).

\bibitem{Slavnov'sReg-B}	T.~D.~Bakeyev and A.~A.~Slavnov, \ModPhysLett{A}{11}{1996}{1539}.


\bibitem{HigherDerivReg-Falcetto}	M.~Asorey and F.~Falceto, \PhysRev{D}{54}{1996}{5290}.


\bibitem{Bars} 				I.~Bars, in \emph{Introduction to Supersymmetry in Particle and Nuclear Physics}, 
							eds. Castan\~{o}s et al.\ (Plenum, New York, 1984) 107.

\bibitem{A&C} 				T.~Appelquist and J.~Carazzone, Phys. Rev. \textbf{D 11} (1975) 2856;
							K.~Symanzik, Comm. Math. Phys. 34 (1973) 7. 
							See also J.~Collins, \emph{Renormalization} (CUP, 1984).

\bibitem{P&S} See \eg\ M.~E.~Peskin \& D.~V.~Schroeder, \emph{An
Intoduction to Quantum Field Theory} (Perseus Books Publishing, L.L.C., 1995).


\bibitem{I&Z} See \eg\ C.~Itzykson and J.~B.~Zuber, \emph{Quantum
Field Theory} (McGraw-Hill, New York, 1980).


\bibitem{Weinberg}	See \eg\ S.~Weinberg, \emph{The Quantum Theory of Fields}
(Cambridge University Press, Cambridge 1996), Vol.~2.



\bibitem{Gribov}			V.~N.~Gribov, \NuclPhys{B}{139}{1978}{1}.

\bibitem{Goldstone}			J.~Goldstone, \emph{Nuovo Cim.} \textbf{19} (1961) 269.



\bibitem{PreReg-Warr}		B.~J.~Warr, \AnnPhys{183}{1988}{1}.

\bibitem{One-loop-UVDivs}	A.~A.~Slavnov, \TheorMathPhys{13}{1972}{1064}; 
							B.~W.~Lee and J.~Zinn-Justin, \PhysRev{D}{5}{1972}{3121}.





\bibitem{beta4-Verm}		T.~van~Ritburgen, J.~A.~M.~Vermaseren and S.~A.~Larin, \PhysLett{B}{400}{1997}{379}.
							

\bibitem{beta4-Other}		M.~Czakon, The Four-Loop QCD Beta-Function and Anomalous Dimensions, 
							\NuclPhys{B}{710}{2005}{485}.

\end{thebibliography}
\end{document}